\documentclass[12pt,a4paper,oneside]{ociamthesis}
\pdfoutput=1
\usepackage{amsmath}
\usepackage{amssymb}
\usepackage{amsfonts}
\usepackage{graphicx} 
\usepackage{fancyhdr}
\usepackage{subfigure}
\usepackage{JournalNames}
\usepackage{paralist}
\usepackage{times}
\usepackage{caption} 
\usepackage{pdfpages}

\usepackage[numbers,sort&compress]{natbib}
\usepackage{hyperref}

\hypersetup{colorlinks=false,pdfpagemode=UseNone,bookmarksopen=true,bookmarks=false,pdfauthor={Joseph Parker},pdfstartview={FitH},pdfpagelayout={OneColumn},pdfborder={0 0 0}}

\newcommand{\g}{{\bf g}}
\renewcommand{\L}{{\cal L}}

\newcommand{\blue}{\color{Blue}}
\newcommand{\z}{\hat{z}}

\renewcommand{\labelenumi}{\arabic{enumi})}

\title{Gyrokinetic simulations of fusion plasmas using a spectral velocity space representation}
\author{Joseph Parker}
\college{Brasenose College}
\renewcommand{\submittedtext}{A thesis submitted for}
\degree{Doctor of Philosophy}
\degreedate{Trinity 2015}

\begin{document}

% Spelling/phrases
\renewcommand{\d}{\mathrm{d}}

% From SGK paper
\newcommand{\alfven}{Alfv\'en}
\newcommand{\ampere}{Amp\`ere}
\newcommand{\apriori}{\emph{a priori}}
\newcommand{\cdf}{complementary distribution function}
\newcommand{\cvk}{Case--Van Kampen}
\newcommand{\collop}{collision operator}
\newcommand{\eg}{\emph{e.g.}}
\newcommand{\etc}{\emph{etc.}}
\newcommand{\Eqn}{Equation}
\newcommand{\eqn}{equation}
\newcommand{\eqm}{equilibrium}
\newcommand{\evp}{eigenvalue problem}
\newcommand{\dop}{Kirkwood}
\newcommand{\fp}{Fokker--Planck}
\newcommand{\FH}{Fourier--Hermite}
\newcommand{\FHH}{Fourier--Hankel--Hermite}
\newcommand{\gcs}{guiding centre space}
\newcommand{\gk}{gyrokinetic}
\newcommand{\gkm}{gyrokinetic-Maxwell}
\newcommand{\Gkm}{Gyrokinetic-Maxwell}
\newcommand{\gkeqn}{gyrokinetic equation}
\newcommand{\gkpot}{gyrokinetic potential}
\newcommand{\hh}{Hankel--Hermite}
\newcommand{\ic}{initial condition}
\newcommand{\ie}{\emph{i.e.}}
\newcommand{\ipc}{inter-processor communication}
\newcommand{\ivp}{initial value problem}
\newcommand{\Kn}{\mathrm{Kn}}
\newcommand{\LB}{L\'enard--Bernstein}
\renewcommand{\LB}{Lenard--Bernstein}
\newcommand{\lbd}{Kirkwood}
\newcommand{\lhs}{left-hand side}
\newcommand{\oneplusoned}{1+1D}
\newcommand{\rhs}{right-hand side}
\newcommand{\tkmk}{tokamak}
\newcommand{\Tkmk}{Tokamak}
\newcommand{\twostream}{two-stream}
\newcommand{\vk}{Van Kampen}
\newcommand{\Vk}{Van Kampen}
\newcommand{\vp}{Vlasov--Poisson}
\newcommand{\vps}{Vlasov--Poisson system}
\newcommand{\VanKampen}{Van Kampen}
\newcommand{\vanKampen}{van Kampen}
\newcommand{\wrt}{with respect to}

\newcommand{\agk}{\textsc{AstroGK}}
\newcommand{\agks}{\textsc{SpectroGK}}
\newcommand{\gstwo}{\textsc{GS2}}
\newcommand{\gene}{\textsc{Gene}}
\newcommand{\sgk}{\textsc{SpectroGK}}
\newcommand{\trinity}{\textsc{Trinity}}

% environments
\newcommand{\chp}{Chapter}
\newcommand{\Fig}{Figure~} % at start of sentence
\newcommand{\fig}{Figure~}
\newcommand{\Figs}{Figures~} % at start of sentence
\newcommand{\figs}{Figures~}
\newcommand{\partref}{Part~}
\newcommand{\referencename}{Ref.~}
\newcommand{\Sec}{Section}
\renewcommand{\sec}{Section}
\renewcommand{\sec}{\S}
\newcommand{\secs}{Sections}
\renewcommand{\secs}{\S\S}
\newcommand{\tab}{Table}
\newcommand{\tbl}{Table}
\newcommand{\thesis}{thesis}

% Bib stuff
\newcommand{\citepos}[1]{\citeauthor{#1}'s (\citeyear{#1})}

% Contentious
\newcommand{\cf}{\nu}
\newcommand{\ff}{\tilde{f}}
\newcommand{\ffs}{\tilde{f}_s}
%\renewcommand{\nu}{\nuup}

% Relations
\newcommand \id{\equiv}
\renewcommand{\O}{{\cal O}}		% Order

% integrals
\newcommand{\intii}{\int_{-\infty}^\infty}
\newcommand{\intio}{\int_{-\infty}^0}
\newcommand{\intoi}{\int_{0}^\infty}
\newcommand{\intotp}{\int_{0}^{2\pi}}
\newcommand{\intpp}{\int_{-\pi}^\pi}

% One off notation

\newcommand{\curl}{\del\times}
\newcommand{\Div}{\del\cdot}
\newcommand{\dee}{\partial}
\newcommand{\del}{\nabla}
\newcommand{\deltab}{\boldsymbol{\delta}}
\newcommand{\dv}{\dee_v}
\newcommand{\fd}[2]{\frac{\mathrm{d}{#1}}{\mathrm{d}{#2}}}
\newcommand{\pd}[2]{\frac{\partial{#1}}{\partial{#2}}}
\newcommand{\pdd}[2]{\frac{\dee^2 #1}{\dee #2^2}}
\newcommand{\sumi}[1]{\sum_{#1=0}^\infty}
\newcommand{\sumii}[1]{\sum_{#1=1}^\infty}
\newcommand{\tfd}[2]{\d #1/\d #2}
\newcommand{\tpd}[2]{\partial{#1}/\partial{#2}}
\newcommand{\tpdd}[2]{\dee^2 #1/\dee #2^2}
\newcommand{\tpds}[2]{\partial_{#2}{#1}}

\newcommand{\Nproc}{N_{\textrm{proc}}}

% unit vectors
\newcommand{\unitx}{\hat{\x}}
\newcommand{\unity}{\hat{\y}}
\newcommand{\unitz}{\hat{\z}}
\newcommand{\Yhat}{\hat{\Y}}
\newcommand{\zhat}{\hat{\z}}
\newcommand{\X}{\boldsymbol{X}}
\newcommand{\Y}{\boldsymbol{Y}}
\newcommand{\V}{\boldsymbol{V}}

% averages
\newcommand{\bra}{\left<}
\newcommand{\ket}{\right>}
\newcommand{\gar}[1]{\bra #1 \ket_{\r}}
\newcommand{\gaR}[1]{\bra #1 \ket_{\R}}
\newcommand{\ga}[2]{\bra #1 \ket_{#2}}
\newcommand{\ta}[1]{\left<#1\right>_{\textrm{turb}}}

%brackets
\newcommand{\lb}{\left\{}  
\newcommand{\rb}{\right\}}
\newcommand{\lp}{\left(}
\newcommand{\rp}{\right)}
\newcommand{\rs}{\right]}

% dot variables
\newcommand{\td}{\dot{\vartheta}}
\newcommand{\Rd}{\dot{\R}_s}
\newcommand{\mud}{\dot{\mu}_s}
\newcommand{\muNd}{\dot{\mu}_{Ns}}
\newcommand{\Ed}{\dot{E}_s}
\newcommand{\ENd}{\dot{E}_{Ns}}

% Distributions
\newcommand{\gfived}{\breve{g}}

% EM fields

\newcommand{\AN}{\A_{N}}
\newcommand{\Apara}{A_{\parallel}}
\newcommand{\ANpara}{A_{N\parallel}}
\newcommand{\Aperp}{\A_{\perp}}
\newcommand{\ANperp}{\A_{N\perp}}
\newcommand{\BN}{\B_{N}}
\newcommand{\Bpara}{B_{\parallel}}
\newcommand{\BNpara}{B_{N\parallel}}
\newcommand{\Bperp}{\B_{\perp}}
\newcommand{\E}{\boldsymbol{E}}
\newcommand{\ENs}{E_{Ns}}
\newcommand{\Epara}{E_{\parallel}}
\newcommand{\varphid}{\delta\varphi}
\newcommand{\varphidN}{\delta\varphi_{N}}
\newcommand{\varphiN}{\varphi_{N}}
\newcommand{\varphihat}{\hat{\varphi}}

% E
\newcommand{\fE}{\widetilde{\E}} % full
\newcommand{\bE}{\bar{\E}} % background
\newcommand{\pE}{\E} % perturbed
\newcommand{\pEN}{\E_N} % perturbed
% B
\newcommand{\fB}{\widetilde{\B}}
\newcommand{\bB}{\bar{\B}} % background
\newcommand{\pB}{\B} % perturbed
\newcommand{\Bpar}{B_{\parallel}}
\newcommand{\pBN}{\B_N} % perturbed
% A
\newcommand{\A}{\boldsymbol{A}}
\newcommand{\fA}{\widetilde{\A}}
\newcommand{\bA}{\bar{\A}} % background
\newcommand{\pA}{\A} % perturbed
\newcommand{\Apar}{A_{\parallel}}
\newcommand{\pAN}{\A_N} % perturbed
\newcommand{\pApara}{A_{\parallel}} % perturbed
\newcommand{\pAperp}{\A_{\perp}} % perturbed
\newcommand{\pANpara}{A_{N\parallel}} % perturbed
\newcommand{\pANperp}{\A_{N\perp}} % perturbed
% varphi
\newcommand{\varphif}{\widetilde{\varphi}}
\newcommand{\varphib}{\bar{\varphi}} % background
\newcommand{\varphip}{\varphi} % perturbed
\newcommand{\varphipN}{\varphi_N} % perturbed

\renewcommand{\u}{\boldsymbol{u}}
\renewcommand{\v}{\boldsymbol{v}}

% Real/Complex
\newcommand{\bbC}{\mathbb{C}}
\newcommand{\bbR}{\mathbb{R}}

% Symbolic  transforms
\newcommand{\FFT}{{\cal F}}
\newcommand{\IFFT}{{\cal F}^{-1}}
\newcommand{\HT}{{\cal H}}
\newcommand{\F}{{\cal F}}
\renewcommand{\Finv}{{\cal F}^{-1}}
\newcommand{\Fmat}{{\mathsf{F}}}
\newcommand{\Fmatinv}{{\mathsf{F}}^{-1}}

% Free energy
\newcommand{\esinv}{{\frak E}} %electrostatic invariant
\newcommand{\ESInvar}{{\frak E}} %electrostatic invariant

% Normalizing quantities
\newcommand{\mNe}{m_{Ne}}
\newcommand{\mNs}{m_{Ns}}
\newcommand{\nref}{n_{s}}
\newcommand{\TNe}{T_{Ne}}
\newcommand{\TNs}{T_{Ns}}
\newcommand{\Tref}{T_{s}}
\newcommand{\vth}{v_{\textrm{th}}}
\newcommand{\vthe}{v_{\textrm{th}e}}
\newcommand{\vthi}{v_{\textrm{th}i}}
\newcommand{\vthr}{v_{\textrm{th}r}}
\newcommand{\vths}{v_{\textrm{th}s}}
\newcommand{\vthNi}{v_{\textrm{th}Ni}}
\newcommand{\vthNe}{v_{\textrm{th}Ne}}
\newcommand{\vthNs}{v_{\textrm{th}Ns}}

% Vlasov Poisson text
\newcommand{\DFT}{discrete Fourier transform}
\newcommand{\ps}{pseudospectral}
\newcommand{\psly}{pseudospectrally}

% Vlasov Poisson resolution
\newcommand{\Nt}{N_{\vartheta}}

% Collisions
\newcommand{\Cmat}{{\mathsf{C}}} % VP paper

% Drift velocities
\newcommand{\vChi}{\v{\chi}}
\newcommand{\vPhi}{\v{\varphi}}

% Lorentz Force
\newcommand{\Fvec}{\boldsymbol{F}}

% Collision operators
\newcommand{\CBoltzmann}{C_{\mathsf{B}}}
\newcommand{\CLandau}{C_{\mathsf{L}}}

% Fluid stuff
\newcommand{\energy}{{\cal E}}
\newcommand{\energyflux}{\boldsymbol{{\cal F}}}
\newcommand{\momflux}{\boldsymbol{\Pi}}
\newcommand{\Itensor}{\boldsymbol{\mathsf{I}}}

%% Hereafter may or may not be used:

\renewcommand{\b}{\boldsymbol{b}}
\newcommand{\B}{\boldsymbol{B}}
\newcommand{\CN}{C_{N}}
\newcommand{\chiNs}{\chi_{Ns}}
\newcommand{\Dx}{\Delta x}
\newcommand{\dfNs}{\delta f_{Ns}}
\newcommand{\dfs}{\delta f_{s}}
\newcommand{\et}{\e_{\vartheta}}
\newcommand{\eperp}{\e_{\perp}}
\newcommand{\fcharge}{\tilde{\varrho}}
\renewcommand{\j}{{\boldsymbol{j}}}
\newcommand{\fcurrent}{\tilde{\boldsymbol{j}}}
\newcommand{\fsa}[1]{\left<\left<#1\right>\right>}
\newcommand{\FNs}{F_{Ns}}
\newcommand{\FNso}{F_{N0s}}
\newcommand{\Fs}{F_{s}}
\newcommand{\Fso}{F_{0s}}
\newcommand{\Fsi}{F_{1s}}
\newcommand{\Fsii}{F_{2s}}
\newcommand{\fNs}{f_{Ns}}
\newcommand{\fs}{f_{s}}
\newcommand{\fso}{f_{0s}}
\newcommand{\fsi}{f_{1s}}
\newcommand{\fsip}{f_{s1p}}
\newcommand{\fsii}{f_{2s}}
\renewcommand{\g}{\bar{g} }
\newcommand{\gkpar}{\varepsilon}
\newcommand{\gks}{g_{s\k}}
\renewcommand{\H}{{\cal H}}
\newcommand{\I}{{\cal I}}
\newcommand{\intt}{\Delta t}
\renewcommand{\intt}{\tilde{\tau}}
\renewcommand{\k}{\boldsymbol{k}}
\newcommand{\kpara}{k_{\parallel}}
\newcommand{\kNperp}{k_{N\perp}}
\newcommand{\Lnr}{L_n^{-1}}
\newcommand{\LTr}{L_T^{-1}}
\newcommand{\muNs}{\mu_{Ns}}
\newcommand{\nablaN}{\nabla_{N}}
\newcommand{\nablaNperp}{\nabla_{N\perp}}
\newcommand{\nablaNpara}{\nabla_{N\parallel}}
\newcommand{\nablaperp}{\nabla_{\perp}}
\newcommand{\nlt}[2]{\lb{#1},{#2}\rb}
\newcommand{\nNs}{n_{Ns}}
\newcommand{\OmegaNs}{\Omega_{Ns}}
\newcommand{\qNs}{q_{Ns}}
\newcommand{\R}{\boldsymbol{R}}
\renewcommand{\r}{\boldsymbol{r}}
\newcommand{\RN}{\R_{N}}
\newcommand{\RNs}{\R_{Ns}}
\newcommand{\RNpara}{\R_{N\parallel}}
\newcommand{\RNperp}{\R_{N\perp}}
\newcommand{\RNsperp}{\R_{Ns\perp}}
\newcommand{\rN}{\r_{N}}
\newcommand{\rNpara}{\r_{N\parallel}}
\newcommand{\rNperp}{\r_{N\perp}}
\newcommand{\rhob}{\boldsymbol{\rho}}
\newcommand{\rhobs}{\boldsymbol{\rho}_s}
\newcommand{\rhoNe}{\rho_{Ne}}
\newcommand{\rhoNs}{\rho_{Ns}}
\newcommand{\rhoNi}{\rho_{i}}
\newcommand{\rhoN}{\boldsymbol{\rho}_N}
\newcommand{\rd}{\dot{\r}}
\newcommand{\tN}{t_{N}}
\renewcommand{\v}{\boldsymbol{v}}
\newcommand{\vN}{\v_{N}}
\newcommand{\VN}{\V_{N}}
\newcommand{\vcut}{V_{\textrm{cut}}}
\newcommand{\vd}{\dot{\v}}
\newcommand{\vD}{\v_D}
\newcommand{\vNpara}{v_{N\parallel}}
\newcommand{\vNperp}{v_{N\perp}}
\newcommand{\vNperpv}{\v_{N\perp}}
\newcommand{\VNpara}{V_{N\parallel}}
\newcommand{\VNperp}{V_{N\perp}}
\newcommand{\VNperpv}{\V_{N\perp}}
\newcommand{\Vpara}{V_{\parallel}}
\newcommand{\vpara}{v_{\parallel}}
\newcommand{\Vperp}{V_{\perp}}
\newcommand{\vperp}{v_{\perp}}
\newcommand{\Vperpv}{\V_{\perp}}
\newcommand{\vperpv}{\boldsymbol{v}_{\perp}}
\newcommand{\xN}{x_{N}}
\newcommand{\x}{\boldsymbol{x}}
\newcommand{\y}{\boldsymbol{y}}
\newcommand{\yN}{y_{N}}
\newcommand{\Z}{\boldsymbol{Z}}
\newcommand{\ZN}{Z_{N}}
\newcommand{\zN}{z_{N}}
\newcommand{\ZNs}{q_{Ns}}

% Hankel transform variables
\newcommand{\kvar}{k'}
\newcommand{\pvar}{p}

% Linear operators
\newcommand{\Lstar}{${\cal L}_g$, ${\cal L}_{\varphi}$, ${\cal L}_A$, ${\cal L}_B$}

\newcommand{\Aparak}{A_{\parallel\k}}
\newcommand{\Bpk}{B_{\parallel\k}}
\newcommand{\Bparak}{B_{\parallel\k}}
\newcommand{\dBp}{B_{\parallel}}
\newcommand{\dhBp}{\hat{B}_{\parallel}}
\newcommand{\gs}{g_{s}}
\newcommand{\hks}{h_{s\k}}
\newcommand{\kperp}{k_{\perp}}
\newcommand{\kperpv}{\k_{\perp}}
\newcommand{\LBr}{L_{B}^{-1}}
\newcommand{\sgn}{\textrm{sgn}}
\newcommand{\stm}{\vthNs}
\newcommand{\varphik}{\varphi_{\k}}
\newcommand{\vDNL}{\v_{D}^{\textrm{NL}}}
\newcommand{\vDNLx}{v_{Dx}^{\textrm{NL}}}
\newcommand{\vDNLy}{v_{Dy}^{\textrm{NL}}}
\newcommand{\vDs}{\v_{Ds}}
\newcommand{\Vparas}{V_{\parallel s}}
\newcommand{\Vperps}{V_{\perp s}}
\newcommand{\ind}{\I_{[0,\vcut]}}

\newcommand{\calD}{\boldsymbol{\cal D}}%{\pmb{\cal F}}
\newcommand{\charc}{characteristic}
\newcommand{\fb}{\mathbf{f}}
\newcommand{\hdexp}{n} % hypercollisional exponent
\newcommand{\hm}{m} % hermite mode
\newcommand{\hc}{\tilde{\cal C}} % hermite cardinal poly
\newcommand{\hcf}{{\cal C}} % hermite cardinal function 
\renewcommand{\Im}{\textrm{Im}}
\renewcommand{\Re}{\textrm{Re}}
\newcommand{\sfrac}[2]{{#1}/{#2}}
\newcommand{\Tmat}{\mathbf{\mathsf{T}}}
\newcommand{\wn}{\tilde{\omega}_n}
\newcommand{\wT}{\tilde{\omega}_T}
\renewcommand{\z}{\hat{\boldsymbol{z}}}

% PJD
\newcommand{\Ex}[1]{(\ref{#1})}
\newcommand{\Mat}{\mathbf{\mathsf{M}}}
\newcommand{\Dmat}{\mathbf{\mathsf{D}}}
\newcommand{\Jmat}{\mathbf{\mathsf{J}}}
\newcommand{\dthree}{\mathcal{I}_{m \ge 3}}

% Hyphenation
\hyphenation{non-dim-en-sion-al-ization non-dim-en-sionalize quasi-neutrality}

\baselineskip=18pt plus1pt

%set the number of sectioning levels that get number and appear in the contents
\setcounter{secnumdepth}{3}
\setcounter{tocdepth}{2}

\maketitle
\pdfoutput=1

\begin{abstract}

Magnetic confinement fusion reactors suffer severely from heat and particle losses through turbulent transport, which has inspired the construction of ever larger and more expensive reactors. 
Numerical simulations are vital to their design and operation, but particle collisions are too infrequent for fluid descriptions to be valid. 
Instead, strongly magnetised fusion plasmas are described by the gyrokinetic equations, a nonlinear integro-differential system for evolving the particle distribution functions in a five-dimensional position and velocity space, and the consequent electromagnetic field. 
Due to the high dimensionality, simulations of small reactor sections require hundreds of thousands of CPU hours on cutting-edge High Performance Computing platforms.

We develop a Hankel--Hermite spectral representation for velocity space that exploits structural features 
of the particle streaming, gyroaveraging, and collision terms in the gyrokinetic system. 
This representation exactly conserves a discrete free energy in the absence of explicit dissipation, while our Hermite hypercollision operator captures Landau damping with as few as ten variables. 
Calculation of the electromagnetic fields also becomes purely local. 
This eliminates all inter-processor communication in, and hence vastly accelerates, searches for linear instabilities.  
We implement these ideas in \mbox{\sgk}, an efficient parallel code.

Turbulent fusion plasmas may dissipate free energy through linear phase mixing to fine scales in velocity space, as in Landau damping, or through a nonlinear cascade to fine scales in physical space, as in hydrodynamic turbulence. 
Using \sgk\ to study saturated electrostatic drift-kinetic turbulence in Cartesian geometry, we find that the nonlinear cascade completely suppresses linear phase mixing at energetically-dominant scales, so the turbulence is fluid-like. 
We use these observations to derive Fourier--Hermite spectra for the electrostatic potential and distribution function, 
and confirm these spectra with \sgk\ simulations.

\end{abstract}
          % include the abstract
\begin{romanpages}
\tableofcontents
\listoffigures
\end{romanpages}

%now include the files of latex for each of the chapters etc
\part{Introduction}
%\chapter{Introduction}
\chapter{Fusion, turbulence, and gyrokinetic theory}

\section{Motivation}

The world needs energy.
Global energy consumption is around $6\times10^{20}$ J per year---an average power of $20$ TW---with usage growing at around 2\% per year.\footnote{Based on the U.S. Energy Information Administration's estimates for the years 1980--2012. Global consumption for 2012 quoted as $528.797\times10^{15}$ Btu. \cite{EIA}}     
Indeed, even though global consumption fell in 2009 as a result of the financial crisis, 
consumption in Asia and Oceania still grew by 5\%---a salient reminder of the future pressures on energy production from increasing consumption and population growth.

Current energy production is dominated by fossil fuels---oil, coal and natural gas---which account for 87\% of global production \cite{BP}.
Producing energy by burning fossil fuels is simple and cheap,
but produces greenhouse gases like carbon dioxide, and other pollutants.
Moreover, reserves are limited.
Estimates are very rough, relying on future consumption and resource discovery, but proven reserves of coal
will last 100 years at current production rates, while oil and gas reserves will last 50 years \cite{BP}.

The other contributions to global energy production are from hydroelectric power (7\%), nuclear fission (4\%) and renewable energies (2\%) \cite{BP}.
Hydroelectric power is severely limited by geography.
Nuclear fission also has its problems, particularly with the need to store radioactive waste, 
with political and security concerns,
and with its susceptibility to meltdown in the case of accident, attack or,
as seen at Fukushima in 2011, natural disaster. 
Renewable energies, like solar or wind, are still in their infancy.
These struggle with cost efficiency, and are only viable in the UK due to government subsidy.\footnote{Renewable energy is subsidized by the U.K. Government at \pounds50 per MWh, compared to subsidies for nuclear fission ($\geq$\pounds33/MWh), gas (\pounds4/MWh), oil (55p/MWh) and coal (20p/MWh) \cite{EAC}.}

A better prospect for clean, safe and large-scale energy production is nuclear fusion.
Nuclear fusion could power the whole world
using only a tiny fraction of the amount of fuel required in nuclear fission or in burning fossil fuels.
Fusion has no carbon dioxide emissions,
no meltdowns,
and no radioactive waste (save decommissioned fusion reactors, which would decay to safe levels after 50 years \citep{Cook01}).
%Instead its product is helium---harmless and useful.
Instead its product is helium, which is not only harmless and useful for applications like magnetic resonance imaging, 
but also a finite commodity which is relatively scarce \cite{Nuttall12}.

These are, unfortunately, the same arguments for fusion that have been made for the last seventy years.
But nuclear fusion's now long-standing reputation as a promising future energy source belies the extraordinary progress made by fusion research.
In the period 1970--2000,
the fusion triple product, 
the accepted measure of fusion performance,
increased by a factor of 10,000---a rate of improvement that outstrips Moore's law for the growth of the number of transistors on a chip \citep{Ikeda10}.
While that technology transformed society with personal computers and laptops and tablets and smartphones,
nuclear fusion's moment is yet to come,
awaiting an elusive further factor of 6 that would take the fusion triple product from the current best performance into the realm of a viable fusion power plant \citep{Ikeda10}.

\section{Nuclear fusion}

What are the methods, and challenges, of nuclear fusion?
In nuclear fusion, we recreate one of the simpler reactions which sustain the sun: deuterium--tritium fusion.
When one nucleus of deuterium and one of tritium fuse, they produce 17.6 MeV of energy divided between the reaction products: a helium nucleus (with 3.5 MeV of kinetic energy) and a neutron (with 14.1 MeV) \citep{Rebhan02}.
Assuming a rate of $10^{21}$ reactions per second, only 10 mg per second of fuel provides 1 GW of power \citep[assuming 30\% reactor efficiency,][]{Maisonnier07}.
That is, 3500 tonnes of fuel per year would power the whole world, 
equivalent to 8400 million tonnes of oil \cite{HighcockPhD}.
Deuterium is readily available in seawater \citep{Ongena08},
and while tritium itself is rare, it may be ``bred'' using lithium and the neutron produced in the reaction.
Lithium too is abundant in seawater \citep{Ongena08}.

We thus have a fuel cycle, and fuel which may be refined from seawater.
What is the challenge?
This: for the deuterium and tritium nuclei to fuse,
they must have enough energy to overcome their mutual electrostatic repulsion.
This requires temperatures of around $10^7$~K.
No known material can withstand this temperature to confine the fuel directly.
Some other method is required.

The most promising method exploits the full ionization of deuterium and tritium atoms at these high temperatures. 
The electrons and ions dissociate to form a plasma,
and in an electromagnetic field experience the Lorentz force
\begin{align}
  \label{eq:LorentzForce}
  \Fvec_s = q_s\left( \E + \frac{\v\times\B}{c} \right),
\end{align}
where $s$ denotes the particle species (ion $i$ or electron $e$),
$q_s$ and $\v$ are the particles' charge and velocity,
$\E$ and $\B$ are the electric and magnetic fields,
and $c$ is the speed of light.
We use Gaussian units throughout this thesis.
A particle in a strong magnetic field (with no electric field, $\E=0$) is constrained to move helically around a magnetic field line, with 
frequency of gyration (cyclotron frequency, or gyrofrequency) $\Omega_s = q_sB/m_sc$, where $m_s$ is the particle mass and $B=|\B|$.
The 
radius of gyration (gyroradius) is $\rho_s=\vperp/\Omega_s$, where $\vperp$ is velocity perpendicular to the magnetic field.
The particle's motion remains centred around this field line, so the particle may be confined if field lines are made to form closed surfaces.
This is only possible for magnetic fields which are topologically a torus.

Thus in magnetic confinement fusion, the plasma is confined with a strong magnetic field designed to form layers of closed toroidal surfaces called flux surfaces.
There are two main classes of magnetic confinement fusion device: \tkmk s (which are axisymmetric about the central axis)
and stellarators (which have more intricate non-axisymmetric shapes).
\Tkmk s are the more common device, so toroidal geometry is prevalent in fusion theory and simulations (and is used in popular codes such as \gstwo\ and \gene, discussed in \chp~\ref{sec:GKMSystem}).

The problem facing fusion devices is that their confinement is not perfect.
Two effects which degrade confinement are collisions and particle drifts.
Collisions cause particles to be knocked from their field line onto another nearby,
and so through multiple collisions, particles can cross flux surfaces and escape from the device.
This process is called ``classical transport''.  
Classical transport alone does not present a problem.
Indeed 
a typical \tkmk\ volume of 10 m$^3$ contains around $10^{22}$ particles, which collide at a rate of roughly $4\times10^{13}$ s$^{-1}$cm$^{-3}$.
Thus each particle experiences collisions at a rate $\tau_c^{-1}\sim 4\times10^{-2}$s$^{-1}$.
Each collision deflects a particle by a distance like the ion gyroradius, typically $\rho_i\sim5\times10^{-4}$m,
so that the random walk diffusion coefficient is $D\sim \rho_i^2\tau_c^{-1}\sim10^{-8}$m$^2$s$^{-1}$.
Assuming a major radius of $L\sim 1$m, the confinement time is $\tau\sim L^2/D\sim 10^8$s.
That is, if affected by collisions alone, a particle at the core takes a few years to leave the machine.

A more significant effect occurs when collisions and particle drifts combine.
Particle drifts are slow motions perpendicular to magnetic field caused by perturbations to the equilibrium electric and magnetic fields in Lorentz's law \eqref{eq:LorentzForce}.
The primary drifts (which we derive in \sec\ref{sec:DerivationOfSlabGyrokineticEquations}) 
are the $\E\times\B$ drift due to perturbations to the electric field caused by motions of the particles,
and the gradient-$B$ and curvature drifts, which arise from inhomogeneities in the background magnetic field \citep{Hastie95}.
%The effect of drifts is incorporated into \tkmk\ design, and can be manipulated such that outwards drifts will, on average, be countered by inwards drifts.
Tokamaks are designed so that outwards drifts will, on average, be 
countered by inwards drifts.
The problem is rather the combination of collisions and drifts.  
A particle may drift outwards, collide with another particle and be knocked onto a new field line where it again drifts outwards.
This process, called ``neoclassical transport'', is much more significant than classical transport.
Neoclassical transport has been extensively studied \citep{HelanderSigmar02}
and good estimates may be derived for its size in various simple fusion devices \citep{Hinton85}.
Indeed, neoclassical transport is controllable,
and even including its effect, we could achieve fusion in a reactor with a minor radius as small as 1m \citep{Galambos95}---a ``table-top \tkmk''.
The reason we are yet to achieve sustainable fusion is transport 
from an altogether different 
and less well-understood phenomenon,
that of plasma turbulence.

\section{Turbulence}
Turbulence is a ubiquitous phenomenon in liquids, gases and plasmas, appearing as an erratic and quickly-evolving mixture of eddies, 
%waves 
vortices
and jets.
In strongly magnetized fusion plasmas,
turbulence manifests itself as fluctuations in the electromagnetic field, and in plasma properties like density and temperature.
It is the field fluctuations which degrade confinement.
Magnetic field perturbations may deform the background magnetic field, causing a very large loss of confinement.
Fortunately, this effect is minimal in the current generation of machines as the particle pressure is never large relative to the magnetic pressure \citep{Colas98}.
Rather the problem is electric field perturbations which induce eddies in the plasma flow that lie across magnetic field lines.
These eddies rapidly convect particles across flux surfaces, leading to turbulent transport an order of magnitude larger than neoclassical transport.

Moreover, turbulence is driven by ion temperature gradients \cite{Horton80,Waltz88,Fonck89,Wootton90,Cowley91,Kotschenreuther95confinement,Carreras97,Dimits00} and electron density and temperature gradients \cite{Dorland00,Gene,Dannert05}.
High pressure at the core of the machine, as required for fusion at a commercially viable rate, necessitates a strong pressure gradient between the core and the edge.
But since pressure is the product of density and temperature, large pressure gradients mean large density and temperature gradients, so we find that turbulence is driven by the very quantities we want to maximize.
This leads to ever-increasing machine sizes, since a larger machine means both that high core pressure can be achieved with shallower gradients, and that particles must be transported further in order to escape the device.
This explains the planned $837$ m$^3$ volume of the ITER \tkmk\footnote{This formerly stood for ``International Thermonuclear Experimental Reactor'', though there is no mention of this in official literature \cite{ITER}. It is widely held that the word ``experimental'' proved unpopular in the context. Now the name is simply \emph{iter}, ``the way'' in Latin \cite{Krommes12}.}
currently under construction in Cadarache, France, over five times the volume of its predecessor, the Joint European Torus (JET), located at Culham, U.K. 
Such large sizes bring a raft of problems.
For one, devices are so expensive as to be only feasible through large international collaborations.
For another, the increased machine size presents engineering problems associated with large heat loads, neutron fluxes and mechanical stresses
on the \tkmk's material wall.

To progress we must not only mitigate turbulence, but understand and control it.
This is now the focus of much experimental and theoretical work.
Experimentalists have created ``transport barriers'', annular regions of laminar flow about the \tkmk\ core \cite{MiuraJT60,deVries09}.
By their suppression of turbulence, these allow much steeper temperature gradients and therefore higher core pressures.
Theoreticians too have studied mechanisms for suppressing turbulence and creating transport barriers, such as shear flow \cite{Highcock12}.

The nature of turbulence means progress with theory is very difficult.
Turbulence is an inherently nonlinear phenomenon.
Thus analytic solutions are rare, and understanding comes from scaling theories and numerical simulations. 
Moreover, there are additional problems particular to strongly magnetized plasma turbulence that do not arise in fluid turbulence.

Firstly, plasma turbulence spans multiple scales in space and time.
The turbulent fluctuations depend on scale, and behaviour changes as the scale first crosses the ion gyroscale $\rho_i$, then the electron gyroscale $\rho_e$.
These scales are well-separated from one another, $\rho_i\gg\rho_e$,
and from the system size $L\gg\rho_i$, the scale of bulk movement of the plasma.
Similarly, the typical timescale for plasma dynamics is well-separated from other timescales, being much slower than the timescale $\Omega_s^{-1}$ for particle gyration,
but much faster than the timescale $\tau$ for transport and the background field evolution.

In addition, plasma turbulence is kinetic. 
That is, it is characterized by effects---like Landau damping discussed in \chp~\ref{sec:ParallelVelocitySpace}---which must be explained in terms of the interaction of particles with the electromagnetic field (a kinetic description),
rather than solely in terms of macroscopic variables, like density, momentum and temperature (a fluid description).
This means that kinetic models must include velocities, as well as positions, as coordinates.
Consequently
kinetic models are higher-dimensional than fluid models,
and thus much more computationally challenging.

\section{Fluid, kinetic and gyrokinetic theory}
We now discuss the mathematical description of gases and plasmas.
We first consider the case of a dilute gas of monatomic, neutral particles. 
The system is described by
the distribution function $\tilde{f}(\x,\v,t)$, the number
density of particles at position $\x$ moving with velocity $\v$ at
time $t$.
It evolves according to the Boltzmann equation
\begin{align}
  \pd{\tilde{f}}{t} + \v\cdot\nabla\tilde{f} = \nu\CBoltzmann[\tilde{f},\tilde{f}],
  \label{eq:BoltzmannEquation}
\end{align}
where $\nu\CBoltzmann$ is the Boltzmann collision operator describing collisions between pairs of particles with frequency $\nu$.
Collisions between three or more particles are negligibly infrequent.
Collisions between pairs of particles conserve mass, momentum, and energy,
so the following three integrals of the collision operator vanish:
	\begin{subequations}
\begin{align}
  \int \d^3\v ~ \CBoltzmann[\tilde{f},\tilde{f}] = 0,
	\\
  \int \d^3\v ~ \v \CBoltzmann[\tilde{f},\tilde{f}] = 0,
	\\
  \int \d^3\v ~ \frac{1}{2}|\v|^2 \CBoltzmann[\tilde{f},\tilde{f}]
  = 0.
\end{align}
  \label{eq:CollisionalConservation}
	\end{subequations}
Other moments of $\CBoltzmann$ typically do not vanish.

Macroscopic quantities of the gas are given by velocity moments of $\tilde{f}$.
The first few are
\begin{subequations}
\begin{align}
  n(\x,t) = \int \d^3\v ~ \tilde{f}(\x,\v,t),
  \\
  \u(\x,t) = \frac{1}{n} \int \d^3\v ~\v \tilde{f}(\x,\v,t),
  \\
  T(\x,t) = \frac{1}{3n} \int \d^3\v ~|\v-\u|^2 \tilde{f}(\x,\v,t),
\end{align}
  \label{eq:MacroscopicMoments}% put comment here as subequations are in text mode, so label inserts space
\end{subequations}
respectively the number density, bulk velocity, and temperature 
in so-called energy units where $T^{1/2}$ is the isothermal (Newtonian) sound speed.
In a fluid description, we replace the Boltzmann equation \eqref{eq:BoltzmannEquation} for the distribution function $\tilde{f}$
with a system of equations for macroscopic quantities, moments of distribution function like $n$, $\u$ and $T$.
%%%Using the definitions \eqref{eq:MacroscopicMoments}, we derive fluid equations by taking moments of the Boltzmann equation \eqref{eq:BoltzmannEquation},
Taking moments of the Boltzmann equation \eqref{eq:BoltzmannEquation} corresponding to the integrals \eqref{eq:CollisionalConservation}, 
we derive three macroscopic conservation laws
\begin{align}
  \pd{(mn)}{t} + \nabla\cdot\left( mn\u \right) = 0, 
  \hspace{1cm}
  \pd{}{t}(mn\u) + \nabla\cdot \momflux = 0, 
  \hspace{1cm}
  \pd{\energy}{t} + \nabla\cdot\energyflux = 0, 
  \label{eq:NearlyEuler}
\end{align}
where $m$ is the particle mass.
The energy density $\energy$, momentum flux $\momflux$, and energy flux $\energyflux$ are defined by
further moments of $\tilde{f}$:
\begin{subequations}
\begin{align}
  \energy = m \int \d^3\v ~ \frac{1}{2}|\v|^2\tilde{f}(\x,\v,t), 
  %\hspace{0.2cm}
  \\
  \momflux = m \int \d^3\v ~  \v\v \tilde{f}(\x,\v,t),
  %\hspace{0.2cm}
  \\
  \energyflux = m \int \d^3\v ~ \frac{1}{2} \v|\v|^2 \tilde{f}(\x,\v,t). 
\end{align}
  \label{eq:EnergiesMomentumDefintions}%
\end{subequations}
The evolution equations for $n$, $\u$, $\energy$ \eqref{eq:NearlyEuler} are not closed as the fluxes $\momflux$ and $\energyflux$ are unknown.
We could find evolution equations for these fluxes by taking further moments of the Boltzmann equation,
\begin{subequations}
\begin{align}
  \pd{\momflux}{t} + \nabla\cdot\left( m\int \d^3\v~ \v\v\v \tilde{f} \right) 
  = \nu m\int \d^3\v ~ \v\v\CBoltzmann[\tilde{f},\tilde{f}]
  \neq 0, 
  \\
  \pd{\energyflux}{t} + \nabla\cdot\left( m\int \d^3\v~ \frac{1}{2}\v\v|\v|^2 \tilde{f} \right) 
  = \nu m\int \d^3\v ~ \frac{1}{2}\v|\v|^2\CBoltzmann[\tilde{f},\tilde{f}]
  \neq 0, 
  \label{eq:NextMomentsEvolution}
\end{align}
\end{subequations}
but then these depend on yet higher unknown moments.
In fact, this is a generic problem---the ``closure problem''---which arises in taking moments of kinetic equations like \eqref{eq:BoltzmannEquation}:
the evolution of the $n$th moment will always depend on the $(n+1)$th moment because of the velocity in the streaming term $\v\cdot\nabla\tilde{f}$.
One may close the system (and thereby produce a fluid model) by defining the $(n+1)$th moment in terms of known quantities.
%One may close the system (and thereby produce a fluid model) by instead defining $\momflux$ and $\energyflux$ in terms of known quantities.
In the simplest model, the Euler equations, one motivates %these expressions 
definitions for $\energy$, $\momflux$ and $\energyflux$
by noting that if collisions in a gas are frequent, the collision frequency $\nu$ is large relative to expected hydrodynamic timescales.
The evolution equations for all moments except $n$, $\u$ and $\energy$ have a collision term on the \rhs.
Thus in the limit of very frequent collisions, the \rhs\ dominates all moment equations.
It is therefore reasonable to set $\CBoltzmann[\tilde{f},\tilde{f}]=0$.
This equation is solved by the Maxwell--Boltzmann distribution
\begin{align}
  \tilde{f}^{(0)}(\x,\v,t) = \frac{n(\x,t)}{[2\pi T(\x,t)]^{3/2}} \exp \left ( - \frac{|\v - \u(\x,t)|^2}{2T(\x,t)} \right) .
  \label{eq:MaxwellBoltzmannDistribution}
\end{align}
Substituting this into \eqref{eq:EnergiesMomentumDefintions} gives
\begin{align}
\begin{split}
  \energy = \frac{1}{2}m|\u|^2 + \frac{3}{2}mnT ,
  \hspace{0.5cm}
 % \\
  \momflux = m n\u\u + mnT \Itensor ,
  \hspace{0.5cm}
  %\\
  \energyflux =  \frac{1}{2}mn\u|\u|^2 + \frac{5}{2}mnT\u , 
\end{split}
  \label{eq:EnergiesMomentumModels}
\end{align}
which on further substitution into \eqref{eq:NearlyEuler} yields the Euler equations for $n$, $\u$ and $T$,
\begin{subequations}
\begin{align}
  \pd{n}{t} + \nabla\cdot\left( n\u \right) = 0, 
  %\hspace{1cm}
  \\
  mn \left( \pd{\u}{t} + \u\cdot \nabla \u \right) + \nabla (mnT) = 0, 
  %\hspace{1cm}
  \\
  \frac{3}{2}mn\left( \pd{T}{t} + \u\cdot \nabla T\right) + mnT\ \nabla\cdot \u = 0.
\end{align}
  \label{eq:EulerEquations}%
\end{subequations}
The Euler equations are in fact the leading order result of the Chapman--Enskog expansion \cite{ChapmanCowling},
where the distribution function in the Boltzmann equation is expanded in powers of the Knudsen number $\Kn$, the ratio of the mean free path to the characteristic system length.
Expanding to higher orders yields further fluid models:
the Navier--Stokes equations at $\O(\Kn)$,
the Burnett equations at $\O(\Kn^2)$
and the
super-Burnett equations at $\O(\Kn^3)$ \cite{Burnett35,Burnett36}.
%Other choices for the moments \eqref{eq:EnergiesMomentumModels} lead to different fluid models, such as the Navier--Stokes equations or the Burnett equations \cite[see \eg][]{Dellar07}.

In a plasma, the relevant kinetic equation is the Fokker--Planck--Landau equation
\begin{align}
  \pd{\tilde{f}_s}{t} + \v\cdot\nabla\tilde{f}_s + \frac{q_s}{m_s}\left( \E + \frac{\v\times\B}{c} \right)\cdot\pd{ {\tilde{f}_s}}{\v} = \sum_{r}\nu_{sr}\CLandau[\tilde{f}_s,\tilde{f}_r],
  \label{eq:FokkerPlanckLandauEquation}
\end{align}
where $\tilde{f}_s$ is the distribution function for species $s$.
Now we have an additional term for the acceleration due to the Lorentz force \eqref{eq:LorentzForce},
and a different collision operator, the Landau operator for collisions between charged particles
mediated by long-range Coulomb interactions (see \sec\ref{sec:LandauCollOpAndProperties}),
where $\nu_{sr}$ is the collision frequency for collisions between species $s$ and $r$.
Macroscopic quantities are still defined by moments of $\tilde{f}$ \eqref{eq:MacroscopicMoments} and \eqref{eq:EnergiesMomentumDefintions}, 
but in addition, moments also give the charge density $\varrho$ and current density $\j$,
\begin{align}
  \varrho = \sum_s q_s \int\d^3 \v ~ \tilde{f},
  \hspace{1cm}
  \j = \sum_s q_s \int\d^3 \v ~ \v\tilde{f}.
	\label{eq:MaxwellSourcesIntro}
\end{align}
The charge and current densities appear as sources in Maxwell's equations,
\begin{subequations}
\label{eq:MaxwellIntro}
\begin{align}
  \nabla\cdot\E = 4\pi \varrho, \label{eq:GausssLawIntro}\\
  \nabla\cdot\B = 0, \label{eq:NoMonopoleIntro}\\
  \pd{\B}{t} = -c\nabla\times\E, \label{eq:FaradaysLawIntro}\\
  \nabla\times\B = \frac{4\pi}{c}\j +\frac{1}{c}\pd{\E}{t}, \label{eq:AmperesLawIntro}
\end{align}
\end{subequations}
which determine the electric and magnetic fields appearing in the Lorentz force term in \eqref{eq:FokkerPlanckLandauEquation}.
Thus the plasma is described by a coupled nonlinear integro-differential equation system \eqref{eq:FokkerPlanckLandauEquation} and \eqref{eq:MaxwellIntro} for $\tilde{f}$ in six dimensional phase space.
Moreover, as mentioned above, this system contains a wide range of scales in both space and time.
This is analytically and numerically intractable.

The wide range of timescales both causes the largest difficulty and points towards a solution.
Estimates for timescales in the JET \tkmk\ regime give
a gyrofrequency $\Omega_i\sim 2\times10^8$ s$^{-1}$,
typical turbulent fluctuations $\omega\sim 2\times10^4$ s$^{-1}$,
and transport rate $\tau^{-1}\sim 2$ s$^{-1}$ \cite{Abel13}.
Supposing one needs ten timesteps to properly resolve ion gyromotion,
one would need to evolve the system for $10^5$ timesteps to see just one oscillation of a typical turbulent fluctuation,
and for $10^9$ timesteps to see any change in macroscopic properties.
To progress, we must relax this timestep restriction.

This is achieved through gyrokinetic theory, which was introduced independently by \citet{RutherfordFrieman68} and \citet{TaylorHastie68} in the 1960's (and is derived and discussed in \chp~\ref{sec:GKMSystem}).
The key idea is to average over the particle gyromotion.
This eliminates the fast timescale $\Omega^{-1}_s$ from the problem, 
and removes the dependence on azimuthal velocity.
Thus the six-dimensional kinetic system for charged particles 
reduces to the five-dimensional gyrokinetic system for the motion of charged rings.
The gyrokinetic system, summarized in \sec\ref{sec:GyrokineticSummary}, is still very expensive computationally.
Simulations of a whole tokamak are not routinely possible (though ``global'' codes do exist, \eg\ \cite{ORB5}) and even flux-tube simulations of a tokamak section \cite{Cowley91} require upwards of hundreds of
thousands of CPU hours each on High Performance Computing platforms.
Because of the high-dimensionality, resolution in each dimension is usually coarse.

It is standard to compute directly with the gyrokinetic equations, or simplified (\eg\ reduced-dimensional) cases of it.
Working with this system, we are guaranteed to include all physical effects which are slower than the gyromotion, 
and all behaviour which depends on velocity space.
This approach is the subject of this thesis, as outlined in the next section. 
There are however also fluid models which are derived from gyrokinetics.
These ``gyrofluid'' models, derived in the 1990's by Hammett and coworkers \cite{Hammett92,Dorland93,HammettBeer93}, 
close the moment equations using a model for Landau damping \cite{HammettPerkins90}.
The computation effort saved by replacing a velocity space grid of $\O(100)$ points with four \cite{Dorland93} or six \cite{HammettBeer93} moments
may be redeployed in increasing spatial resolution, for example.
Alternatively, gyrofluid models have been used in conjunction with transport solvers like \trinity\ \cite{Trinity} 
to provide computationally inexpensive approximations to turbulent properties like heat fluxes.

Gyrofluid models are of interest to us as in the spectral representation of the distribution function (which we shortly introduce) there is a direct correspondance between coefficients of the spectral representation and moments of the distribution function. 
Therefore, by minimizing the velocity space resolution in our methods we are naturally approaching a fluid-like representation of gyrokinetics.
Ultimately it would be interesting to compare our representation (which is entirely based on computational considerations) 
with gyrofluid models (which are based on modelling assumptions),
but this falls outside the scope of this thesis.

\section{Numerical methods for the gyrokinetic equations}

In this thesis we improve numerical methods for the solution of the \gkm\ system.
We employ spectral methods in velocity space to complement the Fourier spectral method typically used in physical space.
In spectral methods, dependent variables---here the distribution function and the electromagnetic fields---are represented using a series expansion in a family of orthogonal functions.
Thus instead of discretizing the system on a finite grid,
we derive a finite set of moment equations for the coefficients of the series expansion, analogous to the moment equations \eqref{eq:EulerEquations} derived previously for a neutral gas.

Such spectral methods are desirable as they efficiently represent the distribution function, minimizing the resolution and memory required.
They are also conservative: unlike grid methods, they introduce no spurious numerical dissipation in the calculation of derivatives, and thus exactly conserve quadratic invariants of the \gkm\ system such as free energy (see \sec\ref{sec:FreeEnergy}). 
%As well as exactly capturing a physical property of the system, this is also favourable for the stability of the numerical method. 
%Moreover since each basis function is characterized by a scale (\eg\ in Fourier methods, each wave has a wavelength), spectral methods a natural 
Finally, spectral methods introduce a change of coordinates (\eg, from space coordinate $\x$ to Fourier wavenumber $\k$)
which can be exploited to make the equations local in phase space. %, and therefore more amenable to parallel computation.
This is particularly useful for computation on multi-processor platforms, because data and operations which are local in phase space are also local to a processor,
and so do not require costly inter-processor communication.
Such considerations are important as it is typically inter-processor communication, not computation, that limits the performance of codes,
particularly for large numbers of processors.

\begin{figure}[tbp]
  \centering
  \subfigure[]{
		\includegraphics[width=0.49\textwidth]{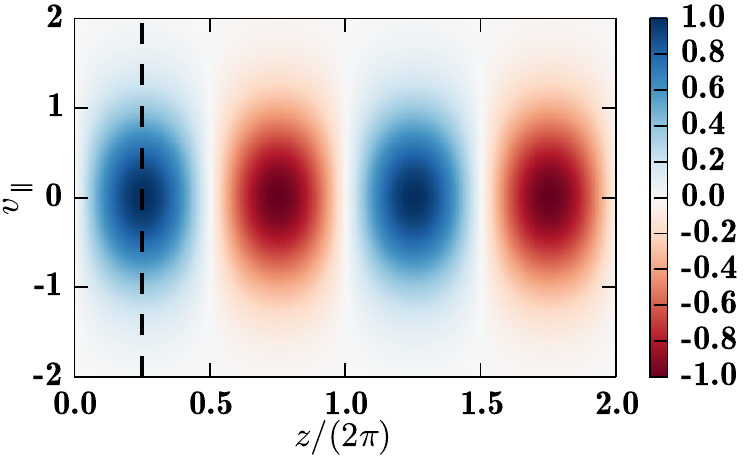}
		\includegraphics[width=0.49\textwidth]{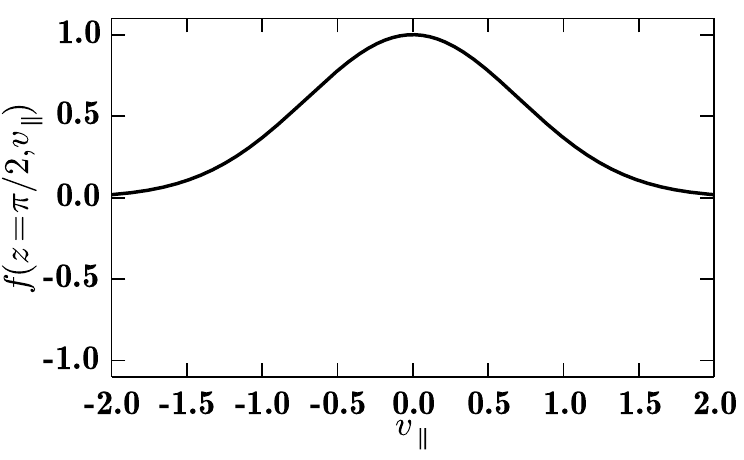}
}
  \subfigure[]{
		\includegraphics[width=0.49\textwidth]{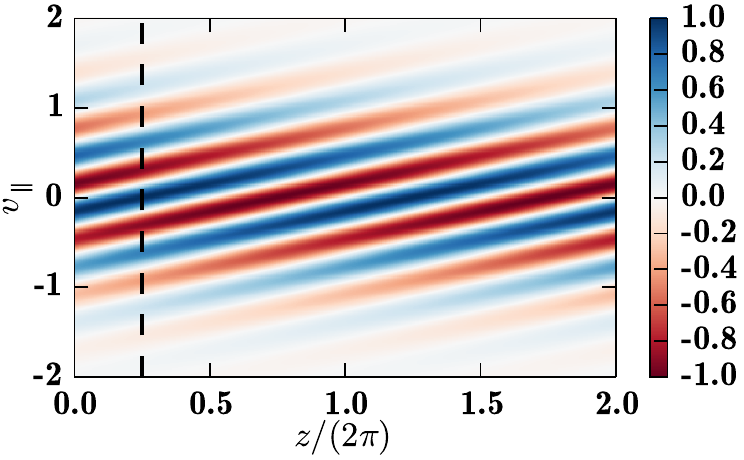}
		\includegraphics[width=0.49\textwidth]{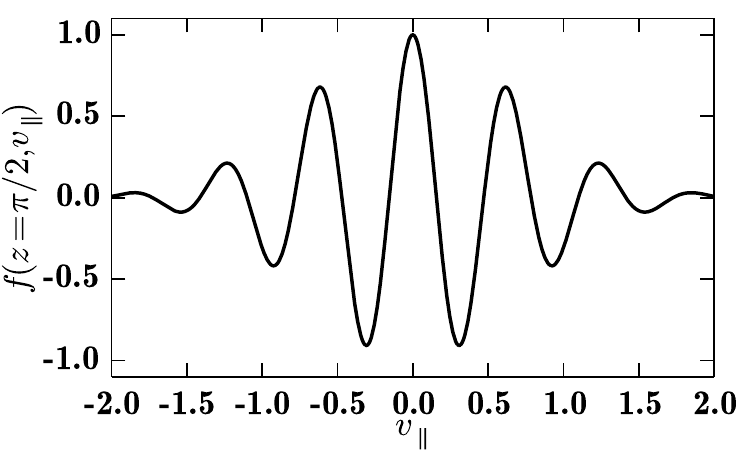}
}
  \subfigure[]{
		\includegraphics[width=0.49\textwidth]{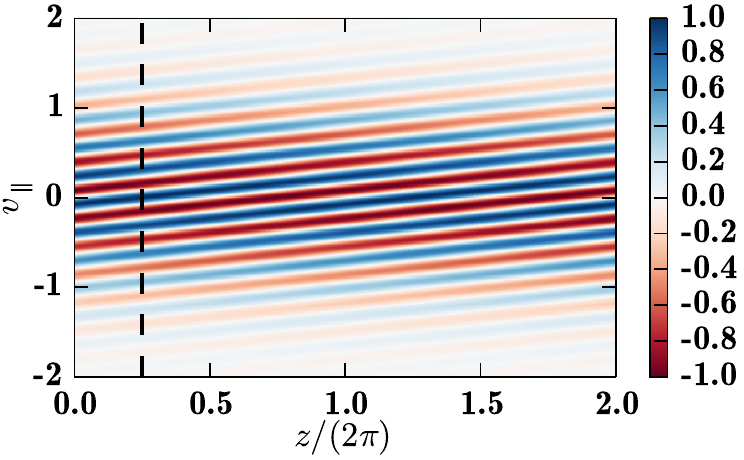}
		\includegraphics[width=0.49\textwidth]{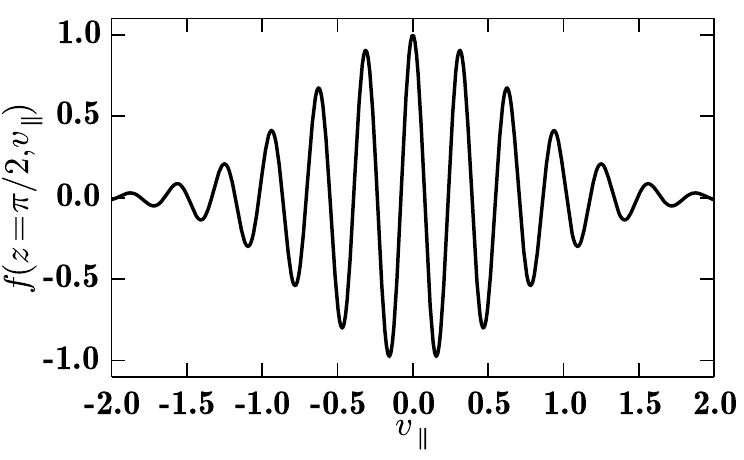}
}
\caption[Solution to the free streaming equation.]{ 
	The distribution function $\tilde{f}(z,\vpara,t)=e^{-\vpara^2}\sin(z-\vpara t)$, which solves the free streaming equation with the initial condition $\tilde{f}(z,\vpara,0)=e^{-\vpara^2}\sin(z)$.
The distribution function is plotted at three times, 
	(a) $t=0$,
	(b) $t=10$, and
	(c) $t=20$.
	The left-hand column shows contours of the distribution function,
	while the right-hand column shows a slice through the distribution function at $z=\pi/2$ (which is marked with a dashed line in the left-hand column).
	The distribution function becomes more sheared with increasing time, and velocity space slices through the distribution function become characterized by finer scales.
\label{fig:intro_free_streaming}
}
\end{figure}

In this thesis we use the Hermite and Hankel spectral representations in parallel and perpendicular velocity space respectively.
Parallel velocity space must be treated very carefully 
in order to capture Landau damping, an important dissipation mechanism
which is
related to the formation of infinitesimally fine scales in the distribution function,
due to the streaming term $\v\cdot\nabla\tilde{f}_s$ in the kinetic equation \eqref{eq:FokkerPlanckLandauEquation}.
To understand this effect, we consider the simplest such case, the free streaming equation
\begin{align}
	\pd{\tilde{f}}{t} + \vpara\pd{\tilde{f}}{z} = 0,
	\label{eq:FreeStreamingEquationIntro}
\end{align}
which has the solution $\tilde{f}(z,\vpara,t)=\tilde{f}(z-\vpara t,\vpara,0)$.
That is, the solution is the initial distribution sheared in phase space, 
as shown in \fig\ref{fig:intro_free_streaming} where we plot $\tilde{f}(z,\vpara,t)=e^{-\vpara^2}\sin(z-\vpara t)$, 
the solution to the free streaming equation with the initial condition $\tilde{f}(z,\vpara,t=0)=e^{-\vpara^2}\sin z$.
The distribution function rapidly oscillates in velocity space,
forming infinitesimally fine velocity space scales as $t\to\infty$.
Although the distribution function itself does not decay,
the rapid oscillation means that moments of the distribution function \wrt\ parallel velocity, 
like the electrostatic potential, do decay.
At any finite time, the distribution function $\tilde{f}$ is continuous and differentiable;
however it is not an exponentially decaying separable solution of \eqref{eq:FreeStreamingEquationIntro}.
Seeking eigenmodes of \eqref{eq:FreeStreamingEquationIntro} by taking a Fourier transform in $z$ and $t$ (with wavenumber $k$ and frequency $\omega$ respectively) we find a continuous spectrum of eigenvalues $\omega=k\vpara$ related to the singular eigenfunctions $\delta(\omega-kv)$.
Moreover, these eigenfunctions are not square integrable.

The behaviour of the free streaming equation is replicated in the Fokker--Planck equation \eqref{eq:FokkerPlanckLandauEquation}.
Here the decay of the electrostatic potential with time is known as Landau damping.
The decay behaviour is different to that of the free streaming equation due to the presence of Lorentz force terms,
but again is ultimately due to the formation of fine scales in velocity space.
The Landau-damped distribution function is derived by solving the kinetic equation via a Laplace transform (as shown in \chp\ \ref{sec:ParallelVelocitySpace}).
As before, infinitesimally fine scales form in velocity space as $t\to\infty$,
so the velocity space moments decay.
Again, the distribution function is not a time eigenmode of the problem; rather the eigenmodes are singular ``Case--Van Kampen'' modes.

This behaviour changes when collisions are introduced.
Collisions provide a velocity space diffusion, a term like $\nu\tpd{^2\tilde{f}_s}{\vpara^2}$ on the \rhs\ of \eqref{eq:FreeStreamingEquationIntro} 
which acts to smooth the fine scale structure in the distribution function.
The eigenmodes of the system are now continuous and differentiable, being obtained from a differential equation in $\vpara$.
Moreover, the Landau-damped distribution function emerges as an eigenmode of the system in the singular limit of vanishing collisions, $\nu\to0$.
The velocity space diffusion continues to have an effect due to the formation of infinitesimal scales in this limit.
Eigenmodes of the strictly collisionless system ($\nu=0$) are still singular. 

This behaviour makes Landau damping difficult to capture in a discrete system where there is necessarily a shortest resolved velocity scale.
With insufficiently strong collisions, the eigenmodes of the discrete system are discrete approximations to the singular solutions of the strictly collisionless equation. 
However, smooth solutions may be found by solving the collisional problem with $\nu$ sufficiently large to smooth the distribution function so that it is resolved for a given grid.
In principle, one may then recover the Landau solution by taking the limit $\nu\to0$ while simultaneously taking the number of grid points to infinity to ensure the solution is always resolved. 

The Hermite representation introduced in \chp\ \ref{sec:ParallelVelocitySpace} provides a convenient description of parallel velocity space.
The $m$th order Hermite polynomial has a characteristic velocity scale $\vth/\sqrt{m}$ so that each expansion coefficient represents a different velocity space scale.
The square of each coefficient represents that scale's contribution to the free energy.
Moreover, the electrostatic potential is proportional to the coefficient of the zeroth order polynomial.
The streaming term $\vpara\partial_z\tilde{f}$ becomes nearest neighbour mode coupling in $m$ that results in the transfer of free energy from low $m$ to high $m$.
Thus Landau damping may be interpreted as the flux of free energy out of the electrostatic potential at $m=0$ and towards high $m$ where it is dissipated by collisions.
This free energy transfer is linear and reversible, but we show in \chp\ \ref{sec:ParallelVelocitySpace} that
the solution of an \ivp\ of the linearized kinetic system approaches an eigenmode where only the forward transfer occurs.
The backward transfer is also observed in numerical simulations where insufficient collisions at the largest retained $m$ result in free energy reflecting back to low $m$ and causing the electrostatic potential to grow.
This numerical phenomenon is called recurrence.

In \chp\ \ref{sec:ParallelVelocitySpace}, we introduce the iterated Kirkwood hypercollisional operator which selectively damps the finest resolved scales in the distribution function, while leaving the details of the free energy transfer unaffected.
Doing so, 
we very efficiently capture Landau damping behaviour and prevent recurrence, retaining only around ten moments in the Hermite expansion.

The Hermite representation is also convenient as it is mostly local in phase space, with the only significant nonlocality coming from the nearest neighbour mode coupling due to the streaming term.
In particular, the parallel velocity space integral in the sources \eqref{eq:MaxwellSourcesIntro} required to calculate the electromagnetic fields from Maxwell's equations \eqref{eq:MaxwellIntro} are coefficients of single Hermite modes, not sums as they would be in a grid discretization. 
This motivates our use of the Hankel transform, which localizes the perpendicular velocity space integral in the gyrokinetic and spatially Fourier transformed version of \eqref{eq:MaxwellSourcesIntro} where there is an additional Bessel function factor in the integrand.
In Fourier--Hankel--Hermite space, the linearized gyrokinetic equations are local but for the Hermite mode coupling, and so may be solved very efficiently as a one-dimensional problem.
While this is no longer true of the nonlinear gyrokinetic system, it too has interesting locality properties which may be exploited,
as we discuss in \chp\ \ref{sec:PerpendicularVelocitySpaceHankelTransform}.

We implement the Fourier--Hermite--Hankel representation in the spectral gyrokinetics code \sgk, described in \chp~\ref{sec:SpectroGK}.
This is an implementation of the \gkm\ system derived in \chp~\ref{sec:GKMSystem},
supporting electromagnetic perturbations and multiple kinetic species.
\sgk\ is based on \gstwo\ and \agk, both grid point codes in velocity space, but shares their well-tested parallelization framework and software infrastructure.
Other advantages of the implementation include exact free energy conservation in the absence of explicit collisions, and the capture of Landau damping through the use of the iterated Kirkwood hypercollisional operator. 
\sgk\ is the main practical outcome of this thesis, a tool ideally suited to studying turbulence in weakly collisional plasmas.
Moreover, \sgk\ is a useful test-bed for new algorithms and optimizations, and the natural starting point in future efforts to develop a spectral toroidal code. 

In \chp s \ref{sec:FreeEnergyFlowAndDissipation} and \ref{sec:ScalingLawsForDriftKineticTurbulence},
we use \sgk\ to study electrostatic drift-kinetic turbulence in Cartesian slab geometry,
a convenient long perpendicular wavelength limit of the \gkm\ system,
which nonetheless captures its important features.
In \chp\ \ref{sec:FreeEnergyFlowAndDissipation},
we describe the turbulent behaviour in terms of competing free energy cascades:
the linear transfer of free energy to fine velocity space scales via phase-mixing, as described in \chp\ \ref{sec:ParallelVelocitySpace},
and the nonlinear transfer to fine physical space scales,
similar to that in hydrodynamic turbulence.
We show that phase space is divided into two regions depending on which cascade is faster, where one cascade dominates the other.
Moreover, we show the surprising result that the nonlinearity excites a transfer of free energy from small to large scales in velocity space
which counteract the forward transfer of free energy from large to small velocity space scales from linear phase-mixing.
Thus where the nonlinearity is dominant, it completely suppresses the transfer of free energy to fine velocity space scales,
and the turbulence in that region of phase space is fluid-like.
In fact, these scales also correspond to the energetically dominant scales in the plasma, 
so that the overall behaviour of the turbulence is fluid-like.

We use this new understanding of drift-kinetic turbulence in \chp\ \ref{sec:ScalingLawsForDriftKineticTurbulence} 
to derive complete scaling laws for the spectra of the electrostatic potential and the distribution function,
and verify these scalings using \sgk\ simulations.

From these two \chp s we have a complete understanding of the behaviour of electrostatic drift-kinetic turbulence in a slab.
Moreover, we have developed the analytical and numerical tools required to study gyrokinetic turbulence in future work.

%%%In \chp~\dots we illustrate the solution of the 1+1D \vps\ with the Fourier--Hermite method using a modified version of the \sgk.
%%%We demonstrate that a velocity space form of the Hou--Li \cite{Hou07} spectral filter prevents recurrence and gives correct results even in regimes
%%%where filamentation and Landau damping are dominant.
%%%We study 
%%%nonlinear Landau damping and the 
%%%\twostream\ 
%%%instability.
%%%The availability of high quality numerical solutions to the \oneplusoned\ \vps\ makes it a good benchmark for novel numerical algorithms, and our 5D gyrokinetic code \sgk\ was readily adapted to solve this system instead.  
%%%
%%%[Free energy flow and dissipation in linear ITG.]

\newcommand{\GKMSystemChapTwo}{\eqref{eq:GyrokineticEquationSummary}--\eqref{eq:MaxwellSummary}}

\chapter{\Gkm\ system}
\label{sec:GKMSystem}

% Listing
\renewcommand{\labelenumi}{\arabic{enumi})}

%%%The gyrokinetic \eqn s are derived (for astrophysical uses) in \cite{AstroGK} and presented in various forms, \eg\ \citet[with useful notes]{Schekochihin09}; \cite{Numata,Plunk,HighcockPhD}.
%%%The purpose of this note is to write the gyrokinetic \eqn s in a form which is spectral in Fourier, Hermite and Hankel space.  
%%%The motivation for this is that this formulation localizes Maxwell's \eqn s (and to some extent the nonlinear term) and therefore minimizes the communication required in a parallel calculation.  

%\section{Introduction}

We begin by deriving the \gkm\ system which models turbulence in a 
wide variety of astrophysical and nuclear fusion plasmas.
Such turbulence spans multiple scales in space and time (see \tab\ \ref{tab:ParameterValues}).
Turbulence occurs on spatial scales comparable with the ion gyroradius $\rho_i$, while macroscopic quantities like densities, bulk velocities and mean temperatures vary over much longer lengths comparable with the system size $L$.
Similarly, turbulent fluctuations have characteristic frequency $\omega$ which is much faster than the rate of evolution of macroscopic properties $\tau^{-1}$, where $\tau$ is the transport time. 
In \sec\ref{sec:SmallScaleAveraging} we introduce the turbulent average to exploit this separation of scales.
This average partitions quantities into a mean part (which evolves slowly on large spatial scales) and a turbulent part (which evolves quickly on small spatial scales).
The equations for each part decouple, allowing for independent solution at different scales.

\begin{table}[tp]
	\begin{center}
	\begin{tabular}{|l|p{2.5cm}|p{2.5cm}|p{2.5cm}|p{2.5cm}|}
		\hline
		&	\multicolumn{2}{|c|}{\Tkmk s} & \multicolumn{2}{|c|}{Astrophysics}  \\
\hline
Parameter & JET & ITER \mbox{\small (projected)} & Solar wind at 1AU & Accretion flow near Sgr A* \\
\hline
$\rho_i$ (m) & $5.1\times10^{-4}$ & $3.2\times10^{-4}$ & $9\times10^4$ & $4\times10^3$ \\
$L$ (m) & 1 & 2 & $\sim 10^8$ & $\sim 10^{11}$ \\
$\vthi$ (m s$^{-1}$) & $8.7\times10^4$  & $7.7\times10^4$  & $9\times10^4$ & $10^{8}$ \\
$\tau^{-1}$ (s$^{-1}$) $^{\textrm{a}}$ & $2.0$ & $0.3$ & &  \\
$\omega$ (s$^{-1}$) $^{\textrm{b}}$ & $2.0\times10^{4}$ & $5.5\times10^{4}$ & $4\times10^{-4}$ & $7\times10^{-4}$ \\
$\Omega_i$ (s$^{-1}$) & $1.7\times10^8$ & $2.4\times10^8$ & 1 & $7\times10^{-2}$ \\
\hline
$\gkpar = \rho_i/L$ & $5\times10^{-4}$ & $2\times10^{-4}$ & $9\times10^{-4}$ & $4\times10^{-9}$ \\
$\beta_i=8\pi n_iT_i/B_0^2$ & $0.025$ & $0.04$ & $5$ & $4$ \\
\hline
  \end{tabular}
  \caption{Parameter values adapted from
		\citet[][\tab~1]{Abel13} (\tkmk s)
		and 
		\citet[][\tab~1]{Schekochihin09} (astrophysics).
		Notes:
		$^\textrm{a}$ Transport time for astrophysical plasmas estimated as bulk velocity $U$ \citep[from][]{Schekochihin09} over the system size $L$.
		$^\textrm{b}$ Plasma frequency estimated for astrophysical plasmas as $\omega\sim \kpara v_A \sim v_A/L$ with $v_A$ the \alfven\ velocity.
		 \label{tab:ParameterValues}}
	\end{center}
\end{table}

Further, the turbulent fluctuations themselves exhibit a disparity of scales.
In a strong magnetic field, the particles gyrate around field lines with gyrofrequency $\Omega_s=q_sB/m_s c$ which for both ions and electrons is much larger than the typical frequency of fluctuations $\omega$.
The turbulence is also spatially anisotropic, with particles streaming along mean fields much faster than they drift across them.
Hence typical wavelengths in the turbulences are much longer parallel to the mean field than perpendicular to it. 
These two properties allow for separation of scales via the ``gyroaverage'', the average over the particle gyration (see \sec\ref{sec:Gyroaverage}), 
a crucial procedure which makes the system tractable.
Gyroaveraging removes the fast cyclotron time scales, 
as well as eliminating the dependence on azimuthal velocity,
so the six-dimensional system for particles 
reduces
to a five-dimensional system for charged rings.

Gyrokinetic theory, which determines the evolution of these charged rings, was introduced independently by 
\citet{RutherfordFrieman68} and \citet{TaylorHastie68}.
These built on the earlier guiding centre approximation 
%(see \eg\ \citet{Alfven50}), 
(see \eg\ \textsc{Alfv\'en \&} \textsc{F\"alt-} \linebreak \textsc{hammer} \cite{Alfven50}), 
which 
applied the idea of studying the motion of charged rings to a single particle rather than a distribution. 
The derivation is greatly simplified by \citeauthor{Catto78}'s \cite{Catto78} introduction of guiding centre coordinates (see also \sec\ref{sec:GyrokineticVariables}).
Gyrokinetic theory was extended to include electromagnetic perturbations by \citet{AntonsenLane80} and \citet{Catto81},
and to incorporate nonlinear effects by \citet{FriemanChen82}.
Finally, gyrokinetic theory was united with neoclassical theory (which describes large scale, non-turbulent fluctuations, see \eg\ \citet{Catto87}) in a single theoretic framework by \citet{Abel13}.

The standard derivation of the \gkm\ system is via an asymptotic expansion in the gyrokinetic parameter $\varepsilon=\rho_i/L$, the ratio of the ion gyroradius to the system size.
All other scale disparities (\eg\ $\omega/\Omega_i$) are related to $\varepsilon$ using the ``$\delta f$-ordering'' 
introduced by \citet{AntonsenLane80} and \citet{FriemanChen82}, 
see \S\ref{sec:DeltaFOrdering}.
Thus every term has a definite size, 
yielding a hierarchy of coupled equations 
to be solved order-by-order in $\varepsilon$.
These equations are not closed as the fast particle gyration always results in a term corresponding to a higher order perturbation to the distribution function.
Gyroaveraging removes these terms yielding closed equations at each order.
Further, turbulent averaging separates equations into mean and turbulent parts
allowing simultaneous derivation of the gyrokinetic equation for turbulent fluctuations and the neoclassical equation for large scale perturbations.

In the above asymptotic approach, energy is conserved by the system overall, but is not conserved at each order.
Indeed, this energy transfer between orders is interpreted as a feature of a multiscale system \citep{Abel13}.
Alternative Hamiltonian derivations of gyrokinetics do conserve energy at each order by retaining terms which are formally small in the asymptotic expansion \cite[see][]{Dubin83,Hahm88,Brizard07,Garbet10}.
However, the gyrokinetic equation itself does conserve free energy, the weighted integral of disturbance amplitudes (see \S\ref{sec:FreeEnergy}), which is related to the Boltzmann entropy (see \S\ref{sec:LandauCollOpAndProperties}).
The free energy is quadratic and is neatly expressed via Parseval's theorem as the sum of squares of coefficients of the Fourier--Hankel--Hermite spectral expansion described in \chp s~\ref{sec:Hypercollisions} to \ref{sec:SpectroGK}.

%%%{\red This formulation results in an addition term, the ``parallel nonlinearity'' proportional to $\delta E_{\parallel}(\tpd{\delta f}{\vpara})$, 
%%%which is absent in the asymptotic derivation \citep{Abel13}.
%%%The importance of this term is contested, with some authors claiming it to be unimportant in simulations \citep{Candy06}, 
%%%while others find it necessary for the description of certain phase-space phenomena \citep{Brizard07}.
%%%We omit the parallel nonlinearity, consistent with its small order; 
%%%this is equivalent to the statement that while small scale structure may form in velocity space,
%%%the formal ordering 
%%%$\tpd{\delta f}{\vpara}\sim \delta f/\vthi$
%%%will not be violated.
%%%However, the parallel nonlinearity is implemented in \sgk, and we encounter it in \chp~\ref{sec:VlasovPoisson} where it is uncontroversially present in the Vlasov--Poisson system.
%%%}

Finally, since gyrokinetic theory was originally developed for \tkmk\ modelling, 
it is commonly presented in toroidal geometry \cite[see][\S3.3]{Abel13}.
However, Cartesian slab geometry typically suffices for astrophysical applications.
As a consequence of neglecting geometry terms, slab gyrokinetics no longer has the distinction between ``trapped'' and ``passing'' particles determined by a particle's magnetic moment (one of the gyrokinetic variables introduced by \citet{Catto78}).
Therefore velocity space in slab gyrokinetics is more commonly expressed in terms of parallel and perpendicular velocity \cite[\eg][]{Howes06,Numata10},
which simplifies the final equations and more naturally describes reduced dimension models 
(see \sec\ref{sec:HankelSpaceStructureGyrokinetics}).

In this chapter,
we derive gyrokinetics 
(\secs\ref{sec:Preliminaries}--\ref{sec:DerivationOfSlabGyrokineticEquations})
following the asymptotic approach in \citet{Abel13}, but simplified for a slab geometry.
Unlike other slab derivations \cite[\eg][]{Howes06}, we explicitly nondimensionalize
so that the gyrokinetic parameter appears in the equations and we work directly with the same normalized quantities as in the \gstwo\ family of codes \citep{GS2,Numata10}. %\footnote{I believe this derivation is unique in taking this approach.}.
The derivation keeps up to second order in the gyrokinetic parameter, corresponding to deriving the equations solved in \sgk.
By retaining the next order, we could derive the transport equations for the evolution of macroscopic quantities.
In \sec\ref{sec:FreeEnergy}
we derive equations for free energy which we use in \chp s \ref{sec:FreeEnergyFlowAndDissipation} and \ref{sec:ScalingLawsForDriftKineticTurbulence} to characterize plasma turbulence.
Finally in \sec\ref{sec:ComputationalFormsOfTheGyrokineticEquations} we derive the various simplified versions of gyrokinetics studied in this thesis by taking limits of the \gkm\ system.

\section{Preliminaries}
\label{sec:Preliminaries}
\subsection{Gyrokinetic--Maxwell equations}
The starting point is the Fokker--Planck \eqn\ %\citep[(A9)]{Howes06}
\begin{align}
  \fd{\ff_s}{t} = \pd{\ff_s}{t} + \v\cdot\nabla \ff_s + \frac{q_s}{m_s}\lp \fE + \frac{\v\times\fB}{c} \rp\cdot \pd{\ff_s}{\v} = \sum_r C_{sr}[\ff_s,\ff_r], %+C_{ss}(f_s,f_s) 
  \label{eq:FokkerPlanckEquation}
\end{align}
which describes the evolution of the distribution function $\ff_s$ of species $s$ with mass $m_s$ and charge $q_s$ in the six-dimensional phase space $(\r,\v)$.
We use Gaussian units, 
with speed of light $c$,
electric field $\fE$
and 
magnetic field $\fB$.
%It states that the distribution is constant on phase space characteristics except for the effect of collisions. 
The operator $C_{sr}$ is the Landau operator, describing collisions between particles of species $s$ and $r$.

The electromagnetic fields $\fE$ and $\fB$ are found via Maxwell's equations
\begin{align}
  \nabla\cdot\fE = 4\pi \fcharge, \label{eq:GausssLaw}\\
  \nabla\cdot\fB = 0, \label{eq:NoMonopole}\\
  \pd{\fB}{t} = -c\nabla\times\fE, \label{eq:FaradaysLaw}\\
  \nabla\times\fB = \frac{4\pi}{c}\fcurrent +\frac{1}{c}\pd{\fE}{t}, \label{eq:AmperesLaw}
\end{align}
where the charge density $\fcharge$ and current $\fcurrent$ are velocity moments of the distribution function
\begin{align}
  \fcharge = \sum_s q_s \int \mathrm{d}^3\v ~ \ff_s ,\label{eq:FullCharge}\\
  \fcurrent = \sum_s q_s \int \mathrm{d}^3\v ~ \v\ff_s. \label{eq:FullCurrent}
  %\label{}
\end{align}
We neglect Debye-scale and relativistic effects,
\begin{align}
  \kperp^2\lambda_{De}^2 \ll 1 \label{eq:Debye},\\
  \vths^2/c^2 \ll 1 \label{eq:NonRelativistic},
  %\label{}
\end{align}
where $\kperp$ is a typical perpendicular wavenumber, $\lambda_{De}=\sqrt{T_e/4\pi n_ee^2}$ is the electron Debye length and $\vths=\sqrt{2T_s/m_s}$ is the thermal velocity, with $n_s$ and $T_s$ the species density and temperature,
and $e$ the electron charge.

%%%{\red We show in \sec\ref{sec:GaussLawQuasineutrality}
%%%that on lengthscales much larger than the electron Debye length, Gauss' law \eqref{eq:GausssLaw} reduces to the quasineutrality condition
%%%\begin{align}
%%%  \fcharge = 0,
%%%  \label{eq:QuasineutralityAsCharge}
%%%\end{align}
%%%while neglecting relativistic effects gives a form of \ampere's law \eqref{eq:AmperesLaw} without the displacement current
%%%\begin{align}
%%%  \nabla\times\fB = \frac{4\pi}{c}\fcurrent. 
%%%  %\label{}
%%%\end{align}
%%%}

Further, we satisfy \eqref{eq:NoMonopole} and \eqref{eq:FaradaysLaw} by introducing the potentials $\varphif$ and $\fA$,
\begin{align}
  \fE &= -\nabla\varphif + \frac{1}{c}\pd{\fA}{t},
  \label{eq:EPotentialForm}
  \\
  \fB &= \nabla\times\fA, 
  \label{eq:BAsVectorPotential}
\end{align}
and we work in the Coulomb gauge, $\nabla\cdot\fA = 0$.

%%%Thus the system we solve comprises the \fp\ equation \eqref{eq:FokkerPlanckEquation} with the fields determined by equations \eqref{eq:QuasineutralityAsCharge}--\eqref{eq:BAsVectorPotential}. 

\subsection{Small scale averaging}
\label{sec:SmallScaleAveraging}

The gyrokinetic equation is derived using the separation of temporal and spatial scales to partition physical quantities into their mean and fluctuating parts.
To do so, we introduce the average over the short turbulent length and time scales, $\ta{\cdot}$,  and write each physical quantity $\widetilde{Q} = \ta{\widetilde{Q}} + Q$, where by construction $\ta{Q}=0$.

%The turbulent average is the combination of a spatial and temporal average.
In space, the macroscopic length scale which characterizes the equilibrium is well-separated from the gyroscale on which fluctuations occur.
We can therefore find an intermediate scale $\ell$ for which
\begin{align}
  \rho \ll \ell \ll L,
%%%  \hspace{1cm}
%%%  \omega^{-1} \ll T \ll \tau
  %\label{}
\end{align}
and define the perpendicular average
\begin{align}
  \ga{F(\r,\v,t)}{\perp} = \frac{1}{\ell^2}\int_{\ell^2} \d^2\r'_{\perp} ~ F(\r'_{\perp},r_{\parallel},\v,t) ,
  %\Bigg / \int_{\ell^2}\mathrm{d}^2\r',
  \label{eq:PerpendicularAverage}
\end{align}
where integration is over a square of side $\ell$ perpendicular to the field line and centred at $\r$.
The perpendicular average varies on the length scales 
$\ell$.
%$\sim\ell$ 
It therefore averages over the turbulent fluctuations on the gyroscale $\rho\ll\ell$, but leaves variation on the macroscale $L\gg\ell$ unaffected.
%(Perhaps tweak this to be the same as the intermediate average and exploit the box periodicity.)

Similarly we find an intermediate timescale $\intt$ between the turbulent and transport time scales
\begin{align}
%%%  \rho \ll \ell \ll L,
%%%  \hspace{1cm}
  \omega^{-1} \ll \intt \ll \tau,
  %\label{}
\end{align}
and define the time average
\begin{align}
  \ga{F(\r,\v,t)}{\intt} = \frac{1}{\intt}\int_{t-\intt/2}^{t+\intt/2} \d t' ~ F(\r,\v,t') .
  %\label{}
\end{align}
Again, this averages over turbulent timescales without affecting quantities that vary on the transport time scales.

The turbulent average is defined as the combination of these two averages
\begin{align}
  \ga{F}{\textrm{turb}} = \ga{\ga{F}{\intt}}{\perp}.
  %\label{}
\end{align}
This allows the separation of all quantities into mean parts and turbulent parts
\begin{align}
  \ff_s = F_s + f_s , \hspace{1cm} F_s = \ta{\ff_s} \label{eq:DistFnSeparation}, \\
  \fE = \bE + \pE , \hspace{1cm} \bE = \ta{\fE}, \\
  \fB = \bB + \pB , \hspace{1cm} \bB = \ta{\fB}, \\
  \fA = \bA + \pA , \hspace{1cm} \bA = \ta{\fA}, \\
  \varphif = \varphib + \varphip , \hspace{1cm} \varphib = \ta{\varphif} .
  %\label{<++>}
\end{align}
The scales of the temporal and spatial variations of a typical quantity $Q$
can then be estimated as
\begin{align}
\begin{split}
  \pd{}{t}\ln\ta{Q}\sim \tau^{-1},
  \hspace{1cm}
  \pd{}{t}\ln Q \sim \omega, \\
  \nabla\ln\ta{\widetilde{Q}}\sim \b\cdot\nabla\ln Q \sim L^{-1},
  \hspace{1cm}
  \nabla_{\perp}\ln Q \sim \rho_i^{-1}.
  \label{eq:GradientSizes}
\end{split}
\end{align}
where $\b$ is the unit vector in the direction of the magnetic field, and $\nabla_{\perp}$ is the gradient in the direction perpendicular to $\b$.
These state that macroscopic time evolution is on the transport timescale $\tau$, while turbulent evolution is on the timescale of plasma dynamics $\omega^{-1}$.
Macroscopic spatial variation occurs over system scales $L$, as do turbulent spatial fluctuations parallel to the field line,
while perpendicular spatial fluctuations occur on scales comparable to the ion gyroradius.

\subsection{Geometry}

We solve for the perturbed distribution function $f_s$ in a Cartesian box with spatial dimension $(x,y,z)\in[0,L_x]\times[0,L_y]\times[0,L_z]$ and velocity space dimension $(\vpara,\vperp)\in\bbR\times [0,\infty)$.
The mean parts of the electromagnetic field, $\bE$ and $\bB$, are imposed.  
For relevant astrophysical and nuclear fusion regimes, there is no mean electric field $\bE=0$, $\varphib=0$,
and the magnetic field $\bB=B_0\b$ is approximately constant.
Specifically, $\bB$ has no explicit time dependence, $\tpd{\bB}{t}=0$.
The vector $\b$ is approximately a unit vector pointing in the $z$-direction.
It is curl-free ($\nabla\times\b=0$) and has a small curvature pointing in the $x$-direction, $\boldsymbol{\kappa}=\kappa\unitx=(\b\cdot\nabla)\b$.\footnote{These conditions are satisfied by any vector $\b=(b_1(x,y),b_2(x,y),1)$ such that
	\begin{inparaenum}[(i)]
	\item $b_1^2+b_2^2\ll 1$, so that $|\b|\approx1$,
	\item $\tpd{b_1}{y}=\tpd{b_2}{x}$, so that $\nabla\times\b=0$; and
	\item $b_1\tpd{b_2}{x}+b_2\tpd{b_2}{y}=0$, so that $\boldsymbol{\kappa}\cdot\unity=0$.
	\end{inparaenum}
	}
	The field strength $B_0$ is assumed to be a small deviation from a constant reference field strength. It is constant along the field line, $\b\cdot\nabla B_0=0$, but has a small linear variation in the $x$-direction, \ie\ $\nabla\ln B_0=-L_B^{-1}\unitx$ where $L_B^{-1}$ is a constant.

	Similarly, we impose perpendicular gradients in density and temperature by expanding the leading order distribution function (which we show in \S\ref{sec:DerivationOfSlabGyrokineticEquations} is a Maxwellian) about a global reference density $n_0$ and temperature $T_0$.  %(see \S\ref{sec:Maxwellian}).
	We again take the perturbation to be constant and pointing in the $x$-direction, $L^{-1}_n\unitx = -\nabla\ln n_0$ and $L^{-1}_T\unitx = -\nabla\ln T_0$.

\begin{figure}[tbp]
  \centering
  \subfigure[\label{fig:Geometry}]{\includegraphics[width=0.49\textwidth]
{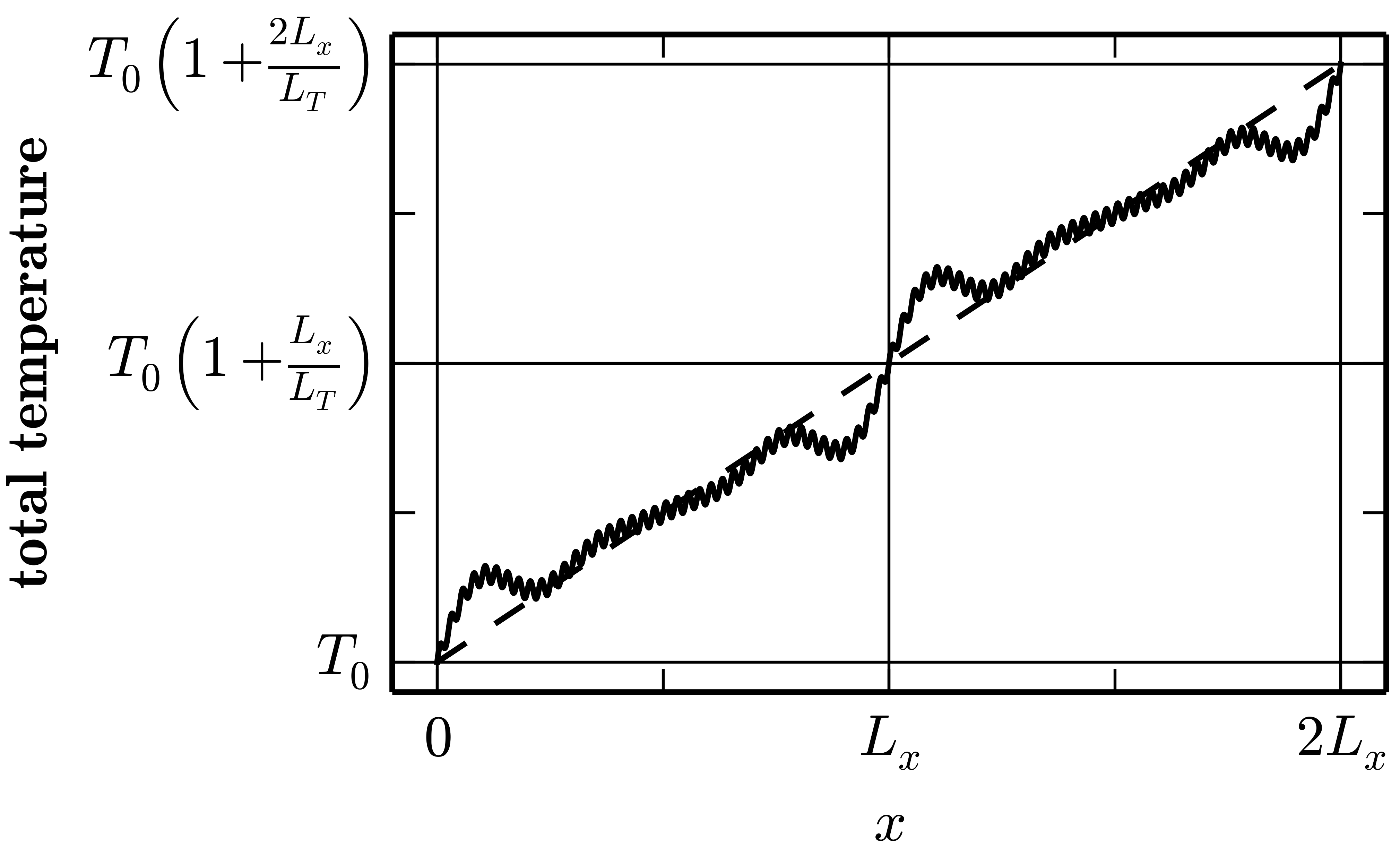}}
\subfigure[\label{fig:Geometry2}]{\includegraphics[width=0.49\textwidth]
{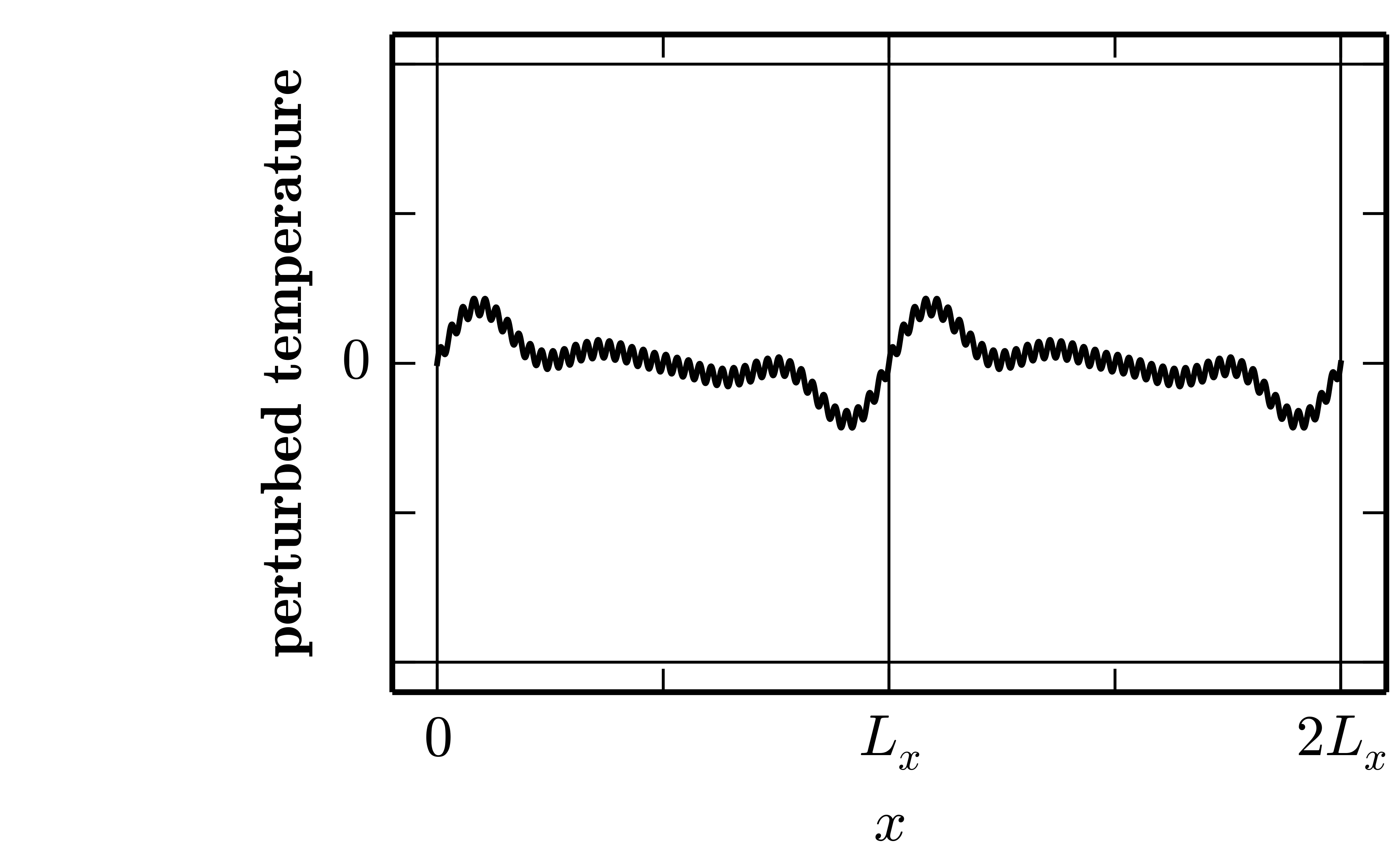}}
\caption[Periodic temperature fluctuations about a macroscopic gradient.]{Schematic representation of periodic temperature fluctuations about a macroscopic temperature gradient. (a) The total temperature (solid) and the macroscopic temperature profile (dashed) over two domain lengths, $[0,L_x]$, $[L_x,2L_x]$. (b) The temperature perturbation about the macroscopic profile.  While the total temperature is not periodic, the perturbed temperature may be periodic for sufficiently large $L_x$. \label{fig:Gradients}  }
\end{figure}

As there are no macroscopic gradients in the $y$- and $z$-directions, we use periodic boundary conditions in $y$ and $z$.
This is computationally convenient, as it permits a Fourier series representation for $f_s$.
It is permissible provided that the correlation length of the turbulence (the characteristic length scale, $\rho_i$) is much shorter than the box length, so that a point in space is not affected by its periodic image.
Moreover, while the gradients in the $x$-direction prevent the whole problem being periodic in $x$, the constant gradients in the density, temperature and magnetic field enter the equations for $f_s$ in a way which does not prevent $f_s$ being periodic.
Therefore we also take $f_s$ to be periodic in $x$. 
This means that there are small periodic perturbations to density, temperature and the electric field superimposed on non-periodic macroscopic gradients, see \fig\ref{fig:Gradients}.

%\subsection{Gyrokinetic ordering}
\subsection{$\boldsymbol{\delta f}$ ordering}
\label{sec:DeltaFOrdering}

We now impose the $\delta f$ ordering introduced by \citet{AntonsenLane80} and \citet{FriemanChen82}.
This orders all quantities (except the transport and collisional timescales) with respect to the small gyrokinetic parameter $\gkpar$,
\begin{align}
  \frac{\omega}{\Omega_i} 
  \sim
  \frac{\fs}{\Fs} \sim
  %\frac{\fsi}{\fso} \sim
  \frac{|\pB|}{|\bB|}\sim
  \frac{e\varphip}{T} 
  %\sim \frac{\nabla_{\parallel}\fs}{\nabla_{\perp}\fs}
  %\sim \frac{\nabla_{\parallel}\fsi}{\nabla_{\perp}\fsi}
  %\sim \frac{\nabla_{\parallel}\varphi}{\nabla_{\perp}\varphi}
  %\sim \frac{\kpara}{\kperp} 
  \sim \frac {\rho_i}{L} \id \varepsilon \ll 1.
  \label{eq:GKOrdering}
\end{align}
The first term states that plasma dynamics are much slower than particle gyromotion.
The second two terms state that the fluctuations in the distribution function and magnetic field are small in amplitude relative to the mean values.
%There is no mean electric field for an analogous ordering for $\delta\E$, so instead the fourth term imposes that the $\E\times\B$ velocity is much smaller than the thermal velocity
As $\bE=0$, there is no analogous ordering for $|\pE|/|\bE|$. 
Instead the fourth term imposes that the $\E\times\B$ velocity is much smaller than the thermal velocity
\begin{align}
  \frac{c|\pE|}{B}\sim \frac{c\varphip}{\rho_s B} \sim \vths \gkpar.
  %\label{}
\end{align}

It follows from \eqref{eq:GKOrdering} that the vector potential fluctuations $\pA$ are very small.
This is because $\bA$ and $\pA$ vary on different perpendicular length scales,
\begin{align}
  \bB = \nabla\times\bA \sim \frac{1}{L}\bA,
  \hspace{1cm}
  \pB = \nabla\times\pA \sim \frac{1}{\rho_s}\pA,
  %\label{<++>}
\end{align}
%$\B \sim \A/L$, $\delta\B \sim \delta\A/\rho_s$, 
so that 
\begin{align}
  \frac{|\pA|}{|\bA|}\sim \frac{\rho_s|\pB|}{L|\bB|}\sim \gkpar^2.
  %\label{}
\end{align}
Consequently the electric field \eqref{eq:EPotentialForm} is primarily electrostatic.

It also follows from the gradients \eqref{eq:GradientSizes} and the $\delta f$ ordering \eqref{eq:GKOrdering} that the variation in turbulent fluctuations is anisotropic, with cross-field variation much faster than variation along the field line
\begin{align}
  \frac{\b\cdot\nabla Q}{|\nabla_{\perp} Q|}\sim \gkpar.
  %\label{}
\end{align}

The $\delta f$ ordering thus gives sizes for all scales except the transport and collisional timescales.
To determine the transport time scale we use the gyro-Bohm estimate for turbulent thermal diffusivity $\chi_{T_s}\sim\rho_s^2\omega$ \citep{Dimits00}.
Then
\begin{align}
  \frac{1}{\tau}\sim\frac{\chi_{T_s}}{L^2}\sim \gkpar^2 \omega \sim \gkpar^3 \Omega_i,
  %\label{}
\end{align}
so that transport is two orders slower than plasma dynamics.
This means time derivatives of the leading-order distribution do not appear at orders we study, so for our purposes $\Fs$ is constant in time. 

We also formally order the collision time 
\begin{align}
  \cf \sim \gkpar \Omega_i.
  %\label{}
\end{align}
We can therefore study plasmas which are either weakly collisional $\cf\sim \omega$ or collisionless $\cf\ll\omega$ by taking a subsidiary ordering for $\cf$ \citep{Howes06}. 

Finally, we expand the mean and fluctuating parts of the distribution function $\ff_s$ \eqref{eq:DistFnSeparation} in powers of the gyrokinetic parameter $\gkpar$,
\begin{align}
  \ff_s = \Fs + \fs ,
  \hspace{1cm}
  \Fs = \Fso + \Fsi + \Fsii + \ldots
  \hspace{1cm}
  \fs = \fsi + \fsii + \ldots
  \label{eq:DistFnExpansion}
\end{align}
where $F_{ns}\sim f_{ns} \sim \gkpar^n\Fso$.
There is no $f_{0s}$ term in the $f_s$ expansion, consistent with $f_s/F_s\sim\gkpar$.

Now every term in the \gkm\ system has a well-defined order \wrt\ $\gkpar$.
In \sec\ref{sec:Nondimensionalization} we nondimensionalize using scales which make this ordering transparent; but first we introduce gyrokinetic variables and the gyroaverage which are crucial for the derivation.

\subsection{Gyrokinetic variables}
\label{sec:GyrokineticVariables}

To derive the \gk\ \eqn, we transform the \fp\ \eqn\ \eqref{eq:FokkerPlanckEquation} from position space coordinates $(\r,\v)$ to guiding centre space coordinates introduced by \citet{Catto78}: guiding centre position $\R_s$, particle energy $E_s$, magnetic moment $\mu_s$, gyroangle $\vartheta$ and sign of parallel velocity $\sigma$. 
These are defined by
\begin{align}
  \R_s = \r - \rhobs , 
  \hspace{1cm}
  E_s = \frac{1}{2}m_sv^2 , \hspace{1cm}
  \mu_s = \frac{m_s\vperp^2}{2B_0}, \hspace{1cm}
  \sigma = \frac{\vpara}{|\vpara|},
  \label{eq:GyrokineticVariables}
\end{align}
where the gyroradius is
\begin{align}
  \rhobs =  \frac{\b\times\v}{\Omega_s}, 
  \label{eq:rhobsDefinition}
\end{align}
%%%\begin{align}
%%%  \R_s = \r - \rhobs , 
%%%  \hspace{1cm}
%%%  \label{eq:R}
%%%  \rhobs =  \frac{\b\times\v}{\Omega_s}, \\
%%%  E_s = \frac{1}{2}m_sv^2 , \\
%%%  \mu_s = \frac{m_s\vperp^2}{2B_0}, \\
%%%  \sigma = \frac{\vpara}{|\vpara|},
%%%  %\label{}
%%%\end{align}
and the gyroangle $\vartheta$ is defined implicitly in terms of parallel and perpendicular velocity 
\begin{align}
  \v = \vpara\b + \vperpv,
  \hspace{1cm}
  \vperpv = \vperp(\cos\vartheta\unitx + \sin\vartheta\unity).
  \label{eq:vCylindric}
\end{align}
%%%\begin{align}
%%%  \eperp = \cos\vartheta\x + \sin\vartheta\y ,
%%%  \hspace{1cm}
%%%  \et = -\sin\vartheta\x + \cos\vartheta\y
%%%  %\label{}
%%%\end{align}
%%%From these we have that
%%%\begin{align}
%%%  \b\times\v = 
%%%  \left | \begin{array}{ccc}
%%%     \x  & 0 & \vperp\cos\vartheta \\
%%%     \y  & 0 & \vperp\sin\vartheta \\
%%%     \b  & 1 & \vpara \end{array} \right |
%%%     = \vperp\et
%%%  %\label{}
%%%\end{align} 
In position space we denote the same gyroangle with $\varTheta$ to emphasize which coordinate system is meant.
Derivatives in $(\r,\vpara,\vperp,\varTheta)$ and $(\R_s,E_s,\mu_s,\vartheta,\sigma_s)$ are related via the chain rule. 
%%%\begin{align}
%%%  \pd{}{\r} = \pd{\R_s}{\r}\cdot\pd{}{\R_s} + \pd{E_s}{\r}\cdot\pd{}{E_s} + \pd{\mu_s}{\r}\cdot\pd{}{\mu_s} + \pd{\vartheta}{\r}\cdot\pd{}{\vartheta}
%%%  \label{<++>}
%%%\end{align}<++>
We only require the relations
\begin{align}
  \pd{}{\R_s} = \nabla , 
  \hspace{1cm}
 \pd{}{\vartheta} =
  \pd{}{\varTheta} - \frac{\vperpv}{\Omega_s}\cdot\nabla .
  \label{eq:ThetaChainRule}
\end{align}

It is also convenient to introduce the gyrokinetic potential
\begin{align}
	\chi_s = \varphi - \v\cdot\A.
  \label{eq:GyrokineticPotentialDefinition}
\end{align}
In the gyrokinetic equation, all information about the electromagnetic field enters as a function of $\chi_s$.
Note that the gyrokinetic potential is independent of species, but will become species-dependent when nondimensionalized in \S\ref{sec:NormalizationElectromagneticFieldGyrokineticPotential}.

\subsubsection{Gyroaverage}
\label{sec:Gyroaverage}
The average over a particle gyration, or gyroaverage, is an important tool that allows us to close the kinetic equation at each order of the asymptotic expansion. 
The gyroaverage is defined
\begin{align}
  %\ga{a(\r)}{\R_s} = \frac{1}{2\pi}\int_0^{2\pi} \d\vartheta ~  a(\R_s+\rhobs),
	\ga{a(\r)}{\R_s} = \frac{1}{2\pi}\int_0^{2\pi} \d\vartheta ~  a\left( \R_s+ \frac{\b\times\v}{\Omega_s} \right),
  %\ga{a(\r,\v)}{\R_s} = \frac{1}{2\pi}\int_0^{2\pi} \d\vartheta ~  a(\r(\R_s,E_s,\mu_s,\vartheta,\sigma),\v(\R_s,E_s,\mu_s,\vartheta,\sigma))
  \label{eq:GyroaverageDefinition}
\end{align}
where $\v$ is a function of $\vartheta$ defined by \eqref{eq:vCylindric}.
Thus the gyroaverage \eqref{eq:GyroaverageDefinition} is a function of the guiding centre variables $\R_s, E_s, \mu_s, \sigma$. 
The analogously-defined gyroaverage 
\begin{align}
  %\ga{a(\R_s)}{\r} = \frac{1}{2\pi}\int_0^{2\pi} \d\varTheta ~  a(\R_s-\rhobs),
	\ga{a(\R_s)}{\r} = \frac{1}{2\pi}\int_0^{2\pi} \d\varTheta ~  a\left( \R_s - \frac{\b\times\v}{\Omega_s} \right),
  %\ga{a(\r,\v)}{\R_s} = \frac{1}{2\pi}\int_0^{2\pi} \d\vartheta ~  a(\r(\R_s,E_s,\mu_s,\vartheta,\sigma),\v(\R_s,E_s,\mu_s,\vartheta,\sigma))
  \label{eq:GyroaverageDefinitionTwo}
\end{align}
is a function of the position space variables $\r,\vperp,\vpara$.

It follows from \eqref{eq:ThetaChainRule} that for an arbitrary function $a(\r)$,
\begin{align}
  \begin{split}
  \ga{\vperpv\cdot\nabla a}{\R_s} 
%%%  = -\Omega_s\ga{(\v\times\b)\cdot \pd{\lp \v\times\b\rp/\Omega_s}{\v} \cdot \nabla a}{\R_s}
%%%  \\
%%%  = \Omega_s\ga{(\v\times\b)\cdot \left.\pd{\r}{\v}\right|_{\R_s} \cdot \nabla a}{\R_s}
%%%  = \Omega_s\ga{(\v\times\b)\cdot \left.\pd{a}{\v}\right|_{\R_s}}{\R_s}
  = -\Omega_s\ga{ \pd{a}{\vartheta}}{\R_s} = 0.
  \end{split}
  \label{eq:GAVPerpDotGradVanishes}
\end{align}

\subsubsection{\fp\ equation in gyrokinetic variables}
The \fp\ equation \eqref{eq:FokkerPlanckEquation} in guiding centre variables is
\begin{align}
  \fd{\ff}{t} = 
  \pd{\ff_s}{t} + \Rd\cdot\pd{\ff_s}{\R_s} + \Ed\pd{\ff_s}{E_s} + \mud\pd{\ff_s}{\mu_s}
  + \td\pd{\ff_s}{\vartheta} 
  = \sum_{s'}C[\ff_s,\ff_{s'}], 
  \label{eq:FokkerPlanckGuidingCentre}
\end{align}
where the dot is the full time derivative along the particle orbit, given by the Vlasov operator
\begin{align}
  \fd{}{t} = \pd{}{t} + \dot{\r}\cdot\nabla + \dot{\v}\cdot\pd{}{\v},
  %\label{}
\end{align}
with $\rd$ and $\vd$ the particle motions
\begin{align}
  \dot{\r} = \v,
  \hspace{1cm}
  \dot{\v} = \frac{q_s}{m_s}\lp -\nabla\varphif - \frac{1}{c}\pd{\fA}{t} + \frac{\v\times\fB}{c} \rp .
  \label{eq:RDotVDot}
\end{align}
We therefore need to evaluate the time derivatives of the gyrokinetic variables \eqref{eq:GyrokineticVariables},
\begin{align}
  \begin{split}
  \label{eq:RDot}
    \Rd &= \dot{\r} + \fd{}{t}\lp \frac{\v\times\b}{\Omega_s}\rp
    \\
    &= \vpara\b + \frac{\v\times((\v\cdot\nabla)\b)}{\Omega_s} 
		\\
		& \hspace{2.5cm}
    + \frac{\b}{\Omega_s}\times\left[ \frac{q_s}{m_s}\nabla\varphif + \frac{q_s}{m_sc}\pd{\fA}{t} - \frac{q_s\v\times\pB}{m_sc} + \v(\v\cdot\nabla\ln B_0)\right] , 
    \end{split}
    \\
  \begin{split}
  \Ed &= m_s\v\cdot\dot{\v} 
      = - q_s \v\cdot\nabla\varphif - \frac{q_s}{c} \v\cdot\pd{\fA}{t} ,
    \end{split}
  \\
  \begin{split}
    \mud &= \frac{m_s}{B_0}\vperpv\cdot\dot{\v}_{\perp} - \frac{\mu_s}{B_0}\dot{B}_0 
    \\
  &= -\frac{q_s}{B_0}\vperpv\cdot\nabla\varphif - \frac{q_s}{cB_0}\vperpv\cdot\pd{\fA}{t} + \frac{q_s}{cB_0}\vperpv\cdot(\v\times\pB) 
   - \mu_s\v\cdot\nabla\ln B_0 ,
  \label{eq:MuDot}
    \end{split}
\end{align}
where we have used \eqref{eq:RDotVDot} for $\dot{\r}$ and $\dot{\v}$.
We have assumed that the magnetic field has spatial dependence but no explicit time dependence, so that
\begin{align}
  \fd{\b}{t} = \left({\v}\cdot\nabla\right)\b,
  \hspace{1cm}
  \fd{\pB}{t} = {\v}\cdot\nabla\pB,
  \hspace{1cm}
  \fd{\Omega_s}{t} = \frac{q_s}{m_sc}\fd{B_0}{t} = \Omega_s\vperpv\cdot\nabla\ln B_0,
\end{align}
where we have also assumed that $B_0$ does not change parallel to the field, $\b\cdot\nabla B_0=0$.
%%%\begin{align}
%%%    \Rd &= \dot{\r} + \fd{}{t}\lp \frac{\v\times\b}{\Omega_s}\rp,
%%%    \hspace{1cm}
%%%  \Ed &= m_s\v\cdot\dot{\v} ,
%%%    \hspace{1cm}
%%%    \mud &= \frac{m_s}{B_0}\vperpv\cdot\dot{\v}_{\perp} - \frac{\mu_s}{B_0}\dot{B}_0 
%%%\end{align}

To find the gyrophase evolution $\dot{\vartheta}$ we take the time derivative of \eqref{eq:vCylindric},
\begin{align}
  \dot{\v}_{\perp} = \frac{\dot{v}_{\perp}}{\vperp}\vperpv + (\b\times\vperpv)\dot{\vartheta}.
  %\label{<++>}
\end{align}
Taking the scalar product of this with $(\b\times\vperpv)$, the first term on the \rhs\ vanishes and we obtain
\begin{align}
  \begin{split}
  \dot{\vartheta} &= \frac{1}{\vperp^2}(\b\times\vperpv)\cdot\dot{\v}_{\perp} 
  \\
  &= \frac{1}{\vperp^2}(\b\times\vperpv)\cdot\left[ \frac{q_s}{m_s}\lp -\nabla_{\perp}\varphif - \frac{1}{c}\pd{\fA_{\perp}}{t} + \frac{(\v\times\fB)_{\perp}}{c} \rp\right] .
\end{split}
  \label{eq:ThetaDot}
\end{align}

We have thus expressed the \fp\ equation in terms of the gyrokinetic variables; we now nondimensionalize and solve order-by-order in the gyrokinetic parameter.

\subsection{Nondimensionalization}
\label{sec:Nondimensionalization}

We now nondimensionalize using scales which respect the $\delta f$ ordering \eqref{eq:GKOrdering}.
We follow the approach used in the \textsc{GS2} family of codes and nondimensionalize \wrt\ a fictitious reference species, denoted with subscript $r$.
The scales are based on those in \citet{HighcockPhD} and are given in \tab s \ref{tab:Normalizing} and \ref{tab:Normalized}.
Thus for our slab system, we obtain the same equations as solved in \agk\ \citep{Numata10}. 

Note that the normalization with respect to a reference species removes constant factors and leads to unexpected quantities.  
For example, the normalized thermal velocity is $\vthNs=\sqrt{\TNs/\mNs}$,
while the unnormalized thermal velocity is the usual $\vths=\sqrt{2T_s/m_s}$. 
Consequently surprising factors of 2 occasionally appear.

\begin{table}[tp]
  \begin{tabular}{lll}
  Gyrokinetic parameter & $\varepsilon = \rho_r/L$ &\\
  Thermal velocity & $\vthr = \sqrt{2T_r/m_r}$ &\\
  Gyrofrequency & $\Omega_r = q_rB_0/m_rc$ & \\
  Gyroradius & $\rho_r = \vthr/\Omega_r$ & \\
  Plasma beta & $\beta_s = 8\pi n_sT_s/B_0^2$ &
  \end{tabular}
  \caption{Normalizing quantities, adapted from \citet{HighcockPhD}. \label{tab:Normalizing}}
\end{table}
\begin{table}[tp]
  \begin{tabular}{lll}
    Equilibrium distribution & $\FNs = F_s/(n_s/\vths^3)$ &\\
    Perturbed distribution & $\fNs =  f_s/(\varepsilon n_s/\vths^3)$ &\\
    Electrostatic potential & $\varphiN = \varphi/(\varepsilon T_r/q_r)$&\\
    Vector potential &  $\AN = \A/(\varepsilon T_r/q_r(\vthr/c))$& \\
    Mean magnetic field & $\b = \bB/B_0$ &\\
    Perturbed magnetic field & $\pBN = \pB/(\varepsilon B_0)$ &\\
    Radial coordinate & $\xN = x/\rho_r$ &\\
    Poloidal coordinate & $\yN = y/\rho_r$ &\\
    Parallel coordinate & $\zN = z/L$ &\\
    Perpendicular gradient & $\nablaNperp = \rho_r\nabla_{\perp}$ & \\ 
    Background gradients & $\nablaNperp = L\nabla_{\perp}$ & \\
    Velocity coordinates & $\vN = \v/\vths$ &\\
    Thermal velocity & $\vthNs = \vths/\vthr=\sqrt{\TNs/\mNs}$ & \\
  Time & $\tN = t/(L/\vthr)$ &\\
  Charge & $\ZNs = q_s/q_r = q_s$ &\\
  Density & $\nNs = n_s/n_r$ &\\
  Mass & $\mNs = m_s/m_r$ & \\
  Temperature & $\TNs = T_s/T_r$ &\\
  Gyrofrequency & $\OmegaNs = \Omega_s/\Omega_r = \ZNs/\mNs$ & \\
  Collision operator & $\CN = C/(\vthr/L)$ &
  \end{tabular}
  \caption{Normalized quantities, adapted from \citet{HighcockPhD}.\label{tab:Normalized}}
\end{table}

In the following subsections we nondimensionalize 
the electromagnetic fields and the 
%\linebreak 
gyrokinetic potential (\S\ref{sec:NormalizationElectromagneticFieldGyrokineticPotential}),
the gyrokinetic variables, their time derivatives and their gyroaverages (\S\ref{sec:NormalizedGyrokineticVariables}),
the background Maxwellian (\S\ref{sec:Maxwellian})
and the Boltzmann response term (\S\ref{sec:BoltzmannResponse}).

\subsubsection{Electromagnetic fields and gyrokinetic potential}
\label{sec:NormalizationElectromagneticFieldGyrokineticPotential}
The normalized electromagnetic fields and gyrokinetic potential are defined by
\begin{align}
    \fB = B_0 \fB_N,
    \hspace{1cm}
    \fE = \pE = \frac{\varepsilon T_r}{q_r\rho_r}\pEN,
    \hspace{1cm}
    \chi_s = \frac{\varepsilon T_r}{q_r} \chi_{Ns} ,
\end{align}
where
\begin{align}
  \label{eq:MagneticFieldNormalized}
    \fB_N &= \b + \frac{\varepsilon}{2}\left( \nabla_{N\perp}\times\pAN \right) + \frac{\varepsilon^2}{2} \left(\nabla_{N\parallel}\times\pAN\right) , 
    \\
  \label{eq:ElectricFieldNormalized}
   \pEN &= -  \nabla_{N\perp} \varphipN - \varepsilon \nabla_{N\parallel}\varphipN - \varepsilon \pd{\pAN}{t_N} ,
    \\
  \label{eq:GyrokineticPotentialNormalized}
  \chi_{Ns} &= \varphipN - \vthNs \vN \cdot \pAN ,
\end{align}
and where the factors of $1/2$ arise due to the choice of the definition of the thermal velocity in the normalization. 

\subsubsection{Gyrokinetic variables}
\label{sec:NormalizedGyrokineticVariables}
The nondimensional gyrokinetic variables are
\begin{align}
%%%  \rNpara &= \RNpara \\
%%%  \rho_r\rNperp &= \rho_r\RNs + \frac{\vths}{\Omega_s} (\b\times\vN)
%%%\id  \rho_r\RNs + \rho_s (\b\times\vN)\\
%%%\rN &= \RNs + \frac{\rho_s}{\rho_r} (\b\times\vN)
%%%= \RNs + \rhoNs\rhoN
\RNs = \rN - \rhoNs\rhoN , \hspace{1cm}
%\rhoN &= \b\times\vN  \\
\ENs = v_N^2, \hspace{1cm} 
\muNs = \vNperp^2 , \hspace{1cm}
\sigma_N = \frac{\vNpara}{|\vNpara|},
\end{align}
and the nondimensional derivatives are 
\begin{align}
  \begin{split}
   \pd{ }{\RNs} = \nabla_N , %\\[1ex]
%%%    \pd{ }{\vNpara} &=   \pd{ }{\VNpara}  \\[1ex]
%%%    \pd{ }{\vNperp} &=   \pd{ }{\VNperp} - \rhoNs \et \cdot\pd{}{\RNs}  \\[1ex]
%%%    \pd{ }{\vartheta} &=   \pd{ }{\Theta} + \rhoNs\VNperpv\cdot\pd{}{\RNs}  \\[1ex]
    \hspace{1cm}
    \pd{ }{\vartheta} =   \pd{ }{\varTheta} - \rhoNs\vNperpv\cdot\nabla_N .
%%%    \pd{ }{\vNperpv} =   \pd{ }{\VNperpv} + \rhoNs \b\times \pd{}{\RNs} .
\end{split}
  %\label{}
\end{align}
%%%Normalized coordinate change now controlled by parameter $\rho$---doing this in dimensional units as is common in literature, more familiar to see $\Omega_s^{-1}$ in that position.  
The nondimensional time derivatives are defined implicitly by
\begin{align}
  \Rd = \vthr\dot{\R}_{Ns}, 
  \hspace{1cm}
  \Ed = \frac{T_s\vthr}{L} \dot{E}_{Ns},
  \hspace{1cm}
  \mud = \frac{T_s\vthr}{B_0L}\dot{\mu}_{Ns},
  \hspace{1cm}
  \td = \frac{\vthr}{L} \dot{\vartheta}_N .
  %\label{}
\end{align}
From these and equations (\ref{eq:RDot}--\ref{eq:MuDot}), \eqref{eq:ThetaDot} we have
\begin{align}
  %%% CHECKED 12/9/2013. 
  %%% replaced rhoNs vthNs with TNs/qNs
  \begin{split}
    \label{eq:RDotNormalized}
    \dot{\R}_{Ns} 
    &= \vthNs\vNpara\b + \varepsilon \frac{\TNs}{\qNs} \vN\times((\vN\cdot\nablaN)\b) 
    \\
%%%    & %\hspace{1cm}
%%%    + \varepsilon \b\times\ls \frac{1}{2\vthNs} \nablaNperp\varphiN 
%%%    - \vN\times\delta\BN
%%%    +  \rhoNs \vN(\vN\cdot\nablaN\ln B_0 ) 
%%%    \rs
%%%    \\
     &+ \varepsilon \b\times\left[ \frac{1}{2} \nablaNperp \chiNs 
    + \frac{1}{2}\lp\vN\cdot\nablaN\rp \pAN
    +  \frac{\TNs}{\qNs} \vN(\vN\cdot\nablaN\ln B_0 ) 
    \right]
    +\O(\varepsilon^2) ,
%%%    \\
%%%    & \hspace{1cm}
%%%    + \b\times\ls  \varepsilon^2 \frac{1}{2\vthNs} \nablaNpara \varphiN 
%%%    + \varepsilon^2 \frac{1}{\vthNs} \pd{\AN}{\tN} 
%%%    \rs
    \end{split}\\
  %%% CHECKED 12/9/2013. Correct as works with zeroth order solution
  \begin{split}
    \dot{E}_{Ns} & = - \frac{\qNs\vthNs}{\TNs} \lp \vN\cdot\nablaNperp\varphipN - \varepsilon \vN\cdot\nablaNpara\varphipN - \varepsilon \vN\cdot\pd{\pAN}{\tN} \rp
    %+\O(\varepsilon^2)   ,
    \end{split}\\
  %%% Changed \v to \vperpv in final term of magnetic moment  16/11/2014. 
  \begin{split}
    \dot{\mu}_{Ns}   &= -\frac{1}{\rhoNs}\vNperpv\cdot\nablaNperp\varphipN - \frac{\varepsilon}{\rhoNs}\vNperpv\cdot\pd{\pAN}{\tN} 
    %%%%%%%%%%%%%%% 
    % Writing in terms of dA,
    %+ 2\vthNs\vNperpv\cdot(\vN\times\delta\BN) 
    % becomes:
    %\\ &
    %\hspace{1.5cm}
    %  + \vthNs\vNperpv\cdot(\vN\times ( \nabla_{N\perp}\times\delta\AN)) 
    %\\ &
    %\hspace{1.5cm}
    %  + \varepsilon\vthNs\vNperpv\cdot(\vN\times ( \nabla_{N\parallel}\times\delta\AN)) 
    % further manipulation gives
  \\ &
  \hspace{1.5cm}
    + \vthNs\vNpara\vNperpv\cdot \nabla_{N\perp}\pANpara
  - \varepsilon\vthNs\vNperpv\cdot[( \vNpara\cdot \nabla_{N} ) \pANperp ] 
    %%%%%%%%%%%%%%% 
  \\ &
  \hspace{1.5cm}
   -\vthNs \mu_{Ns}\vNperpv\cdot\nablaN\ln B_0
    +\O(\varepsilon^2)   ,
    \end{split}
  %\label{}
    \\
    \label{eq:ThetaDotNormalized}
    \dot{\vartheta}_N &= - \frac{1}{\varepsilon}\Omega_{Ns} + \O(1) ,
  %\label{}
\end{align}
where in the magnetic moment we have used \eqref{eq:MagneticFieldNormalized} and written $\pAN = \pANpara\b + \pANperp$.
The gyroaverages of 
%these 
\eqref{eq:RDotNormalized}--\eqref{eq:ThetaDotNormalized}
are
\begin{align}
%%%  \begin{split}
%%%    \ga{ \dot{\R}_{Ns} }{\R_s}
%%%    &= \vNpara\b + \varepsilon  \rhoNs \vN\times((\vN\cdot\nablaN)\b) 
%%%    \\
%%%     &+ \varepsilon \b\times\ls \frac{1}{2\vthNs} \nablaN \chiNs 
%%%    + \lp\vN\cdot\nablaN\rp \AN
%%%    +  \rhoNs \vN(\vN\cdot\nablaN\ln B_0 ) 
%%%    \rs
%%%    \end{split}\\
  \begin{split}
    \ga{ \dot{\R}_{Ns} }{\R_s}
    &= \vthNs\vNpara\b + \varepsilon  \frac{\TNs}{\qNs} \b\times \left[ \vNpara^2 (\b\cdot\nablaN)\b  +\frac{1}{2}\vNperp^2 \nablaN\ln B_0\rs
    \\
    & \hspace{6cm}  
    + \frac{ \varepsilon  }{2} \b\times \nablaNperp \ga{\chiNs}{\R_s} 
    + \O(\varepsilon^2),
    \end{split}\\
  \begin{split}
    \ga{ \dot{E}_{Ns} }{\R_s}  & = 
    - \varepsilon \frac{\qNs\vthNs}{\TNs}\lp \vN\cdot\nablaNpara \ga{\varphipN}{\R_s} + \ga{ \vN\cdot\pd{\pAN}{\tN}}{\R_s}\rp ,
    \end{split}\\
  \begin{split}
    \ga{\dot{\mu}_{Ns}}{\R_s}   &=  - \frac{\varepsilon}{\rhoNs}\ga{\vNperpv\cdot\pd{\pAN}{\tN}}{\R_s} 
    %+ \vthNs\vNpara\vNperpv\cdot \nabla_{N\perp}\delta\ANpara
    - \varepsilon\vthNs\ga{\vNperpv\cdot[( \vNpara\cdot \nabla_{N} ) \pANperp ]}{\R_s}
  \\ &
  \hspace{10.4cm}
   %-\vthNs \mu_{Ns}\vNperpv\cdot\nablaN\ln B_0
    +\O(\varepsilon^2)   ,
    \end{split}
  %\label{}
    \\
    \ga{\dot{\vartheta}_N}{\R_s} &= - \frac{1}{\varepsilon}\Omega_{Ns} + \O(1).
  %\label{}
\end{align}
Notice that many terms vanish due to the property $\ga{\vperpv\cdot\nabla a}{\R_s}=0$, \eqn\ \eqref{eq:GAVPerpDotGradVanishes}.
In particular, taking the gyroaverage reduces the order of the energy and magnetic moment terms from $\ENd\sim\muNd\sim \O(1)$ to $\ga{\ENd}{\R_s}\sim\ga{\muNd}{\R_s}\sim\O(\gkpar)$. 

\subsubsection{Maxwellian}
\label{sec:Maxwellian}
We will also need the Maxwellian
\begin{align}
  f_M(\R_s) = n_s(\R_s)\lp\frac{m_s}{2\pi T_s(\R_s)}\rp^{3/2}\exp\lp -\frac{m_s v^2}{2T_s(\R_s)}\rp,
  %\label{}
\end{align}
and the normalized Maxwellian
\begin{align}
  f_{NM} = \pi^{-3/2}\exp\lp -v_N^2 \rp
  = \pi^{-3/2}\exp\lp -E_N \rp.
  \label{eq:NormalizedMaxwellian}
\end{align}
The density and temperature gradients vanish in this normalization as they are small in $\gkpar$.
Therefore to obtain an expression for the gradient, we expand the density and temperature about their reference values to obtain
\begin{align}
  \pd{f_M}{\R_s} = \left[ \frac{\tpd{\nref}{\R_s}}{\nref} + \lp \frac{m_sv^2}{2\Tref}- \frac{3}{2}\rp \frac{\tpd{\Tref}{\R_s}}{\Tref}\rs f_M(\R_s) .
  \label{eq:MaxwellianGradient}
\end{align}
We show in the next section that the density and temperature gradients must be perpendicular to the magnetic field, 
and without loss of generality we take these to be in the $\X_s$ direction,
\begin{align}
  \Lnr\X_s = -\frac{\tpd{n_s}{\R_s}}{n_s},
  \hspace{1cm}
  \LTr\X_s = -\frac{\tpd{T_s}{\R_s}}{T_s},
  %\label{}
\end{align}
where the gradients of the Maxwellian \eqref{eq:MaxwellianGradient} are assumed constant, and $L_n^{-1}\sim L_T^{-1}\sim L^{-1}$.
Thus the normalized gradient is
\begin{align}
  \pd{f_{MN}}{\R_s} =  -\left[ \omega_n + \lp v_N^2 - \frac{3}{2}\rp \omega_T \rs f_M\X_s ,
  \label{eq:MaxwellianGradientNormalized}
\end{align}
where the normalized density and temperature gradients are
\begin{align}
 \omega_n = L/L_n ,
  \hspace{1cm}
 \omega_T = L/ L_T.
  %\label{}
\end{align}

\subsubsection{Boltzmann response}
\label{sec:BoltzmannResponse}
The Boltzmann response term, $q_sF_{0s}\varphi/T_s$, 
appears as the particular integral in the solution to the \fp\ equation at $\O(1)$, see \eqref{eq:OrderOneSolution}.
To use this solution at the next order we must calculate the full time derivative $\tfd{}{t}(q_sF_{0s}\varphi/T_s)$, which is simplest to do in position space.
However this requires care as gradients and time derivatives of $F_{0s}$ and $\varphi$ vary on different scales.
Consequently terms such as 
\begin{align}
  \begin{split}
  \nabla_{\perp}(\varphip \Fso) &=  \Fso \nabla_{\perp}\varphip + \varphip \nabla_{\perp}\Fso
  %\\ &
  = \varepsilon\frac{n_sT_r}{q_r\vths^3\rho_r}\lp  F_{Ns0} \nabla_{N\perp}\varphipN
  + \varepsilon \varphipN \nabla_{N\perp}F_{Ns0} \rp,
  %\label{}
  \end{split}
\end{align}
are the sums of terms which are at different order in $\gkpar$.
%%%Therefore at any fixed order in $\varepsilon$ it is legitimate to write
%%%\begin{align}
%%%  \nabla_{N\perp}\lp \varphiN F_{Ns0}\rp = 
%%%  F_{Ns0} \nabla_{N\perp}\varphiN .
%%%  %\label{}
%%%\end{align}
Therefore derivatives of products must be expanded before normalization.

With this proviso, we calculate the time derivative of the Boltzmann response
\begin{align}
  \begin{split}
  &\fd{}{t}\lp \frac{q_s\Fso\varphip}{T_s}\rp
	\\
	&
	\hspace{1.5cm} 
	= 
  \lp \pd{}{t} + \v\cdot\nabla + \frac{q_s}{m_s}\lp -\nabla\varphif - \frac{1}{c}\pd{\fA}{t} +\frac{\v\times\fB}{c}\rp\cdot\pd{}{\v}\rp \lp \frac{q_s\Fso\varphip}{T_s}\rp 
  \\
	&
	\hspace{1.5cm} 
  =
  \frac{n_s\vthr}{\vths^3L}\lp \frac{\qNs\vthNs\FNso}{\TNs}\vNperpv\cdot\nablaNperp\varphipN\rp
  \\
  &
	\hspace{2cm} 
	+ \varepsilon \frac{n_s\vthr}{\vths^3L}
  \Bigg(
  \frac{\qNs\FNso}{\TNs}\pd{\varphipN}{\tN}
  +\vthNs\vN\cdot\nablaN\lp\frac{\qNs\FNso}{\TNs}\rp\varphipN
  \\ 
	& 
	\hspace{5cm}
  +\frac{\qNs\vthNs\FNso}{\TNs}\vNpara\b\cdot\nablaN\varphipN
  \\ & \hspace{5cm}
  +\frac{\qNs^2\vthNs}{\TNs^2}\FNso\varphipN\vNperpv\cdot\nablaNperp\varphipN 
  \Bigg)
  + \O(\varepsilon^2) 
  ,
  \label{eq:ddtBoltzmannResponse}
  \end{split}
\end{align}
and its gyroaverage
\begin{align}
  \begin{split}
    \ga{ \fd{}{t}\lp \frac{q_s\Fso\varphip}{T_s}\rp }{\R_s}
  &=
   \varepsilon \frac{n_s\vthr}{\vths^3L}
  \Bigg(
  \frac{\qNs\FNso}{\TNs}\pd{\ga{\varphipN}{\R_s}}{\tN}
  \\ & \hspace{2cm}
  +\vthNs\vNpara\b\cdot\nablaN\lp\frac{\qNs\FNso}{\TNs}\rp\ga{\varphipN}{\R_s}
  \\ & \hspace{2cm}
  +\frac{\qNs\vthNs\FNso}{\TNs}\vNpara\b\cdot\nablaN\ga{\varphipN}{\R_s}
  \Bigg)
  + \O(\varepsilon^2)
  .
  \label{eq:GABoltzmannResponse}
  \end{split}
\end{align}

\section{Derivation of slab gyrokinetic \eqn s}
\label{sec:DerivationOfSlabGyrokineticEquations}

\renewcommand{\AN}{\A}
\renewcommand{\Apara}{A_{\parallel}}
\newcommand{\Apk}{A_{\parallel\k}}
\renewcommand{\Apk}{A_{\parallel\k}}
\renewcommand{\ANpara}{A_{\parallel}}
\renewcommand{\Aperp}{\A_{\perp}}
\renewcommand{\ANperp}{\A_{\perp}}
\renewcommand{\BN}{\B}
\renewcommand{\Bpara}{B_{\parallel}}
\renewcommand{\BNpara}{B_{\parallel}}
\renewcommand{\Bperp}{\B_{\perp}}
\renewcommand{\CN}{C}
\renewcommand{\chiNs}{\chi_{s}}
\newcommand{\dBpk}{B_{\parallel\k}}
\renewcommand{\dfNs}{\delta f_{s}}
\renewcommand{\Epara}{E_{\parallel}}
\renewcommand{\FNs}{F_{s}}
\renewcommand{\FNso}{F_{0s}}
\renewcommand{\fNs}{f_{s}}
\renewcommand{\fs}{f_{s}}
\renewcommand{\k}{\boldsymbol{k}}
\renewcommand{\kNperp}{k_{\perp}}
\renewcommand{\mNe}{m_{e}}
\renewcommand{\mNs}{m_{s}}
\renewcommand{\nablaN}{\nabla}
\renewcommand{\nablaNperp}{\nabla_{\perp}}
\renewcommand{\nablaNpara}{\nabla_{\parallel}}
\renewcommand{\nablaperp}{\nabla_{\perp}}
\renewcommand{\nNs}{n_{s}}
\renewcommand{\OmegaNs}{\Omega_{s}}
\renewcommand{\qNs}{q_{s}}
\renewcommand{\RN}{\R}
\renewcommand{\RNs}{\R_{s}}
\renewcommand{\RNpara}{\R_{\parallel}}
\renewcommand{\RNperp}{\R_{\perp}}
\renewcommand{\RNsperp}{\R_{s\perp}}
\renewcommand{\rN}{\r}
\renewcommand{\rNpara}{\r_{\parallel}}
\renewcommand{\rNperp}{\r_{\perp}}
\renewcommand{\rhob}{\boldsymbol{\rho}}
\renewcommand{\rhobs}{\boldsymbol{\rho}_s}
\renewcommand{\rhoNe}{\rho_{e}}
\renewcommand{\rhoNs}{\rho_{s}}
\renewcommand{\rhoN}{\boldsymbol{\rho}}
\renewcommand{\tN}{t}
\renewcommand{\TNe}{T_{e}}
\renewcommand{\TNs}{T_{s}}
\renewcommand{\vN}{\v}
\renewcommand{\VN}{\V}
\renewcommand{\varphiN}{\varphi}
\renewcommand{\varphidN}{\delta\varphi}
\newcommand{\varphiNk}{\varphi_{\k}}
\renewcommand{\varphiNk}{\varphip_{\k}}
\renewcommand{\vcut}{v_{\textrm{cut}}}
\renewcommand{\vNpara}{v_{\parallel}}
\renewcommand{\vNperp}{v_{\perp}}
\renewcommand{\vNperpv}{\v_{\perp}}
\renewcommand{\VNpara}{V_{\parallel}}
\renewcommand{\VNperp}{V_{\perp}}
\renewcommand{\VNperpv}{\V_{\perp}}
\renewcommand{\vthNe}{v_{\textrm{th}e}}
\renewcommand{\vthNi}{v_{\textrm{th}i}}
\renewcommand{\vthNs}{v_{\textrm{th}s}}
\renewcommand{\xN}{x}
\renewcommand{\y}{\boldsymbol{y}}
\renewcommand{\yN}{y}
\renewcommand{\zN}{z}
\renewcommand{\ZNs}{Z_{s}}
\renewcommand{\ZN}{Z_{s}}

% E
\renewcommand{\pEN}{\pE} % perturbed
% B
\renewcommand{\pBN}{\pB} % perturbed
\newcommand{\pBNpara}{\Bpar} % perturbed
% A
\renewcommand{\pAN}{\pA} % perturbed
\renewcommand{\pANpara}{\pApara} % perturbed
\renewcommand{\pANperp}{\pAperp} % perturbed
% varphi
\renewcommand{\varphipN}{\varphip} % perturbed

We now solve the \fp\ equation \eqref{eq:FokkerPlanckEquation} order-by-order in $\varepsilon$.
From hereon we work in normalized variables, but for presentation drop the subscript $N$.
The \fp\ \eqn\ to $\O(\gkpar^2)$ is 
\begin{align}
  \begin{split}
  %  \varepsilon^2 \pd{\FNs}{\tN} +
  \pd{\fs}{\tN} 
  %\\ &
  &+ \lp \dot{\R}_{0} + \varepsilon \dot{\R}_{1}\rp_{\parallel} \cdot \pd{\ff_s}{\R_s}%\lp \FNs + \fNs\rp 
  %\\ &
  + \lp %\dot{\R}_{0} + 
      \varepsilon \dot{\R}_{1}
      +\varepsilon^2 \dot{\R}_{2}
      \rp_{\perp} \cdot \pd{}{\R_s}\lp \FNs + \frac{1}{\varepsilon} \fNs\rp 
  \\
  &+ \lp \dot{E}_0 +\varepsilon \dot{E}_1 \rp \pd{\ff_s}{E_s}
  %\\ &
  + \lp \dot{\mu}_0 + \varepsilon \dot{\mu}_1\rp \pd{\ff_s}{\mu_s}
 % \\ &
  + \lp \frac{1}{\varepsilon}\dot{\vartheta}_0 + \dot{\vartheta}_1
  + \varepsilon\dot{\vartheta}_2
  \rp \pd{\ff_s}{\vartheta}
	\\  & =  \sum_{s'} C_{ss'}[\ff_s,\ff_{s'}] ,
  \end{split}
  \label{eq:NondimensionalFokkerPlanck}
\end{align}
where $\ffs = \FNs + \fNs$ 
with $\FNs = \Fso + \varepsilon\Fsi+\ldots$ and $\fNs=\varepsilon\fsi+\varepsilon^2\fsii+\ldots$,
the normalized version of the distribution function expansion \eqref{eq:DistFnExpansion}. 
We have also written $\dot{\R}_s=\dot{\R_0}+\gkpar\dot{\R_1}$, where $\dot{\R}_0$ is the zeroth-order contribution to $\dot{R}_s$ and so on. 
Note that $\dot{\R}_0$ has no perpendicular component, $(\dot{\R}_0)_{\perp}=0$.

%\subsection{Solutions at each order}
At $\O(\gkpar^{-1})$ we find that the background distribution function $F_{0s}$ is gyrophase independent.
At $\O(1)$ we find that $F_{0s}$ is Maxwellian, and that the perturbation $f_{1s}$ can be decomposed into the Boltzmann response and the gyrophase-independent distribution function for guiding centres, $h_s$.
At $\O(\gkpar)$ we derive the gyrokinetic equation for $h_s$,
and the neoclassical drift-kinetic equation for $F_{1s}$.
We finish our derivation at $\O(\gkpar)$, but continuing to $\O(\gkpar^2)$ we would obtain the transport equations \citep[see \eg][]{Abel13}.
\paragraph{Order $\O(1/\varepsilon)$: gyrotropy of $\Fso$}
At lowest order we find
\begin{align}
  \begin{split}
  -\OmegaNs  \pd{\Fso}{\vartheta}   = 0,
  %\label{}
  \end{split}
\end{align}
so that $\Fso(\R_s,E_s,\mu_s,t)$ is gyrotropic, \ie\ independent of gyrophase.

\paragraph{Order $\O(1)$}
At next order we find
\begin{align}
  \label{eq:OrderOne}
  \begin{split}
  &  \dot{\R}_{0} \cdot \pd{\Fso}{\R_s}
  %\\ &
  %+ \dot{\R}_{0\perp} \cdot \pd{\delta f_s}{\R_s}
  + \dot{E}_0 \pd{\Fso}{E_s}
  %\\ &
  + \dot{\mu}_0\pd{\Fso}{\mu_s}
 % \\ &
  + \dot{\vartheta}_0 \pd{}{\vartheta}\lp \Fsi + \fsi\rp
  %\\  & 
	=  \sum_{s'} C_{ss'}[\Fso,F_{0s'}] .
  \end{split}
\end{align}
To solve for $\Fso$ we remove the perturbations $\Fsi$ and $\fsi$ by gyroaveraging.
In addition the energy and magnetic moment terms vanish as $\Fso$ (and therefore $\tpd{\Fso}{E_s}$ and $\tpd{\Fso}{\mu_s}$) are gyrotropic, and $\ga{\dot{E}_0}{\R_s}=\ga{\dot{\mu}_0}{\R_s}=0$. 
We treat the remaining terms
\begin{align}
  \begin{split}
  & \vthNs \vpara\b \cdot \pd{\Fso}{\R_s}
= \ga{ \sum_{s'} C_{ss'}[\Fso,F_{0s'}] }{\R_s},
  \label{eq:FirstOrderGyroaveraged}
  \end{split}
\end{align}
as in the proof of Boltzmann's $H$-theorem: 
we multiply by $(1+\ln \Fso)$, integrate over all velocities and take the perpendicular average \eqref{eq:PerpendicularAverage}.
The \lhs\ becomes
\begin{align}
  \begin{split}
  \ga{\int\d^3\v~ (1+\ln \Fso)\vthNs\vpara\b \cdot \pd{\Fso}{\R_s} }{\perp}  
  =
  \ga{\int\d^3\v~ \vpara\b \cdot \pd{}{\R_s}\lp \Fso\ln\Fso\rp }{\perp}  
  \\
  =
	\int_{\ell^2} \d^2\r{\int\d^3\v~ \vpara\b \cdot \pd{}{\R_s}\lp \Fso\ln\Fso\rp }
  =
	\int_{\ell^2} \d^2\r{\int\d^3\v~ \vpara\b \cdot \nabla \lp \Fso\ln\Fso\rp }
  \\
  =
	\int_{\ell^2} \d^2\r{\int\d^3\v~ \nabla\cdot \left[ \vthNs\vpara\b \lp \Fso\ln\Fso\rp \right] }
  = 0 ,
  \end{split}
  %\label{}
\end{align}
where we have used that $\nabla=\tpd{}{\R_s}$ in normalized variables.
The final expression vanishes as the magnetic field is orthogonal to the plane of integration.

The \rhs\ of \eqref{eq:FirstOrderGyroaveraged} becomes 
\begin{align}
	\ga{ \int \d^3\v~ \ga{ \sum_{s'} \ln \Fso C_{ss'}[\Fso,F_{0s'}] }{\R_s} }{\perp} = 0,
  \label{eq:SolveByMaxwellian}
\end{align}
where we have used the fact that the collision operator conserves particle number, 
\linebreak
\ie,\ $\int\d^3\v~C_{ss'}[\Fso,F_{0s'}]=0$.
By Boltzmann's $H$-theorem, \eqref{eq:SolveByMaxwellian} is solved by a local \linebreak Maxwellian \citep{HelanderSigmar02}. 
Substituting this into \eqref{eq:FirstOrderGyroaveraged} and noting that the equation must hold for all $\vpara$, we find that the Maxwellian must have no parallel gradients and no bulk flow \citep{Abel13}, \ie\ it has the form of the Maxwellian introduced in \sec\ref{sec:Maxwellian}.
Substituting \eqref{eq:NormalizedMaxwellian} and \eqref{eq:MaxwellianGradientNormalized} into \eqref{eq:OrderOne} we find
\begin{align}
  %\label{eq:OrderOne}
  \begin{split}
    - \dot{E}_0F_{0s}
  + \dot{\vartheta}_0 \pd{}{\vartheta}\lp F_{1s} + f_{1s}\rp
 =   \frac{\qNs\vthNs}{\TNs} F_{0s}  \vN\cdot\nablaNperp\varphip
  - \Omega_s \pd{}{\vartheta}\lp F_{1s} + f_{1s}\rp
 = 0.
  \end{split}
\end{align}
Using the turbulent average to separate this into mean and fluctuating parts gives
\begin{align}
  %\label{eq:OrderOne}
  \begin{split}
    \Omega_s\pd{F_{1s}}{\vartheta} = 0,
  \end{split}
  \\
  %\label{eq:OrderOne}
  \begin{split}
     \Omega_s \pd{f_{1s}}{\vartheta} 
 %= - \dot{E}_0F_{0s}
 =   \frac{\qNs\vthNs}{\TNs} F_{0s}  \vN\cdot\nablaNperp\varphipN 
 =  - \frac{\qNs\OmegaNs}{\TNs} F_{0s}  \pd{\varphipN}{\vartheta}.
  \end{split}
	\label{eq:df1dtheta}
\end{align}
For the last equality in \eqref{eq:df1dtheta}, we have used the chain rule \eqref{eq:ThetaChainRule}, and the fact that the electrostatic potential $\varphipN(\r)$ is gyrophase independent in position space, $\tpd{\varphi}{\varTheta}=0$, but not in guiding centre space $\tpd{\varphi}{\vartheta}\neq0$.
Integrating \wrt\ $\vartheta$ we find that $\Fsi$ is gyrotropic, and that
\begin{align}
  \label{eq:OrderOneSolution}
  \begin{split}
     f_{1s}
 = h_s - \frac{\qNs}{\TNs} F_{0s} \varphipN  
  \end{split}
\end{align}
where $h_s(\R_s,E_s,\mu_s)$ is the $\vartheta$-independent complementary function.
Thus $\fsi$ is composed of a Boltzmann response $-q_s\Fso\varphipN/\TNs$, and the complementary function $h_s$ which we interpret as the distribution function for guiding centres. 
We find an evolution equation for $h_s$ at next order in $\gkpar$.

\paragraph{Order $\O(\varepsilon)$}
The \fp\ equation at $\O(\varepsilon)$ is
\begin{align}
  \begin{split}
     \pd{ h_{s}}{\tN} 
  %\\ &
   & + \dot{\R}_{0} \cdot \pd{}{\R_s}\lp F_{1s} + h_{s} \rp 
    + \dot{\R}_1\cdot\pd{}{\R_s}\lp F_{0s} + h_{s}\rp
  %\\ &
  \\
  &+ \dot{E}_1  \pd{F_{0s}}{E_s}
  + \lp \dot{E}_0 \pd{}{E_s} + \dot{\mu}_0 \pd{}{\mu_s}\rp \lp F_{1s} + h_{s}\rp
 % \\ &
  + \dot{\vartheta}_0  \pd{}{\vartheta}\lp F_{2s} + f_{2s}\rp
  \\ &
  + \fd{}{t}\lp -\frac{\qNs\Fso}{\TNs}\varphipN\rp
%%%  \\  & 
  =  \sum_{s'} \lp C_{ss'}[F_{0s},F_{1s'}+h_{s'}] + C_{ss'}[F_{1s}+h_{s},F_{0s'}]\rp 
  ,
  \end{split}
\end{align}
where we have separated $\fsi$ into $h_s$ and the Boltzmann response.
Notice that the Boltzmann response, being Maxwellian, vanishes from the collision operator.
For convenience, we evaluate the full time derivative of the Boltzmann response in position space $(\r,\v)$, as given by \eqref{eq:ddtBoltzmannResponse}.
As before we remove higher-order perturbations by gyroaveraging. 
Using the gyrotropy of $F_{0s}$, $F_{1s}$ and $h_s$, we find
\begin{align}
  \begin{split}
   &  \pd{ h_{s}}{\tN} 
      + \vthNs\vpara\b \cdot \pd{}{\R_s}\lp F_{1s} + h_{s} \rp 
     + (\vD +\vChi )     \cdot\pd{}{\R_s}\lp F_{0s} + h_{s}\rp
  %\\ &
     \\ & \hspace{0.5cm} = 
  \frac{\qNs\FNso}{\TNs}\pd{\ga{\chiNs}{\R_s}}{\tN}
  +\ga{ \sum_{s'} \lp C_{ss'}[F_{0s},F_{1s'}+h_{s'}] + C_{ss'}[F_{1s}+h_{s},F_{0s'}]\rp }{\R_s} ,
  \end{split}
  \label{eq:AlmostGKE}
\end{align}
where 
\begin{align}
  \vChi =   \frac{1}{2} \b\times \nablaN \ga{\chiNs}{\R_s}       ,
  %\label{}
\end{align}
is the drift velocity due to the fluctuating gyrokinetic potential \eqref{eq:GyrokineticPotentialDefinition}, and
\begin{align}
  \vD =  
 %\rhoNs\vthNs
 \frac{\TNs}{\qNs}\b\times \left[ \vNpara^2 (\b\cdot\nablaN)\b  +\frac{1}{2}\vNperp^2 \nablaN\ln B_0\rs     ,
  %\label{}
\end{align}
is the guiding-centre drift velocity with terms corresponding to the curvature drift and $\nabla B$ drift respectively.
We have also used the gyroaverage of the Boltzmann response time derivative \eqref{eq:GABoltzmannResponse}.

We separate \eqref{eq:AlmostGKE} into mean and fluctuating parts using the turbulent average.
This gives an equation for the neoclassical distribution function $\Fsi$,
\begin{align}
  \begin{split}
  &   \vthNs \vpara\b \cdot \pd{F_{1s}}{\R_s}
     %+ \lp  \rhoNs \b\times \left[ \vNpara^2 (\b\cdot\nablaN)\b  +\frac{1}{2}\vNperp^2 \nablaN\ln B_0\rs  \rp
       + \vD
     \cdot\pd{F_{0s}}{\R_s}
     %\\ & \hspace{5cm}
  = 
  \ga{ \sum_{s'} \lp C_{ss'}[F_{0s},F_{1s'}] + C_{ss'}[F_{1s},F_{0s'}]\rp }{\R_s},
  \end{split}
\end{align}
and the gyrokinetic equation for the guiding centre distribution $h_s$,
\begin{align}
  \begin{split}
     \pd{ h_{s}}{\tN} 
     & 
   + \lp \vthNs\vpara\b+ \vD +\vChi\rp  \cdot \pd{h_s}{\R_s}
     + \vChi     \cdot\pd{F_{0s}}{\R_s}
  \\  & \hspace{2cm}
  = 
  \frac{\qNs\FNso}{\TNs}\pd{\ga{\chiNs}{\R_s}}{\tN}
  +\ga{ \sum_{s'} \lp C_{ss'}[F_{0s},h_{s'}] + C_{ss'}[h_{s},F_{0s'}]\rp }{\R_s}.
  \end{split}
\end{align}

\subsection{Maxwell's \eqn s}
It remains to express the electromagnetic field in terms of the guiding centre distribution $h_s$.
The fields enter the gyrokinetic equation only through the gyroaveraged gyrokinetic potential $\ga{\chiNs}{\R_s}=\ga{\varphipN-\vths\v\cdot\pAN}{\R_s}$, 
so we relate the potentials $\varphipN$ and $\pAN$ 
to integrals of $h_s$ via Gauss' and \ampere 's laws.

The electromagnetic field is defined by four scalar fields $(\varphipN,A_x,A_y,\Apar)$. 
However, imposing the Coulomb gauge $\nabla\cdot\A=0$, we may describe the electromagnetic field completely using three scalar fields $\varphi$, $\Apar$ and $\Bpar$ as follows.
To leading order, the normalized fields \eqref{eq:MagneticFieldNormalized}, \eqref{eq:ElectricFieldNormalized} are
\begin{align}
  \label{eq:MagneticFieldNormalizedStep}
   \pBN &= \frac{1}{2}\left( \nabla_{\perp}\times\pAN \right) ,
    \\
%  \label{eq:ElectricFieldNormalized}
   \pEN &= -  \nabla_{\perp} \varphipN .
\end{align}
Writing $\pAN=\pANpara\b + \pANperp$, \eqref{eq:MagneticFieldNormalizedStep} becomes
\begin{align}
  \label{eq:PotentialFormOfDeltaB}
  \pBN &= \frac{1}{2}\lp \nabla_{\perp}\pANpara \times \b \rp + \frac{1}{2} \lp \nabla_{\perp}\times \pANperp \rp
= \frac{1}{2}\lp \nabla_{\perp}\pANpara \times \b \rp + \pBNpara\b ,
\end{align}
where $\Bpar=\b\cdot(\nabla_{\perp}\times\pANperp)/2$.
Further, when written in Fourier space, the gyroaveraged gyrokinetic potential $\ga{\chi_s}{\R_s}$ is also a function of only the three scalars $\varphipN$, $\Apar$ and $\Bpar$, as we show in \sec\ref{sec:FourierSpaceRepresentationGKMSystem}.
Thus we close the system by relating $\varphipN$, $\Apar$ and $\Bpar$  
to integrals of $h_s$ via quasineutrality and \ampere 's law.

The scalars $\varphipN$, $\Apar$ and $\Bpar$ are functions of position space $\r$, but Maxwell's equations relate these to velocity space integrals of distribution functions which are functions of \gcs.
%%%The velocity space integrals are therefore evaluated at fixed $\r$,
%%%\begin{align}
%%%  %\int \d^3\v ~ a(\R_s,E_s,\mu_s,\vartheta,\sigma)
%%%  \int \d^3\v ~ a(\R_s,\V_s)
%%%  &\id
%%%  %\int \d^3\v ~ a(\R_s(\r,\v),E_s(\r,\v),\mu_s(\r,\v),\vartheta(\r,\v),\sigma(\r,\v)).
%%%  \int \d^3\v ~ a(\R_s(\r,\v),\V_s(\r,\v)).
%%%\end{align}
%%%The gyroangle integration in this is related to the gyroaverage at fixed $\r$,
%%%\begin{align}
%%%  %\ga{a(\R_s,E_s,\mu_s,\vartheta,\sigma)}{\r}
%%%  \ga{a(\R_s,\V_s)}{\r}
%%%  = 
%%%  %\frac{1}{2\pi}\int_0^{2\pi} \d\vartheta ~ a(\R_s(\r,\v),E_s(\r,\v),\mu_s(\r,\v),\vartheta(\r,\v),\sigma(\r,\v)),
%%%  \frac{1}{2\pi}\int_0^{2\pi} \d\vartheta ~ a(\R_s(\r,\v),\V_s(\r,\v)),
%%%\end{align}
%%%and consequently a gyroaverage of the distribution appears:
%%%\begin{align}
%%%  \int \d^3\v ~ \ga{a(\R_s(\r,\v),E_s(\r,\v),\mu_s(\r,\v),\vartheta(\r,\v),\sigma(\r,\v))}{\r}
%%%  %\label{}
%%%\end{align}
The velocity space integrals are therefore evaluated at fixed $\r$ 
and consequently a gyroaverage of the distribution function appears
\begin{align}
  \begin{split}
%%%  \int \d^3\v ~ a(\R_s,\V_s)
%%%  &\id
  \int \d^3\v ~ a(\R_s(\r,\v),\V_s(\r,\v))
    =
  2\pi\int \d\vpara\d\vperp~\vperp  \ga{a(\R_s,\V_s)}{\r}
  =
  \int \d^3\v ~ \ga{a(\R_s,\V_s)}{\r}.
%%%  = 
%%%  \int \d^3\v ~ a\ga{(\R_s,\V_s)}{\r}
  \end{split}
\end{align}

\subsubsection{Gauss' Law and quasineutrality}\label{sec:GaussLawQuasineutrality}
Gauss' law in normalized units is
\begin{align}
  - \gkpar\lp \lambda_{Dr}/\rho_r\rp^2 \nablaNperp^2\varphipN + \O\lp\gkpar^2\lambda_{Dr}^2/\rho_r^2 \rp
  = \sum_s \qNs\nNs\int \d^3\v ~ \ga{\ffs}{\r} .
  %\label{}
\end{align}
At leading order only the \rhs\ remains, so that
using the leading order distribution function $\FNso=\pi^{-3/2}e^{-v^2}$,
we see that
the plasma is neutral overall,
\begin{align}
  \sum_s \qNs\nNs = 0 .
  %\label{}
\end{align}
Neglecting Debye length effects $\lambda_{Dr}\ll\rho_r$, %\eqref{eq:Debye}, 
the \lhs\ also vanishes at $\O(\gkpar)$.
Further, separating the equation into mean and fluctuating parts using the turbulent average, 
we obtain the equation for the fluctuating part at $\O(\gkpar)$ as
\begin{align}
   \sum_s \qNs\nNs 
   \int\d^3\v \ga{ -\frac{\qNs\varphipN}{\TNs}\FNso + h_s }{\r}  = 0 .
  %\label{}
\end{align}
Integrating the Boltzmann response term and noting that both $\varphipN(\r)$ and $\FNso=\pi^{-3/2}e^{-v^2}$ are independent of $\R_s$,
we obtain the quasineutrality condition
\begin{align}
\sum_s \frac{\qNs^2\nNs}{\TNs}\varphipN
  = \sum_s \qNs\nNs   \int \d^3\vN ~ \ga{h_s}{\r} .
  %\label{}
\end{align}

%%%\subsubsection{Faraday's law}
%%%\begin{align}
%%%  c\nabla\times\E = -\pd{\delta\B}{t}
%%%  %\label{}
%%%\end{align}
%%%True using the potential form.
%%%Writing
%%%\begin{align}
%%%  \A = \Apara\b + \Aperp = \Apara\b + \nabla\lambda\times\b
%%% % \label{}
%%%\end{align}
%%%then $\B$ is 
%%%\begin{align}
%%%  \B &= \nabla\Apara\times\b - \nabla^2\lambda\b = \nabla\Apara\times\b + \delta\Bpara\b
%%%  \label{eq:PotentialFormOfB}
%%%  \\
%%%  \Bperp &= \nabla\Apara\times\b
%%%  \\
%%%  \delta\Bpara\b &= \nabla\times\Aperp
%%%  \label{eq:Bperp}
%%%\end{align}
%%%so the fields are determined by $\varphi$, $\Apara$ and $\delta\Bpara$.
%%%\cite{Howes06} suggests that the above calculation is valid up to $\O(\varepsilon^2)$, but I think this part of the calculation is exact.

\subsubsection{Amp\`ere--Maxwell law}
The \ampere--Maxwell law is 
\begin{align}
  %\nabla\times\fB = \beta_r \sum_s \qNs\nNs\vthNs \fcurrent - \frac{1}{2}\gkpar^2\lp\frac{\vthr^2}{c^2}\rp\pd{\nablaNperp\varphiN}{\tN}
  \nabla\times\fB = \beta_r \sum_s \qNs\nNs\vthNs \fcurrent + \O\lp \gkpar^2\frac{\vthr^2}{c^2}\rp .
  \label{eq:AmpereMaxwellLaw}
\end{align}
Using the definition of the current \eqref{eq:FullCurrent}, this becomes
\begin{align}
%%%  \nabla\times\fB = \beta_r \sum_s \qNs\nNs\vthNs \int\d^3\v~\vN \FNso +\gkpar \beta_r \sum_s \qNs\nNs\vthNs \int\d^3\v ~ \vN ( \Fsi + \fsi )\\
%%%  \nabla\times\fB = \beta_r \sum_s \qNs\nNs\vthNs \int\d^3\v~\vN \FNso +\gkpar \beta_r \sum_s \qNs\nNs\vthNs \int\d^3\v ~ \vN ( \Fsi + h_s - \frac{\qNs}{\TNs}\Fso\varphiN ) \\
  \nabla\times\fB = \gkpar \beta_r \sum_s \qNs\nNs\vthNs \int\d^3\v ~ \ga{\vN ( \Fsi + h_s  )}{\r},
  \label{eq:AmpereMaxwellLawTwo}
\end{align}
correct to $\O(\gkpar)$.
Integrals of the Maxwellian parts of the distribution function have vanished as these are odd in $\v$.
Note that as the plasma is nonrelativistic $\vthr^2/c^2\ll1$, the displacement current is negligible even for very low beta $\beta_r\sim\gkpar^2$.

Separating $\fB$ using the turbulent average gives the fluctuating part
%%%\begin{align}
%%%  \nabla\times\delta\B = \beta_r \sum_s \qNs\nNs\vthNs \int\d^3\v ~ \vN  h_s  
%%%  %\label{}
%%%\end{align}
\begin{align}
  \nabla\times\pBN = \beta_r \sum_s \qNs\nNs\vthNs \int\d^3\v ~ \ga{\vN  h_s}{\r}  
  ,
  \label{eq:AmpereMaxwellLawTurbulent}
\end{align}
where the integral is evaluated at fixed $\r$.
Using \eqref{eq:PotentialFormOfDeltaB} 
then gives
\begin{align}
  -\frac{1}{2}\nabla^2_{\perp}\Apar \b + \nabla\pBNpara\times\b  = \beta_r \sum_s \qNs\nNs\vthNs \int\d^3\v ~ \ga{\vN  h_s}{\r}  
  ,
  \label{eq:PenultimateAmplereMaxwell}
\end{align}
so that the parallel component is
\begin{align}
  -\nabla^2_{\perp}\Apar  = 2\beta_r \sum_s \qNs\nNs\vthNs \int\d^3\v ~ \vNpara\ga{  h_s}{\r}  .
\end{align}
Taking the perpendicular curl of \eqref{eq:PenultimateAmplereMaxwell} gives the perpendicular component
\begin{align}
  -\nabla^2_{\perp}\Bpar  = \b\cdot \left[\nablaNperp \times \beta_r \sum_s \qNs\nNs\vthNs \int\d^3\v ~ \ga{\vNperpv  h_s}{\r}\rs.
  %\label{}
\end{align}

\subsection{Summary}
\label{sec:GyrokineticSummary}
The \gkm\ system consists of the gyrokinetic \eqn
\begin{align}
  \begin{split}
     \pd{ h_{s}}{\tN} 
     & 
   + \lp \vthNs \vpara\b+ \vD +\vChi\rp  \cdot \pd{h_s}{\R_s}
     + \vChi     \cdot\pd{F_{0s}}{\R_s}
  \\  & \hspace{1cm}
  = 
  \frac{\qNs\FNso}{\TNs}\pd{\ga{\chiNs}{\R_s}}{\tN}
  +\ga{ \sum_{s'} \lp C_{ss'}[F_{0s},h_{s'}] + C_{ss'}[h_{s},F_{0s'}]\rp }{\R_s},
  \end{split}
  \label{eq:GyrokineticEquationSummary}
\end{align}
with the gyrokinetic potential
\begin{align}
  \chi_s = \varphipN - \vthNs \v\cdot \pAN ,
  \label{eq:GyrokineticPotentialSummary}
\end{align}
and the drift velocities
\begin{align}
  \vChi =   \frac{1}{2} \b\times \nablaN \ga{\chiNs}{\R_s}       ,
  %\label{}
  \\
  \vD =  
 %\rhoNs\vthNs 
 \frac{\TNs}{\qNs}\b\times \left[ \vNpara^2 (\b\cdot\nablaN)\b  +\frac{1}{2}\vNperp^2 \nablaN\ln B_0\rs     ,
  %\label{}
\end{align}
coupled to the quasineutrality condition,
and the parallel and perpendicular components of \ampere's law
\begin{subequations}
\label{eq:MaxwellSummary}
\begin{align}
\sum_s \frac{\qNs^2\nNs}{\TNs}\varphipN
  = \sum_s \qNs\nNs   \int \d^3\vN ~ \ga{h_s}{\r}  ,
  \label{eq:QuasineutralitySummary}
  \\
  %\label{}
%%%\end{align}
%%%\begin{align}
  -\nabla^2_{\perp}\Apar  = 2\beta_r \sum_s \qNs\nNs\vthNs \int\d^3\v ~ \vNpara\ga{  h_s}{\r},  \\
  -\nabla^2_{\perp}\Bpar  = \b\cdot \left[\nablaNperp \times \beta_r \sum_s \qNs\nNs\vthNs \int\d^3\v ~ \ga{\vNperpv  h_s}{\r}\rs .
  %\label{}
\end{align}
  \end{subequations}

%%%\hrule
%%%
%%%The gyrokinetic \eqn\ is,
%%%\begin{align}
%%%  \pd{h_s}{t} + \vpara\b\cdot\pd{h_s}{\R_s} + \v_D\cdot\lp\pd{h_s}{\R_s}+\pd{f_{0s}}{\R_s}\rp = \frac{q_s}{T_{0s}}\pd{\ga{\chi}{\R_s}}{t}f_{0s} + C[h_s],
%%%\end{align}
%%%where $\chi$ the gyrokinetic potential is,
%%%\begin{align}
%%%  \chi = \varphi - \v\cdot\A.
%%%\end{align}
%%%The drift velocity $\v_D$ is,
%%%\begin{align}
%%%  \v_D = -\Yhat \frac{\vperp^2}{2\Omega_s}L^{-1}_{B_0}
%%%-\Yhat \frac{\vpara^2}{\Omega_s}\kappa
%%%- \pd{\ga{\chi}{\R_s}}{\R_s}\times\frac{\b}{B_0}.
%%%\end{align}
%%%Maxwell's \eqn s are,
%%%\begin{align}
%%%  \sum_s\ls -\frac{q_s^2n_{0s}}{T_{0s}}\varphi + q_s\int\d^3\v \ga{h}{\r}\rs = 0, \\
%%%  -\nabla_\perp^2\Apara = \mu_0\sum_s q_s\int \d^3\v\ \vpara\ga{h}{\r},
%%%\\
%%% B_0 \nabla_\perp \delta\Bpara =  -\mu_0\sum_s q_sm_s\nabla\cdot\int\d^3\v\ \ga{\vperpv\vperpv h}{\r}.
%%%\end{align}

\section{Free energy}
\label{sec:FreeEnergy}

The collisionless \gkm\ system with no background gradients conserves free energy, 
\begin{align}
  W = \int \d^3\r\left[ \sum_s \lp \int \d^3\v \frac{n_{0s}T_{0s}\ga{h_s^2}{\r}}{2 F_0} - \frac{n_{0s}\qNs^2\varphip^2}{2T_{0s}}\rp + \frac{|\pBN|^2}{\beta_r}\rs ,
  \label{eq:FreeEnergySchekochihin}
\end{align}
a quadratic invariant which we show in \chp~\ref{sec:Hypercollisions} is related to Boltzmann entropy.
To write a global budget equation for $W$
we multiply the gyrokinetic equation \eqref{eq:GyrokineticEquationSummary} by $n_sT_{0s}h_s/F_0$, sum over species, and integrate over all velocities and guiding centres,
\begin{align}
  \label{eq:FreeEnergyEquationFirstManipulation}
  \begin{split}
   & \fd{}{\tN} \left( \sum_s\int\d^3\R_s\int\d^3\v \left( \frac{n_sT_{0s}h_s^2}{2F_{0s}} \right) \right)
  - \sum_s\int\d^3\R_s\int\d^3\v  \qNs n_s h_s \pd{\ga{\chiNs}{\R_s}}{\tN}
   \\
     & 
   + \sum_s\int\d^3\R_s\int\d^3\v  \lp \vthNs \vpara\b+ \vD +\vChi\rp  \cdot \pd{}{\R_s} \left( \frac{n_sT_{0s}h_s^2}{2F_{0s}} \right)
   \\ & 
     + \sum_s\int\d^3\R_s\int\d^3\v \frac{n_sT_{0s}h_s}{F_{0s}}  \vChi     \cdot\pd{F_{0s}}{\R_s}
  \\  & % \hspace{1cm}
  = 
    \sum_s\int\d^3\R_s\int\d^3\v \frac{n_sT_{0s}h_s}{F_{0s}} \ga{ \sum_{s'} \lp C_{ss'}[F_{0s},h_{s'}] + C_{ss'}[h_{s},F_{0s'}]\rp }{\R_s}.
  \end{split}
\end{align}
The integrand on the second line is a divergence because the velocities $\vpara\b$, $\vD$ and $\vChi$ are divergence-free.
Therefore the second line vanishes.
The third line is the source of free energy due to background temperature and density gradients, 
\begin{align}
  \label{eq:FreeEnergyInjection}
  \begin{split}
    {\cal T} =  \sum_s\int\d^3\R_s\int\d^3\v \frac{n_sT_{0s}h_s}{F_{0s}}  \vChi     \cdot\pd{F_{0s}}{\R_s}.
  \end{split}
\end{align}
The final line is a collisional sink of free energy,
\begin{align}
  \label{eq:FreeEnergyCollisionalSink}
  \begin{split}
    {\cal C} = \sum_s\int\d^3\R_s\int\d^3\v \frac{n_sT_{0s}h_s}{F_{0s}} \ga{ \sum_{s'} \lp C_{ss'}[F_{0s},h_{s'}] + C_{ss'}[h_{s},F_{0s'}]\rp }{\R_s},
  \end{split}
\end{align}
which is non-positive, owing to Boltzmann's $H$-theorem \citep{Abel13}.

The first line of \eqref{eq:FreeEnergyEquationFirstManipulation} may be written as a single time derivative:
the second term is
\begin{align}
  \begin{split}
    &
\sum_s\int\d^3\R_s\int\d^3\v ~ \qNs n_s h_s \pd{\ga{\chiNs}{\R_s}}{\tN}
  % = \sum_s\int\d^3\r\int\d^3\v~ \qNs\nNs \ga{h_s}{\r}\pd{\chiNs}{\tN}
  %\\ &
%   \\ & \hspace{1.0cm} 
   = \sum_s\int\d^3\r\int\d^3\v~ \qNs\nNs \ga{h_s}{\r}\pd{}{\tN}(\varphipN - \vthNs\v\cdot\pAN)
%%%   \\ & \hspace{0.5cm} 
%%%  = \sum_s\int\d^3\r\int\d^3\v~ \qNs\nNs \ga{h_s}{\r}\pd{\varphipN}{\tN}
%%%  - \sum_s\int\d^3\r\int\d^3\v~ \qNs\nNs\vthNs \ga{\v h}{\r}\cdot\pd{\pAN}{\tN}
   \\ & \hspace{1.0cm} 
   = \fd{}{\tN}\left( \sum_s\int\d^3\r~ \frac{\qNs^2\nNs\varphipN^2}{2\TNs}\right)
  - \frac{1}{\beta_r}\int\d^3\r~ \lp\nabla\times\pBN\rp\cdot\pd{\pAN}{\tN} 
   \\ & \hspace{1.0cm} 
   = \fd{}{\tN}\left( \sum_s\int\d^3\r~ \frac{\qNs^2\nNs\varphipN^2}{2\TNs}
  - \frac{1}{\beta_r}\int\d^3\r~ |\pBN|^2 \right),
  \end{split}
\end{align}
where the first equality uses the identity
$\int \d^3\R_s \ga{h(\R_s)\chi(\r)}{\R_s}= \int \d^3\r \ga{h(\R_s)\chi(\r)}{\r}$,
and the definition $\chi_s = \varphipN - \vthNs \v\cdot \pAN$, % \eqref{eq:GyrokineticPotentialSummary},
the 
second
%third 
equality uses Maxwell's equations \eqref{eq:AmpereMaxwellLawTurbulent} and \eqref{eq:QuasineutralitySummary}, 
and the final equality uses the identity
\begin{align}
  \begin{split}
    \lp\nabla\times\pBN\rp\cdot\pd{\pAN}{\tN}
%%% This is the proof in index notation:
%%%    = 
%%%    \left(\varepsilon_{ijk} \partial_jB_k\right)\lp \tpds{A_i}{\tN}\rp
%%%    = 
%%%     \partial_j \left(\varepsilon_{ijk}B_k \tpds{A_i}{\tN}\right)
%%%    -  B_k \varepsilon_{ijk}\partial_j \left(\tpds{A_i}{\tN}\right)
    & = 
    \nabla\cdot \left( \pBN \times \pd{\pAN}{\tN}\right)
    +  \pBN\cdot\left( \nabla \times \pd{\pAN}{\tN}\right)
    \\ & = 
    \nabla\cdot \left( \pBN \times \pd{\pAN}{\tN}\right)
    + \pd{|\pBN|^2}{\tN}.
  \end{split}
\end{align}
Thus \eqref{eq:FreeEnergyEquationFirstManipulation} becomes the free energy balance equation 
\begin{align}
  \label{eq:FreeEnergyEquation}
  \begin{split}
   & \fd{W}{\tN} 
   + {\cal T}
  = 
  {\cal C},
  \end{split}
\end{align}
showing that free energy \eqref{eq:FreeEnergySchekochihin} is conserved in the absence of collisions and driving gradients.

%%% Do you want to note that VP has an extra term - as does the parallel nonlinearity?

\subsection{Electrostatic invariant}
In addition to free energy, the two-dimensional electrostatic \gkm\ system also conserves the ``electrostatic invariant'' 
\begin{align}
  \label{eq:ElectrostaticInvariant}
  \ESInvar = \int\d^3\r \sum_s \frac{n_sq_s^2}{T_s}\varphi \left(1 - \mathsf{\Gamma}_0\right)\varphi , 
\end{align}
where $\mathsf{\Gamma}_0$ is the operator
$\mathsf{\Gamma}_0\varphi=\int\d^3\v~ \FNso\ga{\ga{\varphi}{\R_s}}{\r}$.
To see this, we take the gyroaverage $\ga{\cdot}{\r}$ of \eqref{eq:GyrokineticEquationSummary}, multiply by $n_sq_s$, sum over species and integrate over all velocity space to give
%%%\begin{align}
%%%  \begin{split}
%%%  &   \pd{ }{\tN} \left( \int\d^3\v  \sum_s n_sq_s \ga{h_{s}}{\r}  \right)
%%%     + \int\d^3\v  \sum_s n_sq_s \ga{ \lp \vthNs \vpara\b+ \vD +\vPhi\rp  \cdot \pd{h_s}{\R_s} }{\r}
%%%  \\  & \hspace{1cm}
%%%     + \int\d^3\v  \sum_s n_sq_s \ga{\vPhi     \cdot\pd{F_{0s}}{\R_s}}{\r}
%%%  = 
%%%  \pd{}{\tN}  \sum_s \frac{n_sq_s^2}{\TNs} \mathsf{\Gamma}_0\varphi,
%%%  \end{split}
%%%  \label{eq:FreeEnergyEquationSecondManipulation}
%%%\end{align}
\begin{align}
  \begin{split}
    &   \pd{ }{\tN} \left(  \sum_s \frac{n_sq_s^2}{T_s} \left(1 - \mathsf{\Gamma}_0\right)\varphi\right)
     + \int\d^3\v  \sum_s n_sq_s \ga{ \lp \vthNs \vpara\b+ \vD +\vPhi\rp  \cdot \pd{h_s}{\R_s} }{\r}
  \\  & \hspace{1cm}
     + \int\d^3\v  \sum_s n_sq_s \ga{\vPhi     \cdot\pd{F_{0s}}{\R_s}}{\r}
  = 0  ,
  \end{split}
  \label{eq:FreeEnergyEquationThirdManipulation}
\end{align}
where $\vPhi=\b\times\nabla\ga{\varphiN}{\R_s}/2$, 
and where we have used quasineutrality \eqref{eq:QuasineutralitySummary} on the first term.
The velocity integral of the collision term vanishes as collisions are mass conserving. 
Multiplying by $\varphi$ and integrating over all position space, we obtain the equation for the electrostatic invariant
\begin{align}
  \begin{split}
    &   \fd{\ESInvar }{\tN} 
     + \int\d^3\r\int\d^3\v  \sum_s n_sq_s\varphi \ga{ \vthNs \vpara\b \cdot \pd{h_s}{\R_s} }{\r}
  = 0  .
  \end{split}
  \label{eq:FreeEnergyEquationFourthManipulation}
\end{align}
%%%Written as a superposition of Fourier modes, the second term is
%%%\begin{align}
%%%  \begin{split}
%%%%%%    \int\d^3\r\int\d^3\v  \sum_s n_sq_s \int\d^3\k \hat{\varphi}_{-\k} e^{-i\k\cdot\r} \ga{ i\kpara' \vthNs \vpara \int\d^3\k' \hat{h}_{s\k'} e^{i\k'\cdot\R_s}  }{\r}
%%%%%%    \\
%%%%%%    \int\d^3\r\int\d^3\v  \sum_s n_sq_s \int\d^3\k \hat{\varphi}_{-\k} e^{-i\k\cdot\r} i\kpara' \vthNs \vpara \int\d^3\k' J_0(\rhoNs\kperp'\vperp) \hat{h}_{s\k'} e^{i\k'\cdot\r}  
%%%%%%    \\
%%%    \sum_s n_sq_s\vthNs  \int\d^3\k\int\d^3\v ~     i\kpara \vpara  J_0(\rhoNs\kperp\vperp) \hat{\varphi}_{-\k}\hat{h}_{s\k} .
%%%  \end{split}
%%%  \label{eq:FreeEnergyEquationFifthManipulation}
%%%\end{align}
Generally the second term is nonzero, so that $\ESInvar$ is not conserved.
However, in the two-dimensional case, there can be no parallel variation, so that $\b\cdot\tpd{h_s}{\R_s}=0$ and the electrostatic invariant is conserved.

%%%It may also be written as a superposition of Fourier modes
%%%\begin{align}
%%%  \begin{split}
%%%    \ESInvar =  \int\d^3\k\int\d^3\v  \sum_s \frac{n_sq_s^2}{T_s}  \left(1 - {\Gamma}_0\right)|\hat{\varphi}_{\k}|^2 ,
%%%  \end{split}
%%%  \label{eq:ElectrostaticInvariantFourier}
%%%\end{align}
%%%where $\Gamma_0=I_0((\rhoNs\kperp)^2/2)\exp(-(\rhoNs\kperp)^2/2)$.

\section{Computational forms of the gyrokinetic equations}
\label{sec:ComputationalFormsOfTheGyrokineticEquations}
In this section we derive the various forms of the \gkm\ system used in this \thesis. 
Firstly in \sec\ref{sec:ComplementaryDistributionFunction} we replace the guiding centre distribution function $h_s$ with the complementary distribution function $g_s$ used for computation in \sgk.  
This yields the most general set of equations implemented in \sgk.
We then derive reduced equations by taking simplifying limits of the \gkm\ system.
All equations we study have adiabatic electrons (derived in \sec\ref{sec:AdiabaticElectrons}) 
and are electrostatic (derived in \sec\ref{sec:ElectrostaticEquations}).
%%%(Throw away curvature and grad B).
Finally in \sec\ref{sec:LinearDriftKinetics} we neglect finite Larmor radius effects to derive the drift kinetic equation
studied in \chp s \ref{sec:Hypercollisions}, \ref{sec:FreeEnergyFlowAndDissipation} and \ref{sec:ScalingLawsForDriftKineticTurbulence}.

\subsection{Complementary distribution function}
\label{sec:ComplementaryDistributionFunction}
The \gkm\ system \eqref{eq:GyrokineticEquationSummary}--\eqref{eq:MaxwellSummary} 
%and the free energy \eqref{eq:FreeEnergySchekochihin}
is written in terms of the guiding centre distribution $h_s$,
as is usual for theoretical discussion \cite[\eg][]{Schekochihin09}.
We now write the system in terms of the 
complementary distribution function
\begin{align}
  g_s = h_s  - \frac{\qNs}{\TNs} \ga{\chiNs}{\R_s}F_{0s}.
  \label{eq:ComplementaryDistributionFunction}
\end{align}
With this, 
the gyrokinetic equation becomes
%%%\begin{align}
%%%  \begin{split}
%%%     \pd{ g_{s}}{\tN} 
%%%  %\\ &
%%%     & +\vthNs \vpara\b \cdot \pd{}{\R_s}\lp g_s + \frac{\qNs\FNso}{\TNs}{\ga{\chiNs}{\R_s}}\rp
%%%     + \vD     \cdot\pd{}{\R_s}\lp F_{0s} + g_s + \frac{\qNs\FNso}{\TNs}{\ga{\chiNs}{\R_s}}  \rp
%%%  %\\ &
%%%  \\  & = 
%%%  \ga{ \sum_{s'} \lp C_{ss'}(F_{0s},h_{s'}) + C_{ss'}(h_{s},F_{0s'})\rp }{\R_s}
%%%  \end{split}
%%%\end{align}
\begin{align}
  \begin{split}
    \pd{g_s}{t} &  
		+ \vthNs\vpara \pd{}{Z_s} \lp g_s + \frac{\qNs}{\TNs}\ga{\chi_s}{\R_s}F_0 \rp  
    + \frac{1}{2} \lp \b \times \pd{\ga{\chi_s}{\R_s}}{\R_s}\rp \cdot \pd{g_s}{\R_s}
  \\ & %\hspace{-1cm}
  - \frac{1}{2} \frac{\TNs}{\qNs}\left[  2\kappa\vpara^2 + L_B^{-1}{\vperp^2}\rs \pd{g_s}{Y_s}
  \\ & %\hspace{-1cm}
 +\frac{1}{2} \left[ - 2\kappa\vpara^2 - L_B^{-1}{\vperp^2}
 +  \omega_n + \lp \vperp^2 + \vpara^2 -\frac{3}{2}\rp \omega_T \rs \pd{\ga{\chi_s}{\R_s}}{Y_s}F_0
  \\ & %\hspace{-1cm}
 = \nu \ga{C[h_s]}{\R_s} ,
  \end{split}
  \label{eq:GyrokineticEquationComplementaryGuidingCentre}
\end{align}
where $\R_s=(X_s,Y_s,Z_s)$.
This is a convenient form for computation, as the gyrokinetic equation now contains only one time derivative term.
Moreover, this form is analogous to the Vlasov equation studied by, among others, \citet{Landau46}, \citet{VanKampen55} and \citet{Case59}.
This allows us to apply their work on the Vlasov equation to the gyrokinetic equation in \chp~\ref{sec:ParallelVelocitySpace}. 

\subsection{Fourier space representation}
\label{sec:FourierSpaceRepresentationGKMSystem}
As the spatial domain is triply periodic,
we express both functions of position and guiding centre space as Fourier series
\begin{align}
  \begin{split}
    g_s(\R_s) = \sum_{\k} e^{i\k\cdot\R_s}\hat{g}_s(\k),
   % \\
    \hspace{1cm}
%%%    \hat{g}_s(\k) = \frac{1}{V}\int_{\mathsf{\Omega}} \d^3\R_s~e^{-i\k\cdot\R_s} g_s(\R_s),
%%%    \\
  \varphi(\r) = \sum_{\k} e^{i\k\cdot\r}\hat{\varphi}(\k),
   % \\
%%%    \hspace{1cm}
%%%    \hat{\varphi}(\k) = \frac{1}{V}\int_{\mathsf{\Omega}} \d^3\r~e^{-i\k\cdot\r} \varphi(\r),
  \end{split}
  \label{eq:FourierSeries}
\end{align}
where $\sum_{\k}$ denotes the sum over all wavevectors $\k=(k_x,k_y,\kpara)^T$.
Wavenumbers $k_x$, $k_y$ and $\kpara$ are related to the lengths $L_x$, $L_y$, $L_z$ which define the periodic box $\mathsf{\Omega}=[0,L_x]\times[0,L_y]\times[0,L_z]$,
with $k_x\in\{0,1/x_0,2/x_0,\ldots\}$ and $x_0=L_x/2\pi$, and similarly in $y$ and $z$.
Where unambiguous, we also denote Fourier components with a subscript $\k$, \eg\ $\varphiNk\id\hat{\varphi}(\k)$.
We derive the inverses to \eqref{eq:FourierSeries} 
using the orthogonality formula
\begin{align}
  \begin{split}
    \frac{1}{V}\int_{\mathsf{\Omega}}\d^3\R_s ~ e^{i(\k-\k')\cdot\R_s} = \delta(\k-\k')
    ,
  \end{split}
  \label{eq:FourierSeriesOrthogonality}
\end{align}
where $V=L_xL_yL_z$ is the volume of the box,
and $\delta$ is the Kronecker delta which is unity if all components of $\k$ and $\k'$ are equal, and zero otherwise.
With this, the respective inverses to \eqref{eq:FourierSeries} are
\begin{align}
  \begin{split}
%%%    g_s(\R_s) = \sum_{\k} e^{i\k\cdot\R_s}\hat{g}_s(\k),
%%%   % \\
%%%    \hspace{1cm}
    \hat{g}_s(\k) = \frac{1}{V}\int_{\mathsf{\Omega}} \d^3\R_s~e^{-i\k\cdot\R_s} g_s(\R_s),
%%%    \\
%%%  \varphi(\r) = \sum_{\k} e^{i\k\cdot\r}\hat{\varphi}(\k),
   % \\
    \hspace{1cm}
    \hat{\varphi}(\k) = \frac{1}{V}\int_{\mathsf{\Omega}} \d^3\r~e^{-i\k\cdot\r} \varphi(\r).
  \end{split}
  \label{eq:FourierSeriesInverse}
\end{align}

The Fourier modes interact very neatly with the gyroaverage \eqref{eq:GyroaverageDefinition},
with gyroaveraging leading to factors of Bessel functions due to the relation \citep{Temme96}
\begin{align}
  J_n(x) = \frac{1}{2\pi}\int_0^{2\pi} \d\vartheta ~  e^{i(n\vartheta-x\sin\vartheta)}.
  %\label{eq:GyroaverageDefinition}
\end{align}
We use this to calculate gyroaverages.
Choosing the direction of $\kperpv = \kperp\unitx$ to simplify the algebra (but not change the phase-independent results),
we have
$\rhobs = \rhoNs\b\times\v = \rho_s\vperp(-\sin\vartheta\unitx+\cos\vartheta\unity)$
and
$\vperpv=\vperp(\cos\vartheta\unitx+\sin\vartheta\unity)$,
so we deduce
\begin{align}
  \begin{split}
  \ga{e^{i\k\cdot\r}}{\R_s} 
  &= \frac{1}{2\pi}\int_0^{2\pi} \d\vartheta ~  e^{i\k\cdot(\R_s+\rhoNs\b\times\v)}
  \\
  &= \frac{1}{2\pi}\int_0^{2\pi} \d\vartheta ~  e^{i\k\cdot\R_s}e^{-i\rhoNs\kperp\vperp\sin\vartheta}
  = J_0(\rhoNs\kperp\vperp)e^{i\k\cdot\R_s},
  \end{split}
  \label{eq:GyroaverageOfFourierMode}
  \\
  \begin{split}
  \ga{\v e^{i\k\cdot\r}}{\R_s} 
  &= \frac{1}{2\pi}\int_0^{2\pi} \d\vartheta ~ \vperp \left( \frac{e^{i\vartheta}+e^{-i\vartheta}}{2}\unitx + \frac{e^{i\vartheta}-e^{-i\vartheta}}{2i}\unity\right) e^{-i\rhoNs\kperp\vperp\sin\vartheta}  e^{i\k\cdot\R_s}
  \\
  &= \frac{i\vperp}{\kperp}J_1(\rhoNs\kperp\vperp) e^{i\k\cdot\R_s}\k\times\b,
  %\label{eq:GyroaverageDefinition}
  \end{split}
\end{align}
and similarly $\ga{e^{i\k\cdot\R_s}}{\r} = J_0(\rhoNs\kperp\vperp)e^{i\k\cdot\r}$, \emph{etc}.
With these we can write the gyrokinetic potential as
\begin{align}
  \begin{split}
  \ga{\chiNs}{\R_s} 
  %= \sum_{\k} \ga{\chiNs}{\R_s,\k} e^{i\k\cdot\R_s}
   %&= \ga{\varphiN - \vthNs\v\cdot\A}{\R_s} 
  &= \ga{\sum_{\k}\left( \varphiNk - \vthNs\vpara\Apk - \vthNs\vperpv\cdot\A_{\perp\k}\right)e^{i\k\cdot\r} }{\R_s} 
  \\
  &= \sum_{\k} \left[ J_0(a_s) \varphiNk - \vthNs\vpara J_0(a_s) \Apk   +  \frac{\TNs}{\qNs}\frac{2\vperp^2 J_1(a_s)}{a_s} \dBpk  \rs e^{i\k\cdot\R_s},
  \end{split}
  \label{eq:GyroaveragedPotentialGuidingCentre}
\end{align}
where $a_s=\rhoNs\kperp\vperp$, and we have used $\A_{\k}=2\B_{\k}\times\kperpv/(i\kperp^2)$ which follows from \eqref{eq:MagneticFieldNormalizedStep}.
We define the (velocity-dependent) Fourier component of the \gkpot
\begin{align}
  \begin{split}
   \ga{\chiNs}{\R_s,\k} 
  =  J_0(a_s) \varphiNk - \vthNs\vpara J_0(a_s) \Apk   +  \frac{\TNs}{\qNs} \frac{2\vperp^2 J_1(a_s)}{a_s} \dBpk .
  \end{split}
  \label{eq:GyroaveragedPotential}
\end{align}
This simplifies further calculation as \eqref{eq:GyroaveragedPotential} is gyrophase independent, so we may find the gyroaverage of $\ga{\chiNs}{\R_s}$ \eqref{eq:GyroaveragedPotentialGuidingCentre} using only the gyroaverage of Fourier modes \eqref{eq:GyroaverageOfFourierMode}.

\subsubsection{\Gkm\ system in Fourier space}
\label{sec:GKMSystemInFourierSpace}

Inserting the Fourier series \eqref{eq:FourierSeries} 
into the gyrokinetic equation \eqref{eq:GyrokineticEquationComplementaryGuidingCentre}
and applying the operator 
$V^{-1}\int_{\mathsf{\Omega}} \d^3\R_s~e^{-i\k\cdot\R_s}$,
we obtain
\begin{align}
  \begin{split}
    \pd{\gks}{t} &  + i\vthNs\vpara\kpara  \lp \gks 
    + \frac{\qNs}{\TNs} \ga{\chi_s}{\R_s,\k}F_0 \rp   
    +\nlt{\ga{\chi_s}{\R_s}}{g_s}_{\k}     
    \\ & %\hspace{4cm}
 + \frac{ik_y}{2} \left[ -2\kappa\vpara^2 - L_B^{-1}\vperp^2 +  \omega_n + \lp \vperp^2 + \vpara^2 -\frac{3}{2}\rp \omega_T \rs \ga{\chi_s}{\R_s,\k} F_0
   \\ & 
    - \frac{ik_y}{2} \frac{\TNs}{\qNs}  \lp 2\kappa\vpara^2+L_B^{-1}{\vperp^2}\rp \gks
 = \nu \ga{C[h_s]}{\R_s,\k} ,
  \end{split}
  \label{eq:GKNumata}
\end{align}
where the nonlinear term is
\begin{align}
\nlt{\ga{\chi_s}{\R_s}}{g_s}_{\k}    
= \frac{1}{2} \sum_{\k'} \b\cdot(\k\times\k') \ga{\chiNs}{\R_s,\k'} g_{\k-\k'}.
  \label{eq:GKSystemNonlinearTeam}
\end{align}
Further, inserting the \cdf\ \eqref{eq:ComplementaryDistributionFunction} 
and gyrokinetic potential \eqref{eq:GyroaveragedPotentialGuidingCentre} 
into Maxwell's equations \eqref{eq:MaxwellSummary},
we obtain
\begin{subequations}
  \label{eq:FieldSolveMoments}
\begin{align}
  &\varphiNk \sum_s \frac{\nNs\qNs^2}{\TNs}(1 - \Gamma_{0s})
  - \dBpk\sum_s \qNs\nNs\Gamma_{1s} = \sum_s \qNs \nNs \int \d^3\v ~\gks J_0(a_s),
  \label{eq:QN}
  \\
  &\Apk\left[ \frac{\kperp^2}{2\beta_r}   + \sum_s \frac{\qNs^2\nNs}{2\mNs}\Gamma_{0s} \rs
  = \sum_s \qNs\nNs \vthNs\int \d^3\v ~\gks \vpara J_0(a_s), 
  \label{eq:AmpPara}
\\
  &\varphiNk\sum_s \nNs\qNs\Gamma_{1s} +
\dBpk \lp \frac{2}{\beta_r} + \sum_s \TNs\nNs\Gamma_{2s}\rp
= - \sum_s \nNs \TNs \int \d^3\v ~\gks \vperp^2 \frac{2J_1(a_s)}{a_s} ,
  \label{eq:MaxBPara}
\end{align}
\end{subequations}
where 
\begin{subequations}
  \label{eq:GammaDefinitions}
\begin{align}
  \Gamma_{0s} 
  = \int \d^3\v ~ J_0(a_s)^2 F_0(v)
  = I_0(b_s)e^{-b_s},
  \\
  \Gamma_{1s}
  = \int  \d^3\v ~ \frac{2\vperp^2J_0(a_s)J_1(a_s)}{a_s} F_0(v)
  = (I_0(b_s)-I_1(b_s))e^{-b_s},
\\
  \Gamma_{2s} 
  = \int \d^3\v ~ \left( \frac{2\vperp^2}{a_s}\right)^2 J_1(a_s)^2 F_0(v)
  = 2 \Gamma_1(b_s),
\end{align}
\end{subequations}
with $b_s = (\rhoNs\kperp)^2/2$ and with $I_0$, $I_1$ the modified Bessel functions \citep{Howes06}.
The equations \eqref{eq:GKNumata}--\eqref{eq:FieldSolveMoments} are solved by \sgk, as discussed in \chp~\ref{sec:SpectroGK}.

Finally, we may write the free energy \eqref{eq:FreeEnergySchekochihin} in terms of the \cdf\ in Fourier space
\begin{align}
  W =
%%%  \sum_{s,\k} \int \d^3\v \frac{n_{0s}T_{0s}|g_{\k}|^2}{2 F_0}
%%%  + \sum_{s,\k}\frac{n_{0s}\qNs^2|\varphiNk|^2}{2T_{0s}}
%%%  - \sum_{\k} \frac{\kperp^2|\Apk|^2}{4\beta_r}
%%%  - \sum_{\k} \frac{|\Bpk|^2}{\beta_r} . 
  \sum_{s,\k} \int \d^3\v \frac{n_{0s}T_{0s}|g_{\k}|^2}{2 F_0}
	+ \sum_{s,\k}\frac{n_{0s}\qNs^2(1-\Gamma_{0s})|\varphiNk|^2}{2T_{0s}}
  + \sum_{\k} \frac{\kperp^2|\Apk|^2}{4\beta_r}
  + \sum_{\k} \frac{|\Bpk|^2}{\beta_r} . 
  \label{eq:FreeEnergyCDF}
\end{align}

\subsection{Simplified equations}

We now derive a series of simplified equation sets used in this thesis
from a number of limits of the
\gkm\ system \eqref{eq:GKNumata}--\eqref{eq:FieldSolveMoments}.
Firstly we show the adiabatic electron closure in \sec\ref{sec:AdiabaticElectrons} which reduces the two species system to a system for just ions.
This assumes electrons are massless and have infinite velocity so that the electron equilibrium distribution instantaneously forms.
We then neglect magnetic field perturbations, deriving the electrostatic equations for ions in \sec\ref{sec:ElectrostaticEquations}.
%This system, studied in \chp~\ref{sec:FreeEnergyFlowAndDissipation}, is the most complicated gyrokinetic system treated in this work.
%%%{\red
%%%Then in \sec\ref{sec:PerpendicularMaxwellian} we assume the distribution function is proportional to a perpendicular Maxwellian, which one may justify at large perpendicular spatial scales $\kperp\rho_i\ll1$. 
%%%This was used by \cite{Hatch13} to study free energy dissipation (see \chp~\ref{sec:FreeEnergyFlowAndDissipation}).
%%%The system also provides an interesting comparison to the four-dimensional (linear) gyrokinetics solved by \sgk\ (see \chp~\ref{sec:SpectroGK}).
%%%}
Finally in \sec\ref{sec:LinearDriftKinetics} we remove finite Larmor radius effects to derive the drift kinetic equation.  
The linearized drift kinetic equation provides the paradigm for Landau damping which we study in \chp~\ref{sec:ParallelVelocitySpace},
while we study the properties of nonlinear drift kinetic turbulence in \chp s \ref{sec:FreeEnergyFlowAndDissipation} and \ref{sec:ScalingLawsForDriftKineticTurbulence}.

\subsubsection{Adiabatic electrons}
\label{sec:AdiabaticElectrons}

We first derive the adiabatic closure, where massless electrons instantaneously assume their equilibrium distribution.
Formally we take the limit $m_e\to0$, $\vthNe\to\infty$ such that $\TNe = \mNe\vthNe^2$ remains constant.
In this limit, finite electron Larmor radius effects are removed as $\rhoNe=\vthe/\Omega_e=T_e/(q_e\vthe)\to0$. 
With the limits
\begin{align}
  \lim_{x\to0} J_0(x) = 
  \lim_{x\to0} \frac{2J_1(x)}{x} =
  \lim_{x\to0} I_0(x) = 
   1,
   \hspace{1cm}
   \lim_{x\to0} I_1(x) = 0,
  %\label{}
\end{align}
the gyrokinetic equation \eqref{eq:GKNumata} for electrons is still
\begin{align}
  \begin{split}
    \pd{g_{\k e}}{t} &  + i\vthNe\vpara\kpara  \lp g_{\k e} 
    + \frac{q_e}{T_e} \ga{\chi_e}{\R_e,\k}F_0 \rp   
    +  \nlt{\ga{\chi_e}{\R_e}}{g_e}_{\k}     
    \\ & %\hspace{4cm}
 + \frac{ik_y}{2} \left[ -2\kappa\vpara^2 - L_B^{-1}\vperp^2 +  \omega_n + \lp \vperp^2 + \vpara^2 -\frac{3}{2}\rp \omega_T \rs \ga{\chi_e}{\R_e,\k} F_0
   \\ & 
   - \frac{ik_y}{2} \frac{T_e}{q_e}  \lp 2\kappa\vpara^2+L_B^{-1}{\vperp^2}\rp g_{\k e}
 = \nu \ga{C[h_e]}{\R_e,\k} ,
  \end{split}
  \label{eq:GKElectrons}
\end{align}
but where the gyrokinetic potential \eqref{eq:GyroaveragedPotential} is now
\begin{align}
\begin{split}
  \ga{\chi_e}{\R_e,\k} 
  & =  \varphiNk - \vthe\vpara \Apk   +  \frac{T_e}{q_e}\vperp^2 \dBpk  .
%  = -\vthe\vpara \Apk  + \O(1).
%  \label{eq:GyroaveragedPotential}
\end{split}
\end{align}
We solve \eqref{eq:GKElectrons}, interpreting $\vthe$ as a large expansion parameter.
At $\O(\vthe^2)$ only the second term in the first parenthesis of \eqref{eq:GKElectrons} remains, giving $\Apk=0$.
Thus $g_{\k e}\sim\ga{\chi_e}{\R_e,\k}\sim \O(1)$, so at $\O(\vthe)$ the only contribution is from the remainder of the streaming term, which gives
\begin{align}
  \begin{split}
     g_{\k e} =
    - \frac{q_e}{T_e}\varphiNk F_0 - \vperp^2\dBpk F_0 .
  \end{split}
  \label{eq:ComplementaryDistributionElectrons}
\end{align}
Maxwell's equations become
\begin{subequations}
\begin{align}
  &\varphiNk \frac{n_iq_i^2}{T_i}(1 - \Gamma_{0i})
  - \dBpk \lp q_i n_i\Gamma_{1i} + q_e n_e \rp = \sum_s \qNs \nNs \int \d^3\v ~\gks J_0(a_s),
  \label{eq:QN_m_to_0}
  \\
%%%  &\Apk\left[ \frac{\kperp^2}{2\beta_r}   
%%%  + \frac{q_e^2n_e}{2m_e} 
%%%  + \frac{q_i^2n_i}{2m_i}\Gamma_{0i} \rs
  & 0 
  = \sum_s \qNs\nNs \vthNs\int \d^3\v ~\gks \vpara J_0(a_s), 
 % \label{eq:AmpPara}
\\
  &\varphiNk \lp n_eq_e +n_iq_i\Gamma_{1i}\rp +
\dBpk \lp \frac{2}{\beta_r} 
+  2T_en_e 
+  T_in_i\Gamma_{2i}\rp
= - \sum_s \nNs \TNs \int \d^3\v ~\gks \vperp^2 \frac{2J_1(a_s)}{a_s} ,
  %\label{eq:MaxBPara}
\end{align}
\end{subequations}
where 
  $\Gamma_{0e} = 1$,
  $\Gamma_{1e} = 1$
  and
  $\Gamma_{2e} = 2$.
The velocity space integrals are found for electrons by inserting $g_{\k e}$ \eqref{eq:ComplementaryDistributionElectrons} and noting $a_e\to0$, resulting in simple integrals of Maxwellians.
Further simplifying by using $n_i=n_e$ and $q_i=-q_e$, we obtain
\begin{subequations}
  \label{eq:MaxwellOneSpecies}
\begin{align}
  \varphiNk \frac{n_iq_i^2}{T_i}\lp 1 + \frac{T_i}{T_e} - \Gamma_{0i}\rp
  - \dBpk q_i n_i \Gamma_{1i}  = q_in_i\int\d^3\v~ g_{i\k} J_0(a_i) ,
  \label{eq:QNOneSpecies}
  \\
  0
  = q_in_i\vthi \int\d^3\v~ \vpara g_{i\k} J_0(a_i) ,
 % \label{eq:AmpPara}
\\
  \varphiNk n_iq_i \Gamma_{1i} +
\dBpk \lp \frac{2}{\beta_r} 
+  T_in_i\Gamma_{2i}\rp
= - n_i T_i \int \d^3\v ~ \vperp^2  g_{i\k}\frac{2J_1(a_i)}{a_i}  .
  %\label{eq:MaxBPara}
\end{align}
\end{subequations}
This is almost the same as the $\Apk=0$, single species case of Maxwell's equations \eqref{eq:FieldSolveMoments},
however \eqn\ \eqref{eq:QNOneSpecies} contains an extra temperature ratio term $(\varphiNk n_iq_i^2/T_i)(T_i/T_e)$ on the \lhs.
The gyrokinetic equation for ions is
\begin{align}
  \label{eq:GKIons}
  \begin{split}
    \pd{g_{\k i}}{t} &  + i\vthi\vpara\kpara  \lp g_{\k i} 
    + \frac{q_i}{T_i} \ga{\chi_i}{\R_i,\k}F_0 \rp   
    +\nlt{\ga{\chi_i}{\R_i}}{g_i}_{\k}     
    \\ & %\hspace{4cm}
 + \frac{ik_y}{2} \left[ -2\kappa\vpara^2 - L_B^{-1}\vperp^2 +  \omega_n + \lp \vperp^2 + \vpara^2 -\frac{3}{2}\rp \omega_T \rs \ga{\chi_i}{\R_i,\k} F_0
   \\ &
    - \frac{ik_y}{2} \frac{T_i}{q_i}  \lp 2\kappa\vpara^2+L_B^{-1}{\vperp^2}\rp g_{\k i}
 = \nu \ga{C[h_i]}{\R_i,\k}. 
  \end{split}
\end{align}
Equations \eqref{eq:MaxwellOneSpecies}--\eqref{eq:GKIons} are solved by \sgk\ when a single ion species is selected.

\subsubsection{Electrostatic equations}
\label{sec:ElectrostaticEquations}
The one species, electrostatic version of gyrokinetics, solved in \chp~\ref{sec:FreeEnergyFlowAndDissipation}, 
is found by setting $\Apar=\Bpar=0$
in \eqref{eq:MaxwellOneSpecies}--\eqref{eq:GKIons}
so that 
$\varphipN$ replaces $\chi_i$ in the gyrokinetic equation.
We also neglect magnetic field inhomogeneities ($\kappa=L_B^{-1}=0$), %and the density gradient ($\omega_n=0$), 
so that
\begin{align}
  \label{eq:GKIonsES}
  \begin{split}
    \pd{g_{\k i}}{t} &  + i\vthi\vpara\kpara  \lp g_{\k i} 
    + \frac{q_i}{T_i} \ga{\varphip}{\R_i,\k}F_0 \rp   
    +\nlt{\ga{\varphip}{\R_i}}{g_i}_{\k}     
    \\ & %\hspace{4cm}
 + \frac{ik_y}{2}  \left[ %-2\kappa\vpara^2 - L_B^{-1}\vperp^2 + 
   \omega_n +
   \lp \vperp^2 + \vpara^2 -\frac{3}{2}\rp \omega_T \rs
   \ga{\varphip}{\R_i,\k} F_0
%%%   \\ & 
%%%    - \frac{ik_y}{2} \frac{T_i}{q_i}  \lp 2\kappa\vpara^2+L_B^{-1}{\vperp^2}\rp g_{\k i}
 = \nu \ga{C[h_i]}{\R_i,\k},
  \end{split}
\end{align}
with the electrostatic potential determined using quasineutrality
\begin{align}
  \varphiNk \frac{n_iq_i^2}{T_i}\lp 1 + \frac{T_i}{T_e} - \Gamma_{0i}\rp
  = q_in_i\int\d^3\v~ g_{i\k} J_0(a_i).
  \label{eq:QNOneSpeciesES}
\end{align}
As quasineutrality alone is sufficient to determine the electrostatic potential, \ampere's law is not considered.
\subsubsection{Drift kinetic equations}
\label{sec:LinearDriftKinetics}
Finally, we derive the drift kinetic equations 
by considering the long perpendicular wavelength limit $\kperp\ll1$.
This removes finite Larmor radius effects so that 
$\ga{\varphip}{\R_i,\k} = \varphiNk$ and
$\Gamma_{0i}=1$.
Further, assuming that $g_{\k i}$ is proportional to a Maxwellian in perpendicular velocity space, we obtain
%%%\begin{align}
%%%  \begin{split}
%%%    \pd{g_{\k i}}{t} &  + i\vthi\vpara\kpara  \lp g_{\k i} 
%%%    + \frac{q_i}{T_i}\varphiNk F_0 \rp   
%%%    \\ & %\hspace{4cm}
%%% + \frac{ik_y}{2} \left[  L_n^{-1} + \lp \vperp^2 + \vpara^2 -\frac{3}{2}\rp L_T^{-1} \rs \varphiNk F_0
%%% = \nu \ga{C[h_i]}{\R_i,\k} .
%%%  \end{split}
%%%\end{align}
\begin{subequations}
\begin{align}
	\begin{split}
    &\pd{g_{\k i}}{t}   + i\vthi\vpara\kpara  \lp g_{\k i} 
    + \frac{q_i}{T_i}\varphiNk F_0^{\parallel} \rp   
 %   \\ & %\hspace{4cm}
    + \frac{ik_y}{2} \left[  \omega_n + \lp \vpara^2 -\frac{1}{2}\rp \omega_T \rs \varphiNk F_0^{\parallel}
			\\
			& \hspace{9cm} + \Big\{ \varphiNk , g_s \Big\}_{\k}
 = \nu C[h_i],
	\end{split}
  \\
   &\varphiNk \frac{n_iq_i^2}{T_i}\lp \frac{T_i}{T_e} \rp
   = q_in_i\intii\d\vpara~ g_{i\k} ,
%  \label{eq:QN}
\end{align}
\label{eq:LinearizedDriftKinetics}
\end{subequations}
where $F_0^{\parallel}(\vpara)=\exp(-\vpara^2)/\sqrt{\pi}$,
and the nonlinear term is
\begin{align}
\Big\{ \varphiNk , g_s \Big\}_{\k}
= \frac{1}{2} \sum_{\k'} \b\cdot(\k\times\k') \varphi_{\k'} g_{\k-\k'}.
  \label{eq:DKSystemNonlinearTeam}
\end{align}
In the development of our collision operator and spectral method in \chp\ \ref{sec:ParallelVelocitySpace}, we study the linearized version of \eqref{eq:LinearizedDriftKinetics} obtained by neglecting this nonlinear term.

\part{Numerical Methods}
\chapter{Parallel velocity space and the Hermite spectral representation}
\label{sec:ParallelVelocitySpace}
\label{sec:Hypercollisions}

% Macro variables
\newcommand{\fulldist}{F}
\newcommand{\oneddist}{\tilde{f}}

% Maths
\renewcommand{\a}{\mathsf{a}}
\newcommand{\at}{\tilde{a}}
\newcommand{\atpm}{\tilde{a}^{\pm}}
\newcommand{\atmp}{\tilde{a}^{\mp}}
\newcommand{\atp}{\tilde{a}^{+}}
\newcommand{\atm}{\tilde{a}^{-}}
\renewcommand{\E}{\hat{E}}
\newcommand{\EL}{E^{\mathrm{(L)}}}
\newcommand{\ER}{E^{\mathrm{(R)}}}
\newcommand{\Ev}{\boldsymbol{E}}
\newcommand{\fouriersum}{\sum_{j=0}^{N_k-1}}
\newcommand{\f}{\hat{f}}
\renewcommand{\I}{{\cal I}}
\newcommand{\indjj}{\I_{jj'}}
\renewcommand{\k}{\boldsymbol{k}}
\renewcommand{\kperpv}{\k_{\perp}}
\newcommand{\N}{{\cal N}}
\newcommand{\n}{\boldsymbol{n}}
\newcommand{\Nmat}{{\mathsf{N}}}
\newcommand{\RE}{{\cal R}}
\newcommand{\sumo}[2]{\sum_{#1=0}^{#2}}
\renewcommand{\vpara}{{v}}
\newcommand{\xperp}{{x}_{\perp}}

% This is CUP's notation
\newcommand{\e}{\textrm{e}}
\renewcommand{\i}{\textrm{i}}
\newcommand{\upi}{\pi}
\newcommand{\Real}{\textrm{Re}}
\newcommand{\Imag}{\textrm{Im}}
% ... overwritten in line with the thesis 
\renewcommand{\e}{e}
\renewcommand{\i}{i}
\renewcommand{\a}{{\bf a}}

\renewcommand{\fouriersum}{\sum_{j=-\Nt}^{\Nt}}

% Text subscripts
%\renewcommand{\B}{\mathrm{B}}
\newcommand{\SV}{\mathrm{SV}}

\newcommand{\vpstilde}{\eqref{eq:TildeSystem}}
\newcommand{\ThreeDVP}{\eqref{eq:ThreeDKinetic}--\eqref{eq:ThreeDPoisson}}
\newcommand{\ctseqns}{\eqref{eq:CtsFourierHermite}--\eqref{eq:CtsFourierHermitePoisson}}
\newcommand{\dvp}{\eqref{eq:DiscreteFourierHermite}}
\newcommand{\vpseqns}{\eqref{eq:OneDSystem}}

% BBBBBBBBBBBBBBBBBBB

\newcommand{\Exac}[1]{(\ref{#1}\textit{-c})}
\newcommand{\dd}[2]{\frac{\mathrm{d} #1}{\mathrm{d} #2}}

\newcommand{\Cthreed}{\widetilde{C}}
\newcommand{\Ft}{\widetilde{F}}

\renewcommand{\blue}{}

% Common phrases and contentious spellings
\renewcommand{\cvk}{Case--Van Kampen}
\renewcommand{\lb}{L\'enard--Bernstein}
\renewcommand{\lb}{Lenard--Bernstein}

%\renewcommand{\nu}{\nuup}

% Maths
\renewcommand{\b}{\mathbf{b}}
\newcommand{\calF}{\boldsymbol{\cal F}}%{\pmb{\cal F}}
\newcommand{\cfe}{\nu}
\newcommand{\crit}{\textrm{crit}}
\newcommand{\etacrit}{\nu_\textrm{crit}}
\newcommand{\etacritobs}{\nu_\textrm{crit}^{\mathrm{obs}}}
\renewcommand{\etacritobs}{\nu_\textrm{crit}}
\newcommand{\Ffived}{\tilde{F}}
\renewcommand{\Im}{\textrm{Im}}
\renewcommand{\kpara}{k}
\newcommand{\ls}{\left[}
\renewcommand{\Re}{\textrm{Re}}
\renewcommand{\vthi}{v_{\mathrm{th}i}}
\renewcommand{\vpara}{v}
\renewcommand{\z}{\hat{z}}

% Modified free energy, E_\mu
\newcommand{\Emu}{\tilde{E}_{\mu}}
\newcommand{\Emuzero}{\tilde{E}_{0}}
\newcommand{\Enomu}{\tilde{E}}  % for use with explicit mu arguement: E(mu) not E_mu

% Change from GENE to GS2 normalization
\newcommand{\distpert}{g}

%\section{Introduction}
\label{sec:intro}
\label{sec:ParallelVelocityIntro}

Having derived the \gkm\ system in \chp~\ref{sec:GKMSystem}, we now turn our attention to its numerical solution.
Due to the high dimensionality, simulations of large domains (like whole \tkmk s) are barely feasible.
Even smaller domains (like flux-tube simulations of a \tkmk\ section) require upwards of hundreds of thousands of CPU hours each.
Even so, resolution is usually coarse. 
For example, 
recent numerical work by \citet{Highcock11} with the gyrokinetic code \textsc{GS2} \cite{GS2} used $64\times32\times14$ points for physical space and $24\times8\times2$ on a pitch angle-energy-sign grid in velocity space.
That is, over 11 million degrees of freedom, but still modest resolution in each dimension.
The velocity space resolution above uses the equivalent of a grid of 16 points in parallel velocity space,
while other recent studies \cite{Dannert04,Kammerer08,Peeters09,PueschelDannertJenko10} 
% Dannet Jenko 2004 Nv = 40 <-- "Solves" recurrence problem with numerical diffusion   
% Kammerer Merz Jenko 2008 GENE linear GK Nv = 64
% PDJ 2010 GENE Nv =32
% Peeters 2009 nonlinear code, uses Nv = 128 in linear benchmarks "rather large" to avoid recurrence problem.  
typically use between 32 to 128 points for parallel velocity space.
Since the problem size is this multiplied by the resolutions in the four other dimensions,
there is a strong motivation to improve the treatment of the parallel velocity degrees of freedom.

In this \chp\ and the next, we study spectral representations for parallel and perpendicular velocity space respectively,
with the goal of deriving a representation for the distribution function which requires little resolution,
but nonetheless can capture important features of the solution (like Landau damping, described shortly).
To investigate the representation of parallel velocity space, 
we use the linearized model for drift-kinetic ions derived in \sec\ref{sec:LinearDriftKinetics},
\begin{subequations}
\begin{align}
  \label{eq:KineticEquationChapter3}
    \pd{g}{t}   + i %\vthi
		\kpara\vpara  \lp g
    + %\frac{q_i}{T_i}
		\varphi F_0 \rp   
    %+ \frac{ik_y}{2} \left[  \omega_n + \lp \vpara^2 -\frac{1}{2}\rp \omega_T \rs \varphiNk F_0^{\parallel}
		+ i \left[  \tilde{\omega}_n + \lp \vpara^2 -\frac{1}{2}\rp \tilde{\omega}_T \rs \varphi F_0
 = \nu C[g],
  \\
   \varphi %\frac{n_iq_i^2}{T_i}\lp \frac{T_i}{T_e} \rp
   = %q_in_i
	 \intii\d\vpara~ g ,
  \label{eq:QuasineutralityChapter3}
\end{align}
\label{e.8}%
\end{subequations}%
where $\tilde{\omega}_n=k_y\omega_n/2$ 
and
$\tilde{\omega}_T=k_y\omega_T/2$.
For brevity we have written the complementary distribution function as $g=g_{\k i}$ and electrostatic potential as $\varphi=\varphiNk$.
We have also dropped the subscript $\parallel$ on the parallel wavenumber $\kpara$, parallel velocity $\vpara$, and parallel Maxwellian $F_0=\exp(-\vpara^2)/\sqrt{\pi}$.
As the system is linear, we may study different wavenumbers independently, so we write the full distribution function as $F(z,v,t)=F_0(v)+e^{ikz}g(v,t)$.
Further, we have chosen ions to be the reference species in the normalization ($r=i$ in \S\ref{sec:Nondimensionalization}), so the normalized charge, temperature, and thermal velocity appearing in \eqref{eq:LinearizedDriftKinetics} are all unity.

This system was recently studied numerically by \citet{PueschelDannertJenko10} 
using the \textsc{Gene} code \cite{Gene}.
For comparison with that work, we choose the parameters $\tilde{\omega}_n=0.3$ and $\tilde{\omega}_T=3.0$,
and consider the parallel wavenumbers $k\in[0,8\alpha_i]$, 
where the constant $\alpha_i=0.34$ is a dimensionless combination of length and velocity scales arising from the normalization used in \textsc{Gene} \cite[see][]{PueschelDannertJenko10}.
This system exhibits a wide range of growing and decaying behaviour in the electrostatic potential $\varphi$,
and is therefore a useful benchmark for numerical methods.
%This system is ideal for studying numerical methods, as it exhibits both the growing and decaying behaviours for $\varphi$ decribed shortly.

\setcounter{subsection}{0}
\subsection{Landau damping}

We begin by describing the analytic solution to \eqref{e.8},
where the presence of a destabilising density gradient $\tilde{\omega}_n$ and temperature gradient $\tilde{\omega}_T$
requires a small amendment to the text-book theory of Landau damping and \cvk\ modes \cite{VanKampenBook}.
We then discuss how this theory changes when velocity space is discretized, as is necessary in numerical solutions.

\begin{figure}[tbp]
  \centering
  \subfigure[\label{fig:contours}]{\includegraphics[width=0.49\textwidth]%,trim=-50 -50 -50 0,clip=true]
%{contour_paper.pdf}}
{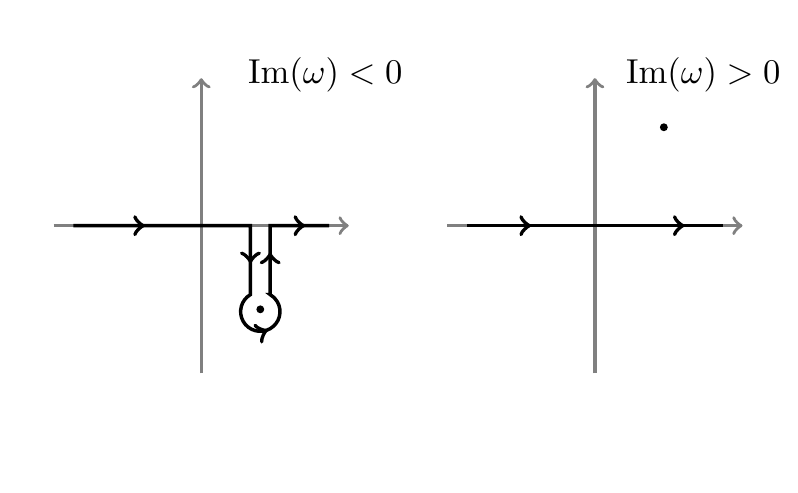}}
\subfigure[\label{f.fsystem}]{\includegraphics[width=0.49\textwidth]
%{fsystem_gr.ps}
%{asystem_gr_bw.ps}
%{asystem_gr_paper.ps}}
{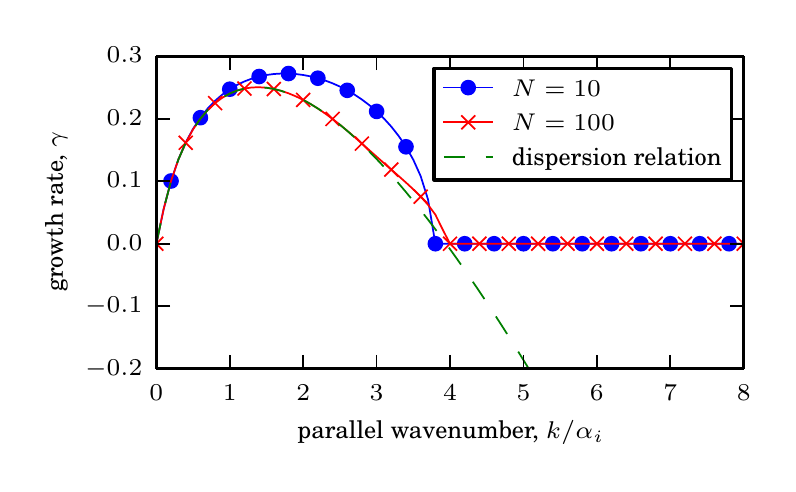}}
\subfigure[\label{fig:VKM}%despite name, real lls on complex part
]{\includegraphics[width=0.49\textwidth]
%{van_kampen_mode.ps}
%{van_kampen_mode_mark.ps}}
%{van_kampen_mode_mark.pdf}}
%{van_kampen_mode_mark_alt.pdf}}
%{van_kampen_mode_mark_alt_imag.pdf}}
{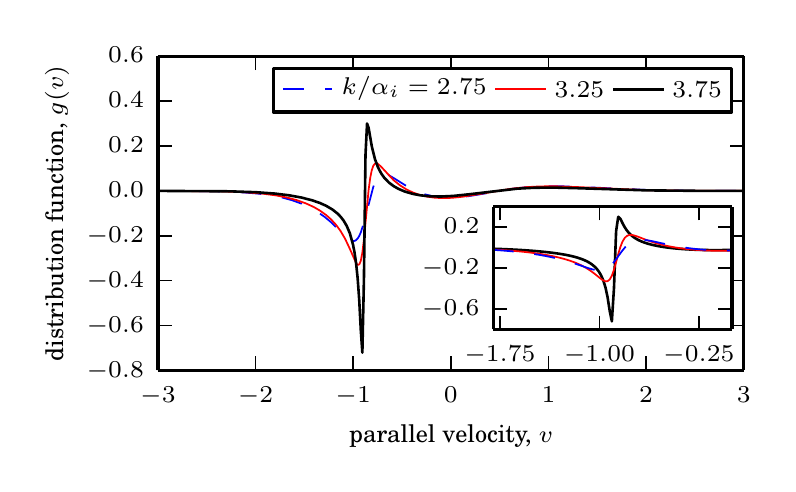}}
\subfigure[\label{fig:dougherty_intro}]{\includegraphics[width=0.49\textwidth]
%{numerical_diffusion_velocity_space_real.ps}}
%{ef_hd_vs_nu_n1_N5000_k410_lls_real.ps}}
%{ef_hd_vs_nu_n1_N5000_k410_lls_real.pdf}}
{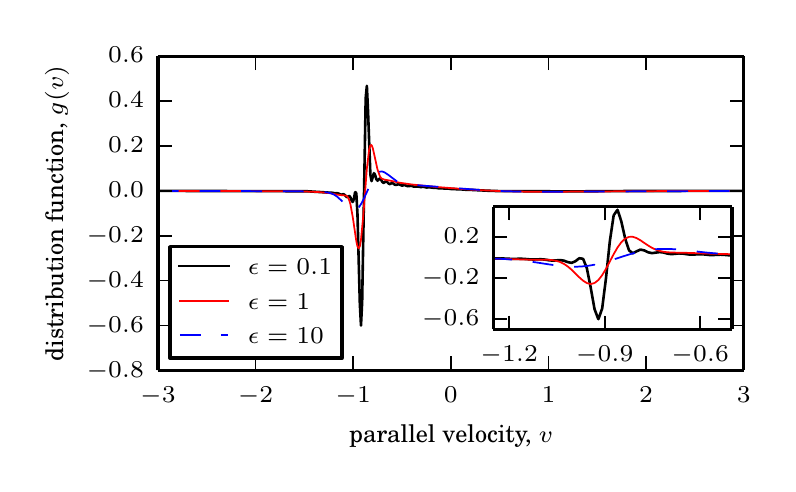}}
%{ef_hd_vs_nu_n1_N5000_k410_lls_imag.pdf}}
%%%\subfigure[\label{fig:DeltaIntro}]{\includegraphics[width=0.49\textwidth]
%%%{delta_fn_hs.pdf}}
\caption[Integration contours for the plasma dispersion function, and growth rates and eigenmodes for a na\"ive velocity space discretization.]{ %Results for the discretization of (\ref{e.8}). 
(a)
Integration contours of the plasma dispersion function $Z(\zeta)$ for $\Im(\omega)\lessgtr 0$.
(b) Growth rate plotted against parallel wavenumber for a discretization with $N$ points on a Gauss--Hermite grid.  
(c)
The fastest growing eigenmode for different wavenumbers.
A \cvk\ mode forms as the growth rate $\gamma\to0^+$ as $k\to k_{\text{crit}}$.  
(d) 
The slowest decaying eigenmode corresponding to $\kpara=4.1\alpha_i$, calculated using the \dop\ operator (\ref{eq:Dougherty_linear_hs}). %As collisionality decreases, the \cvk\ mode forms. 
The distribution develops a boundary layer of width $|\gamma|$ about the resonant velocity $v=\omega_R/k$, and becomes more singular as $\cfe\to0$. 
%The wavenumber is normalized to the connection length, and velocity is normalized to the ion thermal velocity.
\label{fig:Intro_example}}
\end{figure}

Writing \eqref{e.8} in operator notation, we have
\begin{align}
  \pd{\distpert}{t} = - i L_{\cfe} [\distpert],
  \label{eq:ddt_operator_eqn}
\end{align}
where $L_\cfe$ is the operator defined by
\begin{align}\label{eq:Lepsdef}
  L_\cfe [\distpert] =  
  \kpara\vpara \distpert
	+ \ls \tilde{\omega}_n + \wT\lp\vpara^2-\frac{1}{2}\rp  + 
\kpara\vpara  \rs \varphi F_0
  + i\cfe C[\distpert],
\end{align}
and $\varphi$ is related to $\distpert$ by \eqref{eq:QuasineutralityChapter3}.
We first consider the collisionless case ($\cfe=0$) and solve \eqref{eq:ddt_operator_eqn} using Landau's method \cite{Landau46}, as follows.
We take the Laplace transform 
\begin{align}
	\bar{g}(v,p) = \intoi \d t ~ e^{-pt}g(v,t),
	\label{eq:LaplaceTransformDefinition}
\end{align}
of \Ex{eq:ddt_operator_eqn} in time, 
and rearrange to find the transformed distribution function 
\begin{align}
	\bar{\distpert}(v,p) = \frac{\distpert^0(v)}{p+ikv} 
  + \frac{ \ls\wn + \wT\lp\vpara^2-\frac{1}{2}\rp + k \vpara  \rs F_0(v) \bar{\varphi}(p)}{ip - kv},
  %\propto f(t=0)e^{-ivt} + e^{-i\omega t}.
  \label{eq:gbarSoln}
\end{align}
where $\distpert^0(v)=g(v,t=0)$ is the initial distribution.
We have used the property of Laplace transforms that $\overline{\tpd{g}{t}}=p\bar{g}-g^0$. 
We then integrate \eqref{eq:gbarSoln} over all velocities and use the quasineutrality condition \eqref{eq:QuasineutralityChapter3} to
obtain a linear equation for $\bar{\varphi}(p)$.
We solve, and invert the Laplace transform to give
%%%\begin{align}
%%%  \bar{\varphi}(t) =  \frac{1}{D(ip)}\int_{-\infty}^{\infty}\frac{f_0(v)}{p+iv}\,\d v, 
%%%  \label{eq:PhiBar}
%%%\end{align}
%and invert the Laplace transform to give
%%%\begin{align}
%%%  \varphi(t) = \frac{1}{2\pi i} \int_{-i\infty + \sigma}^{i\infty + \sigma}\lp \frac{1}{D(ip)}\int_{-\infty}^{\infty}\frac{f_0(v)}{p+iv}\,\d v \rp e^{pt} \,\d p
%%%  \label{eq:PhiSoln}
%%%\end{align}
\begin{align}
  \varphi(t) = \frac{1}{2\pi i} \int_{-i\infty + \sigma}^{i\infty + \sigma} \d p\  \bar{\varphi}(p)\ e^{pt} ,
  \hspace{1cm}
  \bar{\varphi}(p) = \frac{1}{D(ip)}\int_{-\infty}^{\infty}\d v\ \frac{\distpert^0(v)}{p+ikv},
  \label{eq:PhiSoln}
\end{align}
where the $p$-integral is the Bromwich integral (the Laplace transform inverse) with $\Re(\sigma)$ to the right of all poles in the integrand, and $D(\omega)$ is given by 
\begin{align}
  D( \omega ) = 2+\frac{\wT}{k}\frac{\omega}{k}  + \ls \lp \frac{\omega_n}{k} + \frac{\wT}{k}\lp \lp\frac{\omega}{k}\rp^2-\frac{1}{2}\rp\rp + \frac{\omega}{k}\rs Z\lp\frac{\omega}{k}\rp .
  \label{eq:DispersionRelation}
\end{align}
The plasma dispersion function $Z(\zeta)=\int_{-\infty}^{\infty}\d v\ F_0(v)/(v-\zeta)$ 
has its integration contour along the real line for $\Im(\zeta)>0$ and along the Landau contour for $\Im(\zeta)\leq0$.
The Landau contour is a deformation of the real line which passes below the complex pole at $v=\zeta$, as shown in \fig\ref{fig:contours}.
Properties of the plasma dispersion function, such as asymptotic expansions, are well-known \cite{Fried61,NRL}.
In the long-time limit, $\varphi$ becomes proportional to $e^{p^*t}$ where $p^*$ is the pole of the integrand with largest real part. 
For non-singular initial conditions, this is given by $p^*=-i\omega$, 
where $\omega$ is the solution of $D(\omega)=0$ with largest imaginary part.
Thus, in the long time limit, we have $\varphi\propto e^{-i\omega t}$ (but, as we see shortly, not necessarily $\distpert\propto e^{-i\omega t}$).

Solving $D(\omega)=0$ with no driving or collisions, $\wn=\wT=\cfe=0$, gives Landau damping, the surprising result that the growth rate $\Im(\omega)$ is negative, \ie\ that the electrostatic potential decays despite the system being apparently time-reversible.
With driving, the growth rate plotted in green in \fig\ref{f.fsystem} exhibits growth and decay for different wavenumbers, so provides a useful benchmark for studying numerical methods. 

\newcommand{\hatfc}{\hat{\distpert}_c}
\newcommand{\hatf}{\hat{\distpert}}
The distribution function $\distpert$ 
is also given by Landau's method.  Inverting \eqref{eq:gbarSoln}, we obtain
\begin{align}
  \distpert(v,t) = \frac{1}{2\pi i}\int_{-i\infty + \sigma}^{i\infty+\sigma} \d p\ \ls \frac{\distpert^0(v)}{p+ikv} 
  + \frac{ \ls\wn + \wT\lp\vpara^2-\frac{1}{2}\rp + k \vpara  \rs F_0(v) \bar{\varphi}(p)}{ip - kv}\rs e^{pt}.
  %\propto f(t=0)e^{-ivt} + e^{-i\omega t}.
  \label{eq:fSoln}
\end{align}
The first term in the integrand integrates to $\distpert^0(v)e^{-ikvt}$, which is purely oscillatory and satisfies the particle streaming part of \Ex{eq:ddt_operator_eqn}, $\tpd{\distpert}{t}+ikv\distpert =0$.
The second term is proportional to $\varphi$.
Leaving the $p$ integral alone and instead integrating over $v$ gives
\begin{align}
 \intii\d v\ \distpert(v,t) = \frac{1}{2\pi i}\int_{-i\infty + \sigma}^{i\infty+\sigma}\d p \ \ls \intii \frac{\distpert^0(v)}{p+ikv}\,\d v 
  - D(ip) \bar{\varphi}(p) + \bar{\varphi}(p)\rs e^{pt} = \varphi(t),
  \label{eq:phiSoln}
\end{align}
where using $\bar{\varphi}$ from \Ex{eq:PhiSoln}, we see the contribution $D(ip)\bar{\varphi}$ from the second term in the integrand in \Ex{eq:fSoln} exactly cancels the first, purely oscillatory term. 
Thus even though $\distpert$ oscillates, its integral $\varphi$ can decay without violating the quasineutrality condition.  
This shows that the Landau-damped solution \eqref{eq:fSoln} for $\distpert$ is not an eigenmode of \Ex{eq:ddt_operator_eqn}.

In contrast, for growing modes $\Im(\omega)>0$, the $\varphi$ behaviour dominates \Ex{eq:fSoln} in the long time limit, so $\distpert\propto\varphi$ and the solution is an eigenmode.
It is therefore instructive to consider the %time harmonic Ansatz
separable solutions
\begin{align}
  \distpert(v,t) = \hat{\distpert}(v)\, e^{-i\omega t},
  \hspace{1cm}
  \varphi(t) = \hat{\varphi}\, e^{-i\omega t}.
  \label{eq:wave_ansatz}
\end{align}
The long-time behaviour of the \ivp\ will be determined by the fastest growing or slowest decaying mode, for which $\Im(\omega)$ is largest.
With \eqref{eq:wave_ansatz}, equation \eqref{eq:ddt_operator_eqn} becomes the linear eigenvalue problem
\begin{align}\label{eq:operator_eqn}
  \omega \hat{\distpert} = L_\cfe [\hat{\distpert}],
\end{align}
with the collisionless case being $\omega \hat{\distpert}=L_0[\hat{\distpert}]$ for $\cfe=0$. 
The operator $L_0$ is real, so its eigenvalues occur in complex conjugate pairs.
Therefore the dominant growth rate, the largest imaginary part of any eigenvalue, is non-negative.
The eigenfunctions $\hat{\distpert}$ and $\hat{\varphi}$ are found by solving \Ex{eq:operator_eqn}   
for $\hatfc$, the analytic continuation of $\hatf$ into the complex plane such that $\hatfc(v)=\hatf(v)$ for real $v$,
\begin{align}
  \hatfc(v) = \frac{\ls\wn + \wT\lp\vpara^2-\frac{1}{2}\rp  +  k\vpara  \rs\varphi F_0}{\omega - kv}
  + A\varphi \delta (\omega  - kv).
  \label{eq:KrookDist}
\end{align}
The delta function term arises from dividing \eqref{eq:operator_eqn} by $(\omega-kv)$ because $A(\omega-kv)\delta(\omega-kv)\id 0$, with $A$ an arbitrary constant.
The solution is singular at the point $v=\omega(k)/k$ whose location varies with $k$.
Putting \eqref{eq:KrookDist} into the quasineutrality condition \eqref{eq:QuasineutralityChapter3} yields  
the consistency condition
\begin{align}
  \int_{-\infty}^{\infty}\d v\  \frac{\ls\omega_n + \wT\lp\vpara^2-\frac{1}{2}\rp +  k\vpara  \rs F_0}{\omega - kv} 
  + A\int_{-\infty}^{\infty}\d v\ \delta(\omega - k v)
  = 1,
  \label{eq:ConsistencyCondition}
\end{align}
%an integral \eqn\ for $\omega(k)$,
where the integration contour is the real line in the complex $v$ plane.
When $\omega$ is not real, the second integral vanishes and \Ex{eq:ConsistencyCondition} becomes an integral \eqn\ to determine $\omega(k)=\omega_R+i\gamma$. 
When $\gamma>0$, \eqn\ \Ex{eq:ConsistencyCondition} is $D(\omega)=0$ and the method agrees with the Landau approach.
However, there are no solutions such that $\gamma<0$, which we argue as follows.
The first integral in \Ex{eq:ConsistencyCondition} is a function of $\omega$ that is symmetric in the real axis, so for $\gamma<0$ \eqn\ \Ex{eq:ConsistencyCondition} reduces to $D({\omega_R}-i\gamma)=0$. 
Thus every decaying solution corresponds to a solution growing with rate $-\gamma$, and there are no wavenumbers where the dominant mode has a negative growth rate.

When $\omega$ is real ($\gamma=0$), the second integral remains in \Ex{eq:ConsistencyCondition} and 
the first term is taken as a principal value integral.
There are modes for all real $\omega_R$, with \eqref{eq:ConsistencyCondition} serving to determine $A$.
This results in a continuous spectrum of singular eigenmodes $\hatf_c$ \eqref{eq:KrookDist} with real eigenvalues $\omega_R$, the \cvk\ modes \cite{Case59,VanKampen55}.
Thus the dispersion relation found with eigenmodes agrees with the Landau dispersion relation for growing modes, but gives a zero growth rate for non-growing mode while the Landau growth rate is strictly negative.
This reflects the fact that the Landau-damped solutions are not eigenmodes, so we will not find decay by studying single eigenmodes of the collisionless system.
The eigenmode approach does not contradict Landau's method however; \citet{Case59} showed that the \cvk\ modes are complete, and the distribution for a Landau-damped solution is an infinite superposition of \cvk\ modes. 
Its integral $\varphi$ decays through phase mixing, the interaction of an infinite number of oscillating modes with different frequencies.

The picture changes when $\cfe>0$, as any nonzero degree of collisionality counteracts phase mixing.
Plasma collisions are dominated by glancing encounters due to the long-range Coulomb interactions between charged particles,
which are modelled by \fp-type operators with first and second order velocity space derivatives (see \S\ref{sec:LandauCollOpAndProperties} onwards).
\Eqn\ \Ex{eq:operator_eqn} becomes a second-order differential \eqn\ for $\hat{\distpert}$, and the nature of the eigenvalue problem changes.
Collisions balance particle streaming, regularizing the problem and preventing the formation of structure at infinitesimal scales.
\citet{LenardBernstein} solved the undriven system with a simple \fp\ operator, finding a damped eigenmode solution with a decay rate that tends to the Landau rate in the limit of vanishing collisions. 
\citet{Ng99,Ng04} showed that this system has a discrete spectrum and a complete set of smooth and square integrable eigenmodes, instead of the continuous spectrum of singular eigenmodes in the collisionless system.
A subset of the collisional eigenmodes have eigenvalues that tend to solutions of the Landau dispersion relation $D(\omega)=0$ in the limit of vanishing collision frequency \cite{Ng06}.
As collisionality is reduced, the eigenmodes develop boundary layers of width $|\gamma|$ and become more singular with oscillations of width $\cfe^{-1/4}$ \cite{Ng06}.
Therefore, while Landau's solution is for an \ivp, we may obtain the Landau growth rate from a collisional \evp, and so we consider \evp s for computation convenience.

Many problems, such as nonlinear problems or those in complicated geometry, must be solved numerically, so we discuss the system with discrete velocity space.
In the continuous case, the collisionless limit ($\cfe\to0$) is singular and corresponds to the formation of infinitesimally fine scales in velocity space.
The solution changes from smooth collisional behaviour for $\cfe>0$ to singular collisionless behaviour at $\cfe=0$.
Once discretized, there are no arbitrarily small velocity scales, and the solution must vary continuously with $\cfe$ (because the problem becomes a finite matrix eigenvalue problem).
Now we expect behaviour to change at a finite $\etacrit$ which depends on the resolution of velocity space.
When $\cfe<\etacrit$, the velocity space structure is too fine to be resolved and the behaviour is \cvk-like, while when $\cfe>\etacrit$ the solution is smooth and collisional. 

\subsection{Landau damping in a discrete system}

We discretize \Ex{eq:operator_eqn} and solve the matrix eigenvalue problem
\begin{align}\label{eq:dfdt=MF}
  \omega \hat{\distpert}_j = M_{jl}\hat{\distpert}_l,  
\end{align}
where the matrix $\Mat$ is the discretization of $L_{\cfe}$, and $\hat{\distpert}_j$ is the vector of values of $\hat{\distpert}(v_j)$ on the discretization grid $\{v_0,\ldots,v_N\}$.
There is an implied sum over the repeated index $l$.
Without collisions, $\Mat$ is real so its eigenvalues occur in complex-conjugate pairs, and the discretization cannot capture Landau damping.
\Figs \ref{f.fsystem} and \subref{fig:VKM} show typical behaviour for a discretized collisionless system.  
In \fig\ref{f.fsystem} we plot the growth rate calculated from \Ex{eq:dfdt=MF} where we discretized $\hat{\distpert}$ on the grid of $N$ Gauss--Hermite points described in \sec\ref{sec:HermiteSystem} for $N=10$ and $N=100$.
The numerical growth rate is plotted in blue and red against the exact dispersion relation in dashed green.
The calculation is qualitatively correct for positive growth rates, but requires $\O(100)$ grid points for quantitative agreement.
Regardless of the resolution, the method fails to produce negative growth rates. 
\Fig\ref{fig:VKM} shows the fastest growing mode for three wavenumbers near the critical wavenumber where the growth rate crosses zero to become negative.
As the wavenumber increases, the growth rate tends to zero and the structure in the eigenmodes becomes finer until the grid is unable to resolve them.
(The eigenmode shown is calculated on a higher resolution uniform grid.)
This corresponds to the formation of a \cvk\ mode as the growth rate decreases to zero.
For zero growth rates, the eigenvectors are discretizations of \cvk\ modes: numerical approximations to delta functions located at the discretization grid points.
The eigenvalue spectrum thus becomes dense along the real line in the limit of increasing resolution.

By including a collision term in $\Mat$ in \Ex{eq:dfdt=MF}, we can find decaying eigenmodes for the driven system which approach the correct Landau decay rate, analogous to the eigenmodes for the undriven system found by \citeauthor{LenardBernstein} (see \sec\ref{sec:hypercollisions}).
%In the \ivp\ in \sec\ \ref{sec:init-value-probl}, \fig\ \ref{fig:ivp_contour}(b), we see the initial conditions propagate to fine velocity scales where they are damped, leaving an eigenmode on the long-time limit.
These eigenmodes become more singular as collisionality is reduced.
\Fig\ref{fig:dougherty_intro} shows the slowest decaying eigenmode for $k=4.1\alpha_i$ calculated on a Hermite grid using the \dop\ collision operator introduced in \sec\ref{sec:lb} with different collision frequencies $\cfe$.
Unlike in the continuous system, where any finite collisionality regularizes the problem, some minimum collisionality is required in the discrete case. 
This minimum is related to resolution, and we show in \sec\ref{sec:ParameterChoices} that discrete approximations to $L_\cfe$ based on Hermite polynomials correctly capture decaying modes provided that the number of modes $N$ exceeds some $N_{\crit}(\cfe)$, a critical number needed to capture the roll-over of the Hermite spectrum of the eigenfunction at the point where collisions become dominant.
Crucially $N_\crit\to\infty$ as $\cfe\to0$.
Natural discretizations of $L_0$ lose Landau damping as first $N$ is fixed, then $\cfe\to0$ is taken.
To correctly capture Landau damping, one must set $\cfe>0$ so that $L_\cfe$ has 
a discrete spectrum of square-integrable eigenfunctions, and then choose $N$ to satisfy $N>N_\crit(\cfe)$.

The above criterion $N\geq N_{\crit}(\cfe)$ implies that $N$ must be too large for efficient computation when $\cfe$ is sufficiently small to represent realistic fusion plasmas.
In this \chp\ we investigate the effectiveness of a hypercollisional operator in capturing weakly collisional plasmas, and find that accurate calculations are possible with a few tens of degrees of freedom in velocity space.
An efficient formalism for velocity space in \Ex{e.8} uses the Hermite spectral representation.
Hermite polynomials have been commonly used in kinetic theory of neutral particles 
and of plasmas
as they are orthogonal \wrt\ a Maxwellian weight, and are the eigenfunctions of the linearized Boltzmann collision operator for Maxwell molecules (a convenient model of particles with an inverse 5th power repulsion \cite{Maxwell67}).
Hermite polynomials are also the eigenfunctions of the \lb\ and \dop\ collision operators, two \fp-type collision operators that are suitable for representing weakly collisional plasmas.
We introduce a higher-order hypercollisional operator by iterating the \dop\ operator. This gives numerically accurate results for very low computational cost, (\ie\ it has a low value of $N_{\text{crit}}$) and so allows a coarse resolution to be used.
We also study the Hermite spectrum of the eigenmodes, and investigate the effect of hypercollisions in the initial value problem.
The operator is formulated in Hermite space and so is {simplest} to apply on a Gauss--Hermite velocity grid (see \sec\ref{sec:TruncatedHermiteSeries}). 
However, we demonstrate later that the hypercollisional operator may be used with any velocity space discretization.

\section{Hermite representation}

\label{sec:hermite-expansion}
The use of spectral expansions of the distribution function has a long
history in kinetic theory. It is natural to consider functions of
velocity that are orthogonal \wrt\ the Gaussian weight function that
arises from the Maxwell--Boltzmann equilibrium distribution.
\citet{Burnett35,Burnett36} used a combined expansion in spherical
harmonics and Sonine polynomials to greatly simplify the computation of
the collision integrals that arise in the calculation of the viscosity
and thermal conductivity in the Chapman--Enskog expansion. 
The standard Hermite
polynomials are orthogonal with respect to a Gaussian weight function in one dimension
\citep{AbramowitzStegun} so
\citet{Grad49Note,Grad49Kinetic,GradHandbuch} introduced sets of tensor
Hermite polynomials as a Cartesian alternative to Burnett's expansion.
Both expansions convert an integro-differential kinetic equation into
an infinite hierarchy of partial differential equations in $\x$ and
$t$ for the expansion coefficients.

The same expansion in Hermite polynomials for velocity space, and in Fourier modes for physical space,
was used in early simulations of the \oneplusoned\ \vps, 
such as by \citet{Armstrong67}, \citet{Grant67} and \citet{Joyce71}, albeit
with different forms of dissipation and,  inevitably, much lower resolution than is currently feasible.

However, through disappointment with the available velocity-space resolution \citep{Gagne77}, and with
higher dimensional models becoming computationally feasible, interest turned instead
to particle-in-cell (PIC) methods. These represent the distribution function using a set of
macro-particles located at discrete points ($\x_i, \v_i)$ in phase space, each of which represents many physical ions or electrons \citep{Dawson83,Hockney88,Birdsall04}.
The method exploits the structure of the \lhs\ of the kinetic equation \Ex{eq:FokkerPlanckEquation}
as a derivative along a characteristic in phase space. A PIC method evolves the solution by propagating macro-particles along their characteristics, analogous to the Lagrangian formulation of fluid dynamics.
The representation of the continuous distribution function $\Ft(\x,\v,t)$ by a discrete set of $n$ macro-particles creates
an $O(1/\sqrt{n})$ sampling error, sometimes called ``shot noise'', that creates particular difficulties in the tail
of the distribution where $\Ft$ is much smaller than its maximum value. An information preservation approach
has been developed to alleviate this problem in particle simulations of neutral gases \citep{FanShen01} but it has not
yet been employed for plasmas. 

Instead, more recent  multi-dimensional gyrokinetics codes have returned to Eulerian representations
of velocity space, using fixed grids either for velocities \citep{Gene,GKW} or for energy and pitch angles \citep{GYRO,GS2}.
However Hermite polynomials have been used to develop reduced kinetic \citep{Zocco11} 
and gyrofluid models \citep{Hammett93,Parker95},
as well as the Hermite expansion coefficients being used to characterize velocity space behaviour \citep{Schekochihin14,Kanekar14,Plunk14} in analogy with \textsc{Kolmogorov}'s 1941 theory of hydrodynamic turbulence \cite{Kolmogorov41} expressed using the energy spectrum in Fourier space.
This has reignited interest in using Hermite polynomials for computation in new reduced-dimension \citep{Hatch13,Viriato} gyrokinetics codes, and our fully five-dimensional \sgk\ described in \chp~\ref{sec:SpectroGK}. 
A recent comparison between a PIC code and a Fourier--Hermite code
shows that the latter is much more computationally efficient for the 1+1D Vlasov--Poisson system when high accuracy is desired
\citep{CamporealeDelzannoBergenMoulton13}.

%%%In this \chp\ we use a Hermite representation for the velocity space dependence of the small perturbation $\distpert$ to the distribution function $F(z,v,t)=F_0(v)+\distpert(v,t)e^{ikz}$. 

\subsection{Hermite functions}
\label{sec:hermite_functions}
\label{sec:HermiteFunctions}

We introduce the standard Hermite polynomials $H_m(v)$, and the re-normalized Hermite functions $\phi^m(v)$, defined by
\begin{align}\label{eq:HermFunDef}
  H_m(\vpara) = (-1)^m \e^{\vpara^2}\fd{^m}{\vpara^m}\lp \e^{-\vpara^2}\rp,
  \hspace{1cm}
  \phi^m(\vpara) = \frac{H_m(\vpara)}{\sqrt{2^m m!}}, 
\end{align}
for $m=0,1,2,\ldots$.  The $\phi^m$ are orthonormal with respect to
the weight function $\e^{-v^2}/\sqrt{\upi}$, equal to our
dimensionless Maxwell--Boltzmann distribution. Introducing the dual
Hermite functions $\phi_m(v)=\e^{-v^2}\phi^m(v)/\sqrt{\pi}$ thus
establishes the bi-orthonormality conditions
\begin{align}\label{eq:orthogonality}
  \int_{-\infty}^{\infty} \d v \, \phi_n(v)\phi^m(v) =
   \delta_{nm} \mbox{ for } m,n \ge 0.
\end{align}
Each $\phi_m$ satisfies the velocity space boundary conditions $\phi_m(v)\to0$ as $v\to\pm\infty$.
They form a complete set for functions $\distpert(v)$ that are analytic in a strip centred on the real axis and satisfy the
decay condition $|\distpert(v)| < c_1 \exp{(-c_2v^2/2)}$ for some constants $c_1 > 0$ and $c_2 > 1$ \citep{Holloway96,Boyd01}
so we may expand the perturbed distribution function as
\begin{align}
	\distpert(v) = \sum_{m=0}^{\infty} g_m \phi_m(v), 
	\hspace{1cm}\mbox{ with }\hspace{1cm} 
	g_m = \int_{-\infty}^{\infty} \d v \, \distpert(v) \phi^m(v). \label{eq:GradF}
\end{align}
This is an asymmetrically weighted Hermite expansion, in the terminology of \citet{Holloway96}. 
The symmetric expansion uses the orthogonal functions 
$\psi_m(v) = e^{-v^2/2} \phi^m(v) \pi^{-1/4}$.
%$\psi_m(v) = \exp(-v^2/2) \phi^m(v) \pi^{-1/4}$.
In kinetic theory problems one may choose the width of the Gaussian weight function in the Hermite orthogonality
condition to differ from the width of the Maxwell--Boltzmann distribution \citep{Tang93,Schumer98}, but \citet{LeBourdiecEtAl06} found that taking the two Gaussians to have the same width gives optimal accuracy. 
The expansion \Ex{eq:GradF} then coincides with the expansion
originally introduced by \citet{Grad49Note,Grad49Kinetic,GradHandbuch}.

The first few expansion coefficients in \eqref{eq:GradF} have physical interpretations, 
\begin{gather}\label{eq:macroQ}
  \begin{split}
    \distpert_0 &= \intii \d v~ \distpert(v) = n_i, %  \rho, 
    \\ 
    \distpert_1 &= \intii \d v~ \sqrt{2}v\distpert(v)  = \sqrt{2} n_i u_i, % \rho u, 
    \\
    \distpert_2 &= \intii \d v~ \sqrt{2}\lp v^2 - \frac{1}{2}\rp
    \distpert(v) 
    %= 2\sqrt{2}{\cal E} - \frac{1}{\sqrt{2}} n_i , %\rho, 
    %= 2\sqrt{2}{\cal E} - \sqrt{2} n_iT_i , %\rho, 
    = \sqrt{2}(2{\cal E}_i - n_iT_i) , %\rho, 
  \end{split}
\end{gather}
where $n_i$, %$\rho$, 
$u_i$, and $\mathcal{E}_i$ are the perturbations due to $\distpert$ in the ion fluid density, fluid velocity, and energy density respectively, and $T_i=1/2$ is the background \eqm\ temperature in our dimensionless variables.
These expressions are of particular importance for efficient computation, since integrals required to find the current and charge densities in Maxwell's equations \eqref{eq:MaxwellSummary} are now single Hermite coefficients, rather than integrals approximated as sums over velocity space grids.

In contrast to low $m$ moments, coefficients for higher $m$ are associated with fine scales and kinetic, not fluid, moments:
the asymptotic behaviour of the symmetrically weighted Hermite functions for large $m$ is \cite{AbramowitzStegun,Olver10:18117},
%$\psi_m(v) \sim A_m \cos(v \sqrt{2m} - m\upi/2)$ for constants $A_m$. 
\begin{align}
  \psi_m(v)\sim A_m \cos(v\sqrt{2m}-m\pi/2) ,
	\hspace{1cm}\mbox{ with }\hspace{1cm} 
  A_m = \frac{2^{1-m/2}\Gamma(m)^{1/2}}{\pi^{1/4}m^{1/2}\Gamma(m/2)} .
  \label{eq:OlverLimits}
\end{align}
Hermite functions with
larger $m$ thus represent progressively finer scale structures in $v$. 
Hermite functions with adjacent $m$ values satisfy the recurrence relation
\begin{align}\label{eq:rr}
 \vpara\phi_m(\vpara) = \sqrt{\frac{m+1}{2}}\phi_{m+1}(\vpara) + \sqrt{\frac{m}{2}}\phi_{m-1}(\vpara).
\end{align}
This may be written as a matrix \eqn\ using the Jacobi matrix $\Jmat$,
\begin{align}
  v\phi^m(v) = \sum_{n=0}^{\infty} J_{mn}\phi^{n}(v),
  \label{eq:rrJacobi}
\end{align}
where $\Jmat$ is infinite, symmetric, and tridiagonal with nonzero entries $J_{m,m-1}=J_{m-1,m}=\sqrt{m/2}$ \cite{Gil07}. 
Derivatives of Hermite functions may be expressed as
\begin{align}
  \fd{\phi_m}{v} = -\sqrt{2(m+1)}\phi_{m+1},
	\hspace{1cm}\mbox{ and }\hspace{1cm} 
  \fd{\phi^m}{v} = \sqrt{2m}\phi^{m-1}.
  \label{eq:HermiteDerivatives}
\end{align}
The relations \eqref{eq:rr} and \eqref{eq:HermiteDerivatives} allow us to develop a pure spectral discretisation in velocity space. Once initial conditions have been expressed as coefficients in a Hermite expansion, we do not need
a collocation grid in $v$ analogous to that required to compute the nonlinear terms in $\x$ space (see \sec\ref{sec:5DCodeGKE}). % and \sec\ref{sec:DiscretizedSystem}).
We later use initial conditions that have simple expressions as truncated Hermite expansions, such as the Maxwellian in \chp~\ref{sec:FreeEnergyFlowAndDissipation}, but generically one must evaluate the Hermite transform of the initial conditions.

\begin{figure}[tbp]
  \centering
  \subfigure[\label{fig:DeltaIntro}]{\includegraphics[width=0.49\textwidth]
{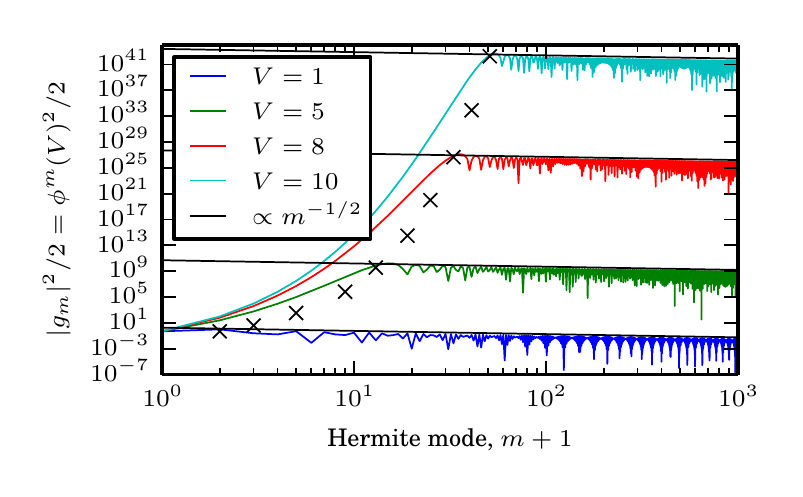}}
%{weighted_hermitec_functions_paper.pdf}}
\subfigure[\label{fig:weighted_hermitec_functions}]{\includegraphics[width=0.49\textwidth]
{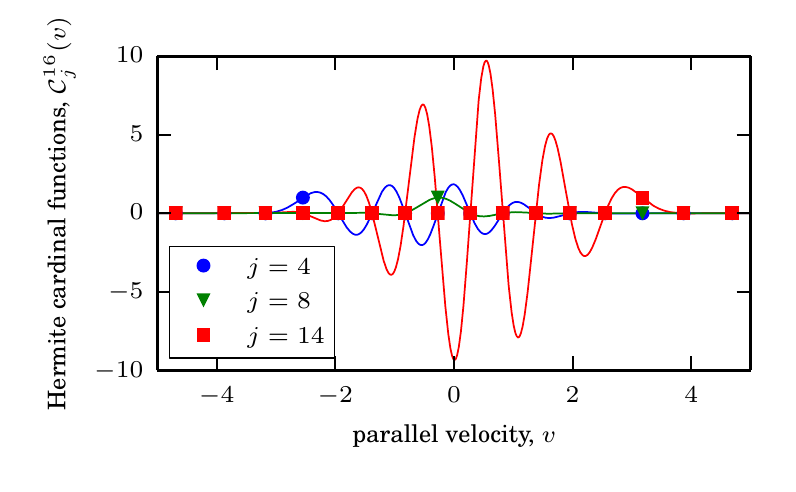}}
\caption[Hermite spectra for delta functions, and the Hermite cardinal functions.]{(a) Hermite spectra of $\delta(v-V)$. Black lines show $m^{-1/2}$ decay, and crosses give a lower bound for the position of the maxima $m=V^2/2-1$. (b) Three examples of 16th order Hermite cardinal functions, with markers located at Hermite grid points. Each function is zero at all grid points but one, and one at the other. \label{fig:hermitec_and_delta}}
\end{figure}

Finally, we shall find the Hermite expansions of delta functions useful.
Combining the two expressions in \eqref{eq:GradF} gives
 \begin{align}
  \distpert(v) = \sum_{m=0}^\infty \distpert_m\phi_m(v) = \int_{-\infty}^\infty\d V\ \lp\sum_{m=0}^\infty\phi^m(V)\phi_{m}(v)\rp \distpert(V),
\end{align}
from which we deduce the resolution of the identity formula
\begin{align}
  \delta(v-V) = \sum_{m=0}^\infty \phi^m(V)\phi_m(v).
  \label{eq:HermiteDelta}
\end{align}
Therefore the expansion coefficients for the delta function $\delta(v-V)$ are $\distpert_m=\phi^m(V)$. 
The spectrum $|\distpert_m|^2/2$, plotted in \Fig\ref{fig:DeltaIntro}, increases super-exponentially until a maximum is reached, and then decays like $m^{-1/2}$, which follows from \Ex{eq:OlverLimits} and Stirling's formula \cite{AbramowitzStegun}
\begin{align}
  \log \Gamma(m+1) =
	\log m! = m\log m -m + \frac{1}{2}\log(2\pi m)  + \O(m^{-1}).
  %\label{}
\end{align}
If the maximum is at $m^*$, then 
\begin{align}
  \left|\phi^{m^*}(V)\right| > 
  \left|\phi^{m^*+1}(V)\right| >
  \left| \sqrt{\tfrac{2}{m^*+1}}V\phi^{m^*}(V)\right| - \left|\sqrt{\tfrac{m^*}{m^*+1}}\phi^{m^*-1}(V)\right|,
  \label{eq:PhiInequalityStep}
\end{align}
where the first inequality is by definition of a maximum, and the second is due to the recurrence relation \eqref{eq:rr} and the triangle inequality.
Rearranging \eqref{eq:PhiInequalityStep} gives
\begin{align}
  \lp \sqrt{\tfrac{2}{m^*}}|V|
  - \sqrt{\tfrac{m^*+1}{m^*}}\rp
  \left|\phi^{m^*}(V)\right|  %\label{}
  < \left|\phi^{m^*-1}(V)\right|
  < \left|\phi^{m^*}(V)\right|,
\end{align}
which yields the bound $m^* > V^2/2 - 1$. This is also plotted in \Fig\ref{fig:DeltaIntro}.

\subsection{Truncated Hermite series}
\label{sec:TruncatedHermiteSeries}
We approximate the distribution function perturbation \Ex{eq:GradF} by the truncated Hermite series
\begin{align}\label{eq:HermExpTrunc}
  \distpert(\vpara,t) =\sum_{m=0}^{N}\distpert_m(t)\phi_m(\vpara),
\end{align}
which is a polynomial of degree $N$ multiplied by $F_0$.
We expect $\distpert$ to decay as $e^{-v^2}$, so this expansion
 provides a spectrally accurate approximation
to (\ref{eq:GradF}) using a convenient bi-orthonormal basis.

%Analogously to \Ex{eq:AnDef}, we calculate $a_n$ as sums of $v_j$ via Gauss--Hermite quadrature \cite{StoerBulirsch}
We calculate moments via the Gauss--Hermite quadrature formula \cite{StoerBulirsch}
\begin{align}\label{eq:GHQuad}
  %\int_{-\infty}^{\infty} p(v) F_0(v) \d v = \sum\limits_{j=1}^{N}\frac{p(v_{j})}{N\lp\phi^{N-1}\lp v_{j}\rp\rp^2},
  \int_{-\infty}^{\infty} p(v) F_0(v) \d v = \sum\limits_{j=0}^{N}w_j^N p(v_{j}),
  \quad w_j^N = \ls N \phi^{N}\lp v_{j}\rp^2\rs^{-1},
\end{align}
where the quadrature points $v_j$ are the $N+1$ roots of
the Hermite polynomial $H_{N+1}(v)$.
Equation \eqref{eq:GHQuad} holds exactly for all polynomials $p(v)$ with degree $2N+1$ or
less. 
Setting $p(v) = \phi^m(v)\phi^n(v)$ and using the bi-orthogonality condition \eqref{eq:orthogonality} establishes that Hermite polynomials of degree at most $N$ form a finite orthogonal set:
\begin{align}\label{eq:DiscOrth}
  %\int_{-\infty}^{\infty} \phi^m(v)\phi^n(v) F_0(v) \d v = 
  %\sum\limits_{j=1}^{N}\frac{\phi^m(v_j)\phi^n(v_j)}{N\lp\phi^{N-1}\lp v_{j}\rp\rp^2} = \delta_{nm}, \quad n,m\in\{0,1,\ldots,N-1\}.
  \sum\limits_{j=0}^{N}w^N_j\phi^m(v_j)\phi^n(v_j) = \delta_{nm}, 
	\hspace{1cm}\mbox{ for }\hspace{1cm} 
	n,m\in\{0,1,\ldots,N\}.
\end{align}
Dual to this is the discrete resolution of the identity formula, obtained by multiplying \Ex{eq:DiscOrth} by $\phi^m(v_l)$ and summing over $m$,
\begin{align}%\label{eq:GHQuad}
  %\sum\limits_{n=0}^{N-1}\frac{\phi^n(v_j)\phi^n(v_l)}{N\lp\phi^{N-1}\lp v_{j}\rp\rp^2} = \delta_{jl}, \quad j,l\in\{1,2,\ldots,N\}.
  \sum\limits_{m=0}^{N}w^N_j \phi^m(v_j)\phi^m(v_l) = \delta_{jl}, 
	\hspace{1cm}\mbox{ for }\hspace{1cm} 
	j,l\in\{0,1,\ldots,N\}.
\end{align}
These formulae motivate the introduction of the Hermite cardinal polynomials
\begin{align}
  %\hc^N_j(v) =  \sum\limits_{n=0}^{N-1}\frac{\phi^n(v_j)\phi^n(v)}{N\lp\phi^{N-1}\lp v_{j}\rp\rp^2},
  \hc^N_j(v) =  \sum\limits_{m=0}^{N}w_j^N\phi^m(v_j)\phi^m(v),
  \label{eq:hermitec}
\end{align}
polynomials of degree $N$ that are analogous to Whittaker's sine cardinal function for 
\linebreak
Fourier series \cite{Whittaker28,McNameeStengerWhitney71,Stenger81} in that $\hc^N_j(v_l)=\delta_{jl}$. %for $v_l$ the roots of the Hermite polynomial $H_{N}(v)$. 
Hermite cardinal polynomials lie in the span of Hermite polynomials up to degree $N$, being the Lagrange interpolation polynomials for the Gauss--Hermite quadrature points. 
We also define the Hermite cardinal functions
\begin{align}
  \hcf^N_j(v) = \frac{F_0(v)}{F_0(v_j)}\hc^N_j(v).
  %\label{}
\end{align}
These also satisfy $\hcf^N_j(v_l)=\delta_{jl}$ for Hermite roots, but decay like a Maxwellian as $v\to\pm\infty$. 
Three examples of $\hcf^{16}_j(v)$ are plotted in \fig\ref{fig:hermitec_and_delta}(b).
With these, our truncated expansion \Ex{eq:HermExpTrunc} becomes
\begin{align}\label{eq:HermExpTrunc1}
  \distpert(\vpara,t) =\sum_{m=0}^{N}\distpert_m(t)\phi_m(\vpara)
  = \sum_{j=0}^{N} \distpert(v_j,t)\hcf^{N}_j(v),
\end{align}
and we may express our truncated approximation of $\distpert$ either in terms of the coefficients $\distpert_m(t)$ or the function values $\distpert(v_j,t)$. 
As \Ex{eq:HermExpTrunc1} shows, this truncation is equivalent to a discretization of $\distpert$ on the grid of $N+1$ Gauss--Hermite
quadrature points $v_j$ with coefficients $\distpert_m$ related to the function values $\distpert(v_j)$ through an invertible linear transformation.

The optimal method for calculating the Hermite roots $v_j$ uses a truncation of the Jacobi matrix from \Ex{eq:rrJacobi}. 
The leading $N+1\times N+1$ submatrix $\Jmat_{N}$ of $\Jmat$ has the property that its characteristic polynomial is proportional to the Hermite function $\phi^{N+1}(v)$ \cite{GautschiBook,Totik05}. 
Therefore the roots of the Hermite polynomial $\phi^{N+1}(v)$ may be readily computed as the eigenvalues of the symmetric tridiagonal matrix $\Jmat_{N}$ using the QR algorithm \cite{TrefethenBau}. 

\subsection{Hermite system}
\label{sec:hermite-system}
\label{sec:HermiteSystem} % my autocomplete only picks up single words

Substituting the truncated expansion (\ref{eq:HermExpTrunc}) into (\ref{eq:ddt_operator_eqn}) 
and projecting onto each of the $\phi_m$ in turn gives the system
\begin{gather}  \label{eq:asystem}
  \pd{\distpert_m}{t} + \frac{i}{\kpara}\lp \wn \delta_{m0} 
 + \frac{\wT}{\sqrt{2}}\delta_{m2}\rp \, \distpert_0
+ \frac{i}{\sqrt{2}}\delta_{m1} \, \distpert_0
+i\lp\sqrt{\frac{m+1}{2}}\distpert_{m+1} + \sqrt{\frac{m}{2}}\distpert_{m-1}\rp = 0,
\end{gather}
where we have neglected collisions and eliminated $\varphi = \distpert_0$ using the quasineutrality condition.
We have also rescaled time by $t\mapsto kt$ to simplify the algebra in subsequent sections.
The recurrence relation \Ex{eq:rr} allows the streaming term $v\distpert$ in  (\ref{eq:Lepsdef}) to be expressed as a coupling between mode $m$ and modes $m \pm 1$. 
The electrostatic response and the driving from background gradients appear only in the first few equations for $m \in \{0,1,2\}$.
\Eqn\ \Ex{eq:asystem} is the problem treated by \citet{Ng99} in their analytic work, but generalized to include density and temperature gradients. 

Equation \Ex{eq:asystem} may be rewritten as the matrix equation
\begin{align}\label{eq:amatrix}
  \pd{\distpert_m}{t} = -iM_{mn}\distpert_n,
\end{align}
for the %$N+1$ by $N+1$ 
matrix
\newcommand{\0}{ {\color{white} 0 } }
\begin{align}\label{eq:matrix_L}
	\begin{split}
  \Mat 
  &= \lp
   \begin{array}{cccccc}
   \frac{\wn}{\kpara} & \frac{1}{\sqrt{2}} &  \0 & \0  & \0 & \0  \\  
   \frac{1%\tau_e
}{\sqrt{2}} + \frac{1}{\sqrt{2}}& \0  & 1  & \0  & \0 & \0 \\
   \frac{\wT}{\kpara\sqrt{2}} & 1  &  \0 & \sqrt{\frac{3}{2}} & \0 & \0 \\
   \0  & \0  & \sqrt{\frac{3}{2}}  & \0   & \ddots &\0 \\
   \0 &\0 &\0 & \ddots & \0 &\sqrt{\frac{N}{2}} \\
   \0 &\0 &\0 & \0     & \sqrt{\frac{N}{2}} & \0
   \end{array}
   \rp 
	 \\
  &= \Jmat_N 
  + \lp
   \begin{array}{cccccc}
     \frac{\wn}{\kpara} & \phantom{\frac{1}{\sqrt{2}}} &  \0 & \0  & \0 & \0  \\  
   \frac{1%\tau_e
   }{\sqrt{2}} & \0  & \phantom{1}  & \0  & \0 & \0 \\
\frac{\wT}{\kpara\sqrt{2}} & \phantom{1}  &  \0 & \phantom{\sqrt{\frac{3}{2}}} & \0 & \0 \\
\0  & \0  & \phantom{\sqrt{\frac{3}{2}}}  & \0   & \phantom{\ddots} &\0 \\
\0 &\0 &\0 & \phantom{\ddots} & \0 & \phantom{\sqrt{\frac{N}{2}}} \\
\0 &\0 &\0 & \0     & \phantom{\sqrt{\frac{N}{2}}} & \0
   \end{array}
   \rp .
	\end{split}
\end{align}
This is %$\Jmat_N$, 
the $N+1$th order leading submatrix of the Jacobi matrix for Hermite polynomials $\Jmat$ in \Ex{eq:rrJacobi}, with three additional non-zero elements in the first column due to the driving and the electrostatic
response.
Thus $\Mat$ is a rank-one update to $\Jmat_N$, that is $\Mat=\Jmat_N + xy^T$ for vectors $x=(\wn/k,1/\sqrt{2},\wT/k,0,\dots,0)^T$ and $y^T=(1,0,\dots,0)$.  Theory exists for inverting matrices of this form \cite{ShermanMorrison49} and for finding eigenvalues in special cases such as when $\Mat$ is symmetric ($x=y$) \cite{Bunch78} or when $x$ or $y$ are eigenvectors of $\Jmat_N$ \cite{Zhou11}, 
but it does not seem possible to exploit this structure for the asymmetric eigenvalue problem with a general rank-one update.

%%%We also use a truncated Hermite series \Ex{eq:HermExpTrunc} to approximate $f$. 
%%%This is equivalent to setting $a_{N+1}$ and all higher modes to zero, and so corresponds to truncating \Ex{eq:matrix_L} after the first $N+1$ rows and columns.  

\begin{figure}[tb]
  \centering
%  \subfigure[Growth rate against parallel wavenumber for the system (\ref{eq:amatrix}). $N$ is the number of terms included in the series (\ref{eq:HermExpTrunc}).  \label{f.asystem}]{\includegraphics[width=0.49\textwidth]{asystem_gr.ps}}
  \subfigure[\label{fig:EFasystem}]{
    \includegraphics[width=0.48\textwidth]
%{asystem_hs_k=2.ps}}
{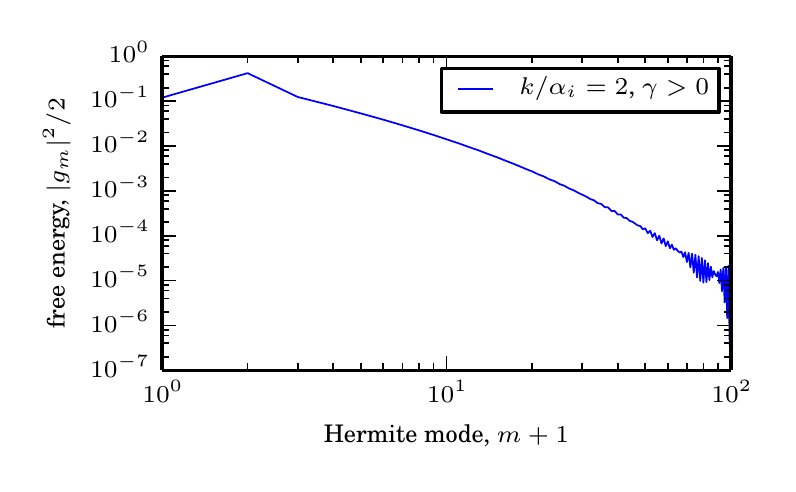}}
  \subfigure[\label{fig:LargeKspec}]{
    \includegraphics[width=0.48\textwidth]
%{asystem_hs_k=6.ps}}
{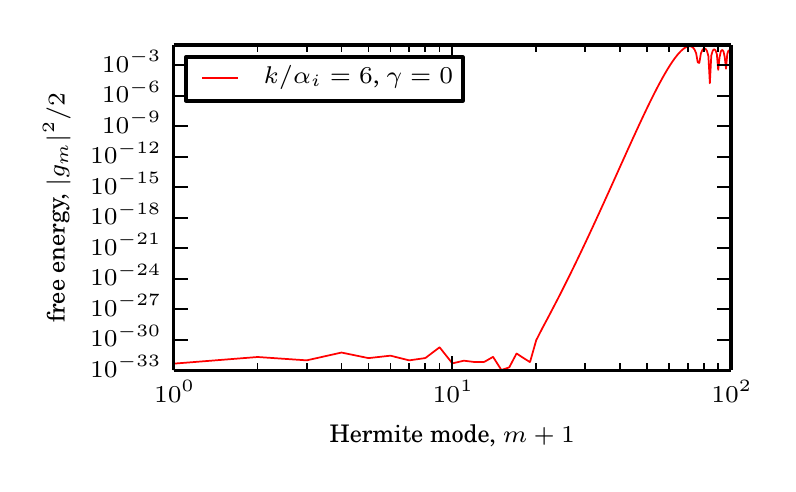}}
%\caption{Hermite spectra for two values of the parallel wavenumber, $\kpara=2$ and $\kpara=6$. $N=100$.\label{fig:EFasystem}}
\caption[Hermite spectra of growing and decaying eigenmodes.]{Hermite spectra for two values of the parallel wavenumber, (a) $\kpara/\alpha_i=2$ and (b) $\kpara/\alpha_i=6$, calculated using the truncation (\ref{eq:HermExpTrunc}) with $N=100$. Spectra are plotted against $m+1$ to not omit the zeroth mode on the logarithmic scale.\label{fig:hermite_truncated}}
\end{figure}

The growth rates shown in \fig\ref{f.fsystem} are calculated by finding the eigenvalues of the truncated matrix $\Mat$. As explained above, this is equivalent to discretizing $\distpert(v)$ on a Gauss--Hermite grid in velocity space.
However we may now explain these results in terms of the truncated representation of $\distpert$ (\ref{eq:HermExpTrunc}).
Putting $m=N$ in \Ex{eq:amatrix} gives
\begin{align}
  \sqrt{\frac{N+1}{2}}\distpert_{N+1} + \sqrt{\frac{N}{2}}\distpert_{N-1}% = -\frac{\omega}{\alpha_i\kpara} \, a_{N},
= \omega \distpert_{N},
\label{eq:lastrow}
\end{align}
but a simple truncation sets  $\distpert_{N+1}=0$ and is only valid when $ \distpert_{N+1}$ is negligibly small.\footnote{More generally, the coefficient $\distpert_{N+1}$ in (\ref{eq:lastrow}) may be approximated by a linear combination of the coefficients $\distpert_0,\distpert_1,\ldots,\distpert_{N}$.
This assumes that the smallest scale structure may be extrapolated from larger scales, an
approach studied by Smith \cite{Smith}.}
To see if this is so we plot the Hermite spectrum for $N=100$ and two wavenumbers $\kpara/\alpha_i=2$ and $\kpara/\alpha_i=6$ in \fig\ref{fig:hermite_truncated}. 
The spectrum for $\kpara/\alpha_i=2$ decays %as $n^{-1/2}$ 
from $\distpert_m \sim 1 $ at small $m$ to $\distpert_m \sim 10^{-3}$ at $m=100$, so the last few terms retained are roughly a thousand times smaller than the first few terms.
The spectrum for $\kpara/\alpha_i=6$ does not decay.
The $\distpert_m$ are initially negligible, but grow rapidly for $12\lesssim m \lesssim 40$ and then saturate for $m \gtrsim 40$. 
The spectrum is similar to the growing part of the spectra of delta functions shown in \fig\ref{fig:DeltaIntro}. 
We may expect this as $k=6\alpha_i$ exceeds $k_{\textrm{crit}}$, the critical wavenumber at which the \cvk\ mode forms (see \fig\ref{fig:VKM}). 
Therefore the truncation is valid for growing modes, but not for decaying modes.

\subsection{Hermite representation of a Case--Van Kampen mode}

The behaviour of the Hermite spectrum for $k/\alpha_i=6$ may be understood by considering the free streaming \eqn 
%%%\begin{align}
%%%  \pd{\distpert}{t} + v \pd{\distpert}{z} = 0,
%%%\end{align}
\begin{align}\label{eq:free_streaming}
  \pd{\distpert}{t} + ikv\distpert = 0 ,
 %(v - \omega)f  = 0 .
  %\label{}
\end{align}
the kinetic \eqn\ in \Ex{eq:KineticEquationChapter3} with no driving from background gradients, collisions, or electrostatic response. 
%%%Seeking a plane wave solution as in \Ex{eq:SpatialFourierMode} 
%%%and rescaling time as in \sec\ref{sec:HermiteSystem} 
%%%we obtain
Seeking time periodic solutions proportional to $e^{-i\omega t}$ and rescaling time as in \sec\ref{sec:HermiteSystem} gives
\begin{align}%\label{eq:free_streaming}
 % \pd{f}{t} + vf = 0 .
  (v - \omega)\hat{\distpert}  = 0 .
  %\label{}
\end{align}
This \eqn\ admits the generalized eigenfunctions $\hat{\distpert} = \delta(v-\omega)$, which are equivalent to the \cvk\ modes in the original system. 
Indeed the collisionless and undriven form of \Ex{e.8} may be reduced to \Ex{eq:free_streaming} using a different integral transform in place of the Fourier transform \cite{MorrisonShadwick94}.

The calculations in \sec\ref{sec:hermite-system} hold for \eqn~(\ref{eq:free_streaming}) if we omit the electrostatic response term and set $\wn=\wT=0$. 
Now $\Mat=\Jmat_N$ %in (\ref{eq:matrix_L}) 
and the truncated matrix eigenvalue problem becomes,
\begin{align}
\label{eq:everyrow}
  \sqrt{\frac{m+1}{2}}\distpert_{m+1} + \sqrt{\frac{m}{2}}\distpert_{m-1} = \omega \, \distpert_{m},
\end{align}
for $m=0,\ldots,N-1$. The last row is modified because $\distpert_{N+1} \equiv 0$ due to the truncation. 
The matrix 
%%%\begin{equation}
%%%  \Jmat_{N} = \begin{pmatrix}
%%%   \0 & \frac{1}{\sqrt{2}} &  \0 & \0  & \0 & \0 & \0 \\  
%%%   \frac{1}{\sqrt{2}}& \0  & 1  & \0  & \0 & \0 & \0 \\
%%%   \0 & 1  &  \0 & \sqrt{\frac{3}{2}} & \0 & \0 & \0 \\
%%%   \0  & \0  & \sqrt{\frac{3}{2}}  & \0  & \sqrt{2} & \0 & \0 \\
%%%   \0  & \0  & \0  &\sqrt{2} & \0 & \ddots & \0 \\
%%%   \0 &\0 &\0 &\0 & \ddots & \0 & \sqrt{\frac{N}{2}}  \\
%%%   \0 &\0 &\0 &\0 & \0 & \sqrt{\frac{N}{2}} 
%%%   \end{pmatrix},
%%%\end{equation}
$\Jmat_N$ is the $N+1$th order leading submatrix of the Jacobi matrix for Hermite polynomials $\Jmat$ in \Ex{eq:rrJacobi}. %, $\Mat_{N+1}=\Jmat_{N+1}$.  
%For general orthogonal polynomials, Jacobi submatrices $\Jmat_{N+1}$ are used to calculate the grids and weights for quadrature rules as these depend on the eigenvalues and eigenvectors of $\Jmat_{N+1}$, and finding eigenvalues of symmetric tridiagonal matrices is efficient and stable \cite{Gautschi85,TrefethenBau}. 
%The characteristic polynomial of $\Jmat_{N+1}$ is proportional to the ${N+1}$th orthogonal polynomial, so that the eigenvalues of $\Jmat_{N+1}$ are the roots of that polynomial.  
The characteristic polynomial of $\Jmat_{N}$ is proportional to $\phi^{N+1}(v)$, so the eigenvalues of $\Jmat_{N}$ are the roots of $\phi^{N+1}(v)$.
This may also be seen by noting that the relation \Ex{eq:everyrow} coincides with the recurrence relation (\ref{eq:rr}) for $m=0,\ldots,N-1$, so 
$\phi^m(v)$ is a candidate eigenvector with components $(\phi^0(v),\ldots,\phi^N(v))$
and eigenvalue $v=\omega$. 
To satisfy the last row of the matrix equation we need
$\phi^{N+1}(v)=0$, which determines the eigenvalues as
the $N+1$ Gauss--Hermite quadrature points $v_j$ that satisfy $\phi^{N+1}(v_j) = 0$.
The distribution functions corresponding to the eigenvectors are
\begin{align}
  \distpert_j(v) = \sum_{m=0}^{N}\phi^m(v_j)\phi_m(v)
  %=  \ls w^{N+1}_j\rs^{-1} C^{N+1}_j(v)F_0(v)
  %=  (N+1)\ls\phi^N(v_j)\rs^2C^{N+1}_j(v)F_0(v)
  , \label{eq:CVKeigenfunc}
\end{align}
which are approximations of the delta functions $\delta(v-v_j)$ %in \eqn\ \Ex{eq:HermitecDelta}, %derived in \sec\ \ref{sec:hermite_functions}.
such that in the limit $N\to\infty$ we recover the Hermite representation of a delta function \Ex{eq:HermiteDelta}.
Moreover the Gauss--Hermite quadrature points expand to fill the real line because they spread outwards proportional to $\sqrt{N}$
as $N \to \infty$, and their spacing decreases as $1/\sqrt{N}$ \cite{BoydBook}. 
Thus the spectra of the matrices $\Jmat_N$ approach a continuous spectrum as $N \to \infty$, and the
corresponding eigenvectors approach the  Case--Van Kampen modes.

\section{Collisions}
\label{sec:collisions}

As we have seen in \sec\ref{sec:HermiteSystem},
discrete representations of collisionless dynamics require regularization to 
prevent infinitesimally fine scales forming in velocity space. It is natural to do this with physical collision operators, \ie\ to study weakly collisional plasmas.
The relevant collision operator is the Landau operator \cite{Landau36} which contains the long-range Coulomb potential rather than the 
%short-range Lenard-Jones \cite{LennardJones24} or 
Maxwell molecule inverse 5th power potentials \cite{Maxwell67} used in the kinetic theory of uncharged particles.
The Landau operator is not amenable to simple computation, 
so in \sec\ref{sec:lb} we consider model collision operators which share the properties of the Landau operator discussed in \sec\ref{sec:LandauCollOpAndProperties}. 
With the Coulomb potential, collisions are predominantly glancing, so velocity changes are small. 
This allows a Taylor expansion of the distribution function which leads to first and second-order velocity space derivatives in the Landau operator.   
Thus the operator produces diffusion in velocity space, which is effective at removing fine scales in the distribution.\footnote{It is also possible to remove fine scales using a velocity-dependent collision frequency \cite[\eg][]{Cercignani66,Struchtrup97}, but we do not discuss this.}
In Hermite space this corresponds to damping higher modes which represent smaller scales,
and provides a mechanism for $\distpert_{N+1}$ and higher to be made negligible.
However this requires that we resolve collisional scales, which we show in \sec\ref{sec:lb} to be prohibitively expensive.
To circumvent this we use the separation of scales of the Hermite system (\ref{eq:asystem}): driving and the Boltzmann response are represented by low moments, and only particle streaming appears in the high moments. 
Free energy cascades from low moments to high moments via particle streaming, similar to the turbulent cascade through wavenumbers in hydrodynamic theory. Here however, the cascade is linear and reversible. Still, we find for our system that changes to the collision model do not affect the low moments, nor properties determined by these, such as the growth rate. Therefore in \sec\ref{sec:hypercollisions} we introduce a hypercollisional 
operator with high order derivatives in velocity space to artificially increase the scales on which collisions act and to make damping severe over a narrow range of high modes.
This leaves low modes unaffected, so the correct growth rate is found with fewer terms in the Hermite expansion.

This approach is similar to the hyperviscosity used in spectral methods for physical space in hydrodynamic turbulence \citep[\eg][]{%BasdevantLegrasSadournyBeland81,
PassotPouquet88,CeruttiMeneveauKnio00}
and plasma simulations \cite[\eg][]{PueschelDannertJenko10}.
A higher order term, typically $-(-\nabla^2)^n$ for some integer $n$, operating on the velocity or vorticity confines dissipation to a small range of high wavenumbers. %, with the majority of the spectrum being the correctly-behaving inertial range.
The approach of damping velocity space with a hypercollisional operator has also been considered before \cite[\eg][]{Joyce71,Knorr74,Shoucri77}.
These works use hypercollisions to prevent recurrence, the spurious reappearance of free energy in low moments after finite time,  %---a manifestation of which is the zero growth found in the high wavenumbers in \fig~\ref{f.fsystem}---
rather than as a means to reduce the necessary resolution in velocity space.

\subsection{The Landau collision operator and its properties}
\label{sec:LandauCollOpAndProperties}

\newcommand{\vbar}{\bar{\v}}

Particles interacting via the long-range Coulomb potential experience many small deflections due to glancing collisions. 
Expanding the distribution function in the Boltzmann binary collision operator in
a Taylor expansion for small velocity changes,
and cutting off the divergent contributions at small and large scales 
gives the Landau operator
\begin{align}\label{eq:Kulsrud05_Landau_coll_op}%\label{eq:Landau_coll_op}
  C[F] = \pd{}{\v}
\cdot\int K(\v-\v')\cdot\lp F(\v')\pd{}{\v}F(\v) - F(\v)\pd{}{\v'}F(\v')\rp\,\d\v'.
\end{align}
Here $K(\v-\v')=(I-\hat{\vbar}\hat{\vbar})/|\vbar|$, where $\vbar=\v-\v'$ is the closing velocity, and $\hat{\vbar}$ a unit vector in that direction.
The collision frequency is $\cfe = 2\pi n_i q_i^4\log\Lambda/(m_i^2\vth^3)$ where $\log\Lambda$ is the Coulomb logarithm that absorbs the effects of the cutoffs, and $n_i$, $q_i$, $m_i$ and $\vthi$ are the ion number density, charge, mass and thermal velocity (the characteristic velocity scale).
\Eqn~\Ex{eq:Kulsrud05_Landau_coll_op} is for ion--ion collisions, the relevant term in \eqn~\Ex{e.8}. 
Collisions between the adiabatic electrons and the ion perturbation are neglected, being smaller by a factor of the square root of the mass ratio $\sqrt{m_e/m_i}\approx1/40$ \cite{Hazeltine03}. 
The Landau collision operator (\ref{eq:Kulsrud05_Landau_coll_op}) may be rewritten in the \fp\ form
\begin{align}\label{eq:fp_form}
    C[F] = \pd{}{\v}\cdot\lp \calD [F]\cdot\pd{}{\v} F - \calF[F]F\rp,
\end{align}
using the diffusion tensor $\calD$ and friction vector $\calF$, 
\begin{align}
\calD[F] = \int K(\v-\v')F(\v')\,\d\v',
~~~~~
\calF[F] = \int K(\v-\v')\cdot\pd{F(\v')}{\v'}\,\d\v'.
\end{align}
The $F$ dependence of $\calD$ and $\calF$ makes them inconvenient expressions to work with.
Inserting approximations for $\calD$ and $\calF$ into the \fp\ form \Ex{eq:fp_form} leads to useful model collision operators. 
The properties of the Landau operator that model operators should share are often given \citep[\eg][]{Cercignani,Anderson07} as the model operator
\begin{inparaenum}[(i)]
  \item representing the effect of small-angle collisions,
  \item vanishing when applied to any Maxwellian distribution, 
  \item conserving mass, momentum and energy, 
    and
  \item driving the distribution function to a Maxwellian as $t\to\infty$.
\end{inparaenum}

The first property holds for \fp\ operators due to the presence of velocity derivatives in \eqref{eq:fp_form}. 
We only use the global Maxwellian $F_0$, so the second property is %simply 
$C[F_0]=0$.
Conservation of mass, momentum and energy correspond respectively to the zeroth, first and second velocity moments of the collision operator vanishing,
\begin{align}\label{eq:cons_ints}
  \int_{-\infty}^\infty C[F]\,\d v = 
  \int_{-\infty}^\infty vC[F]\,\d v =
  \int_{-\infty}^\infty v^2C[F]\,\d v = 0.
\end{align}

\newcommand{\smallpert}{\distpert}

The final property imposes a bound on the behaviour of $F$ at long times.  We express this using a convenient measure of entropy relative to the Maxwellian $F_0$,
\begin{align}
  \label{eq:ParallelVelocityRelativeEntropy}
  R[F|F_0] \id\int_{-\infty}^{\infty}F\log \lp\frac{F}{F_0}\rp - F + F_0~\d v.
\end{align}
This relative entropy was introduced for studying the Boltzmann \eqn\ \cite[\eg][]{Pauli00,Bardos93}, in particular for establishing rigorous 
hydrodynamic limits \citep{Bardos93,Lions01,Golse04}.
It is also used to establish the existence and long-time attracting properties of steady solutions of the Vlasov--Poisson system \citep{Bouchut93,Dolbeault99} and in other plasma applications \citep{Krommes94,Hallatschek04,Zocco11}.
Expanding for small perturbations $\smallpert=F-F_0\ll F_0$,
\begin{align}\label{eq:freeenergymotivation}
  R[F|F_0] = \frac{1}{2}\int_{-\infty}^{\infty}\d v~\frac{\smallpert^2}{F_0} + \O(\smallpert^3),
\end{align}
so $R$ is positive definite and quadratic in small perturbations.
The leading order term defines the free energy of the perturbed distribution for linearized theory,
which may also be expressed in terms of the Hermite expansion coefficients \eqref{eq:GradF} as
\begin{align}\label{eq:freeenergy}
  W_f \id \frac{1}{2}\int_{-\infty}^{\infty}\frac{\smallpert^2}{F_0}\,\d v 
=\frac{1}{2}\sum_{n,m=0}^{\infty}\int_{-\infty}^{\infty}\distpert_n \distpert_m\phi_n\phi^m\,\d v  
= \frac{1}{2}\sum_{m=0}^{\infty}|\distpert_m|^2 .
\end{align}
This is also a statement of Parseval's theorem for expansions in Hermite functions.
The total free energy for the system \eqref{e.8} is  $W=W_f+W_{\varphi}$, where $W_\varphi = |\varphi|^2/2=|\distpert_0|^2/2$ is the free energy of the electric field.
The free energy $W$ is conserved in the collisionless system with no temperature gradient ($\cfe=\wT=0$), as shown in \sec\ref{sec:EigenmodeFormationIVP}.
The Hermite spectrum $E_m\id |\distpert_m|^2/2$ (previously plotted in \figs\ref{fig:hermitec_and_delta} and \ref{fig:hermite_truncated}) is thus the contribution to $W_f$ from each Hermite mode.

Relative entropy $R$ differs from the Boltzmann entropy,
\begin{align}
  H[F] \id \int_{-\infty}^\infty F\log F\,\d v,
\end{align}
only by terms which are linear in number density and energy density.
Using the Maxwellian\\ $F_0=(n_0/(2\pi T_0)^{-1/2})\exp(-v^2/2T_0)$,
 \begin{align}\label{eq:relent}
   \begin{split}
     R[F|F_0] &=\int_{-\infty}^{\infty}F\log F - F\log F_0 - F + F_0~\d v
 %&= H[F]- n_i\log n_0 + \frac{1}{2}n_i\log(2\pi T_0) + \frac{{\cal E}}{T_0} - n_i + n_0, \\
     \\ &
 = H[F]+ \frac{{\cal E}}{T_0}  + n_0  - n_i\log \lp \frac{ e n_0 }{(2\pi T_0)^{1/2}}\rp       ,
   \end{split}
 \end{align}
 where $n_i$ and ${\cal E}$ are the number density and energy density associated with the total distribution $F$.
Thus $T_0R$ is analogous to the Gibbs free energy density at a point in space with temperature $T_0$ and pressure $n_0T_0$, plus a chemical potential term to account for the possibility of local changes in particle number.  
Relative entropy $R$ is also the plasma part of the general grand canonical ensemble density \cite{Hallatschek04}.
 When the collision operator conserves $n_i$ and ${\cal E}$, the behaviour of $R$ and $H$ under collisions is the same,
\ie\ $\tfd{H}{t}=\tfd{R}{t}$,
 and we may recast statements about the entropy in terms of the relative entropy.
In particular, the plasma is driven to \eqm\ by collisions if and only if
\begin{align}\label{eq:dRdt}
  \fd{R}{t} = \int_{-\infty}^{\infty} C[F]\log\frac{F}{F_0} \,\d v \leq 0.
\end{align}

\subsection{The \lb\ and Kirkwood collision operators}
\label{sec:lb}
We study two \fp\ type 
model collision operators 
where $\calD$ and $\calF$ are approximated.
\citet{LenardBernstein} proposed $\calD =1/2$ (the \eqm\ temperature in our dimensionless variables) and $\calF=-\v$, which in one dimension gives
\begin{align}\label{eq:LBcoll}
  C\ls F\rs = \pd{}{v}\lp v F + \frac{1}{2}\pd{F}{v}\rp . %\id \L[F_1].
\end{align}
The Boltzmann entropy decreases under \lb\ collisions %(and some other operators in \fp\ form) 
% See the eqn before 7 in Desvillettes.  There is a specific form.  I can't get hold of reference [1] though.
\cite{Desvillettes01},
\begin{align}\label{eq:LBR}
  \fd{H}{t} = \int_{-\infty}^{\infty} C[F]\log\frac{F}{F_0} \,\d v 
= -\frac{1}{2}\int_{-\infty}^{\infty}F\left| \pd{}{v}\lp\log\frac{F}{F_0}\rp\right|^2\,\d v \leq 0,
\end{align}
with equality only if $F=F_0$. 
%We give a linearized version of this result for the \dop\ operator in \sec\ref{sec:hypercollisions}.
The eigenfunctions and eigenvalues are the Hermite functions
\begin{align}\label{eq:LBeigenfn}
  C[\phi_m] = -  m \phi_m, ~~~ \textrm{for all} ~ ~  m\geq0.%~~~ m\in\N \setminus \{0,2\},
\end{align}
The \lb\ operator preserves the leading order Maxwellian, $C[F_0]=0$, which follows from $F_0 = \phi_0$.
Thus $C[F]=C[\smallpert]$, and expanding the perturbed distribution function as $\smallpert=\sum_m \distpert_m\phi_m$, we investigate the conservation properties by calculating the integrals (\ref{eq:cons_ints}).
Mass is conserved since
\begin{align}\label{eq:lbmass}
  \intii C[F]\,\d v = -\intii \sum_m m\distpert_m \phi_m\,\d v = -\sum_m m\distpert_m \delta_{m0} = 0,
\end{align}
but the \lb\ operator conserves neither momentum nor energy,
\begin{align}\label{eq:lbmomen}
  \intii vC[F]\,\d v = -\frac{\distpert_1}{\sqrt{2}}, ~~~~~~~~~~
  \intii v^2C[F]\,\d v = -2\sqrt{2} \distpert_2.
\end{align}
The \dop\ operator \cite{Kirkwood46} restores conservation of momentum and energy by incorporating local values of the fluid velocity $u$ and temperature $T$ into $\calF = v-u$ and $\calD=T$, giving
\begin{align}
  \label{eq:Dougherty}
  C[F] =  \pd{}{v}\lp (v-u[F])F + T[F]\pd{F}{v}\rp,
\end{align}
where
\begin{align}
n_i[F] = \intii F\,\d v,
~~~~~
  u[F] = \frac{1}{n_i}\intii vF\,\d v, 
~~~~~
T[F] = \frac{1}{n_i}\intii (v-u)^2F\,\d v,
\end{align}
so that the \dop\ operator is a nonlinear integro-differential operator.

\subsection{The linearized \dop\ collision operator}
\label{sec:LinearizedCollisionOperators}

Formally in \eqn\ \Ex{eq:KineticEquationChapter3} we require the linearized form of the Landau operator \Ex{eq:Kulsrud05_Landau_coll_op} which describes collisions between the ion Maxwellian $F_0$ and a perturbation $\smallpert$ \cite{HelanderSigmar},
%%%\begin{align}\label{eq:Landau_coll_op}
%%%  C[F_1] = \pd{}{\v}
%%%\cdot\int K(\v-\v')\cdot\lp F_0(\v')\pd{}{\v}F_1(\v) + F_1(\v')\pd{}{\v}F_0(\v)  - F_1(\v)\pd{}{\v'}F_0(\v') -F_0(\v)\pd{}{\v'}F_1(\v')  \rp\,\d\v'.
%%%\end{align}
\begin{align}\label{eq:Landau_coll_op}
  C[\smallpert] = \pd{}{\v}
  \cdot\int K(\v-\v')\cdot F_0(\v)F_0(\v')\ls \pd{}{\v}\lp \frac{\smallpert(\v)}{F_0(\v)}\rp - \pd{}{\v'}\lp\frac{\smallpert(\v')}{F_0(\v')}\rp  \rs\,\d\v'.
\end{align}
We replace this with the linearized \dop\ operator
\begin{align}
  \label{eq:Dougherty_linear}
  \L[\smallpert] =  \pd{}{v}\lp \frac{1}{2}\pd{\smallpert}{v} + v\smallpert + T[\smallpert]\pd{F_0}{v} - u[\smallpert]F_0\rp.
  %\id \L[F_1]
\end{align}
%which extends the \lb\ operator with two additional terms that restore momentum and energy conservation.
In Hermite space this linearized \dop\ operator is 
\begin{align}
  \label{eq:Dougherty_linear_hs}
  \L[\smallpert] = -\sum_{m=0}^\infty m\distpert_m\phi_m  \dthree ,  %% + \nu a_1\phi_1 + 2\nu a_2\phi_2, 
\end{align}
where the indicator function $\dthree$ is defined by
\begin{equation}
\dthree = \begin{cases} 1 & \mbox{ for } m \ge 3, \\
0 & \mbox { otherwise}. \end{cases}
\end{equation}
The eigenfunctions are the Hermite functions $\phi_m$, each with corresponding eigenvalue $-m\dthree$. Thus $\phi_0$, $\phi_1$ and $\phi_2$ all have eigenvalue zero,
which establishes the required conservation properties.

%\subsection{Growth rate with \dop\ collisions}
With (\ref{eq:Dougherty_linear_hs}) the collisional version of the moment system (\ref{eq:asystem}) is
\begin{align}\label{eq:lb_amended}
  \begin{split}
    \pd{\distpert_m}{t} + \frac{i}{\kpara}\lp \wn
    \delta_{n0} + \frac{\wT}{\sqrt{2}}\delta_{n2}\rp \distpert_0
+    \frac{i}{\sqrt{2}}\delta_{m1} \distpert_0
   +i\lp\sqrt{\frac{m+1}{2}}\distpert_{m+1} + \sqrt{\frac{m}{2}}\distpert_{m-1}\rp 
\\
=    - \cf m \distpert_m \dthree.
  \end{split}
\end{align}
Writing this in the form of \Ex{eq:amatrix}, there are now imaginary entries on the diagonal of the matrix.
As it is complex, its eigenvalues do not necessarily occur in conjugate pairs, and so the method can capture negative growth rates.

\begin{figure}[tb]
  \centering
  \subfigure[\label{f.LB_nu}]{\includegraphics[width=0.48\textwidth]
%{gr_lb_nu_mark.ps}%{gr_lb_nu.ps}
%{gr_lb_nu_mark.pdf}
{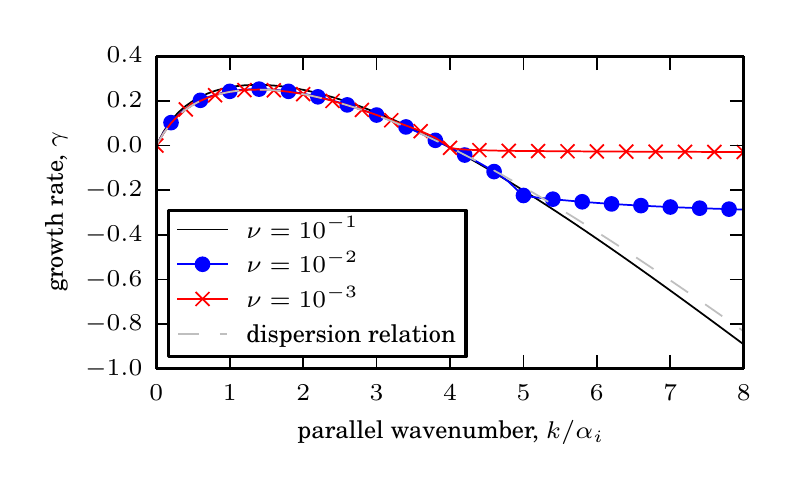}
}
\subfigure[\label{fig:LB_N}]{\includegraphics[width=0.48\textwidth]
%{gr_lb_N.ps}
%{gr_lb_N_mark.ps}
%{gr_lb_N_mark.pdf}
{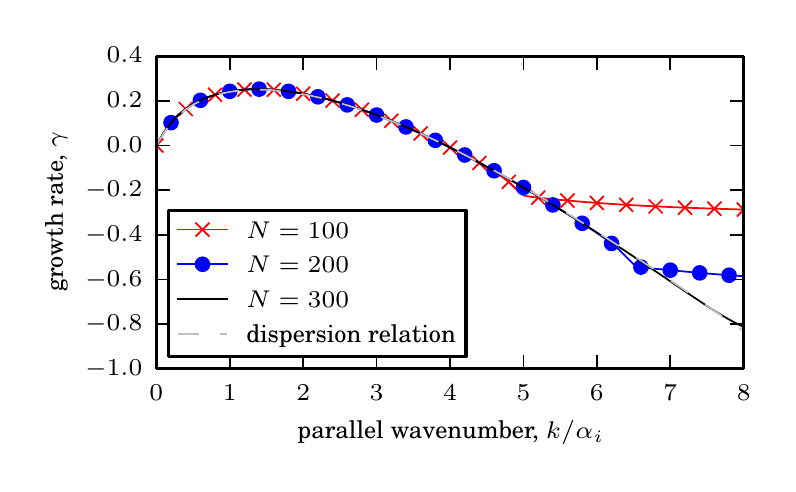}
} 
\subfigure[\label{fig:lb_hs_k2}]{\includegraphics[width=0.48\textwidth]
  %{lb_hs_k=2.ps}} 
  %{lb_hs_k=2.pdf}} 
  {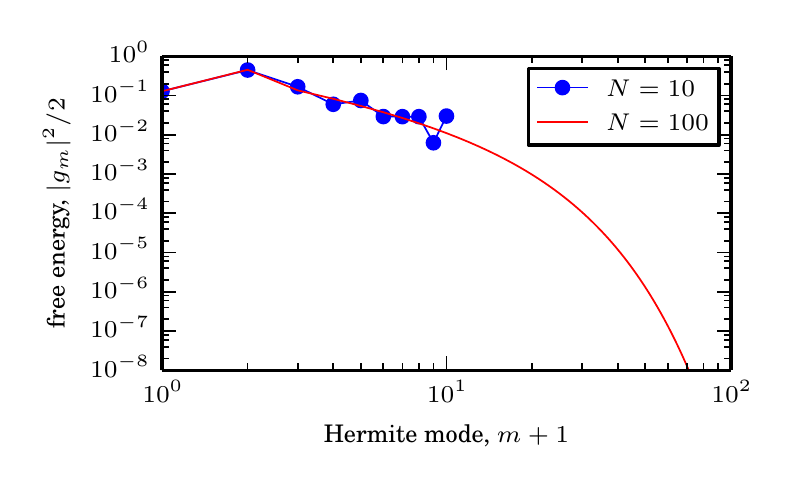}} 
  \subfigure[\label{fig:lb_hs_k6}]{\includegraphics[width=0.48\textwidth]
  %{lb_hs_k=6.ps}}
  %{lb_hs_k=6.pdf}}
  {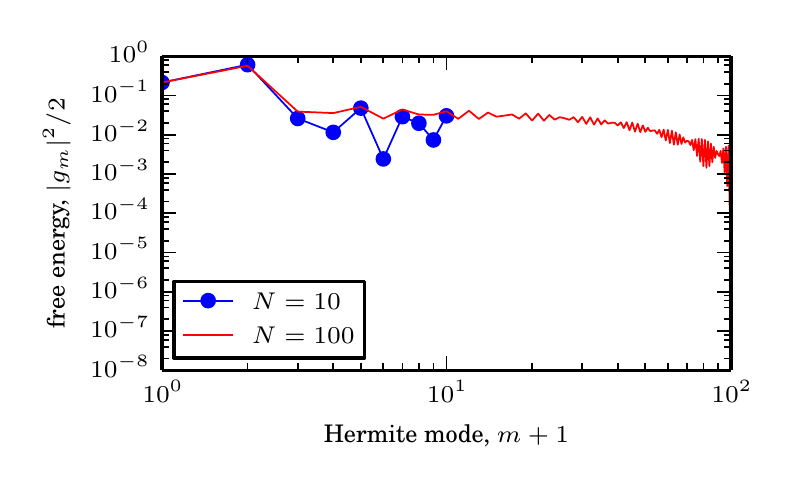}} 
  \caption[Growth rates and Hermite spectra calculated with Kirkwood collisions.]{\lbd\ collisions. The growth rate is plotted for (a) varying collision frequency $\cf$ and fixed truncation point $N=100$, and (b) varying truncation point and fixed collision frequency $\cf=10^{-2}$.
Spectra are plotted for truncations $N=10$, $N=100$ and wavenumbers (c) $k/\alpha_i=2$, (d) $k/\alpha_i=6$.}
\label{fig:LB}
\end{figure}

\Fig\ref{fig:LB} shows the growth rate plotted against parallel wavenumber for different collision frequencies $\cf$ and numbers of Hermite functions $N$.
\Fig\ref{f.LB_nu} shows 
that when the collision frequency is small, the growth rate is similar to the collisionless case and negative growth rates are not correctly captured.
However when the collision frequency is large enough to correctly capture the qualitative growth rate behaviour, the growth rate is too large in the low wavenumbers and too small in the high wavenumbers.
The correct rate is found by increasing $N$, but \fig\ref{fig:LB_N} shows this requires hundreds of terms.

\Figs \ref{fig:lb_hs_k2} and \subref{fig:lb_hs_k6} show the Hermite spectra for two truncations with $\cf=10^{-2}$ and a fixed parallel wavenumber,
corresponding to points on the red curve in \fig\ref{fig:LB_N}.
As in \fig\ref{fig:hermite_truncated}, the $N=100$ calculation is correct for 
$k/\alpha_i=2$ and incorrect for %at the higher wavenumber 
$k/\alpha_i=6$.
Now, however, both spectra decay, %with increasing $n$
even though the $k/\alpha_i=6$ spectrum is not negligible at the truncation point.
The behaviour of the $N=10$ spectra is similar to the $N=100$ spectra. 
This shows we can truncate at different points without qualitative changes to the spectra, but also that we must use a large number of terms to resolve collisional scales. 
%We therefore introduce a collision operator which gives greater control over the damping of the spectrum at different modes $n$.

\subsection{Hypercollisions}
\label{sec:hypercollisions}
To reduce the number of Hermite modes required to resolve collisional scales we introduce a hypercollisional operator that leaves most of the spectrum unaffected but severely damps a narrow range of the highest modes.  
This is defined by
\begin{align}\label{eq:DHDC}
  C[\phi_m] = - (m/N)^\hdexp \phi_m \dthree.
\end{align}
It differs from the \dop\ collision operator \Ex{eq:Dougherty_linear_hs} by the power $\hdexp$, and the replacement of $m$ by $m/N$.
This is a convenient rescaling of $\cf$ so that the damping strength depends on relative position within the truncated 
series, not absolute mode number.
The highest mode is damped by $\cfe$ for all $N$, and the parameter $\hdexp$ determines how sharply the damping changes in Hermite space.
A similar operator, $C[\phi_m]=-m(m-1)(m-2)\phi_m$, which also conserves mass, momentum and energy, was proposed by \citeauthor{CamporealeDelzannoBergenMoulton13} \cite{CamporealeDelzannoBergenMoulton13}.
The latter does not have a simple definition as a differential operator in velocity space, but can be constructed from the \lb\ operator \cite{CamporealeDelzannoBergenMoulton13}.
In contrast, the hypercollisional operator \eqref{eq:DHDC} may be defined using the Kirkwood operator $\L$ \Ex{eq:Dougherty_linear}, analogously to the definition of hyperviscosity as a power of a Laplacian,
\begin{align}\label{eq:HDop}
  C[f] \id -\frac{(-1)^\hdexp}{N^\hdexp} \L^\hdexp[f].
\end{align}
The hypercollision operator inherits some properties from the \dop\ operator.
The equilibrium distribution is unaffected by collisions as $\L[F_0]=0$, and the Hermite functions $\phi_m$ are the eigenfunctions.
Hypercollisions conserve mass, momentum or energy.
The free energy $W_f$ \eqref{eq:freeenergy} does not increase under collisions, which is sufficient for a linearized theory:
differentiating \Ex{eq:freeenergy} \wrt\ time gives
\begin{align}
    \fd{W_f}{t} 
= -\sum_{m=3}^{N} \lp\frac{m}{N}\rp^\hdexp |\distpert_m|^2, 
\end{align}
so that relative entropy $R$, and therefore entropy $H$, decrease to leading order in small perturbations $\smallpert\ll F_0$.

With hypercollisions, equation (\ref{eq:asystem}) becomes,
\begin{align}\label{eq:asystemHD}
\begin{split}
  \pd{\distpert_m}{t} + \frac{i}{\kpara}\lp \wn
    \delta_{m0} + \frac{\wT}{\sqrt{2}}\delta_{m2}\rp \distpert_0
 +\frac{i}{\sqrt{2}}\delta_{m1}\distpert_0
 +i\lp\sqrt{\frac{m+1}{2}}\distpert_{m+1} + \sqrt{\frac{m}{2}}\distpert_{m-1}\rp 
 \\
 = -\cf \lp\frac{m}{N}\rp^\hdexp \distpert_m \dthree.
\end{split}
\end{align}
As with Kirkwood collisions, the corresponding matrix is not purely imaginary, resulting in negative growth rates. 

\begin{figure}[tb]
  \centering 
  \subfigure[\label{f.GR_HD}]{\includegraphics[width=0.49\textwidth]
    %{hd-asystem_gr.ps}} 
    {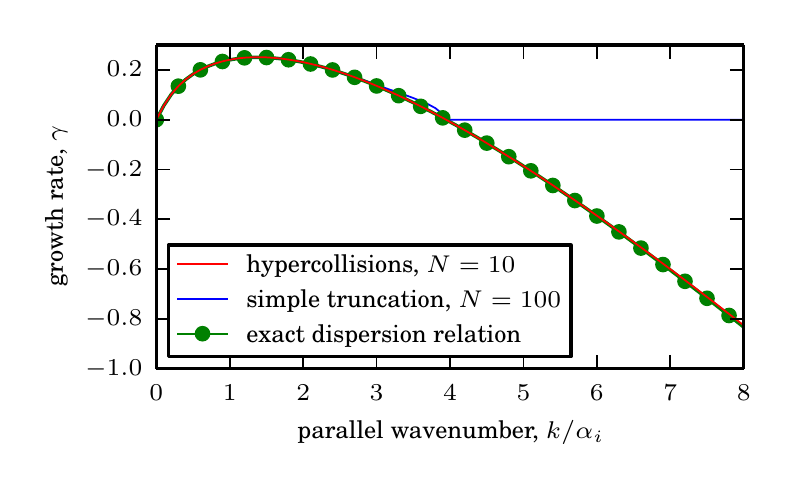}} 
    \subfigure[\label{f.GR_HD_error}]{\includegraphics[width=0.49\textwidth]
    %{normalized_error.ps}} 
    {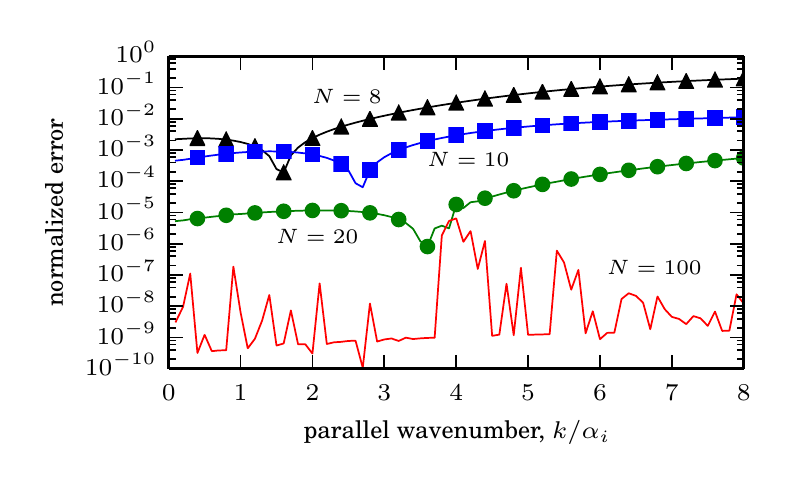}} 
    \subfigure[\label{fig:HSk2}]{\includegraphics[width=0.49\textwidth]
    %{hd_hs_k2.ps}}
    {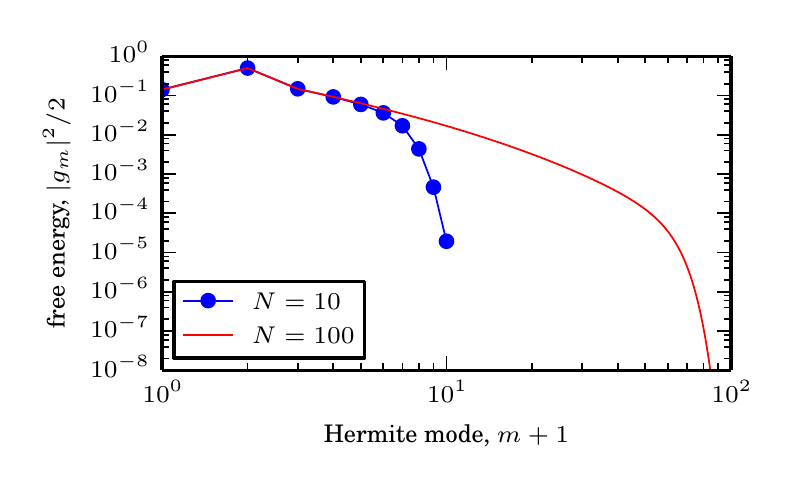}}
    \subfigure[\label{fig:HSk6}]{\includegraphics[width=0.49\textwidth]
  %{hd_hs_k6.ps}}
  {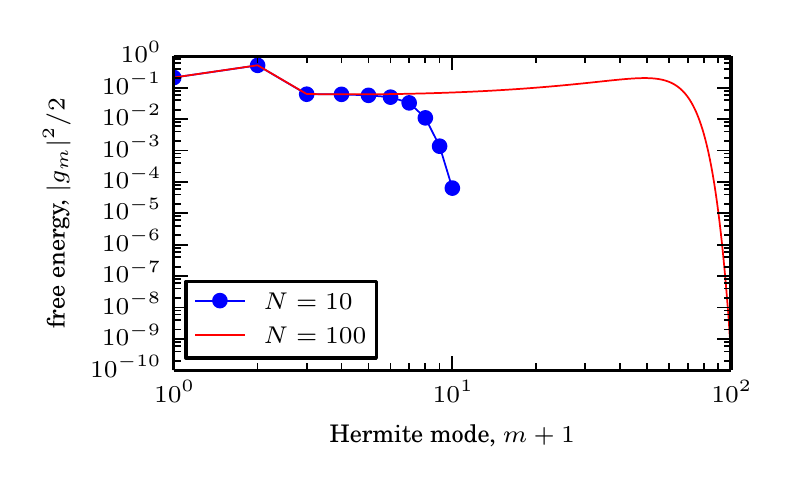}}
\caption[Growth rates and Hermite spectra calculated with hypercollisions.]{Results for hypercollisions, $\cf=16$, $\hdexp=6$: 
(a) the growth rate against parallel wavenumber, and
(b) the normalized error in the calculation %The peaks near $\kpara=1$ are due to the exact growth rate changing sign.
The last two figures show Hermite spectra for (c) $\kpara/\alpha_i=2$ and (d) $\kpara/\alpha_i=6$.
}
\end{figure}

\Fig\ref{f.GR_HD} shows that with hypercollisions the growth rate may be captured accurately for all wavenumbers with as few as 10 modes.
The quality of the approximation depends on the parameters $\cf$ and $\hdexp$, but is quite robust to changes (a property we investigate in \sec\ref{sec:ParameterChoices}).
\Fig\ref{f.GR_HD_error} shows the normalized error plotted for various values of $N$.
This error is defined as
\begin{align}\label{eq:NormErr}
  \mathrm{Normalized\ Error} = \frac{|\gamma_{\textrm{calc}}-\gamma_{\textrm{exact}}|}{|\gamma_{\textrm{exact}}|},
\end{align}
where $\gamma_{\textrm{calc}}$ is the computed growth rate, and $\gamma_{\textrm{exact}}$ is the growth rate from the dispersion relation \eqref{eq:DispersionRelation}.
The normalized error decreases as $N$ increases, but tends to increase as $\kpara$ increases.
The troughs in the first three lines correspond to a change of sign in the numerator of (\ref{eq:NormErr}).
%The smaller peaks %in all lines around $k=4$ correspond to the zero in the numerator.
The error %of the growth rate calculated with hypercollisions 
is below one percent with as few as 10 terms in the Hermite expansion.

\Figs\ref{fig:HSk2} and \subref{fig:HSk6} show that high modes of the hypercollision spectra are now damped to zero before the series is truncated.
This is true for all truncations as the highest mode is always damped by $-\cfe \distpert_N$. %both $N=10$ and $N=100$ due to the fact damping depends on $(n/N)$. 
Again the behaviour of the low modes is independent of $N$ with the $N=10$ spectrum agreeing with $N=100$ spectrum before its highest modes are damped.
Before damping, the $k/\alpha_i=6$ spectrum increases in magnitude with increasing $m$, showing that the distribution would have fine structure in velocity space but for the effect of collisions and truncation.

\section[Theoretical Hermite spectra and the hypercollision plateau]{Theoretical Hermite spectra and the hypercollision\\ plateau}

We now study the Hermite spectrum of the hypercollisional system \Ex{eq:asystemHD}. 
We derive a partial differential equation for the free energy density $E_{\hm}=|\distpert_{\hm}|^2/2$ as a function of continuous $(t,\hm)$ space.   
Solving for eigenmodes in time, we find a theoretical expression for the Hermite spectrum in terms of the growth rate and hypercollision parameters. 

We also plot the calculated growth rate against the collision frequency $\cfe$ and find a ``hypercollision plateau'', an interval in $\cfe$ where the growth rate is accurately calculated, independent of the precise value of $\cfe$.
Using the theoretical spectrum, we find the ends of the plateau and thereby derive inequalities for the parameter values in the hypercollision operator. 

\subsection{Theoretical Hermite spectra}
\label{sec:theoretical_hermite_spectra}
\label{sec:TheoreticalHermiteSpectra}

To derive the theoretical Hermite spectrum we follow the approach of \citet{Watanabe04} and \citet{Zocco11}, and multiply the kinetic equation (\ref{eq:asystemHD}) by $\bar{g}_{\hm}$ (where overbar denotes the complex conjugate) to obtain the difference \eqn
\begin{align}\label{eq:difference_eqn}
  \begin{split}
		\bar{\distpert}_{\hm}\pd{\distpert_{\hm}}{t} +i\lp\sqrt{\frac{\hm+1}{2}}\distpert_{\hm+1}\bar{\distpert}_{\hm} + \sqrt{\frac{\hm}{2}}\bar{\distpert}_{\hm}\distpert_{\hm-1}\rp 
		= -\cf \lp\frac{\hm}{N}\rp^\hdexp |\distpert_{\hm}|^2 ,
  \end{split}
\end{align}
for $\hm>2$. 
The substitution $b_{\hm} = i^{\hm}\distpert_{\hm}$ gives the equation
\begin{align}\label{eq:bm}
  \begin{split}
		b_{\hm}\pd{b_{\hm}}{t} + \lp\sqrt{\frac{\hm+1}{2}}b_{\hm+1}b_{\hm} - \sqrt{\frac{\hm}{2}}b_{\hm}b_{\hm-1}\rp 
		= -\cf \lp\frac{\hm}{N}\rp^\hdexp b_{\hm}^2,
  \end{split}
\end{align}
which supports real-valued solutions $b_{\hm}$. 
Defining $\Gamma_{\hm+1/2}=b_{\hm+1}b_{\hm}\sqrt{(\hm+1)/2}$, 
the mode coupling term may be interpreted as a finite difference approximation to a derivative in $\hm$ considered as a continuous variable,
\begin{align}
  \begin{split}\label{eq:flux}
		\frac{\Gamma_{\hm+1/2} - \Gamma_{\hm-1/2}}{(\hm+1/2)-(\hm-1/2)} \approx \pd{\Gamma_{\hm}}{\hm}.
  \end{split}
\end{align}
We interpret $\Gamma_{\hm}$ as the free energy flux through mode numbers, the same concept as the flux that cascades through wavenumber space in Kolmogorov's hydrodynamic turbulence theory \cite{Kolmogorov41}.
The approximation is valid in the large $\hm$ limit, in which $\Gamma_{\hm}\approx b_{\hm}^2\sqrt{\hm/2}$, provided $b_{\hm+1}\approx b_{\hm}$.
It is known from the theory of finite difference approximations for advective partial differential equations 
that \eqref{eq:bm} also supports ``alternating'' or ``parasitic'' solutions for which $b_{\hm+1}\approx -b_{\hm}$,
and the flux propagates in the opposite direction
\cite{MesingerArakawa76,Strikwerda04}.
We return to this point when studying the \ivp\ in \sec\ref{sec:ivp}.

In the large $\hm$ limit, (\ref{eq:bm}) may be rewritten as an evolution \eqn\ for the free energy density, $E_{\hm}=|\distpert_{\hm}|^2/2=b_{\hm}^2/2$, in continuous $(\hm,t)$ space,
\begin{align}\label{eq:EmPDE}
	\pd{E_{\hm}}{t} + \pd{}{\hm}\lp \sqrt{2\hm} E_{\hm}\rp =  -2\cf \lp\frac{\hm}{N}\rp^\hdexp E_{\hm}.
\end{align}
Introducing the variables $\mu=\sqrt{2\hm}$ and $\Emu = \mu E_{\hm}$,
and
multiplying (\ref{eq:EmPDE}) by $\mu$ gives
\begin{align}\label{eq:EmuPDE}
  \lp\pd{}{t}+\pd{}{\mu}\rp\Emu = -2\cf \mu^{2\hdexp}(2N)^{-\hdexp}\Emu.
\end{align}
Considering eigenmodes in time we have $\dee \distpert_{\hm}/\dee t = -i\omega \distpert_{\hm}$ and $\d\Emu/\d t = 2\gamma \Emu$, where $\gamma=\Im(\omega)$. Substitution into (\ref{eq:EmuPDE}) yields
\begin{align}
  \fd{\Emu}{\mu} = -2\gamma \Emu -2\cf \mu^{2\hdexp}(2N)^{-\hdexp}\Emu.
\end{align}
Solving, and rewriting the solution $\Emu$ using the original variables gives the Hermite spectrum of the eigenmode as
\begin{align}\label{eq:theoreticalspectra}
	E_{\hm} = \frac{C}{\sqrt{2\hm}}\exp\lp -\frac{\gamma}{|\gamma|}\lp\frac{\hm}{\hm_\gamma}\rp^{1/2} - \lp\frac{\hm}{\hm_c}\rp^{\hdexp+1/2}\rp,
\end{align}
where the growth rate cutoff $\hm_\gamma$ and the collisional cutoff $\hm_c$ are defined by
\begin{align}\label{eq:mc}
  \hm_\gamma = \frac{1}{8\gamma^2}, ~~~~~~~~~~
  \hm_c = \ls \frac{N^\hdexp\lp \hdexp +1/2\rp}{\cf\sqrt{2}}\rs^{1/(\hdexp+1/2)}.
\end{align}
The spectrum has three parts: an underlying $\hm^{-1/2}$ decay, and two exponential factors due to the growth rate and collisions.
Typically $\hm_\gamma<\hm_c$ so collisions dominate at the largest $\hm$. 
Modes above the collisional cutoff, $\hm>\hm_c$, are strongly damped, with the steepness of the cutoff increasing with increasing $\hdexp$.
Between cutoffs, $\hm_\gamma<\hm<\hm_c$, the behaviour depends on 
the sign of $\gamma$, with the spectrum growing for $\gamma<0$ and  
decaying for $\gamma>0$.
The $\hm^{-1/2}$ behaviour is observed in the modes below both cutoffs. 
As $\gamma\to0$, the spectrum becomes like $\hm^{-1/2}$, like that of a delta function, as found in \S\ref{sec:HermiteFunctions}.

\begin{figure}[tb]
  \centering \subfigure[]{\includegraphics[width=0.49\textwidth]
    %{theoretical_spectrum_k2.ps}} 
    {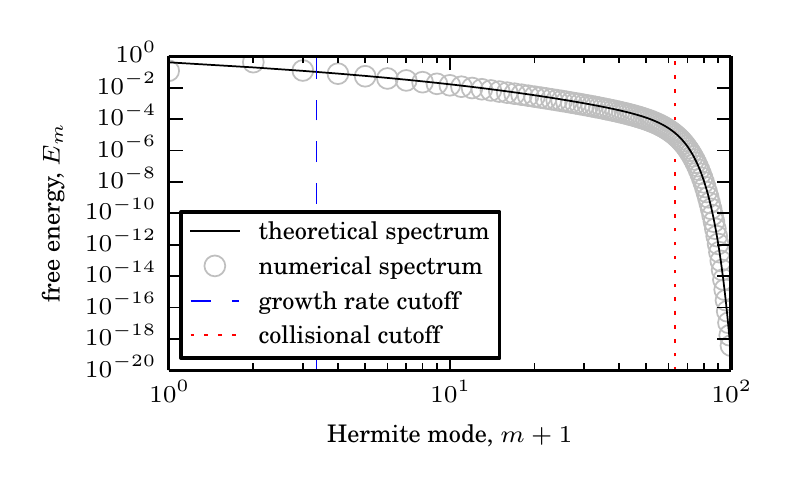}} 
    \subfigure[\label{fig:theoreticalspectra_b}]{\includegraphics[width=0.49\textwidth]
      %{theoretical_spectrum_k6.ps}}
      {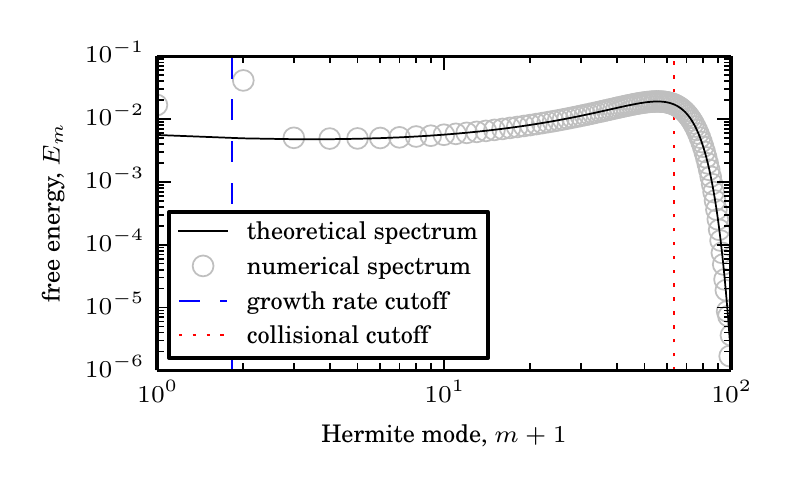}}
\caption[Calculated Hermite spectra against theoretical spectra]{Calculated Hermite spectra with hypercollisions against theoretical spectra for $\hdexp=6$, $\cfe=10$ and two wavenumbers (a) growing mode $k/\alpha_i=2$ and (b) decaying $k/\alpha_i=6$. The two vertical lines are the growth rate and collisional cutoffs. }
\label{fig:theoreticalspectra}
\end{figure}

The theoretical spectra (\ref{eq:theoreticalspectra}) for $\kpara/\alpha_i=2$ and
$\kpara/\alpha_i=6$ are plotted in \fig\ref{fig:theoreticalspectra} along with 
corresponding spectra calculated as eigenvectors of \Ex{eq:asystemHD}, and the collisional and growth rate cutoffs.
The growth rate $\gamma$ in (\ref{eq:theoreticalspectra}) is not determined in this derivation of the spectrum, so we use the growth rate calculated in the discrete eigenvalue calculation \Ex{eq:asystemHD}.
Since the theoretical and numerical spectra are both eigenfunctions, their overall normalizations are arbitrary. 
We have aligned the theoretical spectrum to the numerical spectra using a least squares fit for the normalization.
The theoretical spectra for both growth and decay show remarkable agreement with the numerical spectra.
This is particularly striking as the theoretical spectra are derived in the large $\hm$ limit, but hold for modes as small as $\hm=2$.
The collisional cutoff also gives an accurate guide for the point at which hypercollisions become dominant.

%\subsection{Hypercollisional plateau}
\subsection{Hypercollisional plateau against collisionality}
\label{sec:ParameterChoices}
\label{sec:HypercollisionalPlateau}

\begin{figure}[tbp]
  \centering
  \subfigure[\label{fig:PlateauVsStrength}]{\includegraphics[width=0.49\textwidth]{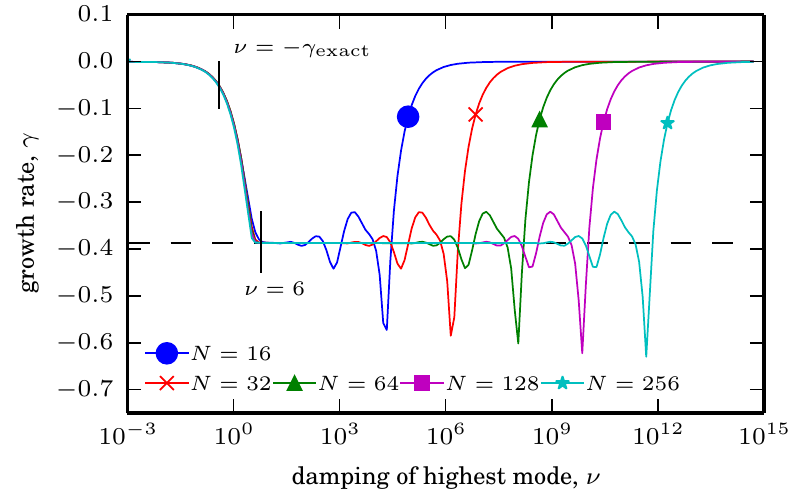}}
    \subfigure[\label{fig:PlateauVsFrequency}]{\includegraphics[width=0.49\textwidth]{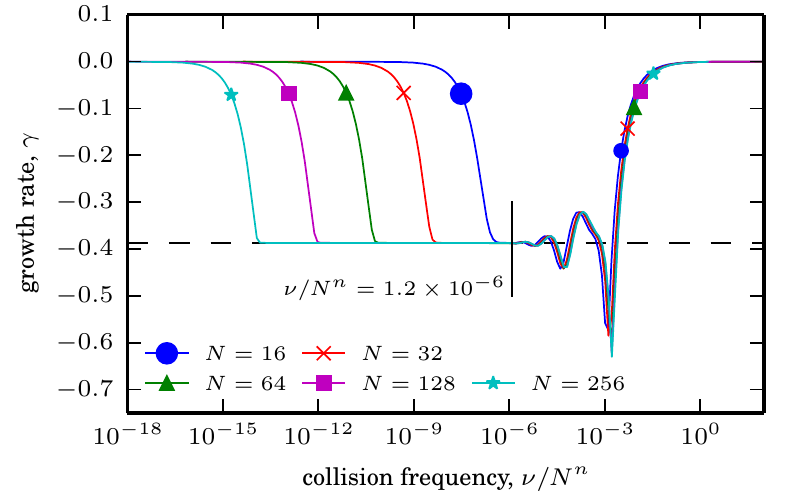}}
    \subfigure[\label{fig:DPlateauDNuVsStrength}]{\includegraphics[width=0.49\textwidth]{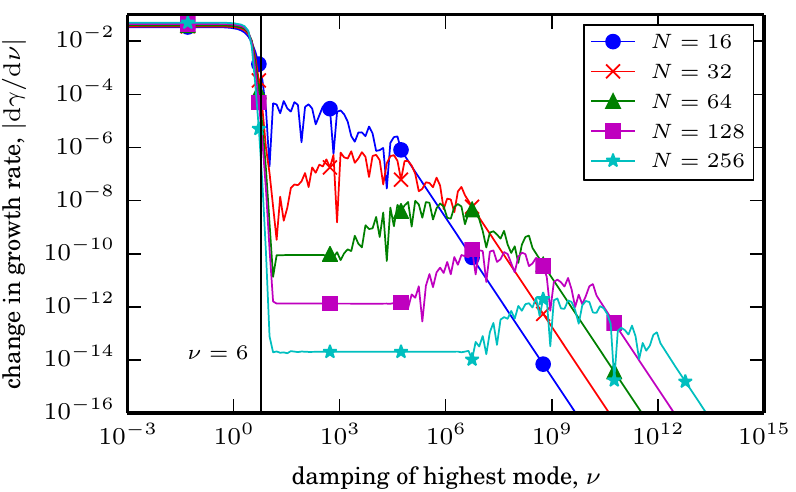}}
  \subfigure[\label{fig:DPlateauDNuVsFrequency}]{\includegraphics[width=0.49\textwidth]{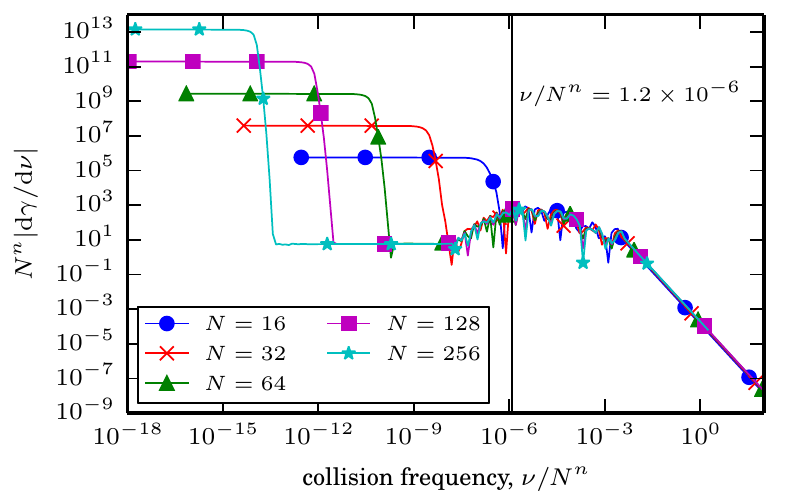}}
  \caption[The hypercollision plateau for different resolutions.]{\label{fig:gr_vs_coll}\label{fig:GrVsColl}
  The hypercollision plateau for $\hdexp=6$.
  (a,b) The calculated growth rate for different resolutions plotted against (a) the damping of the highest mode, and (b) the hypercollision frequency. 
  There is a plateau where the calculated growth rate equals the value obtained from the dispersion relation \eqref{eq:DispersionRelation}, indicated with the dashed line. 
  In (a), the small vertical lines indicate estimates for the start of the plateau: ($\cfe=-\gamma_{\mathrm{exact}}$) from the lower bound \eqref{eq:EtaCrit}, and ($\cfe=6$) by inspection.
  In (b), the vertical line indicates the end of the plateau obtained by inspection.
	The end is at $\cfe/N^{\hdexp}\approx 1.2\times10^{-6}$ which corresponds to $\hm_\crit=10$ (see \eqref{eq:EtaOnNCrit}). 
\label{fig:dplateaudnu}
  (c) The derivative of the calculated growth rate $|\tfd{\gamma}{\cfe}|$ against $\cfe$. 
	(d) The derivative of the calculated growth rate $|\tfd{\gamma}{(\cfe/N^{\hdexp})}|$ against $\cfe/N^{\hdexp}$. 
  In (c,d) there is also a plateau that begins at fixed $\cfe$ and ends at fixed $\cfe/N^\hdexp$, but appears smaller than the plateau in the growth rate due to the logarithmic scale.
}
\end{figure}

We now study how the calculation of the growth rate depends on the parameters $\cf$, $\hdexp$, and $N$ in the hypercollisional operator.
In particular, we wish to choose $\cf$ and $\hdexp$ to minimize the necessary resolution $N$.
We first fix $N$ and $\hdexp$, and plot the growth rate of the $k/\alpha_i=6$ decaying mode against the damping of the highest mode, $\cf$, in \fig\ref{fig:PlateauVsStrength}, and against the collision frequency, $\cf/N^\hdexp$, in \fig\ref{fig:PlateauVsFrequency}.
We find that the decay rate is underestimated until some critical value $\cf_{\mathrm{crit}}$ (marked with the vertical line $\cfe=6$ in \fig\ref{fig:PlateauVsStrength}).
Then there is a plateau where the correct value is calculated for a range of $\cf$, before the decay rate becomes incorrect as $\cf$ becomes too large.
This pattern is repeated for different values of $N$. 
As $N$ increases, $\cf_{\mathrm{crit}}$ is unchanged so that $\cf_{\mathrm{crit}}/N^\hdexp$, the minimum collision frequency required for correct capture of the growth rate, decreases.
However the right-hand end of the plateau %where the solution becomes over-damped 
is at a fixed $\cf/N^\hdexp$ ($\approx1.2\times10^{-6}$, marked in \fig\ref{fig:PlateauVsFrequency})
so the range of feasible values of $\cf$ increases with increasing $N$. 

We find the location of the plateau by calculating $\tpd{\gamma}{\cfe}$ via matrix perturbation theory \cite{Hinch91}. 
We write \eqref{eq:asystemHD} as the \evp\
\begin{align}
  %\omega\a = (M - i\cfe(\hm/N)^\hdexp\dthree) \a,
  \omega\a = (M - D) \a,
  \label{eq:EigenvalueEquationVectors}
\end{align}
where $D$ is a diagonal matrix whose entries are $D_{mm}=i\cfe(\hm/N)^\hdexp\dthree$, and $\a=(\distpert_\hm)$ is the eigenvector of Hermite coefficients for the dominant eigenvalue $\omega$.
Taking the $\cfe$-derivative, we obtain
\begin{align}
  \pd{\omega}{\cfe}\a + \omega\pd{\a}{\cfe} =  - i\lp\frac{\hm}{N}\rp^\hdexp\dthree \a  + \ls M - i\cfe\lp\frac{\hm}{N}\rp^\hdexp\dthree \rs \pd{\a}{\cfe}   ,
  %\label{eq:}
\end{align}
so that left-multiplying by the adjoint eigenvector $\b^*$ and rearranging gives
\begin{align}
  -i\pd{\omega}{\cfe}
  =  -\frac{\b^*\ls \lp\frac{\hm}{N}\rp^\hdexp\dthree \rs\a}{\b^*\a} 
	=  -\frac{\sum_{\hm=3}^Nb_{\hm}^* \lp\frac{\hm}{N}\rp^\hdexp \distpert_{\hm}}{\sum_{\hm=0}^Nb_{\hm}^*\distpert_{\hm}} 
  .
  \label{eq:dwdeta}
\end{align}
The $\tpd{\a}{\cfe}$ terms cancel since $\b^*\omega= \b^*(M -i\cfe(\hm/N)^{\hdexp}\dthree)$, the adjoint equation of \eqref{eq:EigenvalueEquationVectors}.
A set of parameters $\{\cfe,\hdexp,N\}$ is in the plateau when $\tpd{\omega}{\cfe}$ is small. 
In \figs\ref{fig:DPlateauDNuVsStrength} and \subref{fig:DPlateauDNuVsFrequency} we plot $\tpd{\gamma}{\cfe}=\Re\lp-i\tpd{\omega}{\cfe}\rp$ against $\cfe$ and against $\cfe/N^\hdexp$. 
There is a plateau where $\tpd{\gamma}{\cfe}$ is small and constant, 
and, as with the growth rate plateau, the plateau begins at fixed $\cfe$ and ends at fixed $\cfe/N^\hdexp$.
As it is plotted on a logarithmic scale, the plateau in $\tpd{\gamma}{\cfe}$ appears to begin later and end sooner than the plateau in the growth rate.
Therefore \figs\ref{fig:DPlateauDNuVsStrength} and \subref{fig:DPlateauDNuVsFrequency} give conservative values for the extent of the plateau.

\begin{figure}[tbp]
  \subfigure[\label{fig:ExpectedMVsStrength}]{\includegraphics[width=0.49\textwidth]{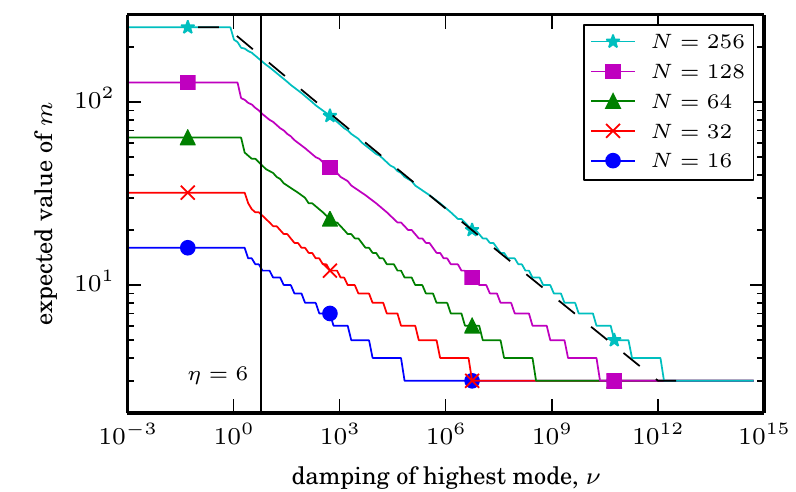}}
  \subfigure[\label{fig:ExpectedMVsFrequency}]{\includegraphics[width=0.49\textwidth]{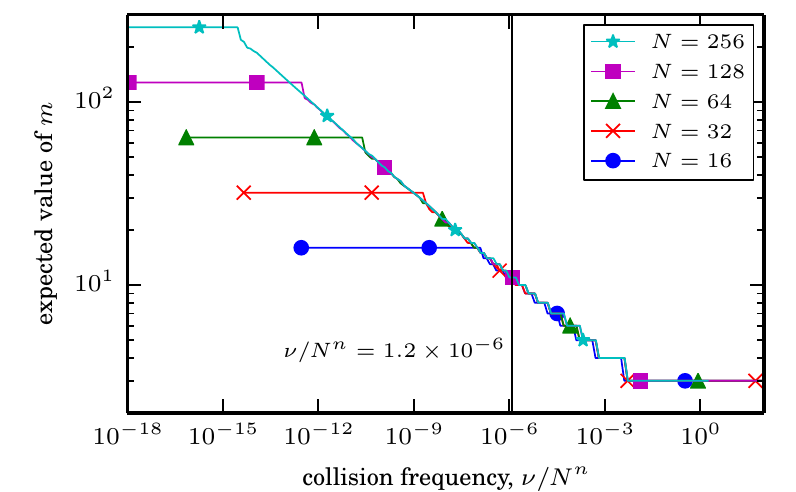}}
   
	\caption[Location of the largest term in the free energy dissipation spectrum.]{Location of the largest term in the numerator of \eqref{eq:dwdeta} against (a) $\cfe$ and (b) $\cfe/N^\hdexp$.
  The dashed line in (a) gives the theoretical shape of the $N=256$ curve from solving \eqref{eq:LocationPolynomials}.
}
\end{figure}

Equation \eqref{eq:dwdeta} shows that $\tpd{\omega}{\cfe}$ is a sum with contributions from each Hermite mode $\hm=3,\dots,N$. 
The location of the largest term in this sum is also a good indicator of the extent of the plateau.
In \figs\ref{fig:ExpectedMVsStrength} and \subref{fig:ExpectedMVsFrequency} we plot the $\hm$-value corresponding to the largest term in the sum $\sum_{\hm=3}^Nb_{\hm}^* \lp\frac{\hm}{N}\rp^\hdexp \distpert_{\hm}$, the numerator in \eqref{eq:dwdeta}. 
To the left of the plateau, this is at $\hm=N$,
it then decreases algebraically until it reaches $\hm=3$. 
\Fig\ref{fig:ExpectedMVsStrength} shows that the plateau begins with $\hm$ slightly smaller than $N$ at $\cfe=6$,
while 
\Fig\ref{fig:ExpectedMVsFrequency} shows that the end of the plateau $\cfe/N^{\hdexp}\approx1.2\times10^{-6}$ corresponds to $\hm\approx10$ for all resolutions.

We may relate the largest term in \eqref{eq:dwdeta} to the dissipation of free energy by collisions (discussed in \sec\ref{sec:EigenmodeFormationIVP}) as follows.
Since the eigenvector $\a$ and adjoint eigenvector $\b$ both satisfy the theoretical spectrum \eqref{eq:theoreticalspectra},
we may approximate the sum
\begin{align}
\begin{split}
  \sum_{\hm=3}^Nb_\hm^* \lp\frac{\hm}{N}\rp^\hdexp \distpert_\hm
  & \approx \sum_{\hm=3}^N \lp\frac{\hm}{N}\rp^\hdexp  |\distpert_\hm|^2
  \\
	& = \sum_{\hm=3}^N  \lp\frac{\hm}{N}\rp^{\hdexp}  \frac{2C}{\sqrt{2\hm}}\exp\lp -\frac{\gamma}{|\gamma|}\lp\frac{\hm}{\hm_\gamma}\rp^{1/2} - \lp\frac{\hm}{\hm_c}\rp^{\hdexp+1/2}\rp.
  \label{eq:FreeEnergyDissipationSpectrum}
\end{split}
\end{align}
The \rhs\ of this is the free energy dissipated by collisions (see ${\cal C}$ \eqref{eq:FreeEnergyCollisionalSinkChp3}).
Thus the largest term in \eqref{eq:FreeEnergyDissipationSpectrum} is the Hermite mode which is most affected by collisions.

From \eqref{eq:FreeEnergyDissipationSpectrum}
we may find an approximate expression for the location of the largest term, and use this to derive inequalities for the extent of the plateau.
Treating $\hm$ as a continuous variable and taking the $\hm$-derivative of \eqref{eq:FreeEnergyDissipationSpectrum},
we find that the location of the largest term $\hm$ satisfies
\begin{align}
  \frac{\cfe}{N^\hdexp}\hm^{\hdexp+1/2} + \gamma \hm^{1/2} - \frac{\hdexp-1/2}{\sqrt{2}} = 0.
  \label{eq:LocationPolynomials}
\end{align}
The numerical solution of this equation (constrained between 3 and $N$) is plotted for the $N=256$ case in dashed black in \fig\ref{fig:ExpectedMVsStrength} and is in excellent agreement with the observed largest value.
To derive inequalities for the extent of the plateau, we solve \eqref{eq:LocationPolynomials} approximately by taking balances of two of the terms. 
The first term is positive while the other two are negative, so we balance the first term with the larger of the other two.
This is too crude to give quantitative agreement, but we shall derive qualitative behaviour.
With $\hdexp=6$, $\gamma=-0.387$, the second term is larger than the third when $\hm\gtrsim100$.
For sufficiently large $N$, this is the case at the left-hand end of the plateau.
Thus balancing the first two terms in \eqref{eq:LocationPolynomials}, we find that the largest term is located at
\begin{align}
  \hm = \lp\frac{-\gamma}{\cfe}\rp^{1/\hdexp} N,
  \label{eq:LargestTermApproximateLocation}
\end{align}
provided $\gamma<0$.
For the parameters to give a point in the plateau, the dominant $\hm$ must satisfy $\hm<N$, so dividing \eqref{eq:LargestTermApproximateLocation} by $N$ we derive a lower bound for $\cfe$,
\begin{align}
  \cfe > -\gamma \id \etacrit.
  \label{eq:EtaCrit}
\end{align}
Plotting this in \fig\ref{fig:PlateauVsStrength}, we see that this underestimates the beginning of the plateau; in fact we have already observed that the plateau begins at $\etacritobs=6$.
Substituting $\etacritobs=6$ into \eqref{eq:LargestTermApproximateLocation}, we deduce that rather than $\hm<N$, we must have $\hm\lesssim 0.6 N$.
That is, even the smallest collision frequency must yield a spectrum where $\sim40\%$ of the modes are after the most damped mode.

At the right-hand end of the plateau, $\hm<100$ and the first and third terms in \eqref{eq:LocationPolynomials} balance, giving
\begin{align}
  \hm = \ls\frac{N^\hdexp(\hdexp-1/2)}{\cfe\sqrt{2}}\rs^{1/(\hdexp+1/2)} .
  \label{eq:RightHandEndSolution}
\end{align} 
\Fig\ref{fig:PlateauVsFrequency} shows that the plateau ends at a critical value of the collision frequency, $(\cfe/N^\hdexp)_\crit$, 
so rearranging \eqref{eq:RightHandEndSolution} we obtain the inequality 
\begin{align}
\frac{\cfe}{N^\hdexp} <   \frac{(\hdexp-1/2)}{\hm_\crit^{\hdexp+1/2}\sqrt{2}} = \lp \frac{\cfe}{N^\hdexp}\rp_\crit,
  \label{eq:EtaOnNCrit}
\end{align} 
where $\hm_\crit$ is a critical value of the peak of the Hermite spectrum defined implicitly by the equality in \eqref{eq:EtaOnNCrit}.
Taking the observed value of $(\cfe/N^\hdexp)_\crit\approx1.2\times10^{-6}$ from \fig\ref{fig:PlateauVsFrequency}, we obtain $\hm_\crit=10$ (which also agrees well with the value given by \fig\ref{fig:ExpectedMVsFrequency}).
This shows that the even the largest collision frequency must leave around 10 modes unaffected.

We use the inequalities for the two ends of the plateau to estimate the minimum possible resolution $N$.  
At the minimum resolution, the left and right ends of the plateau coincide
so we combine the conditions $\hm\geq \hm_\crit =10$
and $\hm\lesssim 0.6 N$ 
to obtain the lower bound $N\gtrsim 17$.
Again, this is a conservative estimate---the correct growth rate was calculated in \sec\ref{sec:hypercollisions} with $N=10$---however for $N\lesssim17$, there is no guarantee of a range of $\cfe$ values that give the correct result.

The existence of $\cfe_{\mathrm{crit}}$ gives a discrete analog of the work of \citet{Ng99,Ng04}.
There, Landau-damped modes are found to be the eigenmodes of the  collisional problem in the limit of vanishing collisions;
however the eigenfunctions of the purely collisionless problem are the undamped \cvk\ modes.
In our work \cvk\ modes are found without collisions, and Landau damped modes are found with hypercollisions provided $\cfe>\cfe_{\mathrm{crit}}$.
This means to take the limit of vanishing collisions $\cfe/N^\hdexp\to0$ we must also take the limit of infinite resolution $N\to\infty$. 
%\Fig\ \ref{fig:gr_vs_coll} shows it is not possible to take the limit of vanishing collisions in the discrete setting without also taking the limit of infinite resolution.
%That is, the correct value is found for the growth rate in the limit of zero collision frequency if this limit is attained by increasing $N$ rather than decreasing $\cf$.
%%%Indeed when we have enough resolution to capture the collisional cutoff, $n_c<N$, \eqn\ (\ref{eq:abs_coll}) also shows the collision frequency is bounded below for finite $N$
%%%\begin{align}
%%%  \frac{\cf}{N^\hdexp} > \frac{\hdexp+1/2}{\sqrt{2}N^{\hdexp+1/2}}.
%%%\end{align}
%and the collision frequency can only tend to zero in the limit $N\to\infty$.

\begin{figure}[tbp]
  \centering
  \subfigure[\label{fig:GrVsCollVsMHigh}]{\includegraphics[width=.49\textwidth]
      {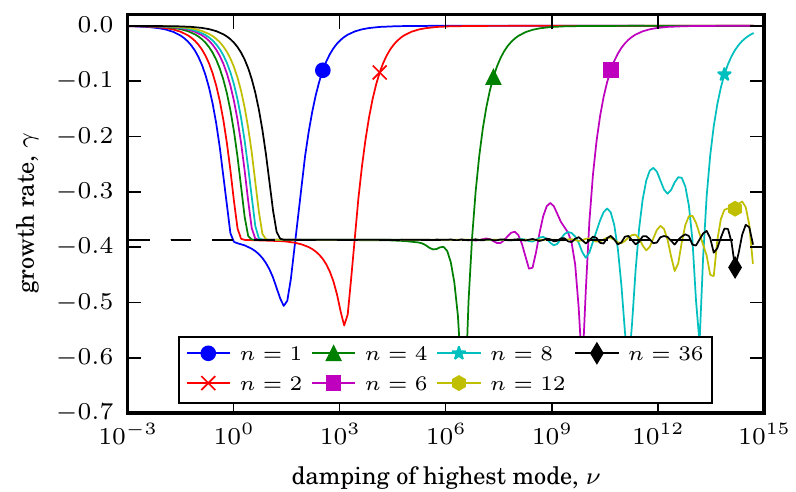}}
      \subfigure[\label{fig:GrVsCollVsMLow}]{\includegraphics[width=.49\textwidth]
      %{plateau_gr_against_m_collision_strength_16.pdf}}
    {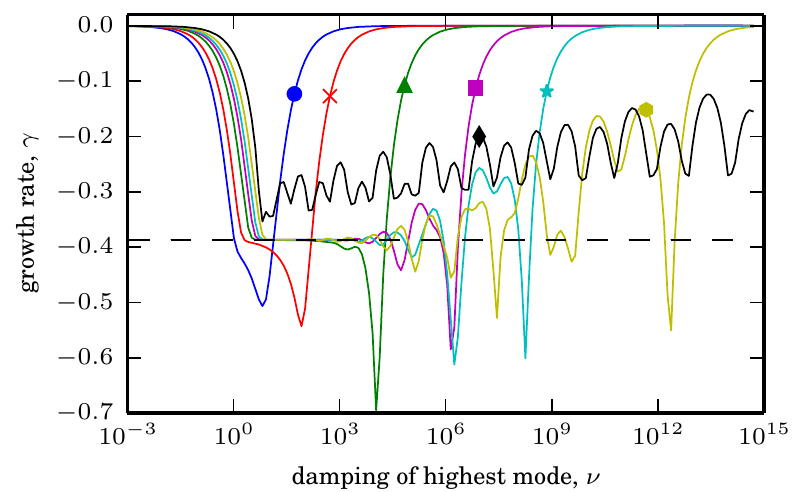}}
  \caption[The hypercollision plateau for different hypercollision exponents.]{\label{fig:GrVsCollVsM}
  The calculated growth rate for different hypercollision exponents plotted against the damping of the highest mode for different resolutions (a) $N=128$, and (b) $N=32$. The $\hdexp=1$ corresponds to the Kirkwood operator, $\hdexp=6$ to the hypercollision operator used in \sec\ref{sec:collisions}, and $\hdexp=36$ to the velocity space version of the Hou--Li filter \cite{Hou07} used in \citet{Parker14}.
}
\end{figure}

\subsection{Hypercollisional plateau against hypercollision exponent}
\label{sec:PlateauVsM}
We now study the plateau as a function of hypercollision exponent $\hdexp$.
In \fig\ref{fig:GrVsCollVsM} we plot the calculated growth rate against the damping of the highest mode for different $\hdexp$ at resolutions $N=32$ and $N=128$.
The overall pattern is very similar to \fig\ref{fig:PlateauVsStrength}, 
with the plateau beginning at a nearly constant $\cfe$ and ending at fixed $(\cfe/N^\hdexp)$.
Thus the plateau extends further with increasing $\hdexp$.
This is seen in the plot for high resolution, $N=128$, in \fig\ref{fig:GrVsCollVsMHigh}.
Two features of this plot are notable.
Firstly, there is effectively no plateau for the Kirkwood collision operator, $\hdexp=1$, indicating it is not appropriate for use as an operator for providing numerical dissipation (as opposed to a physically-motivated collision operator).
Secondly, the behaviour at the right-hand end of the plateau changes with increasing $\hdexp$.
For low $\hdexp\lesssim4$, the growth rate decreases further at the right-hand end of the plateau, before returning monotonically to zero.
At higher $\hdexp\gtrsim6$, the growth rate beyond the end of the plateau oscillates about the correct value before returning to zero.
\Fig\ref{fig:GrVsCollVsMHigh} suggests that one should increase $\hdexp$ indefinitely,
with even very sharp cutoffs like $\hdexp=36$ (which corresponds to the velocity space version of the Hou--Li filter \cite{Hou07} employed in \citet{Parker14}) performing very well.
However this is only the case for large $N$.
\Fig\ref{fig:GrVsCollVsMLow} shows the plateaus for $N=32$.  
The low values of $\hdexp$ still have effectively no plateau, but now large $\hdexp$ also do not have a plateau.
With even smaller resolutions down to around $N=16$, only $\hdexp\in\{4,6,8\}$ retain any plateau.
Thus for simulations with limited resolution we should choose $\hdexp\in\{4,6,8\}$ for which there is a robust plateau. 

\subsection{Convergence with hypercollision parameters}

\begin{figure}[tbp]
  \centering
  \subfigure[]%\label{fig:GrVsCollVsMHigh}]
  %{\includegraphics[width=0.48\textwidth]{images/hypercollisions/convergence_against_m/error_nu_10.eps}}
  {\includegraphics[width=0.48\textwidth]{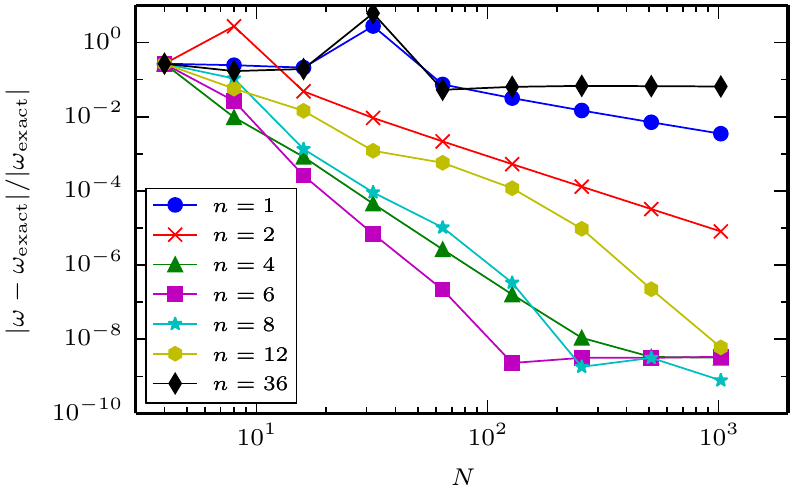}}
      \subfigure[]%\label{fig:GrVsCollVsMLow}]
      %{\includegraphics[width=0.48\textwidth]{images/hypercollisions/convergence_against_m/error_nu_10_hdexp.eps}}
      {\includegraphics[width=0.48\textwidth]{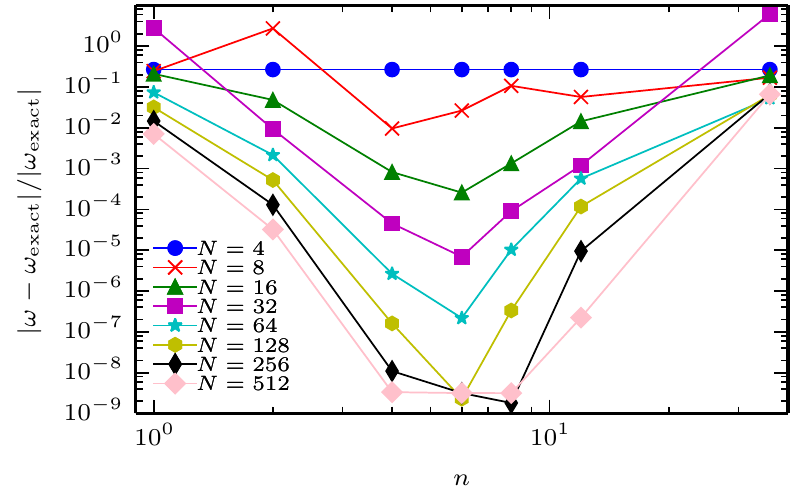}}
    \subfigure[]%\label{fig:GrVsCollVsMLow}]
      %{\includegraphics[width=0.48\textwidth]{images/hypercollisions/convergence_against_m/error_nu_100.eps}}
      {\includegraphics[width=0.48\textwidth]{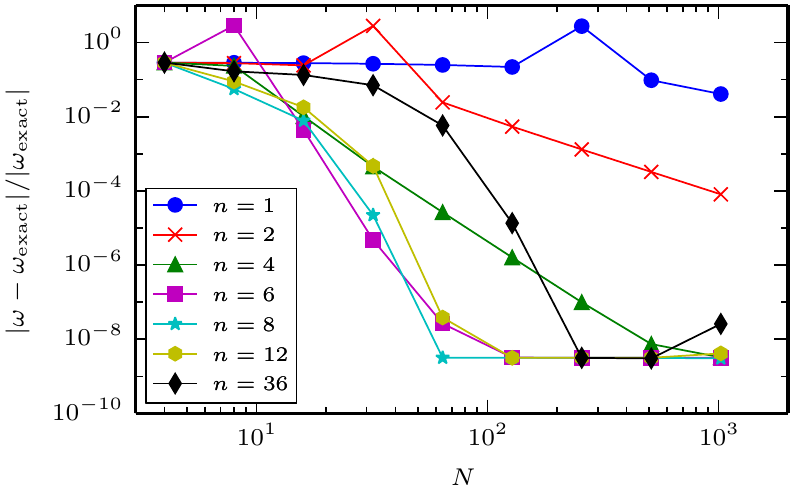}}
    \subfigure[]%\label{fig:GrVsCollVsMLow}]
    %{\includegraphics[width=0.48\textwidth]{images/hypercollisions/convergence_against_m/error_nu_100_hdexp.eps}}
    {\includegraphics[width=0.48\textwidth]{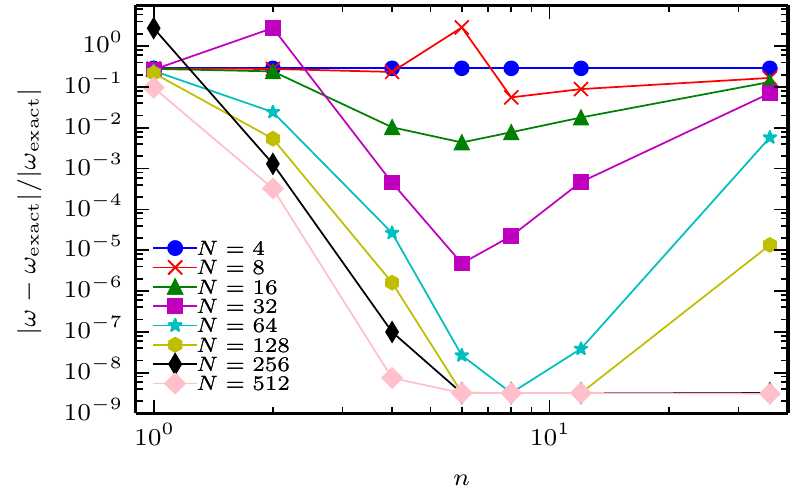}}
  \caption[Convergence of growth rate calculation.]{\label{fig:ConvergenceWithM}
  Convergence to the complex frequency with $N$ and $\hdexp$ at $k/\alpha_i=6$ and fixed $\cfe$.
  (a) Convergence against $N$ for different $\hdexp$ with $\cfe=10$ and (b) the same data plotted against $\hdexp$ for each $N$.
  (c,d) Corresponding plots with $\cfe=100$.
}
\end{figure}

Finally we consider convergence of the calculated complex frequency $\omega$ to the exact frequency from the dispersion relation $\omega_{\textrm{exact}}$.
We plot the normalized error $|\omega-\omega_{\textrm{exact}}|/|\omega_{\textrm{exact}}|$ at fixed $\cfe$ against $N$ and against $\hdexp$ in \fig\ref{fig:ConvergenceWithM}.
In \fig\ref{fig:ConvergenceWithM}(a) we plot the error against $N$ for $\cfe=10$ and various $\hdexp$.
This shows algebraic convergence with $N$, with fastest convergence for $\hdexp\approx6$.
In \fig\ref{fig:ConvergenceWithM}(b) we plot the same data against $\hdexp$.
This shows that convergence is fastest for the values $\hdexp\in\{4,6,8\}$ corresponding to the region with a robust plateau in \sec\ref{sec:PlateauVsM}.
However, there is almost no convergence for small ($\hdexp=1$) or large ($\hdexp\geq36$) values of $\hdexp$.

\Fig\ref{fig:ConvergenceWithM}(c,d) correspond to \Fig\ref{fig:ConvergenceWithM}(a,b), but now with $\cfe=100$.
These exhibit similar behaviour, but now with increased collision strength, larger values of $\hdexp$ yield convergence.

\section{Initial value problem}
\label{sec:ivp}

We now return to the free energy equation \eqref{eq:EmuPDE} and solve it as an \ivp\ by the method of characteristics.
This allows us to explain recurrence in terms of free energy flow in Hermite space.
We also show that an eigenmode forms in the long time limit of a sufficiently damped system.

\subsection{Solution by method of characteristics}

\begin{figure}[tb]
  \centering
%\subfigure{\includegraphics[trim=1cm 0.5cm 3.5cm 0.5cm,width=0.49\textwidth,clip]{contour_recurrence.png}}
%\subfigure{\includegraphics[trim=1cm 0.5cm 3.5cm 0.5cm,width=0.49\textwidth,clip]{contour.png}}
  \subfigure[]{\includegraphics[width=0.48\textwidth]{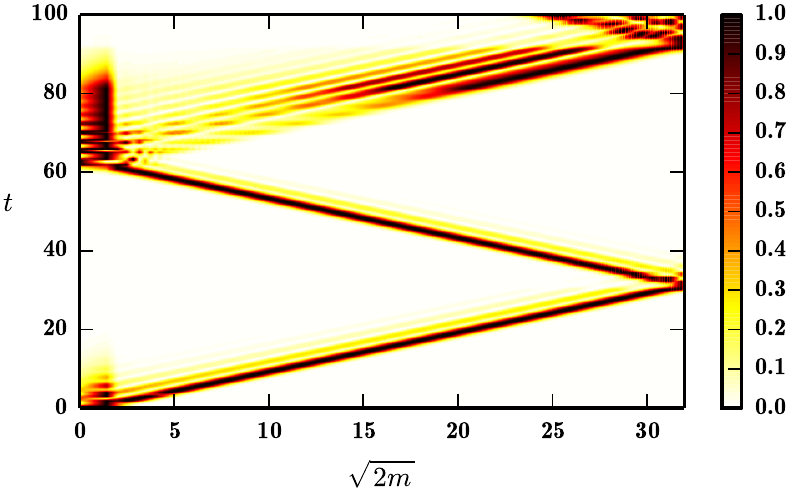}}
  \subfigure[]{\includegraphics[width=0.48\textwidth]{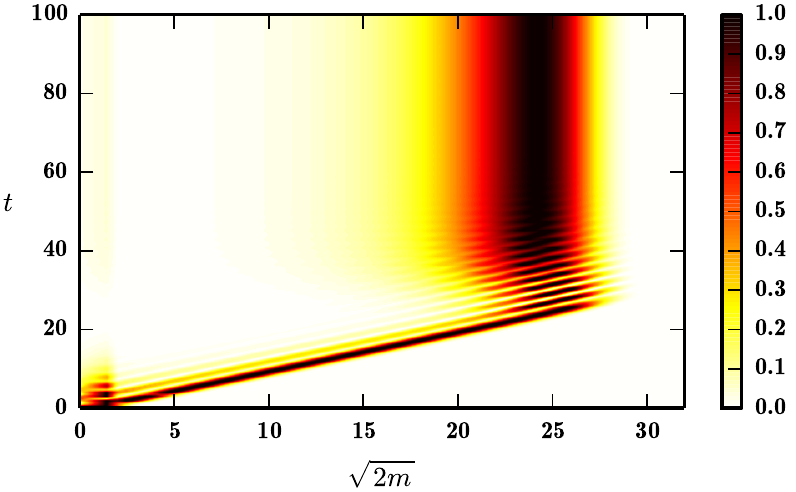}}
  %{\small (a)\hspace{.47\textwidth}(b)\hspace{.47\textwidth}}
  \caption[Free energy propagation with and without recurrence.]{Normalized Hermite spectra for the initial value problem. At each time level the spectrum is normalized by the maximum value of $|\distpert_\hm|^2/2$ at that time. Dark areas show relatively large values of free energy. We use the variable $\mu=\sqrt{2\hm}$ so that characteristics are straight lines with gradient one. (a) With no damping the free energy reflects off the highest mode and propagates back to the lower modes. (b) With hypercollisions free energy is damped before reaching the highest mode. An eigenmode similar to \fig\ref{fig:theoreticalspectra_b} forms. \label{fig:ivp_contour}}
\end{figure}

The characteristic \eqn s for \eqref{eq:EmuPDE} are,
\begin{align}\label{eq:char_eqn}
  \fd{\Emu}{\mu} = -\frac{2\cf \mu^{2\hdexp}}{(2N)^{\hdexp}}\Emu,
  \hspace{1cm}
  \fd{\mu}{t} = 1,
\end{align}
showing that free energy propagates outwards along the straight line characteristics $\mu=\mu_0+t$.
The solution of \eqref{eq:char_eqn} is
\begin{align}\label{eq:mu>t}
	\Emu = \Emuzero(\mu-t)\exp\lp \frac{\cf}{(\hdexp+1/2)(2N)^{\hdexp}}\lp(\mu-t)^{2\hdexp+1}-\mu^{2\hdexp+1}\rp\rp,
%  \hat{E} = \hat{E}_0(\mu-t)\exp\lp \frac{\nu(\mu-t)^{2n+1}}{(n+1/2)(2N)^n}-\frac{\nu\mu^{2n+1}}{(n+1/2)(2N)^n}\rp,
~~~~~
\text{for} 
~~~
\mu \geq t,
\end{align}
where $\Emuzero(\mu)=\Enomu(\mu,t=0)$.
Without collisions, %$\cf=0$, 
the initial values are conserved along characteristics, while with collisions the $\Emu$ decay at a $\mu$-dependent rate.
We only have initial data for non-negative mode numbers, $\hm\in[0,N]$ or $\mu\in[0,\sqrt{N}]$.
The \charc s $t = \mu - \mu_0$ from these data only cover the region in phase space where $\mu\geq t$.
In the region $\mu < t$, we still have \charc s with $\tfd{\mu}{t}=1$, but no knowledge of the value $\Emu$ along these. 
Moreover in deriving \eqn~(\ref{eq:EmuPDE}) we neglected the driving and Boltzmann response terms in $\hm=0,1,2$, and took the limit of large $\hm$.
Therefore to find the solution inside the \charc\ cone $\mu< t$ we must solve (\ref{eq:asystemHD}) numerically.  
The solutions for a collisionless and collisional case are shown in \fig\ref{fig:ivp_contour}.

We can use the solution (\ref{eq:mu>t}) to investigate the recurrence phenomena seen in Hermite simulations of collisionless initial value problems \cite{Grant67}.
Since instabilities are driven on large scales, the initial conditions are dominated by low Hermite modes.
These propagate to higher $\hm$ along the characteristics 
%$t = \sqrt{2n} - \sqrt{2n_0}$ 
%$n=\frac{1}{2}\lp t+\sqrt{2n_0}\rp^{2}$
%$\hm=\lp t/\sqrt{2}+\sqrt{\hm_0}\rp^{2}$
$\hm^{1/2}=\hm_0^{1/2} + t/\sqrt{2}$
and are undamped as $\cf=0$.
When the propagation reaches $\hm=N$, the calculation becomes invalid as the condition applied is $\distpert_{N+1}=0$. 
This contradicts the assumption that $\distpert_\hm$ is slowly varying in $\hm$, and results in $\distpert_{N}$ taking an incorrect value.  Through mode coupling, this error propagates back to lower modes.
Assuming the initial conditions are only in $\distpert_0$, the propagation of initial conditions to the highest mode takes a time $t = \sqrt{2N}$ and so limiting the velocity space resolution also limits the validity of the solution in time. 
This is seen in \fig\ref{fig:ivp_contour}(a) where
without sufficiently strong collisions, free energy reflects back into low modes causing an incorrect solution.
This gives an explanation for the scaling of the time until recurrence with total number of modes used that was observed in early simulations \cite{Armstrong67,Grant67}. 

\Eqn\ \Ex{eq:EmuPDE} only has outward-propagating characteristics, while \fig\ref{fig:ivp_contour}(a) shows both outwards and inward propagating characteristics.  We may also describe inward-propagating characteristics in this framework.  
The derivation of \Ex{eq:EmuPDE} depends on the modelling assumption that $b_{\hm}=i^{\hm}\distpert_{\hm}$ is slowly varying, so that $b_{\hm+1}\approx b_{\hm}$ and (\ref{eq:flux}) gives a flux.
This leads to the outward characteristics $t = \mu - \mu_0$.
However one may equally use $c_{\hm}=(-i)^{\hm}\distpert_{\hm}$, assuming instead $c_{\hm+1}\approx c_{\hm}$.
This reverses the sign on the flux and results in inward characteristics $t = \mu_0 - \mu$. 
This provides the mechanism for reflected free energy to propagate towards low modes if collisions provide insufficient damping in the region of the truncation
point.
Putting $c_{\hm}$ into \eqn~(\ref{eq:difference_eqn}) and proceeding as before leads to
\begin{align}
  \lp\pd{}{t}-\pd{}{\mu}\rp\Emu = -2\cf \mu^{2\hdexp}(2N)^{-\hdexp}\Emu,
\end{align}
(with $\Emu = \mu c^2_{\hm}/2$),
like \eqn~(\ref{eq:EmuPDE})
but with 
%$(\tpd{}{t}-\tpd{}{\mu})$ 
$(\partial_{t}-\partial_{\mu})$ 
replacing 
%$(\tpd{}{t}+\tpd{}{\mu})$.
$(\partial_{t}+\partial_{\mu})$.
Solving this gives
\begin{align}
    \Emu = \Emuzero(\mu+t)\exp\lp \frac{\cf}{(\hdexp+1/2)(2N)^\hdexp}\lp\mu^{2\hdexp+1}-(\mu+t)^{2\hdexp+1}\rp\rp,
\end{align}
with \charc s $\mu=\mu_0-t$, which is the second, inwards \charc\ in %\fig\ref{fig:ivp_contour}(a).
\linebreak Figure \ref{fig:ivp_contour}(a).

In contrast, when collisions are suitably effective, no free energy is reflected and the eigenmode with largest growth rate is found as the solution in the large time limit, as shown in \fig\ref{fig:ivp_contour}(b).

\subsection{Wave analysis of the \ivp}
We can capture both inwards and outwards \charc s in a single framework using an approach adapted from the wave analysis of finite difference schemes \cite{Trefethen82,Whitham11,Strikwerda04}.
%\Eqn\ \Ex{eq:bm} is first order in time so has waves travelling in one direction only, however we can find .
We put the wave Ansatz
\begin{align}
	b_{\hm}=\hat{b}(\hm)\exp(i(k\hm-\omega(k,\hm)t))
  %\label{}
\end{align}
into \Ex{eq:bm} and, neglecting collisions, find the frequency
    \begin{align}
      \omega = \sqrt{2(\hm+1)}\ \frac{\hat{b}(\hm+1)}{\hat{b}(\hm)}\frac{1}{2i}\lp e^{ik} - \frac{\hat{b}(\hm-1)\sqrt{\hm}}{\hat{b}(\hm+1)\sqrt{\hm+1}}e^{-ik}\rp.
      \label{}
    \end{align}
  Seeking a real frequency, we decompose this as
    \begin{align}
      \omega = \sqrt{2(\hm+1)}\ \frac{\hat{b}(\hm+1)}{\hat{b}(\hm)} \sin k,
      \hspace{1cm}
      \frac{\hat{b}(\hm-1)\sqrt{\hm}}{\hat{b}(\hm+1)\sqrt{\hm+1}} = 1,
      \label{}
    \end{align}
    where the second term is two independent recurrence relations, one each for even and odd $\hm$.
  Solving with a power series we find
    \begin{align}
      \hat{b}(\hm) = b_{\pm}\lp \hm^{-1/4} - \frac{5}{16}\hm^{-5/4} + \O\lp \hm^{-9/4}\rp\rp,
      %\label{}
    \end{align}
    where $b_+$ and $b_-$ are two undetermined constants corresponding to the even and odd series respectively.
    For the amplitude of $\omega$ to vary smoothly with $\hm$, we need $b_+ = \pm b_-$. 
    Thus there are two cases for $\hat{b}(\hm)$, one where the sign changes between $\hm$ and $\hm+1$, and one where it does not. 
    This gives two cases for the frequency
    \begin{align}
      \omega = \pm\sqrt{2(\hm+1)}\lp 1- \frac{1}{4\hm}\rp\sin k + \O(\hm^{-3/2}).
      \label{}
    \end{align}
    where the $\pm$ sign is coupled to the sign in $b_+=\pm b_-$.
    The sign may be absorbed into a change in $k$, so we consider only the case where $b_+=b_-$ and $\omega$ takes the plus sign.

    In the \ivp, the initial conditions in the low moments cause an outgoing wave.
    For the boundary condition $b_{N+1}=0$ to be satisfied, the outgoing wave must cancel with a reflected wave at the boundary, so we seek a solution as the sum of two waves:
    \begin{align}
      b_\hm = \hat{b}(\hm)e^{i(k\hm-\omega(k) t)}  + \alpha \hat{b}(\hm)e^{i(k^*\hm-\omega(k^*) t)}.
      \label{eq:WaveSumAnsatz}
    \end{align}
    For $b_{N+1}=0$ to hold at all times, we require $\omega(k)=\omega(k^*)$, so for non-trivial solutions $k^*=\pi-k$.
    Then $b_{N+1}=0$ becomes an \eqn\ for $\alpha$, giving $\alpha = (-1)^{N}e^{2ik(N+1)}$ and
    \begin{align}
      b_\hm = \hat{b}(\hm)e^{i(k\hm-\omega t)} + (-1)^{N+\hm}e^{2ik(N+1)} \hat{b}(\hm)e^{-i(k\hm+\omega t)}.
      %\label{}
    \end{align}
    Thus the reflected wave has the same frequency and minus the wavenumber of the original wave.  
    The amplitude of the reflected wave oscillates with $\hm$, analogous to the oscillation $c_\hm=(-1)^\hm b_\hm$ for the inward wave in the previous theory. 

    Collisions are simple to include. Retaining the \rhs\ of \Ex{eq:bm}, the frequency is modified to
    \begin{align}
      \omega = \sqrt{2(\hm+1)}\lp 1- \frac{1}{4\hm}\rp\sin k  - i\cfe \lp\frac{\hm}{N}\rp^n + \O(\hm^{-3/2}).
      \label{}
    \end{align}
The algebra follows as before, except each wave solution is now damped by a factor of $\exp\lp -\cfe \lp \hm/N\rp^mt\rp$.
An incoming wave is still needed to satisfy the boundary condition, but in practice it is not seen if the outward-travelling wave has been damped sufficiently, as in \fig\ref{fig:ivp_contour}(b).

\subsection{Eigenmode formation in the \ivp}
\label{sec:EigenmodeFormationIVP}

The free energy balance also gives insight into the \ivp.
As noted in \sec\ref{sec:LandauCollOpAndProperties}, the total free energy is 
$W=W_f+W_{\varphi}$ where $W_f=\sum_{m=0}^{\infty}|g_m|^2/2$ and $W_{\varphi}=|\varphi|^2/2$ are the free energy contributions from the distribution function and the electrostatic potential respectively.
Evolution equations for $W_f$ and $W_{\varphi}$ are 
derived by multiplying \eqref{eq:asystemHD} by $\bar{\distpert}_m$, and adding the resulting equation to its complex conjugate,
giving an equation for $\tpd{|\distpert_\hm|^2}{t}$.
Summing these equations over $\hm$ gives an equation for $\tfd{W_f}{t}$, while the $\hm=0$ case gives an equation for $\tfd{W_\varphi}{t}$:
\begin{align}
  \fd{W_f}{t} + {\cal T} + {\cal F} = {\cal C},
  \hspace{1cm}
  \fd{W_\varphi}{t} - {\cal F} = 0,
	\label{eq:FreeEnergyEvolutions}
\end{align}
where
\begin{align}
	{\cal T} = \Re\lp \frac{i\tilde{\omega}_T}{k}\varphi\intii \lp v^2-\frac{1}{2}\rp \bar{\distpert}~ \d v\rp
	= \Re\lp \frac{i\tilde{\omega}_T}{k\sqrt{2}} \distpert_0\bar{\distpert}_2 \rp,
\end{align}
is the free energy source due to the temperature gradient,
\begin{align}
  \label{eq:FreeEnergyCollisionalSinkChp3}
  {\cal C} = \Re\lp \cfe \intii \frac{\bar{\distpert}C[\distpert]}{F_0} ~ \d v\rp
  = \cfe \sum_{\hm=3}^{N} \lp\frac{\hm}{N}\rp^\hdexp |\distpert_\hm|^2,
\end{align}
is the free energy sink due to collisions
and 
\begin{align}
	{\cal F} = \Re\left( i\sqrt{2}\bar{\distpert}_1\distpert_0\right),
	%\label{<++>}
\end{align}
is the free energy flux out of $\hm=0$.
Summing the two equations in \eqref{eq:FreeEnergyEvolutions} yields the conservation equation
\begin{align}
  \label{eq:FreeEnergyConservationChp3}
  \fd{}{t}\left( W_f + W_{\varphi}\right) + {\cal T}  = {\cal C}.
\end{align}
As noted in \S\ref{sec:LandauCollOpAndProperties}, free energy is conserved in the absence of driving and collisions.

Introducing the time-integrated sources and sinks,
\begin{align}
  W_T = \int_0^t {\cal T}(t') ~\d t,
  \hspace{1cm}
  W_C = \int_0^t {\cal C}(t') ~\d t,
\end{align}
equation \eqref{eq:FreeEnergyConservationChp3} becomes the conservation law
\begin{align}
%  \label{eq:}
  \fd{}{t}\left( W_f + W_{\varphi} +W_T - W_C \right) = 0 .
\end{align}

\begin{figure}[tb]
  \centering
\subfigure[\label{fig:W}
]{\includegraphics[width=0.49\textwidth,trim=0.0cm 0.0cm 0.0cm 0.0cm,clip=true]
{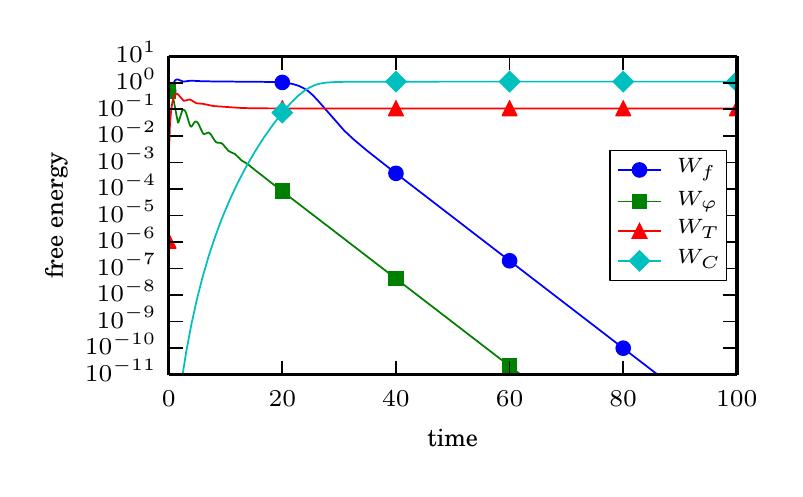}}
\subfigure[\label{fig:dWdt}
]{\includegraphics[width=0.49\textwidth,trim=0.0cm 0.0cm 0.0cm 0.0cm,clip=true]
{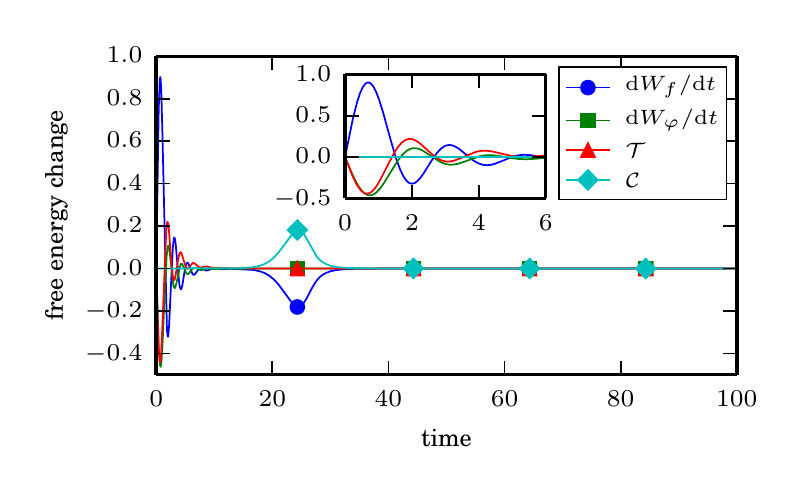}}
\subfigure[%\label{fig:contours}
]{\includegraphics[width=0.49\textwidth,trim=0.0cm 0.0cm 0.0cm 0.0cm,clip=true]
{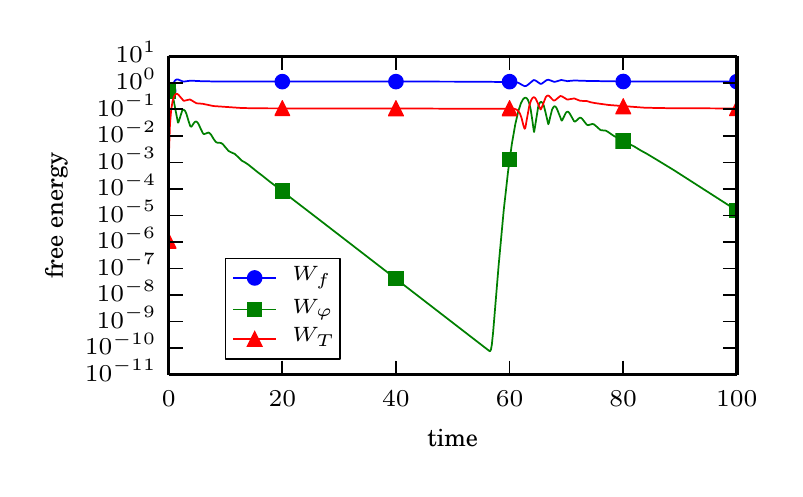}}
\subfigure[%\label{fig:contours}
]{\includegraphics[width=0.49\textwidth,trim=0.0cm 0.0cm 0.0cm 0.0cm,clip=true]
{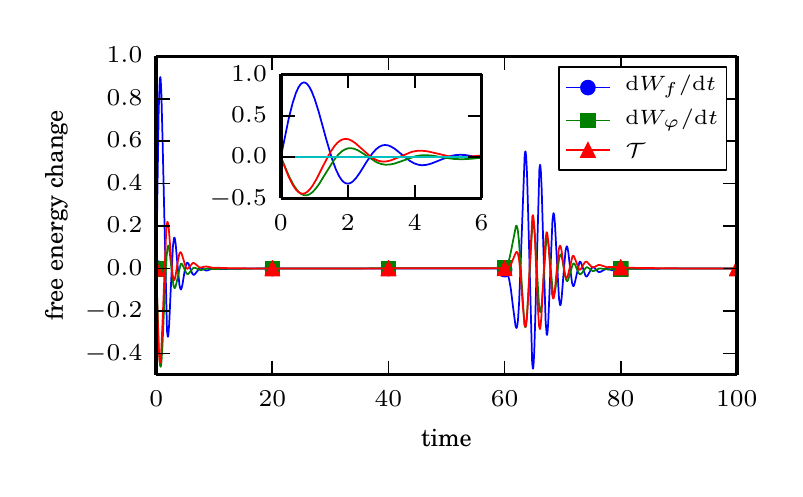}}
\caption[Time traces for free energies and their respective time derivatives.]{ 
  Time traces for (a) free energies and (b) their respective time derivative for the \ivp\ for $k/\alpha_i=6$ and with hypercollisions $\cfe=10$, $\hdexp=6$.
  (c,d) Corresponding plots with no collision operator.
\label{fig:FreeEnergyTimeTrace}
}
\end{figure}

In \fig\ref{fig:FreeEnergyTimeTrace}(a) and (b) we plot the free energies and their derivatives against time for an initial value simulation of the decaying mode $k/\alpha_i=6$ with hypercollisions ($\cfe=10, \hdexp=6$) and resolution $N=512$, corresponding to \fig\ref{fig:ivp_contour}(b).
After an initial transient lasting until $t=10$, the system enters a state reminiscent of the linear Landau solution \eqref{eq:fSoln} and \eqref{eq:phiSoln}, where the free energy of the electric field $W_{\varphi}$ decays exponentially while the free energy of the distribution $W_f$ remains constant.
This corresponds to the initial propagation of free energy out to high moments, as seen for early times in \fig\ref{fig:ivp_contour}.
On reaching the highest modes, the free energy in the distribution function is damped by collisions, as shown by the increase in ${\cal C}$ and corresponding decrease in $\tfd{W_f}{t}$ at around $t=25$ in \fig\ref{fig:FreeEnergyTimeTrace}(b).
Thereafter both $W_\varphi$ and $W_f$ decay exponentially with the linear damping rate, showing that an eigenmode solution has formed.

For comparison, in \fig\ref{fig:FreeEnergyTimeTrace}(c) and (d) we also plot the time traces for the system without collisions, corresponding to \fig\ref{fig:ivp_contour}(a).
After the initial transient, this system enters the Landau regime and remains in this regime until the recurrence time.
This corresponds to the time taken for free energy to return to the lowest modes, having reflected from the effective hard-wall boundary that results from setting $\distpert_{N+1}=0$ at the highest mode at around $t=25$, rather than being damped as in \fig\ref{fig:FreeEnergyTimeTrace}(b).
Note that, except during the transient and the recurrence, $W_f$ remains constant and $W_\varphi$ decays at the correct linear rate.
Note also that collisions have no effect on $W_\varphi$ because $C[\phi_0]=0$, so the values of $W_\varphi$ in the two simulations are the same until the recurrence time.
Thus one can use the strictly collisionless system to determine collisionless behaviour, but the validity of the simulation is limited in time.

%%%\subsection{Landau damping as free energy flow}
%%%Finally, we can use the theoretical spectrum \eqref{eq:theoreticalspectra} to determine the dependence of free energy $W_f$,
%%%collisional dissipation ${\cal C}$ 
%%%and the free energy flux $\Gamma_m\approx|\distpert_m|^2\sqrt{m/2}$ (\sec\ref{sec:TheoreticalHermiteSpectra})
%%%on the hypercollision parameters.
%%%Substituting the underlying algebraic spectrum $|\distpert_m|^2/2\sim C/\sqrt{2m}$ into $W_f$ \eqref{eq:freeenergy}
%%%and approximating the sum with an integral from $m=1$ to $m=m_c=[N^n(n+1/2)/\nu]^{(1/(2n+1))}$ (\ie, where the spectrum becomes strongly damped),
%%%we obtain
%%%\begin{align}
%%%	W_f \sim \int_1^{m_c} \d m ~ \sqrt{2}Cm^{-1/2} \sim 2\sqrt{2}C\left(\frac{N^n(n+1/2)}{\nu}\right)^{\frac{1}{2n+1}}.
%%%	%\label{}
%%%\end{align}
%%%As one might expect, 
%%%the free energy contained in the distribution function increases with increasing resolution $N$ and decreases with increasing collision frequency $\nu$.
%%%
%%%Using the same approach, we approximate the collisional dissipation \eqref{eq:FreeEnergyCollisionalSink} as
%%%\begin{align}
%%%	{\cal C} \sim \nu \int_1^{m_c} \d m ~ \left(\frac{m}{N}\right)^n\frac{\sqrt{2}C}{\sqrt{m}}
%%%	\sim \frac{\nu C}{N^n(n+1/2)} = C.
%%%	%\label{}
%%%\end{align}
%%%Thus the rate of collisional dissipation is independent of resolution and all hypercollision parameters.

\section{Arbitrary velocity space grids}

The hypercollision operator is simple to implement on a Gauss--Hermite grid in velocity space, and in Hermite space, since Hermite functions are its eigenfunctions and damping is represented by a diagonal matrix $\Dmat$ in a Hermite basis. We can also apply the same hypercollision operator on arbitrary grids in velocity space, such as the parallel velocity grid $v_{\parallel}$ used in \cite{Peeters09} or the pitch-angle/energy grid $(v_\perp^2/v^2,v^2)$ (with $v\id\vperp^2+v_\parallel^2$ and $\vperp$ the perpendicular velocity) used in \cite{Numata10,GYRO}, only now the collision matrix is not diagonal.  
To construct the collision matrix, we map distribution function values on the supplied velocity grid to Hermite coefficients
with a matrix $\Tmat^{-1}$,
apply the hypercollision matrix $\Dmat$, and map
back to distribution function values at grid points with $\Tmat$. That is, we compute
\begin{align}\label{eq:fdamped_matrix_eqn} 
  %C[\fb] = \Tmat \Dmat \Tmat^{-1}\fb
	C[\boldsymbol{\distpert}] = \Tmat \Dmat \Tmat^{-1}\boldsymbol{\distpert}
\end{align}
where $\boldsymbol{\distpert}=(\distpert(v_1),\dots,\distpert(v_N))$ is the vector of distribution function values on the velocity space grid, and the diagonal hypercollision matrix in Hermite space has
coefficients 
\begin{align}
  D_{\hm l} = -\cf\lp\frac{\hm}{N}\rp^\hdexp\delta_{\hm l} \dthree.
\end{align} 
The expansion of the distribution function as $\distpert(v) = \sum_{\hm=0}^N\distpert_\hm\phi_\hm(v)$
determines the transformation matrix $\Tmat$ through
\begin{align}
  \distpert(v_j) = \sum_{\hm=0}^N \distpert_\hm \phi_\hm(v_j),  \quad T_{j\hm} = \phi_\hm(v_j),
\end{align}
and we invert $\Tmat$ numerically to find $\Tmat^{-1}$. It is only
necessary to perform this calculation once if the velocity grid does
not change in time, and computing $C[f]$ is then equivalent to
multiplication by a constant dense matrix.
\begin{figure}
% must generate growth rate figure on Maths Inst versions of python
  \centering \subfigure[] {\includegraphics[width=.49\textwidth]
    %{f-system_damped_growth_rate,N=16_paper.ps}} 
    {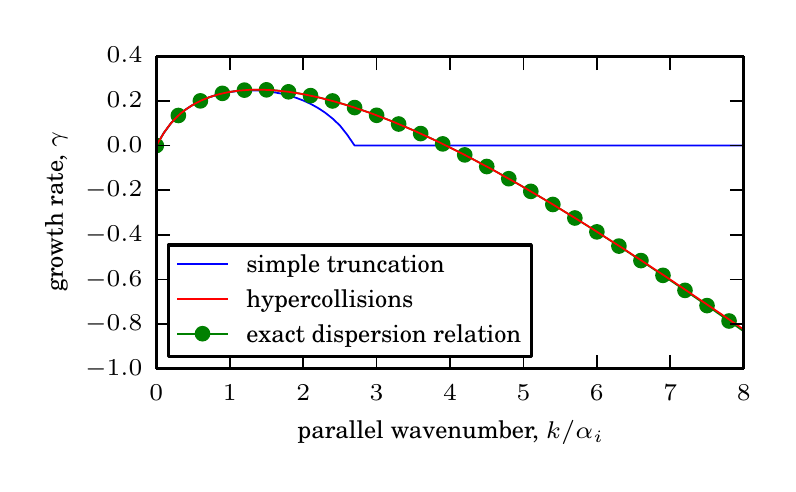}} 
    \subfigure[]{\includegraphics[width=.49\textwidth]
      %{f-system_damped_eigenfunctions,N=16_paper.ps}}
      %{f-system_damped_eigenfunctions,N=16_paper.pdf}
      {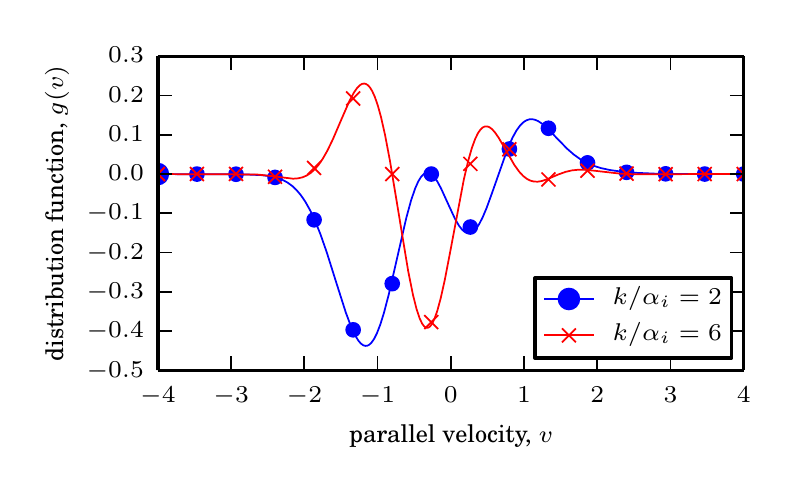}
    }
    \caption[Hypercollisional damping performed on a uniform grid.]{Hypercollisional damping performed on a 16 point uniform grid in velocity space: (a) growth rate and (b) eigenfunctions for wavenumbers $k/\alpha_i=2$ and $k/\alpha_i=6$. The calculated function values are marked and interpolated with a 15th order polynomial.}
  \label{f.uniform_hc}
\end{figure}
The growth rate and eigenfunctions calculated with hypercollisions on
a uniform velocity space grid with 16 points %using this approach 
are shown in \fig\ref{f.uniform_hc}.

\section{Summary}

In this \chp\ we have introduced the Hermite spectral representation for parallel velocity space.
Expanding the distribution function in terms of polynomials which are orthogonal \wrt\ the Maxwell--Boltzmann distribution is a relatively old idea from the kinetic theory of neutral gases dating from \citet{Burnett35} in the 1930's,
with Hermite polynomials themselves being introduced by \citet{Grad49Note,Grad49Kinetic} in the 1940's.
Hermite polynomials were also used in early plasma simulations \cite{Grant67,Armstrong67,Joyce71}
and are still valuable in modern multi-processor simulations.
Their main advantage is the neat structure of the \gkm\ equations when written in Fourier--Hermite space.
There is very little mode coupling in the kinetic equation, only nearest neighbour coupling from the streaming term.
Moreover, the integrals over parallel velocity space for the source terms in Maxwell's equations become evaluations of single Hermite coefficients, rather than sums over velocity space grids.
This allows very efficient solution of linear problems.

We have also studied the hypercollision operator.
This too is a relatively old idea, dating at least to the 1970's \cite{Joyce71},
though it appears only to have been used as an \emph{ad hoc} method to prevent recurrence in calculating Landau-damped solutions.
We showed that a suitable hypercollision operator gives accurate growth rates with around ten to twenty degrees of freedom
(compared to the $\O(100)$ for grid methods)
while also capturing Landau damping (which collisionless methods cannot do at all).
We studied the hypercollision operator parameters and showed that there is a ``plateau'' in parameter space, a region where the calculation gives an accurate growth rate and is insensitive to changes in resolution, collision frequency, and hypercollision exponent.
This means the hypercollision operator is very well suited to practical use.

We have also derived an analytic expression for the Hermite spectra of eigenmodes.
The underlying $m^{-1/2}$ spectrum is the same as that of a delta function, showing that infinitesimally fine velocity space scales would develop, were it not for hypercollisions which exponentially damp the finest scales.
To find this spectrum, we derived a partial differential equation for free energy and its flux through Hermite space.
We showed that phase mixing transfers free energy from large to small velocity space scales, analogously to the free energy cascade through Fourier wavenumbers in hydrodynamic turbulence.
In this case, however, the transfer is linear and reversible.
Indeed, in collisionless \ivp s we observe a reverse flux from small to large scales, resulting from the reflection of free energy at the highest resolved mode.
This forwards and backwards transfer of free energy will be an important mechanism in our study of the nonlinear drift kinetic system in 
%\partref\ref{sec:PartResults}.
\chp s \ref{sec:FreeEnergyFlowAndDissipation} and \ref{sec:ScalingLawsForDriftKineticTurbulence}.
There however it will not be the unphysical reflection from the highest mode which causes the flux of free energy to large scales;
it will be the result of the interaction with the $\boldsymbol{E}\times\B$ nonlinearity.

% put notation back\dots
\renewcommand{\vpara}{{v}_{\parallel}}
\renewcommand{\kpara}{{k}_{\parallel}}

\renewcommand{\b}{\boldsymbol{b}}

\chapter{Perpendicular velocity space and the Hankel transform}
\label{sec:PerpendicularVelocitySpaceHankelTransform}

%We now consider the perpendicular velocity space representation.
In \chp\ \ref{sec:ParallelVelocitySpace}, we introduced the Hermite representation for parallel velocity space 
which minimized mode coupling in the \gkm\ system.
In particular, 
the Hermite representation reduces the integrals over parallel velocity space to find the charge and current densities in Maxwell's equations to single Hermite coefficients.
We wish to achieve something similar with the more complicated integrals in perpendicular velocity space.
In \chp\ \ref{sec:GKMSystem}, we derived the \gkm\ system for $h_s$, the distribution function for charged rings (see summary \S\ref{sec:GyrokineticSummary}).
The distribution function $h_s$ is a function of guiding centre position $\R_s=\r-\rhobs$, where $\r$ is the position space coordinate and $\rhobs=\Omega^{-1}_s\b\times\v$ is the gyroradius.
In contrast, the electromagnetic field variables $\varphi$, $\Apar$, and $\Bpar$ are functions of position space $\r$.
Therefore, Maxwell's equations for determining these from $h_s$ using the quasineutrality condition and \ampere's law \eqref{eq:MaxwellSummary},
\begin{subequations}
\label{eq:MaxwellSummaryHankelChapter}
\begin{align}
\sum_s \frac{\qNs^2\nNs}{\TNs}\varphipN
  = \sum_s \qNs\nNs   \int \d^3\vN ~ \ga{h_s}{\r} ,
  \label{eq:QuasineutralitySummaryHankelChapter}
  \\
  -\nabla^2_{\perp}\Apar  = 2\beta_r \sum_s \qNs\nNs\vthNs \int\d^3\v ~ \vNpara\ga{  h_s}{\r},  \\
  -\nabla^2_{\perp}\Bpar  = \b\cdot \left[\nablaNperp \times \beta_r \sum_s \qNs\nNs\vthNs \int\d^3\v ~ \ga{\vNperpv  h_s}{\r}\rs ,
\end{align}
  \end{subequations}
	contain gyroaverages in the integrands to account for the change of variables from $\R_s$ to $\r$. 
With the Fourier representation introduced in \sec\ref{sec:FourierSpaceRepresentationGKMSystem}, 
	the gyroaverages becomes multiplications by Bessel functions.
%%%	; specifically
%%%	$\ga{g_s}{\R_s}\mapsto J_0(\rhoNs\kperp\vperp)\hat{g}$
%%%	and
%%%	$\ga{g_s}{\R_s} \mapsto ({i\vperp}/{\kperp})J_1(\rhoNs\kperp\vperp) \hat{g}_s\kperpv\times\boldsymbol{b}$,
%%%	where $\hat{g}_s$ is the Fourier transform of $g_s$ and $\kperp$ is the perpendicular wavenumber $\kperp=\sqrt{k_x^2+k_y^2}$.
	Inserting the complementary distribution function $g_s = h_s  - ({\qNs} \ga{\chiNs}{\R_s}F_{0s}/\TNs$) 
	into \eqref{eq:MaxwellSummaryHankelChapter} gives
\begin{subequations}
  \label{eq:FieldSolveMomentsHankelChapter}
\begin{align}
  &\varphiNk \sum_s \frac{\nNs\qNs^2}{\TNs}(1 - \Gamma_{0s})
  - \dBpk\sum_s \qNs\nNs\Gamma_{1s} = \sum_s \qNs \nNs \int \d^3\v ~\gks J_0(a_s),
  \label{eq:QNHankelChapter}
  \\
  &\Apk\left[ \frac{\kperp^2}{2\beta_r}   + \sum_s \frac{\qNs^2\nNs}{2\mNs}\Gamma_{0s} \rs
  = \sum_s \qNs\nNs \vthNs\int \d^3\v ~\gks \vpara J_0(a_s), 
  \label{eq:AmpParaHankelChapter}
\\
  &\varphiNk\sum_s \nNs\qNs\Gamma_{1s} +
\dBpk \lp \frac{2}{\beta_r} + \sum_s \TNs\nNs\Gamma_{2s}\rp
= - \sum_s \nNs \TNs \int \d^3\v ~\gks \vperp^2 \frac{2J_1(a_s)}{a_s} ,
  \label{eq:MaxBParaHankelChapter}
\end{align}
\end{subequations}
where 
  $\Gamma_{0s} = I_0(b_s)e^{-b_s}$,
  $\Gamma_{1s} = (I_0(b_s)-I_1(b_s))e^{-b_s}$
	and
  $\Gamma_{2s} = 2 \Gamma_1(b_s)$,
with $a_s=\rho_s\kperp\vperp$ and $b_s = (\rhoNs\kperp)^2/2$.
Thus while the integrands in \eqref{eq:FieldSolveMomentsHankelChapter} contain only simple factors in parallel velocity space, 
they contain Bessel function factors in perpendicular velocity space.

To find the fields, we therefore need to evaluate the perpendicular velocity space integral
\begin{align}
  %\g(\kvar) %= {\cal H}_{\pvar} \gks 
	\int_0^\infty \d\vperp~ \vperp J_0(\rho_s\kperp\vperp)g(\vperp) ,
  \label{eq:ContinuousHankelTransformAtK}
\end{align}
for (\ref{eq:FieldSolveMomentsHankelChapter}a,b), and the integral
\begin{align}
	\int_0^\infty \d\vperp~ \vperp^2 J_1(\rho_s\kperp\vperp)g(\vperp) ,
  \label{eq:ContinuousHankelTransformAtKdB}
\end{align}
for \eqref{eq:MaxBParaHankelChapter}.
We focus on \eqref{eq:ContinuousHankelTransformAtK} since this is always needed to find $\varphiNk$ and $\Apk$.
The integral \eqref{eq:ContinuousHankelTransformAtKdB} is only needed when $\dBpk\neq0$, and we treat this later in \sec \ref{sec:DiscreteHankelTransform}.

We will see in \sec\ref{sec:ContinuousHankelTransform} that the integral \eqref{eq:ContinuousHankelTransformAtK} is the zeroth order Hankel transform of $g$,
\begin{align}
  \g(\pvar) %= {\cal H}_{\pvar} \gks 
  %\tilde{g}(\pvar) %= {\cal H}_{\pvar} \gks 
  = \int_0^\infty \d\vperp~ \vperp J_0(\pvar\vperp)g(\vperp) ,
  \label{eq:ContinuousHankelTransform}
\end{align}
evaluated at the point $p=\rho_s\kperp$ \citep{Davies02}.
This shows that the phase space line $p=\rho_s\kperp$ is of particular significance for computing the electromagnetic field: we may find $\varphiNk$ and $\Apk$ by knowing $\g$ {only along the line} $p=\rho_s\kperp$, rather than everywhere in phase space.
This is similar to the idea that finding $\varphiNk$ and $\Apk$ only requires the zeroth and first Hermite moments.

In this Chapter we represent $g$ using the Hankel transform $\g$ to exploit this phase space locality.
We discuss two approaches to solving the \gkm\ system in Hankel space:
(1) using a purely spectral method with the grid $p_i=\rho_s\kperp=\rho_s\sqrt{k_x^2+k_y^2}$ implied by the uniform Fourier grids $k_x$, $k_y$;
and (2)
a pseudospectral method using grids of scaled Bessel roots in $p$ and $\vperp$.
For the former, we use the continuous Hankel transform \eqref{eq:ContinuousHankelTransform}
whose properties are introduced in \sec\ref{sec:ContinuousHankelTransform}.
This is equivalent to representing the distribution function $g$ as an infinite superposition of zeroth-order Bessel functions with different arguments.
For the latter, we introduce the discrete Hankel transform in \sec\ref{sec:DiscreteHankelTransform}, which is equivalent to representing $g$ as the finite sum of zeroth-order Bessel functions with different arguments.
Both the continuous and discrete Hankel transforms prove useful in computation, but when each is used depends on the properties 
of the particular system being solved.
In \sec\ref{sec:HankelSpaceStructureGyrokinetics} we discuss the Hankel space structure of gyrokinetics to inform the different Hankel space implementations in \sgk.
Finally in \sec\ref{sec:HankelTreatmentsPerpendicularVelocitySpace} we present the two approaches 
to perpendicular velocity space which are implemented in \sgk:
the spectral (Galerkin) and pseudospectral (collocation) approaches,
which are analogous respectively to the expansion coefficient and grid point value approaches discussed in \chp~\ref{sec:ParallelVelocitySpace}.

\section{Continuous Hankel transform}
\label{sec:ContinuousHankelTransform}

%%%The continuous (zeroth order) Hankel transform is defined \citep{Davies02}
%%%\begin{align}
%%%  \g(\pvar) %= {\cal H}_{\pvar} \gks 
%%%  %\tilde{g}(\pvar) %= {\cal H}_{\pvar} \gks 
%%%  = \int_0^\infty \d\vperp~ \vperp J_0(\pvar\vperp)g(\vperp) .
%%%  \label{eq:ContinuousHankelTransform}
%%%\end{align}
The continuous Hankel transform \eqref{eq:ContinuousHankelTransform} is defined provided that the integral
\begin{align}
  \intoi \d\vperp\ \vperp^{1/2}|g(\vperp)| ,
  %\label{eq:ContinuousHankelTransform}
\end{align}
exists.
It is sufficient for $g(\vperp)$ to be piecewise continuous and of bounded variation in every finite subinterval of $(0, \infty)$ %``satisfies Direchlet's condition of limited fluctuation on $[0,\infty]$'' 
\citep{FiskJohnson87}.

The Hankel transform \eqref{eq:ContinuousHankelTransform} is its own inverse,
\begin{align}
  g(\vperp) = \intoi \d\pvar~ \pvar J_0(\pvar\vperp) \g(\pvar)  ,
  \label{eq:ContinuousHankelInverse}
\end{align}
which follows from the orthogonality relation for Bessel functions:
\begin{align}
  \intoi \d\vperp ~ \pvar \vperp J_0(\pvar\vperp)J_0(q\vperp)  = \delta(\pvar-q) .
  \label{eq:BesselFunctionOrthogonality}
\end{align}
Therefore the inverse Hankel transform \eqref{eq:ContinuousHankelInverse} represents $g$ as the superposition of zeroth order Bessel functions with different arguments.

The Hankel transform \eqref{eq:ContinuousHankelTransform} is the radial part of the two dimensional Fourier transform, 
and is therefore equivalent to the two dimensional Fourier transform of an axisymmetric function \cite{Piessens00}.
In particular, since the distribution function is gyrophase independent, the Hankel transform is the Fourier transform of $g$ in perpendicular velocity space.
The line $p=\rho_s\kperp$ is then the set of phase space points where the Fourier wavenumbers for perpendicular physical space and perpendicular velocity space are equal, for the species used to define the normalization (for which $\rho_s=1$ in our units).

\section{Discrete Hankel transform}
\label{sec:DiscreteHankelTransform}

We also consider the discrete Hankel transform, which we introduce via a quadrature rule on a grid of scaled Bessel function roots,
based on a result due to \citet{Frappier93}.
As well as the discrete Hankel transform,
this gives us a quadrature rule to use for other perpendicular velocity integrals which are not (zeroth-order) Hankel transforms,
such as in the perpendicular \ampere's law for $\Bpar$ \eqref{eq:MaxBPara}
and in the free energy \eqref{eq:FreeEnergyCDF}.
The quadrature rule is based on the formula
\begin{align}
  \label{eq:Frappier}
  \intoi \d\vperp ~ \vperp f(\vperp) = \frac{2}{\tau^2}\sum_{n=1}^{\infty} \frac{1}{J_1^2(j_n)} f( j_n/\tau ) ,
\end{align}
where $j_n$ are the roots of the zeroth-order Bessel function, so that $J_0(j_n)=0$, and $\tau>0$ is a free scaling parameter.
This formula holds for functions $f$ such that $f(\vperp)=\O\left(\vperp^{-\delta}\right)$ for $\delta>2$ as $\vperp\to\infty$,
and such that for every $\epsilon>0$, there exists a constant $A$ such that $|f(z)|\leq A \exp(2\epsilon |z|)$ as $|z|\to\infty$, for complex $z$ \citep[Theorem 1]{Frappier93}.
These are satisfied by typical distribution functions.

Motivated by the Gaussian decay of the Maxwell--Boltzmann distribution, 
we assume $f(\vperp)$ is negligible for $\vperp$ greater than some velocity space cutoff $\vcut$, and construct a velocity grid using the first $N_p$ scaled Bessel roots, $v_n = j_n/\tau = j_n \vcut / j_{N_p+1}$, with $\tau=j_{N_p+1}/\vcut$.
With \eqref{eq:Frappier}, this yields the quadrature rule
\begin{align}
  %\intoi \d\vperp ~ \vperp f(\vperp)
  %= 
  \int_0^{\vcut} \d\vperp ~ \vperp f(\vperp)
   = \frac{2}{\tau^2}\sum_{n=1}^{N_p} \frac{1}{J_1^2(j_n)} f( v_n ) 
   = \sum_{n=1}^{N_p} w_n f( v_n ) ,
  \label{eq:QuadratureRule}
\end{align}
with the weights $w_n = 2\vcut^2/(j^2_{N_p+1}J_1^2(j_n))$.
We may use this to evaluate the free energy integrals \eqref{eq:FreeEnergyCDF} by setting $f(\vperp) = |g(\vperp)|^2/F^{\perp}_0(\vperp)$,
and the integrals for charge and current density in Maxwell's equations \eqref{eq:FieldSolveMoments} by setting $f(\vperp)$ equal to 
$J_0(\kperp \vperp) g(\vperp)$ or $\vperp J_1(\kperp \vperp) g(\vperp)$.

\begin{figure}[tbp]
  \centering
%\subfigure[\label{fig:contours}]
{\includegraphics[width=0.49\textwidth]
{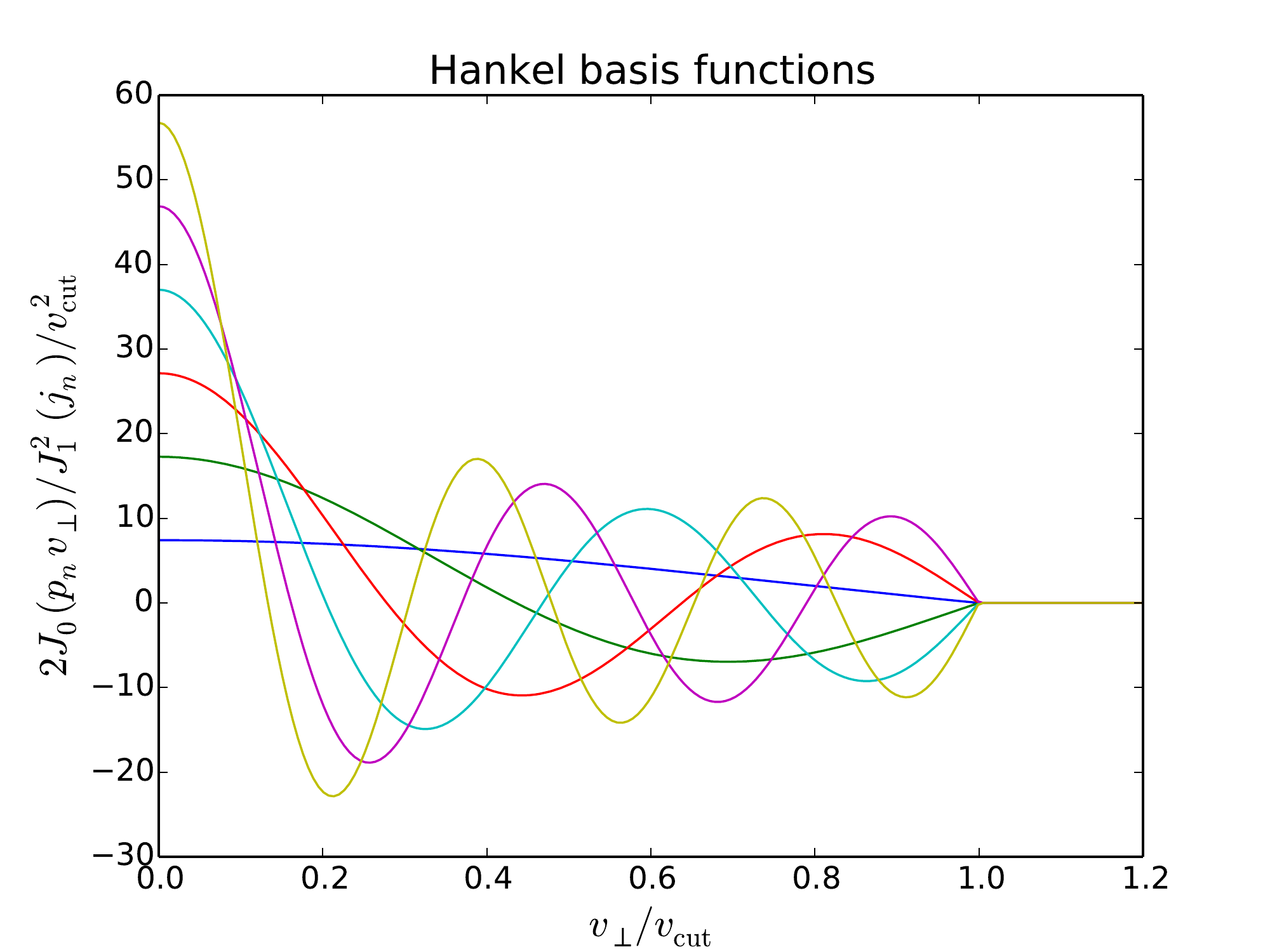}}
%%%\subfigure[\label{f.fsystem}]{\includegraphics[width=0.49\textwidth]
%%%{asystem_gr_paper.pdf}}
%%%\subfigure[\label{fig:VKM}%despite name, real lls on complex part
%%%]{\includegraphics[width=0.49\textwidth]
%%%{van_kampen_mode_by_complex_lls.pdf}}
%%%\subfigure[\label{fig:dougherty_intro}]{\includegraphics[width=0.49\textwidth]
%%%{ef_hd_vs_nu_n1_N5000_k410_lls_imag.pdf}}
\caption{ 
  The first six Hankel basis functions.
\label{fig:BesselBasis}
}
\end{figure}

The quadrature rule \eqref{eq:QuadratureRule} is consistent with the discrete Hankel transform given by \citet{FiskJohnson87},
which we now derive.
Setting $f(\vperp)=J_0(\kperp \vperp) g(\vperp)$ in \eqref{eq:QuadratureRule}, we obtain the semi-discrete Hankel transform
\begin{align}
  \bar{g}(p)
  %= \intoi
  = \int_0^{\vcut}
  \d\vperp ~ \vperp J_0(p\vperp) g(\vperp)
   = \sum_{n=1}^{N_p} w_n J_0(pv_n)g( v_n ) ,
  \label{eq:SemiDiscreteHankelTransform}
\end{align}
the Hankel transform which is discrete in $\vperp$, but continuous in $p$.
Further, restricting \eqref{eq:SemiDiscreteHankelTransform} to the grid $p_m=j_m/\vcut$, we obtain the discrete Hankel transform 
\begin{align}
  \g(p_m) 
  %&= {\cal H}g(v_m) 
  = \sum_{n=1}^{N_p} w_n J_0(p_mv_n) g(v_n)
  = \frac{\vcut^2}{j_{N_p+1}^2}\sum_{n=1}^{N_p} \frac{2J_0(p_mv_n)}{J_1^2(j_n)} g(v_n).
  \label{eq:DiscreteHankelPair}
\end{align}
This $p$-grid is chosen to exploit the discrete orthogonality relation \citep{FiskJohnson87}
\begin{align}
   \sum_{l=1}^{N_p}\frac{J_0(j_mj_l/j_{N_p+1})J_0(j_nj_l/j_{N_p+1})}{J_1^2(j_l)} =  \frac{J^2_1(j_m)j_{N_p+1}^2}{4}  \delta_{nm}.
  \label{eq:DiscreteOrth}
\end{align}
With the discrete Hankel variable $p_n=j_n/\vcut$
and the velocity grid $v_n=j_n\vcut/j_{N_p+1}$,
\eqref{eq:DiscreteOrth} gives the two orthogonality relations
\begin{align}
   \sum_{l=1}^{N_p}\frac{J_0( p_m v_l)J_0( p_n v_l)}{J_1^2(j_l)} 
= \sum_{l=1}^{N_p}\frac{J_0(p_l v_m )J_0(p_l v_n)}{J_1^2(j_l)} 
= \frac{J^2_1(j_m)j_{N_p+1}^2}{4} \delta_{nm}.
  \label{eq:DiscreteOrthVP}
\end{align}
Multiplying \eqref{eq:DiscreteHankelPair} by $2J_0(p_mv_l)/(\vcut^2J_1^2(j_m))$, summing over $m$
and using \eqref{eq:DiscreteOrthVP},
we obtain the exact inverse transform to the discrete Hankel transform \eqref{eq:DiscreteHankelPair},
\begin{align}
  \label{eq:DiscreteHankelPairBack}
  g(v_l) 
  = \frac{1}{\vcut^2}\sum_{m=1}^{N_p} \frac{2J_0(p_mv_l)}{J_1^2(j_m)} \g(p_m)
  = \frac{j_{N_p+1}^2}{\vcut^4}\sum_{m=1}^{N_p} w_m J_0(p_mv_l) \g(p_m).
\end{align}
As with the Hermite representation \eqref{eq:HermExpTrunc},
we may think of \eqref{eq:DiscreteHankelPairBack}
either a discretization of $g$ on a velocity space grid,
or as a truncated spectral expansion.
In \eqref{eq:DiscreteHankelPairBack},
the basis functions are $2J_0(p_mv_l)/(\vcut^2J_1^2(j_m))$. 
The continuous $\vperp$ version of the basis functions,
$2J_0(p_m\vperp)/(\vcut^2J_1^2(j_m))\I_{[0,\vcut]}$
(where $\I_{[0,\vcut]}=1$ if $\vperp\in[0,\vcut]$ and $0$ otherwise),
are plotted in \fig\ref{fig:BesselBasis}.
The nonzero part of these functions oscillates with characteristic wavelength $2\pi/p_m$ so that higher $p_m$ correspond to more quickly oscillating functions.

The discrete Hankel transforms \eqref{eq:DiscreteHankelPair} and \eqref{eq:DiscreteHankelPairBack} form a self-inverse pair up to a constant factor
\begin{align}
  \g(p_m) = \frac{\vcut^2}{j_{N+1}^2}\sum_{n=1}^{N_p}h_{mn}g(v_n),
  \hspace{1cm}
  g(v_n) = \frac{1}{\vcut^2}\sum_{m=1}^{N_p}h_{nm}\g(p_m),
  \label{eq:DiscreteHankelSummary}
\end{align}
where $h_{mn}$ are the coefficients of the transform matrix:
\begin{align}
  h_{mn}= \frac{2J_0(j_nj_m/j_{N_p+1}) }{J_1^2(j_n)} .
  \label{eq:DiscreteHankelMatrix}
\end{align}

\section{Hankel space structure of gyrokinetics}
\label{sec:HankelSpaceStructureGyrokinetics}

To allow comparison of different perpendicular velocity space methods,
we now study the structure of the \gkm\ system in Hankel space. 
It suffices to consider the collisionless, single ion species, electrostatic \gkm\ system ($\Apar=\Bpar=0$) with no driving ($L_T^{-1}=L_n^{-1}=0$) and no inhomogeneities in the magnetic field ($\kappa=L_B^{-1}=0$).
%In \chp~\ref{sec:SpectroGK}, we show that all the effects neglected here\footnote{Bar the curvature drift?} may also be treated purely spectrally.
All the effects neglected here (bar the gradient-$B$ term) may also be treated purely spectrally,
\ie\ using the expansion coefficients without introducing a collocation grid in $\vperp$.
Taking the Hermite transform of \eqref{eq:GKIonsES} and denoting Fourier--Hermite expansion coefficients by $g_{\k m}$, 
the ion gyrokinetic equation becomes
\begin{align}
  \begin{split}
    \pd{g_{\k m}(\vperp)}{t} 
 & 
 +   i\vthNi\kpara \left(\sqrt{\frac{m+1}{2}}g_{\k,m+1}(\vperp) + \sqrt{\frac{m}{2}} g_{\k,m-1}(\vperp)\right)
      \\ &
      \hspace{-1.6cm}
  + i \frac{\qNs}{\sqrt{\mNs\TNs}} \frac{\kpara}{\sqrt{2}}\delta_{m1}  
  J_0(\rhoNi\kperp\vperp) F_0^{\perp}(\vperp)   {\varphiNk} 
  +  \left\{ \ga{\varphiN}{\R_i} , g_{m}(\R_i,\vperp) \rb_{\k}  
  %+ \intii\d\vpara \phi^m(\vpara) \lb \ga{\varphiN}{\R_s} , g_{i} \rb_{\k}  
    = 
    0,
  \end{split}
  \label{eq:ReducedSpectralGKEqn}
\end{align}
and the quasineutrality condition becomes
\begin{align}
  \varphiNk \frac{n_iq_i^2}{T_i}\lp 1 + \frac{T_i}{T_e} - \Gamma_{0i}\rp
  &= 2\pi q_in_i\int\d\vperp~\vperp J_0(\rhoNi\kperp\vperp) g_{\k,m=0}(\vperp)
  \label{eq:ReducedQuasineutralityIntegral}
  \\
  &= 2\pi q_in_i\ \bar{g}_{\k,m= 0}(p=\rhoNi\kperp) .
  \label{eq:ReducedQuasineutrality}
\end{align}
Here to obtain \eqref{eq:ReducedQuasineutrality}
we have noted that the integral 
%in the quasineutrality condition 
on the \rhs\ of \eqref{eq:ReducedQuasineutralityIntegral}
is in the form of a Hankel transform \eqref{eq:ContinuousHankelTransform} with argument $p=\rhoNi\kperp$.
The electrostatic 
\linebreak
potential 
is therefore determined using only the single Fourier--Hankel--Hermite mode
\linebreak
$\bar{g}_{\k,m=0}(p=\rhoNi\kperp)$, \ie, a single $(p,m)$ dual velocity space point for each Fourier wavenumber $\k$. 
Thus the system greatly simplifies if this mode decouples from other modes.   

Taking the continuous Hankel transform of \eqref{eq:ReducedSpectralGKEqn}, we obtain the gyrokinetic equation
\begin{align}
  \begin{split}
    \pd{\bar{g}_{\k m}(p)}{t} 
 & 
 +   i\vthNi\kpara \left(\sqrt{\frac{m+1}{2}}\bar{g}_{\k,m+1}(p) + \sqrt{\frac{m}{2}} \bar{g}_{\k,m-1}(p)\right)
      \\ &
      %\hspace{-0.8cm}
  + i \frac{\qNs}{\sqrt{\mNs\TNs}} \frac{\kpara}{\sqrt{2}}\delta_{m1}  
  \frac{1}{2\pi}\exp\left( -(\rhoNi\kperp)^2/2\right) I_0\left( (\rhoNi\kperp)^2/2\right)  {\varphiNk} 
      \\ &
      %\hspace{-0.8cm}
      + \intoi \d\vperp\ \vperp J_0(p\vperp) \left\{ \ga{\varphiN}{\R_i} , g_{m}(\R_i,\vperp) \rb_{\k}  
   %+ \intoi \d\vperp\ \vperp J_0(p\vperp)\intii\d\vpara\ \phi^m(\vpara) \lb \ga{\varphiN}{\R_i} , g_i \rb_{\k}  
    = 
    0,
  \end{split}
  \label{eq:ReducedHankelGKEqn}
\end{align}
for $\bar{g}$ defined in \eqref{eq:ContinuousHankelTransform}.
This shows that the  Hankel and Fourier modes are coupled only through the nonlinear term,
while the Hermite modes are coupled only through the streaming term.
In the linearized system, the Fourier--Hankel modes decouple, meaning that the system reduces to a one-dimensional problem in Hermite space, parameterized by the wavenumber $\k$ and Hankel mode $p=\rhoNi\kperp$.
As established in \chp\ \ref{sec:Hypercollisions}, the one dimensional Hermite system may be solved very efficiently.

In the nonlinear system, the Fourier--Hermite modes remain coupled.
The nonlinear system has been studied in Hankel space in the simplified case with $\kpara=0$
\citep{Tatsuno09,Tatsuno10,Plunk10,PlunkTatsuno11}.
This so-called ``two-dimensional gyrokinetics'' (\ie\ the two perpendicular spatial dimensions plus perpendicular velocity space)
has two significant differences from five-dimensional gyrokinetics \eqref{eq:ReducedQuasineutrality} and \eqref{eq:ReducedHankelGKEqn}.
Firstly, as parallel velocity space is neglected, there is no free energy transfer via parallel phase mixing and therefore no Landau damping.
Secondly, while collisionless five-dimensional gyrokinetics conserves free energy \eqref{eq:FreeEnergySchekochihin},
two-dimensional gyrokinetics conserves both free energy and the electrostatic invariant \eqref{eq:ElectrostaticInvariant},
see also \cite{Plunk10}. 
Free energy cascades forwards (as in five-dimensional gyrokinetics), while the electrostatic invariant has either a forwards or inverse cascade depending on the distribution of free energy in phase space \citep{PlunkTatsuno11}.
Despite these differences, the nonlinear term, 
and therefore the perpendicular free energy transfer mechanism,
is the same in two- and five-dimensional gyrokinetics,
and it is reasonable to expect some properties of the two-dimensional system to be replicated in five dimensions.

Studies of two-dimensional gyrokinetics highlight the significance of the phase space line $p=\rho_i\kperp$,
beyond its meaning as the only Hankel modes required to determine the electromagnetic field.
\citet{Plunk10} showed that Fourier and Hankel modes are coupled via a three mode interaction which is most significant near the line $p=\rho_i\kperp$ (see \eqref{eq:KIntegral} below).
\citet{Plunk10} also derived spectra for the free energy:
\begin{align}
  \begin{split}
    W(\rho_i\kperp,p) \propto
    \begin{cases}
      (\rho_i\kperp)^{-2}p^{-1/3} , \hspace{1cm} \rho_i\kperp\gg p \gg 1,\\
      p^{-2}(\rho_i\kperp)^{-1/3} , \hspace{1cm}  p\gg \rho_i \kperp \gg 1,
    \end{cases}
  \end{split}
\end{align}
showing that these decay away from the line $p=\rho_i\kperp$.
\citet{Tatsuno10} confirmed these spectra numerically,
and moreover showed that free energy transfer, 
while in principle nonlocal,
is in effect local
with all significant transfer confined to a narrow band about $p=\rho_i\kperp$.

These simulations were carried out using the code \agk\ \citep[see also \sec\ref{sec:AstroGK}]{Numata10},
which uses a Fourier representation in perpendicular space,
and a pitch angle-energy-sign grid in velocity space.
To resolve the fine velocity space structure, these simulations used up to 16384 points in velocity space, corresponding to $N_p\sim\O(100)$.\footnote{In this case, \agk\ is unnecessarily resolving both parallel and perpendicular velocity space, when only perpendicular velocity space is required.
This is because perpendicular velocity cannot be expressed using only one of pitch angle and energy.
}
The Hankel representation was used only for post-processing, rather than for computation;
however the results suggest properties which may be exploited in computations,
and we now discuss different approaches to performing these calculations in Hankel space.

\section[Alternative Hankel treatments of perpendicular velocity space]{Alternative Hankel treatments of perpendicular\\velocity space}
\label{sec:HankelTreatmentsPerpendicularVelocitySpace}

We compare three approaches to solving the \gkm\ system which arise from writing these equations in different forms in Hankel and perpendicular velocity space. 
We do this to determine which is the most efficient method, and where our knowledge of the Hankel space structure of gyrokinetics can lead to computational savings.
Ultimately we ask: is the system better treated by discretizing \eqref{eq:ReducedSpectralGKEqn} and \eqref{eq:ReducedQuasineutralityIntegral} in perpendicular velocity space, or by discretizing \eqref{eq:ReducedQuasineutrality} and \eqref{eq:ReducedHankelGKEqn} in Hankel space?
For each, what is the best discretization grid to use?

To measure efficiency we consider the computational complexity and \ipc\ involved in calculating terms in the equations.
Large complexity is caused by sums over phase space, while communication is caused by nonlocality in phase space. 
It transpires that the calculation of the nonlinear term dominates both complexity and communication, 
due to the distributed Fourier and Hankel transforms.
We follow the traditional numerical analysis approach of quantifying complexity,
but by minimizing the number of Fourier and Hankel transforms, we will be minimizing 
both complexity and communication.
%We also discuss the discretization of the integral in the quasineutrality condition---this is nontrivial in both $p$ and $\vperp$, and may also lead to \ipc.
%The Hermite coupling term also leads to \ipc, but as the coupling is nearest-neighbour and the term contains no sum, this calculation is far dominated by the calculation of nonlinear term and the quasineutrality integral.

The nonlinear term has the structure of a Fourier convolution, and so is most efficiently calculated pseudospectrally 
using the Fast Fourier Transform \citep{Cooley65,Orszag70}.
This requires $\O(N_X\log_2 N_X)$ operations, where $N_X=N_xN_yN_z$ is the total number of spatial grid points. %\footnote{In fact, the Fourier transforms are performed as the sequence of three one-dimensional Fourier transforms, requiring slightly more operators, $\O(N_X(\log_2N_x)(\log_2N_y)(\log_2N_z))$, but allowing more flexibility to reduce inter-processor communication, see \sec\ref{sec:DataLayoutParallelization}.}.
The discrete Hankel transform can also appear in the calculation of the nonlinear term in Hankel space.
The Hankel transform \eqref{eq:ContinuousHankelTransform} is equivalent to the radial part of a two-dimensional Fourier transform \citep{Piessens00},
but while there is work on polar Fourier transforms \cite[\eg][]{Averbuch06}, there is no exact radial Fast Fourier Transform.
Similarly there are efficient approximations to the continuous Hankel transform for large numbers of points \citep{Siegman77,Ogata05} but as we expect to have relatively few points in perpendicular velocity, we may also calculate the discrete transform \eqref{eq:DiscreteHankelPair} using the matrix multiplication \eqref{eq:DiscreteHankelMatrix}, for which highly efficient implementations are available (\eg\ BLAS \cite{Lawson79}).

\begin{figure}[tbp]
  \centering
%\subfigure[\label{fig:contours}]
  \subfigure[]{\includegraphics[width=0.49\textwidth]
{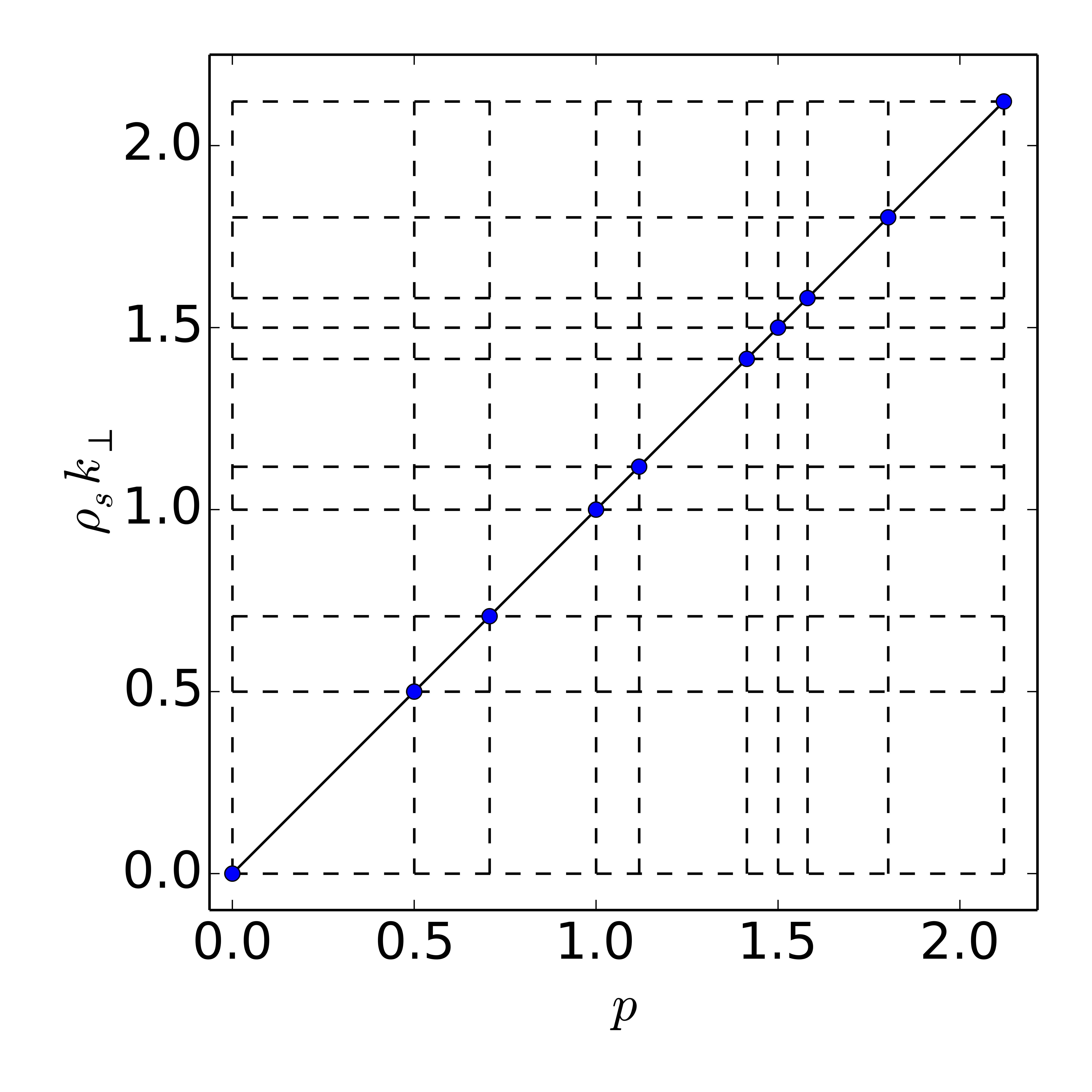}}
\subfigure[\label{fig:HankelInterpolationGrid}]{\includegraphics[width=0.49\textwidth]
{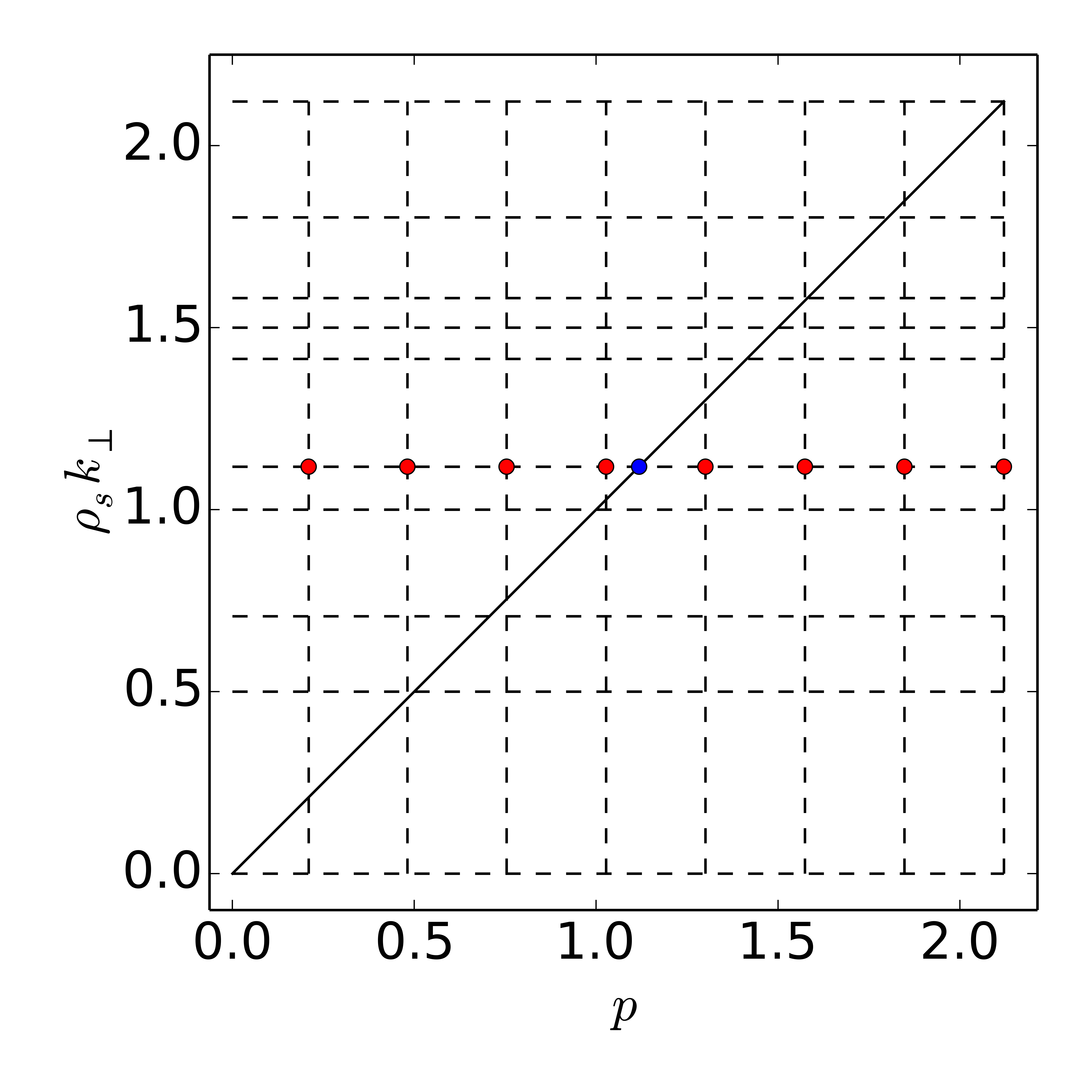}}
\caption[Possible discretizations in Hankel space.]{ 
  Possible discretizations in Hankel space.
  The diagonal line is $p=\rho_s\kperp$, the line along which the electrostatic potential is calculated.
  (a) The Hankel grid in $p$ is chosen to comprise the same points as the $\rho_s\kperp$ points. The electrostatic potential is found as a single function evaluation (at the blue points), but the Hankel grid will contain many points.
  (b) An arbitrary grid is used for Hankel space. The point $p=\rho_s\kperp$ is not in general a grid point. The electrostatic potential is found by interpolating the value of $\bar{g}_{i0}(\k,\rho_i\kperp)$ at the blue point, from the values $\bar{g}_{i0}(\k,p_n)$, the red points. 
\label{fig:SchematicPGrids}
}
\end{figure}
We are also interested in the calculation of the electrostatic potential from the quasineutrality condition \eqref{eq:ReducedQuasineutralityIntegral}.
In Hankel space, the electrostatic potential is proportional to a single \FHH\ mode, and in the linearized system, 
the lack of coupling lets us evolve this mode alone.
In the full system, mode coupling means that we need a grid in Hankel space.
To take advantage of the calculation of the electrostatic potential as a single mode, we would need a grid point at every $p=\rho_s\kperp$---a Hankel space grid with $\O(N_xN_y)$ points!
The alternative is either to neglect modes that are far from $p=\rho_s\kperp$,
or to use a smaller grid in $p$, and evaluate $\bar{g}_0(\k,p=\rho_s\kperp)$ via interpolation as a sum of $\bar{g}_0(\k,p_n)$.
These two approaches are shown schematically in \fig\ref{fig:SchematicPGrids}.
A discretization on a generic $\vperp$ grid will also lead to the evaluation of the electrostatic potential via a sum, but now through a quadrature rule.
These summations will have complexity far smaller than that of the calculation of the nonlinear term, but will result in (modest) inter-processor communication, as discussed in \sec\ref{sec:FieldSolve}.

We now derive the three approaches to the nonlinear term:
a discretization in $\vperp$,
a pseudospectral approach,
and a purely spectral approach on a grid in $p$.
The discretization grids are not chosen \apriori, but must be determined as part of the method.

\subsection{Discretization in velocity space}
\label{sec:NonlinearTermVelocitySpace}

The nonlinear term in $\vperp$--Fourier--Hermite space from equation \eqref{eq:ReducedSpectralGKEqn} is
\begin{align}
  \begin{split}
%%%    \intii\d\vpara\ \int_{\bbR^3}\d^3\R_s\ e^{-i\k\cdot\R_s} \phi^m(\vpara) \lb \ga{\varphiN}{\R_s} , g_s \rb  
%%%   \\ = \intii\d\vpara\ \int_{\bbR^3}\d^3\R_s\ e^{-i\k\cdot\R_s} \phi^m(\vpara) \left( \pd{\ga{\varphiN}{\R_s}}{X}\pd{g_s}{Y} - \pd{\ga{\varphiN}{\R_s}}{Y}\pd{g_s}{X} \right)
%%%   \\ = \int_{\bbR^3}\d^3\R_s\ e^{-i\k\cdot\R_s} \phi^m(\vpara) \left( \pd{\ga{\varphiN}{\R_s}}{X}\pd{g_{m}}{Y} - \pd{\ga{\varphiN}{\R_s}}{Y}\pd{g_{m}}{X} \right)
%%%   \\ = \int_{\bbR^3}\d^3\R_s\ e^{-i\k\cdot\R_s} \left( \pd{\ga{\varphiN}{\R_s}}{X}\pd{g_{m}}{Y} - \pd{\ga{\varphiN}{\R_s}}{Y}\pd{g_{m}}{X} \right)
		\int_{\bbR^3}\d^3\tilde{\k} ~ ~ \b\cdot(\tilde{\k}\times\k) J_0(\rho_i\tilde{k}_\perp\vperp)\hat{\varphiN}(\tilde{\k}) \hat{g}_{m}(\k-\tilde{\k},\vperp) ,
  \end{split}
  %\label{eq:ReducedSpectralGKEqn}
\end{align}
a Fourier convolution which may be written as the product of Fourier transforms
\begin{align}
  \begin{split}
%%%    \intii\d\vpara\ \int_{\bbR^3}\d^3\R_s\ e^{-i\k\cdot\R_s} \phi^m(\vpara) \lb \ga{\varphiN}{\R_s} , g_s \rb  
%%%   \\ = \intii\d\vpara\ \int_{\bbR^3}\d^3\R_s\ e^{-i\k\cdot\R_s} \phi^m(\vpara) \left( \pd{\ga{\varphiN}{\R_s}}{X}\pd{g_s}{Y} - \pd{\ga{\varphiN}{\R_s}}{Y}\pd{g_s}{X} \right)
%%%   \\ = \int_{\bbR^3}\d^3\R_s\ e^{-i\k\cdot\R_s} \phi^m(\vpara) \left( \pd{\ga{\varphiN}{\R_s}}{X}\pd{g_{m}}{Y} - \pd{\ga{\varphiN}{\R_s}}{Y}\pd{g_{m}}{X} \right)
%%%   \\ = \int_{\bbR^3}\d^3\R_s\ e^{-i\k\cdot\R_s} \left( \pd{\ga{\varphiN}{\R_s}}{X}\pd{g_{m}}{Y} - \pd{\ga{\varphiN}{\R_s}}{Y}\pd{g_{m}}{X} \right)
   \FFT \left( \IFFT\left( ik_xJ_0(\rho_i\kperp\vperp) \hat{\varphiN}\right) \IFFT\left( ik_y \hat{g}_m \right) - \IFFT\left( ik_yJ_0(\rho_i\kperp\vperp) \hat{\varphiN}\right) \IFFT\left( ik_x \hat{g}_m \right) \right) . 
  \end{split}
  \label{eq:vperpNonlinearTerm}
\end{align}
There is no coupling in $\vperp$--Hermite space, but the convolution couples Fourier modes.
The five Fourier transforms dominate the work.
Three of the Fourier transforms are of distribution-sized arrays and require $\O(N_XN_mN_p\log_2(N_X))$ operations,
while the two transforms of $J_0(\kperp\vperp)\hat{\varphi}$ only require $\O(N_XN_p\log_2(N_X))$ operations
since $\hat{\varphi}$ has no $m$-dependence.

Although the lack of $\vperp$-coupling in the nonlinear term allows an arbitrary treatment of $\vperp$,
we need to introduce a quadrature grid to approximate the perpendicular velocity integral to find $\hat{\varphi}$ from the quasineutrality condition.

\subsection{Discretization in Hankel space}
\label{sec:NonlinearTermHankelSpace}
\begin{figure}[tbp]
  \centering
%\subfigure[\label{fig:contours}]
{\includegraphics[width=0.49\textwidth]
{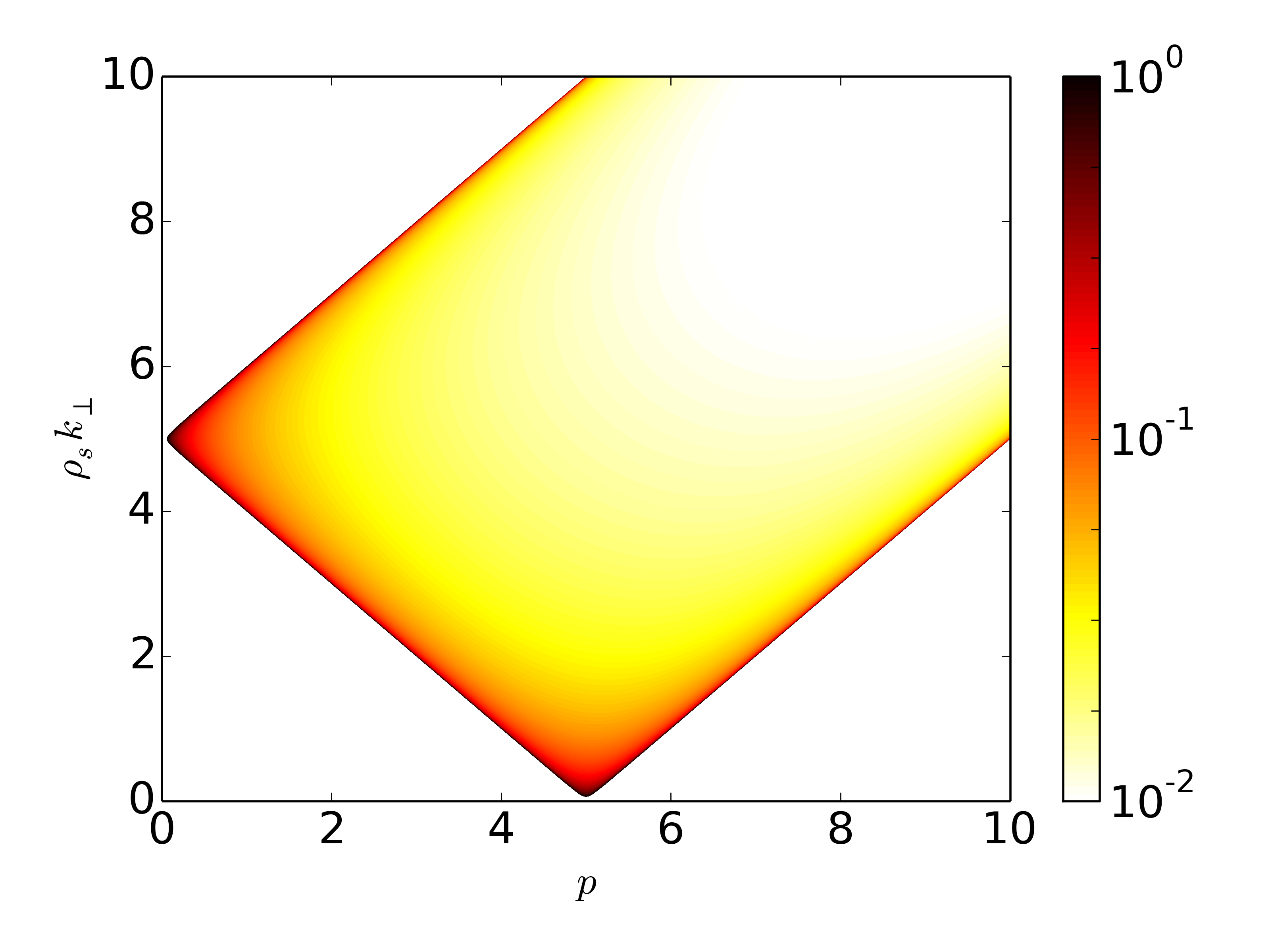}}
\caption{ 
  Contour of the integral $K(\kperp,p,5)$ from \eqref{eq:KIntegral}.
\label{fig:KIntegral}
}
\end{figure}

The nonlinear term in Hankel space (from equation \ref{eq:ReducedHankelGKEqn}) is
\begin{align}
  \begin{split}
		\HT \FFT \Big\{ & \IFFT\left( ik_xJ_0(\rho_i\kperp\vperp) \hat{\varphiN}\right) \IFFT\left( ik_y\HT^{-1} \bar{g}_m \right) 
		\\
		& \hspace{3cm}
		- \IFFT\left( ik_yJ_0(\rho_i\kperp\vperp) \hat{\varphiN}\right) \IFFT\left( ik_x\HT^{-1} \bar{g}_m \right) \Big\} ,
%%%    \intoi \d\vperp\ \vperp J_0(p\vperp) \int_{\bbR^3}\d^3\k'\ \b\cdot(\k'\times\k) J_0(\kperp'\vperp)\hat{\varphiN}(\k') \hat{g}_{m}(\k-\k',\vperp) 
%%%   \\ = 
%%%   \intoi \d\vperp\ \vperp J_0(p\vperp) \int_{\bbR^3}\d^3\k'\ \b\cdot(\k'\times\k) J_0(\kperp'\vperp)\hat{\varphiN}(\k') \intoi \d q\ q J_0(q\vperp) \bar{g}_{m}(\k-\k',q) 
%%%   \\ = 
%%%  \intoi \d q\ q  \int_{\bbR^3}\d^3\k'\ \b\cdot(\k'\times\k) \intoi \d\vperp\ \vperp J_0(\kperp'\vperp)J_0(p\vperp)J_0(q\vperp) \hat{\varphiN}(\k') \bar{g}_{m}(\k-\k',q) 
%%%   \\ = 
%%%  \intoi \d q\ q  \int_{\bbR^3}\d^3\k'\ \b\cdot(\k'\times\k) K(\kperp',p,q) \hat{\varphiN}(\k') \bar{g}_{m}(\k-\k',q) 
%%%   \\ = 
%%%   \intoi \d q\ q  \FFT \left( \IFFT\left( ik_x K(\kperp,p,q)\hat{\varphiN}\right) \IFFT\left( ik_y \bar{g}_m \right) - \IFFT\left( ik_yK(\kperp,p,q) \hat{\varphiN}\right) \IFFT\left( ik_x \bar{g}_m \right) \right)
  \end{split}
  \label{eq:NonlinearTermPseudoSpectral}
\end{align}
where we have substituted in the definition $\bar{g}=\HT \hat{g}$ with $\HT$ the continuous Hankel transform \eqref{eq:ContinuousHankelTransform}.
The pseudospectral approach is to discretize this in Hankel space on the grid of Bessel roots $p_n=j_n/\vcut$ introduced in \sec\ref{sec:DiscreteHankelTransform}
and use the discrete Hankel transform \eqref{eq:DiscreteHankelSummary}
instead of ${\cal H}$.
In addition to the five Fourier transforms in \eqref{eq:vperpNonlinearTerm}, 
there are now three Hankel transforms each of complexity $\O(N_p^2N_XN_m)$.
Thus this direct pseudospectral approach is not useful.
However \citet{Plunk10} showed that the integrals in the Hankel transforms in \eqref{eq:NonlinearTermPseudoSpectral} may be rearranged to give
\begin{align}
  \begin{split}
%%%  \HT \FFT \left( \IFFT\left( ik_xJ_0(\kperp\vperp) \hat{\varphiN}\right) \IFFT\left( ik_y\HT \bar{g}_m \right) - \IFFT\left( ik_yJ_0(\kperp\vperp) \hat{\varphiN}\right) \IFFT\left( ik_x\HT \bar{g}_m \right) \right) ,
%%%    \intoi \d\vperp\ \vperp J_0(p\vperp) \int_{\bbR^3}\d^3\k'\ \b\cdot(\k'\times\k) J_0(\kperp'\vperp)\hat{\varphiN}(\k') \hat{g}_{m}(\k-\k',\vperp) 
%%%   \\ = 
%%%   \intoi \d\vperp\ \vperp J_0(p\vperp) \int_{\bbR^3}\d^3\k'\ \b\cdot(\k'\times\k) J_0(\kperp'\vperp)\hat{\varphiN}(\k') \intoi \d q\ q J_0(q\vperp) \bar{g}_{m}(\k-\k',q) 
%%%   \\ = 
%%%  \intoi \d q\ q  \int_{\bbR^3}\d^3\k'\ \b\cdot(\k'\times\k) \intoi \d\vperp\ \vperp J_0(\kperp'\vperp)J_0(p\vperp)J_0(q\vperp) \hat{\varphiN}(\k') \bar{g}_{m}(\k-\k',q) 
%%%   \\ = 
%%%  \intoi \d q\ q  \int_{\bbR^3}\d^3\k'\ \b\cdot(\k'\times\k) K(\kperp',p,q) \hat{\varphiN}(\k') \bar{g}_{m}(\k-\k',q) 
%%%   \\ = 
   \intoi \d q\ q  \FFT \Big[ \IFFT\big( ik_x & K(\rho_i\kperp,p,q)  \hat{\varphiN}\big)  \IFFT\big( ik_y \bar{g}_m(\k,q) \big) 
     \\ 
  & - \IFFT\left( ik_yK(\rho_i\kperp,p,q) \hat{\varphiN}\right) \IFFT\left( ik_x \bar{g}_m(\k,q) \right) \Big],
  \end{split}
  %\label{eq:ReducedSpectralGKEqn}
\end{align}
where 
\begin{align}
  \begin{split}
    K(\rho_i\kperp,p,q) = \intoi \d \vperp\ \vperp J_0(\rho_i\kperp\vperp)J_0(p\vperp)J_0(q\vperp) ,
  \end{split}
  \label{eq:KIntegral}
\end{align}
is a known integral:
if the scalars $\rho_i\kperp$, $p$, $q$ can form the three sides of a triangle, then $K=1/(2\pi\Delta)$ where $\Delta$ is the area of that triangle; otherwise $K=0$.
The example of $K(\rho_i\kperp,p,5)$ is  
plotted in \fig\ref{fig:KIntegral}. 
The integral is zero for $q\not\in ( |\rho_i\kperp-p|,\rho_i\kperp+p )$;
otherwise, it is largest as $q$ approaches edge of that interval (but zero when $q=|\rho_i\kperp-p|$ or $q=\rho_i\kperp+p$.
Thus we may rewrite the nonlinear term as
\begin{align}
  \begin{split}
%%%    \int_{|\kperp-p|}^{\kperp+p} \d q\ q  \int_{\bbR^3}\d^3\k'\ \b\cdot(\k'\times\k) K(\kperp',p,q) \hat{\varphiN}(\k') \bar{g}_{m}(\k-\k',q) 
%%%   \\ = 
   \int_{|\rho_i\kperp-p|}^{\rho_i\kperp+p} \d q\ q  \FFT \Big[ \IFFT\big( ik_x & K(\rho_i\kperp,p,q)  \hat{\varphiN}\big)  \IFFT\big( ik_y \bar{g}_m(\k,q) \big) 
     \\ 
  & - \IFFT\left( ik_yK(\rho_i\kperp,p,q) \hat{\varphiN}\right) \IFFT\left( ik_x \bar{g}_m(\k,q) \right) \Big] .
  \end{split}
  \label{eq:NonlinearTermReducedRange}
\end{align}
This shows that some localization is preserved from the linearized case.
In the linearized case we only need to evolve the Hankel mode $p=\rho_i\kperp$.
In this nonlinear case, we need to evolve all Hankel modes, but for each $p$ and $\rho_i\kperp$, only the subset of modes $( |\rho_i\kperp-p|,\rho_i\kperp+p )$ contributes to the nonlinear term.
Of course, to evolve the $\rho_i\kperp+p$ mode, we need to know modes up to $2\rho_i\kperp+p$, \etc, so in this fashion all modes are coupled. 
However at a given $p$ and a given timestep, only a subset of modes are required to exactly calculate the nonlinear term.

\begin{figure}[tbp]
  \centering
%\subfigure[\label{fig:contours}]
{\includegraphics[width=0.49\textwidth]
%{images/kpgridmasked.png}}
{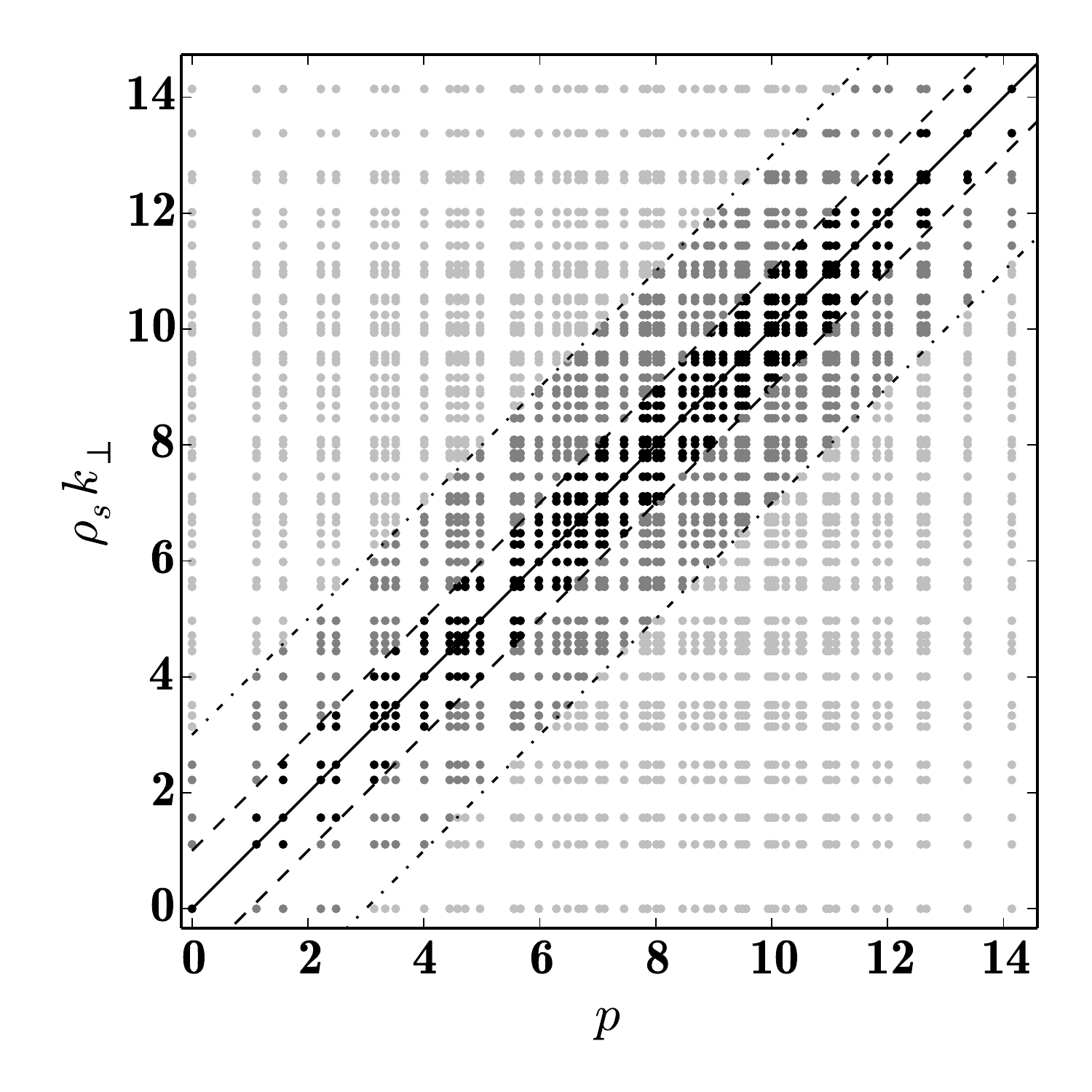}}
\caption[A $\kperp$--$p$ grid masked around the line $p=\rho_i\kperp$.]{ 
  A $\kperp$--$p$ grid masked around the line $p=\rho_i\kperp$. The grid is formed taking the same uniform 100 point grid in $k_x$ and $k_y$, which yields 53 unique $\kperp=\sqrt{k_x^2+k_y^2}$ points. The Hankel grid is chosen such that $p=\rho_i\kperp$, yielding a total array of $53^2=2809$ points. The black and dark grey points are in the intervals $p\in[\rho_i\kperp-1,\rho_i\kperp+1]$ and $p\in[\rho_i\kperp-3,\rho_i\kperp+3]$ respectively.
 There are 457 black points and 1289 dark grey points, so restricting the calculation to these regions reduces the number of Hankel modes by $84\%$ and $54\%$ respectively.
\label{fig:kpgridmasked}
}
\end{figure}

As before, the three distribution-function-sized Fourier transforms have complexity 
$\O(N_pN_mN_X\log_2(N_X))$.
However now we must evaluate the transforms of $K\hat{\varphi}$ in \eqref{eq:NonlinearTermReducedRange} for all $p$ and $q$, so that the 
two transforms of the electrostatic potential are  increased by factor of $N_p$ to $\O(N_p^2N_X\log_2(N_X))$.
Thus this method is also not viable, unless the resolution is such that $N_p$ is smaller than the $\O(100)$ points required in the velocity grid method in \sec\ref{sec:NonlinearTermVelocitySpace}. 
Such a reduction may be achieved by using the observed localization properties in Hankel space,
namely, the localization of free energy flow about the diagonal $p=\rho_i\kperp$ \citep{Tatsuno10} and decay in the spectrum away from the diagonal \citep{Plunk10,Tatsuno10}.
These properties suggest that we may neglect modes away from the diagonal.
\Fig\ref{fig:kpgridmasked} shows an example of this for $\kperp$--$p$ space with $\kperp$ and $p$ discretized on the same grid.
Restricting the simulation to points within a band about $p=\rho_i\kperp$ dramatically reduces the resolution and therefore complexity.
For example,
restricting the simulation to the band $p\in[\rho_i\kperp-3,\rho_i\kperp+3]$ reduces the number of Hankel modes by $54\%$, 
while restricting it to the band $p\in[\rho_i\kperp-1,\rho_i\kperp+1]$ reduces the number of modes by $84\%$.
Moreover, the width of the band is a parameter which takes the system from a full kinetic simulation to a reduced model.

\subsection{\sgk\ discretization}
The fully spectral approach described in \sec\ref{sec:NonlinearTermHankelSpace} is implemented in \sgk, but is not used for the simulations in this thesis.
The reason for this is that simulations of the $\rho_i\kperp<1$ regime (like that presented in \S\ref{sec:FreeEnergyConservation}) has limited perpendicular phase mixing, so the resolution requirements in $\vperp$ are modest.
Instead it is more efficient to work on a grid in perpendicular velocity, calculating the nonlinear term as \eqref{eq:vperpNonlinearTerm}.
We discretize on the grid of scaled Bessel roots $v_n=j_n\vcut/j_{N_p+1}$ as described in \sec\ref{sec:DiscreteHankelTransform},
with the electrostatic potential calculated from \eqref{eq:ReducedQuasineutrality} by setting the integrand to be $f(\vperp)=J_0(\rho_i\kperp \vperp) g_m(\k,\vperp)$ in the quadrature rule \eqref{eq:QuadratureRule},
\begin{align}
  %\intoi \d\vperp ~ \vperp f(\vperp)
  %= 
  \bar{g}_m(\k,\rho_i\kperp)
  =
  \int_0^{\vcut} \d\vperp ~ \vperp J_0(\rho_i\kperp \vperp) g_m(\k,\vperp)
   = \sum_{n=1}^{N_p} w_n J_0(\kperp v_n) g_m(\k, v_n) .
  \label{eq:QuasineutralityIntegralVSpace}
\end{align}
By discretizing on this grid, 
the method is completely equivalent to the pseudospectral method \eqref{eq:NonlinearTermPseudoSpectral}
which solves for $\bar{g}_m(\k,p_n)$ on the grid $p_n=j_n/\vcut$.
In this case however, the calculation of $\bar{g}_m(\k,\rho_s\kperp)$ for the quasineutrality condition may be viewed as an interpolation formula,
as follows.
Using \eqref{eq:DiscreteHankelPairBack} to substitute $\bar{g}_m(\k,p_l)$ for $g_m(\k,v_n)$, the \rhs\ of \eqref{eq:QuasineutralityIntegralVSpace}
becomes
\begin{align}
  \bar{g}_m(\k,\rho_i\kperp)
  =
  \int_0^{\vcut} \d\vperp ~ \vperp J_0(\rho_i\kperp \vperp) g_m(\k,\vperp)
%%%  = \sum_{l=1}^{N_p} \left[ \sum_{n=1}^{N_p} \frac{4}{\vcut^2}\frac{J_0(\rho_s\kperp v_n)J_0(p_lv_n)}{J_1^2(j_n)J_1^2(j_l)}\right] \bar{g}_m(\k, p_l) 
%%%  = \sum_{l=1}^{N_p} \hat{w}_l(\rho_s\kperp) \bar{g}_m(\k, p_l) ,
  = \sum_{l=1}^{N_p} \hat{w}_l(\rho_i\kperp) \bar{g}_m(\k, p_l) ,
  \label{eq:HankelInterpolationFormula}
\end{align}
where
\begin{align}
  \hat{w}_l(\rho_i\kperp) =  \sum_{n=1}^{N_p} \frac{4}{\vcut^2}\frac{J_0(\rho_i\kperp v_n)J_0(p_lv_n)}{J_1^2(j_n)J_1^2(j_l)},
  \label{eq:QuasineutralityIntegralHankel}
\end{align}
which is an interpolation formula for $\bar{g}_m(\k,\rho_i\kperp)$ using data points $\bar{g}_m(\k,p_n)$, as in \fig\ref{fig:HankelInterpolationGrid}.
Note the similarity of \eqref{eq:QuasineutralityIntegralHankel} %$\hat{w}_l(\rho_s\kperp)$ 
to the discrete orthogonality relation \eqref{eq:DiscreteOrthVP}.
Indeed, for grid points $\rho_i\kperp=p_{n}$, we have $\hat{w}_l(\rho_i\kperp)=\delta_{ln}$ and the sum reduces to the single term $\bar{g}_m(\k,p_{n})$.

Therefore, performing the calculation on a grid in $\vperp$ and approximating the velocity space integral in the quasineutrality condition as a velocity space sum
is equivalent to performing the calculation on a grid in $p$ and evaluating the integral as a sum of Hankel modes.
Moreover, the calculation on a velocity grid is more efficient as it does not entail the calculation of discrete Hankel transforms.
In this thesis we perform the calculation in Fourier--Hermite--$\vperp$ space, only using the discrete Hankel transform \eqref{eq:DiscreteHankelPair} to output data in Hankel space.

%%%\section{Notes}
%%%
%%%What should you do about choosing a $p$ grid in the multispecies case?  Also, for simplicity should you set $\rho_i=1$?  (This obscures the first question\dots) 
%%%
%%%Note that the Hankel transform is a different idea to the Fourier--Bessel series (standard REF for this?) which has order of Bessel functions are variable rather than argument.

\chapter{\sgk}
\label{sec:SpectroGK}

% These definitions are transplanted from the Vlasov Poisson paper to accompany text that is moved earlier
\newcommand{\toab}{third-order Adams--Bashforth}
\newcommand{\tsop}{\mathsf{L}}
\renewcommand{\tsop}{{\cal L}}
\renewcommand{\tsop}{{\cal T}}
\renewcommand{\tsop}{{\cal A}}
\newcommand{\VP}{Vlasov--Poisson}
\newcommand{\VPS}{Vlasov--Poisson system}

%%%\section{Introduction}

We now describe \sgk, the code in which we implement the spectral methods described in \chp s \ref{sec:ParallelVelocitySpace} and \ref{sec:PerpendicularVelocitySpaceHankelTransform}, and which we use in our study of the \gkm\ system in \chp s \ref{sec:FreeEnergyFlowAndDissipation} and \ref{sec:ScalingLawsForDriftKineticTurbulence} (and the Vlasov--Poisson system in Appendix \ref{sec:VPPaper} and Ref.\ \cite{Parker14}). 
We present the \gkm\ system in the exact form solved by \sgk\ in \sec\ref{sec:SpectralGKMSystemRecap}, describe the \sgk\ algorithm in \sec\ref{sec:AlgorithmDescription}, and verify the code and study its performance in \sec\ref{sec:CodeVerification}.

\section{Spectral \gkm\ system}
\label{sec:SpectralGKMSystemRecap}

\sgk\ solves the \gkm\ system using a Fourier representation for guiding centre space and
a Hermite expansion in parallel velocity space.
The treatment of perpendicular velocity space depends on the particular system being solved.
For the linearized system with no parallel magnetic field perturbations, $\Bpar=0$,
\sgk\ solves the system spectrally using a Hankel representation in perpendicular velocity space.
Both $\varphi$ and $\Apar$ are obtained from a single \FHH\ mode with Hankel point $p=\rho_s\kperp$.
Since there is no Hankel mode coupling, it is only necessary to evolve this single Hankel mode for each perpendicular Fourier wavenumber.
As Fourier modes also decouple in the linearized system, this problem is in effect a one-dimensional problem in Hermite space parameterized by $\k$, and hence $p=\rho_s\kperp$.  
In practice, \sgk\ solves this system on a four-dimensional \FH\ grid, allowing simultaneous solution for many Fourier modes.
This ``four-dimensional'' spectral system is described in \sec\ref{sec:4DCode}.
Otherwise, \sgk\ solves the five-dimensional \gkm\ system with Fourier--Hermite modes, and a discretization of perpendicular velocity space using the grid of scaled Bessel roots discussed in \chp\ \ref{sec:PerpendicularVelocitySpaceHankelTransform}.

In \chp\ \ref{sec:GKMSystem}, we derived the \gkm\ system in wave vector $\k$ and velocity $(\vpara, \vperp)$ coordinates.
In this \chp, we solve the version of this system for the complementary distribution function $g$ expressed in Fourier space, which is summarized in \S\ref{sec:GKMSystemInFourierSpace}.  
The system comprises the gyrokinetic equation
\begin{align}
  \begin{split}
    \pd{\gks}{t} &  + i\vthNs\vpara\kpara  \lp \gks 
    + \frac{\qNs}{\TNs} \ga{\chi_s}{\R_s,\k}F_0 \rp   
    +\left\{{\ga{\chi_s}{\R_s}},g_s\right\}_{\k}
    \\ & %\hspace{4cm}
 + \frac{ik_y}{2} \left[ -2\kappa\vpara^2 - L_B^{-1}\vperp^2 +  \omega_n + \lp \vperp^2 + \vpara^2 -\frac{3}{2}\rp \omega_T \rs \ga{\chi_s}{\R_s,\k} F_0
   \\ & 
    - \frac{ik_y}{2} \frac{\TNs}{\qNs}  \lp 2\kappa\vpara^2+L_B^{-1}{\vperp^2}\rp \gks
 = \nu \ga{C[h_s]}{\R_s,\k} ,
  \end{split}
  \label{eq:GKNumataChp5}
\end{align}
where the nonlinear term is
\begin{align}
  \left\{{\ga{\chi_s}{\R_s}},g_s \right\}_{\k}
= \frac{1}{2} \sum_{\k'} \b\cdot(\k\times\k') \ga{\chiNs}{\R_s,\k'} g_{s,\k-\k'},
  \label{eq:GKSystemNonlinearTeamChp5}
\end{align}
and the gyrokinetic potential is
\begin{align}
  \begin{split}
   \ga{\chiNs}{\R_s,\k} 
  =  J_0(a_s) \varphiNk - \vthNs\vpara J_0(a_s) \Apk   +  \frac{\TNs}{\qNs} \frac{2\vperp^2 J_1(a_s)}{a_s} \dBpk ,
  \end{split}
  \label{eq:GyroaveragedPotentialChp5}
\end{align}
and Maxwell's equations
\begin{subequations}
  \label{eq:FieldSolveMomentsChp5}
\begin{align}
  &\varphiNk \sum_s \frac{\nNs\qNs^2}{\TNs}(1 - \Gamma_{0s})
  - \dBpk\sum_s \qNs\nNs\Gamma_{1s} = \sum_s \qNs \nNs \int \d^3\v ~\gks J_0(a_s),
  \label{eq:QNChp5}
  \\
  &\Apk\left[ \frac{\kperp^2}{2\beta_r}   + \sum_s \frac{\qNs^2\nNs}{2\mNs}\Gamma_{0s} \rs
  = \sum_s \qNs\nNs \vthNs\int \d^3\v ~\gks \vpara J_0(a_s), 
  \label{eq:AmpParaChp5}
\\
  &\varphiNk\sum_s \nNs\qNs\Gamma_{1s} +
\dBpk \lp \frac{2}{\beta_r} + \sum_s \TNs\nNs\Gamma_{2s}\rp
= - \sum_s \nNs \TNs \int \d^3\v ~\gks \vperp^2 \frac{2J_1(a_s)}{a_s} ,
  \label{eq:MaxBParaChp5}
\end{align}
\end{subequations}
where $a_s=\rho_s\kperp\vperp$, and 
$\Gamma_{0s}$,
$\Gamma_{1s}$, and
$\Gamma_{2s}$ are known functions of $\kperp$ given in \eqref{eq:GammaDefinitions}.

\subsection{Full five-dimensional system}
\label{sec:5DCode}

In the general case, \sgk\ solves the \gkm\ system in Fourier--
\linebreak
Hermite--$\vperp$ space,
using the representation
\begin{align}
  g_{s\k}(v_n,\vpara,t) = \sum_{m=0}^{N_m-1} \phi_m(\vpara) \gfived_{sm\k}(v_n,t).
\end{align}
where the
$\phi_m$ are the Hermite basis functions defined in \eqref{eq:HermFunDef}, 
\begin{align}%\label{eq:HermFunDef}
	\phi_m(\vpara) = \frac{H_m(\vpara)}{\sqrt{2^m m!}}\frac{e^{-\vpara^2}}{\sqrt{\pi}}, 
  \hspace{1cm}
  H_m(\vpara) = (-1)^m \e^{\vpara^2}\fd{^m}{\vpara^m}\lp \e^{-\vpara^2}\rp,
\end{align}
and
$v_n=j_n\vcut/j_{N_p+1}$ is the $n$th point in the perpendicular velocity space grid, with $j_n$ the $n$th root of the Bessel function $J_0$, and $\vcut$ the velocity space cutoff, a parameter to be chosen.
For convenience, we will usually suppress the time argument $t$.

\subsubsection{Gyrokinetic equation}
\label{sec:5DCodeGKE}
Taking the Hermite transform of the gyrokinetic equation \eqref{eq:GKNumataChp5}
gives
\begin{align}
  \begin{split}
    \pd{\gfived_{sm\k}}{t} 
    + {\cal L}_g\gfived_{sm\k}
    + {\cal L}_{\varphi} \varphiNk
    + {\cal L}_{A} \Apk
    + {\cal L}_{B} \dBpk
    + {\cal N}[\ga{\chiNs}{\R_s},g_s(\R_s,v_n,\vpara)] 
    = 
    \nu {\cal C}[h_s],
  \end{split}
  \label{eq:SpectralGKEqn}
\end{align}
where \Lstar\ are the linear operators defined by
\begin{subequations}
  \label{eq:LinearOperatorsFiveD}
\begin{align}
  \begin{split}
    {\cal L}_g\gfived_{sm\k}
    = 
  &   i\stm\kpara \lp\sqrt{\frac{m+1}{2}}\gfived_{s,m+1,\k} + \sqrt{\frac{m}{2}} \gfived_{s,m-1,\k}\rp ,
%%%  \\ &  - ik_y\frac{\TNs}{\qNs}   \kappa \lp \frac{1}{2}\sqrt{(m+1)(m+2)}\bar{g}_{m+2} + \frac{2m+1}{2}\bar{g}_m + \frac{1}{2}\sqrt{m(m-1)}\bar{g}_{m-2} \rp
%%%  %+ \frac{ik_y}{2L_B}\frac{\qNs}{\mNs}   \frac{1}{p}\fd{}{p}\lp p\fd{\bar{g}_m}{p}\rp
%%%  - \frac{ik_y\TNs}{2L_B\qNs}\intoi \d\vperp~ \vperp^3 J_0( p\vperp) g_m(\vperp) ,
  \end{split}
  \label{eq:Lg}
\end{align}
\begin{align}
  \begin{split}
    {\cal L}_{\varphi}{\varphiNk} = 
    i \left\{\frac{\qNs}{\sqrt{\mNs\TNs}} \frac{\kpara}{\sqrt{2}}\delta_{m1}  
%%%    -  k_y\kappa \lp\frac{\delta_{m2}}{\sqrt{2}}+\frac{\delta_{m0}}{2}\rp
    + \frac{k_y}{2L_n}\delta_{m0} 
    + \frac{k_y}{2L_T}\lp \frac{\delta_{m2}}{\sqrt{2}}- \delta_{m0}\rp  \rb \I_1  {\varphiNk} 
%%%  +\delta_{m0} \frac{ik_y}{2} \lp L_T^{-1} - L_B^{-1}   \rp  \I_3  {\varphiNk} ,
  +\delta_{m0} \frac{ik_y}{2L_T}  \I_3  {\varphiNk} ,
  \end{split}
  \label{eq:ESGroup2}
\end{align}
\begin{align}
  \begin{split}
    {\cal L}_A{\Apk} 
    = 
    &
    -i\stm\left\{ \frac{\qNs}{\sqrt{\mNs\TNs}}\kpara \lp\frac{\delta_{m2}}{\sqrt{2}}+\frac{\delta_{m0}}{2}\rp 
%%%    -  k_y\kappa \lp\frac{\sqrt{3}\delta_{m3}}{2} + \frac{3\delta_{m1}}{2\sqrt{2}}\rp
    + \frac{k_y}{2\sqrt{2}L_n}\delta_{m1} 
    +  \frac{\sqrt{3}k_y}{4L_T}\delta_{m3}   \rb \I_1  \Apk 
%%%    - \frac{ik_y}{2\sqrt{2}}\delta_{m1} \stm\lp L_T^{-1} - L_B^{-1}   \rp  \I_3  \Apk ,
    \\ &
    - \frac{ik_y}{2\sqrt{2}}\delta_{m1} \stm L_T^{-1}   \I_3  \Apk ,
  \end{split}
  \label{eq:AGroup2}
\end{align}
\begin{align}
  \begin{split}
    {\cal L}_B\dBpk = 
   &
    i\frac{\TNs}{\qNs} \left\{ \frac{\qNs}{\sqrt{\mNs\TNs}}\frac{\kpara}{\sqrt{2}}\delta_{m1}  
%%%    -  k_y\kappa \lp\frac{\delta_{m2}}{\sqrt{2}}+\frac{\delta_{m0}}{2}\rp
    + \frac{k_y}{2L_n}\delta_{m0} 
    + \frac{k_y}{2L_T}\lp \frac{\delta_{m2}}{\sqrt{2}}- \delta_{m0}\rp  \rb 2\I_2  \dBpk 
%%%    +\delta_{m0} \frac{ik_y}{2}\frac{\TNs}{\qNs} \lp L_T^{-1} - L_B^{-1}   \rp  2\I_4  \dBpk ,
\\
&
    +\delta_{m0} \frac{ik_y}{2}\frac{\TNs}{\qNs}  L_T^{-1}  2\I_4  \dBpk .
  \end{split}
  \label{eq:BGroup2}
\end{align}
\end{subequations}
The perpendicular velocity space dependence of the ${\cal L}_{\varphi}$, ${\cal L}_{A}$ and ${\cal L}_{B}$ operators is entirely contained in the four coefficients
%%%\begin{align}
%%%\begin{split}
%%%\I_1 = J_0(k'\vperp)F^{\perp}_0(\vperp),
%%%\hspace{1cm}
%%%\I_2 = \frac{\vperp}{k'} J_1(k'\vperp)F^{\perp}_0(\vperp),
%%%%\hspace{1cm}
%%%\\
%%%\I_3 = \vperp^2J_0(k'\vperp)F^{\perp}_0(\vperp),
%%%\hspace{1cm}
%%%\I_4 = \frac{\vperp^3}{k'}J_1(k'\vperp)F^{\perp}_0(\vperp),
%%%\end{split}
%%%\end{align}
\begin{align}
\begin{split}
\I_1 = J_0(\rho_s\kperp v_n)F^{\perp}_0(v_n),
\hspace{1cm}
\I_2 = \frac{v_n}{\rho_s\kperp} J_1(\rho_s\kperp v_n)F^{\perp}_0(v_n),
%\hspace{1cm}
\\
\I_3 = v_n^2J_0(\rho_s\kperp v_n)F^{\perp}_0(v_n),
\hspace{1cm}
\I_4 = \frac{v_n^3}{\rho_s\kperp}J_1(\rho_s\kperp v_n)F^{\perp}_0(v_n),
\end{split}
\end{align}
where the perpendicular Maxwellian is $F^{\perp}_0(v_n) = \exp(-v_n^2)/\pi$.

%%%and ${\cal C}$ is collision term which we discuss in \sec\ref{sec:Recurrence}.

%%%\begin{align}
%%%  {\cal N}[\ga{\chiNs}{\R_s},g_s] = \intii\d\vpara \phi^m(\vpara) \lb \ga{\chiNs}{\R_s} , g_s \rb_{\k} , 
%%%  \label{eq:NonlinearTermSpectral}
%%%\end{align}
%%%\subsubsection{Nonlinear term}
\label{sec:NonlinearTerm}
The nonlinear term ${\cal N}[\ga{\chiNs}{\R_s},g_s]$ is the Fourier--Hermite transform of \eqref{eq:GKSystemNonlinearTeamChp5}, the Poisson bracket of the distribution function and the gyroaveraged \gkpot, discretized in perpendicular velocity space:
\begin{align}
  \begin{split}
  {\cal N}[\ga{\chiNs}{\R_s},g_s] 
 & = 
 \frac{1}{2} \intii\d\vpara \phi^m(\vpara)
  \\
  & \hspace{0.5cm}
  {\cal F}
  \Bigg[ {\cal F}^{-1}\lp k_x \ga{\chiNs}{\R_s,\k}\rp {\cal F}^{-1}\lp k_y\sum_{m'=0}^{N_m-1} \gfived_{sm'\k}(v_n)\phi_{m'}(\vpara) \rp
\\ 
& \hspace{1cm}
 -
 {\cal F}^{-1}\lp k_y \ga{\chiNs}{\R_s,\k}\rp {\cal F}^{-1}\lp k_x \sum_{m'=0}^{N_m-1} \gfived_{sm'\k}(v_n)\phi_{m'}(\vpara) \rp\Bigg].
  \end{split}
  \label{eq:NonlinearTerm}
\end{align}
Here $\F$ denotes the three-dimensional discrete Fourier transform,
and we have inserted the Hermite expansion of the distribution function.
We may decompose 
${\cal N}[\ga{\chiNs}{\R_s},g_s]  = {\cal N}_{\varphi} + {\cal N}_A + {\cal N}_B$,
where ${\cal N}_{\varphi}$, ${\cal N}_A$ and ${\cal N}_B$ correspond to the $\varphi$, $\Apara$ and $\Bpara$ terms in $\chiNs$ \eqref{eq:GyroaveragedPotentialChp5}.
These represent the $\Ev\times\B$ drift, particle streaming along the perturbed magnetic field, and the gradient-$B$ drift respectively.
The velocity enters these three terms via factors of $J_0(a_s)$, $\vpara J_0(a_s)$, and $\vperp J_1(a_s)$ respectively multiplying $\gfived_{smk}$.
The parallel velocity integral in \eqref{eq:NonlinearTerm} can thus be calculated analytically.
Inserting $\ga{\chiNs}{\R_s,\k}$ into \eqref{eq:NonlinearTerm}
and performing the integrals then
gives
\begin{subequations}
\begin{align}
  \begin{split}
  %{\cal N}[\ga{\varphi}{\R_s},g_s] 
    {\cal N}_{\varphi}
 &
 = 
  \frac{1}{2}{\cal F}
  \Bigg[ {\cal F}^{-1}\lp k_x J_0(\rho_s \kperp v_n)\varphi_{\k}\rp {\cal F}^{-1}\lp k_y \gfived_{sm\k}(v_n) \rp
\\ 
& \hspace{3cm}
 -
 {\cal F}^{-1}\lp k_y J_0(\rho_s\kperp v_n)\varphi_{\k}\rp {\cal F}^{-1}\lp k_x \gfived_{sm\k}(v_n) \rp\Bigg],
  \end{split}
  \label{eq:NonlinearTermPhi}
\end{align}
\begin{align}
%  \\
  \begin{split}
 %& {\cal N}[\ga{\Apara}{\R_s},g_s] 
  {\cal N}_A
%%% \\ 
%%% & \hspace{0.5cm}
 & = 
-\frac{1}{2}\stm  {\cal F}
  \Bigg[ {\cal F}^{-1}\lp k_x J_0(\rho_s\kperp v_n)A_{\parallel\k}\rp
\\    & \hspace{2.5cm}  
{\cal F}^{-1}\lp k_y \lp \sqrt{\frac{m+1}{2}}\gfived_{s,m+1,\k}(v_n) + \sqrt{\frac{m}{2}}\gfived_{s,m-1,\k}(v_n)\rp \rp
\\ 
& \hspace{1.5cm}
 - {\cal F}^{-1}\lp k_y J_0(\rho_s\kperp v_n)A_{\parallel\k}\rp
\\ 
& \hspace{2.5cm}
 {\cal F}^{-1}\lp k_x \lp \sqrt{\frac{m+1}{2}}\gfived_{s,m+1,\k}(v_n) + \sqrt{\frac{m}{2}}\gfived_{s,m-1,\k}(v_n)\rp \rp
\Bigg],
  \end{split}
  \label{eq:NonlinearTermApara}
\end{align}
\begin{align}
%%%  \\
  \begin{split}
  %{\cal N}[\ga{\Bpara}{\R_s},g_s] 
  {\cal N}_B
 &
 = 
 \frac{\TNs}{\qNs} {\cal F} % half of two cancels
 \Bigg[ {\cal F}^{-1}\lp \frac{k_x v_n J_1(\rho_s\kperp v_n)}{\kvar} B_{\parallel\k}\rp {\cal F}^{-1}\lp k_y \gfived_{sm\k}(v_n) \rp
\\ 
& \hspace{3cm}
 -
 {\cal F}^{-1}\lp \frac{k_y v_n J_1(\rho_s\kperp v_n)}{\kvar} B_{\parallel\k}\rp {\cal F}^{-1}\lp k_x  \gfived_{sm\k}(v_n) \rp\Bigg],
  \end{split}
  \label{eq:NonlinearTermBpara}
\end{align}
\end{subequations}
so that we obtain moments $\gfived_m$ in the terms proportional to $\varphi$ and $\Bpara$, and the moments $\gfived_{m-1}$ and $\gfived_{m+1}$ via Hermite mode coupling \eqref{eq:rr} in the term proportional to ${A}_{\parallel}$.  
%%%In practice only the $x$- and $y$-Fourier transforms need repeated evaluation, due to the factors of $k_x$ and $k_y$ in the functions to be transformed.
%%%Therefore we take the $z$-Fourier transform only at the start and end of the calculation of the nonlinear term, and work with functions in $(k_x,k_y,z,m,\vperp)$ space.
%%%Excluding the $z$-Fourier transforms, the algorithm is very similar to that in \agk, but calculated in Hermite space rather than on a grid in parallel velocity space. 

\subsubsection{Maxwell's equations}
Integrals in the field solve \eqref{eq:FieldSolveMomentsChp5} %\eqref{eq:M0}, \eqref{eq:M1}, \eqref{eq:M2} 
are evaluated by replacing $\int\d\vperp~\vperp\ldots$ with $\sum_{n=1}^{N_p}w_n\ldots$,
yielding
\begin{subequations}
  \label{eq:Maxwell5D}
\begin{align}
  \varphiNk \sum_s \frac{\nNs\qNs^2}{\TNs}(1 - \Gamma_{0s})
  - \dBpk\sum_s \qNs\nNs\Gamma_{1s} = 2\pi \sum_s \qNs \nNs  \sum_{n=1}^{N_p}w_nJ_0(\rho_s\kperp v_n) \gfived_{s,0,\k}(v_n) ,
  \label{eq:Maxwell5DQN}
\end{align}
\begin{align}
  \Apk\left[ \frac{\kperp^2}{2\beta_r}   + \sum_s \frac{\qNs^2\nNs}{2\mNs}\Gamma_{0s} \right]
    = \sqrt{2}\pi \sum_s \qNs\nNs \vthNs \sum_{n=1}^{N_p}w_n J_0(\rho_s\kperp v_n) \gfived_{s,1,\k}(v_n)  ,
  %\label{eq:AmpPara}
\end{align}
\begin{align}
\begin{split}
&  \varphiNk\sum_s \nNs\qNs\Gamma_{1s} +
\dBpk \lp \frac{2}{\beta_r} + \sum_s \TNs\nNs\Gamma_{2s}\rp
\\ 
& \hspace{5cm}
= - 2\pi \sum_s  \nNs \TNs \sum_{n=1}^{N_p}w_n  v_n \frac{2J_1(\rho_s\kperp v_n)}{\rho_s\kperp} \gfived_{s,0,\k}(v_n)  ,
  %\label{eq:MaxBPara}
\end{split}
\end{align}
\end{subequations}
where 
  $\Gamma_{0s} = I_0(b_s)e^{-b_s}$,
  $\Gamma_{1s} = (I_0(b_s)-I_1(b_s))e^{-b_s}$,
  and
  $\Gamma_{2s} = 2 \Gamma_1(b_s)$
with 
$b_s = (\rho_s\kperp)^{2}/2$. 
%$b_s = (\rhoNs\kperp)^2/2$. 

\subsection{Reduced four-dimensional system}
\label{sec:4DCode}
The calculation of $\varphi$ and $\Apar$ from Maxwell's equations \eqref{eq:FieldSolveMomentsChp5} only requires the Hankel mode $p=\rhoNs\kperp$, so we may neglect all other Hankel modes if the \gkeqn\ for $\g(p=\rhoNs\kperp)$ evolves independently of other modes.
Taking the Hankel transform of \eqref{eq:SpectralGKEqn}, shows that Hankel mode coupling only occurs in the nonlinear term. 
This is because whenever $g$ appears in linear terms, it is not multiplied by a function of $\vperp$, so we may use the orthogonality of Bessel functions \eqref{eq:BesselFunctionOrthogonality};
however, when $g$ appears in the nonlinear term, it is multiplied by a function of $\vperp$ and we may no longer use orthogonality.
Therefore in linear simulations with $\Bpar=0$, %$L_{B}^{-1}=0$, 
\agks\ does not initialize a $p$-space grid, but instead evolves $\g(p=\rho_s\kperp)$ only for each $(\k,m)$ grid point.
Formally, this is equivalent to taking the continuous Hankel transform of \eqref{eq:GKNumataChp5} at $p=\rhoNs\kperp$ while using the Ansatz %\eqref{eq:FourDimensionalGAnsatz}
\begin{align}
  g_{s\k}(\vperp,\vpara,t) = \sum_{m=0}^{N_m-1} \phi_m(\vpara) \intoi \d p \ p J_0\lp p\vperp\rp\g_{sm\k}(p,t),
\end{align}
and only solving for $\g_{sm\k}(p=\rhoNs\kperp,t)$.
The electromagnetic fields are found from \eqref{eq:FieldSolveMomentsChp5} by evaluating the zeroth and first Hermite coefficients.
The Fourier modes also decouple in both the gyrokinetic equation and the field solve, so that the system is effectively one-dimensional in Hermite space and parametrized by $\k$.

This method does not find the whole distribution function $\g_{sm\k}(p,t)$, but only the part needed to find the electromagnetic fields.
However, it is exact, and valid for generic initial conditions that have been projected onto $p=\rhoNs\kperp$.
In this case \agks\ is significantly faster than \agk\ due to the reduced dimensionality and reduced inter-processor communication.

\subsubsection{Gyrokinetic equation}
Taking the Hankel--Hermite transform of the linearized \gkeqn\ \eqref{eq:GKNumataChp5} we obtain
\begin{align}
  \begin{split}
    \pd{\bar{g}_{sm\k}}{t} 
    + {\cal L}_g\bar{g}_{sm\k}
    + {\cal L}_{\varphi} \varphiNk
    + {\cal L}_{A} \Apk
    + {\cal L}_{B} \dBpk
    = 
    \nu {\cal C}[h_s],
  \end{split}
  \label{eq:SpectralGKEqnHankel}
\end{align}
where \Lstar\ are the linear operators \eqref{eq:LinearOperatorsFiveD}, %\eqref{eq:Lg}--\eqref{eq:BGroup2}, 
but now with coefficients 
\begin{subequations}
\begin{align}
%%%  {\cal I}_1 &= \frac{1}{\pi}\intoi\d\vperp~ \vperp J_0(\kvar\vperp)J_0(\pvar\vperp) e^{-\vperp^2} 
%%%  = \frac{1}{2\pi} \exp\lp -\frac{\kvar^2+\pvar^2}{4}\rp I_0 \lp \frac{\kvar\pvar}{2}\rp  ,
%%%  \label{eq:I1}
%%%  \\ 
%%%  {\cal I}_2 &= \frac{1}{\pi\kvar}\intoi \d\vperp~\vperp^2 J_1(\kvar\vperp)J_0(\pvar\vperp) e^{-\vperp^2} 
%%%  =  \frac{1}{4\pi}  \exp\lp -\frac{\kvar^2+\pvar^2}{4}\rp \left[ I_0 \lp \frac{\kvar\pvar}{2}\rp - \frac{\pvar}{\kvar}I_1 \lp \frac{\kvar\pvar}{2}\rp\rs   ,
%%%  \\ 
%%%  \label{eq:I3}
%%%  {\cal I}_3 &= \frac{1}{\pi}\intoi \d\vperp~\vperp^3 J_0(\kvar\vperp)J_0(\pvar\vperp) e^{-\vperp^2} 
%%%= \frac{1}{2\pi}   \exp\lp -\frac{\kvar^2+\pvar^2}{4}\rp \left[ \lp 1 - \frac{\kvar^2+\pvar^2}{4}\rp I_0 \lp \frac{\kvar\pvar}{2}\rp + \frac{\kvar\pvar}{2}I_1 \lp \frac{\kvar\pvar}{2}\rp\rs , \\ 
%%%\begin{split}
%%%  {\cal I}_4 &= \frac{1}{\pi\kvar}\intoi \d\vperp~\vperp^4 J_1(\kvar\vperp)J_0(\pvar\vperp) e^{-\vperp^2}
%%%  \\ & \hspace{1cm} =  \frac{1}{2\pi} \exp\lp -\frac{\kvar^2+\pvar^2}{4}\rp \left[ \lp 1 - \frac{\kvar^2+2\pvar^2}{8}\rp I_0 \lp \frac{\kvar\pvar}{2}\rp 
%%%  + \frac{\pvar}{\kvar}\lp \frac{3\kvar^2+\pvar^2}{8} - 1\rp I_1 \lp \frac{\kvar\pvar}{2}\rp - \frac{\pvar^2}{8}  I_2 \lp \frac{\kvar\pvar}{2}\rp \rs ,
%%%  \label{eq:I4}
%%%\end{split}
%%%\\
  %\label{}
  {\cal I}_1 &= \frac{1}{\pi}\intoi\d\vperp~ \vperp J_0(\kvar\vperp)^2 e^{-\vperp^2} 
  = \frac{1}{2\pi} \exp\lp -\frac{\kvar^2}{2}\rp I_0 \lp \frac{\kvar^2}{2}\rp  ,
  \label{eq:I1}
  \\ 
  \begin{split}
  {\cal I}_2 &= \frac{1}{\pi\kvar}\intoi \d\vperp~\vperp^2 J_0(\kvar\vperp) J_1(\kvar\vperp)e^{-\vperp^2} 
  \\& =  \frac{1}{4\pi}  \exp\lp -\frac{\kvar^2}{2}\rp \left[ I_0 \lp \frac{\kvar^2}{2}\rp - I_1 \lp \frac{\kvar^2}{2}\rp\rs   ,
  \end{split}
  \\ 
  \begin{split}
  \label{eq:I3}
  {\cal I}_3 &= \frac{1}{\pi}\intoi \d\vperp~\vperp^3 J_0(\kvar\vperp)^2 e^{-\vperp^2} 
\\ & = \frac{1}{2\pi}   \exp\lp -\frac{\kvar^2}{2}\rp \left[ \lp 1 - \frac{\kvar^2}{2}\rp I_0 \lp \frac{\kvar^2}{2}\rp + \frac{\kvar^2}{2}I_1 \lp \frac{\kvar^2}{2}\rp\rs , 
  \end{split}
  \\ 
\begin{split}
  {\cal I}_4 &= \frac{1}{\pi\kvar}\intoi \d\vperp~\vperp^4 J_0(\kvar\vperp) J_1(\kvar\vperp)e^{-\vperp^2}
  \\ & =  \frac{1}{2\pi} \exp\lp -\frac{\kvar^2}{2}\rp \left[ \lp 1 - \frac{3\kvar^2}{8}\rp I_0 \lp \frac{\kvar^2}{2}\rp 
  + \lp \frac{\kvar^2}{2} - 1\rp I_1 \lp \frac{\kvar^2}{2}\rp - \frac{\kvar^2}{8}  I_2 \lp \frac{\kvar^2}{2}\rp \rs ,
  \label{eq:I4}
\end{split}
  %\label{}
\end{align}
\end{subequations}
where $\kvar=\rhoNs\kperp$.
Integrals $\I_1$ and $\I_3$ appear in \cite{Gradshteyn07}, while $\I_2$ and $\I_4$ are given by
\begin{align}
  \I_2 = -\frac{1}{\kvar}\pd{\I_1}{\kvar},
  \hspace{1cm}
  \I_4 = -\frac{1}{\kvar}\pd{\I_3}{\kvar}.
  \label{}
\end{align}

\subsubsection{Maxwell's \eqn s}
\label{sec:SpectralMaxwell}
The first two moments of $g$, (\ref{eq:FieldSolveMomentsChp5}a,b) are single Hankel--Hermite modes, 
so that the quasineutrality condition and the parallel \ampere's law become
\begin{subequations}
  \label{eq:Maxwell4D}
\begin{align}
  \varphiNk \sum_s \frac{\nNs\qNs^2}{\TNs}(1 - \Gamma_{0s})
  - \dBpk\sum_s \qNs\nNs\Gamma_{1s} = 2\pi \sum_s \qNs \nNs \g_{s,0,\k}(p=\rhoNs\kperp) ,
  \label{eq:M0HH}
  \\
  \Apk\left[ \frac{\kperp^2}{2\beta_r}   + \sum_s \frac{\qNs^2\nNs}{2\mNs}\Gamma_{0s} \rs
    = \sqrt{2}\pi\sum_s \qNs\nNs \stm \g_{s,1,\k}(p=\rhoNs\kperp) 
 ,
  \label{eq:M1HH}
\end{align}
which determine $\varphi$ and $\Apar$ when $\Bpar=0$.

We note also that the second moment of $g$ (\ref{eq:FieldSolveMomentsChp5}c) %\eqref{eq:M2} 
is the derivative \wrt\ $p$ of a single mode,
so the perpendicular \ampere's law is
\begin{align}
  \varphiNk\sum_s \nNs\qNs\Gamma_{1s} +
\dBpk \lp \frac{2}{\beta_r} + \sum_s \TNs\nNs\Gamma_{2s}\rp
= 4\pi \sum_s \frac{\nNs \TNs}{\rhoNs\kperp}\left.\pd{\g_{s,0,\k}}{p}\right|_{p=\rhoNs\kperp} .
  \label{eq:M2HH}
\end{align}
\end{subequations}
Thus the four-dimensional method may be extended to include cases where $\Bpar\neq0$ by also evolving $\left.\tpd{\g_{sm\k}}{p}\right|_{p=\rhoNs\kperp}$ via the derivative \wrt\ $p$ of the Hankel-transformed gyrokinetic equation.
%%%\Eqn\ \eqref{eq:SpectralGKEqnHankel} with Maxwell's \eqn s\ \Ex{eq:Maxwell4D} are an exact spectral formulation of the \gkm\ system.
%%%This is an attractive, sparse representation.
%%%In Hermite space, the electromagnetic field only acts on the lowest four moments,
%%%and the moment hierarchy is coupled only through particle streaming.
%%%The Fourier wavenumber $\k$ and Hankel mode $p$ are parameters of the system.  
%%%Moreover the fields are found from Maxwell's \eqn s using only the first two Hermite moments and the value of $\g$ and $\tpd{\g}{p}$ along the line $p=\rhoNs\kperp$. 
%%%%%%\Eqn\ \eqref{eq:SpectralGKEqn} with Maxwell's \eqn s\ \Ex{eq:Maxwell4D} are an exact spectral formulation of the \gkm\ system.
%%%%%%This is an attractive, sparse representation.
%%%%%%In Hermite space, the electromagnetic field only acts on the lowest four moments.
%%%%%%The moment hierarchy is coupled through the particle streaming, curvature and nonlinear terms.
%%%%%%The Fourier wavenumber $\k$ and Hankel mode $p$ are almost parameters of the system.  
%%%%%%In Hankel space, the only mode coupling is through the magnetic curvature integral in \eqref{eq:Lg} and the nonlinear term \eqref{eq:NonlinearTermSpectral}; Fourier wavenumbers are coupled only through the nonlinear term.
%%%%%%Moreover the fields are found from Maxwell's \eqn s using only the first two Hermite moments and the value of $\g$ and $\tpd{\g}{p}$ along the line $p=\rhoNs\kperp$. 
%%%

%\section{Implementation}
\section{Algorithm description}
\label{sec:AlgorithmDescription}

We now discuss details of the \sgk\ algorithm.
For both cases described in \sec\ref{sec:SpectralGKMSystemRecap},
the gyrokinetic equation may be written schematically as
\begin{align}
  %\pd{a_{jm}}{t} = \tsop_{jm}
  \fd{\a}{t} = \tsop\left[\a\right] , 
  \label{eq:dadtIsAa}
\end{align}
where 
$\a$ denotes the set of expansion coefficients. 
Given an algorithm for forming the \rhs, we use the 
third-order Adams--Bashforth scheme described in \sec\ref{sec:TimeIntegration}
to advance the coefficients in time. 
To form $\tsop$, %the timestep operator, 
we must 
determine the electromagnetic field,
calculate the nonlinear term, and control the growth of fine scales in physical and velocity space due to the nonlinear term and particle streaming respectively. 
These are discussed in \secs\ref{sec:FieldSolve}--\ref{sec:Recurrence}.
We describe the parallelization and communication patterns of the code in \sec\ref{sec:DataLayoutParallelization}.
But first we give a brief overview of \agk, the code which is the starting point for \sgk\ development.

\subsection{\agk}
\label{sec:agk}
\label{sec:AstroGK}
\newcommand{\gagk}{g^{\mathsf{AGK}}}

\agk\ is a slab gyrokinetics code, implemented as a simplified version of the \tkmk\ gyrokinetics code \gstwo\ 
that neglects effects due to toroidal geometry.
Both \agk\ and \gstwo\ are publicly available and widely used \cite{GS2}; 
\agk\ (which is documented in \cite{Numata10}) is used to study astrophysical plasmas %\citep[\eg][]{Tatsuno09,Tatsuno10,PlunkTatsuno11,TenBarge13}
\citep[\eg][]{Howes08PRL,PlunkTatsuno11,TenBarge13}
while \gstwo\ is used for \tkmk\ plasmas
\citep[\eg][]{Highcock11,BarnesEtal11}.

\agk\ solves the \gkm\ system \GKMSystemChapTwo,
using the modified complementary distribution function 
$\gagk_s = h_s - (q_s\FNso/\TNs)\ga{\varphi - \vperpv\cdot\Aperp}{\R_s}$,
rather than $g_s=h_s-(q_s\FNso/\TNs)\ga{\chiNs}{\R_s}$ \eqref{eq:ComplementaryDistributionFunction} as used in \sgk.
It uses a Fourier representation for perpendicular space, 
and a discretization on a uniform grid in parallel space.
The latter is inherited from \gstwo's toroidal geometry, and has no particular merit in a slab code. 
The parallel spatial derivatives in the streaming term $(\vpara\tpds{g}{z})$
and the electrostatic response $(\FNso\tpds{\varphi}{z})$
are evaluated by an upwinding compact finite difference scheme.
In velocity space, \agk\ discretizes on grids in energy $E=\vperp^2+\vpara^2$ and pitch angle $\lambda=\vperp^2/E$, with a separate variable for sign of parallel velocity $\sigma=\vpara/|\vpara|$.
A grid of Gauss--Legendre quadrature points is used for pitch angles, 
while a composite grid of Gauss--Laguerre and Gauss--Legendre quadrature points is used for energy.
This grid yields spectral convergence for the integrals used to determine the electromagnetic field from Maxwell's equations (see \cite{Numata10}, \sec3.1 and \sec4.3).
Splitting the integral range in two allows for changes of variables which remove singularities in the integrand at $E=0$ and $E=\infty$.
%See AGK footnote on p4.

The nonlinear term is calculated \psly\ as in \sec\ref{sec:NonlinearTermVelocitySpace},
replacing the convolution in $\kperpv$ with the product of inverse Fourier transforms.
This is efficient as the nonlinear term is otherwise local in the remaining coordinates $z$, $\vperp=\sqrt{\lambda E}$, $\vpara=\sqrt{(1-\lambda)E}$ and $\sigma$.

For time integration,
\agk\ separates the operator ${\cal A}$ in \eqref{eq:dadtIsAa} into linear and nonlinear parts, 
finding the evolution of the distribution function due to each part separately.
The nonlinear part is integrated using the explicit \toab, as discussed in \sec\ref{sec:TimeIntegration}.
The linear part is integrated using \citepos{Kotschenreuther95} implicit Green's function method
(described in \cite{Numata10}, \sec3.3.1)

\agk\ is implemented in Fortran 90 with parallelization achieved using MPI.
A copy of the three-dimensional electromagnetic field is kept on each processor, 
while the five-dimensional distribution function is divided between processors.
The distribution function is stored as a three-dimensional array with the first two dimensions holding parallel space $z$ 
and the sign of parallel velocity $\sigma$
respectively.
The remaining coordinates are concatenated in the final dimension.
The order of this final dimension is determined by a ``layout'' input parameter, 
a five-character string comprising 
\texttt{l} (pitch angle, $\lambda$),
\texttt{e} (energy $E$),
\texttt{x} and \texttt{y} (perpendicular wavenumbers $k_x$, $k_y$)
and
\texttt{s} (species $s$).
The parallelization is over the final dimension so that finite differences in $z$
are always local to each processor,
as are the operations on the same $|\vpara|$.
Generically, all other operations are distributed.
However, by carefully choosing layout and resolution, 
one can arrange for some operations to be local.
For example,
with the layout \texttt{lexys},
velocity space may be made local to each processor by choosing $N_xN_yN_s$ to be divisible by the number of processors $\Nproc$.
Then the sums approximating the velocity space integrals in the field solve become local, and the electromagnetic field is calculated with no \ipc.
Similarly with \texttt{xyles}, choosing $N_{\lambda}N_EN_s$ to be divisible by $\Nproc$ ensures that Fourier space is local, and therefore Fourier transforms are calculated without \ipc.

\sgk\ is also written in Fortran 90 and reuses the code base of \agk\ so far as possible.
It inherits memory management and communication in perpendicular Fourier transforms,
but is otherwise rewritten for Hankel, Hermite, and parallel Fourier space.
\sgk\ is however compatible with \agk\ (and \textsc{GS2}), using the same input parameter names (meaning Fortran namelist input files are transferable), and the same compilation procedures and Makefiles.
%%%New approach to linear terms and field solve.
%%%Also implemented Hankel transform.
%%%\sgk\ inherits the MPI communication routines from \agk, and these were used to implement necessary communication in field solve, particle streaming usw.

%%%The most important part of \sgk\ that is left unaltered from \agk\ is the perpendicular Fourier transforms.
%%%In most parts of $\sgk$ communication has been 

\subsection{Time integration}
\label{sec:TimeIntegration}
\label{sec:Timestepping}
\sgk\ computes the approximate solution of \eqref{eq:dadtIsAa}
using the explicit third-order Adams--Bashforth scheme
\begin{align}
  \label{eq:AB3}
  \a^{i+1} &= \a^{i} + \Delta t\left( \frac{23}{12} \tsop\left[ \a^i \right] 
  - \frac43 \tsop\left[\a^{i-1}\right] 
  + \frac{5}{12} \tsop\left[ \a^{i-2}\right] \right) ,
\end{align} 
where 
$\a^i$ denotes the expansion coefficients at the $i$th time level, 
and $\Delta t$ is the timestep.
\sgk\ also implements a variable time-spacing version of this formula that allows the timestep to change during execution. 
In nonlinear runs, the size of the timestep is varied automatically to satisfy the CFL condition for the nonlinear drift velocity $\vDNL$, exactly as in \agk.
In every iteration the timestep $\Delta t$ is made to satisfy
\begin{align}
  \Delta t \leq C_{\textrm{CFL}} \min\lp \frac{\Delta x}{\max \vDNLx} , \frac{\Delta y}{\max\vDNLy}\rp,
  %\label{}
\end{align}
where $\Delta x = 2\pi/k_{x,\max}$, $\Delta y = 2\pi/k_{y,\max}$, and $C_{\textrm{CFL}}\leq1$ is an input parameter.

\newcommand{\dt}{\Delta t}
\newcommand{\omegamax}{\omega_\mathrm{max}}
\newcommand{\omegatyp}{\omega_\mathrm{typ}}

The Adams--Bashforth scheme \eqref{eq:AB3} is stable and accurate for
non-dissipative wave phenomena, with third-order accuracy in amplitude
and wave speed. It is well-suited to problems like ours where
the calculation of $\tsop$ dominates the computation work.
\citet{Durran91,Durran99} defines the ``efficiency factor'' of a numerical integration
scheme to be the maximum stable timestep for an oscillatory test problem
divided by the number of evaluations of $\tsop$ per timestep.  By this
measure Adams--Bashforth is the most efficient third-order scheme. 
While it has a smaller stable timestep than other schemes such as the
Runge--Kutta family, it only requires one $\tsop$ evaluation per timestep.

The main disadvantage of the \toab\ scheme is that it requires the two previous values
$\a^{i-1}$ and $\a^{i-2}$ to advance from $\a^i$ to $\a^{i+1}$. We must amend the scheme for the first and second
timesteps, for which fewer previous values are available. We use the explicit
Euler method for the first timestep, and the second-order
Adams--Bashforth method, which requires only one previous value, for
the second timestep.

In principle, this reduces the global time accuracy to only second order,
since the first Euler timestep alone contributes an $\O(\dt^2)$ global error. However,
for typical simulation parameters and initial conditions, this error is in fact smaller than the error accumulated
over the subsequent third-order Adams--Bashforth timesteps (see \cite{Parker14}/Appendix \ref{sec:VPPaper}). 
The fastest timescales
in the system arises from parallel streaming with frequencies $\omega = k \sqrt{m/2}$. Let
$\omega_0$ be the typical frequency of the modes that are nonzero in the initial conditions, 
$\omegatyp$ a typical frequency for the dominant modes in the subsequent evolution, and $\omegamax = k_\mathrm{max} \sqrt{m_\mathrm{max}/2}$ the highest frequency present in the simulation. The accumulated amplitude error after a time $T$ is then
\begin{align}
E = \frac{T}{\dt} \frac{3}{8} ( \omegatyp \dt)^4 + \frac{1}{4} (\omega_0 \dt)^3 + \frac{1}{2} (\omega_0 \dt)^2,
\end{align}
using the amplitude errors for the explicit Euler and Adams--Bashforth methods \cite[][Table 2.2]{Durran99}.

The maximum stable timestep is $\dt = \alpha / \omegamax$, with $\alpha \approx 0.723$ for the third order
Adams--Bashforth method \citep{Durran99}. The error after $n$ characteristic times ($T = n/\omegatyp$) using this timestep is
thus
\begin{align}
E = \frac{3}{8} n \alpha^3 ( \omegatyp/\omegamax)^3 + \frac{1}{4} \alpha^3 ( \omega_0 / \omegamax)^3+ \frac{1}{2} \alpha^2 ( \omega_0 / \omegamax)^2,
\end{align}
which is dominated by the first term whenever $\omegatyp > ( \omega_0^2 \omegamax)^{1/3} n^{-1/3}$.
The error due to the first explicit Euler timestep is thus negligible if the initial conditions contain only slowly evolving modes (as do ours) and the system is not heavily over-resolved so that $\omegatyp$ is not small compared with $\omegamax$ in the sense made precise by the
previous inequality.

In \sgk, we use the Adams--Bashforth scheme for all terms on the \rhs\ of the gyrokinetic equation.
It would be possible to implement an implicit scheme for the linear parts of the gyrokinetic equation, such as Kotschenreuther's Green's function method \citep{Kotschenreuther95} that is used in \agk.
This allows larger timesteps at the cost of higher complexity.
However we favour an explicit approach so that we may exploit phase space locality and reduced dimensionality in the linear Fourier--Hermite--Hankel system.
In the nonlinear case, the nonlinear term must be treated explicitly, so that the timestep is limited by the CFL condition. 
In this case it is preferable to use an explicit timestepping algorithm for the linear terms too, as an implicit method gives an increase in complexity without an increase in the timestep.

\subsection{Calculation of the electromagnetic field}
\label{sec:FieldSolve}

In order to form the operator $\tsop$,
we need to calculate 
the quantities $\varphi$, $\Apar$ and $\Bpar$
that determine
the electromagnetic field.
For each wave vector $\k$, 
Maxwell's equations, \eqref{eq:Maxwell5D} or \eqref{eq:Maxwell4D}, may be written as
\begin{align}
\begin{pmatrix}
  c_1 & 0        & c_2 \\
  0        & c_3 & 0  \\
  c_4 & 0 & c_5 \\
\end{pmatrix} 
\begin{pmatrix}
 \varphi \\
 \Apar \\
 \Bpar
\end{pmatrix} 
= 
\begin{pmatrix}
 d_1 \\
 d_2 \\
 d_3
\end{pmatrix} 
,
\label{eq:FieldSolveSchematic}
\end{align}
where the first row is the quasineutrality condition, and the second and third rows are the parallel and perpendicular components of \ampere's law respectively.
In the general five-dimensional version of the code, the coefficients $d_1$, $d_2$, $d_3$ are proportional to 
coefficients of single Fourier--Hermite modes summed over perpendicular velocities (see \ref{eq:Maxwell5D}).
In the four-dimensional version, the coefficients $d_1$, $d_2$, $d_3$ are proportional to coefficients of a single Fourier--Hermite--Hankel mode \eqref{eq:Maxwell4D}.
The coefficients $c_1$, $c_2$, $c_3$, $c_4$ and $c_5$ are known constants which are the same in the four- and five-dimensional versions of the code.
For reference, these constants are tabulated in \tab~\ref{tab:FieldSolveConstants} for the multispecies case, 
and in \tab~\ref{tab:FieldSolveConstantsOneSpecies} for a single ion species with adiabatic electrons.

\renewcommand{\kvar}{\rho_s\kperp}
\renewcommand{\arraystretch}{1.75}
\begin{table}[tp]
  \begin{center}
  \begin{tabular}{|c|c|c|}
    \hline
		Coef. & 5D \eqref{eq:Maxwell5D} & 4D \eqref{eq:Maxwell4D}  \\
    \hline
    $c_1$ & \multicolumn{2}{|c|}{$\sum_s n_sq_s^2(1-\Gamma_{0s})/T_s$} \\
    $c_2$ & \multicolumn{2}{|c|}{$-\sum_s n_sq_s\Gamma_{1s}$} \\
    $c_3$ & \multicolumn{2}{|c|}{$\kperp^2/(2\beta_r)  + \sum_s q_s^2n_s\Gamma_{0s}/(2m_s)$} \\
    $c_4$ & \multicolumn{2}{|c|}{$\sum_sn_sq_s\Gamma_{1s}$} \\
    $c_5$ & \multicolumn{2}{|c|}{$2/\beta_r  + \sum_s n_sT_s\Gamma_{2s}$} \\
    \hline
    $d_1$ & $2\pi \sum_s \qNs \nNs  \sum_{n=1}^{N_p}w_nJ_0(\kvar v_n) g_{s0}(\k,v_n)$ 
    & $2\pi \sum_s \qNs \nNs \g_{s0}(\k,p=\kvar)$ \\
    $d_2$ & $\sqrt{2}\pi \sum_s \qNs\nNs \vthNs \sum_{n=1}^{N_p}w_n J_0(\kvar v_n) g_{s1}(\k,v_n)$
    & $\sqrt{2}\pi\sum_s \qNs\nNs \stm \g_{s1}(\k,p=\kvar)$ \\
    $d_3$ & $- 2\pi \sum_s  \nNs \TNs \sum_{n=1}^{N_p}w_n  v_n \frac{2J_1(\kvar v_n)}{\kvar} g_{s0}(\k,v_n)$ 
    & $4\pi \sum_s \frac{\nNs \TNs}{\kvar}\left.\pd{\g_{s0}}{p}\right|_{p=\kvar}$
    \\[1ex]
    \hline
  \end{tabular}
  \end{center}
\caption[Coefficients in the field solve for the multispecies case.]{Coefficients in the field solve \eqref{eq:FieldSolveSchematic} for the multispecies case. \label{tab:FieldSolveConstants}}
\end{table}

\begin{table}[tp]
  \begin{center}
  \begin{tabular}{|c|c|c|}
    \hline
    Coef. & 5D \eqref{eq:Maxwell5D} & 4D \eqref{eq:Maxwell4D}  \\
    \hline
    $c_1$ & \multicolumn{2}{|c|}{$n_iq_i^2(1 + T_i/T_e -\Gamma_{0i})/T_i$} \\
    $c_2$ & \multicolumn{2}{|c|}{$-n_iq_i\Gamma_{1i}$} \\
    $c_3$ & \multicolumn{2}{|c|}{$0$} \\
    $c_4$ & \multicolumn{2}{|c|}{$n_iq_i\Gamma_{1i}$} \\
    $c_5$ & \multicolumn{2}{|c|}{$2/\beta_r  + n_iT_i\Gamma_{2i}$} \\
    \hline
    $d_1$ & $2\pi q_i n_i  \sum_{n=1}^{N_p}w_nJ_0(\kvar v_n) g_{i0}(\k,v_n)$ 
    & $2\pi  q_i n_i \g_{i0}(\k,p=\kvar)$ \\
    $d_2$ & $\sqrt{2}\pi  q_in_i \vthNi \sum_{n=1}^{N_p}w_n J_0(\kvar v_n) g_{i1}(\k,v_n)$
    & $\sqrt{2}\pi q_in_i \vthNi \g_{i1}(\k,p=\kvar)$ \\
    $d_3$ & $- 2\pi  n_i T_i \sum_{n=1}^{N_p}w_n  v_n \frac{2J_1(\kvar v_n)}{\kvar} g_{i0}(\k,v_n)$ 
    & $4\pi  \frac{n_i T_i}{\kvar}\left.\pd{\g_{i0}}{p}\right|_{p=\kvar}$
    \\[1ex]
    \hline
  \end{tabular}
  \end{center}
\caption[Coefficients in the field solve for a single ion species with adiabatic electrons.]{Coefficients in the field solve \eqref{eq:FieldSolveSchematic} for a single ion species with adiabatic electrons. \label{tab:FieldSolveConstantsOneSpecies}}
\end{table}
Thus the electromagnetic field variables are calculated at every Fourier wavenumber using
\begin{align}
 \varphi &= (c_5d_1 - c_2d_3) / ( c_1c_5 - c_2c_4) ,  \\
 \Apar   &= d_2 / c_3 , \\
 \Bpar &= (c_4d_1 - c_1d_3) / ( c_1c_5 - c_2c_4) .
\end{align}
In electrostatic simulations we only solve the quasineutrality condition,
\begin{align}
 \varphi = d_1 / c_1,
\end{align}
and neglect \ampere's law.

In order to perform these calculations, we must evaluate $d_1$, $d_2$ and $d_3$ from the distribution function, 
which in general requires \ipc.
The four-dimensional case is simple.
At every Fourier wavenumber $\k$, the field is calculated on the processor which holds the point $(\k,m=0,p=\kvar)$;
this requires no communication.\footnote{The Hermite modes $m=0$ and $m=1$ are always on the same processor for a given $(\k,v_n)$; see \sec\ref{sec:DataLayoutParallelization}.}
Each processor sets the electromagnetic fields to zero, except for at the points which that processor calculated.
The fields are then broadcast to all processors via an MPI ``sum all reduce'' operation.

The five-dimensional case is similar, only now a sum over perpendicular velocity is required.
The electromagnetic field is calculated by the processor which holds the point $(\k,m=0,p=v_1)$.
Processors with other velocities at the Fourier--Hermite mode $(\k,m=0)$ send their contribution to the sums in $d_1$, $d_2$, $d_3$ to that processor.
Once calculated, the electromagnetic field is broadcast to all processors as in the four-dimensional case.

\renewcommand{\kvar}{k'}

\subsection{Nonlinear term and dealiasing}
\label{sec:Nonlinear}

As described in \sec\ref{sec:NonlinearTerm} we calculate the
nonlinear term \eqref{eq:NonlinearTerm} using fast Fourier transforms
to convert between expansion coefficients at a discrete set of wavenumbers %$k_j=2\upi j/L$ 
and
function values at a uniformly spaced set of collocation points. % $ z_l = lL / N_k$.
We use the FFTW library \citep{FFTW3} which implements unnormalized \DFT s, %\ie\ the analogue of \eqref{eq:DFTs} without the factor $1/N_k$ in the definition of $\Fmat_{jl}$, for which the forward transform of $\e^{\i k_jz_l}$ has magnitude $N_k$ rather than unity. 
\ie\ in one dimension the pair of operators 
\begin{align}
  \label{eq:DFTs}
  \Fmat_{jl} = \sumo{l}{N_z-1} ~ \e^{-\i k_jz_l} 
  ,
  \hspace{1cm}
  \Fmatinv_{ln} = \sumo{n}{N_z-1} ~ \e^{\i k_nz_l} ,
\end{align}
with wavenumbers $k_n=2\pi n/L_z$ and grid points $z_l=lL_z/N_z$, as in \sec\ref{sec:FourierSpaceRepresentationGKMSystem}.

This \ps\ approach reduces the  cost of computing the nonlinear term 
to $\O(N_VN_X\log N_X)$ instead of $\O(N_VN_X^2)$, 
where $N_X=N_xN_yN_z$ is the total spatial resolution 
and $N_V=N_mN_p$ is the total velocity resolution.
However, it creates a source of error due to aliasing between
different wavenumbers.  The product of two truncated Fourier series
that occurs in the nonlinear term \eqref{eq:NonlinearTerm} %\eqref{eq:DiscreteNonlinearTerm} 
contains terms involving $\e^{\i(k_n+k_{n'})z_l}$. However, only terms with $|k_n+k_{n'}|\le k_{N_z}$ can be represented on the collocation grid.
The remaining terms are omitted from the original convolution \eqref{eq:GKSystemNonlinearTeamChp5}, %\eqref{eq:ModifiedNonlinearTerm}, 
but on the collocation points $z_l$ they are indistinguishable from Fourier modes with
wavenumbers $k_n+k_{n'} \mp k_{N_z}$. This aliasing of wavenumbers outside the
truncation onto retained wavenumbers creates errors in the coefficients of the 
retained wavenumbers. It may be removed by setting the coefficients for the highest
third of wavenumbers to zero before the nonlinear term is
calculated \citep{Orszag71,Boyd01}. This ``two-thirds rule'' removes all spurious
contributions from aliasing for quadratic nonlinearities, such as
the Fourier convolution \eqref{eq:GKSystemNonlinearTeamChp5}. % \eqref{eq:ModifiedNonlinearTerm}. 
All
wavenumbers with $|k_n+k_{n'}|>k_{N_z}$ now map onto wavenumbers that
are eliminated by the filter.

The abrupt transition from unmodified Fourier coefficients to those
that are set to zero acts like a reflecting boundary condition for the
typical Kolmogorov-like nonlinear cascade of disturbances towards larger
wavenumbers. This will distort the higher resolved wavenumbers unless
their Fourier coefficients are already negligible at the two-thirds
rule cut-off point. 
In \agk, this is ensured by applying hyperviscosity in the perpendicular Fourier directions.
In its simplest form, 
hyperviscosity is the term
\begin{align}
  \label{eq:Hyperviscosity}
  -\nu_{\mathsf{hypervisc}} ( \kperp^2/ k_{\perp\max}^2 )^4 g_m(\k,\vperp),
\end{align}
added to the \rhs\ of \eqref{eq:SpectralGKEqn}.
Other viscosity models, such as one based on the nonlinear Smagorinsky eddy viscosity \citep{BelliThesis}, % S 6.2.1
are also implemented in \agk.
As the perpendicular hyperviscosity implementation is closely related to the distributed perpendicular Fourier transform, this routine is inherited by \sgk.

In parallel Fourier space (which is a new implementation)
we employ the smooth non-reflecting cut-off provided
by the \citet{Hou07} filter that multiplies the expansion coefficients by
\begin{align}
  \label{eq:HouLiFilter}
  \exp\lp -36(|\kpara|/k_{\parallel\mathrm{max}})^{36}\rp ,
\end{align}
at each timestep.
This filter is highly selective in wavenumber space, yet it
provides sufficient dissipation to replace both the
two-thirds rule dealiasing, and the parallel hyperviscous dissipation usually
employed to prevent an accumulation of free energy at the finest
resolved scales.
For the same resolution, this smooths the distribution while keeping 12--15\% more modes than the 2/3 aliasing rule \citep{Hou07}.

%%%\subsubsection{Fourier transforms and filtering}
%%%
%%%The perpendicular Fourier transforms in $x$ and $y$ are implemented in \agk\ as two 1D transforms using FFTW \cite{FFTW3}. 
%%%Although this has higher complexity---$\O( (N_x\log N_x)(N_y\log N_y))$ instead of $\O(N_xN_y\log (N_xN_y))$---separating the transforms allows each to be performed locally to processor, with the resulting saving in communication overcoming the increase in work.
%%%Thus in \agk\ the perpendicular Fourier transforms are a single routine combined with memory redistributions and Fourier dealising by the 2/3 rule.
%%%We use the same routines for \sgk's perpendicular Fourier transforms.
%%%
%%%
%%%
%%%\subsubsection{Hankel transform}
%%%The Hankel transform \eqref{eq:ContinuousHankelTransform} is equivalent to the radial part of a two-dimensional Fourier transform \cite{Piessens00}.
%%%However while there is work on polar Fourier transforms \cite[\eg][]{Averbuch06} there is no exact radial FFT.
%%%Similarly there are efficient approximations to the continuous Hankel transform for large numbers of points \cite{Ogata05,Siegman77} but as we expect to have relatively few points in perpendicular velocity, we calculate the discrete transform \eqref{eq:DiscreteHankelPair} by the matrix multiplication \eqref{eq:DiscreteHankelMatrix}.

\subsection{Recurrence and velocity space dissipation}
\label{sec:Recurrence}
As shown in \chp~\ref{sec:Hypercollisions}, the shear in phase space due to the particle streaming term $\vpara\tpds{g}{z}$ tends to form fine scale
structures in velocity space, which, 
for any discretization of velocity space, become unresolvable after some finite, resolution-dependent time.  
The $\vpara\tpds{g}{z}$ term creates the linear coupling between coefficients with adjacent $m$ values in the gyrokinetic equation.
We showed in \chp\ \ref{sec:ParallelVelocitySpace}
that this coupling may be interpreted as a propagation of disturbances towards larger $m$ values that continues until they encounter
the truncation condition $g_{N_m}=0$. 
As with the two-thirds rule mentioned above, this condition appears as a reflecting boundary condition that creates disturbances propagating in the reverse direction towards low $m$ values. 
Disturbances eventually reach $m=0$ to create a spurious increase in the amplitude of the electric field, a phenomenon known as recurrence \citep{Grant67}.

The recurrence may be suppressed using collision terms 
on the right hand side of the gyrokinetic equation, 
chosen to ensure that the coefficients reach negligibly small values before $m$ reaches $N_m$. 
Various collision operators are implemented in \sgk:
the %\textsc{Lenard--Bernstein} (\citeyear{LenardBernstein}) 
\citet{LenardBernstein}
and \citet{Kirkwood46} 
collision operators
\begin{align}
  C_{\mathsf{LB}}[g] &= -\nu  m g, 
\label{eq:LenardBernsteinSGK}
\\
  C_{\mathsf{K}}[g] &=  -\nu  m \mathcal{I}_{\{ m \ge 3 \}} g,
\label{eq:KirkwoodSGK}
\end{align}
(where $\mathcal{I}_{\{ m \ge 3 \}}$ denotes the indicator function)
which were discussed in \sec\ref{sec:lb} and \sec\ref{sec:LinearizedCollisionOperators},
and the Kirkwood hypercollisional operator 
\begin{align}
  C^H_{\mathsf{K}}[g] = - \nu (m/N_m)^{\alpha}\mathcal{I}_{\{ m \ge 3 \}} g ,  \label{eq:HyperColl}
\end{align}
which was discussed in \sec\ref{sec:hypercollisions}.
In these, $\nu$ and $\alpha$ are input parameters.
In addition, 
the velocity-space analogue of the \textsc{Hou--Li} (\citeyear{Hou07}) filter 
\begin{align}
  \label{eq:HouLiM}
  \exp\lp -36(m/(N_m-1))^{36}\rp
\end{align}
is also implemented.  
We discuss this filter in Appendix \ref{sec:VPPaper}/\referencename\cite{Parker14}. %\sec\ref{sec:RecurrenceVlasovPoisson}.

Each of these operators is trivial to apply as all are known functions of $m$ multiplying the distribution function expansion coefficients.
Thus no communication and minimal computation are required.

\subsection{Data layout and parallelization scheme}
\label{sec:DataLayoutParallelization}

In \agks\ the distribution function $g$ has size $N_xN_yN_zN_mN_pN_s$ and the electromagnetic field variables $\varphi$, $\Apar$, and $\Bpar$ have size $N_xN_yN_z$. 
A copy of these scalars is kept on each processor, while the much larger distribution function is divided between processors.
\agks\ inherits its data layout from \agk\ which is described in \sec\ref{sec:AstroGK} and Ref.\ \cite[][\sec3.4]{Numata10}.
Recall that %in \agk\ 
the distribution is stored in a three dimensional array.
The first two dimensions correspond to the parallel space coordinate $z$ and the sign of parallel velocity respectively.
The remaining coordinates are concatenated in the third dimension.
The array is distributed by keeping the first two dimensions local, but splitting the third between processors.

For compatibility, \sgk\ also uses the first dimension for the parallel space dependence.
In the two elements of the second dimension it stores pairs of Hermite modes.
This minimizes communication in the particle streaming term as a Hermite mode is always on processor with at least one neighbour.
This is also convenient for the field solve, as the array containing the two necessary Hermite modes $m=0$ and $m=1$ is referenced by a single index. 
A reference to these Hermite pairs is stored in the third dimension concatenated with Hankel and perpendicular Fourier modes.\footnote{Using \agk\ layout terminology, \sgk\ uses the energy coordinate $\texttt{e}$ to reference the Hankel mode and the lambda index $\texttt{l}$ to reference Hermite pairs. Thus if \texttt{sgn=0,1} gives the index of the second array, the $m$th Hermite mode satisfies $m=\texttt{sgn}+2\texttt{l}$.  As in \agk, $\texttt{x}$, $\texttt{y}$ and $\texttt{z}$ refer to spatial coordinates and $\texttt{s}$ labels the species.  }

In \sgk\ coordinates, there are three terms which potentially require inter-processor communication. 
Firstly, the linear operator $\L_g$ \eqref{eq:Lg} requires neighbouring Hermite modes for streaming. %, and all Hankel modes for the integral if $L_B^{-1}\neq0$.
Secondly, the perpendicular velocity space integrals in Maxwell's \eqn s %\eqref{eq:M0sum}--\eqref{eq:M2sum} 
require sums over all Hankel modes, or over all perpendicular velocity grid points. 
Finally, the Hankel transforms and perpendicular Fourier transforms in the nonlinear term require sums across $k_x$, $k_y$, and $p$. 

In linear runs, there is no communication across Fourier space and the most efficient layouts are where the Hermite mode is most local. 
%If the linear run requires Hankel space ($L_B^{-1}\neq0$), then this should be the next most local.
%Communication is then limited to sending a processor's contribution to the field velocity integrals %\eqref{eq:M0sum}--\eqref{eq:M2sum} 
%using its local data.
%Alternatively 
If velocity space is made entirely local by making the number of Hermite mode pairs to be divisible by the number of processors, \ie, by choosing $N_m/(2\Nproc)\in\mathbb{Z}$, then \sgk\ recognizes this and performs all calculations locally.

In nonlinear runs there is a compromise between keeping velocity space local, which is best for the field solve and linear terms, and keeping physical space local, which is most efficient for the Fourier transforms in the nonlinear term. 
%%%In our experience, the most efficient runs have Hermite space most local followed by Hankel space.
%%%Then the code is limited by the distributed two-dimensional Fourier transforms in the nonlinear term. 
In our experience, it is usually most efficient to perform the Fourier transforms locally, \ie\ to choose parameters such that $N_mN_p/(2\Nproc)\in\mathbb{Z}$.

\section{Code verification}
\label{sec:CodeVerification}
\newcommand{\omegat}{\tilde{\omega}}

\sgk\ and \agk\ both solve the \gkm\ system \GKMSystemChapTwo, but discretize using different coordinates; therefore we expect agreement between the codes in the limit of infinite resolution.
The change from \agk's grid in $z$ to \sgk's Fourier wavenumbers is straightforward, with the difference being that \sgk\ captures resolved Fourier modes exactly, while \agk\ suffers finite-differencing errors.
We therefore focus on velocity space behaviour.
We repeat some of the linear and nonlinear tests used in Ref.\ \cite{Numata10} to test \agk.
This both verifies \sgk\ and helps determine optimal parameters, particularly for the Hankel space representation which contains the undetermined velocity space cutoff $\vcut$. 

\subsection{Linear ITG}
\newcommand{\rhokperp}{\kperp'}
We first compare the growth rate for the electrostatic ion temperature gradient instability obtained from a \sgk\ initial value problem to the value 
given by the dispersion relation.
This is a case where \sgk\ can run in four dimensions.
The four-dimensional code is exact in Fourier--Hankel space (with solutions parameterized by the Fourier wavenumber $\k$, $p=\rho_s\kperp$), 
so this is a test solely of the Hermite space discretization.
We also repeat the test with the five-dimensional code to verify that version's treatment of the linear terms.

The linear electrostatic dispersion relation is found via Landau's Laplace transform method, analogous to that described in \chp~\ref{sec:ParallelVelocitySpace}.
We take the Laplace transform in time of \eqref{eq:GKIonsES} with transform variable $\hat{p}$ and with $\L_A=\L_B={\cal N}=0$.
We then divide by $\hat{p}+i\vths\vpara\kpara$, apply the operator $\intii\d\vpara\intoi\d\vperp~\vperp J_0(\rho_s\kperp\vperp)$,
and use the quasineutrality condition \eqref{eq:QNOneSpeciesES} to obtain the dispersion relation
%%%\begin{align}
%%%  \begin{split}
%%%   &  \frac{n_iq_i^2}{T_i} \left(1+\frac{T_i}{T_e}-\Gamma_{0i}\right) 
%%%  =  - q_in_i \Bigg[ \frac{q_i}{T_i}   
%%%  \lp 1 + \omegat Z(\omegat)\rp \I_1(\rho_i\kperp)
%%%  \\  & \hspace{2cm} 
%%%  + \frac{k_y}{2\vthNi\kpara}\left\{\lp\Lnr - \frac{3}{2}\LTr\rp Z(\omegat)+\LTr Z(\omegat) + \omegat Z(\omegat)\rb \I_1(\rho_i\kperp) 
%%%  \\ &  \hspace{2cm} 
%%%  + \frac{k_y\LTr}{2\vthNi\kpara} Z(\omegat) \I_3(\rho_i\kperp) \Bigg],
%%%  \label{eq:ExactDR}
%%%\end{split}
%%%\end{align}
\begin{align}
  \begin{split}
   &  \frac{n_iq_i^2}{T_i} \left(1+\frac{T_i}{T_e}-\Gamma_{0i}\right) 
  =  - q_in_i \Bigg[ \frac{q_i}{T_i}   
  \lp 1 + \omegat Z(\omegat)\rp \I_1(\rho_i\kperp) + \frac{k_y\omega_T}{2\vthNi\kpara} Z(\omegat) \I_3(\rho_i\kperp)
  \\  & \hspace{2cm} 
  + \frac{k_y}{2\vthNi\kpara}\left\{\lp\omega_n - \frac{3}{2}\omega_T\rp Z(\omegat)+\omega_T Z(\omegat) + \omegat Z(\omegat)\rb \I_1(\rho_i\kperp) 
   \Bigg],
  \label{eq:ExactDR}
\end{split}
\end{align}
where $\omegat = \omega/(\vthNi\kpara)$ with $\omega$ the complex frequency of the solution in the long-time limit, and $\I_1$ and $\I_3$ are given by \eqref{eq:I1} and \eqref{eq:I3}.
The parallel velocity integral leads to factors of the plasma dispersion function 
$Z(\omega) = \pi^{-1/2}\int \d\vpara~{e^{-\vpara^2}}/({\vpara - \omega})$,
as in \eqref{eq:DispersionRelation}.
We solve \eqref{eq:ExactDR} numerically for $\omega(\k)$ and compare the result to that from an \ivp\ with the four-dimensional code.
This is a test for the Hermite representation as the $\k$ and $p$ dependence is treated exactly.

We can also derive a dispersion relation mimicking the five-dimensional code.
We suppose the distribution function is only known on $\vperp$-grid points,
so that rather than taking the continuous Hankel transform and using the quasineutrality condition \eqref{eq:QNOneSpeciesES},
we take the discrete Hankel transform $\sum_{n=1}^{N_p}w_nJ_0(\rho_i\kperp v_n)$ and use the discrete quasineutrality condition \eqref{eq:Maxwell5DQN}.
With this approach, we obtain
\begin{align}
  \begin{split}
   &  \frac{n_iq_i^2}{T_i} \left(1+\frac{T_i}{T_e}-\Gamma_{0i}\right) 
  =  - q_in_i \Bigg[ \frac{q_i}{T_i}   
    \lp 1 + \omegat Z(\omegat)\rp \sum_{n=1}^{N_p} w_k J_0^2(\rho_i\kperp v_n) F_0^{\perp}(v_n) 
  \\  & \hspace{0.8cm} 
  + \frac{k_y}{2\vthNi\kpara}\left\{\lp\omega_n - \frac{3}{2}\omega_T\rp Z(\omegat)+\omega_T Z(\omegat) + \omegat Z(\omegat)\rb \sum_{n=1}^{N_p} w_k J_0^2(\rho_i\kperp v_n) F_0^{\perp}(v_n) 
  \\ &  \hspace{0.8cm} 
  + \frac{k_y\omega_T}{2\vthNi\kpara} Z(\omegat) \sum_{n=1}^{N_p} w_k v_n^2J_0^2(\rho_i\kperp v_n) F_0^{\perp}(v_n)  \Bigg],
  \label{eq:ApproxDR}
\end{split}
\end{align}
which is the same as the exact dispersion relation \eqref{eq:ExactDR} but with $\I_1$ and $\I_3$ approximated as sums.
Therefore the linear behaviour in Hankel space is captured if the quadrature rule suitably approximates the integral $\I_1$ and $\I_3$.
We therefore use these functions to quantify the accuracy in \sec\ref{sec:5dTesting}.

\subsubsection{Four-dimensional ITG}
\label{sec:4DLinearTest}

\begin{figure}[tb]
  \centering
  \subfigure[]{\includegraphics[trim=3.7cm 9cm 4cm 10cm,width=0.49\textwidth,clip]{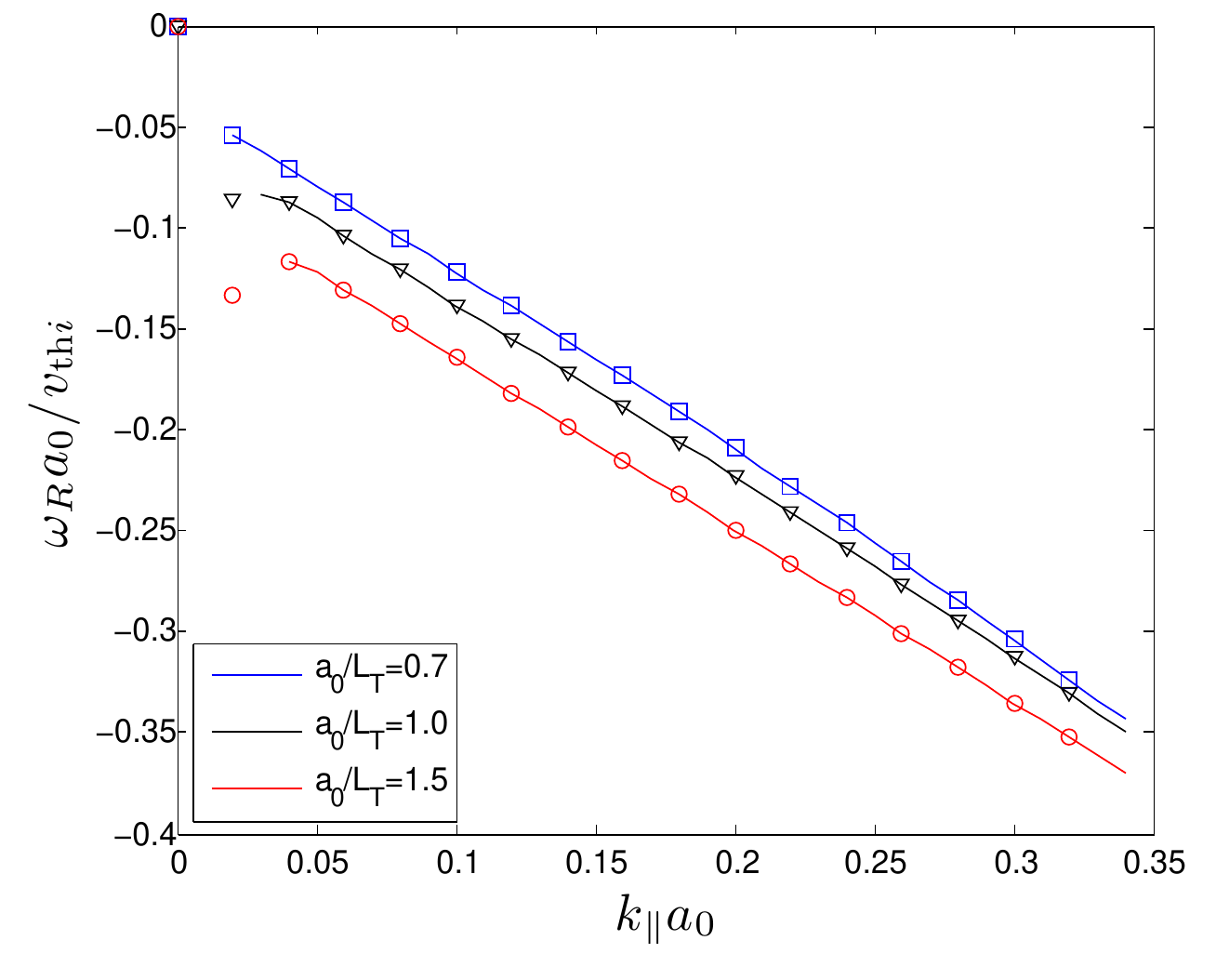}}
  \subfigure[\label{fig:42Growth}]{\includegraphics[trim=3.7cm 9cm 4cm 10cm,width=0.49\textwidth,clip]{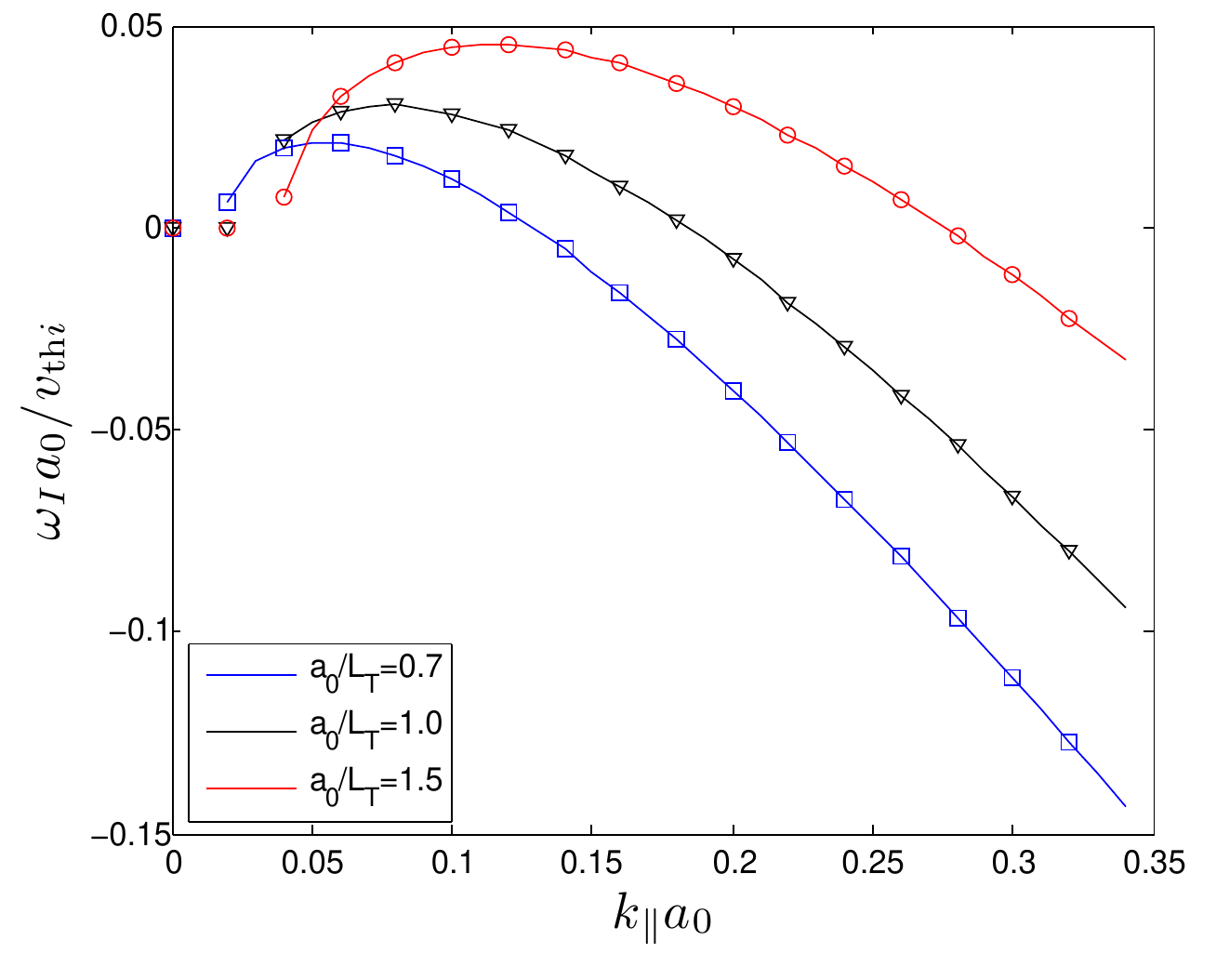}}
  \caption[The dispersion relation for the ITG instability.]{The dispersion relation for the linear slab ITG instability. The marks show values calculated by \agks\ for (a) the frequency and (b) the growth rate against parallel wavenumber, with lines showing the exact dispersion relation.  These plots are the same as the \agk\ results in \citet[\fig 4]{Numata10} for growing modes.  Unlike \agk, \agks\ correctly calculates negative growth rates in the right-hand plot due to the use of the hypercollisional operator \eqref{eq:HyperColl}.  \label{fig:42}}
\end{figure}
In \fig\ref{fig:42} we plot the linear frequency and growth rate obtained from a four-dimensional \ivp\ against $\kpara$ for a range of temperature gradients with $(k_x\rho_i,k_y\rho_i)=(0,1)$ and $T_{0e}=T_{0i}$.
The \ivp\ gives perfect mode-by-mode agreement with the dispersion relation for $k_x$, $k_y$, $\kpara$ with as few as 12 Hermite modes.
Importantly, \sgk\ also captures Landau damping due to the hypercollision operator developed in \chp~\ref{sec:ParallelVelocitySpace},
exactly matching the negative growth rates in the right-hand plot.

\subsubsection{Five-dimensional ITG}
\label{sec:5dTesting}
The four-dimensional linear calculation does not solve for the perpendicular velocity structure.
The method only finds the amplitude of the single Hankel mode $\g(p=\rho_s\kperp)$ that is required to determine the electrostatic potential, while the $\vperp$ structure depends on the superposition of all modes $\g(p)$.
We therefore perform five-dimensional linear calculations to ensure the code correctly represents $\vperp$ structure, and to study its behaviour when discrete perpendicular velocity space is included.     
We show that the linear growth rate obtained in the five-dimensional code converges to the value obtained from the four-dimensional code.
We also show convergence behaviour of the quadrature rule \eqref{eq:QuasineutralityIntegralVSpace}
used the field solve.

\begin{figure}[tb]
  \centering
  %\subfigure[\label{fig:GRConvergenceWithGridSpacing}]{\includegraphics[]{images/sgk/negrid_growth_rate_convergence/negrid_growth_rate_convergence_vcut_on_negrid.eps}}
  \subfigure[\label{fig:GRConvergenceWithGridSpacing}]{\includegraphics[]{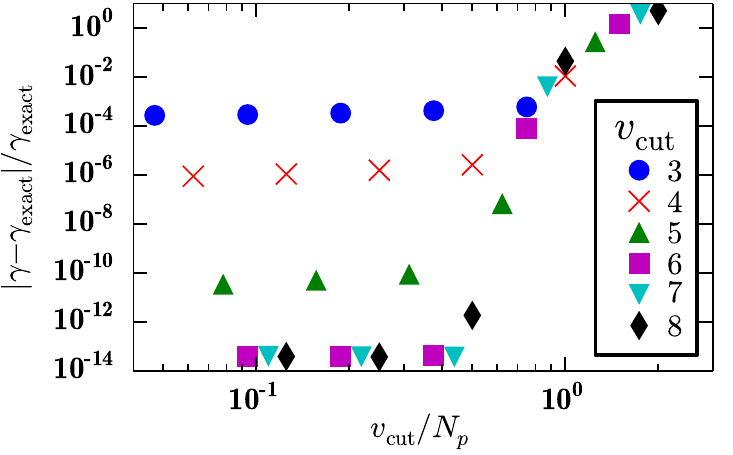}}
  %\subfigure[\label{fig:GRConvergenceWithVcut}]{\includegraphics[]{images/sgk/negrid_growth_rate_convergence/negrid_growth_rate_convergence_vcut.eps}}
  \subfigure[\label{fig:GRConvergenceWithVcut}]{\includegraphics[]{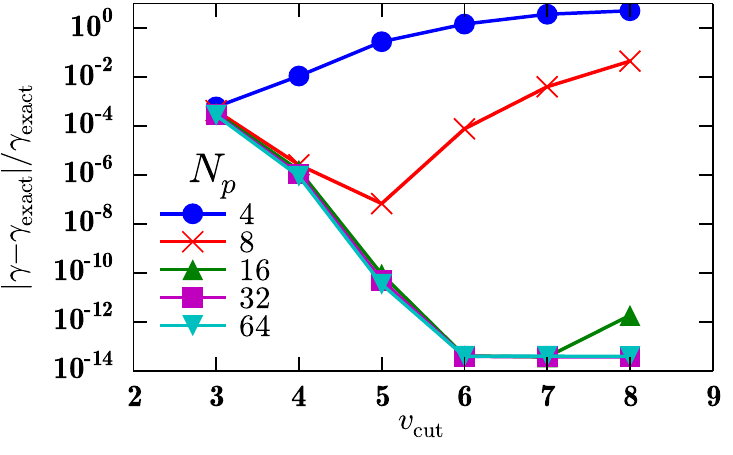}}
  %\subfigure[]{\includegraphics[]{images/sgk/negrid_growth_rate_convergence/negrid_growth_rate_convergence.eps}}
  \caption[Convergence of linear growth rates with perpendicular velocity resolution.]{Convergence of linear growth rates with perpendicular velocity resolution $N_p$ and with velocity space cutoff $\vcut$. %  \label{fig:42}
}
\end{figure}

The perpendicular velocity space grid $v_n = j_n\vcut/j_{N_p+1}$ depends on two parameters: 
$N_p$ the number of grid points,
and $\vcut$ the largest perpendicular velocity (in units of $\vths$) captured by the discretization.
Naturally we want to take $\vcut\to\infty$, but this coarsens the $\vperp$ resolution for a fixed $N_p$. 
In fact, convergence behaviour depends on the ratio $\vcut/N_p$, 
which is the approximate grid spacing, since the Bessel roots $j_n$ are approximately linearly spaced.
In \fig\ref{fig:GRConvergenceWithGridSpacing} we plot the error in the calculation of the growth rate by the five-dimensional code ($\gamma$) compared to the four-dimensional code ($\gamma_{\mathrm{exact}}$) for typical parameters $(\rho_ik_x,\rho_ik_y,R\kpara,R/L_T)=(0,1,0.1,0.7)$ corresponding to a point on the blue curve in \fig\ref{fig:42Growth}.
The growth rate converges super-algebraically with decreasing grid spacing,
until reaching some $\vcut$-dependent minimum error, beyond which there is no further convergence. % with decreasing grid spacing.
This minimum error decreases with increasing $\vcut$, and reaches machine precision for $\vcut\geq6$
at around $\vcut/N_p\approx 0.4$.

For practical reasons, we are typically limited to a modest, fixed resolution, $N_p$,
and must choose a value of $\vcut$.
As the error decreases with decreasing grid spacing $\vcut/N_p$ we generally wish to minimize $\vcut$, 
but doing so for $\vcut<6$ leads to increases in error due to the minimum error for each $\vcut$. %seen in \fig\ref{fig:GRConvergenceWithGridSpacing}.
Consequently there is an optimal value of $\vcut$ for each $N_p$,
as shown in \fig\ref{fig:GRConvergenceWithVcut} where the error is plotted against $\vcut$.
For $N\geq16$, the optimal choice is $\vcut=6$, while for smaller resolutions the optimal choice of $\vcut$ decreases.

As well as verifying the five-dimensional linear calculation,
this test also demonstrates that
the approximate dispersion relation \eqref{eq:ApproxDR} converges 
to the exact dispersion relation \eqref{eq:ExactDR}.
This shows that the sum-approximations to ${\cal I}_1$ and ${\cal I}_3$ which appear in \eqref{eq:ApproxDR} (and in the field solve) must also converge.
However the quantities $\I_1$ and $\I_3$ contain Bessel functions which oscillate more rapidly as $\rho_i\kperp$ increases, and we have only shown convergence at $\rho_i\kperp=1$.
This is sufficient for drift kinetics for which $\rho_i\kperp\lesssim1$,
but for future sub-Larmor scale work it is instructive to study these approximations as a function of $\rho_i\kperp$.

\begin{figure}[tb]
  \centering
%%%  \subfigure[\label{fig:testFunctionVcut6}]{\includegraphics[trim=3.7cm 10.0cm 4.5cm 10.4cm,width=0.49\textwidth,clip]{images/sgk/test_function_vcut_6_2.pdf}}
%%%  \subfigure[\label{fig:PhiErrorVcut4}]{\includegraphics[trim=3.7cm 10.0cm 4.5cm 10.4cm,width=0.49\textwidth,clip]{images/sgk/phi_error_vcut_4_2.pdf}}
%%%  \subfigure[\label{fig:PhiErrorVcut6}]{\includegraphics[trim=3.7cm 10.0cm 4.5cm 10.4cm,width=0.49\textwidth,clip]{images/sgk/phi_error_vcut_6_2.pdf}}
%%%  \subfigure[\label{fig:phi_error_negrid_32}]{\includegraphics[trim=3.7cm 10.0cm 4.5cm 10.4cm,width=0.49\textwidth,clip]{images/sgk/phi_error_negrid_32_mid_2.pdf}}
  % v2
%%%  \subfigure[\label{fig:testFunctionVcut6}]{\includegraphics[trim=0.0cm 0.0cm 0.0cm 0.0cm,width=0.49\textwidth]{images/sgk/function_vcut6.png}}
%%%  \subfigure[\label{fig:PhiErrorVcut4}]{\includegraphics[trim=0.0cm 0.0cm 0.0cm 0.0cm,width=0.49\textwidth]{images/sgk/error_vcut4.png}}
%%%  \subfigure[\label{fig:PhiErrorVcut6}]{\includegraphics[trim=0.0cm 0.0cm 0.0cm 0.0cm,width=0.49\textwidth]{images/sgk/error_vcut6.png}}
%%%  \subfigure[\label{fig:phi_error_negrid_32}]{\includegraphics[trim=0.0cm 0.0cm 0.0cm 0.0cm,width=0.49\textwidth]{images/sgk/error_Np32.png}}
  % v3
  %\subfigure[\label{fig:testFunctionVcut6}]{\includegraphics[trim=0.0cm 0.0cm 0.0cm 0.0cm,width=0.49\textwidth]{images/sgk/test_field_solve/int1_function_vcut6.eps}}
  \subfigure[\label{fig:testFunctionVcut6}]{\includegraphics[trim=0.0cm 0.0cm 0.0cm 0.0cm,width=0.49\textwidth]{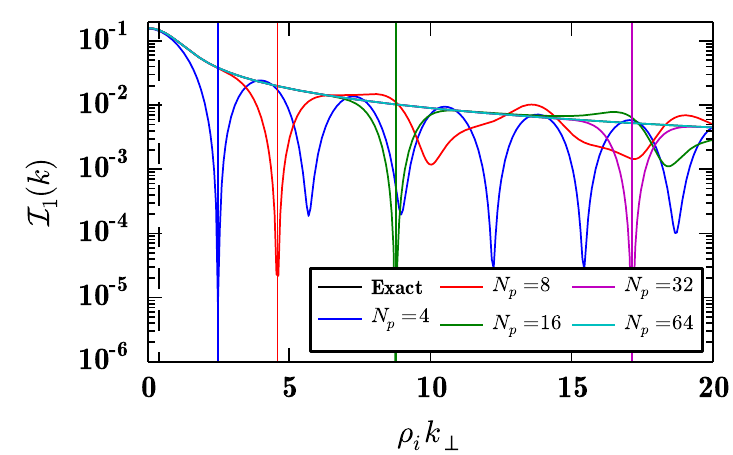}}
  %\subfigure[\label{fig:PhiErrorVcut4}]{\includegraphics[trim=0.0cm 0.0cm 0.0cm 0.0cm,width=0.49\textwidth]{images/sgk/test_field_solve/int1_error_vcut4.eps}}
  \subfigure[\label{fig:PhiErrorVcut4}]{\includegraphics[trim=0.0cm 0.0cm 0.0cm 0.0cm,width=0.49\textwidth]{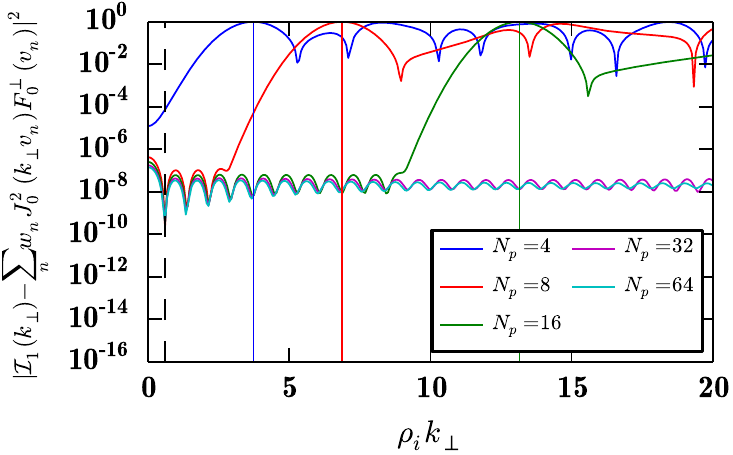}}
  %\subfigure[\label{fig:PhiErrorVcut6}]{\includegraphics[trim=0.0cm 0.0cm 0.0cm 0.0cm,width=0.49\textwidth]{images/sgk/test_field_solve/int1_error_vcut6.eps}}
  \subfigure[\label{fig:PhiErrorVcut6}]{\includegraphics[trim=0.0cm 0.0cm 0.0cm 0.0cm,width=0.49\textwidth]{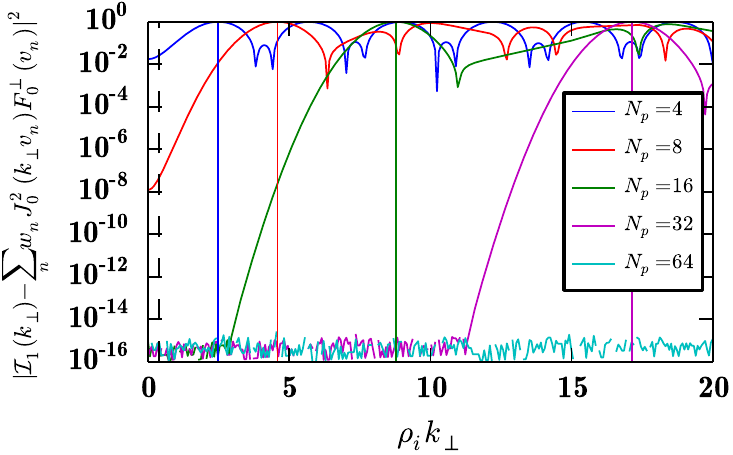}}
  %\subfigure[\label{fig:phi_error_negrid_32}]{\includegraphics[trim=0.0cm 0.0cm 0.0cm 0.0cm,width=0.49\textwidth]{images/sgk/test_field_solve/int1_error_Np32.eps}}
  \subfigure[\label{fig:phi_error_negrid_32}]{\includegraphics[trim=0.0cm 0.0cm 0.0cm 0.0cm,width=0.49\textwidth]{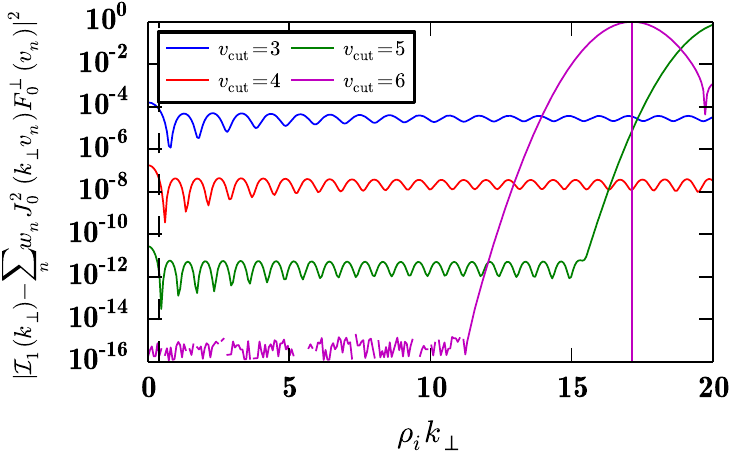}}
  \caption[Testing the Hankel transform at points off the $p$-grid.]{Testing the Hankel transform at points off the $p$-grid. Equation \eqref{eq:phiApproximation} is used to approximate $\I_1(\rho_i\kperp)$ in the field solve. (a) The approximation to $\I_1(\rho_i\kperp)$ with $\vcut=6$ and different $N_p$ .
  (b--c) The error in the approximation against $N_p$ for fixed $\vcut$, (b) $\vcut=4$, (c) $\vcut=6$.
  (d) The error with $N_p=32$ and different $\vcut$.\label{fig:phiScan}}
\end{figure}

We consider the error in the approximation of $\I_1$ by
the quadrature rule
\eqref{eq:QuasineutralityIntegralVSpace},
\begin{align}
  %\intoi \d\vperp ~ \vperp f(\vperp)
  %= 
  %\bar{g}(\kvar)
  \I_1(\kperp)
  %=   \int_0^{\vcut} \d\vperp ~ \vperp J_0(\kperp \vperp) g_m(\k,\vperp)
  =   \int_0^{\infty} \d\vperp ~ \vperp J_0^2(\rho_i\kperp \vperp) F_0^{\perp}(\vperp)
  \approx \sum_{n=1}^{N_p} w_n J_0^2(\rho_i\kperp v_n) F^{\perp}_0(v_n) .
  %\label{eq:DiscretePhiIntegral}%\tag{tag:QNInt}
  \label{eq:phiApproximation}%\tag{tag:QNInt}
\end{align}
As in the previous test, convergence depends on the grid spacing $\vcut/N_p$, 
with a minimum error depending on $\vcut$.
However, since the approximation is in the form of a Hankel transform with transform variable $\rho_i\kperp$,
the range in $\rho_i\kperp$ for which the approximation converges depends on the Hankel grid $p_n=j_n/\vcut$ rather than the velocity space grid.

We plot the error of the approximation \eqref{eq:phiApproximation}
in \fig\ref{fig:phiScan}; results for $\I_2$, $\I_3$ and $\I_4$ are similar.
\Fig\ref{fig:testFunctionVcut6} shows approximations to $\I_1(\rho_i\kperp)$ (itself under the cyan curve)
for various resolutions $N_p$ and $\vcut=6$.
The function $\I_1$ and all approximations to it are positive for all $\kperp$.
The dashed black line indicates the smallest grid point $p_1=j_1/\vcut$, the same for all resolutions.
The coloured vertical indicate the highest grid point $p_{N_p}=j_{N_p}/\vcut$ for each resolution.
Since Bessel roots $j_n$ are approximately linearly spaced, $p_{N_p}$ increases roughly linearly with $N_p$.
The approximation \eqref{eq:phiApproximation} is valid between the lowest and highest grid points, but it also converges below the lowest grid point.
At the highest grid point the approximation tends to zero and this leads to a deterioration before this point. 

This is more clearly seen in the error plots, \fig\ref{fig:PhiErrorVcut4}--\subref{fig:phi_error_negrid_32}.
The error for a given $\vcut$ and $N_p$ has some flat characteristic level. 
However the error increases rapidly as $\rho_i\kperp$ approaches the end of the $p$-grid range, where the error becomes $\O(1)$. 
As before, the error decreases with decreasing grid spacing $\vcut/N_p$. 
But, in addition, the range of convergence in $\rho_i\kperp$ also increases with decreasing $\vcut/N_p$, since the end of the $p$-grid is located at $p_{N_p}=j_{N_p}/\vcut$ which increases roughly linearly with $N_p/\vcut$.

Again, we consider the question of how to choose $\vcut$ given a fixed resolution.
For fixed $N_p$, the error decreases with increasing $\vcut$, provided that the $\rho_i\kperp$ is in the resolution range for all $\vcut$.
For example, in \fig\ref{fig:phi_error_negrid_32} the error decreases with increasing $\vcut$ for $\rho_i\kperp\in(j_1/\vcut,j_{32}/\vcut)$, but increases sharply for $\rho_i\kperp$ above $j_{32}/\vcut$.
Therefore, for fixed resolution, the choice of $\vcut$ depends on the range of $\rho_i\kperp$ to be resolved.
Unfortunately, the point where the error starts to increase does not seem to be simply related to $\vcut$, so must be determined for each choice of $N_p$ and $\vcut$.
However, it is only necessary to consider minimizing the error at the largest $\rho_i\kperp$ to be resolved. 
To see this, suppose we must resolve wavenumbers up to $\rho_i\kperp=15$ with fixed resolution $N_p=32$.
Then from \fig\ref{fig:phi_error_negrid_32} we see it is optimal to choose $\vcut=5$, as this is the value which yields the lowest uniform error.
Since increases in the error occur at the end of the domain, $\vcut=5$ is also the value which minimizes the error at the point $\rho_i\kperp=15$.

\subsection{Free energy conservation}
\label{sec:FreeEnergyConservation}

We now illustrate \sgk's free energy conservation properties.  
Recall from \S\ref{sec:FreeEnergy} that the \gkm\ system conserves free energy $W$ \eqref{eq:FreeEnergySchekochihin} in the absence of driving and dissipation.
The Fourier--Hermite--$\vperp$ spectral representation (\sec\ref{sec:5DCode}) also conserves free energy.
This is a consequence of the orthogonality of the Fourier and Hermite basis functions, and of the fact that the linear operators 
$\L_{\varphi}$, $\L_{A}$ and $\L_{B}$ in the gyrokinetic equation \eqref{eq:SpectralGKEqn}
are each proportional to $F^{\perp}_0(v_n)$.
We may therefore replace the perpendicular velocity space integral operator $\int\d\vperp\ \vperp\ldots$ with its discrete approximation $\sum_{n=1}^{N_p} w_n\ldots$
which preserves all the qualitative properties needed in the derivation in \sec\ref{sec:FreeEnergy}.
The resulting conservation equation is 
\begin{align}
  \label{eq:FreeEnergyEquationChp5}
  \begin{split}
    \fd{}{\tN} 
		\lp \sum_s \lp W_{gs} + W_{\varphi s} \rp \rp
   + {\cal T}
  = 
  {\cal C} + 
  {\cal D},
  \end{split}
\end{align}
where the free energy $W=\sum_s \lp W_{gs} + W_{\varphi s} \rp$ has been written in terms of contributions from the electrostatic potential and distribution function for each species,
\begin{align}
	W_{gs} = \sum_{m,n,\k} n_sT_s\frac{w_n |\gfived_{sm\k}(v_n)|^2}{2F^{\perp}_0(v_n)}
	\hspace{1cm}
	\mathrm{and}
	\hspace{1cm}
	W_{\varphi s} =  \sum_{\k} \frac{n_sq_s^2}{2T_s} \lp 1-\Gamma_{0s}(\kperp) \rp |\varphi_{\k}|^2,
	\label{eq:FreeEnergyContributions}
\end{align}
and the source of free energy from the temperature gradient \eqref{eq:FreeEnergyInjection} is
\begin{align}
  %\label{eq:FreeEnergyInjection}
  \begin{split}
		{\cal T} = \Re\lp \sum_{s,n,\k} n_sT_s\frac{ik_y}{L_T 2\sqrt{2}} w_nJ_0(\rhoNs\kperp v_n)\varphi_{\k}g^*_{s,m=2,\k}(v_n) \rp.
  \end{split}
\end{align}
We have written the collision operator in \eqref{eq:FreeEnergyCollisionalSink} as the sum of a hyperviscous term \eqref{eq:Hyperviscosity}, 
giving dissipation
\begin{align}
	{\cal D} = - \sum_{n,\k,s}\sum_{m=3}^{N_m} \nu_v (\kperp / k_{\perp\max})^8  n_sT_s w_n |\gfived_{sm\k}(v_n)|^2,
	%\label{}
\end{align}
and a hypercollisional term \eqref{eq:HyperColl}, giving dissipation
\begin{align}
	{\cal C} = - \sum_{n,\k,s}\sum_{m=3}^{N_m} \nu_c (m/N_m)^6  n_sT_sw_n |\gfived_{sm\k}(v_n)|^2.
	%\label{}
\end{align}

\begin{figure}[tp]
  \centering
	\subfigure[]{\includegraphics[width=0.49\textwidth]{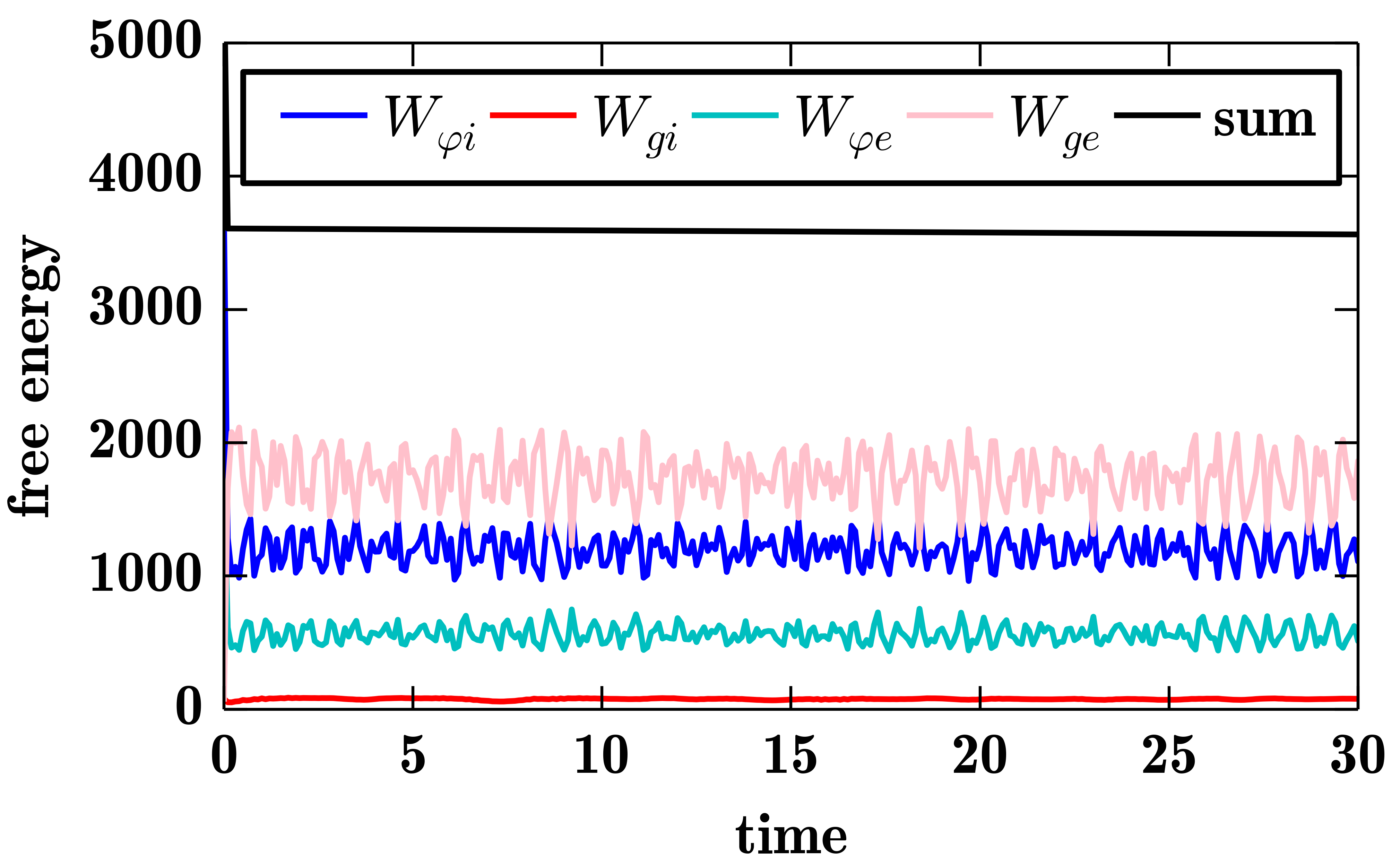}}
	\subfigure[]{\includegraphics[width=0.49\textwidth]{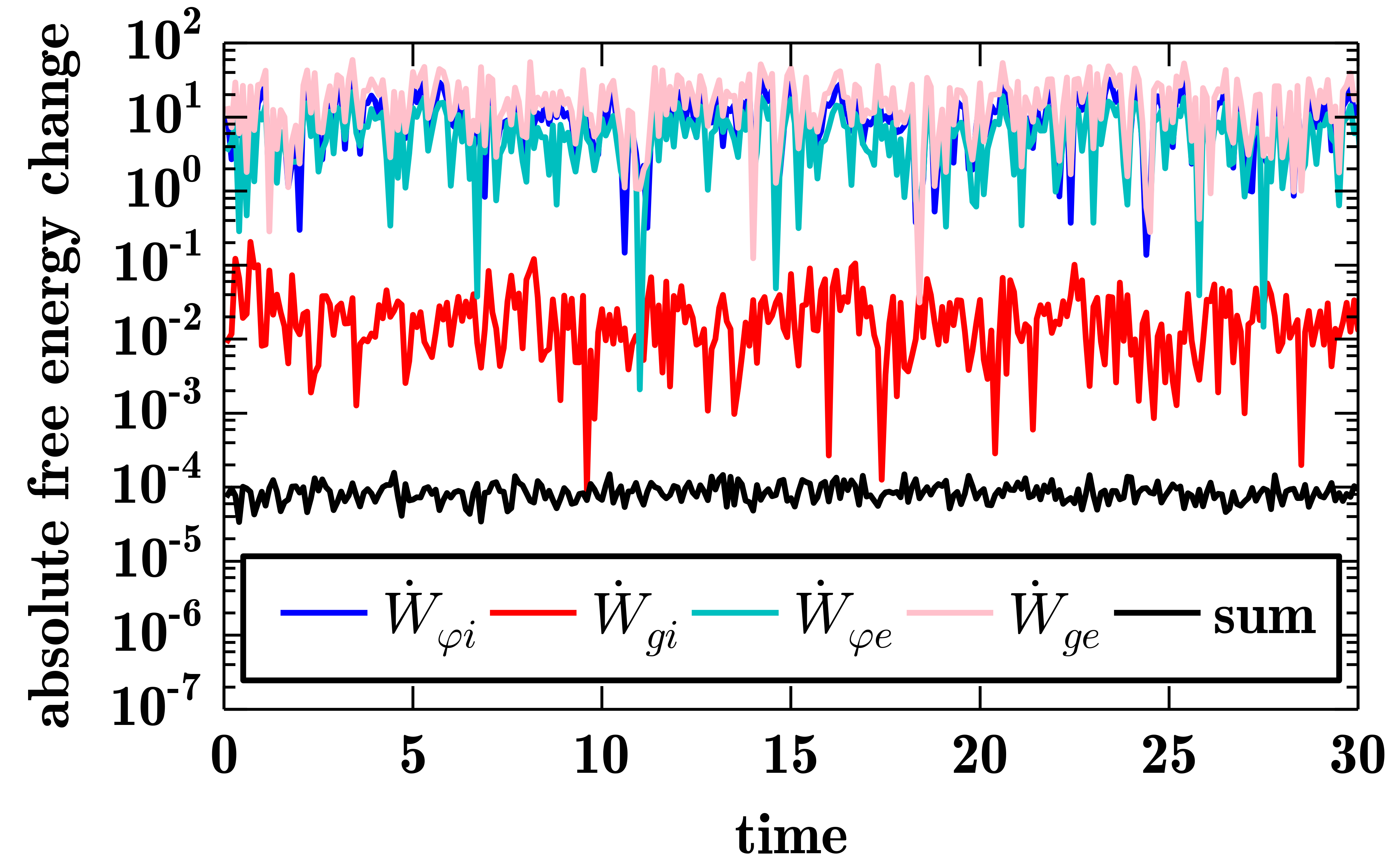}}
	\caption[Free energy conservation in a five-dimensional, two-species, nonlinear, electrostatic gyrokinetic simulation.]{
Free energy conservation in a five-dimensional, two-species, nonlinear, electrostatic gyrokinetic simulation.
		(a) The four contributions to the free energy, and their sum.
		The sum should be constant, but is only approximately constant due to the temporal truncation error.
		(b) Absolute values of the time derivatives of the free energy contributions in (a), and their sum.
		The sum should vanish, but does not due to temporal truncation error.
		\label{fig:FreeEnergyBalanceTwoSpecChp5}
	}
\end{figure}

We first show free energy conservation for a two species ion-electron plasma with mass ratio $m_e/m_i=1/1836$ and no driving (${\cal T}=0$) or dissipation (${\cal C}={\cal D}=0$).
From \eqref{eq:FreeEnergyEquationChp5}, the free energy would be constant in a continuous time formulation.
In \fig\ref{fig:FreeEnergyBalanceTwoSpecChp5}(a) we plot the contributions to the free energy, $W_{gi}$, $W_{ge}$, $W_{\varphi i}$ and $W_{\varphi e}$, and their sum $W$.
The sum $W$ is indeed nearly constant, but decreases slightly due to the dissipative truncation error in the third order Adams--Bashforth timestepping algorithm \cite{Durran99}.
In \fig\ref{fig:FreeEnergyBalanceTwoSpecChp5}(b) we plot the absolute values of the time derivatives of the contributions to the free energy 
$\dot{W}_{gi}$, $\dot{W}_{ge}$, $\dot{W}_{\varphi i}$ and $\dot{W}_{\varphi e}$
(approximated by first-order finite differences), 
and their sum $\dot{W}$, which would vanish in a continuous time formulation.
While the sum $\dot{W}$ does not vanish, it is two orders of magnitude smaller than its smallest contributing term, $\dot{W}_{gi}$.
Moreover, the magnitude of the sum decreases linearly with decreasing timestep, which is consistent with truncation error in the first order finite difference approximation.

\begin{figure}[tp]
  \centering
	\subfigure[]{\includegraphics[width=0.49\textwidth]{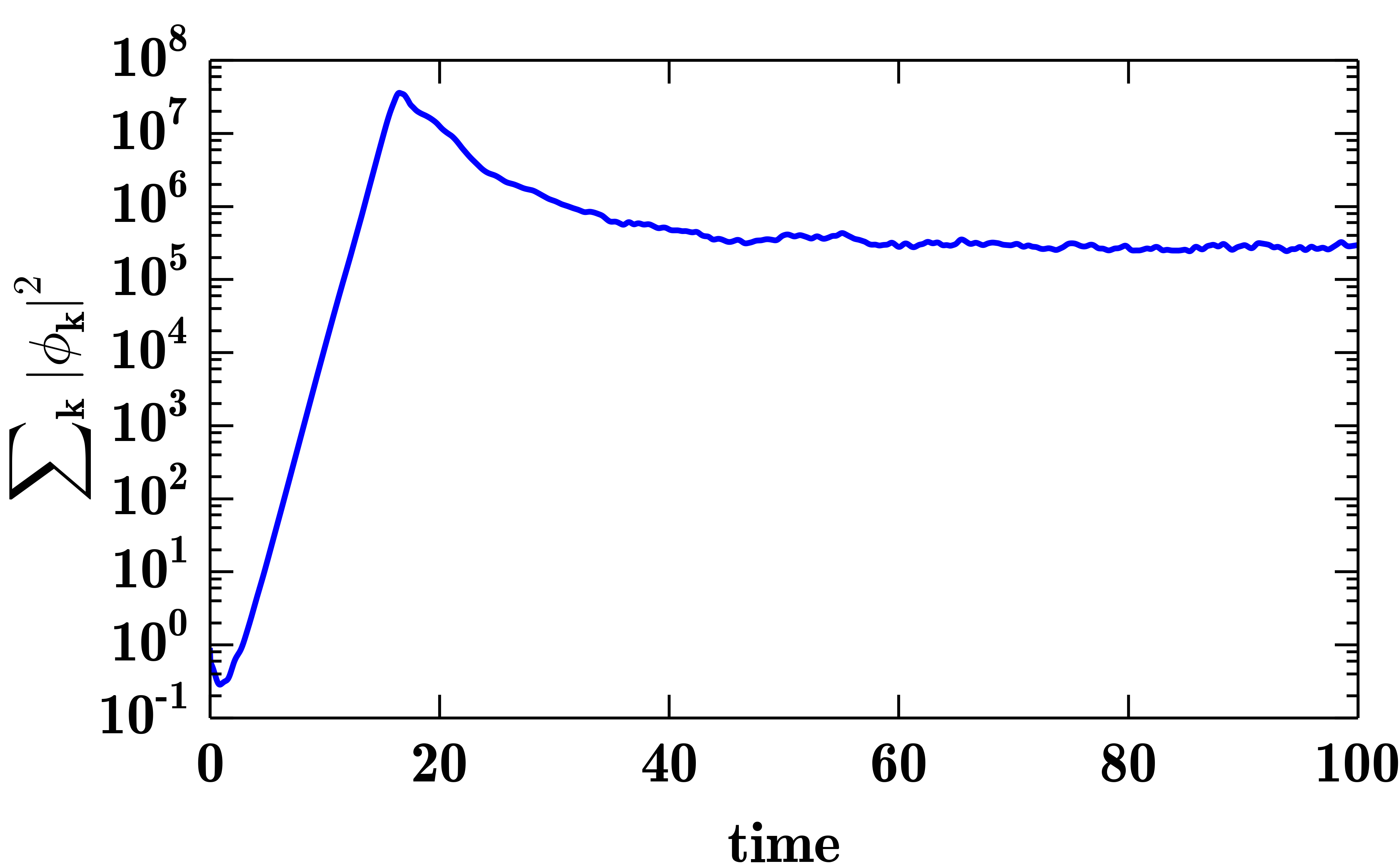}}
	\subfigure[]{\includegraphics[width=0.49\textwidth]{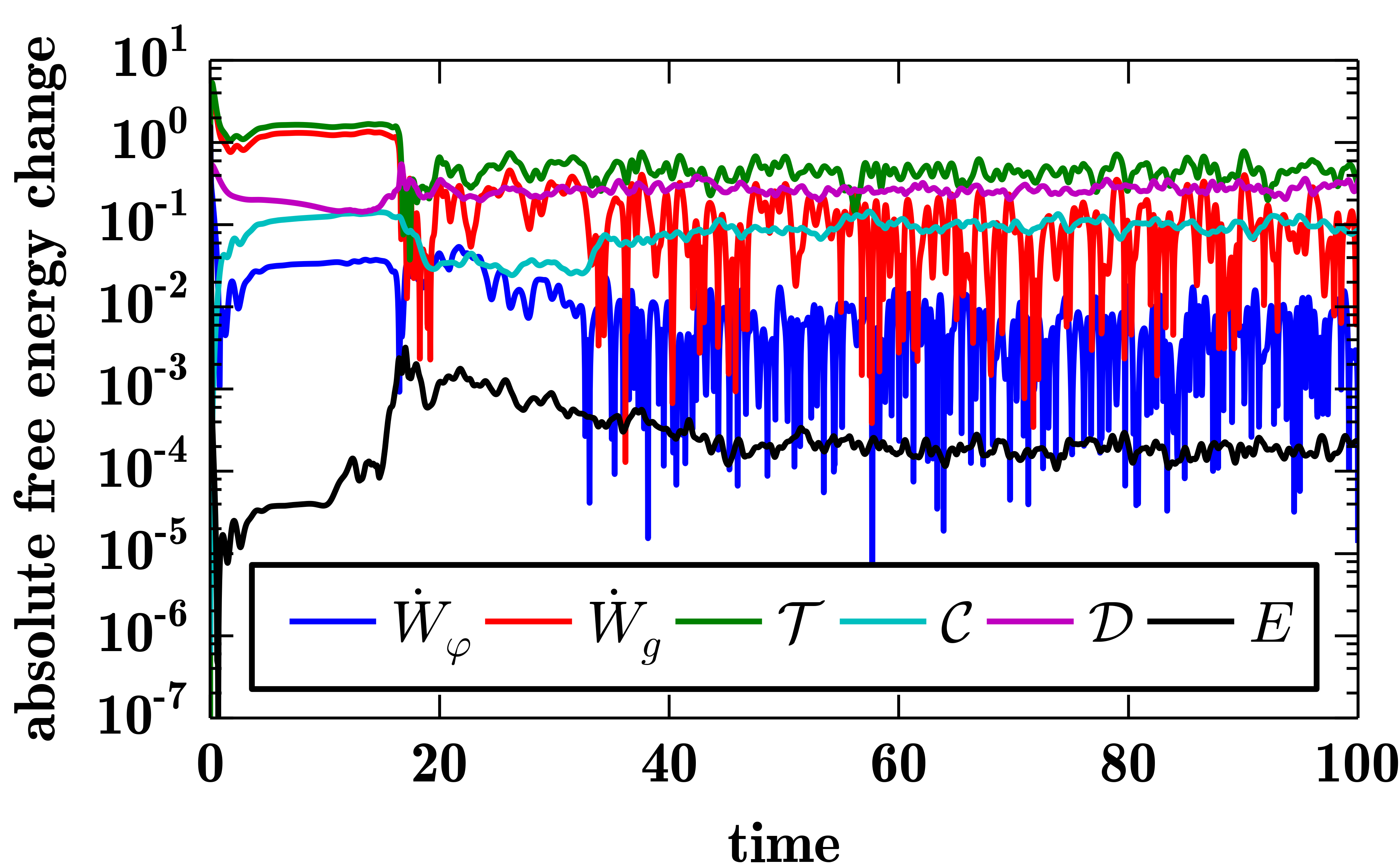}}
	\subfigure[]{\includegraphics[width=0.49\textwidth]{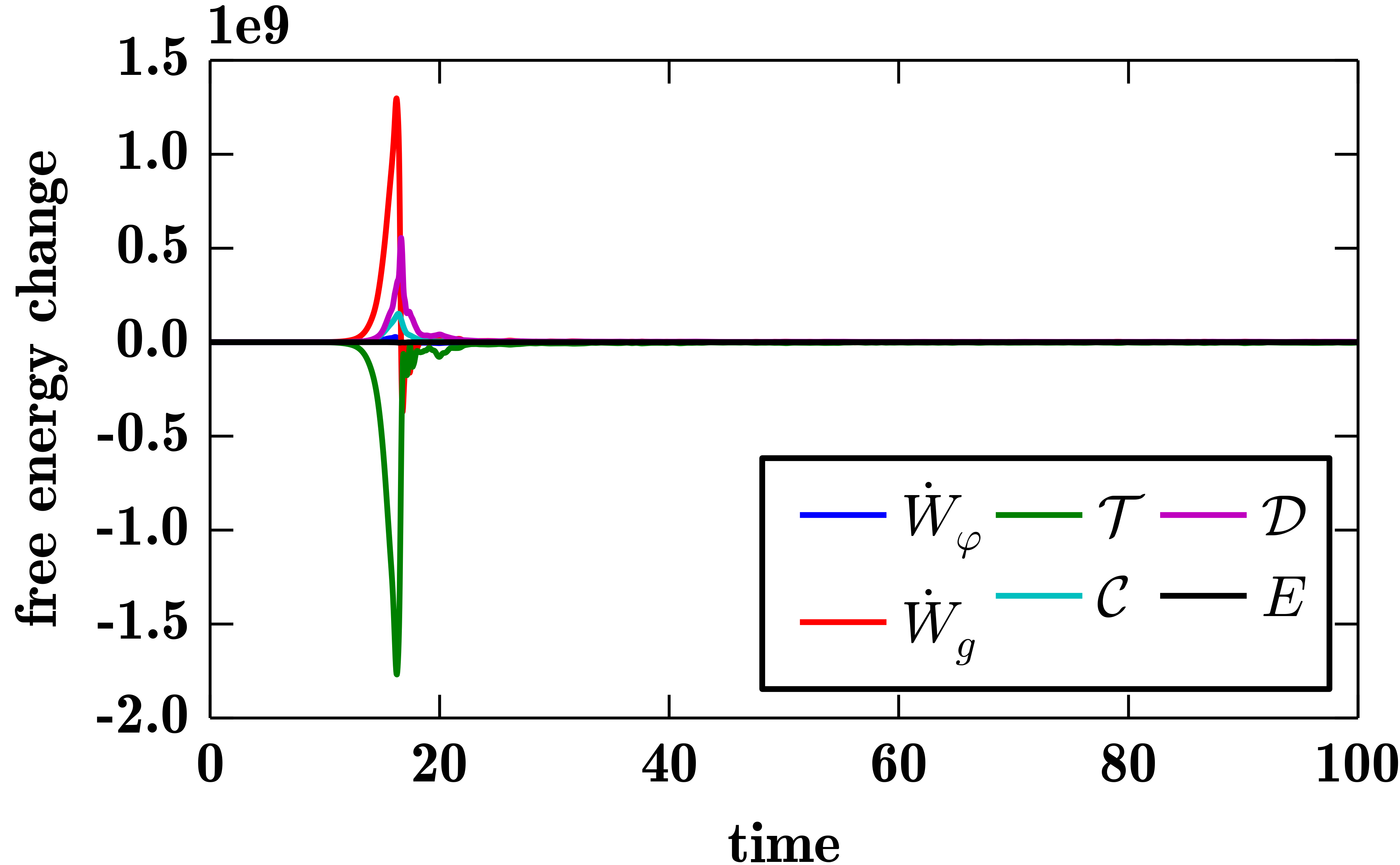}}
	\subfigure[]{\includegraphics[width=0.49\textwidth]{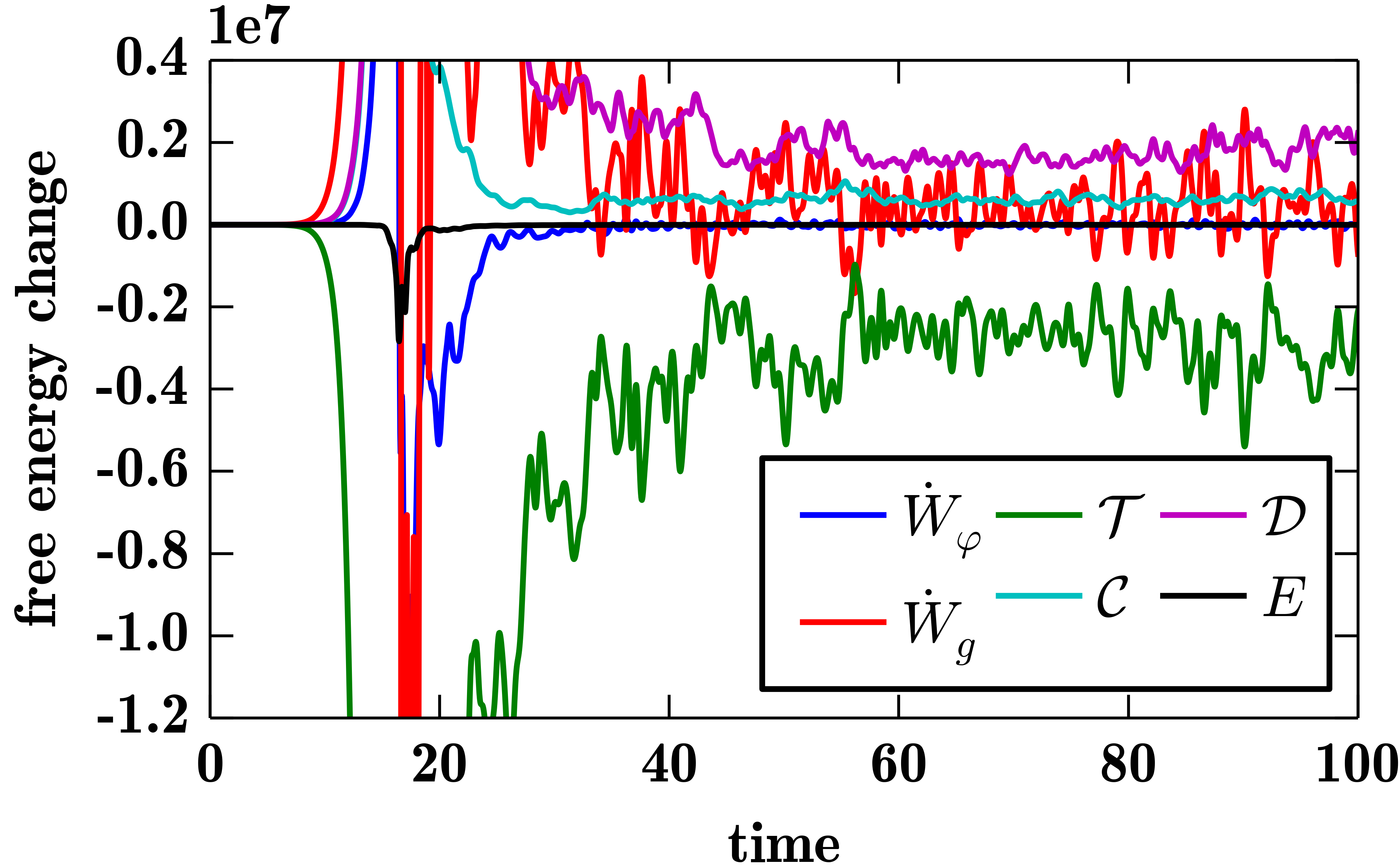}}
	\caption[Free energy conservation in five-dimensional, nonlinear, electrostatic ITG-driven turbulence.]{
Free energy conservation in five-dimensional, nonlinear, electrostatic ITG-driven turbulence.
		(a) Summed electrostatic potential against time.
		(b) Absolute values of time-derivatives of free energy.
		(c) Time-derivatives of free energies on a linear scale.
		(d) A close-up of (c).
		\label{fig:FreeEnergyBalanceChp5}
	}
\end{figure}

We also show free energy conservation for a simulation of nonlinear electrostatic ITG-driven turbulence with kinetic ions and adiabatic electrons.
Turbulence is driven with a temperature gradient $\omega_T=100$, 
and free energy is dissipated by hyperviscosity with coefficient $\nu_v=1$ and by hypercollisionality with coefficient $\nu_c=10$.
Now there is only one species ($s=i$), and to account for the adiabatic electrons, the free energy contribution $W_{\varphi i}$ \eqref{eq:FreeEnergyContributions} becomes
\begin{align}
	W_{\varphi i} =  \sum_{\k} \frac{n_iq_i^2}{2T_i}\lp 1+\frac{T_i}{T_e}-\Gamma_{0i}(\kperp) \rp |\varphi_{\k}|^2,
	\label{eq:FreeEnergyContributionsAdiabatic}
\end{align}
and the free energy balance \eqref{eq:FreeEnergyEquationChp5} becomes
\begin{align}
  \label{eq:FreeEnergyEquationAdiabaticChp5}
  \begin{split}
    \fd{}{\tN} 
		\lp W_{gi} + W_{\varphi i} \rp 
   + {\cal T}
  = 
  {\cal C} + 
  {\cal D},
  \end{split}
\end{align}

In \fig\ref{fig:FreeEnergyBalanceChp5}(a) we plot the electrostatic potential against time,
and in \fig\ref{fig:FreeEnergyBalanceChp5}(b), (c) and (d) we plot the terms in the free energy balance \eqref{eq:FreeEnergyEquationAdiabaticChp5} against time. 
We also plot the error in the free energy balance $E=\dot{W}_{gi}+\dot{W}_{\varphi i}+{\cal T}-{\cal C}-{\cal D}$.
As with \fig\ref{fig:FreeEnergyBalanceTwoSpecChp5}(b), the free energy is not conserved due to the truncation error in the Adams--Bashforth third order timestepping algorithm.
However as before, the error $E$ is two orders of magnitude smaller than its smallest contributing term, and decreases with decreasing timestep.

\subsection{Antenna driving}
\begin{figure}[tp]
  \centering
  \includegraphics[trim=4.8cm 9.4cm 4.2cm 10.12cm,width=0.49\textwidth,clip]{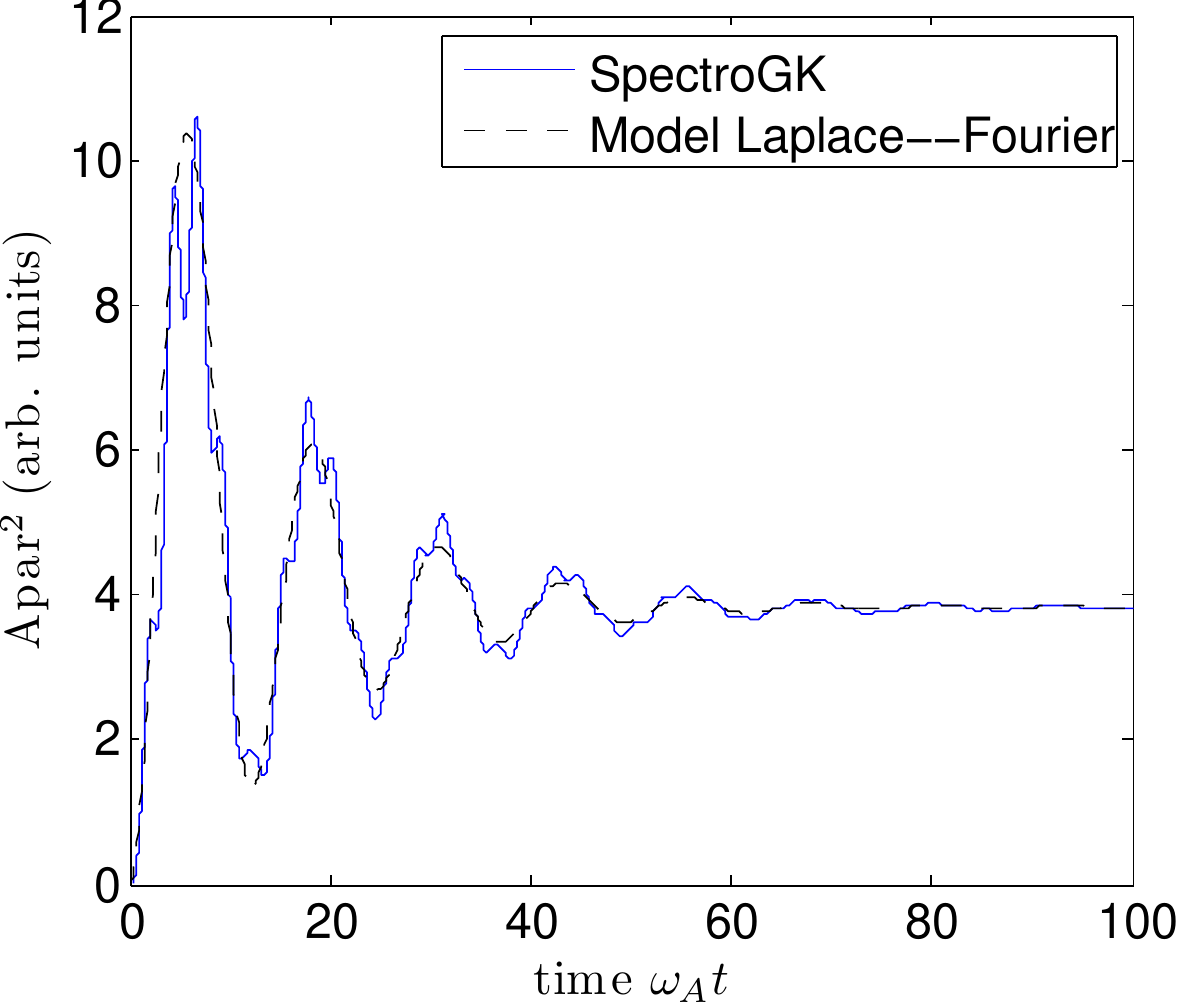}
  \caption{An \alfven\ wave driven by an external parallel current. \label{fig:Antenna}}
\end{figure}

In addition to previous unforced linear waves, we may also drive \alfven\ waves in \sgk\ by adding an external current $A_{\textrm{ext}}=A_{\parallel0}e^{-i\omega_0t}$ to the \lhs\ of the parallel \ampere's law \eqref{eq:AmpPara}.
In \fig\ref{fig:Antenna} we plot the linear system with $\dhBp=0$ driven with the same parameters as the example in \citet[][\fig\ 1]{Numata10}.
We also plot the Laplace--Fourier analytical solution given by \citet{Numata10}.
This is a scenario where \sgk\ significantly outperforms \agk. 
Where \agk\ uses $(N_z,N_{\lambda},N_E)=(32,8,32)=8192$ points to achieve comparable accuracy, \sgk\ uses a single Fourier--Hankel mode and requires around 12 Hermite modes. 

\subsection{Orszag--Tang vortex problem}

\begin{figure}[tp]
  \centering
  %\subfigure[]{\includegraphics[width=0.49\textwidth,trim=0.cm 0.0cm 0.0cm 0.0cm,clip]{images/sgk/AGK_phi_both_128x128x4x1_t31_id16.eps}}
  \subfigure[]{\includegraphics[width=0.49\textwidth,trim=0.cm 0.0cm 0.0cm 0.0cm,clip]{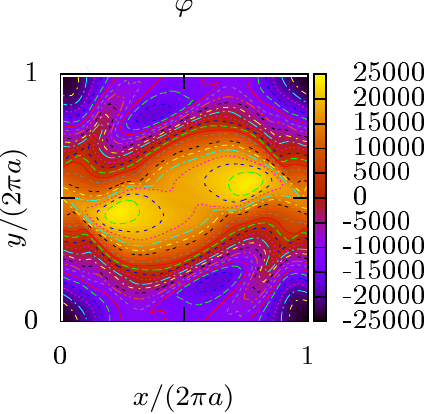}}
  %\subfigure[]{\includegraphics[width=0.49\textwidth,trim=0cm 0.0cm 0.0cm 0.0cm,clip]{images/sgk/SGK_phi_both_128x128x16x1_t31_id17.eps}}
  \subfigure[]{\includegraphics[width=0.49\textwidth,trim=0cm 0.0cm 0.0cm 0.0cm,clip]{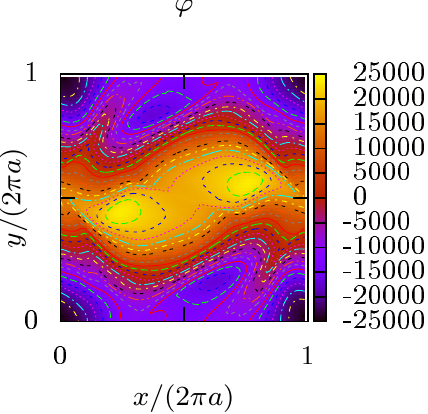}}
  %\subfigure[]{\includegraphics[width=0.49\textwidth,trim=0.0cm 0.0cm 0.0cm 0.0cm,clip]{images/sgk/AGK_apar_both_128x128x4x1_t31_id16.eps}}
  \subfigure[]{\includegraphics[width=0.49\textwidth,trim=0.0cm 0.0cm 0.0cm 0.0cm,clip]{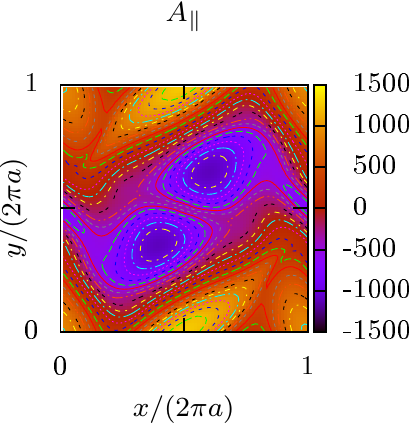}}
  %\subfigure[]{\includegraphics[width=0.49\textwidth,trim=0cm 0.0cm 0.0cm 0.0cm,clip]{images/sgk/SGK_apar_both_128x128x16x1_t31_id17.eps}}
  \subfigure[]{\includegraphics[width=0.49\textwidth,trim=0cm 0.0cm 0.0cm 0.0cm,clip]{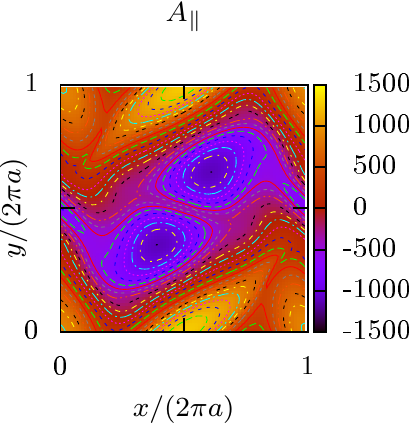}}
    \caption[Contour plots for the Orszag--Tang test problem.]{Contour plots of (a,b) the electrostatic potential $\varphi$ and (c,d) the parallel component of the vector potential $\Apar$ for the Orszag--Tang test problem.  Contours (a,c) were computed by \agk, while (b,d) were computed by \sgk.
			Large contour values are due to \agk's normalization.
      \label{fig:OrszagTang}
}
\end{figure}

To validate \sgk\ in a nonlinear electromagnetic setting, 
we consider the Orszag--Tang vortex problem \citep{OrszagTang79},
a standard test problem for MHD codes.
\citet{OrszagTang79} solved the incompressible reduced MHD equations for flux and stream functions in the plane perpendicular to the mean magnetic field.
\citet{Schekochihin09} showed that in the long wavelength limit $\rho_i\kperp\ll1$, the \gkm\ system with $\Bpar=0$ reduces to incompressible reduced MHD
with a flux function $\varphi/B_0$ and stream function $\Apar$.
Therefore \citet[\S4.5]{Numata10} validated \agk\ by simulating this regime and comparing the results to those from a reduced MHD code.

We repeated this simulation in \sgk\ using the same parameters.
\agk\ and \sgk\ share the same routines for perpendicular Fourier space, and as the parallel direction is neglected, the codes only differ in their perpendicular velocity space implementations---\sgk's discretization on a grid in perpendicular velocity space, 
and \agk's pitch angle-energy-sign discretization (which cannot be separated into parallel and perpendicular velocity).
The simulations are in excellent agreement.
In \fig\ref{fig:OrszagTang} we show the contours of the stream and flux function after one \alfven\ time for \agk\ and \sgk\ simulations.
%%%\begin{align}
%%%  \pd{}{t}\nabla_{\perp}^2\phi  + \left\{ \phi , \nabla_{\perp}^2 \phi\right\} 
%%%  = \frac{1}{\mu_0n_{0i}m_i} + \left\{ \psi , \nabla_{\perp}^2 \psi\right\} + \frac{\mu}{n_{0i}m_i} \nabla^4_{\perp} \phi
%%%  \\
%%%  \pd{\psi}{t}  + \left\{ \phi , \psi\right\} 
%%%  =  \frac{\eta}{\mu} \nabla^2_{\perp} \psi,
%%%  \label{eq:OrszagTang}
%%%\end{align}
%%%where $\mu$ is the viscosity and $\eta$ is the resistivity.

\section{Performance}
\label{sec:Performance}
Finally, we study the performance of \sgk\ measured in timesteps executed per second of wall clock time.
We compare the observed and theoretical complexity of the algorithm in single-processor problems, and determine the weak and strong scalings of the code's parallel performance.
We do this by running nonlinear, collisionless, antenna-driven problems of varying resolution.  
For comparison with \agk, we choose the same problem parameters as in Ref.\ \cite{Numata10}: $\beta=T_{0i}/T_{0e}=n_{0i}/n_{0e}=-q_i/q_e=1$, $m_i/m_e=1836$ and antenna driving at $\rho_i\kperp=1$, the smallest wavenumber in the simulation.

\subsection{Single processor scaling}
\begin{figure}[tb]
  \centering
  \includegraphics[]{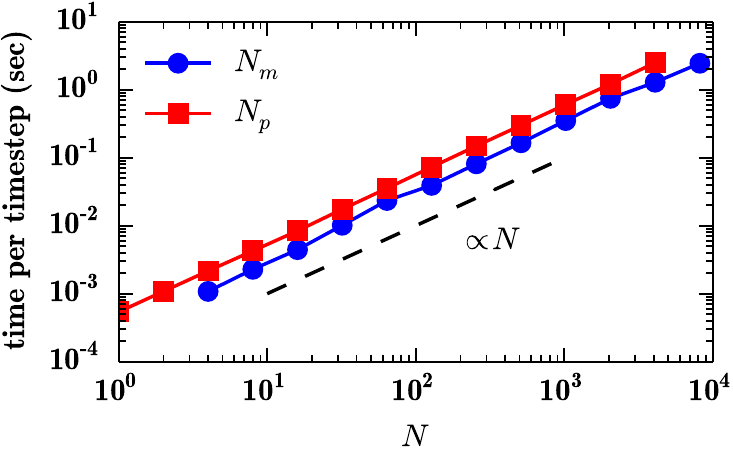}
  \includegraphics[]{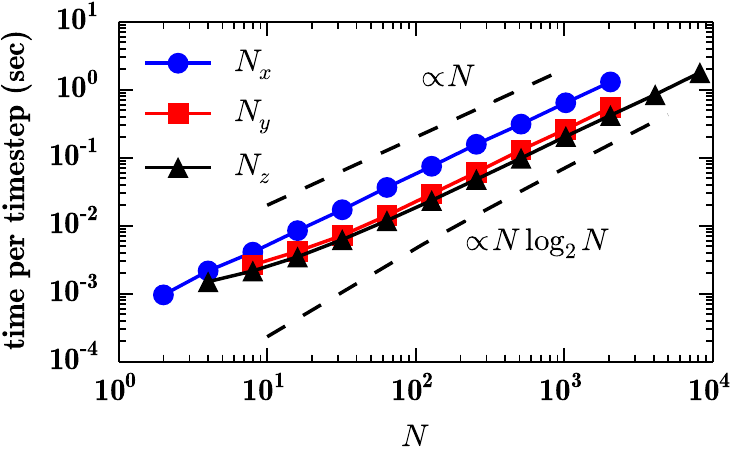}
  \caption[\sgk\ single processor scaling.]{Single processor scaling on a 1.90GHz Intel i3-4030U CPU. The scaling with spatial resolution follows $N\log N$ due to the fast Fourier transforms. The scaling is linear with respect to perpendicular velocity and Hermite space resolution, as expected.   \label{fig:SingleProcessorScaling}}
\end{figure}
We determine the scaling of the \sgk\ algorithm with resolution using small nonlinear runs on a single processor, eliminating any loss in performance due to inter-processor communication.
We take a base resolution of $(N_x,N_y,N_z,N_m,N_p,N_s) = (4,4,8,4,2,2)$ and successively double the resolution of one dimension while keeping the resolution in the other directions fixed. 
The time per timestep is measured and the scalings for each dimension are plotted in \fig\ref{fig:SingleProcessorScaling}.  
The scaling with the spatial resolutions $N_x$, $N_y$ and $N_z$ follow $N\log N$, as expected since the Fast Fourier Transform in the nonlinear term dominates the work.
The algorithm is linear with respect to Hankel and Hermite space, which we observe.

\subsection{Parallel performance}
Parallel performance is measured by a code's agreement with the weak and strong scalings. 
In the weak scaling, the problem size and number of processors are increased simultaneously such that the work done by each processor remains constant.
Ideally, the wall clock time per timestep should remain constant as the number of processors increases.
In the strong scaling, the number of processors is increased but the problem size is held fixed.
Thus the number of points per processor is inversely proportional to the number of processors, and the ideal scaling is a linear decrease in wall clock time per timestep.

\begin{figure}[tp]
  \centering
  %\subfigure[\label{fig:StrongScaling}]{\includegraphics[trim=0.0cm 0.0cm 0.0cm 0.0cm,width=0.49\textwidth,clip]{images/sgk/sgk_strong_scaling.png}}
  %\subfigure[\label{fig:StrongScaling}]{\includegraphics[]{images/sgk/sgk_strong_scaling.eps}}
  \subfigure[\label{fig:StrongScaling}]{\includegraphics[]{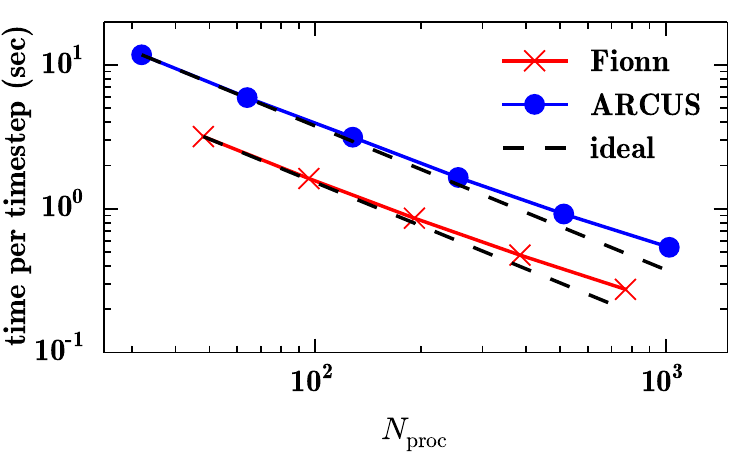}}
  %\subfigure[\label{fig:WeakScaling}]{\includegraphics[]{images/sgk/weak_scaling.eps}}
  \subfigure[\label{fig:WeakScaling}]{\includegraphics[]{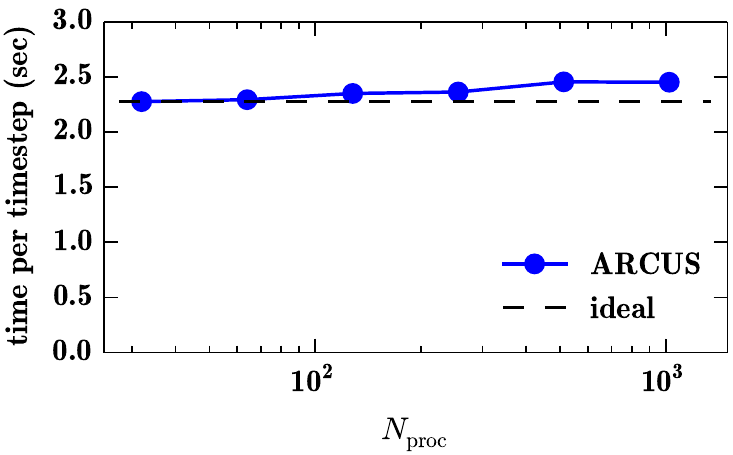}}
  \caption[\sgk\ strong and weak scalings.]{(a) \sgk\ strong scaling for the 2.4GHz Intel Ivy Bridge Xeon processors on the Fionn machine at the Irish Centre for High-End Computing (ICHEC) and for the 2.0GHz Intel Sandy Bridge Xeon processors on the ARCUS machine at the University of Oxford Advanced Research Computing (ARC) facility, both using Infiniband interconnects. (b) Weak scaling on the ARCUS machine.
    \label{fig:Scaling}} 
\end{figure}

The weak and strong scalings for \sgk\ are shown in \fig\ref{fig:Scaling}.
Both scalings are reasonably good, but the performance degrades as the number of cores approaches $10^3$.

\part{Results}
\label{sec:PartResults}
\renewcommand{\B}{\boldsymbol{B}}
\newcommand{\C}{{\cal C}}
\renewcommand{\d}{\mathrm{d}}
\newcommand{\dTpara}{\delta T_{\parallel}}
\newcommand{\dTparahat}{\delta \hat{T}_{\parallel}}
\newcommand{\Ds}{D_{\textrm{s}}}
\newcommand{\Dnl}{D_{\textrm{nl}}}
\renewcommand{\E}{\boldsymbol{E}}
\renewcommand{\g}{\hat{g}_{m}}
\newcommand{\gp}{\hat{g}_{m+1}}
\newcommand{\gm}{\hat{g}_{m-1}}
\renewcommand{\ga}[2]{\left<#1\right>_{#2}}
\renewcommand{\I}{{\cal I}}
\renewcommand{\k}{\boldsymbol{k}}
\renewcommand{\kpara}{k_{\parallel}}
\newcommand{\ppara}{p_{\parallel}}
\newcommand{\qpara}{q_{\parallel}}
\renewcommand{\kperp}{k_{\perp}}
\renewcommand{\kperpv}{\k_{\perp}}
\renewcommand{\mNs}{m_{s}}
\newcommand{\mcut}{m_{\mathrm{cut}}}
\renewcommand{\N}{{\cal N}}
\renewcommand{\qNs}{q_{s}}
\renewcommand{\R}{\boldsymbol{R}}
\renewcommand{\r}{\boldsymbol{r}}
\newcommand{\reference}{Ref.~}
\renewcommand{\rhs}{right-hand side}
\newcommand{\rperp}{\r_{\perp}}
\newcommand{\Source}{{\cal S}}
\renewcommand{\sgn}{{\mathrm{sgn}}}
\renewcommand{\stm}{\sqrt{\frac{T_{0s}}{m_s}}}
\newcommand{\T}{{\cal T}}
\newcommand{\taus}{\tau_{\mathrm{s}}}
\newcommand{\tausi}{\tau_{\mathrm{s}}^{-1}}
\newcommand{\taunl}{\tau_{\mathrm{nl}}}
\newcommand{\taunli}{\tau_{\mathrm{nl}}^{-1}}
\renewcommand{\tfd}[2]{\mathrm{d}#1/{\mathrm{d}#2}}
\renewcommand{\tpd}[2]{\partial#1/{\partial#2}}
\renewcommand{\TNs}{T_{0s}}
\renewcommand{\u}{\boldsymbol{u}}
\newcommand{\uperp}{\boldsymbol{u}_{\perp}}
\newcommand{\uperphat}{\hat{\boldsymbol{u}}_{\perp}}
\newcommand{\upara}{u_{\parallel}}
\newcommand{\uparahat}{\hat{u}_{\parallel}}
\renewcommand{\vpara}{v_{\parallel}}
\newcommand{\vparahat}{\hat{v}_{\parallel}}
\renewcommand{\vperp}{v_{\perp}}
\renewcommand{\vperpv}{\boldsymbol{v}_{\perp}}
\renewcommand{\vth}{v_{\mathrm{th}}}
\renewcommand{\vthi}{v_{\mathrm{th}i}}
\renewcommand{\z}{\boldsymbol{z}}

\newcommand{\deez}{\delta z}
\newcommand{\deev}{\delta \vpara}
\newcommand{\ellc}{\delta z^*}
\renewcommand{\ellc}{\taunli\sqrt{m}}
\renewcommand{\ellc}{\taunl/\sqrt{m}}
\newcommand{\pcl}{\delta z^*}

\newcommand{\Wfluid}{W_{\textrm{fluid}}}
\newcommand{\Wkinetic}{W_{\textrm{kin}}}
\renewcommand{\ta}[1]{\left\langle#1\right\rangle}
\newcommand{\indthree}{{\cal I}_{m\geq3}}

%\chapter{Free energy flow and dissipation in saturated ITG turbulence}
\chapter[Suppression of Landau damping in saturated ITG-driven turbulence]{Suppression of Landau damping in saturated ion temperature gradient driven turbulence}
\label{sec:FreeEnergyFlowAndDissipation}

\emph{This chapter is based on \citet{Parker15ITG}.}

\section{Introduction}%
%
%EGH 
%Turbulence can 
%Gyrokinetic turbulence (in three spatial dimensions) can be thought of as
%a mechanism by which turbulent
%free energy injected 
%at a given scale 
%(that is, a given point in phase space)
%by, for example, an instability, 
%is transferred to some other point in phase space where this 
%turbulent free energy can dissipated (thermalized) by particle collisions.
%can be thought of as a combination of two separate processes:
%
Gyrokinetic turbulence (turbulence in weakly collisional, strongly magnetized plasmas) is a phenomenon
which is widely studied, owing to its ubiquitous occurrence both in magnetic-confinement-fusion experiments \cite{Conner94,Doyle07,Garbet10}
and in astrophysical settings \cite{Howes06,Schekochihin09}.
Like incompressible Navier--Stokes turbulence, gyrokinetic turbulence may be described as the injection, cascade to small scales, and dissipation of some quadratic invariant (in gyrokinetics, free energy).
%Like all turbulence, gyrokinetic turbulence is composed of a set of interacting fluctuations which have the effect of
%transferring the free energy injected by some kind of large scale stirring (by, for example, an instability of the plasma like the ion temperature gradient (ITG) instability \cite{steve, others}) to a smaller scale
%
%Like all turbulence, gyrokinetic turbulence is composed of three things: a mechanism for injecting free energy
%into the plasma by some kind of stirring,
%%(by, for example, a plasma instability), 
%a mechanism for dissipating this energy into heat,
%and, if there is a scale separation between the two processes,
%a mechanism for transferring this energy between them \cite{scheckochintome}.
%%The first of these can be, for example, an instability of the plasma 
%%like the ion temperature gradient (ITG) instability.
%%The second of these is typically collisions between particles.
%%%(typically by collisions between particles)
%%%energy from 
%%
%Where the injection of energy occurs on spatial scales greater than the ion Larmor radius,
%%two processes are at work withi
%%gyrokinetic turbulence can be 
%two separate processes can be identified within this transfer mechanism:
On spatial scales larger than the ion Larmor radius, 
gyrokinetic turbulence
%(in three spatial dimensions) 
%contains two cascades of turbulent energy.
incorporates two mechanisms for dissipating into heat
the free energy injected into the system (\eg, by a plasma instability).
The first is a fluid-like nonlinear cascade which transfers free energy from large spatial scales
(the injection, or outer, scales) to 
smaller, sub-Larmor, spatial scales 
(where the free energy is dissipated 
eventually
by collisions \cite{Schekochihin09,Tatsuno09}).
The second is
parallel phase mixing,
the linear process 
studied in \chp\ \ref{sec:ParallelVelocitySpace}
which transfers free energy from the fluid moments (density, parallel flow and temperature) 
to the kinetic moments
by creating perturbations to the velocity distribution
on ever finer scales (a velocity space cascade),
%of particle velocities
%on ever finer scales in velocity space, 
%parallel velocity scales where it is 
perturbations which are 
dissipated either by collisions, or
%in a collisionless plasma,
by Landau damping \cite{Hammett92,Watanabe06}.
In a linear plasma, the  transfer and dissipation of free energy occurs at a constant rate which is independent of collision frequency \cite{Zocco11,Kanekar14}.

%%%The macroscopic properties of the turbulence (like transport) are directly affected by the interaction of these two cascades.
%%%%To understand macroscopic properties like heat fluxes and collisional dissipation, it is vital to understand the interaction of the two cascades.
%%%In the linearized gyrokinetic system, parallel phase mixing yields a velocity-space cascade with a shallow spectrum that dissipates free energy at a rate independent of collisions \cite{Zocco11,Kanekar14}.
%%%However, \citet{Howes08} showed that imposing this linear damping rate on the nonlinear system
%%%leads to an exponentially-decaying spatial spectrum,
%%%not the empirically-observed algebraic spectrum;
%%%to derive that spectrum, \citet{BarnesEtal11} neglected parallel phase mixing entirely.
%%%However \citet{HatchEtAl13} showed that a significant proportion of injected free energy does cascade %and dissipate 
%%%in velocity space,
%%%albeit with a steeper spectrum than in the linear case.
%%%These observations suggest a complicated relationship between parallel phase mixing and the nonlinear cascade;
%%%there is as yet no complete picture of free energy flow and dissipation in phase space.

The macroscopic properties of the turbulence (like transport) are directly affected by the interaction of these two cascades;
yet while each is understood in isolation, how they interact is not clear.  
Recent work reveals some disquieting observations.
Firstly, the fluid-like theory for the nonlinear cascade by \citeauthor{BarnesEtal11} \cite{BarnesEtal11} is in excellent agreement with the empirically-observed power law spectra of the electrostatic potential, 
but in its derivation it is necessary to neglect free energy transfer by phase mixing, contrary to the expected behaviour from linear plasma kinetic theory.
Attempts to include the constant flux of free energy into velocity space lead to incorrect, exponentially-decaying spectra, as illustrated by \citet{Howes08} and \citet{Podesta10} (while discussing a different problem).
However, a significant proportion of injected free energy does indeed cascade and dissipate 
in velocity space \cite{Watanabe06},
albeit with a slower transfer rate than in the linear case,
and with a dissipation rate that depends on collision frequency \cite{Hatch13}.
These observations suggest a complicated relationship between parallel phase mixing and the nonlinear cascade;
there is as yet no complete picture of free energy flow and dissipation in phase space.

In this \chp, 
we provide this picture by describing the interaction between phase mixing and the nonlinear cascade
in electrostatic drift kinetic turbulence, a convenient limit of gyrokinetics.
We show that the net transfer of free energy from fluid to kinetic modes is strongly inhibited in a turbulent plasma, relative to a linear plasma. 
This is due to a stochastic version of the plasma echo \cite{Gould67,Malmberg68} where the nonlinearity excites ``anti-phase-mixing'' modes
which transfer free energy from small to large scales, leading to statistical cancellation of free energy transfer.
The significance of this effect depends on the relative sizes of the phase-mixing and nonlinear terms,
and we identify regions of wavenumber space where either the echo effect dominates, or where there is phase-mixing at the usual linear rate.
The plasma's energy-containing scales lie within the former echo-dominated region.
Therefore there is very little net transfer of free energy to fine velocity-space scales via linear phase-mixing,
and consequently Landau damping is strongly suppressed as a dissipation mechanism.

\section{Drift kinetics}%
  We study electrostatic ion-temperature-gradient driven drift kinetic turbulence in an unsheared slab with kinetic ions and adiabatic electrons.
	The drift kinetic equation for ions is 
\begin{equation}
  \begin{split}
		%\fd{g}{t}
		 \pd{g}{t} 
		+ \vpara\nabla_{\parallel} \left( g + \varphi F_0 \right)
		+ \uperp\cdot\nabla_\perp g
		= C[g] + \chi.
    \label{eq:GyrokineticEquation}
  \end{split}
\end{equation}
Here $g=n_i^{-1}\int\d^2\vperpv~\delta f$ is the perturbed distribution function integrated over perpendicular velocity space, 
with $n_i$ the background ion density;
$\varphi=Ze\phi/T_i$ with electrostatic potential $\phi$, electron charge $e$, ratio of ion to electron charge $Z$, and background ion temperature $T_i$;
$F_0(\vpara/\vth)=e^{-\vpara^2/\vth^2}/\sqrt{\pi}$ is the one-dimensional Maxwellian, with $\vth$ the ion thermal velocity;
$\uperp=(\rho_i\vth/2)\hat{\z}\times\nabla_{\perp}\varphi$ is the $\E\times\B$ velocity with ion gyroradius $\rho_i$, and $\hat{\z}$ the unit vector in the direction of the magnetic field line;
the perpendicular directions are denoted $x$ and $y$.
The source term from the ion temperature gradient in the negative $x$-direction is 
\begin{equation}
  \begin{split}
		\chi=-\frac{\rho_i\vth}{2L_T}\pd{\varphi}{y}\left( \frac{\vpara^2}{\vth^2}-\frac{1}{2}\right) F_0,
    \label{eq:Chi}
  \end{split}
\end{equation}
where $1/L_T=-\tfd{\log T_i}{x}$ is the constant imposed macroscopic temperature gradient.
Particle collisions $C[g]$ are described shortly.
The electrostatic potential is given by the quasineutrality condition
\begin{equation}
  \begin{split}
		\varphi = \alpha \int_{-\infty}^{\infty}\d\vpara ~  g , 
		~ ~ ~ ~ ~ 
		\alpha = ZT_e / T_i . 
    \label{eq:Quasineutrality}
  \end{split}
\end{equation}

The drift kinetic system conserves free energy 
\begin{align}
	W = \int\d^3\r\ \frac{\varphi^2}{2\alpha} + \int\d^3\r \int_{-\infty}^{\infty}\d\vpara ~ \frac{g^2}{2F_0},
%	\label{<++>}
\end{align}
except for injection by the ion temperature gradient and dissipation by collisions:
\begin{equation}
  \begin{split}
		\fd{W}{t} = \int\d^3\r\int_{-\infty}^{\infty}\! \d \vpara ~ \frac{g\chi}{F_0} + \int\d^3\r\int_{-\infty}^{\infty}\!\d\vpara~ \frac{gC[g]}{F_0}.
    \label{eq:TotalFreeEnergyConservation}
  \end{split}
\end{equation}

\section{Simulations}%
  We study the saturated nonlinear state of drift kinetic turbulence.
We solve equations (\ref{eq:GyrokineticEquation}--\ref{eq:Quasineutrality}) with {\sc SpectroGK},  %\cite{SGK}, a spectral code which exactly captures  free energy transfers.
which was described in \chp\ \ref{sec:SpectroGK}.
We use a Fourier representation in space, with 256 parallel wavenumbers, and 128 wavenumbers in each perpendicular direction, with 8th order hyperviscosity \eqref{eq:Hyperviscosity} cutting off the distribution function at $\kperp\rho_i\sim1$.
  In parallel velocity space, we use the Hermite representation 
	$g(\vpara)=\sum_{m=0}^{\infty}g_mH_m(\vparahat)F_0(\vparahat)/\sqrt{2^mm!}$ 
  for Hermite polynomials 
	$H_m(\vparahat)=e^{\vparahat^2}(-\mathrm{d}/\mathrm{d}\vparahat)^me^{-\vparahat^2}$
	with ${\vparahat} = \vpara/\vth$.
%We use Kirkwood's momentum- and energy-conserving version of the Lenard--Bernstein operator \cite{Kirkwood46,Lenard58} 
%with collision frequency $\nu$ and the indicator function $I_3=1$ for $m\geq3$ and $I_3=0$ otherwise.
%Hypercollisionality and hyperviscosity are controlled by the parameters $\nu_c$, $\nu_v$ respectively.
%For dissipation and dealiasing in the parallel spatial direction, we use the Hou--Li filter \cite{Hou07}.
%The basis functions of the spectral representation are eigenfunctions of all dissipation operators.
	For large $m$, Hermite polynomials behave like $H_mF_0^{1/2}/\sqrt{2^mm!} \sim \cos(\vparahat\sqrt{2m}-m\pi/2)$
	so that each $m$ represents a scale in velocity space.
Low $m$ represent ``fluid'' quantities:
$g_0=\varphi/\alpha$, 
$g_1=\sqrt{2}\upara/\vth$,
$g_2=\dTpara/T_i\sqrt{2}$,
where $\upara$ is the parallel bulk velocity and $\dTpara$ is the perturbed parallel temperature.
In contrast, $g_m$ for $m\geq3$ are ``kinetic'' modes representing finer scales in velocity space.
For collisions we use the Kirkwood operator %\cite{Kirkwood46} 
(the momentum and energy conserving version of the Lenard--Bernstein operator,   %\cite{LenardBernstein}
see \S\ref{sec:lb}
):
$C[g_m]=-m\nu g_m\indthree$, where $\indthree=1$ if $m\geq3$ and is zero otherwise.
In these simulations we use 256 Hermite modes,
regularizing with 6th order hypercollisions $C[g_m]=-\nu_hm^6g_m\indthree$.
As the linear growth rate in drift kinetics increases indefinitely with increasing $\kperp$, we damp the temperature gradient term by a factor $\sim e^{-100(\kperp/k_{\perp\max})^2}$ to separate the free energy injection and dissipation scales. %\alpha ~ 1000
Consequently, we use very large temperature gradients, here presenting results for $R/L_T=1600$.
However due to the exponential factor this is representative of much smaller, but still above marginal, temperature gradients.

We solve fully spectrally, but for presentation consider equations which are spectral in the parallel directions only.
The drift kinetic equation \eqref{eq:GyrokineticEquation} becomes
\begin{equation}
  \begin{split}
  \pd{\g}{t} 
	+ i\kpara\vth \left(\sqrt{\frac{m+1}{2}}\gp + \sqrt{\frac{m}{2}} \gm \right)
	+ \sum_{\ppara+\qpara=\kpara} \hat{\u}_{\perp}(\ppara)\cdot\nabla_{\perp}\g(\qpara)
	\\
	+ \frac{i\kpara\vth}{\sqrt{2}}\hat{\varphi} \delta_{m1}
    = 
    -\nu m \g\indthree  + \hat{\chi},
    \label{eq:GyrokineticEquationFourierHermite}
  \end{split}
\end{equation}
where 
%%%$g_m=g_m(\rperp,\kpara)$,
%%%$\hat{\varphi}=\hat{\varphi}(\rperp,\kpara)$,
%%%$\hat{\u}_{\perp}=(\rho_i\vth/2)\hat{\z}\times\nabla_{\perp}\hat{\varphi}$,
$\hat{\chi}=-(\rho_i\vth/2\sqrt{2}L_T)(\tpd{\hat{\varphi}}{y})\delta_{m2}$, 
$\delta$ is the Kronecker delta, 
and a hat denotes a function in $(\rperp,\kpara)$ space.
Equation \eqref{eq:GyrokineticEquationFourierHermite} neatly contains the two cascades present in the turbulence: the fluid cascade due to the nonlinearity $(\uperphat\cdot\nabla_{\perp}\hat{g})$,
and linear streaming (the mode coupling in $m$). 

\section{Free energy transfer}%
We first study the transfer of free energy between fluid and kinetic modes.
We write the free energy as 
\begin{align}
	W=\int\d^2\rperp\sum_{k\parallel}(\Wfluid+\Wkinetic),
%	\label{}
\end{align}
where 
\begin{align}
	\Wfluid= \frac{(1+\alpha)|\hat{\varphi}|^2}{2\alpha^2} + \frac{|\uparahat|^2}{\vth^2} + \frac{|\dTparahat|^2}{4T_i^2},
	%\label{}
\end{align}
is the free energy contained in fluid modes
and 
\begin{align}
	\Wkinetic=\frac{1}{2}\sum_{m=3}^{\infty}|\g|^2 ,
%	\label{}
\end{align}
is the free energy in kinetic modes.
To derive evolution equations for these, we multiply \eqref{eq:GyrokineticEquationFourierHermite} by $\g^*$ (the complex conjugate of $\g$) 
and add the resulting equation to its complex conjugate to obtain 
\begin{equation}
  \begin{split}
	\fd{}{t} \left( \frac{|\g|^2}{2}\right)
	+  \Gamma_m - \Gamma_{m-1} 
	+	\Im\left( \frac{\kpara\vth}{\sqrt{2}}\g^*\hat{\varphi}\delta_{m1}\right)
    = 
    -\nu m |\g|^2\indthree  + \Re \left(\g^*\hat{\chi}\right) ,
    \label{eq:FreeEnergyModeByMode}
  \end{split}
\end{equation}
where $\tfd{}{t}$ is the convective derivative $\tpd{}{t}+\uperp\cdot\nabla_{\perp}$ expressed in $(\rperp,\kpara)$ space,
and $\Gamma_m=\kpara\vth\sqrt{(m+1)/2}\ \Im(\gp^*\g)$ is the free energy transfer from mode $m$ to $m+1$ \cite{Watanabe04}.

Summing \eqref{eq:FreeEnergyModeByMode} separately for $m=0,1,2$ and $m\geq3$, we obtain 
\begin{equation}
  \begin{split}
		\fd{\Wfluid}{t} = \Source - \T, \hspace{1cm}
		\fd{\Wkinetic}{t} = \T - \C,
    \label{eq:FreeEnergyConservation}
  \end{split}
\end{equation}
where 
$\Source=\Re\left(\dTparahat^* \hat{u}_x\right)/2T_iL_T$ is the free energy source due to the temperature gradient
and
$\C=\nu\sum_{m=3}^{\infty}m|\g|^2$ is the free energy sink due to collisions. % (or Landau damping as $\nu\to0^+$),
%and 
The term 
$\T=\kpara\vth\sqrt{3}\ \Im\left(\hat{g}^*_3\dTparahat\right)/2T_i =\Gamma_2$
is the transfer of free energy from fluid to kinetic modes
%Only $\T$ transfers free energy between fluid and kinetic modes.
%The transfer $\T$ is 
due to streaming $\vpara\nabla_{\parallel}g$ in \eqref{eq:GyrokineticEquation}.
This is the only effect which transfers free energy between fluid and kinetic modes.
Streaming is linear and reversible, so $\T$ may be positive or negative; however analytic theory for linear Landau damping, which sets $\T\equiv\T_L=|\kpara|\vth\sqrt{3/2}|\hat{g}_2|^2$ (\ie\ $\Gamma_m\equiv\Gamma^L_m=|\kpara|\vth\sqrt{(m+1)/2}|\g|^2$), is in excellent agreement with numerically calculated spectra (see \cite{Zocco11,Kanekar14}, or \chp\ \ref{sec:ParallelVelocitySpace}). 

\begin{figure}
  \centering
  \includegraphics[]{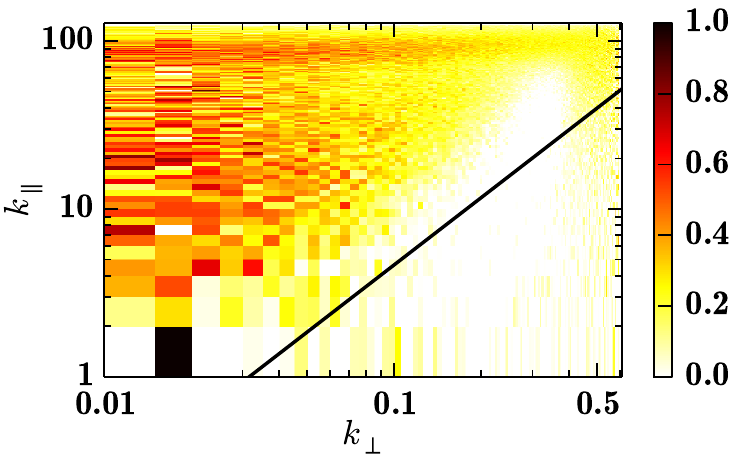}
	\caption[Free energy transfer from fluid to kinetic modes in saturated turbulence, normalized to its value in a linear plasma.]{%}
	\label{fig:FluxSuppressionM2} Free energy transfer from fluid to kinetic modes in saturated turbulence, normalized to its value in a linear plasma, $\ta{\T}/\T_L$.
		The diagonal black line marks the critical balance determined from $\varphi$.
		Free energy transfer to fine velocity-space scales is strongly suppressed and hence the turbulence is fluid-like.
  }
\end{figure}
We now consider saturated plasma turbulence.
Here the time average (denoted $\ta{\cdot}$) of the free energy is constant, $\ta{\tfd{\Wfluid}{t}}=\ta{\tfd{\Wkinetic}{t}}=0$,
so that $\ta{\T}=\ta{\Source}=\ta{\C}\geq0$. 
For phase mixing (and Landau damping) to play a similar role in turbulent plasma as in the linear case, one would expect $\T$ to be similar to the value from the linear case $\ta{\T}\approx {\T_L}$.
In \fig\ref{fig:FluxSuppressionM2} we plot the ratio $\ta{\T}/\T_L$ for saturated drift kinetic turbulence. 
Across all physical scales, the transfer is strongly suppressed from its value in the linear case.
Moreover, across a large range of physical scales, the transfer is completely suppressed, $\ta{\T}\approx 0$.
These have two important consequences.
Firstly, the fluid and kinetic modes are, statistically, very nearly energetically decoupled. 
Secondly, collisional dissipation $\ta{\C}=\ta{\T}$ is also strongly suppressed, so collisions are far less effective as a dissipation mechanism.

%\newsection{Forward and backward propagating modes}%
\section{Phase-mixing and anti-phase-mixing modes}%
This suppression of free energy transfer via linear phase mixing is a nonlinear, kinetic effect.
To understand its mechanism,  we decompose the distribution function in terms of 
$\g^+$ and $\g^-$, propagating modes in Hermite space:
the ``phase-mixing mode'', $\g^+$, propagates forwards from low to high $m$,
while 
the ``anti-phase-mixing mode'', $\g^-$, {propagates backwards} from high to low $m$ \cite{Kanekar14,Parker14,Schekochihin15}.
These are the two modes already observed in the linear problem, see \fig\ref{fig:ivp_contour}(a).
The decomposition is
\begin{align}
\g=\lp i\ \sgn\ \kpara\rp^{-m}\left[ \g^+ + (-1)^m \g^- \right],
	%\label{}
\end{align}
where 
\begin{align}
\g^{\pm}=\frac{1}{2}\lp \pm i\ \sgn\ \kpara\rp^m\lp\g \pm i\ \sgn(\kpara) \gp \rp.
	%\label{}
\end{align}
%With these, the drift kinetic equation for $m\geq3$ becomes
%%%\begin{equation}
%%%  \begin{split}
%%%		\pd{\g^{\pm}}{t} 
%%%		\pm \sqrt{2}|\kpara|\vth m^{1/4}\pd{}{m}\left( m^{1/4}\g^{\pm}\right) + \nu m\g^{\pm} 
%%%  \\
%%%    = 
%%%		-\!\!\!\!\!\sum_{\ppara+\qpara=\kpara}\!\!\!\!\! \uperphat(\ppara)\cdot\nabla_{\perp}\left[ \delta^+_{\kpara\qpara}\g^{\pm}(\qpara) + \delta^-_{\kpara\qpara}\g^{\mp}(\qpara)\right],
%%%    \label{eq:GyrokineticEquationPlusMinus}
%%%  \end{split}
%%%\end{equation}
With these, 
the free energy contributions are approximately $|\g|^2\approx|\g^+|^2+|\g^-|^2$ \cite{Schekochihin15},
with $\g^{\pm}$ evolving as
\begin{equation}
  \begin{split}
	&	\pd{}{t}\left( \frac{|\g^{\pm}|^2}{2}\right)
		\pm \frac{|\kpara|\vth}{\sqrt{2}} \pd{}{m}\left( \sqrt{m}|\g^{\pm}|^2\right) + \nu m|\g^{\pm}|^2
  \\
  &   
\hspace{0.5cm}
		= -\Re\Big(\!\!\!\!\!\sum_{\ppara+\qpara=\kpara}\!\!\!\!\! [\g^{\pm}(\kpara)]^*\uperphat(\ppara)\ \cdot
%%%  \\
%%%  & \hspace{2.5cm} 
	\nabla_{\perp}\left[ \delta^+_{\kpara\qpara}\g^{\pm}(\qpara) + \delta^-_{\kpara\qpara}\g^{\mp}(\qpara)\right]\Big),
%%%    \label{eq:FreeEnergyEquationPlusMinus}
    \label{eq:GyrokineticEquationPlusMinus}
  \end{split}
\end{equation}
for $m\geq3$.
Here $\delta^{\pm}_{\kpara\qpara}=\left[1\pm\sgn(\kpara\qpara)\right]/2$,
which is one if $\kpara$ and $\qpara$ have the same sign, and zero otherwise.
We have also introduced a derivative approximation for the finite difference in linear streaming which is valid 
as the modes $\g^{\pm}$ are smooth in the sense that $\g^{\pm}\approx \hat{g}^{\pm}_{m+1}$ \cite{Schekochihin15}.
In terms of $\g^{\pm}$, the $\E\times\B$ velocity is $\uperphat=\alpha\rho_i\vth\hat{\z}\times\nabla_{\perp}(\hat{g}^+_0+\hat{g}^-_0)/2$.
Further, the normalized free energy transfer from $m$ to $m+1$ is
\begin{equation}
  \begin{split}
		\bar{\Gamma}_m 
		= \frac{\Gamma_m}{\Gamma^L_m} 
		&= \frac{\kpara\vth\sqrt{(m+1)/2}\ \Im(\gp^*\g)}{\kpara\vth\sqrt{(m+1)/2}\ |\g|^2}
%%%		\\ 		&
		\approx
		\frac{|\g^+|^2-|\g^-|^2}{|\g^+|^2+|\g^-|^2}.
    \label{eq:FreeEnergyTransferNormalized}
  \end{split}
\end{equation}
From this expression it follows that the suppression of free energy transfer, $\bar{\Gamma}_m<1$, can only be due to the presence of anti-phase-mixing modes, $\g^-\neq0$.

Let us now consider the linear and nonlinear cases in terms of $\g^{\pm}$.
In the linear case, we neglect the \rhs\ of \eqref{eq:GyrokineticEquationPlusMinus}.
Taking an initial disturbance at large velocity space scales (low $m$) that propagates forwards only, we seek solutions with $\g^-=0$.\footnote{Any $\g^-$ present in the initial conditions ``reflects'' off the hard-wall-like boundary condition at $m=0$ and becomes forward propagating within one streaming time.}
Solving the differential equation in $m$ for $|\g^+|^2$, we find
\begin{equation}
  \begin{split}
		%|\g^+|^2=\frac{A(\k)}{\sqrt{m}}\exp\left( \left(\frac{m}{m_c}\right)^{3/2}\right), 
		|\g^+|^2=\frac{A(\kpara)}{\sqrt{m}}e^{\left(-{m}/{m_c}\right)^{3/2}}, 
		\hspace{0.7cm}
		m_c = \left(\frac{3|\kpara|\vth}{2\sqrt{2}\nu}\right)^{2/3},
    \label{eq:GPlusLinearSolution}
  \end{split}
\end{equation}
where $A$ is a constant of integration, and $m_c$ is the collisional cutoff \cite{Zocco11,Kanekar14}.
For $m<m_c$, $|\g^+|^2$ has the $m^{-1/2}$ spectrum of linear Landau damping, while for $m>m_c$, the spectrum is strongly damped.
As $\g^-=0$, the normalized transfer \eqref{eq:FreeEnergyTransferNormalized} is $\bar{\Gamma}_m=1$ everywhere.

\begin{figure}
  \centering
  \includegraphics[]{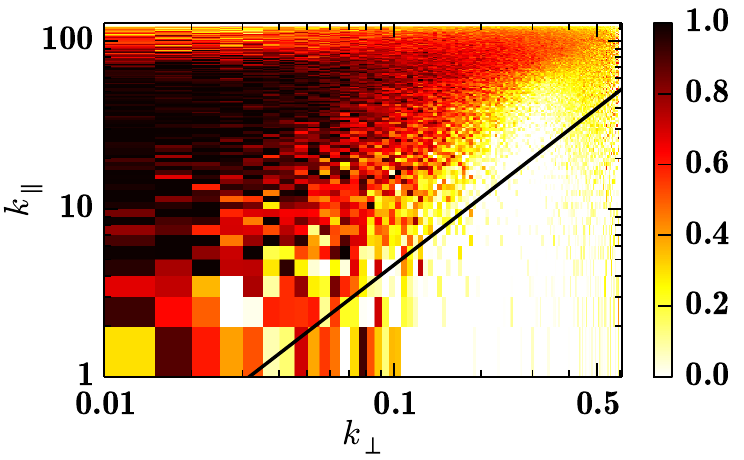}
	\caption[Free energy transfer from mode $100$ to $101$ in saturated turbulence,\\ normalized to its value in a linear plasma.]{%}
    \label{fig:FluxSuppressionM100} Free energy transfer from mode $100$ to $101$ in saturated turbulence, normalized to its value in a linear plasma,
		$\bar{\Gamma}_{100}$.
		The diagonal black line marks the critical balance determined from $\varphi$.
		Free energy transfer is completely suppressed at the energy-containing scales $\taunli\gtrsim\tausi$, and hence Landau damping is strongly suppressed.
  }
\end{figure}

To consider nonlinear drift kinetics, we reinstate the nonlinear term in \eqref{eq:GyrokineticEquationPlusMinus}.
%From \eqref{eq:FreeEnergyTransferNormalized} we see that the flux suppression observed in \fig\ref{fig:FluxSuppressionM2} requires $\g^-\neq0$; but how are these modes excited from initial conditions with $\g^-=0$?
%In \eqref{eq:GyrokineticEquationPlusMinus} we see that the this must be via the nonlinear term: $\g^+$ couples with the $g^+_0$ in $\uperp$ to contribute to the $\tpd{\g^-}{t}$ equation.
Now even with $\g^-=0$ in the initial conditions, the nonlinear term acts as a source in the ``$-$'' equation (the $\hat{g}^+_0$ in $\uperphat$ couples to the two $\g^-$ terms).
This causes energy in the $\g^-$ modes to increase, with the result that $\bar{\Gamma}_m<1$, as in \fig\ref{fig:FluxSuppressionM2}.
The effect is even clearer in the inertial range of $m$, away from driving and dissipation scale effects,
as we see in \fig\ref{fig:FluxSuppressionM100} where we plot $\bar{\Gamma}_{100}$.
Now there are clear regions in wavenumber space where free energy transfer is as in the linear case, $\bar{\Gamma}_{100}=1$,
and where free energy transfer is completely suppressed, $\bar{\Gamma}_{100}=0$.

\section{Critical balance}%
%
%To bound these regions of wavenumber space, we compare the sizes of the phase-mixing and nonlinear terms.
To understand the mechanism that leads to these distinct regions of wavenumber space, we compare the sizes of the phase-mixing and nonlinear terms.
We quantify these by comparing the timescales for the two cascades, the characteristic streaming rate $\tausi\sim \kpara\vth$ and the nonlinear eddy turnover rate $\taunli\sim(\vth/R)(\kperp\rho_i)^{4/3}$ \cite{BarnesEtal11}. 
When $\tausi\gg\taunli$, streaming dominates the nonlinearity, and the problem is essentially linear with $\bar{\Gamma}_m=1$. 
When $\tausi\ll\taunli$, the nonlinearity dominates streaming and flux is suppressed.
In \fig\ref{fig:FluxSuppressionM100} we also plot the line of critical balance, $\taunli\sim\tausi$, with the constant of proportionality determined by the critical balance of the electrostatic potential $\varphi$.
The critical balance line is in good agreement with the boundary of complete suppression $\bar{\Gamma}_m=0$, indicating that free energy transfer is suppressed wherever the nonlinear eddy turnover rate is comparable with or faster than the linear streaming rate $\taunli\gtrsim\tausi$.

\begin{figure}
  \centering
	\subfigure[]{\includegraphics[width=0.49\textwidth]{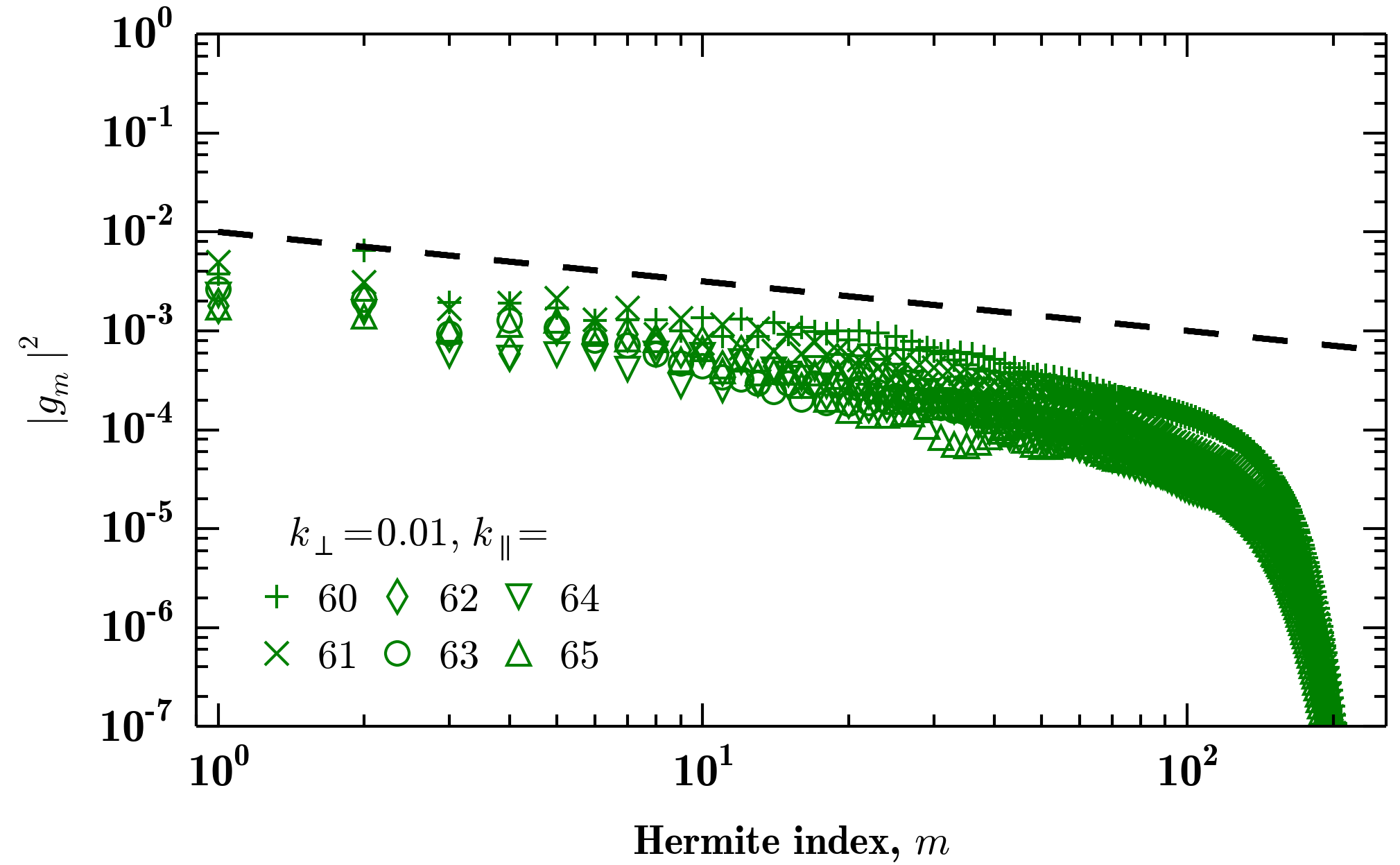}}
	\subfigure[]{\includegraphics[width=0.49\textwidth]{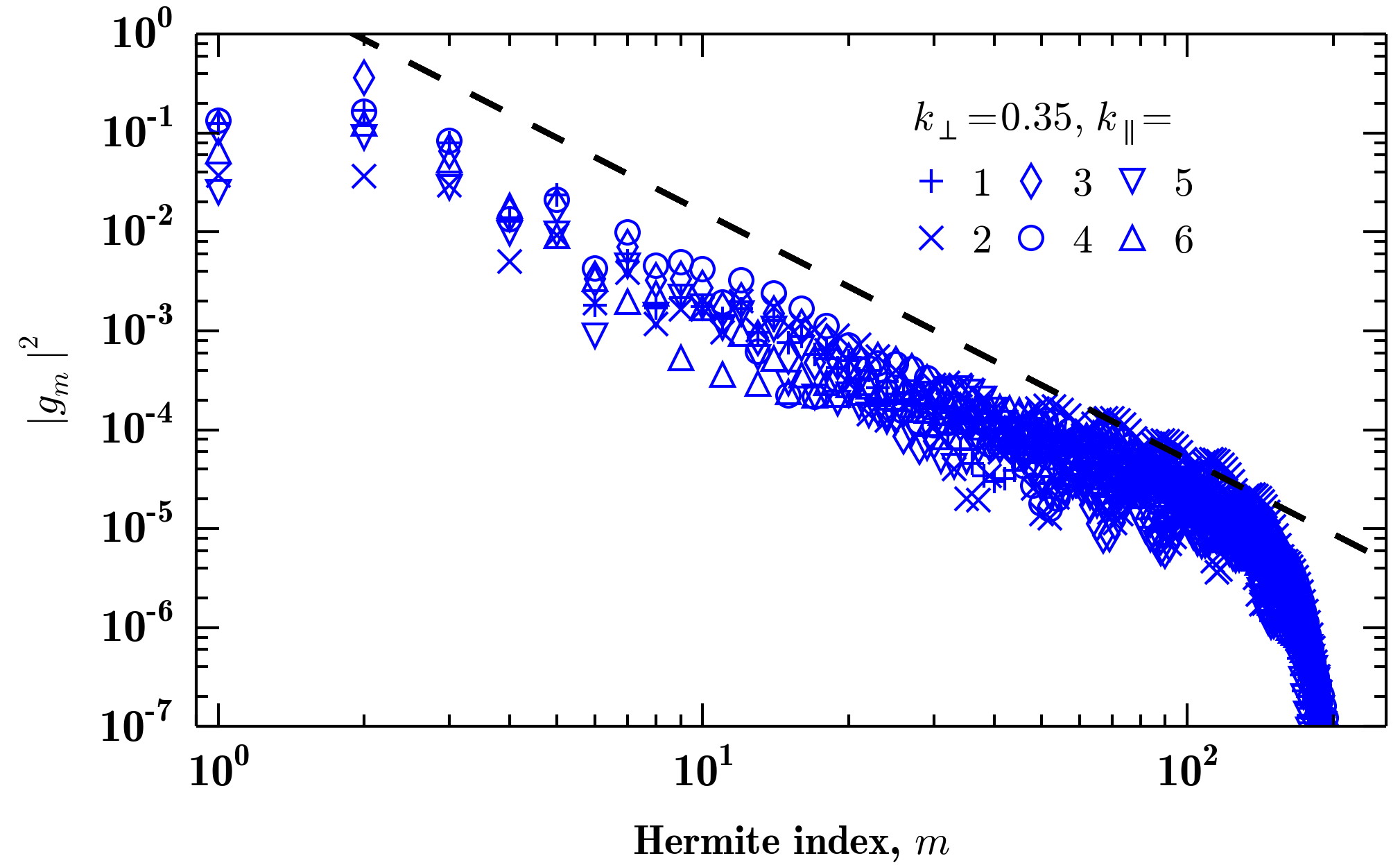}}
	\caption[Hermite spectra at fixed $\k$ in the linear streaming dominated and\\ nonlinearity dominated regions]{%}
    \label{fig:HermiteSpectra} Hermite spectra at fixed $\k$ in (a) the linear streaming dominated region ($m^{-1/2}$) and (b) the nonlinearity dominated region ($m^{-5/2}$).
  }
\end{figure}
\section{Hermite spectra and dissipation}%
The different free energy transfer behaviours in the phase-mixing and the nonlinearity dominated regions give rise to two different Hermite spectra, as plotted in \fig\ref{fig:HermiteSpectra}.
In the phase-mixing-dominated region ($\Gamma_m=1$), we observe the $m^{-1/2}$ spectrum \eqref{eq:GPlusLinearSolution} obtained by neglecting the nonlinear term.
In the nonlinearity dominated region ($\Gamma_m=0$), we observe the $m^{-5/2}$ spectrum recently predicted by \citet{Schekochihin15} and derived in \chp\ \ref{sec:ScalingLawsForDriftKineticTurbulence}.

These spectra exhibit different dissipation behaviours.
The spectrum in the phase-mixing region dissipates free energy at the usual Landau damping rate
  \begin{equation}
		%D_{\textrm{s}} \sim 
		\Ds \sim 
		\int_{3}^{m_c}\d m ~  \nu m m^{-1/2} 
		%\sim \frac{|\kpara|\vth}{\sqrt{2}}
		\sim |\kpara|\vth,
  \end{equation}
which remains finite as $\nu\to0^+$.
In contrast, the dissipation rate in the nonlinear region is
  \begin{equation}
		\Dnl \sim \int_{3}^{m_c}\d m ~  \nu m m^{-5/2} 
		\sim \nu^{4/3} (|\kpara|\vth)^{-1/3},
  \end{equation}
	so $\Dnl\to0$ as $\nu\to0^+$.
	Landau damping is thus suppressed in the nonlinear region.

	Finally, one important property of critical balance is that most of the free energy is contained in modes for which $\taus/\taunl\sim (\kperp\rho_i)^{4/3}/(\kpara R) \gtrsim 1$ (\cite{BarnesEtal11}, or note, \eg, that in \fig\ref{fig:HermiteSpectra} the amplitudes of the spectra are much larger for the nonlinear region). 
	This is exactly the nonlinearity-dominated region where $\bar{\Gamma}_m=0$ and $|g_m|^2\sim m^{-5/2}$.
	Therefore the total dissipation in collisions tends to zero as $\nu\to0^+$, and the vast majority of free energy cascades to dissipation at fine physical space scales.
	
	These spectra and dissipation patterns are in accordance with the earlier work of \citet{Hatch13} which deduced similar properties based on the Hermite spectra summed over all $\k$.
	However we have shown that the Hermite spectrum has different behaviours in different regions of Fourier space.

	\section{Discussion}%
	In this \chp\ we have shown that linear phase mixing and the nonlinear cascade, two effects which one might have expected to be independent,
	are in fact strongly interacting.
The nonlinear term excites anti-phase-mixing modes, suppressing the net transfer of free energy into kinetic modes in the inertial range in $m$.
This has both theoretical and practical implications. 
  Theoretically our results profoundly change our understanding of the way that free energy is cascaded and dissipated in phase space.
  As there is only a small net free energy flux out of fluid modes in the inertial range, it is legitimate to neglect parallel streaming
  when deriving physical space %results. 
	spectra from Kolmogorov arguments, as in Ref.~\cite{BarnesEtal11}. % PJD
Moreover, the Hermite spectrum at dominant scales is a steep $m^{-5/2}$ power law which dissipates no free energy via Landau damping as $\nu\to0^+$. 
Therefore almost all free energy cascades to sub-Larmor scales.  
The steep spectrum also means that free energy dissipation is not independent of collisionality.
This has the important practical implication that enlarged collision frequencies cannot necessarily be used to compensate for low $\vpara$ resolution in weakly collisional simulations.
Conversely, this work suggests a possible refinement to existing gyrofluid models \cite{Hammett92,Dorland93} through incorporating free energy flux conditions into the gyrofluid closure.

Finally, this work has focused on ion-temperature-gradient-driven drift kinetic turbulence.
However, its conclusions depend on inertial range physics which do not depend on details of the injection scale.
The approach presented here will be applicable to other kinetic systems where a nonlinearity interacts with particle streaming.
Indeed, the similar suppression of streaming has already been observed due to a different nonlinearity in the Vlasov--Poisson system \cite[see also Appendix A]{Parker14}.

\newcommand{\Ephi}{E_{\varphi}}
\newcommand{\cst}{\mathrm{constant}}

\chapter{Phase space spectra for drift-kinetic turbulence}
\label{sec:ScalingLawsForDriftKineticTurbulence}

In the previous \chp, we showed that the transfer of free energy via linear streaming is different in different regions of phase space.
When the linear streaming term dominates the nonlinear term, free energy is transferred at the same rate as in the linearized system.
However, when the nonlinearity dominates the streaming, the transfer to fine velocity scales is suppressed, and free energy cascades purely in Fourier space.
In \S\ref{sec:DerivationOfScalings}, we use these observations to construct a theory for the phase space spectra in drift-kinetic turbulence.
In \S\ref{sec:ObservedSpectra}, we verify these spectra using simulations of ITG-driven turbulence.
However, there is nothing in the theory which is specific to ITG turbulence,
and it should apply equally to the inertial range of electron temperature gradient driven turbulence \cite{Dorland00,Gene},
or indeed generic electrostatic drift-kinetic turbulence driven at long perpendicular wavelengths.

\section{Derivation of scalings}
\label{sec:DerivationOfScalings}
\citet*{Schekochihin15} gives a detailed and careful derivation of the scaling laws for drift-kinetic turbulence.
In this \Sec, we give a concise derivation following their approach.

The spectra are derived making three assumptions.
Firstly, the temperature gradient is large, $R/L_T\gg1$.
This ensures scale separation between the energy injection scale and the Larmor (\ie\ dissipation) scale, 
and therefore allows an inertial range in $\kperp$ to develop.
The second assumption is that the turbulence is isotropic in the perpendicular plane, so that $k_x\sim k_y\sim \kperp$.
This assumption was introduced by \citet{BarnesEtal11} based on the conjecture that strong ITG turbulence generates zonal flows, 
modes with $k_y=\kpara=0$ and $k_x$ much smaller than scales characteristic of the inertial range in $\kperp$ (see \eg\ \fig\ref{fig:PhiOneDSpecta}).
Counterintuitively, zonal flows tend to isotropise structures in the $xy$ plane by shearing apart anisotropies.
The shearing rate $S_{\mathrm{ZF}}$ is comparable with the inverse nonlinear time $\taunli$ \cite{Cowley91,Rogers00},
so that wavenumbers satisfy $k_x\sim k_y(S_{\mathrm{ZF}}\taunli)\sim k_y$.
The third assumption is ``critical balance'', a causality argument discussed in \sec\ref{sec:SpectrumForTheElectrostaticPotential}.
Critical balance is used to relate parallel and perpendicular length scales;
no relation can be found using dimensional analysis, 
since the natural ratio $\rho_i/L$ has been formally taken to zero in deriving the gyrokinetic (and drift-kinetic) equations.
Perpendicular isotropy and critical balance will be verified in \sec\ref{sec:ObservedSpectra}.

\subsection{Spectrum for the electrostatic potential}
\label{sec:SpectrumForTheElectrostaticPotential}
We begin by deriving the spectrum of the electrostatic potential, defined as
\begin{align}
	\Ephi(\kpara,\kperp) = 2\pi \kperp \langle |\varphik|^2\rangle ,
	\label{eq:EphiDefinition}
\end{align}
where as before $\varphik$ denotes the Fourier coefficients of $\varphi$, 
and $\langle\cdot\rangle$ denotes a time average over the saturated turbulent state.
We assume that $\Ephi(\kpara,\kperp)$ will have different scalings in the two regions of wavenumber space discussed in \chp~\ref{sec:FreeEnergyFlowAndDissipation}: 
the advection-dominated region ($\taunl\lesssim\taus$) and the streaming-dominated region ($\taus\lesssim\taunl$).
Since we will make local cascade arguments, we assume separate power law spectra in the two regions:
\begin{align}
	\Ephi(\kpara,\kperp) = 
	\begin{cases}
		\kpara^a \kperp^b, 
		& \taus\lesssim\taunl,\\[2ex]
		\kpara^c \kperp^d, 
		& \taus\gtrsim\taunl,
	\end{cases}
	\label{eq:EphiPrototype}
\end{align}
where $a$, $b$, $c$ and $d$ are constants to be determined.

We first find the spectra in the advection-dominated region using critical balance, 
a causality argument introduced for astrophysical magnetohydrodynamic turbulence \cite{Goldreich95,Goldreich97,Boldyrev05}
which has since been used for strong turbulence in other wave-supporting systems \cite{Cho04,Schekochihin09,Nazarenko11}.
Critical balance states that two points along the mean field can be correlated only if the time taken for information to pass between them (the streaming time $\taus\sim(\kpara\vth)^{-1}$) is less than the time taken for turbulence to decorrelate that information in the perpendicular plane (the nonlinear time $\taunl\sim(\kperp^2\varphi)^{-1}$).
Thus the turbulence is uncorrelated in the parallel direction for $\taunl\lesssim\taus$,
and there $\Ephi(\kpara,\kperp)$ has the spectrum of uncorrelated perturbations (white noise), $\Ephi(\kpara,\kperp)\sim\kpara^0\kperp^d$.
That is, $c=0$ in \eqref{eq:EphiPrototype}.
Critical balance also implies that most of the free energy is contained in the scales $\taunl\lesssim\taus$.

We determine the exponent $d$ from a constant flux argument in the advection-dominated region.
As seen in \fig\ref{fig:FluxSuppressionM2}, there is very little free energy flux due to linear streaming in $\taunl\lesssim\taus$.
We therefore argue that free energy has a constant flux through wavenumber space
\begin{align}
	\frac{W}{\taunl} \sim 
	\frac{\varphi^2}{\taunl} \sim \kperp^2\varphi^3 \sim \cst,
	\hspace{1cm}
	\implies
	\hspace{1cm}
	\varphi \sim \kperp^{-2/3},
	\label{eq:ConstFluxArgument}
\end{align}
where we have used the nonlinear time $\taunl\sim(\kperp^2\varphi)^{-1}$.
We thus determine the nonlinear time in terms of $\kperp$ only: $\taunl\sim (\kperp^2\varphi)^{-1} \sim \kperp^{-4/3}$. % with $\varphi^2\sim \int\d\kpara\d\kperp~\kperp|\varphik|^2 \sim \kperp^{-4/3}$.
We also determine the one-dimensional perpendicular spectrum, defined as $\Ephi^{\perp}(\kperp)\id \int \d \kpara ~ \Ephi(\kpara,\kperp)$.
By critical balance, this is dominated by the energy-containing scales with $\taunl\lesssim\taus$, that is, $\kpara\lesssim\kperp^{4/3}$:
\begin{align}
	\Ephi^{\perp}(\kperp) \id 
	\int \d \kpara~ \Ephi(\kpara,\kperp)
	\sim 
	\int_0^{\kperp^{4/3}}\d\kpara~ \kpara^0\kperp^d
	\sim 
	\kperp^{d+\frac{4}{3}}.
	\label{eq:OneDSpectrumOne}
\end{align}
However, using the definition of the one-dimensional spectrum and the constant free energy flux argument \eqref{eq:ConstFluxArgument}, we may also deduce
\begin{align}
	\Ephi^{\perp}(\kperp) \id 
	\int \d \kpara~ 2\pi\kperp\langle |\varphik|^2\rangle
	\sim 
	\frac{\int\d\kpara\d\kperp~ 2\pi\kperp\langle |\varphik|^2\rangle}{\kperp}
	\sim 
	\frac{\varphi^2}{\kperp}
	\sim
	\kperp^{-\frac{7}{3}},
	\label{eq:OneDSpectrumTwo}
\end{align}
so that comparing \eqref{eq:OneDSpectrumOne} and \eqref{eq:OneDSpectrumTwo} determines $d=-11/3$.

We still need two conditions to determine $a$ and $b$, the remaining powers in \eqref{eq:EphiPrototype}.
Firstly we use the standard result based on the Corsin invariant in two-dimensional turbulence (and derived in \cite[][Appendix A]{Schekochihin15}) that spectra in a homogeneous isotropic system decay like $\kperp^3$ for long perpendicular wavelengths. 
We therefore have $b=3$.
The remaining exponent $a$ is found by imposing continuity in the spectrum across $\taunl\sim\taus$, so that
\begin{align}
	\kpara^0\kperp^{-11/3} \sim \kpara^a\kperp^3 \sim \kperp^{\frac{4a}{3}+3}
	\hspace{1cm}
	\implies
	\hspace{1cm}
	a = -5 .
	%	\label{}
\end{align}

\subsection{Spectrum for the distribution function}

We now derive spectra for the distribution function.
These are more complicated since they depend on $m$, as well as $\kperp$ and $\kpara$.
Moreover, there are two sets of spectra, one each for the forward and backward propagating modes $\tilde{g}^{\pm}_m$ introduced in \chp~\ref{sec:FreeEnergyFlowAndDissipation}.

The spectrum of the electrostatic potential $\Ephi$ is divided into two regions along the critical balance line $\taunl\sim\taus$.
Critical balance also appears in the distribution function spectra, but in addition there is a new dependence on the ``phase-mixing threshold'', $\kpara\sim \sqrt{m}\kperp^{4/3}$.
The phase-mixing threshold is like a critical balance where the velocity scale in the streaming time, $\vth$, has been replaced by the velocity scale of the $m$th Hermite mode, $\vth/\sqrt{m}$ (see \eqref{eq:OlverLimits}).

To find the distribution function spectra, we introduce the new function 
\begin{align}
	\tilde{f}(\kpara,\kperp,s) = m^{1/4}
	\begin{cases}
		\tilde{g}^+_m(\kpara,\kperp), 
		& \kpara\geq0, \\[2ex]
		\tilde{g}^-_m(\kpara,\kperp),
		& \kpara<0,
	\end{cases}
	%\label{}
\end{align}
where $s=\sqrt{m}$.
On substitution into the gyrokinetic equation, we find $\tilde{f}$ satisfies
\begin{align}
	\pd{\tilde{f}}{t} + \frac{\kpara\vth}{\sqrt{2}}\pd{\tilde{f}}{s} = 
	- \sum_{\ppara} \uperp(\ppara)\cdot\nabla_{\perp} \tilde{f}(\kpara-\ppara),
	\label{eq:ftilde}
\end{align}
and the mean squared amplitude
$F=\langle|\tilde{f}|^2\rangle$
satisfies
\begin{align}
	\pd{F}{t} + \frac{\kpara\vth}{\sqrt{2}}\pd{F}{s}
	= - 2\ \Re\left( \sum_{\ppara}\tilde{f}^*(\kpara)\uperp(\ppara)\cdot\nabla\tilde{f}(\kpara-\ppara)\right).
	%\label{}
\end{align}
In this equation, both phase-mixing and anti-phase-mixing are represented as propagation along the characteristics $s=\kpara\vth (t-t_0)/\sqrt{2}$.
The generation of echo flux 
(\ie\ conversion from $\tilde{g}^+_m$ to $\tilde{g}^-_m$)
is now represented by mode coupling between positive and negative parallel wavenumbers via the nonlinear term.

We first consider the spectrum of forward propagating modes, 
\begin{align}
	E^+_m(\kpara,\kperp)=2\pi\kperp\langle|\tilde{g}^+_m|^2\rangle.
%	\label{}
\end{align}
Free energy streams linearly along characteristics until coupled to other wavenumbers by the nonlinear term.
Therefore, linear streaming lasts for at most one nonlinear time,
\begin{align}
	\frac{s\sqrt{2}}{\kpara\vth}  = (t-t_0) \lesssim \taunl \sim \kperp^{-4/3}
	\hspace{1cm}
	\implies
	\hspace{1cm}
	s \lesssim \frac{\kpara}{\kperp^{4/3}}.
	\label{eq:ConstFluxArgumentEplus}
\end{align}
In this region $\tilde{f}$, inherits its spectrum $E_f\id 2\pi\kperp\langle |\tilde{f}|^2\rangle = \sqrt{m}E^+_m$ from low $s$, 
that is, from $\Ephi$ at $\kpara\lesssim \kperp^{4/3}$, 
so that
\begin{align}
	E_f = \sqrt{m}E_m^+ \sim \Ephi \sim \kpara^{-5}\kperp^3.
	%\label{}
\end{align}
Thus we have the spectrum in the streaming-dominated region:
\begin{align}
E_m^+(\kpara,\kperp)\sim \kpara^{-5}\kperp^3m^{-1/2}
\hspace{1cm}
\mathrm{in}
\hspace{1cm}
\sqrt{m}\kperp^{4/3}\lesssim \kpara.
	\label{eq:EplusInStreaming}
\end{align}
%%%As there is no echo, no backwards streaming and expect $E^-_m \ll E^+_m$.
%%%
%%%Passive scalar turbulence $b=3$, and therefore $a=-5$.
%%%

To find the spectrum in the advection dominated region $\kpara \lesssim \kperp^{4/3}$,
we note that 
there is significant echo flux, 
so we may neglect linear streaming and treat \eqref{eq:ftilde} as a fluid-like equation for each $s$.
Repeating the constant free energy flux argument used for $\Ephi$ \eqref{eq:ConstFluxArgument}, we obtain
%%%\begin{align}
%%%	\frac{\tilde{f}^2(s)}{\taunl} \sim \kperp^2\varphi \tilde{f}^2 \sim \mathrm{fn}(s),
%%%	\hspace{1cm}
%%%	\implies
%%%	\hspace{1cm}
%%%	E^{+\perp}_m(\kperp) \sim \kperp^{-7/3},
%%%	%\label{}
%%%\end{align}
%%%where $\mathrm{fn}(s)$ denotes a function of $s$ only, and we have used $\varphi\sim\kperp^{-2/3}$.
%%%Unlike for $\Ephi$, the flux is now an unknown function of $s$, and so the scaling with $m$ is undetermined.
%%%This yields $E^{+}_m\sim \kpara^0\kperp^{-11/3}m^{-\sigma}$ (for unknown exponent $\sigma$) in $\kpara\lesssim \kperp^{4/3}$.
%%%
\begin{align}
	\frac{\tilde{f}^2}{\taunl} \sim \kperp^2\varphi \tilde{f}^2 \sim \mathrm{fn}(s),
	\hspace{1cm}
	\implies
	\hspace{1cm}
	\tilde{f}^2 \sim \kperp^{-4/3},
	%\label{}
\end{align}
where $\mathrm{fn}(s)$ denotes a function of $s$ only, and we have used $\varphi\sim\kperp^{-2/3}$.
As in \eqref{eq:OneDSpectrumTwo}, we use this to find the one-dimensional perpendicular spectrum,
\begin{align}
	E^{+\perp}_m(\kperp) \id 
	\int \d \kpara~ 2\pi\kperp\langle |\tilde{g}^+_m|^2\rangle
	\sim 
	\frac{\int\d\kpara\d\kperp~ E^+_m(\kpara,\kperp)}{\kperp}
	\sim 
	\frac{\tilde{f}^2}{\kperp}
	\sim
	\kperp^{-\frac{7}{3}},
	\label{eq:OneDSpectrumThree}
\end{align}
Unlike for $\Ephi$, the flux is now an unknown function of $s$, and so the scaling with $m$ is undetermined.
This yields the scaling the advection-dominated region:
\begin{align}
E^{+}_m(\kpara,\kperp) \sim \kpara^0\kperp^{-11/3}m^{-\sigma}
\hspace{1cm}
\mathrm{in}
\hspace{1cm}
\kpara\lesssim \kperp^{4/3},
	\label{eq:EplusInAdvection}
\end{align}
where the exponent $\sigma$ is unknown.

We now know $E^+_m(\kpara,\kperp)$ in the streaming region, $\sqrt{m}\kperp^{4/3}\lesssim\kpara$ \eqref{eq:EplusInStreaming} and, except for the $m$-dependence, in the advection-dominated region, $\kpara\lesssim\kperp^{4/3}$ \eqref{eq:EplusInAdvection}.
We now need the spectrum in the intermediate region between these two.
To summarize,
\begin{align}
	E^+_m(\kpara,\kperp) = 
	\begin{cases}
		\kpara^0 \kperp^{-11/3} m^{-\sigma}, 
		& \kpara\lesssim\kperp^{4/3},\\[2ex]
		\kpara^{a'} \kperp^{d'} m^{-\sigma'}, 
		& \kperp^{4/3}\lesssim\kpara\lesssim\sqrt{m}\kperp^{4/3},\\[2ex]
		\kpara^{-5} \kperp^{3} m^{-1/2}, 
		& \kpara\gtrsim\sqrt{m}\kperp^{4/3},
	\end{cases}
	\label{eq:EPlusPrototype}
\end{align}
for unknown exponents $a'$, $d'$, $\sigma$ and $\sigma'$.
Continuity of the spectrum across $\kpara\sim\kperp^{4/3}$ and $\kpara\sim\sqrt{m}\kperp^{4/3}$ gives the three constraints
\begin{align}
	a' = -\frac{11+3d'}{4},
	\hspace{1cm}
	\sigma' = 3+\frac{a'}{2},
	\hspace{1cm}
	\sigma=\sigma',
	\label{eq:IntermediateRegionCoeffs}
\end{align}
where one of the four matching conditions is repeated.
To determine $d'$, and hence the other scalings, we consider the free energy cascade in the intermediate region.
Unlike previously, where the cascade is local in all wavenumbers, the rapid decay of $\varphik$ with parallel wavenumber ($\Ephi\sim\kpara^{-5}$) 
%in $\kpara\gtrsim\kperp^{4/3}$ 
means the dominant coupling cannot be local in parallel wavenumber space,
as shown in \citet[][Appendix B]{Schekochihin15}.
Instead the dominant cascade is local in $\kperp$ but nonlocal in $\kpara$.
Thus the free energy cascades in $\kperp$ only, so that
\begin{align}
	\frac{\tilde{f}^2}{\taunl} \sim \kperp^2\varphi \tilde{f}^2 \sim \mathrm{fn}(s,\kpara),
	\hspace{1cm}
	\implies
	\hspace{1cm}
	E^{+}_m(\kpara,\kperp) \sim \kperp^{-7/3},
	%\label{}
\end{align}
where now it is the two-dimensional spectrum that behaves as $\kperp^{-7/3}$, not the one-\linebreak dimensional spectrum as in \eqref{eq:OneDSpectrumThree}.
We therefore have $d'=-7/3$, and so from \eqref{eq:IntermediateRegionCoeffs}, we have $a'=-1$ and $\sigma=\sigma'=5/2$.

Finally, we determine the spectrum of backwards propagating modes.
The spectrum in the advection region is determined in exactly the same way as the forward spectrum \eqref{eq:EplusInAdvection}:
advection completely dominates streaming, and free energy cascades locally in Fourier space, 
yielding the spectrum 
\begin{align}
E_m^-(\kpara,\kperp)\sim \kpara^0\kperp^{-11/3}m^{-\sigma''}
\hspace{1cm}
\mathrm{in}
\hspace{1cm}
\kpara\lesssim\kperp^{4/3}.
	\label{eq:EminusAdvection}
\end{align}
However, unlike the $E^+_m$ spectrum, the $E^-_m$ spectrum only has one region for $\kpara\gtrsim\kperp^{4/3}$.
This is because there is no coupling of parallel wavenumbers in $\kpara\gtrsim\kperp^{4/3}$, even in the intermediate region where the nonlinear term is non-negligible.
Thus there is no coupling between the $\kpara>0$ and $\kpara<0$ modes which generates the backwards propagating modes in \eqref{eq:ftilde}. 
Therefore all we know in $\kpara\gtrsim\kperp^{4/3}$, is that the spectrum decays like $\kperp^3$ in the limit $\kperp\to0$,
and we have 
\begin{align}
E^-_m(\kpara,\kperp)\sim \kperp^3\kpara^{a''}m^{-\sigma'''}
\hspace{1cm}
\mathrm{in}
\hspace{1cm}
\kperp^{4/3}\lesssim\kpara.
	\label{eq:EminusAlsoAdvection}
\end{align}
The remaining coefficients are all found by ensuring the spectrum in \eqref{eq:EminusAdvection} and \eqref{eq:EminusAlsoAdvection} is continuous across $\kpara\sim \kperp^{4/3}$, yielding $a''=-5$, $\sigma''=\sigma'''=5/2$. 

In summary, we have derived the spectrum for the forward propagating modes,
\begin{align}
	E^+_m(\kpara,\kperp) = 
	\begin{cases}
		\kpara^0 \kperp^{-11/3} m^{-5/2}, 
		& \kpara\lesssim\kperp^{4/3},\\[2ex]
		\kpara^{-1} \kperp^{-7/3} m^{-5/2}, 
		& \kperp^{4/3}\lesssim\kpara\lesssim\sqrt{m}\kperp^{4/3},\\[2ex]
		\kpara^{-5} \kperp^{3} m^{-1/2}, 
		& \kpara\gtrsim\sqrt{m}\kperp^{4/3},
	\end{cases}
	\label{eq:Eplusm}
\end{align}
backwards propagating modes,
\begin{align}
	E^-_m(\kpara,\kperp) = 
	\begin{cases}
		\kpara^0 \kperp^{-11/3} m^{-5/2},
		& \kpara\lesssim\kperp^{4/3},\\[2ex]
		\kpara^{-5} \kperp^3 m^{-5/2}, 
		& \kpara\gtrsim\kperp^{4/3},
	\end{cases}
	\label{eq:Eminusm}
\end{align}
and for the electrostatic potential,
\begin{align}
	\Ephi(\kpara,\kperp) = 
	\begin{cases}
		\kpara^0 \kperp^{-11/3}, 
		& \kpara\lesssim\kperp^{4/3},\\[2ex]
		\kpara^{-5} \kperp^3, 
		& \kpara\gtrsim\kperp^{4/3} .
	\end{cases}
	\label{eq:Ephi}
\end{align}
%%%

\section{Observed spectra}
\label{sec:ObservedSpectra}

We now verify the above scaling laws using data from the same simulation introduced in \chp~\ref{sec:FreeEnergyFlowAndDissipation}.
We first verify the assumptions used to derive the spectra---isotropy and critical balance---and then present evidence for the scaling laws for $\Ephi$ and $E^{\pm}_m$.

\subsection{Assumptions}

\begin{figure}
	\centering
	\subfigure[]{\includegraphics{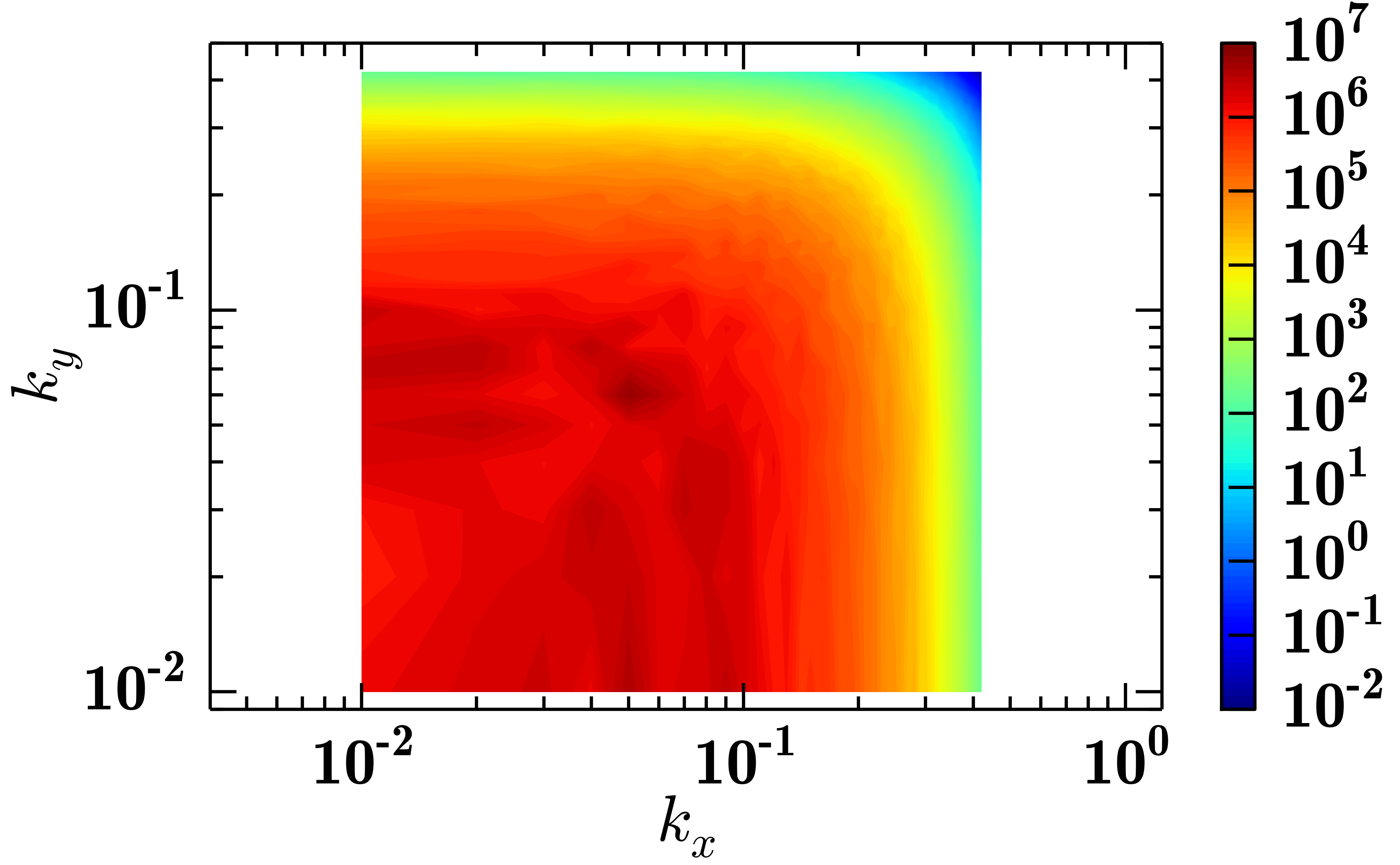}}
	\subfigure[]{\includegraphics{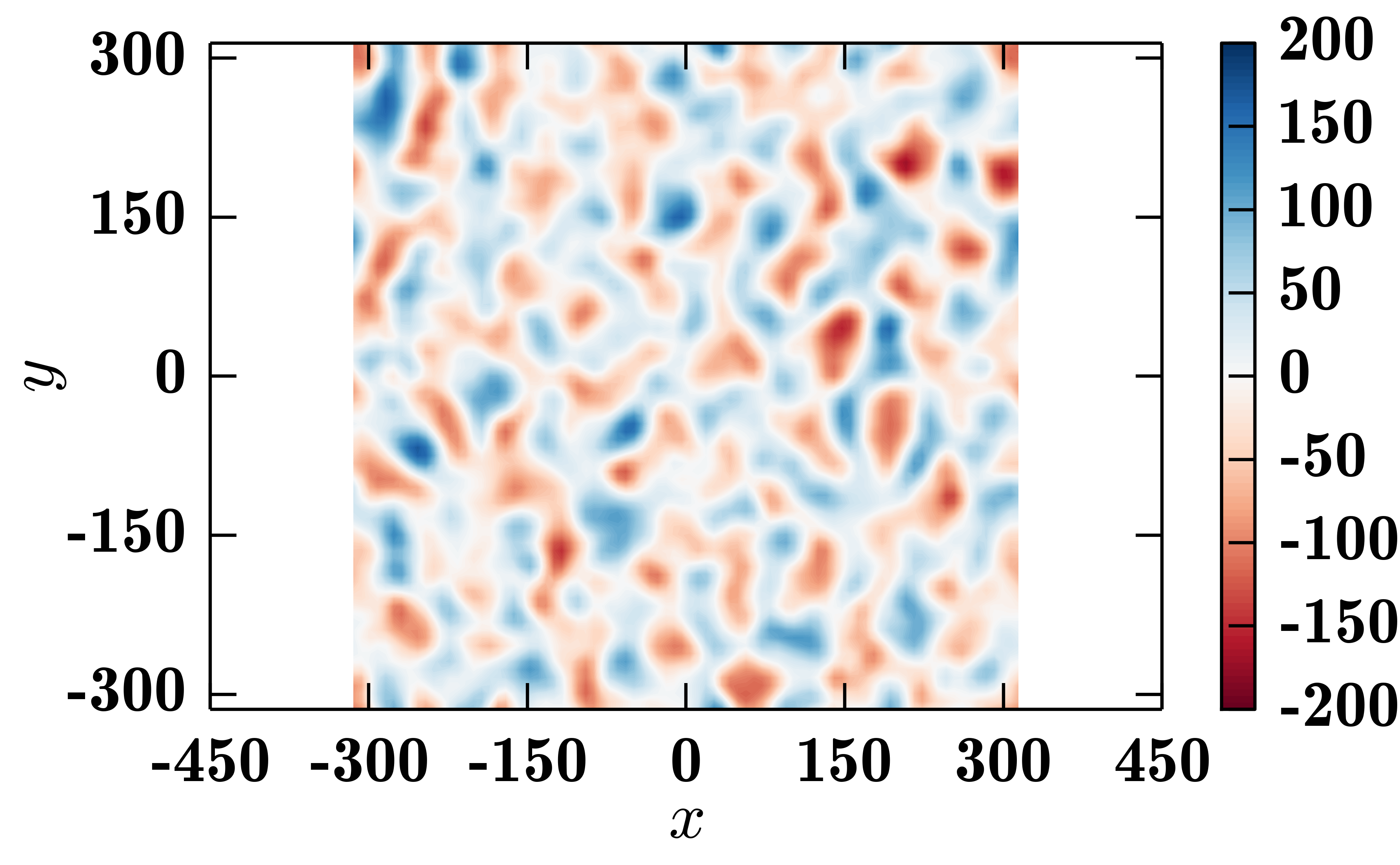}}
\caption[Isotropy of turbulence in the perpendicular plane.]{The turbulence is isotropic in the perpendicular plane. 
	(a) The spectrum of the electrostatic potential summed over $\kpara$, against $k_x$ and $k_y$; and 
	(b) the electrostatic potential summed over $z$, against $x$ and $y$.
	In both plots there is no preferred direction.}
	\label{fig:Isotropy}
\end{figure}
We begin by verifying that the turbulence is isotropic. 
In \fig\ref{fig:Isotropy}(a) we plot the electrostatic potential's spectrum summed over $\kpara$ against $k_x$ and $k_y$, 
and in \fig\ref{fig:Isotropy}(b) we plot the electrostatic potential averaged over $z$ against $x$ and $y$. 
The plots show no distinction between the two perpendicular directions.
The same behaviour is found throughout phase space.

\begin{figure}
	\centering
	\subfigure[]{\includegraphics{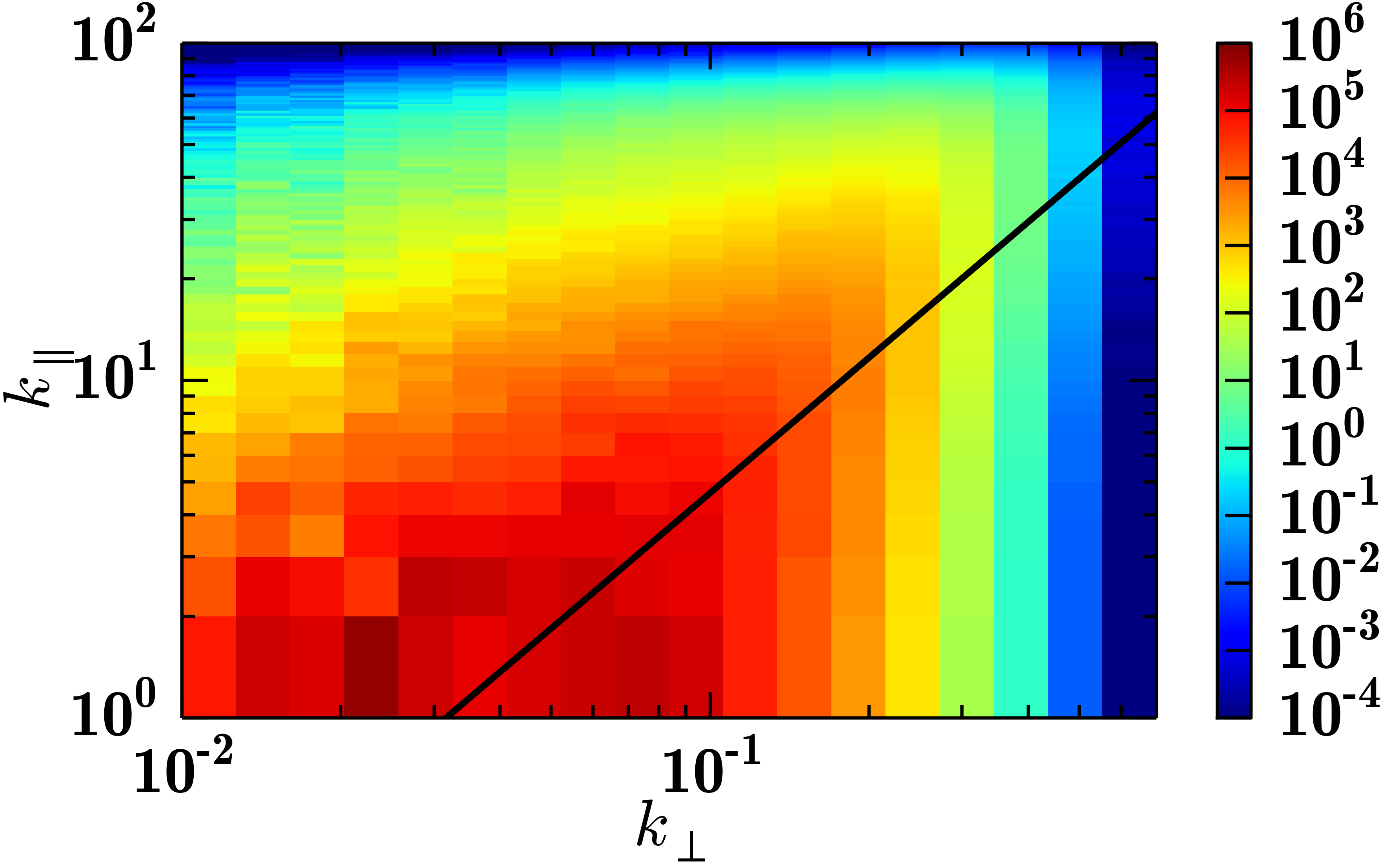}}
	%\subfigure[]{\includegraphics{images/scaling_laws/h_020_196/compensated_spectrum_phi2ta_vs_kz_summed_kperp_id_196.png}}
	%\subfigure[]{\includegraphics{images/scaling_laws/h_020_196/spectrum_phi2ta_vs_kz_summed_kperp_id_196.png}}
	%\subfigure[]{\includegraphics{images/scaling_laws/h_020_196/phi2ta_kperp_vs_kz_norm_to_max_fixed_kz_id_196.png}}
	\subfigure[]{\includegraphics{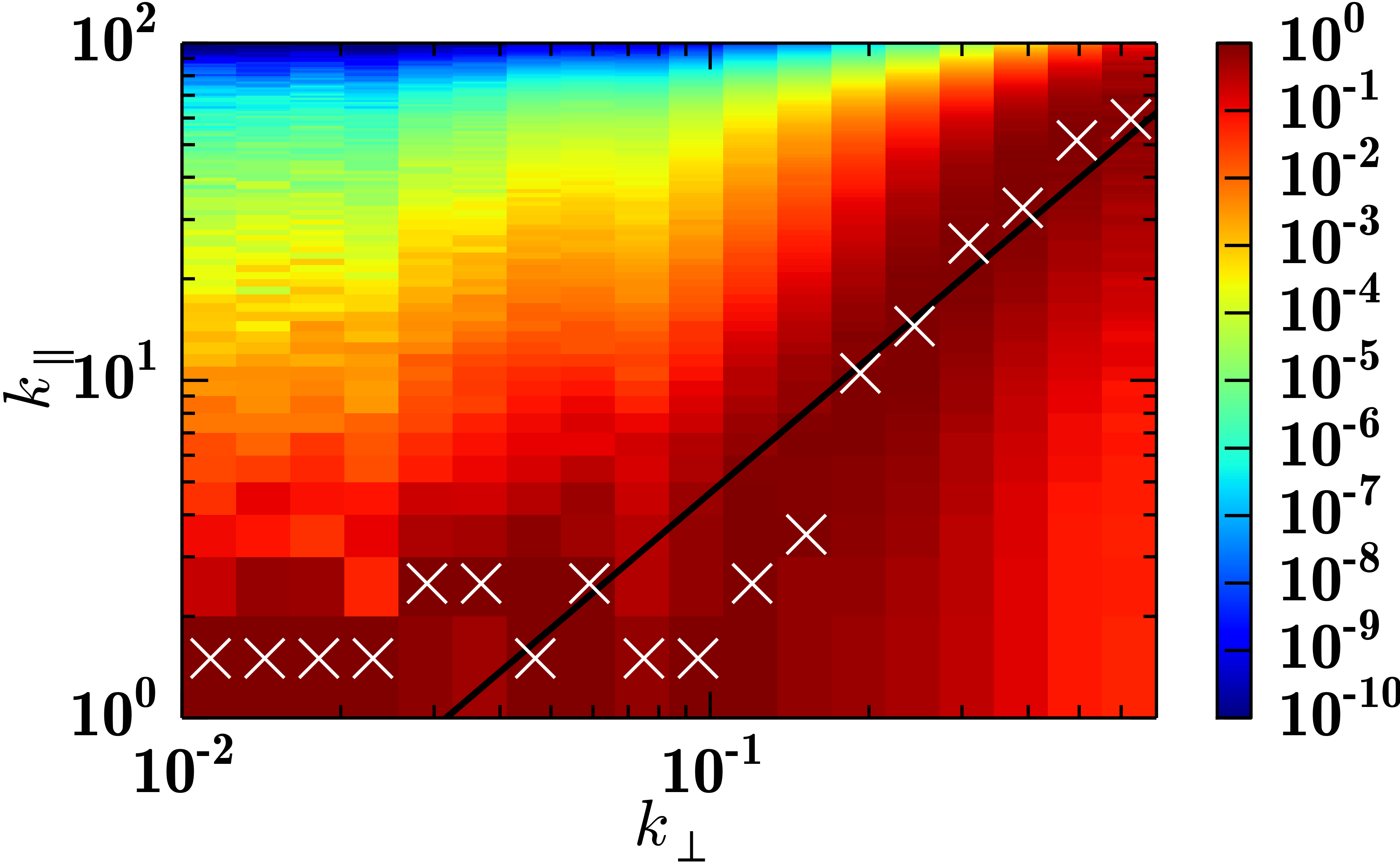}}
\caption[Critical balance for the electrostatic potential.]{Critical balance for the electrostatic potential.
		(a) The electrostatic potential against $\kpara$ and (binned) $\kperp$.
		(b) The same plot as (a), but normalized to the largest value for each fixed $\kperp$ (column).
		This selects the dominant $\kpara$ scale for each $\kperp$, which are marked with crosses.  The black line is the line of critical balance, 
		$\kpara=100\kperp^{4/3}$, a fit to these dominant scales.}
	\label{fig:CriticalBalanceESPot}
\end{figure}
We next verify critical balance. 
As noted in \sec\ref{sec:SpectrumForTheElectrostaticPotential}, the consequence of the critical balance causality argument is that the dominant $\kpara$ modes should satisfy $\kpara\sim\kperp^{4/3}$.
In \fig\ref{fig:CriticalBalanceESPot}(a) we plot the amplitude of the electrostatic potential against $\kpara$ and (binned) $\kperp$.
To show the dominant scales, in \fig\ref{fig:CriticalBalanceESPot}(b) we plot the amplitude of the electrostatic potential normalized to its maximum value at each fixed $\kperp$, that is, $|\varphik|^2/\max_{\kpara}|\varphik|^2$.
We also plot the line $\kpara= 100\kperp^{4/3}$ where the constant of proportionality was chosen to fit the dominant modes.

\subsection{Electrostatic potential}

\begin{figure}
	\centering
	%\subfigure[]{\includegraphics{images/scaling_laws/h_020_196/spectrum_phi2ta_vs_kz_summed_kperp_id_196.png}}
	%\subfigure[]{\includegraphics{images/scaling_laws/h_020_196/spectrum_phi2ta_vs_kperp_summed_kz_id_196.png}}
	\subfigure[]{\includegraphics{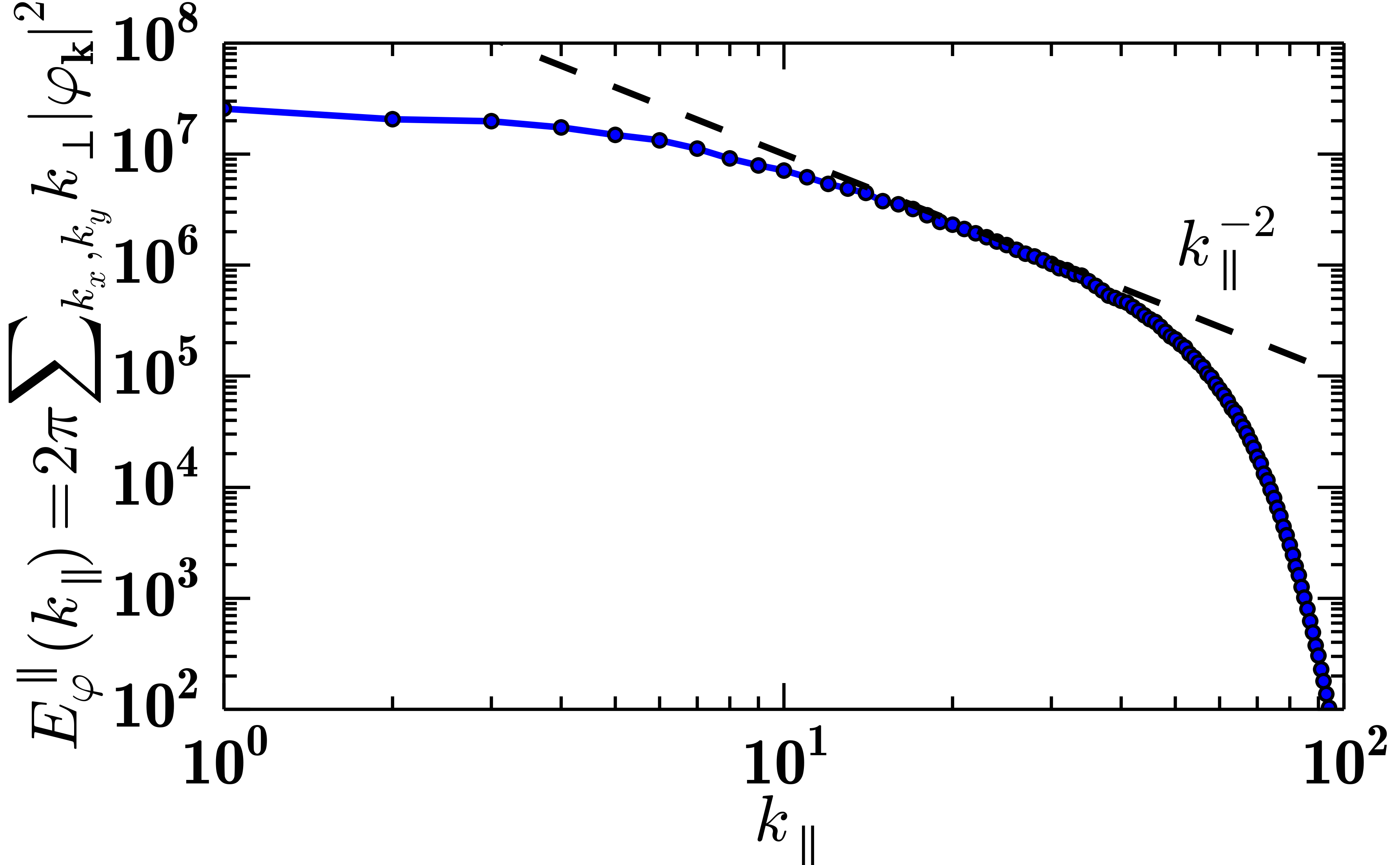}}
	\subfigure[]{\includegraphics{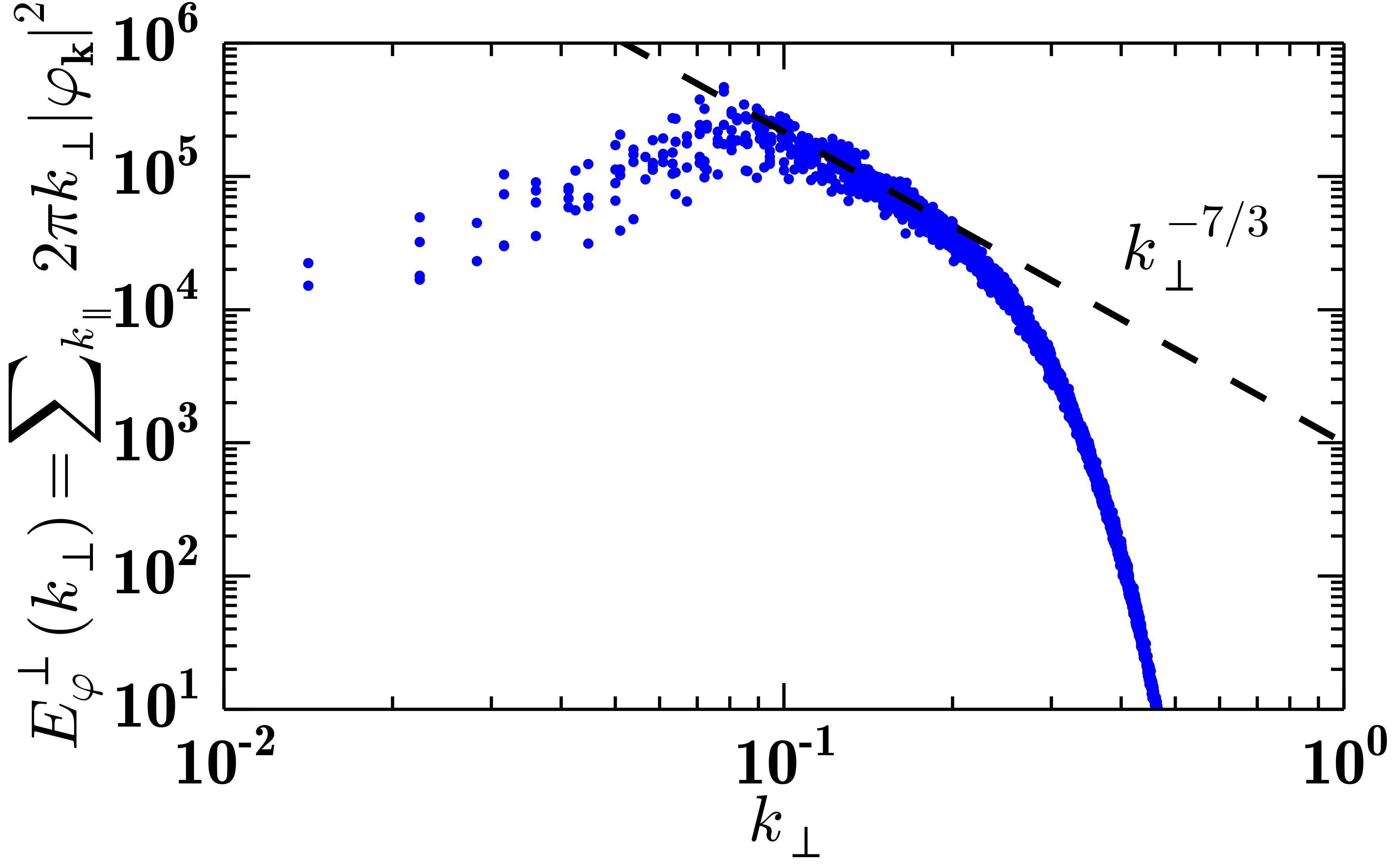}}
	\subfigure[]{\includegraphics{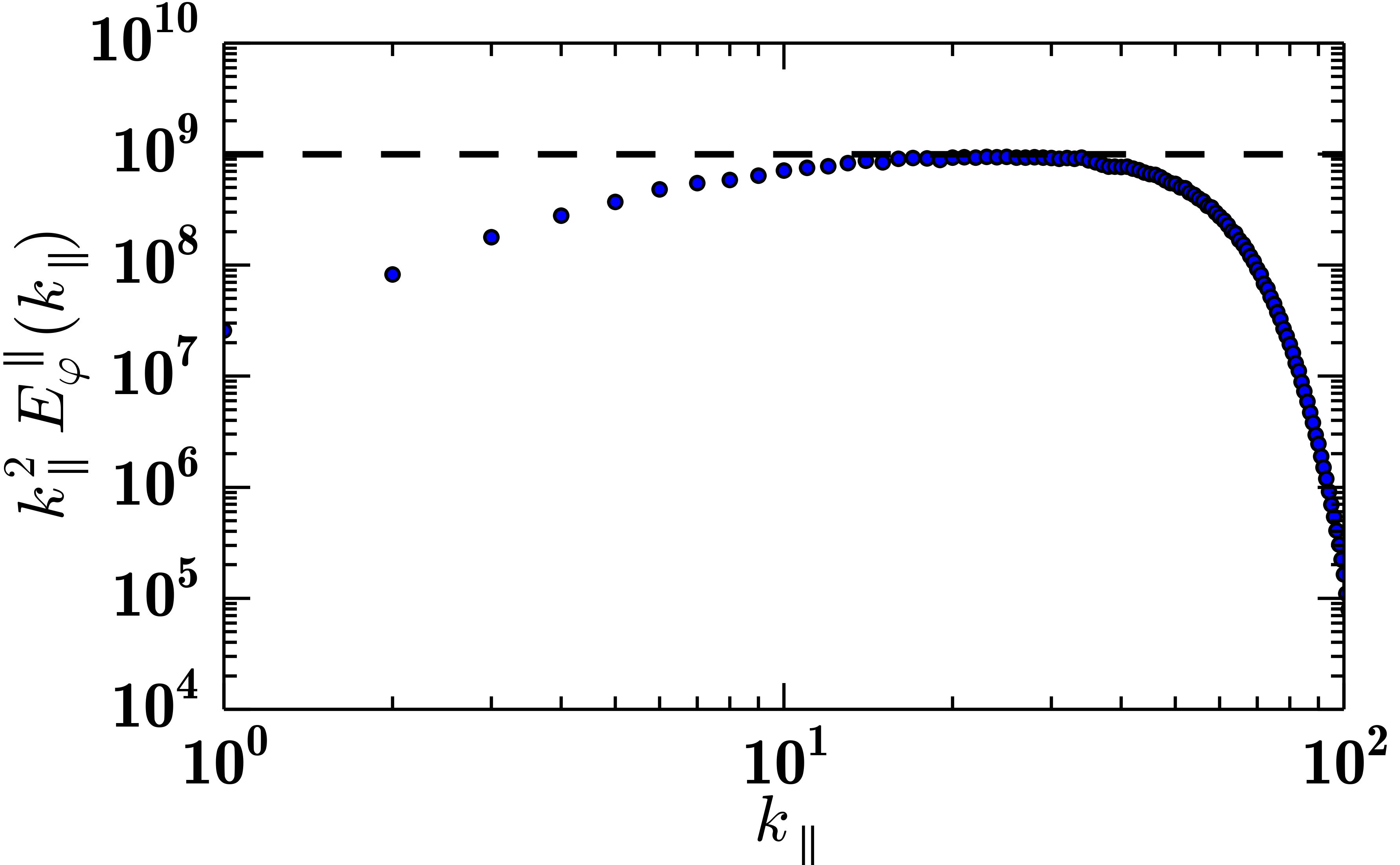}}
	\subfigure[]{\includegraphics{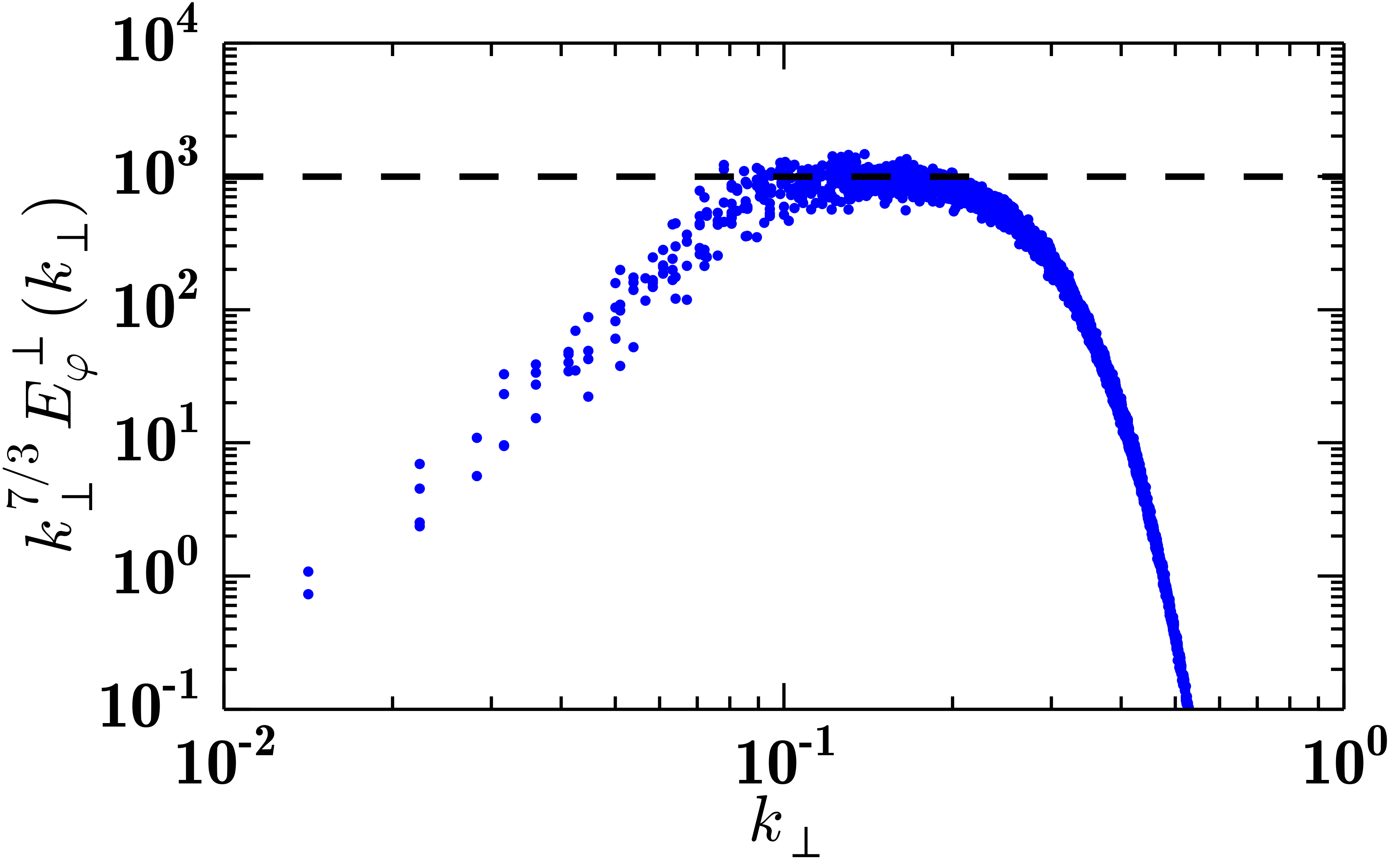}}
	\caption[Summed spectra for the electrostatic potential.]{The spectra of the electrostatic potential $\Ephi(\kpara,\kperp)$, summed over one dimension:
		(a) $\Ephi^{\parallel}(\kpara)=2\pi\sum{}_{k_x,k_y}\kperp \ta{|\varphi_{\k}|^2}$, showing the expected $\kpara^{-2}$ scaling; 
		and (b) $\Ephi^{\perp}(\kperp)=2\pi\sum{}_{\kpara}\kperp \ta{|\varphi_{\k}|^2}$, showing the expected $\kperp^{-7/3}$.
	Corresponding compensated spectra are plotted in (c) and (d).
	}
	\label{fig:PhiOneDSpecta}
\end{figure}

Having established isotropy, 
we now show spectra for the electrostatic potential
as functions of $\kpara$ and $\kperp$.
We begin by defining the one-dimensional spectra
\begin{align}
	\Ephi^{\parallel}(\kpara) = \int\d\kperp~ \Ephi(\kpara,\kperp),
	\hspace{1cm}
	\Ephi^{\perp}(\kperp) = \int\d\kpara~ \Ephi(\kpara,\kperp).
	\label{eq:OneDSpectraEPhi}
\end{align}
Putting the theoretical spectrum \eqref{eq:Ephi}
into these definitions 
and integrating
over the\linebreak energy-containing advection-dominated region $\kpara\lesssim\kperp^{4/3}$
gives
\begin{align}
	\Ephi^{\parallel}(\kpara) \sim \int_{\kpara^{3/4}}^{\infty}\d\kperp~ \kperp^0\kperp^{-11/3} \sim \kpara^{-2},
	\hspace{1cm}
	\Ephi^{\perp}(\kperp) \sim \int_0^{\kperp^{4/3}}\d\kpara~ \kperp^0\kperp^{-11/3} \sim \kperp^{-7/3}.
	\label{eq:OneDSpectraEPhiTheory}
\end{align}
In \fig\ref{fig:PhiOneDSpecta} we plot the discrete spectra
\begin{align}
	\Ephi^{\parallel}(\kpara) = \sum_{k_x,k_y} \Ephi(\kpara,\kperp),
	\hspace{1cm}
	\Ephi^{\perp}(\kperp) = \sum_{\kpara} \Ephi(\kpara,\kperp),
	\label{eq:OneDSpectraEPhiApprox}
\end{align}
which coincide with \eqref{eq:OneDSpectraEPhi} on taking $\Ephi(\kpara,\kperp)$ to be the piecewise constant function obtained from simulations in a finite box.
%%%The parallel spectrum is in excellent agreement with its theoretical scaling.
%%%The perpendicular scaling has an inertial range with the correct $\kperp^{-7/3}$ scaling for lower wavenumbers, but the spectrum decays too quickly at the higher wavenumbers $\kperp>\dots$.
%%%This decay is not due to dissipation which only affects wavenumbers $\kperp>\dots$.
%%%Instead, this effect can be understood by considering the critical balance plot, \fig\dots.
%%%The dominant scales on this plot are marked with a black line.
%%%The dominant scales are entirely contained within the $\kpara$, so that summing over $\kperp$ will always include the dominant scales in the one-dimensional spectrum.
%%%In contrast, the black line only intersects $\kpara$ for $\kpara\in[\dots,\dots]$; 
%%%the dominant scales only contribute to $\Ephi^{\perp}$ in this range,
%%%and indeed we see this matches the inertial range observed in \fig\dots.
%%%We therefore have the counterintuitive result that to improve the perpendicular spectrum, we must increase the parallel resolution;
%%%doing so, we would include more of the dominant modes in the $\kperp$ spectrum before reaching the $\kpara$ dissipation.
Both spectra are in excellent agreement with their theoretical scaling \eqref{eq:OneDSpectraEPhiTheory}.

\begin{figure}
	\centering
	%\subfigure[]{\includegraphics{images/scaling_laws/h_020_196/rainbow_spectrum_phi2ta_vs_kperp_aligned_1_id_200.png}}
	%\subfigure[]{\includegraphics{images/scaling_laws/h_020_196/rainbow_spectrum_phi2ta_vs_kperp_aligned_2_id_200.png}}
%%%	\subfigure[]{\includegraphics{images/scaling_laws/h_020_196/rainbow_spectrum_phi2ta_vs_kperp_aligned_1_id_200.pdf}}
%%%	\subfigure[]{\includegraphics{images/scaling_laws/h_020_196/rainbow_spectrum_phi2ta_vs_kperp_aligned_2_id_200.pdf}}
	%\subfigure[]{
	%}
	%\subfigure[]{
		\includegraphics{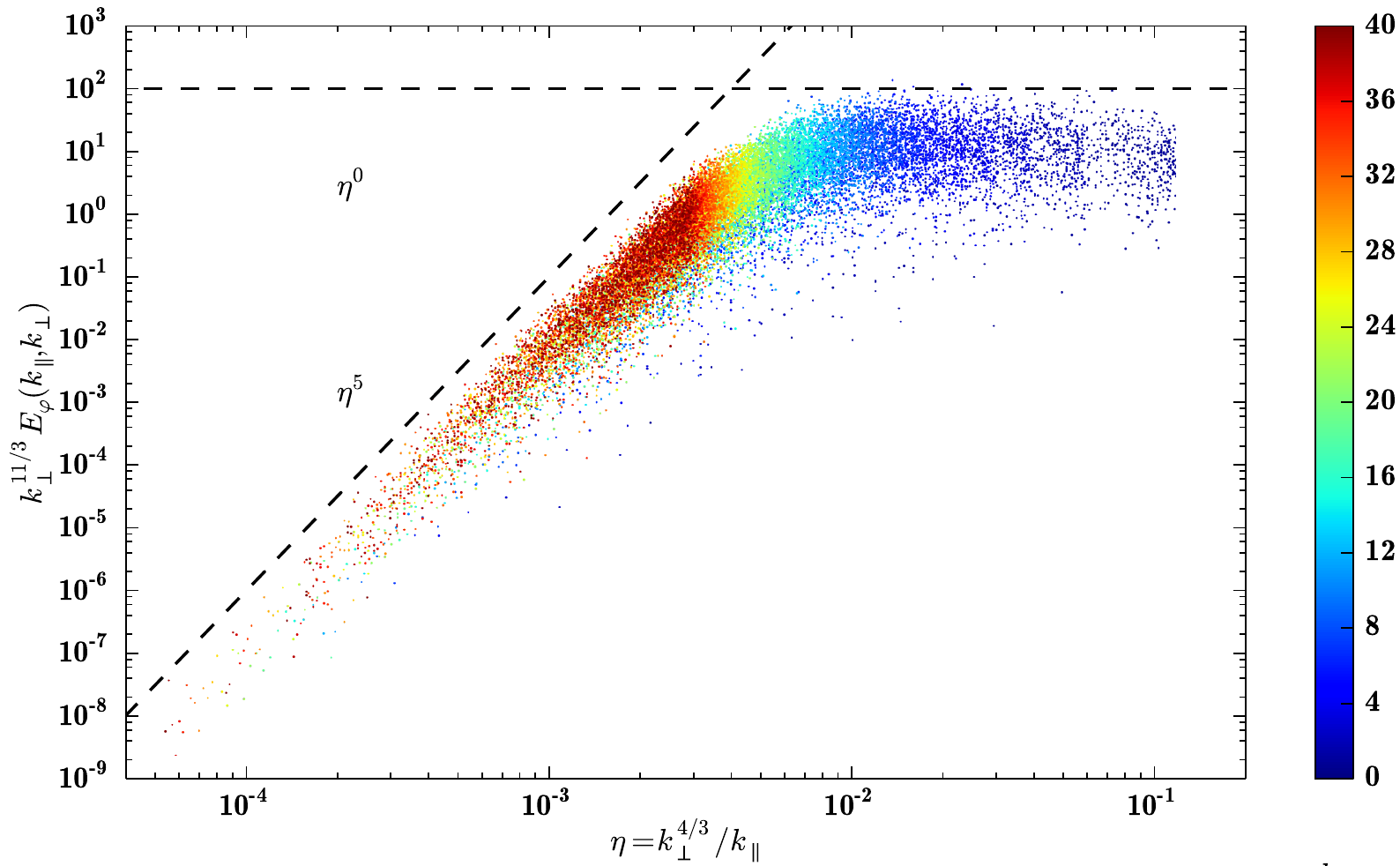}
		\includegraphics{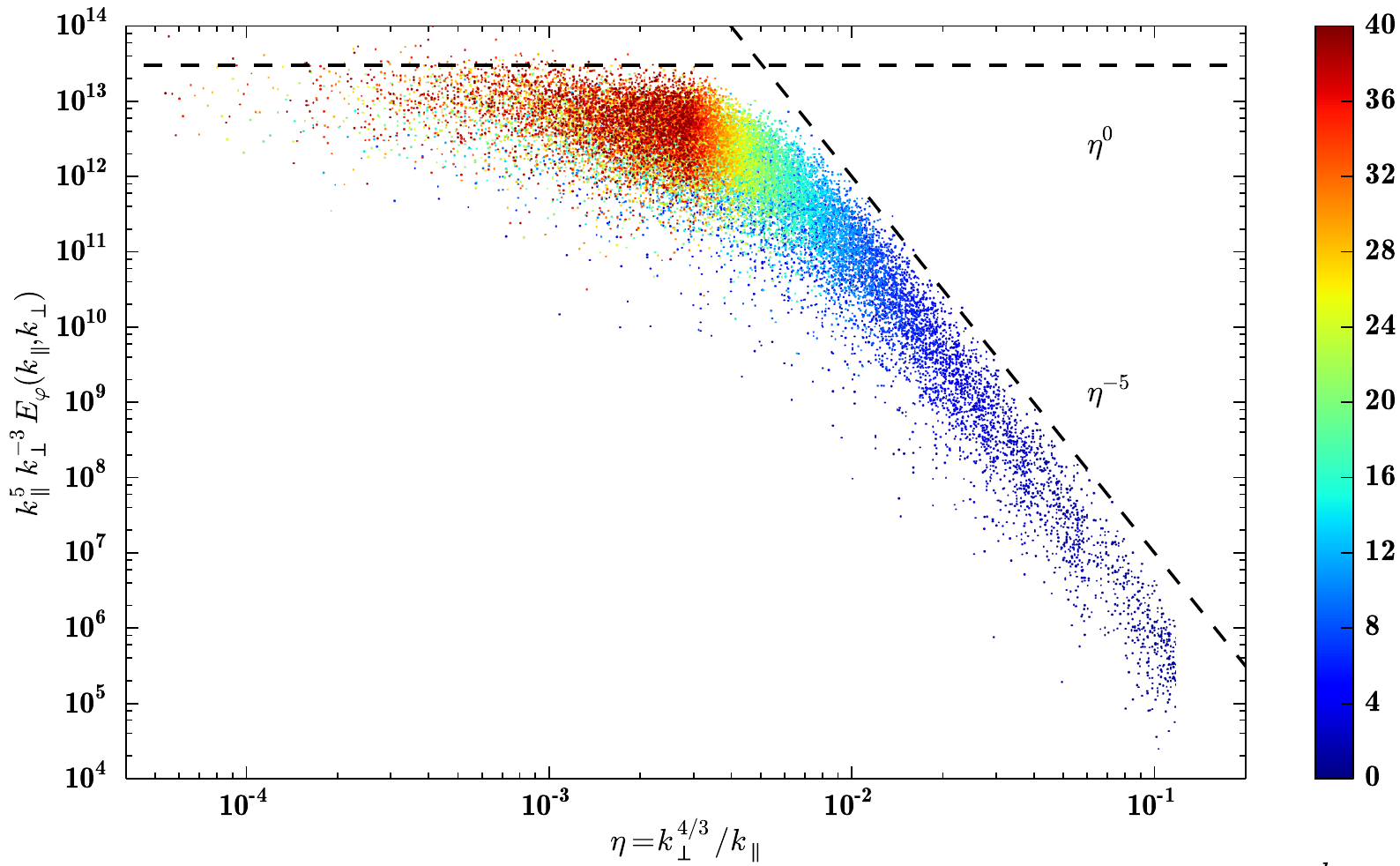}
	%}
	\caption[Spectrum $\Ephi$ as a function of $\eta=\kperp^{4/3}/\kpara$.]
	{The two-dimensional spectrum $\Ephi(\kpara,\kperp)$ as a function of the single variable $\eta=\kperp^{4/3}/\kpara$.
	The whole spectrum collapses onto a single line, as predicted in \eqref{eq:EphiSimilarityAlt} and \eqref{eq:EphiSimilarity}.}
	\label{fig:PhiTwoDSpecta}
\end{figure}

We also plot the two-dimensional spectrum $\Ephi(\kpara,\kperp)$ directly.
Rewriting \eqref{eq:Ephi} in terms of a single variable $\eta=\kperp^{4/3}/\kpara$, we obtain
\begin{align}
	\kperp^{11/3}	\Ephi(\kpara,\kperp) = 
	\begin{cases}
		\cst, 
		& 1\lesssim \eta,\\[2ex]
	 \eta^{5}, 
		& 1\gtrsim \eta,
	\end{cases}
	\label{eq:EphiSimilarityAlt}
\end{align}
and
\begin{align}
	\kpara^5\kperp^{-3}	\Ephi(\kpara,\kperp) = 
	\begin{cases}
	 \eta^{-5}, 
		& 1\lesssim \eta,\\[2ex]
		\cst, 
		& 1\gtrsim \eta.
	\end{cases}
	\label{eq:EphiSimilarity}
\end{align}
Plotting these in \fig\ref{fig:PhiTwoDSpecta},
we see two dimensional spectrum collapses onto a single line with the expected scalings in $\eta$.
This demonstrates that all points in the electrostatic potential satisfy the expected scalings.

\subsection{Distribution function}

We now verify the scaling laws for the spectra of the forwards and backwards propagating modes, $E^{+}_m$ \eqref{eq:Eplusm} and $E^-_m$ \eqref{eq:Eminusm}.
In the advection-dominated region, $\kpara\lesssim\kperp^{4/3}$, the spectra have the same scalings,
and indeed for there to be no free energy flux, we must have $E^+_m\approx E^-_m$.
However, in $\kpara\gtrsim \kperp^{4/3}$ the spectra differ. 
The forwards spectrum has two regions:
a phase-mixing dominated region $\kpara\gtrsim\sqrt{m}\kperp^{4/3}$,
and an intermediate region $\kperp^{4/3}\lesssim \kpara \lesssim \sqrt{m}\kperp^{4/3}$ which bridges between the advection-dominated and phase-mixing dominated regimes.
This region becomes larger with increasing $m$.
In contrast, the spectrum for backwards propagating modes only has one scaling in $\kpara\gtrsim \kperp^{4/3}$.

\subsubsection{Distribution function contours and normalized free energy transfer}

\begin{figure}
	\centering
	\subfigure[]{
		\includegraphics{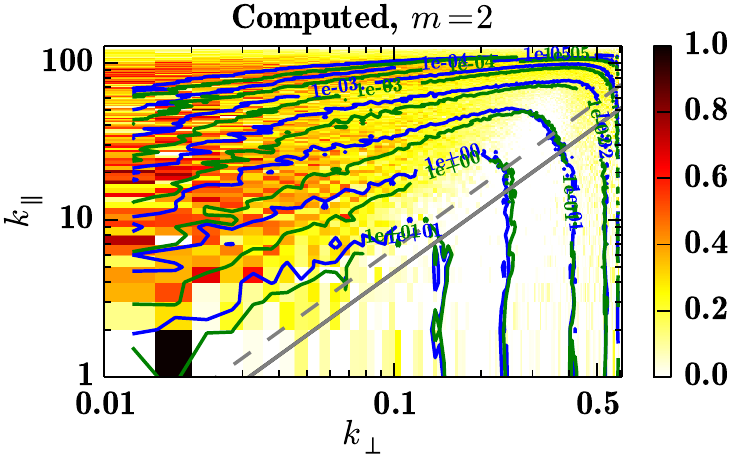}%}
	%\subfigure[]{
		\includegraphics{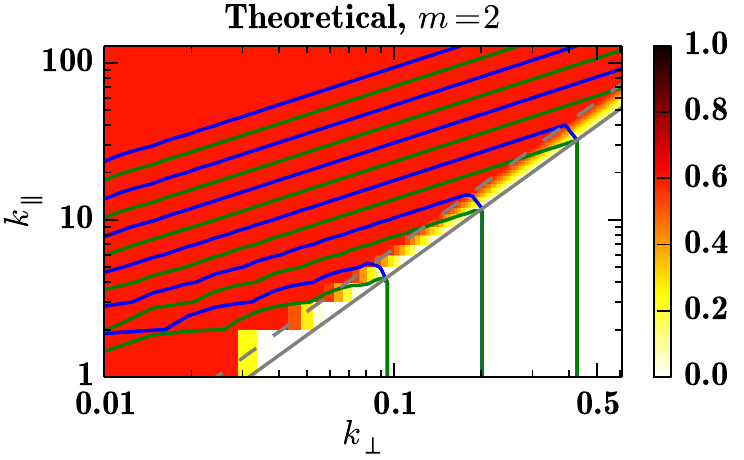}}
	\subfigure[]{
		\includegraphics{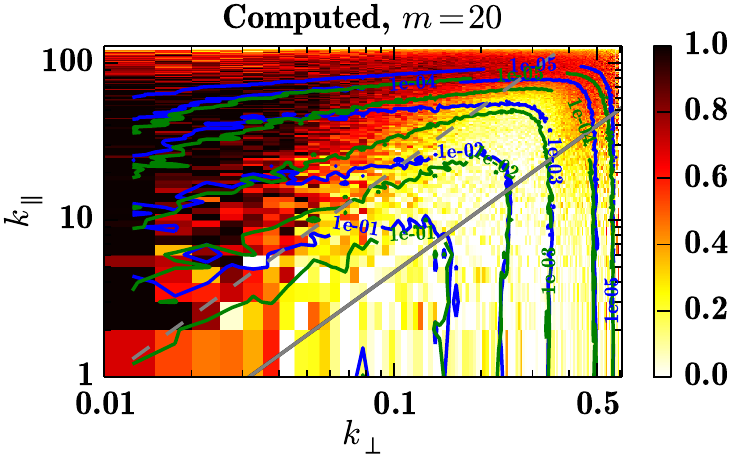}%}
	%\subfigure[]{
		\includegraphics{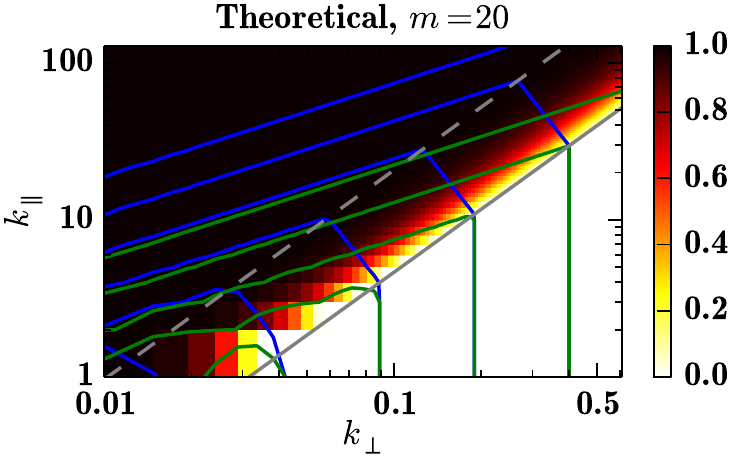}}
	\subfigure[]{
		\includegraphics{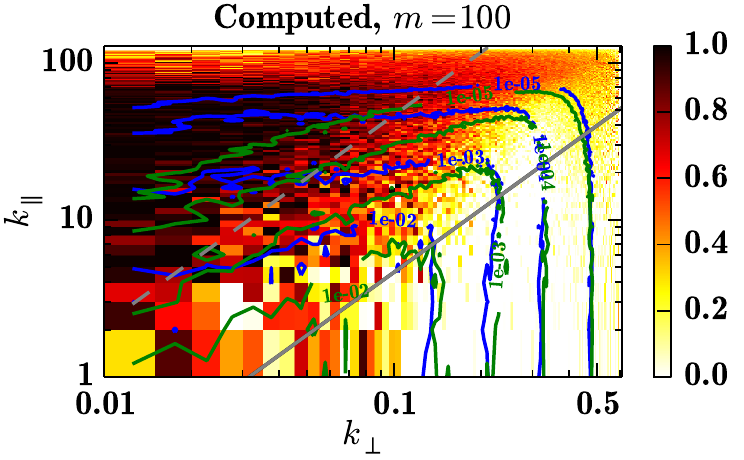}%}
	%\subfigure[]{
		\includegraphics{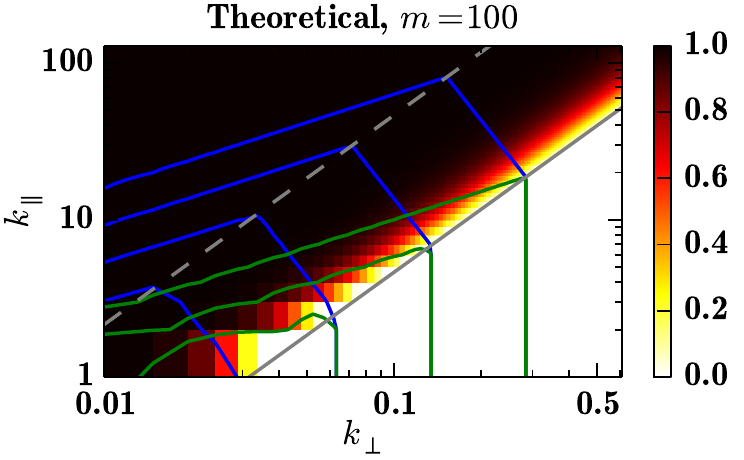}}
	\caption[Contours of $E^{\pm}_m$ and the normalized free energy flux.]{Contours of the spectra $E^+_m$ (blue) and $E^-_m$ (green), superimposed on the normalized free energy transfer $\bar{\Gamma}_m \approx (E^+_m-E^-_m)/(E^+_m+E^-_m)$ %\eqref{eq:FreeEnergyTransferNormalized} 
	for fixed $m$: (a) $m=2$, (b) $m=20$, (c) $m=100$.
	Computed spectra are shown on the left, while theoretical spectra from \eqref{eq:Eplusm} and  \eqref{eq:Eminusm} are shown on the right.
In the advection-dominated region $\kpara\lesssim 100\kperp^{4/3}$ (below the critical balance line in solid grey),
contours of $E^+_m$ and $E^-_m$ coincide and there is no free energy transfer, $\bar{\Gamma}_m\approx0$.
	Above the critical balance line, the contours separate. 
	Contours of $E^-_m$ slope downwards, while contours of $E^+_m$ continue upwards, before turning downwards at the phase-mixing threshold, $\kpara\lesssim 100\sqrt{m}\kperp^{4/3}$, the dashed line.
	As $E^+_m$ and $E^-_m$ are not equal, there is a free energy transfer, $\bar{\Gamma}_m\neq0$.
	For most $m$, the separation between contours is large, so that $\bar{\Gamma}_m\approx1$. 
	For small $m$ however the separation is small, and $\bar{\Gamma}_m<1$.
	%Note that the plots of the theoretical spectra use the same resolution as in simulations to indicate discretization effects at small wavenumbers.
	\label{fig:EplusminusContours}
}
\end{figure}

In \fig\ref{fig:EplusminusContours} we plot contours of $E^+_m$ (in white) and $E^-_m$ (in red) against $\kperp$ and $\kpara$ at $m=2$, $m=20$ and $m=100$.
The contours are superimposed on the normalized free energy transfer
\begin{equation}
  \begin{split}
		\bar{\Gamma}_m 
		= \frac{\Gamma_m}{\Gamma^L_m} 
		&= \frac{\kpara\vth\sqrt{(m+1)/2}\ \Im(\gp^*\g)}{\kpara\vth\sqrt{(m+1)/2}\ |\g|^2}
%%%		\\ 		&
		=\frac{|\g^+|^2-|\g^-|^2}{|\g^+|^2+|\g^-|^2}
		\approx \frac{E^+_m-E^-_m}{E^+_m+E^-_m}.
 %   \label{eq:FreeEnergyTransferNormalized}
  \end{split}
\end{equation}
The plots in the left-hand column show computed spectra, while plots in the right-hand column are produced using the theoretical spectra \eqref{eq:Eplusm} and \eqref{eq:Eminusm}. %, with $\bar{\Gamma}\approx (E^+_m - E^-_m)/(E^+_m + E^-_m)$ \cite{Schekochihin15}.
The line of critical balance $\kpara=100\kperp^{4/3}$ that we determined earlier from the electrostatic potential is marked in black.
The expected behaviour is shown by the theoretical spectra in the right-hand column.
In the advection-dominated region $\kpara \lesssim 100\kperp^{4/3}$,
the spectra are equal, $E^+_m = E^-_m$, so $\bar{\Gamma}_m = 0$.
Above the critical balance line, the contours separate with the contours of $E^-_m$ immediately sloping downwards.
In contrast, the contours of $E^+_m$ slope upwards until reaching the phase-mixing threshold, $\kpara=100\sqrt{m}\kperp^{4/3}$, and then slope downwards with the same gradient as $E^-_m$.
Above the critical balance line, $E^+_m$ is much larger than $E^-_m$ so $\bar{\Gamma}_m\approx 1$.
As the theoretical spectra are discontinuous, all corners are sharp. 
Therefore the contours separate rapidly at the critical balance line and there is a rapid transition from $\bar{\Gamma}_m=0$ to $\bar{\Gamma}_m\approx 1$.
Notice that the theoretical spectra at small $m$ (\fig\ref{fig:EplusminusContours}(b)) also predicts suppression of free energy transfer everywhere in phase space.
This is because for small $m$ the contours do not fully separate so that $\bar{\Gamma}_m$ is a constant that is noticeably less than one.

The computed spectra (in the left-hand column) replicate the key properties of the theoretical spectra, only now all transitions are slower since the computed spectra are smooth.
As before, the spectra coincide in the advection-dominated region, and separate at the line of critical balance.
The $E^-_m$ spectrum slopes downwards with the same gradient as the theoretical spectrum.
The $E^+_m$ also has the correct downwards slope, but without the marked upwards slope that appears in the theoretical spectrum. 
However this is the only difference in their behaviours, which are otherwise in excellent agreement.

We now verify the spectral exponents by plotting $E^+_m$ and $E^-_m$ against one of $\kperp$, $\kpara$ and $m$ 
in \figs\ref{fig:EplusminusVsKperp}, \ref{fig:EplusminusVsKpara} and \ref{fig:EplusminusVsM} respectively, holding the other two variables fixed.
In each Figure, we plot the spectra for $E^+_m$ and $E^-_m$ in the left- and right-hand columns respectively and, where possible, use the same axis limits for both plots.

\subsubsection{$E^{\pm}_m$ versus $\kperp$}

\begin{figure}
	\centering
	\subfigure[]{\includegraphics{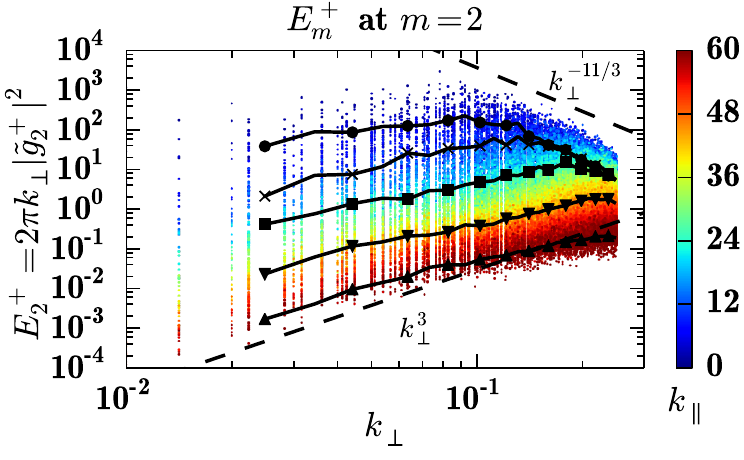}%}
	%\subfigure[]{
	\includegraphics{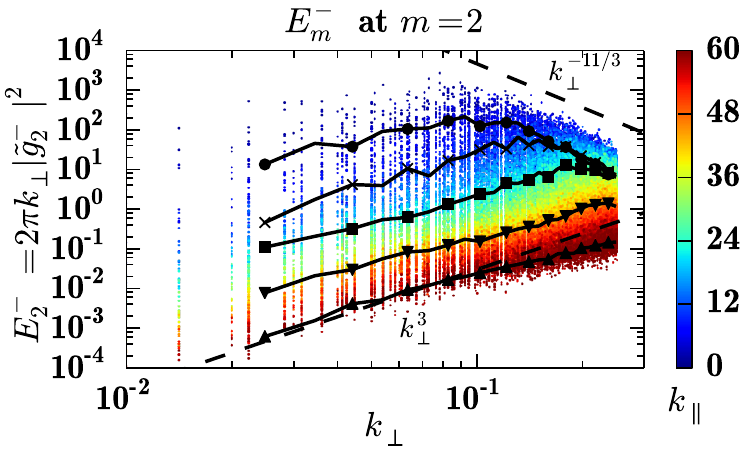}}
	%\subfigure[]{\includegraphics{images/scaling_laws/h_020_196/rainbow_spectrum_gplus2ta_vs_kperp_fixed_kz_m_20_id_200.pdf}}
	%\subfigure[]{\includegraphics{images/scaling_laws/h_020_196/rainbow_spectrum_gminus2ta_vs_kperp_fixed_kz_m_20_id_200.pdf}}
	\subfigure[]{\includegraphics{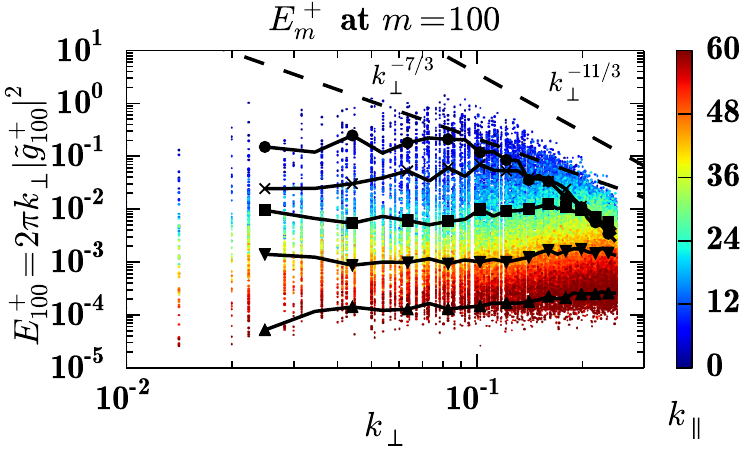}%}
	%\subfigure[]{
	\includegraphics{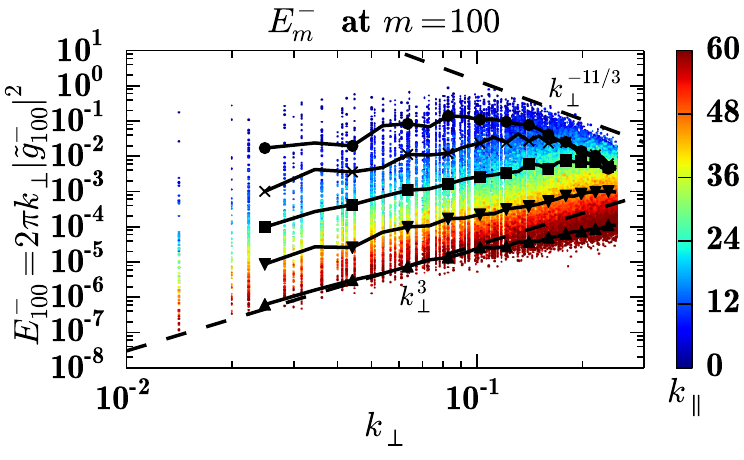}}
	\subfigure[]{\includegraphics{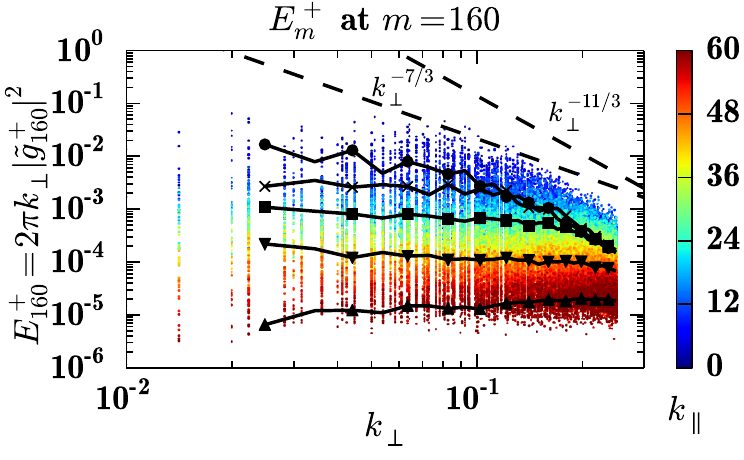}%}
	%\subfigure[]{
	\includegraphics{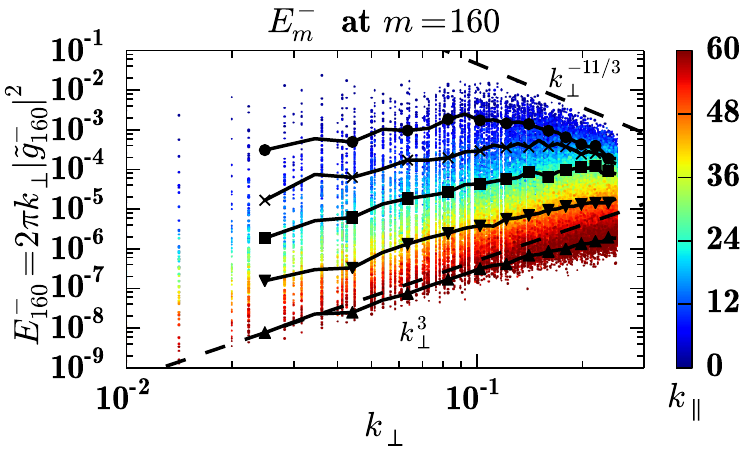}}
	\caption[Spectra of $E^{\pm}_m$ against $\kperp$ at fixed $\kpara$ and $m$.]{Spectra of $E^+_m$ (left column) and $E^-_m$ (right column) against $\kperp$ at fixed $\kpara$ and $m$: (a) $m=2$, (b) $m=100$, (c) $m=160$. 
		Values of $\kpara$ are shown in the colour bar. 
		The five black lines on each plot are bin-averaged spectra at fixed $\kpara$: from top to bottom, $\kpara=4$, $\kpara=10$, $\kpara=20$, $\kpara=40$ and $\kpara=60$.
		The $E^-_m$ spectrum \eqref{eq:Eminusm} has the same scaling in Fourier space for each $m$, so each row shows the same behaviour but with a different overall amplitude.  
For $\kpara\gtrsim\kperp^{4/3}$, $E^-_m\sim\kperp^3$, while for  $\kpara\lesssim\kperp^{4/3}$, $E^-_m\sim\kperp^{-11/3}$.
The positive spectrum \eqref{eq:Eplusm} also scales as $E^+_m\sim\kperp^{-11/3}$ for $\kpara\lesssim\kperp^{4/3}$, but now the $E^+_m\sim\kperp^3$ scaling is confined to $\kpara\gtrsim\sqrt{m}\kperp^{4/3}$. 
Between these regions, $\kperp^{4/3}\lesssim\kpara\lesssim\sqrt{m}\kperp^{4/3}$, we see $E^+_m\sim\kperp^{-7/3}$ develop in the high $m$ plots.
	\label{fig:EplusminusVsKperp}
}
\end{figure}

In \fig\ref{fig:EplusminusVsKperp}, we plot $E^+_m$ and $E^-_m$ against $\kperp$ at fixed $\kpara$ and $m$.
%These are scatter plots as the distribution function is computed on a $(k_x, k_y)$ grid, from which we construct $\kperp=\sqrt{k_x^2+k_y^2}$. 
The colour indicates the value of $\kpara$.  
The black lines are spectra at a fixed $\kpara$ plotted against bin-averaged $\kperp$. 
From top to bottom, the $\kpara$ values for these spectra are $\kpara=4$, $\kpara=10$, $\kpara=20$, $\kpara=40$ and $\kpara=60$.
The three rows show plots for $m=2$, $m=100$ and $m=160$ respectively.

First consider the $E^-_m$ spectra (right column).
The theoretical spectrum \eqref{eq:Eminusm} has two scalings, a spectrum like $\kperp^3$ for $\kpara\gtrsim\kperp^{4/3}$ and like $\kperp^{-11/3}$ for $\kpara\lesssim\kperp^{4/3}$.
While the overall spectrum decreases like $m^{-5/2}$, the shape of the spectrum in Fourier space does not change with $m$.
These properties are observed in the plots.
The spectra are all alike, but decrease in magnitude with $m$.
All spectra behave like $\kperp^3$ at low $\kperp$,
and a there is a good $\kperp^{-11/3}$ scaling at high $\kperp$ for the $\kpara=4$ and $\kpara=10$ spectra.
Moreover, the maximum points of these spectra agree well with their expected positions on the critical balance line, $\kpara=100\kperp^{4/3}$.

The $E^+_m$ spectrum \eqref{eq:Eplusm} is similar to $E^-_m$ at low $m$, but at high $m$ develops an additional $\kperp^{-7/3}$ spectrum in the intermediate region $\kpara^{3/4}m^{-3/8}\lesssim \kperp \lesssim \kpara^{3/4}$.
We see this in the calculated spectra in the left-hand column.
At $m=2$ (\fig\ref{fig:EplusminusVsKperp}(a)) the spectrum is similar to that of $E^-_2$, with two distinct scalings $\kperp^3$ and $\kperp^{-11/3}$.
As $m$ increases, we no longer see the $\kperp^3$ behaviour at low $\kperp$, but rather see a $\kperp^{-7/3}$ spectrum develop.

\subsubsection{$E^{\pm}_m$ versus $\kpara$}

\begin{figure}
	\centering
%%%	\subfigure[]{\includegraphics{images/scaling_laws/h_020_196/spectrum_gplus2ta_vs_kz_fixed_kperp_m_1_id_200.png}}
%%%	\subfigure[]{\includegraphics{images/scaling_laws/h_020_196/spectrum_gminus2ta_vs_kz_fixed_kperp_m_1_id_200.png}}
%%%	\subfigure[]{\includegraphics{images/scaling_laws/h_020_196/spectrum_gplus2ta_vs_kz_fixed_kperp_m_10_id_200.png}}
%%%	\subfigure[]{\includegraphics{images/scaling_laws/h_020_196/spectrum_gminus2ta_vs_kz_fixed_kperp_m_10_id_200.png}}
	\subfigure[]{\includegraphics{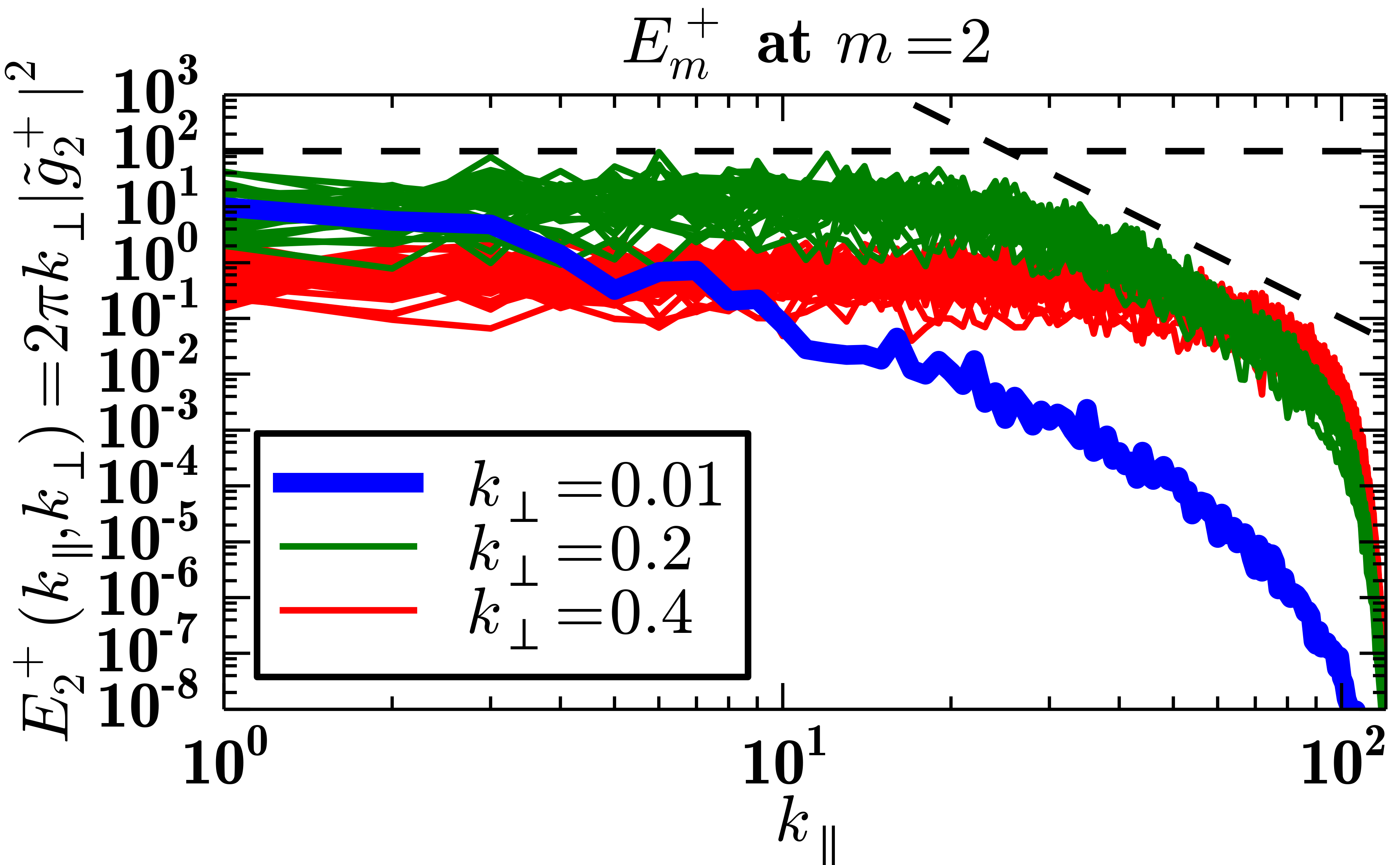}%}
	%\subfigure[]{
	\includegraphics{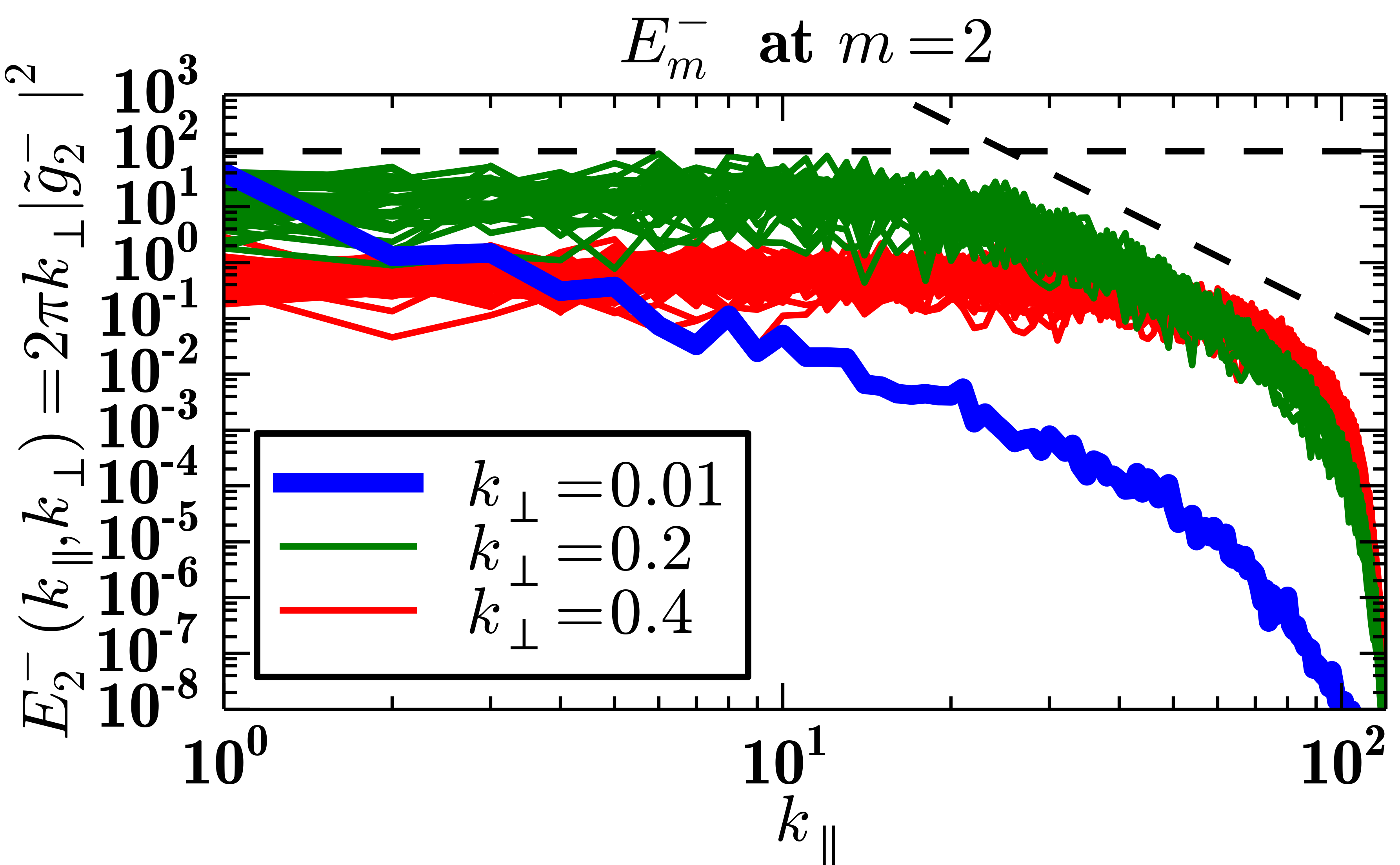}}
	\subfigure[]{\includegraphics{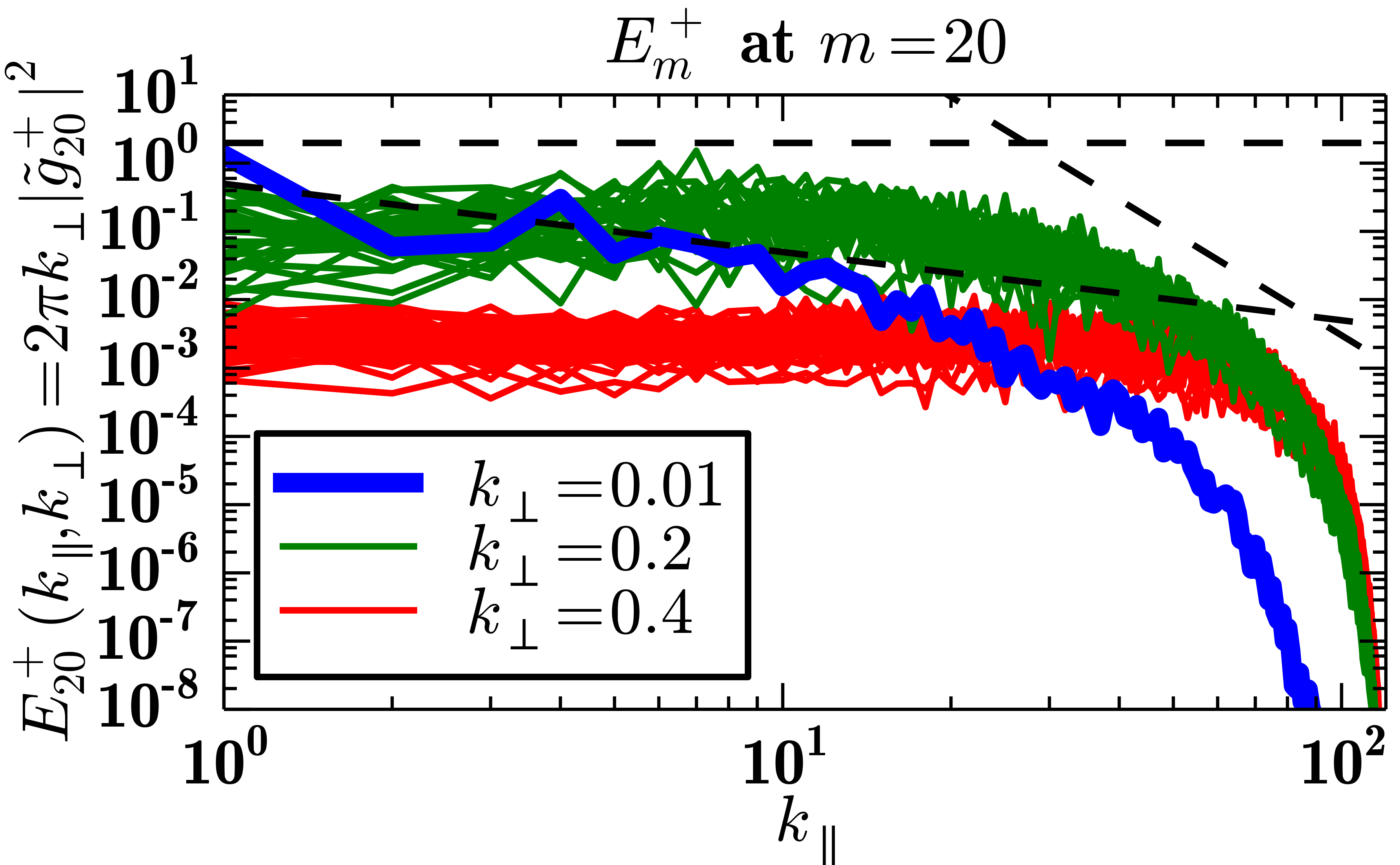}%}
	%\subfigure[]{
	\includegraphics{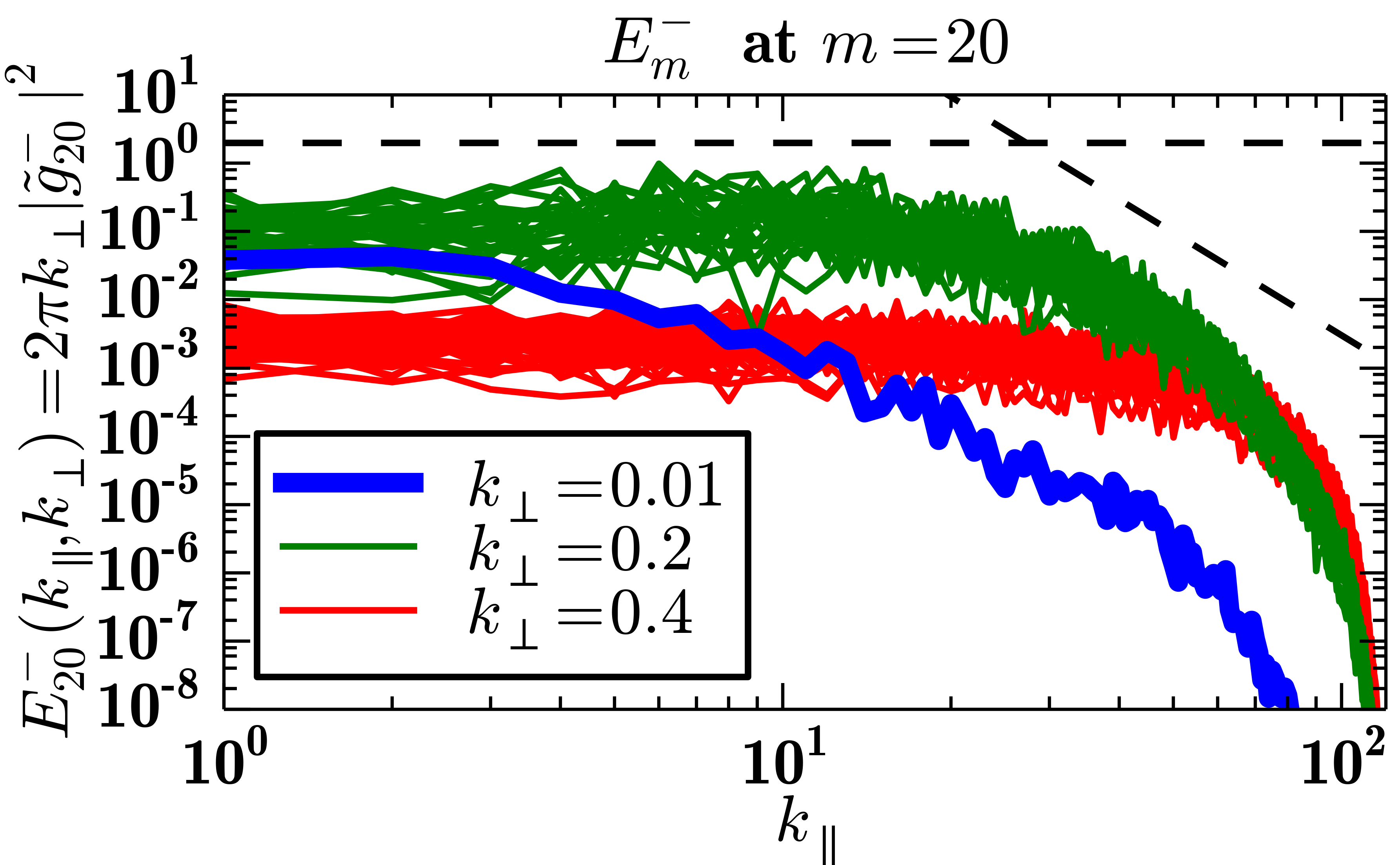}}
	\subfigure[]{\includegraphics{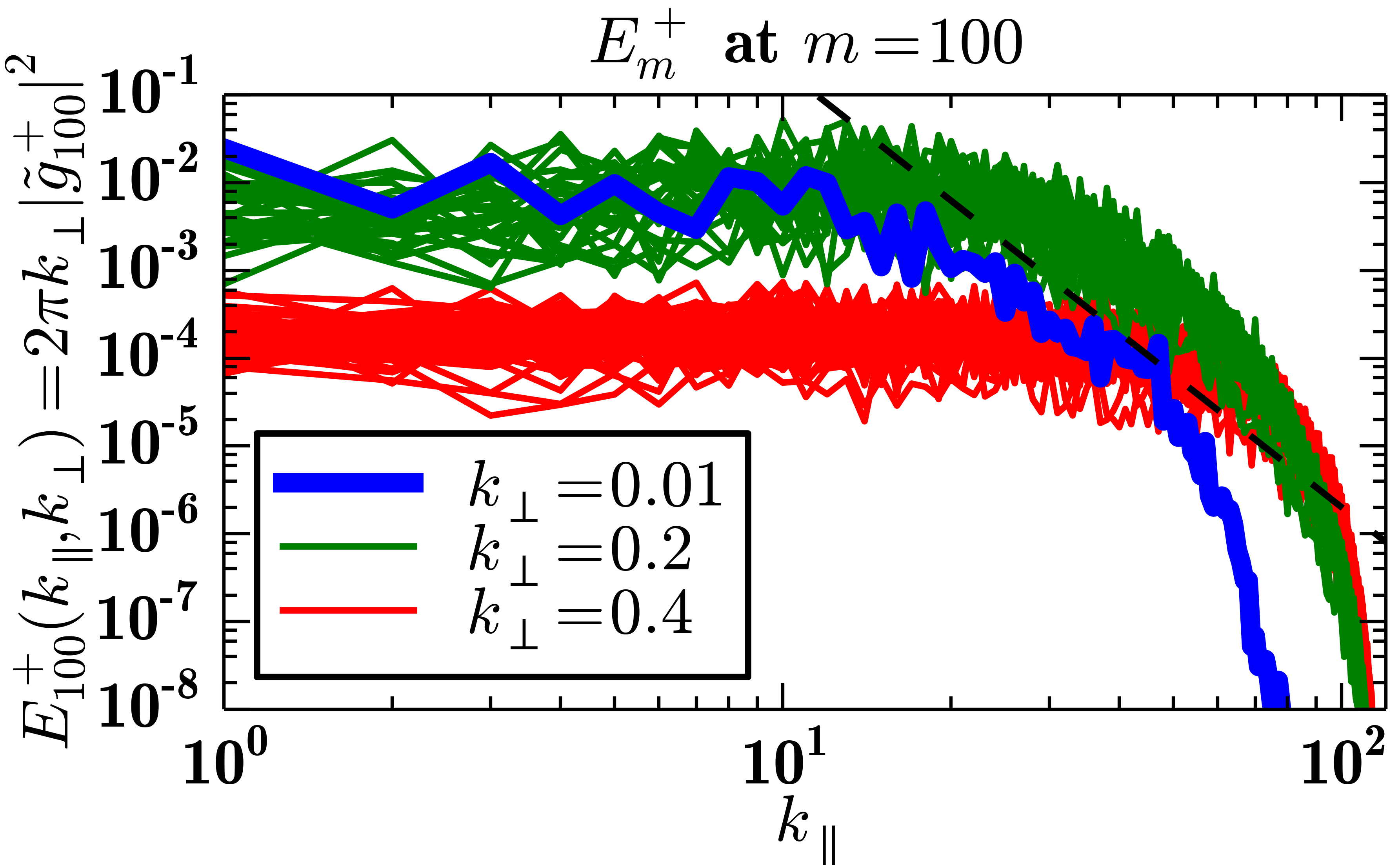}%}
	%\subfigure[]{
	\includegraphics{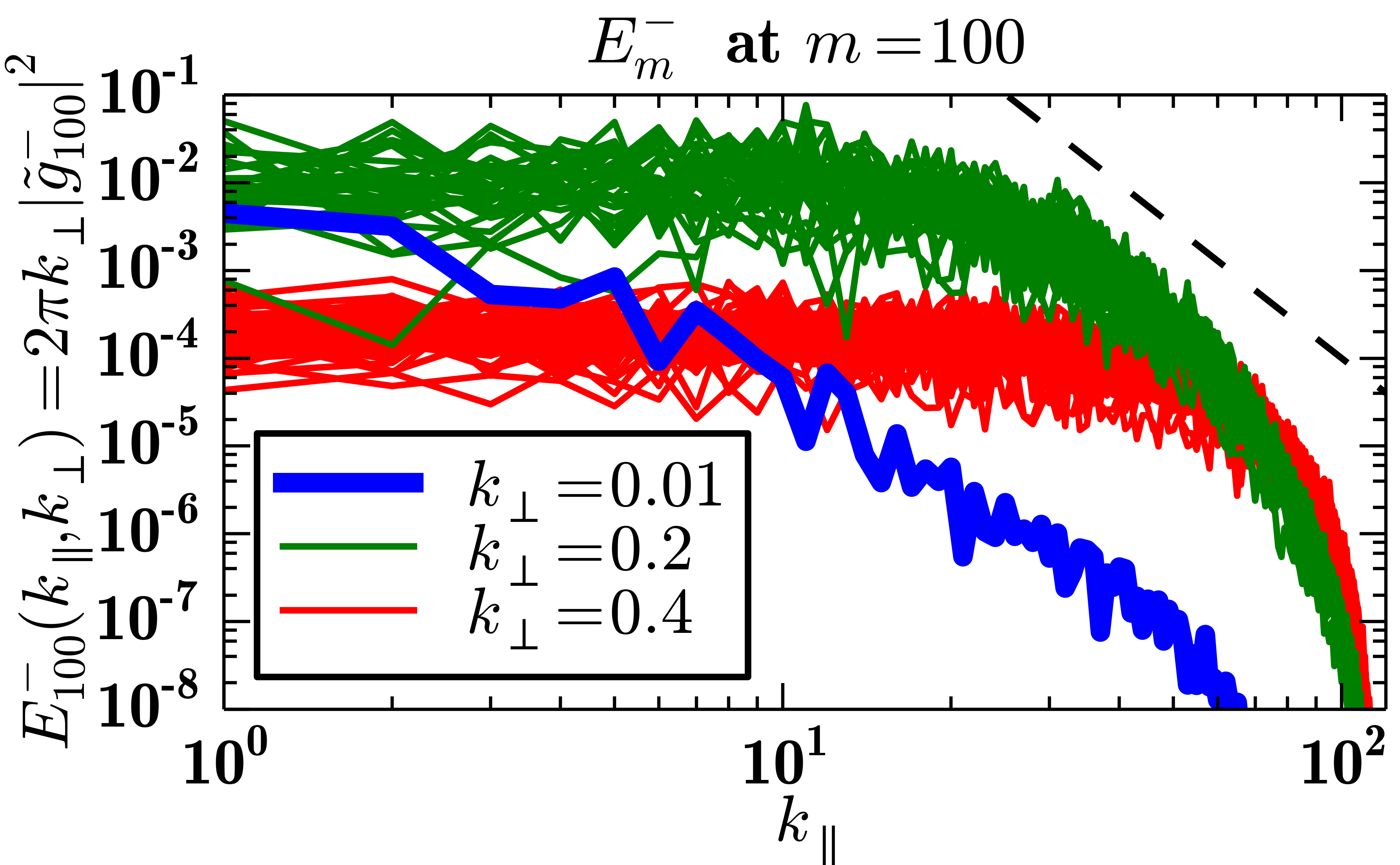}}
	\caption[Spectra of $E^{\pm}_m$ against $\kpara$ at fixed $\kperp$ and $m$.]{Spectra of $E^+_m$ (left column) and $E^-_m$ (right column) against $\kpara$ at fixed $\kperp$ and $m$: (a) $m=2$, (b) $m=20$, (c) $m=100$. 
	  Colour denotes selections of modes with $k_x$, $k_y$ such that $\sqrt{k_x^2+k_y^2}$ approximately equals the given $\kperp$.
		The negative spectrum $E^-_m$ \eqref{eq:Eminusm} has two scalings, shown by the dashed guidelines: 
		$E^-_m\sim\kpara^{-5}$ in $\kpara\gtrsim\kperp^{4/3}$ and $E^-_m\sim\kpara^{0}$ in $\kpara\lesssim\kperp^{4/3}$. 
The positive spectrum \eqref{eq:Eplusm} has three regions:
$E^+_m\sim\kpara^{-5}$ in $\kpara\gtrsim\sqrt{m}\kperp^{4/3}$, 
$E^+_m\sim\kpara^{-1}$ in $\kperp^{4/3}\lesssim\kpara\lesssim\sqrt{m}\kperp^{4/3}$, 
and $E^+_m\sim\kpara^{0}$ in $\kpara\lesssim\kperp^{4/3}$. 
		\label{fig:EplusminusVsKpara}
}
\end{figure}

We next plot the spectra against $\kpara$ at fixed $\kperp$ and $m$ in \fig\ref{fig:EplusminusVsKpara}.
Colour denotes different $\kperp$. 
In each \fig we plot a selection of modes with different $k_x$ and $k_y$ with $\sqrt{k_x^2+k_y^2}$ approximately equal  to the desired $\kperp$.
For $\kperp=0.01$, we plot the mode $(k_x,k_y)=(0,0.01)$, 
while for $\kperp=0.2$ and $\kperp=0.4$, we plot modes with wavenumbers in the range $[0.2,0.21]$ and $[0.4,0.41]$ respectively.
As before, different rows show different $m$: panel (a) shows $m=2$, (b) shows $m=20$, and (c) shows $m=100$. 

The $E^-_m$ spectra (right column) exhibit the expected behaviour: each spectrum is flat for $\kpara\lesssim\kperp^{4/3}$, 
and has a $\kpara^{-5}$ decay for $\kpara\gtrsim\kperp^{4/3}$.
As with \fig\ref{fig:EplusminusVsKperp}, only the overall amplitude of each spectrum depends on $m$, and the three \figs in the right column
are essentially the same plots but with different amplitudes.

The $E^+_m$ spectra also have $\kpara^0$ and $\kpara^{-5}$ scalings at low and high $\kpara$ respectively,
as shown in \Fig\ref{fig:EplusminusVsKpara}.
In addition, there should also be a $\kpara^{-1}$ scaling in the intermediate region $\kperp^{4/3}\lesssim \kpara \lesssim \sqrt{m}\kperp^{4/3}$.
\Figs\ref{fig:EplusminusVsKpara}(b) and (c) are not inconsistent with this scaling, but the spectra are too noisy to provide convincing evidence.

\subsubsection{$E^{\pm}_m$ versus $m$}

\begin{figure}
	\centering
	%\subfigure[]{\includegraphics{images/scaling_laws/h_020_196/spectrum_gplus2ta_vs_m_fixed_kperp_kz_without_dominant_id_200.png}}
	%\subfigure[]{\includegraphics{images/scaling_laws/h_020_196/spectrum_gminus2ta_vs_m_fixed_kperp_kz_without_dominant_id_200.png}}
	\subfigure[]{\includegraphics{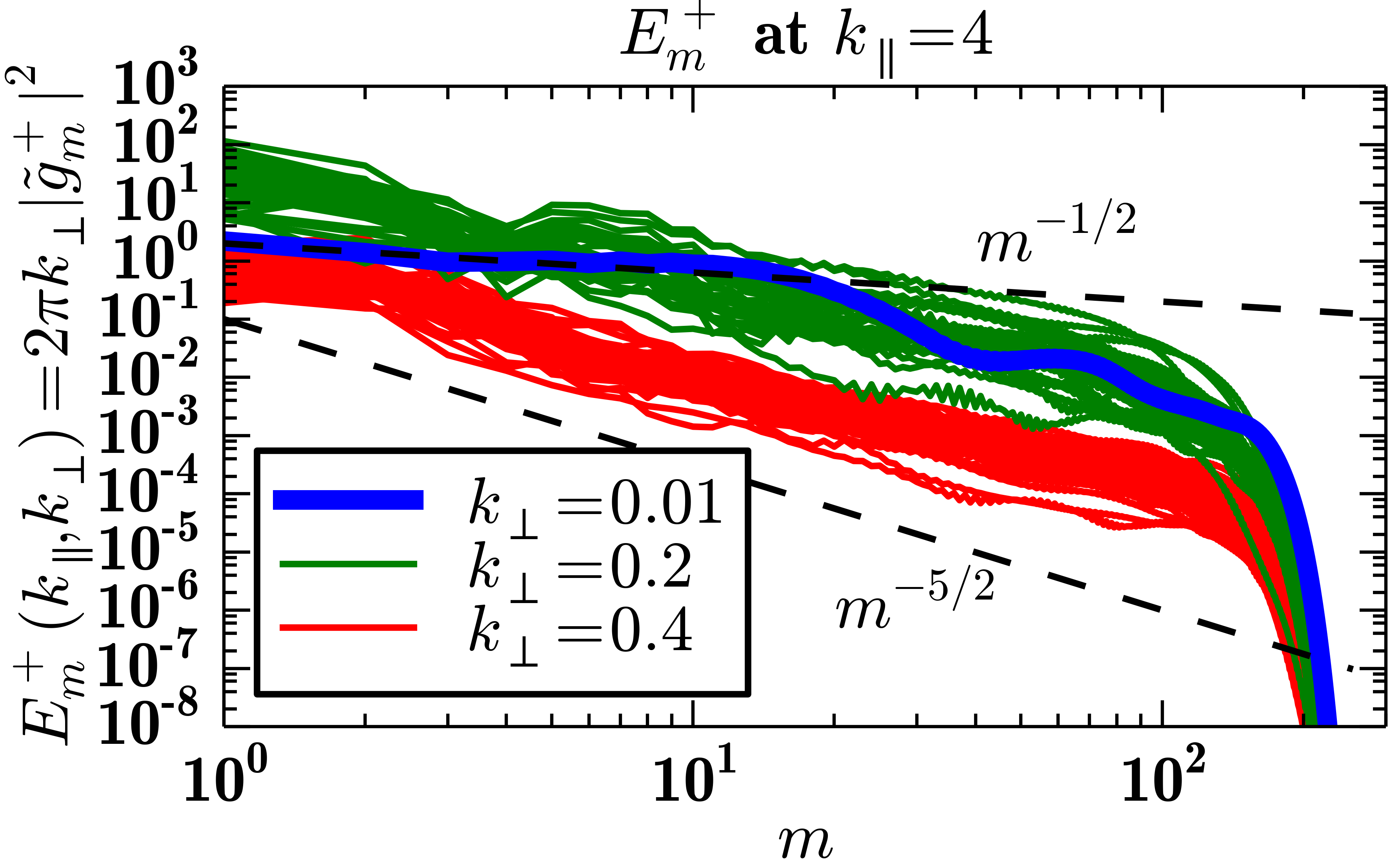}%}
	%\subfigure[]{
	\includegraphics{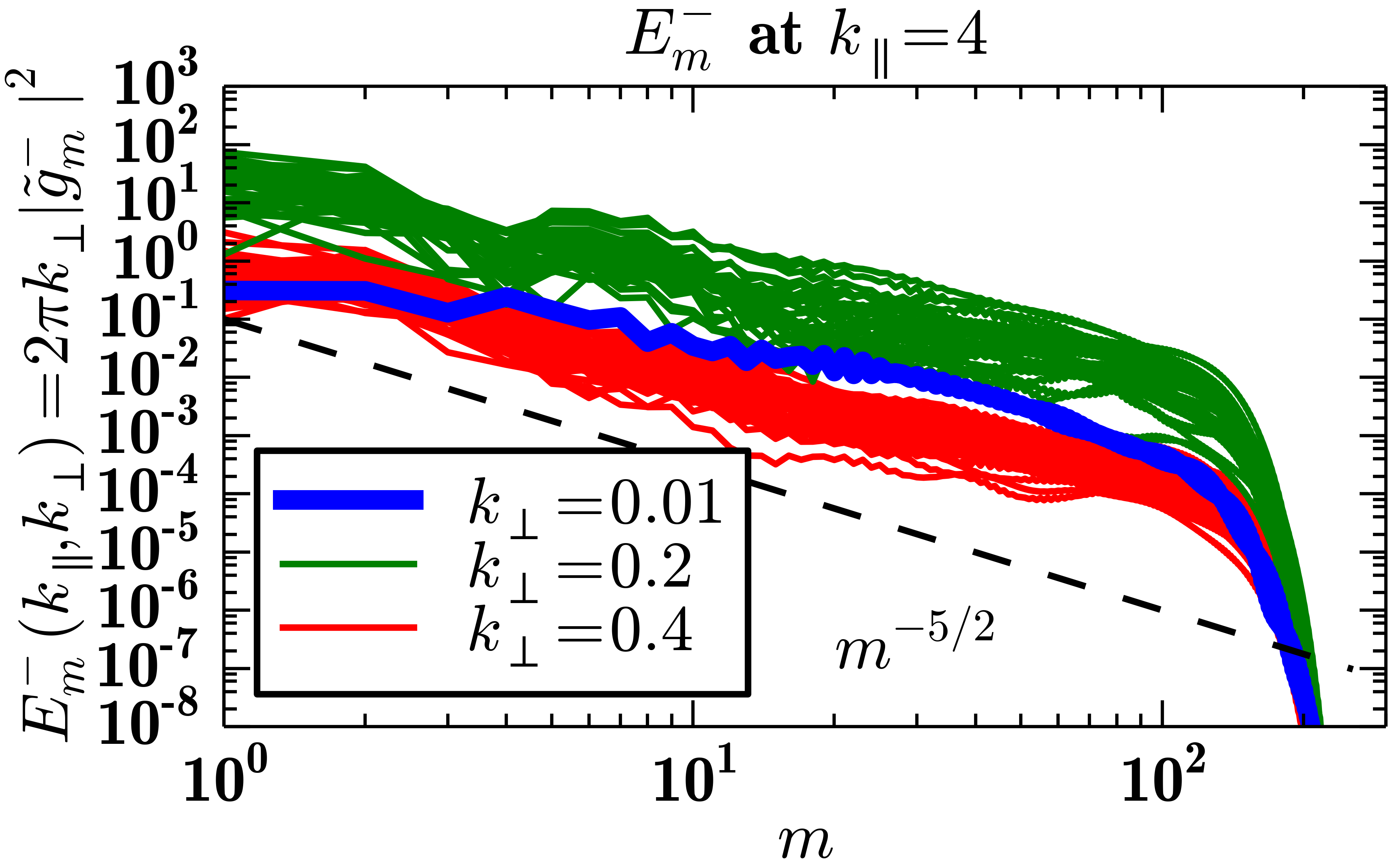}}
	\subfigure[]{\includegraphics{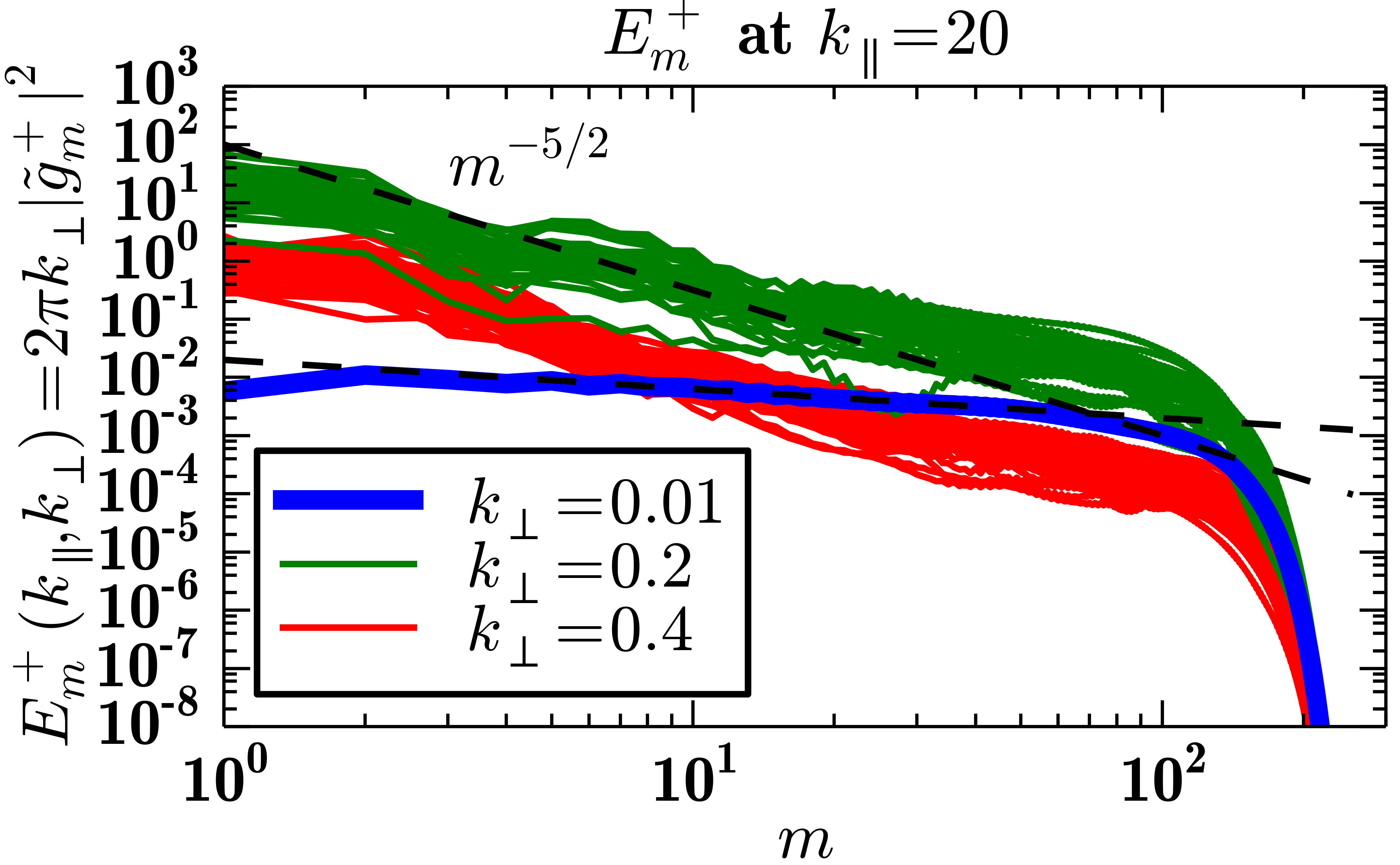}%}
	%\subfigure[]{
	\includegraphics{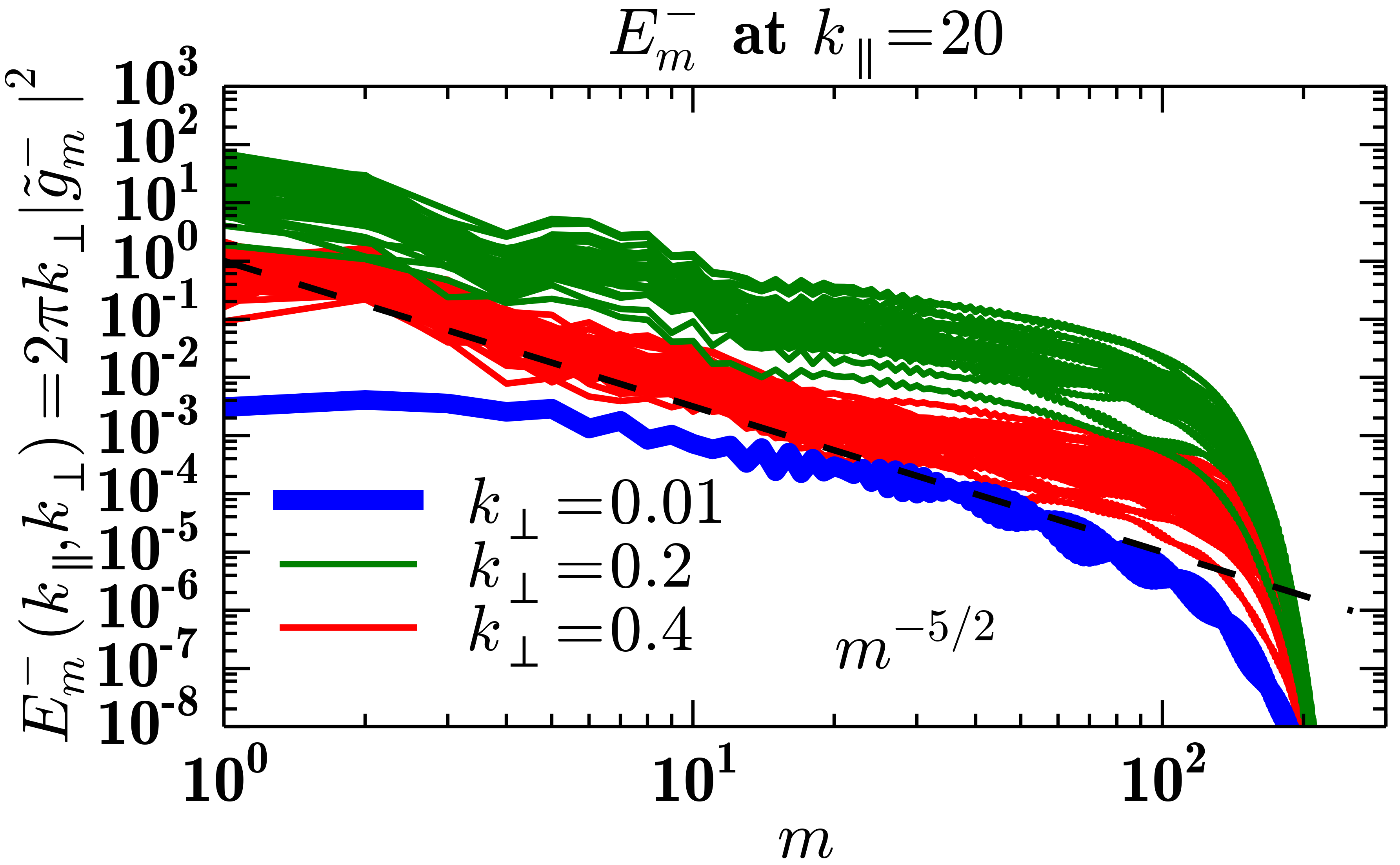}}
	\subfigure[]{\includegraphics{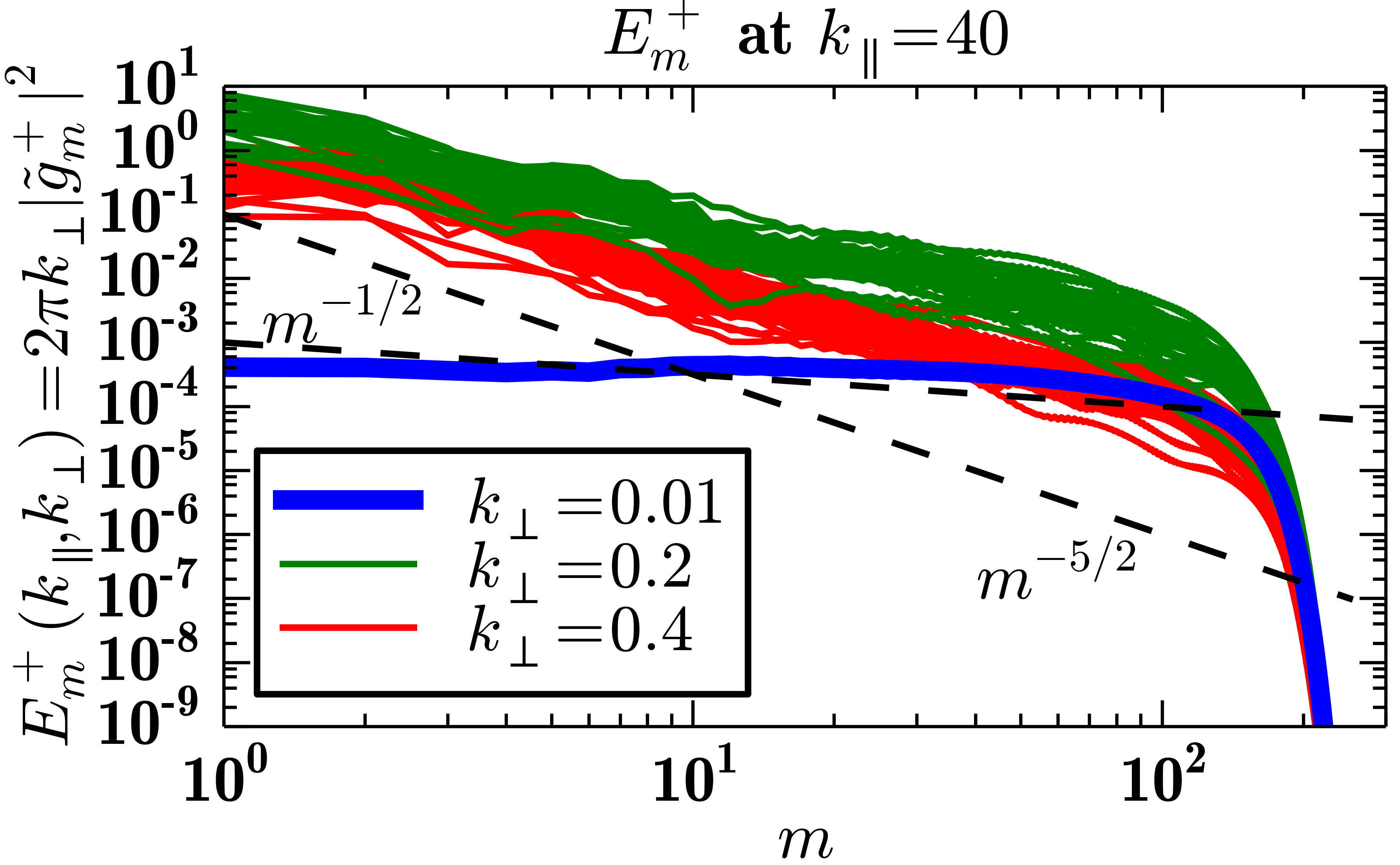}%}
	%\subfigure[]{
	\includegraphics{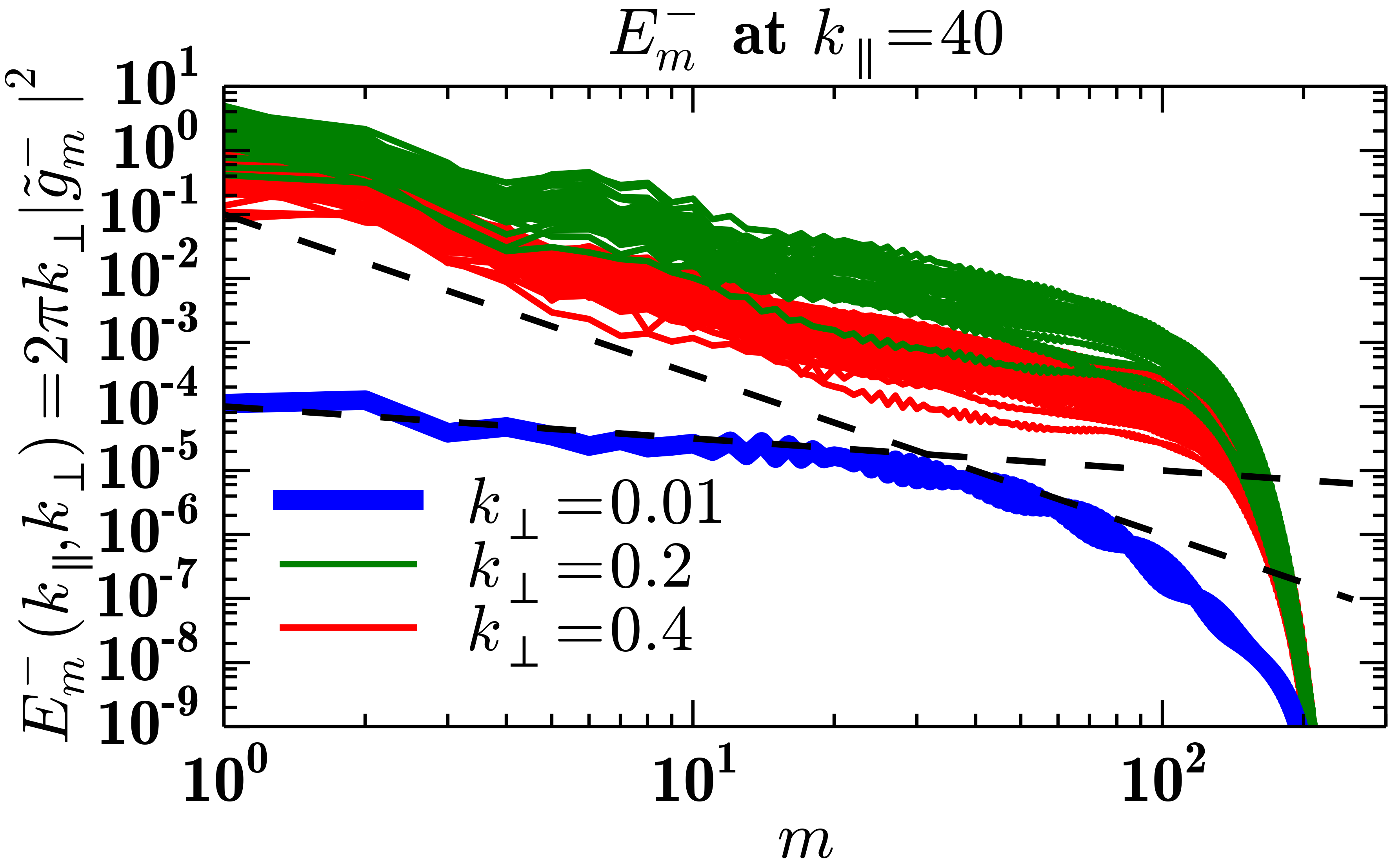}}
	\caption[Spectra of $E^{\pm}_m$ against $m$ at fixed $\kpara$ and $\kperp$.]{Spectra of $E^+_m$ (left column) and $E^-_m$ (right column) against $m$ at fixed $\kperp$ and $\kpara$: (a) $\kpara=4$, (b) $\kpara=20$, (c) $\kpara=40$. 
		  Colour denotes selections of modes with $k_x$, $k_y$ such that $\sqrt{k_x^2+k_y^2}$ approximately equals the given $\kperp$.
The positive spectrum $E^+_m$ \eqref{eq:Eplusm} has two scalings shown by the dashed guide lines:
$E^+_m\sim m^{-5/2}$ in $\kpara\lesssim\sqrt{m}\kperp^{4/3}$, and
$E^+_m\sim m^{-1/2}$ in $\kpara\gtrsim\sqrt{m}\kperp^{4/3}$. 
The negative spectrum $E^-_m$ \eqref{eq:Eminusm} has a universal $m^{-5/2}$ scaling;
however we observe regions where $E^-_m\sim m^{-1/2}$ in $\kpara\gtrsim\sqrt{m}\kperp^{4/3}$.
This may be an artifact from $E^+_m$, as the decomposition into forwards and backwards propagating modes is only asymptotic at large $m$ and so less accurate at low $m$.
	\label{fig:EplusminusVsM}
}
\end{figure}

In \fig\ref{fig:EplusminusVsM} we plot the spectra $E^{\pm}_m$ against $m$.
As before, we group together by colour spectra which have the same $\kperp$ (but different $k_x$ and $k_y$).
We show data for the perpendicular wavenumbers $\kperp=0.01$, $\kperp=0.2$ and $\kperp=0.4$.
Each row of figures is for a different parallel wavenumber: panel (a) shows $\kpara=4$, (b) shows $\kpara=20$, and (c) shows $\kpara=40$.

As expected from \eqref{eq:Eplusm}, the $E^+_m$ spectrum (left-hand column) has a $m^{-5/2}$ scaling in the intermediate and advection-dominated regions, and a $m^{-1/2}$ scaling in the phase-mixing region.

The $E^-_m$ spectrum should have a universal $m^{-5/2}$ scaling.  
This is seen almost everywhere in the right-hand column of \fig\ref{fig:EplusminusVsM}.
However a $m^{-1/2}$ spectrum is also observed above the phase-mixing threshold $\kpara\gtrsim\sqrt{m}\kperp^{4/3}$, at high $\kpara$, low $m$, and low $\kperp$. 
This may be due to the fact that the decomposition into forwards and backwards propagating modes $\tilde{g}^{\pm}_m$ is 
only asymptotic for large $m$, and hence 
less accurate at low $m$.
Thus the scaling could be an artifact from the $E^+_m\sim m^{-1/2}$ spectrum in the phase-mixing region,
particularly since $E^+_m\gg E^-_m$ in this region.

\subsubsection{$E^{-}_m$ versus $\eta=\kperp^{4/3}/\kpara$}

\begin{figure}
	\centering
%%%	\subfigure[]{\includegraphics{images/scaling_laws/h_020_196/spectrum_gminus2ta_vs_kz_summed_kperp_m_id_200.png}}
%%%	\subfigure[]{\includegraphics{images/scaling_laws/h_020_196/spectrum_gminus2ta_vs_kperp_summed_kz_id_200.png}}
%%%	%\subfigure[]{\includegraphics{images/scaling_laws/h_020_196/compensated_spectrum_gminus2ta_vs_kperp_summed_kz_id_200.png}}
%%%	%\subfigure[]{\includegraphics{images/scaling_laws/h_020_196/compensated_spectrum_gminus2ta_vs_kz_summed_kperp_m_id_200.png}}
%%%	\subfigure[]{\includegraphics{images/scaling_laws/h_020_196/spectrum_gminus2ta_vs_m_summed_kperp_kz_id_200.png}}
%%%	\subfigure[]{\includegraphics{images/scaling_laws/h_020_196/spectrum_gminus2ta_vs_m_fixed_kperp_kz_id_200.png}}
%%%	%\subfigure[]{\includegraphics{images/scaling_laws/h_020_196/rainbow_spectrum_gminus2ta_vs_kperp_aligned_2_id_200.png}}
	%\subfigure[]{
		\includegraphics{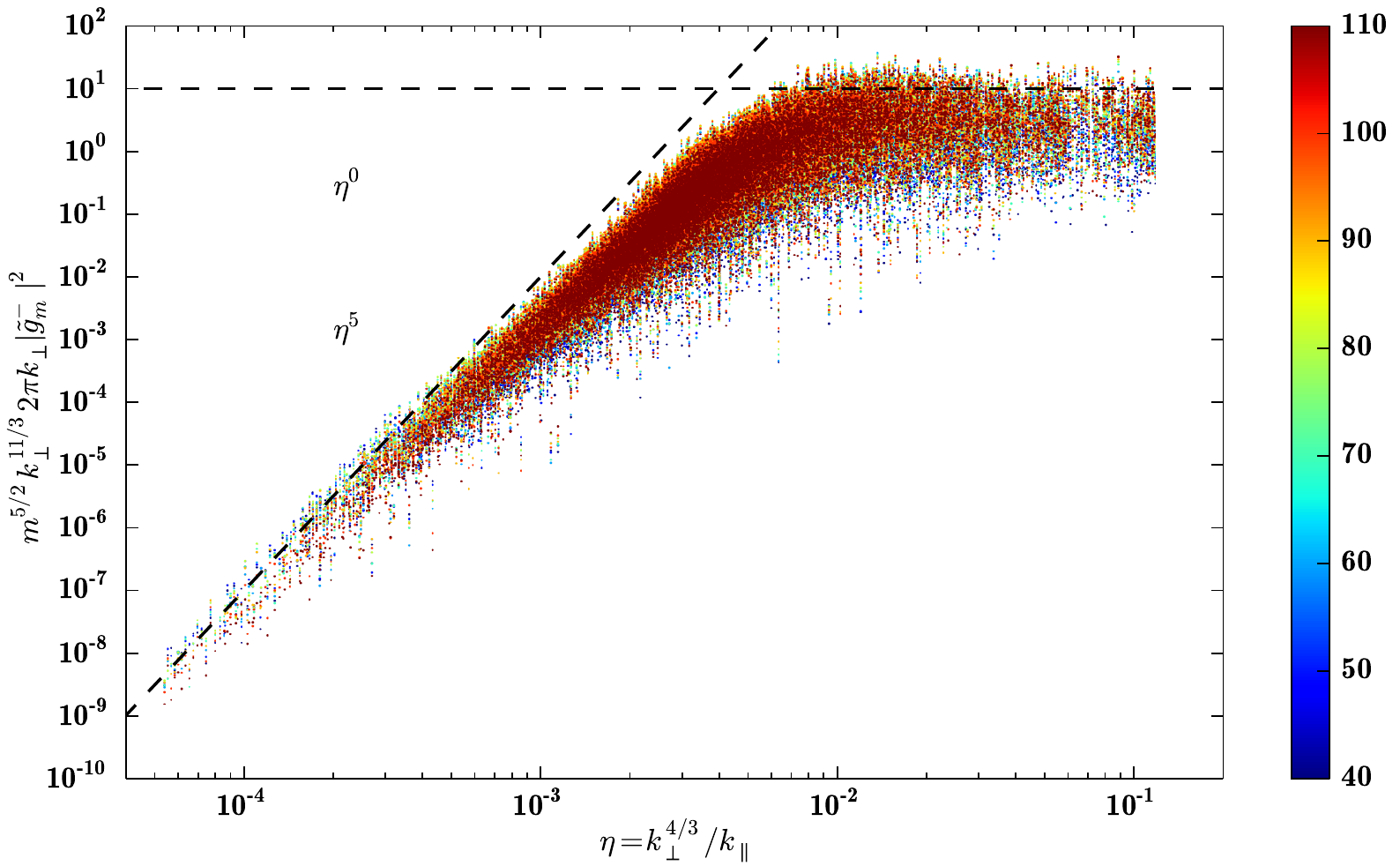}
		\includegraphics{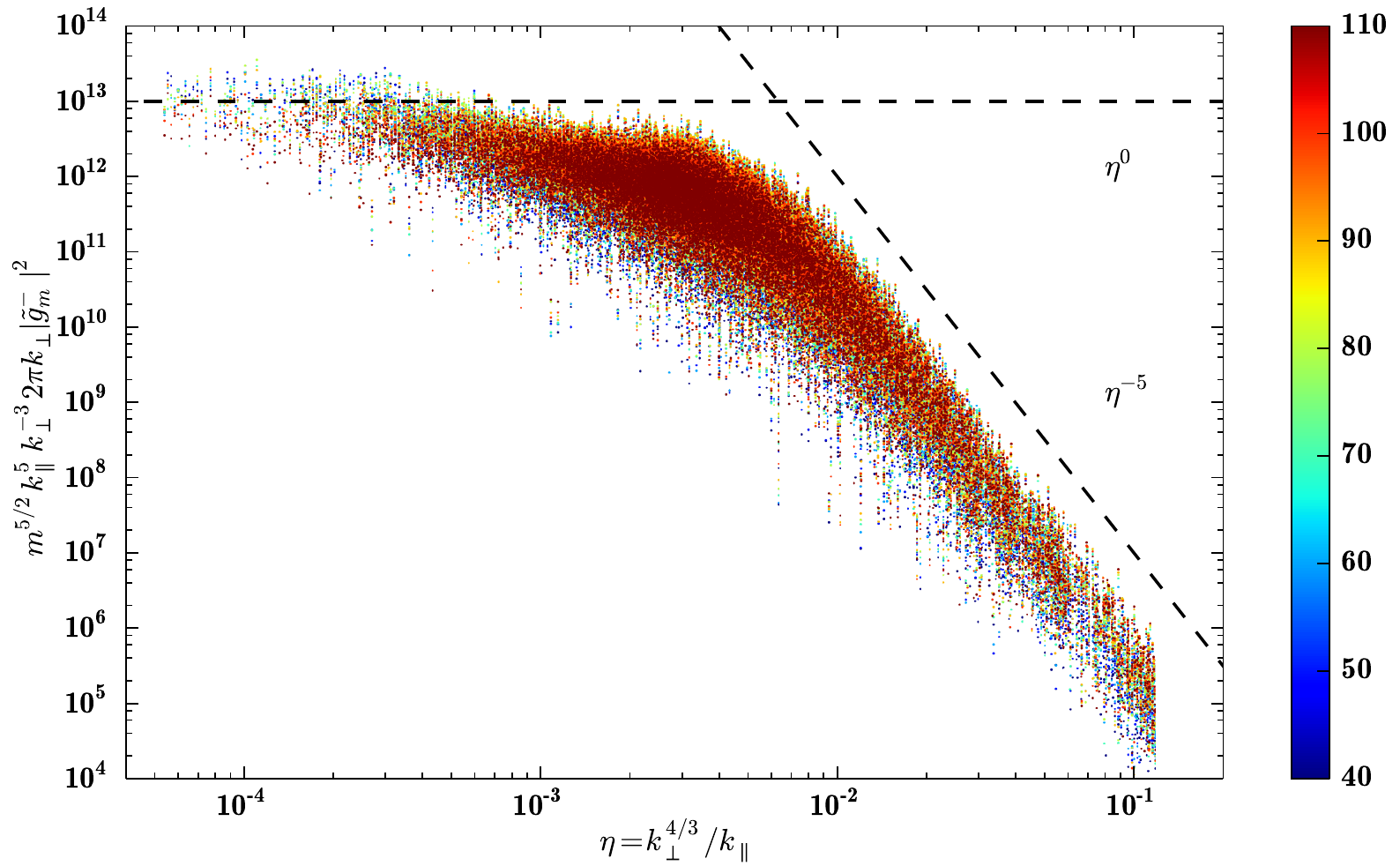}
	%}
%%%	\subfigure[]{\includegraphics{images/scaling_laws/h_020_196/rainbow_spectrum_gminus2ta_vs_all_aligned_3_id_200.png}}
		\caption[Spectrum of $E^-_m$ against $\eta=\kperp^{4/3}/\kpara$.]{
Spectrum of $E^-_m$ against $\eta=\kperp^{4/3}/\kpara$.
These are similar spectra to those for $\Ephi$ in \fig\ref{fig:PhiTwoDSpecta}, but now include all of velocity space.
There is a collapse onto a single line for $m\gtrsim 40$, indicating that for these $m$, the distribution function satisfies the scaling law \eqref{eq:Eminusm}. 
	\label{fig:EminusSpectra}
}
\end{figure}

Finally, we note from equations \eqref{eq:Eminusm} and \eqref{eq:Ephi} 
that the spectrum of backwards propagating modes $E^-_m=2\pi\kperp\langle|\tilde{g}^-_m|^2\rangle$ 
is simply related to the spectrum of $\varphi$ by $E^-_m=m^{-5/2}\Ephi$.
%%%Therefore the backwards spectrum summed over all $m$, $\sum_mE^-_m$, will have the same scalings as $\Ephi$ in Fourier space.
%%%In particular, it has the same one-dimensional spectra as \eqref{eq:OneDSpectraEPhiTheory},
%%%\begin{align}
%%%	\sum_m E^{-\parallel}_m(\kpara) \sim \kpara^{-2},
%%%	\hspace{1cm}
%%%	\sum_m E^{-\perp}_m(\kperp) \sim  \kperp^{-7/3},
%%%	\label{eq:OneDSpectraEminusTheory}
%%%\end{align}
Therefore, like $\Ephi$ in \eqref{eq:EphiSimilarity}, $E^-_m$ can be expressed in terms of the single variable $\eta=\kperp^{4/3}/\kpara$ as
\begin{align}
	\kperp^{11/3}	m^{5/2}	E^-_m(\kpara,\kperp) = 
	\begin{cases}
		\cst, 
		& 1\lesssim \eta,\\[2ex]
	 \eta^{5}, 
		& 1\gtrsim \eta,
	\end{cases}
	\label{eq:EminusmSimilarity}
\end{align}
and
\begin{align}
	\kpara^5\kperp^{-3}	m^{5/2} E^-_m(\kpara,\kperp) = 
	\begin{cases}
	 \eta^{-5}, 
		& 1\lesssim \eta,\\[2ex]
		\cst, 
		& 1\gtrsim \eta.
	\end{cases}
	\label{eq:EminusSimilarityAlt}
\end{align}
%%%We plot these along with the Hermite spectrum, 
%%%\begin{align}
%%%	\sum_{\k}E^{-}_m(\kpara,\kperp) \sim m^{-5/2},
%%%	\label{eq:EminusmHermite}
%%%\end{align}
%%%in \fig\ref{fig:EminusSpectra}.
%%%The Fourier spectra in \fig\ref{fig:EminusSpectra}(a) and (b) are in good agreement with their theoretical values \eqref{eq:OneDSpectraEminusTheory}
%%%with inertial ranges at the same wavenumbers as in $\Ephi$, \fig\ref{fig:PhiOneDSpecta}.
%%%The Hermite spectrum in \fig\ref{fig:EminusSpectra}(c) is too shallow, being a better fit for $m^{-3/2}$ than $m^{-5/2}$.
%%%This reflects the Hermite spectra for fixed Fourier wavenumbers which individually have $m^{-3/2}$ scalings.
We plot this in \fig\ref{fig:EminusSpectra} with colours denoting different $m$ in the range $[40,110]$,
\ie\ above the decomposition artifacts at low $m$, but below the dissipation range.
The distribution function collapses onto a single line, indicating that the computed spectrum satisfies the scaling law \eqref{eq:Eminusm}.

\subsubsection{Summary}

In summary, 
the results presented support the theoretical spectra \eqref{eq:Eplusm}, \eqref{eq:Eminusm} and \eqref{eq:Ephi}  derived by \citet{Schekochihin15}.
The assumptions of the derivation---isotropy and critical balance---are satisfied.
The observed $\Ephi$ spectrum is in strong agreement with the theoretical spectrum \eqref{eq:Ephi}. 
%%%Indeed the two-dimensional spectrum may be plotted in terms of the single variable $\eta=\kperp^{4/3}/\kpara$.
Indeed, the two-dimensional spectrum collapses onto a single line as a function of the similarity variable $\eta=\kperp^{4/3}/\kpara$, showing that the whole spectrum is in agreement with the expected scalings.
The distribution function spectra $E^{\pm}_m$ are also in good agreement with the theoretical spectra \eqref{eq:Eplusm} and \eqref{eq:Eminusm}.
The spectrum $E^-_m$ for negative modes is in very good agreement with the scalings,
and the whole spectrum collapses onto the expected function of a single variable \eqref{eq:EminusmSimilarity}.
The spectrum $E^+_m$ for positive modes is expected to have three distinct scaling regions.
The results do not contradict this picture, but we find no clear evidence for the intermediate region.
This is most likely because computed spectra transition smoothly between the different scaling regions rather than changing sharply as the theoretical spectra have been assumed to do.
To see the intermediate region, we would therefore require more resolution than available in these already well-resolved simulations.
Overall however these results are good evidence for the scalings proposed by \citet{Schekochihin15}.

\chapter{Conclusion}

In this thesis we have developed the theory of a fully spectral Fourier--Hankel--Hermite representation for 
%slab gyrokinetics,
the coupled gyrokinetic-Maxwell equations that describe a strongly magnetised fusion plasma in Cartesian slab geometry,
implemented the resulting equations in \sgk,
a code in Fortran 90 + MPI designed for efficiency on modern High Performance Computing platforms,
and used simulations with \sgk\ to fundamentally change our understanding of saturated drift-kinetic turbulence.

After deriving the \gkm\ system in \chp\ \ref{sec:GKMSystem},
we studied parallel velocity space in \chp\ \ref{sec:ParallelVelocitySpace}.
The most important effect in this dimension is Landau damping, 
a mechanism by which perturbations to the electric field decay in the absence of explicit dissipation.
Landau damping is caused by particle streaming, 
the phase space shear 
$\vpara\partial_z\distpert$
which causes infinitesimally fine scales to form in the distribution function.
While the distribution function itself does not decay, its velocity space moments, like the electrostatic potential $\varphi$, do decay due to integrating over the distribution
function's fine scale structure.
Landau damping is derived via a Laplace transform in time,
so the distribution function is not a time eigenmode of the system.
Rather, the time eigenmodes are singular non-decaying functions, called the \cvk\ modes, which form a continuous spectrum.

The velocity space behaviour changes with the inclusion of any amount of collisions.
\fp-type collision operators, like the \lb\ and \dop\ operators discussed in \S\ref{sec:lb},
contain velocity space diffusion which smooths the finest scales in the distribution function.
Time eigenmodes of the collisional system are now square integrable and smooth, unlike the eigenmodes of the collisionless system.
Moreover, 
as finer and finer scales form in velocity space, the effect of velocity space diffusion remains finite, even as the diffusivity tends to zero.
Landau's choice of contour (see \fig\ref{fig:contours}) captures this limit without explicitly introducing collisions.

The velocity space behaviour changes again when the system is discretized,
as discretization imposes a finest resolvable velocity scale.
Now discrete analogues of smooth solutions are found only if collisions are sufficiently strong to make the solution resolvable on the discretization grid;
otherwise the solutions are discrete approximations to the singular \cvk\ modes.
The Landau-damped solution is still found in the limit of vanishing collisions,
but only in the simultaneous limit of infinite resolution.

The Hermite representation introduced in \chp\ \ref{sec:ParallelVelocitySpace} provides a neat formalism for parallel velocity space.
The $m$th order Hermite polynomial has a characteristic velocity scale $\vth/\sqrt{m}$, 
so that each coefficient in the expansion represents a different velocity space scale.
The square of each coefficient represents that scale's contribution to the free energy.
Moreover, the electrostatic potential is proportional to the zeroth order coefficient.
Collisions, which act preferentially at fine velocity scales, are represented by a damping of the high $m$ coefficients.
The streaming term becomes a nearest neighbour mode-coupling in $m$.
The streaming results in a transfer of free energy in Hermite space, analogous to the nonlinear cascade of free energy due to the $\u\cdot\nabla\u$ term in hydrodynamic turbulence.
Unlike the nonlinear cascade, however, this transfer is linear and reversible.
The linear Landau damping solution corresponds to a forward cascade of free energy towards infinitesimally fine scales. 
The forwards direction is set by the presence of some small collisionality that dissipates free energy at sufficiently fine scales.
In contrast, the backwards cascade is only observed in the linear \gkm\ system which has been discretized (has a finite range of $m$) and has an inadequate treatment of the cut-off at the highest $m$.
Indeed, the backwards cascade appears in recurrence,
the unphysical scenario where free energy reflects from the finest resolved scales and propagates backwards towards large scales,
ultimately causing the damped electric field to suddenly grow. 

In \chp\ \ref{sec:ParallelVelocitySpace}, we develop the theory of the hypercollisional operator---an iterated version of the Kirkwood collision operator
(the momentum and energy conserving version of the Lenard--Bernstein operator).
It is particularly simple to apply in Hermite space, as the Hermite functions are its eigenfunctions.
The hypercollisional operator selectively damps the high $m$ which represent fine velocity space scales, ensuring that there is no reflection at the highest retained $m$ and thus no recurrence.
With this operator, accurate Landau growth rates are calculated, even with modest resolution.
Moreover, the operator exhibits a ``plateau'' in parameter space, a region where the calculated result is accurate and insensitive to changes in collision frequency, hypercollision exponent and resolution. 
This not only makes the operator of great practical value---it is easy to find suitable parameter values---but it also provides a numerical analog of the plateau found in the analytical solution of the collisionless limit of the weakly collisional problem.

Determining the electrostatic potential is straightforward in Fourier--Hermite space 
as integrals over parallel velocity space become 
evaluations of the coefficients of
single modes
due to the orthogonality of the basis functions.
Trying to achieve the same phase space localization motivates the use of the Hankel representation in perpendicular velocity.
In \chp\ \ref{sec:PerpendicularVelocitySpaceHankelTransform} we show that Maxwell's equations are local in Fourier--Hankel--Hermite space,
that is, $\varphi$, $\Apar$ and $\Bpar$ are determined from single modes.
Moreover in the linearized \gkm\ system there is no coupling at all between Fourier and Hankel modes;
the equations reduce to the one-dimensional problem in Hermite space studied in \chp\ \ref{sec:ParallelVelocitySpace}, parameterized by $\k$ and $p=\rho_i\kperp$.

In the nonlinear \gkm\ system there is coupling of Fourier and Hankel modes through the nonlinear term.  
However there remains an interesting localization of phase space, where the evolution of any one Fourier--Hankel mode depends on a limited range of other Fourier--Hankel modes.
We discussed methods of exploiting this property in gyrokinetics, but concluded that these would not improve upon pseudospectral methods for the drift-kinetic problem studied in \chp s \ref{sec:FreeEnergyFlowAndDissipation} and \ref{sec:ScalingLawsForDriftKineticTurbulence}. %\partref\ref{sec:PartResults}.

In \chp\ \ref{sec:SpectroGK} we tested and verified \sgk.
\sgk\ is the main practical outcome of this thesis.
It is a versatile code which runs and scales across many systems, from laptops to High Performance Computing platforms.
Because of its spectral approach, it makes efficient use of memory, and exactly conserves free energy transfers between modes in the absence of explicit dissipation.
\sgk\ has proven useful for studying Vlasov--Poisson turbulence, as in \citet{Parker14}, and for drift-kinetic turbulence, as in \chp s \ref{sec:FreeEnergyFlowAndDissipation} and \ref{sec:ScalingLawsForDriftKineticTurbulence}.

In drift-kinetic turbulence, there are two mechanisms by which free energy injected at large scales (by forcing or an instability like a temperature gradient) can be transferred to dissipation.
Firstly, free energy may be transfered linearly by phase-mixing to dissipation by collisions at fine parallel velocity space scales, %(or Landau damping at infinitesimal scales), 
as discussed in \chp\ \ref{sec:ParallelVelocitySpace}.
Alternatively, it may cascade nonlinearly in Fourier space to dissipation by viscosity at fine physical scales, as in hydrodynamic turbulence.
Each cascade was understood in isolation, but it was unclear how they interacted.
In particular, \citet{BarnesEtal11} had derived the spectrum for the electrostatic potential observed in simulations, but to do so had to neglect the transfer of free energy to fine velocity space scales.
This is despite other authors observing fine velocity space structure, and so some free energy is indeed present at fine scales \cite{Watanabe04,Hatch13}.

In \chp\ \ref{sec:FreeEnergyFlowAndDissipation} we showed the startling result that in saturated nonlinear turbulence, the nonlinear term inhibits, and at energetically-dominant scales completely suppresses, the transfer of free energy by linear phase-mixing.
It does this by exciting backwards propagating modes, like those observed in recurrence in \chp\ \ref{sec:ParallelVelocitySpace}.
Here however the effect is entirely physical and self-generated by the turbulence, rather than by an artificial numerical boundary condition.
The suppression of net free energy flux has two important consequences.
Firstly, the turbulence is largely fluid, with free energy at the dominant scales only cascading to fine scales in physical space.
Secondly, as little free energy reaches fine scales in velocity space, Landau damping is strongly suppressed as a dissipation mechanism.

In \chp\ \ref{sec:ScalingLawsForDriftKineticTurbulence}, we derived scaling laws for the spectra of the distribution function\linebreak and electrostatic potential in drift-kinetic turbulence, verifying these scalings with\linebreak \sgk.
The most significant feature of these results is that the spectra of the electrostatic potential and the backwards propagating modes may both be written in terms of the single variable $\eta=\kperp^{4/3}/\kpara=\taus/\taunl$.
Thus the ratio of timescales not only determines whether linear or nonlinear behaviour is dominant in a region of phase space, 
it is itself a similarity variable for much of the solution.

\section{Future work}
There are a number of exciting directions in which we will take this work.
Firstly, in %\partref\ref{sec:PartResults} 
\chp s \ref{sec:FreeEnergyFlowAndDissipation} and \ref{sec:ScalingLawsForDriftKineticTurbulence}
we studied the essentially four-dimensional case of drift-kinetic turbulence.
It is natural to next study five-dimensional gyrokinetic turbulence, \ie\ extend the spatial domain to include sub-Larmor scales.
In addition to parallel phase-mixing and the nonlinear cascade in Fourier space, at sub-Larmor scales there is also a nonlinear cascade in perpendicular velocity space.
We will extend the scaling theory derived in \chp\ \ref{sec:ScalingLawsForDriftKineticTurbulence} to incorporate this cascade.
As we noted in \chp\ \ref{sec:PerpendicularVelocitySpaceHankelTransform}, there already exists a scaling theory for ``two-dimensional'' gyrokinetic turbulence (turbulence in the perpendicular plane only, with no parallel phase-mixing).
This theory could be immediately applied to sub-Larmor scales in five-dimensional gyrokinetics, 
provided there is significant suppression of net free energy transfer to fine parallel velocity space scales.
To determine if this is the case requires simulations; 
but these simulations would be very costly due to the increased size of Fourier space.
However it seems likely that the same mechanism as in drift kinetics---the nonlinear term exciting backwards propagating modes by coupling parallel wavenumbers---should also apply at sub-Larmor scales.
Indeed, the only difference in the nonlinear term between drift-kinetics and gyrokinetics is a factor of $J_0(\rho_i\kperp\vperp)$ from the gyroaverage of the electrostatic potential (see equation \eqref{eq:NonlinearTermPhi}).
This will change the regions of phase space where there is suppression of free energy transfer, but not the underlying mechanism.
Indeed, in this thesis we have developed the analytical framework required for determining the phase space structure of gyrokinetics,
and we will be able to make deductions about free energy transfer and dissipation in gyrokinetics as we did for drift kinetics in \chp\ \ref{sec:FreeEnergyFlowAndDissipation}. 

We will also investigate fluid models for the perpendicular velocity cascade.
In \chp\ \ref{sec:PerpendicularVelocitySpaceHankelTransform}, we discussed the importance of the phase space line $p=\rho_i\kperp$;
it is the peak of the spectrum in two-dimensional gyrokinetics, and is likely also the peak in five-dimensional gyrokinetics.
By omitting grid points further away from $p=\rho_i\kperp$ in $(\kperp,p)$ phase space than some distance $r$, 
we derived a family of fluid-like models parameterized by $r$.
We will investigate if these low resolution (small $r$) models replicate the behaviour of the full kinetic system ($r\to\infty$).
If so, the $\O(100)$ perpendicular velocity space grid points needed in computing gyrokinetic turbulence may be reduced to just a few Hankel modes.
This is a novel approach to producing fluid models, and we will compare the resulting family of models to existing gyrofluid models.

Further, we will use the suppression of net free energy transfer as a parallel velocity space closure condition for fluid models.
In parallel velocity space, we may think of a fluid model as replacing the $\O(100)$ ``kinetic'' (high $m$) Hermite modes with a closure which defines $g_3$ (the first kinetic mode) in terms of the fluid modes $g_0$, $g_1$ and $g_2$.
In its crudest form, one might imagine imposing no net free energy flux in the advection region and free energy flux as in the linear case in the phase-mixing region.
That is, at each timestep setting $|g_3|=|g_2|$ with phases that satisfy $\Im(g_3^*g_2)=0$ for $\kpara\lesssim A\kperp^{4/3}$, and $\Im(g_3^*g_2)=1$ for $\kpara\gtrsim A\kperp^{4/3}$,
where $A$, the constant in the line of critical balance, is determined from full kinetic simulations.
This closure reduces parallel velocity space from the $\O(100)$ modes of the kinetic system to a three mode fluid system.

Taken together, these two velocity space models may retain the key features of the gyrokinetic system, while offering a reduction in resolution of $\O(10^4)$.

Finally, in this thesis we have studied electrostatic turbulence in slab geometry.
To make this work applicable for fusion, we need to include both electromagnetic effects and toroidal geometry.
\sgk\ can treat electromagnetic problems in slab geometry (as described in \chp\ \ref{sec:SpectroGK}), 
but the Hankel space/perpendicular velocity space representation needs to be amended to treat the trapped and passing particles which arise in toroidal geometry.

\appendix
\part{Appendices} 
%now enable appendix numbering format and include any appendices
\chapter%[Fourier--Hermite spectral representation for the Vlasov--Poisson system in the weakly collisional limit \normalfont{-- J. T. Parker and P. J. Dellar}]
{Fourier--Hermite spectral representation for the Vlasov--Poisson system in the weakly collisional limit\\ \normalfont{J. T. Parker and P. J. Dellar}}
\label{sec:VPPaper}

%\includepdf[pages=-]{ParkerDellar15_a4.pdf}
This paper, Ref. [145], is omitted in the arXiv version.

\chapter%[Irreversible energy flow in forced Vlasov dynamics \normalfont{-- G. G. Plunk and J. T. Parker}]
{Irreversible energy flow in forced Vlasov dynamics\\ \normalfont{G. G. Plunk and J. T. Parker}}

%\includepdf[pages=-,scale=0.8]{PlunkParker14.pdf}

This paper, Ref.~[94], is omitted in the arXiv version.

%next line adds the Bibliography to the contents page
\addcontentsline{toc}{chapter}{Bibliography}
\bibliography{thesis}      

\begin{thebibliography}{187}
\expandafter\ifx\csname natexlab\endcsname\relax\def\natexlab#1{#1}\fi

\bibitem[{\scshape {{U}.{S}. {E}nergy {I}nformation
  {A}dministration}}(2015){\scshape {{U}.{S}. {E}nergy {I}nformation
  {A}dministration}}]{EIA}
{\scshape {{U}.{S}. {E}nergy {I}nformation {A}dministration}} (2015)
  \emph{International Energy Statistics}.
  \url{http://www.eia.gov/cfapps/ipdbproject}, accessed: {14-1-2015}.

\bibitem[{\scshape {BP}}(2014){\scshape {BP}}]{BP}
{\scshape {BP}} (2014) \emph{BP Statistical Review of World Energy}.
  \url{http://www.bp.com/en/global/corporate/about-bp/energy-economics/statistical-review-of-world-energy.html}.

\bibitem[{\scshape {Environmental Audit Committee}}(2013){\scshape
  {Environmental Audit Committee}}]{EAC}
{\scshape {Environmental Audit Committee}} (2013) \emph{Energy subsidies}.
  House of Commons, Ninth Report of Session 2013-14, paper 61.

\bibitem[{\scshape Cook {\scshape et~al.}}(2001){\scshape Cook, Marbach, {Di
  Pace}, Girard \& Taylor}]{Cook01}
{\scshape Cook, I., Marbach, G., {Di Pace}, L., Girard, C. \& Taylor, N.~P.}
  (2001) Safety and environmental impact of fusion. {\em European Fusion
  Development Agreement (EFDA) Report EFDA-S-RE-1. EUR (01) CCE-FU\ FTC\/} .

\bibitem[{\scshape Nuttall {\scshape et~al.}}(2012){\scshape Nuttall, Clarke \&
  Glowacki}]{Nuttall12}
{\scshape Nuttall, W.~J., Clarke, R.~H. \& Glowacki, B.~A.} (2012) Resources:
  Stop squandering helium. {\em Nature\/} {\bfseries 485}, 573--575.

\bibitem[{\scshape Ikeda}(2010){\scshape Ikeda}]{Ikeda10}
{\scshape Ikeda, K.} (2010) {ITER} on the road to fusion energy. {\em Nuclear
  Fusion\/} {\bfseries 50}, 014002.

\bibitem[{\scshape Rebhan \& Van~Oost}(2002){\scshape Rebhan \&
  Van~Oost}]{Rebhan02}
{\scshape Rebhan, E. \& Van~Oost, G.} (2002) Thermonuclear burn criteria. {\em
  Fusion Science and Technology\/} {\bfseries 41}, 15--26.

\bibitem[{\scshape Maisonnier {\scshape et~al.}}(2007){\scshape Maisonnier,
  Campbell, Cook, Di~Pace, Giancarli, Hayward, Puma, Medrano, Norajitra,
  Roccella et~al.others}]{Maisonnier07}
{\scshape Maisonnier, D., Campbell, D., Cook, I., Di~Pace, L., Giancarli, L.,
  Hayward, J., Puma, A.~L., Medrano, M., Norajitra, P., Roccella, M. et~al.}
  (2007) Power plant conceptual studies in {E}urope. {\em Nuclear Fusion\/}
  {\bfseries 47}, 1524.

\bibitem[{\scshape Highcock}(2012){\scshape Highcock}]{HighcockPhD}
{\scshape Highcock, E.~G.} (2012) The zero-turbulence manifold in fusion
  plasmas. Doctoral thesis, University of Oxford.

\bibitem[{\scshape Ongena \& Van~Oost}(2008){\scshape Ongena \&
  Van~Oost}]{Ongena08}
{\scshape Ongena, J. \& Van~Oost, G.} (2008) Energy for future
  centuries-prospects for fusion power as a future energy source. {\em Fusion
  Science and Technology\/} {\bfseries 53}, 3--15.

\bibitem[{\scshape Hastie}(1995){\scshape Hastie}]{Hastie95}
{\scshape Hastie, R.~J.} (1995) Plasma particle dynamics. In {\em Plasma
  Physics: An Introductory Course\/}. Cambridge: Cambridge University Press.

\bibitem[{\scshape Helander \& Sigmar}(2002)]{HelanderSigmar02}
{\scshape Helander, P. \& Sigmar, D.~J.} (2002) {\em Collisional Transport in
  Magnetized Plasmas\/}. Cambridge: Cambridge University Press.

\bibitem[{\scshape Hinton \& Wong}(1985){\scshape Hinton \& Wong}]{Hinton85}
{\scshape Hinton, F.~L. \& Wong, S.~K.} (1985) Neoclassical ion transport in
  rotating axisymmetric plasmas. {\em Physics of Fluids\/} {\bfseries 28},
  3082--3098.

\bibitem[{\scshape Galambos {\scshape et~al.}}(1995){\scshape Galambos,
  Perkins, Haney \& Mandrekas}]{Galambos95}
{\scshape Galambos, J.~D., Perkins, L.~J., Haney, S.~W. \& Mandrekas, J.}
  (1995) Commercial tokamak reactor potential with advanced tokamak operation.
  {\em Nuclear Fusion\/} {\bfseries 35}, 551.

\bibitem[{\scshape Colas {\scshape et~al.}}(1998){\scshape Colas, Zou, Paume,
  Chareau, Guiziou, Hoang, Michelot \& Gr\'esillon}]{Colas98}
{\scshape Colas, L., Zou, X.~L., Paume, M., Chareau, J.~M., Guiziou, L., Hoang,
  G.~T., Michelot, Y. \& Gr\'esillon, D.} (1998) Internal magnetic fluctuations
  and electron heat transport in the tore supra tokamak: Observation by
  cross-polarization scattering. {\em Nuclear Fusion\/} {\bfseries 38}, 903.

\bibitem[{\scshape Horton \& Estes}(1980){\scshape Horton \& Estes}]{Horton80}
{\scshape Horton, W. \& Estes, R.~D.} (1980) Fluid simulation of ion pressure
  gradient driven drift modes. {\em Plasma Physics\/} {\bfseries 22}, 663.

\bibitem[{\scshape Waltz}(1988){\scshape Waltz}]{Waltz88}
{\scshape Waltz, R.~E.} (1988) Three-dimensional global numerical simulation of
  ion temperature gradient mode turbulence. {\em Physics of Fluids\/}
  {\bfseries 31}, 1962--1967.

\bibitem[{\scshape Fonck {\scshape et~al.}}(1989){\scshape Fonck, Howell,
  Jaehnig, Roquemore, Schilling, Scott, Zarnstorff, Bush, Goldston, Hsuan,
  Johnson, Ramsey, Schivell \& Towner}]{Fonck89}
{\scshape Fonck, R.~J., Howell, R., Jaehnig, K., Roquemore, L., Schilling, G.,
  Scott, S., Zarnstorff, M.~C., Bush, C., Goldston, R., Hsuan, H., Johnson, D.,
  Ramsey, A., Schivell, J. \& Towner, H.} (1989) Ion thermal confinement in the
  enhanced-confinement regime of the {TFTR} tokamak. {\em Physical Review
  Letters\/} {\bfseries 63}, 520--523.

\bibitem[{\scshape Wootton {\scshape et~al.}}(1990){\scshape Wootton, Carreras,
  Matsumoto, McGuire, Peebles, Ritz, Terry \& Zweben}]{Wootton90}
{\scshape Wootton, A.~J., Carreras, B.~A., Matsumoto, H., McGuire, K., Peebles,
  W.~A., Ritz, C.~P., Terry, P.~W. \& Zweben, S.~J.} (1990) Fluctuations and
  anomalous transport in tokamaks. {\em Physics of Fluids B\/} {\bfseries 2},
  2879--2903.

\bibitem[{\scshape Cowley {\scshape et~al.}}(1991){\scshape Cowley, Kulsrud \&
  Sudan}]{Cowley91}
{\scshape Cowley, S.~C., Kulsrud, R.~M. \& Sudan, R.} (1991) Considerations of
  ion-temperature-gradient-driven turbulence. {\em Physics of Fluids B\/}
  {\bfseries 3}, 2767--2782.

\bibitem[{\scshape Kotschenreuther {\scshape et~al.}}(1995){\scshape
  Kotschenreuther, Dorland, Beer \& Hammett}]{Kotschenreuther95confinement}
{\scshape Kotschenreuther, M., Dorland, W., Beer, M.~A. \& Hammett, G.~W.}
  (1995) Quantitative predictions of tokamak energy confinement from
  first-principles simulations with kinetic effects. {\em Physics of Plasmas\/}
  {\bfseries 2}, 2381--2389.

\bibitem[{\scshape Carreras}(1997){\scshape Carreras}]{Carreras97}
{\scshape Carreras, B.~A.} (1997) Progress in anomalous transport research in
  toroidal magnetic confinement devices. {\em IEEE Transactions on Plasma
  Science\/} {\bfseries 25}, 1281--1321.

\bibitem[{\scshape Dimits {\scshape et~al.}}(2000){\scshape Dimits, Bateman,
  Beer, Cohen, Dorland, Hammett, Kim, Kinsey, Kotschenreuther, Kritz
  et~al.others}]{Dimits00}
{\scshape Dimits, A.~M., Bateman, G., Beer, M.~A., Cohen, B.~I., Dorland, W.,
  Hammett, G.~W., Kim, C., Kinsey, J.~E., Kotschenreuther, M., Kritz, A.~H.
  et~al.} (2000) Comparisons and physics basis of tokamak transport models and
  turbulence simulations. {\em Physics of Plasmas\/} {\bfseries 7}, 969.

\bibitem[{\scshape Dorland {\scshape et~al.}}(2000){\scshape Dorland, Jenko,
  Kotschenreuther \& Rogers}]{Dorland00}
{\scshape Dorland, W., Jenko, F., Kotschenreuther, M. \& Rogers, B.~N.} (2000)
  Electron temperature gradient turbulence. {\em Physical Review Letters\/}
  {\bfseries 85}, 5579--5582.

\bibitem[{\scshape Jenko {\scshape et~al.}}(2000){\scshape Jenko, Dorland,
  Kotschenreuther \& Rogers}]{Gene}
{\scshape Jenko, F., Dorland, W., Kotschenreuther, M. \& Rogers, B.~N.} (2000)
  Electron temperature gradient driven turbulence. {\em Physics of Plasmas\/}
  {\bfseries 7}, 1904--1910.

\bibitem[{\scshape Dannert \& Jenko}(2005){\scshape Dannert \&
  Jenko}]{Dannert05}
{\scshape Dannert, T. \& Jenko, F.} (2005) Gyrokinetic simulation of
  collisionless trapped-electron mode turbulence. {\em Physics of Plasmas\/}
  {\bfseries 12}, 072309.

\bibitem[{\scshape {ITER Organization Website}}(2015){\scshape {ITER
  Organization Website}}]{ITER}
{\scshape {ITER Organization Website}} (2015) \url{http://www.iter.org/},
  accessed: {16-1-2015}.

\bibitem[{\scshape Krommes}(2012){\scshape Krommes}]{Krommes12}
{\scshape Krommes, J.~A.} (2012) The gyrokinetic description of microturbulence
  in magnetized plasmas. {\em Annual Review of Fluid Mechanics\/} {\bfseries
  44}, 175--201.

\bibitem[{\scshape Miura \& {JT-60 Team}}(2003){\scshape Miura \& {JT-60
  Team}}]{MiuraJT60}
{\scshape Miura, Y. \& {JT-60 Team}} (2003) {Study of improved confinement
  modes with edge and/or internal transport barriers on the Japan Atomic Energy
  Research Institute Tokamak-60 Upgrade (JT-60U)}. {\em Physics of Plasmas\/}
  {\bfseries 10}, 1809--1815.

\bibitem[{\scshape de~Vries {\scshape et~al.}}(2009){\scshape de~Vries,
  Joffrin, Brix, Challis, Cromb\'e, Esposito, Hawkes, Giroud, Hobirk,
  L\"onnroth, Mantica, Strintzi, Tala, Voitsekhovitch \& {JET-EFDA Contributors
  to the Work Programme}}]{deVries09}
{\scshape de~Vries, P.~C., Joffrin, E., Brix, M., Challis, C.~D., Cromb\'e, K.,
  Esposito, B., Hawkes, N.~C., Giroud, C., Hobirk, J., L\"onnroth, J., Mantica,
  P., Strintzi, D., Tala, T., Voitsekhovitch, I. \& {JET-EFDA Contributors to
  the Work Programme}} (2009) Internal transport barrier dynamics with plasma
  rotation in {JET}. {\em Nuclear Fusion\/} {\bfseries 49}, 075007.

\bibitem[{\scshape Highcock {\scshape et~al.}}(2012){\scshape Highcock,
  Schekochihin, Cowley, Barnes, Parra, Roach \& Dorland}]{Highcock12}
{\scshape Highcock, E.~G., Schekochihin, A.~A., Cowley, S.~C., Barnes, M.,
  Parra, F.~I., Roach, C.~M. \& Dorland, W.} (2012) Zero-turbulence manifold in
  a toroidal plasma. {\em Physical Review Letters\/} {\bfseries 109}, 265001.

\bibitem[{\scshape Chapman \& Cowling}(1991)]{ChapmanCowling}
{\scshape Chapman, S. \& Cowling, T.~G.} (1991) {\em {The Mathematical Theory
  of Non-Uniform Gases}\/}. Cambridge: Cambridge University Press.

\bibitem[{\scshape Burnett}(1935){\scshape Burnett}]{Burnett35}
{\scshape Burnett, D.} (1935) The distribution of velocities in a slightly
  non-uniform gas. {\em Proceedings of the London Mathematical Society\/}
  {\bfseries s2-39}, 385--430.

\bibitem[{\scshape Burnett}(1936){\scshape Burnett}]{Burnett36}
{\scshape Burnett, D.} (1936) The distribution of molecular velocities and the
  mean motion in a non-uniform gas. {\em Proceedings of the London Mathematical
  Society\/} {\bfseries s2-40}, 382--435.

\bibitem[{\scshape Abel {\scshape et~al.}}(2013){\scshape Abel, Plunk, Wang,
  Barnes, Cowley, Dorland \& Schekochihin}]{Abel13}
{\scshape Abel, I.~G., Plunk, G.~G., Wang, E., Barnes, M., Cowley, S.~C.,
  Dorland, W. \& Schekochihin, A.~A.} (2013) Multiscale gyrokinetics for
  rotating tokamak plasmas: fluctuations, transport and energy flows. {\em
  Reports on Progress in Physics\/} {\bfseries 76}, 116201.

\bibitem[{\scshape Rutherford \& Frieman}(1968){\scshape Rutherford \&
  Frieman}]{RutherfordFrieman68}
{\scshape Rutherford, P.~H. \& Frieman, E.~A.} (1968) Drift instabilities in
  general magnetic field configurations. {\em Physics of Fluids\/} {\bfseries
  11}, 569--585.

\bibitem[{\scshape Taylor \& Hastie}(1968){\scshape Taylor \&
  Hastie}]{TaylorHastie68}
{\scshape Taylor, J.~B. \& Hastie, R.~J.} (1968) Stability of general plasma
  equilibria---{I} formal theory. {\em Plasma Physics\/} {\bfseries 10},
  479--494.

\bibitem[{\scshape Jolliet {\scshape et~al.}}(2007){\scshape Jolliet, Bottino,
  Angelino, Hatzky, Tran, Mcmillan, Sauter, Appert, Idomura \& Villard}]{ORB5}
{\scshape Jolliet, S., Bottino, A., Angelino, P., Hatzky, R., Tran, T.~M.,
  Mcmillan, B.~F., Sauter, O., Appert, K., Idomura, Y. \& Villard, L.} (2007) A
  global collisionless {PIC} code in magnetic coordinates. {\em Computer
  Physics Communications\/} {\bfseries 177}, 409--425.

\bibitem[{\scshape Hammett {\scshape et~al.}}(1992){\scshape Hammett, Dorland
  \& Perkins}]{Hammett92}
{\scshape Hammett, G.~W., Dorland, W. \& Perkins, F.~W.} (1992) {Fluid models
  of phase mixing, Landau damping, and nonlinear gyrokinetic dynamics}. {\em
  Physics of Fluids B\/} {\bfseries 4}, 2052.

\bibitem[{\scshape Dorland \& Hammett}(1993){\scshape Dorland \&
  Hammett}]{Dorland93}
{\scshape Dorland, W. \& Hammett, G.~W.} (1993) Gyrofluid turbulence models
  with kinetic effects. {\em Physics of Fluids B\/} {\bfseries 5}, 812--835.

\bibitem[{\scshape Hammett {\scshape et~al.}}(1993){\scshape Hammett, Beer,
  Dorland, Cowley \& Smith}]{HammettBeer93}
{\scshape Hammett, G.~W., Beer, M.~A., Dorland, W., Cowley, S.~C. \& Smith,
  S.~A.} (1993) Developments in the gyrofluid approach to tokamak turbulence
  simulations. {\em \ppcf\/} {\bfseries 35}, 973.

\bibitem[{\scshape Hammett \& Perkins}(1990){\scshape Hammett \&
  Perkins}]{HammettPerkins90}
{\scshape Hammett, G.~W. \& Perkins, F.~W.} (1990) Fluid moment models for
  {L}andau damping with application to the ion-temperature-gradient
  instability. {\em Physical Review Letters\/} {\bfseries 64}, 3019--3022.

\bibitem[{\scshape Barnes {\scshape et~al.}}(2010){\scshape Barnes, Abel,
  Dorland, G\"orler, Hammett \& Jenko}]{Trinity}
{\scshape Barnes, M., Abel, I.~G., Dorland, W., G\"orler, T., Hammett, G.~W. \&
  Jenko, F.} (2010) Direct multiscale coupling of a transport code to
  gyrokinetic turbulence codes. {\em Physics of Plasmas\/} {\bfseries 17},
  056109.

\bibitem[{\scshape Schekochihin {\scshape et~al.}}(2009){\scshape Schekochihin,
  Cowley, Dorland, Hammett, Howes, Quataert,  \& Tatsuno}]{Schekochihin09}
{\scshape Schekochihin, A.~A., Cowley, S.~C., Dorland, W., Hammett, G.~W.,
  Howes, G.~G., Quataert, E.,  \& Tatsuno, T.} (2009) Astrophysical
  gyrokinetics: kinetic and fluid turbulent cascades in magnetized weakly
  collisional plasmas. {\em \astrojss\/} {\bfseries 182}, 310--377.

\bibitem[{\scshape Alfv{\'e}n \& F\"althammar}(1950)]{Alfven50}
{\scshape Alfv{\'e}n, H. \& F\"althammar, C.-G.} (1950) {\em {Cosmical
  Electrodynamics}\/}. Oxford: Clarendon Press.

\bibitem[{\scshape Catto}(1978){\scshape Catto}]{Catto78}
{\scshape Catto, P.~J.} (1978) Linearized gyro-kinetics. {\em Plasma Physics\/}
  {\bfseries 20}, 719.

\bibitem[{\scshape Antonsen \& Lane}(1980){\scshape Antonsen \&
  Lane}]{AntonsenLane80}
{\scshape Antonsen, T.~M. \& Lane, B.} (1980) Kinetic equations for low
  frequency instabilities in inhomogeneous plasmas. {\em Physics of Fluids\/}
  {\bfseries 23}, 1205--1214.

\bibitem[{\scshape Catto {\scshape et~al.}}(1981){\scshape Catto, Tang \&
  Baldwin}]{Catto81}
{\scshape Catto, P.~J., Tang, W.~M. \& Baldwin, D.~E.} (1981) Generalized
  gyrokinetics. {\em Plasma Physics\/} {\bfseries 23}, 639.

\bibitem[{\scshape Frieman \& Chen}(1982){\scshape Frieman \&
  Chen}]{FriemanChen82}
{\scshape Frieman, E.~A. \& Chen, L.} (1982) Nonlinear gyrokinetic equations
  for low-frequency electromagnetic waves in general plasma equilibria. {\em
  \PhysFluids\/} {\bfseries 25}, 502--508.

\bibitem[{\scshape Catto {\scshape et~al.}}(1987){\scshape Catto, Bernstein \&
  Tessarotto}]{Catto87}
{\scshape Catto, P.~J., Bernstein, I.~B. \& Tessarotto, M.} (1987) {Ion
  transport in toroidally rotating tokamak plasmas}. {\em Physics of Fluids\/}
  {\bfseries 30}, 2784--2795.

\bibitem[{\scshape Dubin}(1983){\scshape Dubin}]{Dubin83}
{\scshape Dubin, D. H.~E.} (1983) {Nonlinear gyrokinetic equations}. {\em
  Physics of Fluids\/} {\bfseries 26}, 3524.

\bibitem[{\scshape Hahm {\scshape et~al.}}(1988){\scshape Hahm, Lee \&
  Brizard}]{Hahm88}
{\scshape Hahm, T.~S., Lee, W.~W. \& Brizard, A.} (1988) {Nonlinear gyrokinetic
  theory for finite-beta plasmas}. {\em Physics of Fluids\/} {\bfseries 31},
  1940.

\bibitem[{\scshape Brizard \& Hahm}(2007){\scshape Brizard \& Hahm}]{Brizard07}
{\scshape Brizard, A.~J. \& Hahm, T.~S.} (2007) Foundations of nonlinear
  gyrokinetic theory. {\em Reviews of Modern Physics\/} {\bfseries 79}, 421.

\bibitem[{\scshape Garbet {\scshape et~al.}}(2010){\scshape Garbet, Idomura,
  Villard \& Watanabe}]{Garbet10}
{\scshape Garbet, X., Idomura, Y., Villard, L. \& Watanabe, T.~H.} (2010)
  Gyrokinetic simulations of turbulent transport. {\em Nuclear Fusion\/}
  {\bfseries 50}, 043002.

\bibitem[{\scshape {Howes} {\scshape et~al.}}(2006){\scshape {Howes}, {Cowley},
  {Dorland}, {Hammett}, {Quataert} \& {Schekochihin}}]{Howes06}
{\scshape {Howes}, G.~G., {Cowley}, S.~C., {Dorland}, W., {Hammett}, G.~W.,
  {Quataert}, E. \& {Schekochihin}, A.~A.} (2006) {Astrophysical Gyrokinetics:
  Basic Equations and Linear Theory}. {\em \astroj\/} {\bfseries 651},
  590--614.

\bibitem[{\scshape Numata {\scshape et~al.}}(2010){\scshape Numata, Howes,
  Tatsuno, Barnes \& Dorland}]{Numata10}
{\scshape Numata, R., Howes, G.~G., Tatsuno, T., Barnes, M. \& Dorland, W.}
  (2010) {AstroGK: Astrophysical gyrokinetics code}. {\em \JCompPhys\/}
  {\bfseries 229}, 9347--9372.

\bibitem[{\scshape Dorland {\scshape et~al.}}(2009){\scshape Dorland, Highcock,
  Barnes, Hammett, Numata, Tatsuno, Roach, Colyer, Baumgaertel \&
  Dickinson}]{GS2}
{\scshape Dorland, W., Highcock, E.~G., Barnes, M., Hammett, G.~W., Numata, R.,
  Tatsuno, T., Roach, C., Colyer, G., Baumgaertel, J. \& Dickinson, D.} (2009)
  Gyrokinetic simulations project. See http://gyrokinetics.sourceforge.net/.

\bibitem[{\scshape Landau}(1946){\scshape Landau}]{Landau46}
{\scshape Landau, L.~D.} (1946) {On the Vibrations of the Electronic Plasma}.
  {\em \JPhysUSSR\/} {\bfseries 10}.

\bibitem[{\scshape van Kampen}(1955){\scshape van Kampen}]{VanKampen55}
{\scshape van Kampen, N.~G.} (1955) On the theory of stationary waves in
  plasmas. {\em Physica\/} {\bfseries 21}, 949--963.

\bibitem[{\scshape Case}(1959){\scshape Case}]{Case59}
{\scshape Case, K.~M.} (1959) Plasma oscillations. {\em Annals of Physics\/}
  {\bfseries 7}, 349--364.

\bibitem[{\scshape Temme}(1996)]{Temme96}
{\scshape Temme, N.~M.} (1996) {\em Special Functions: An Introduction to the
  Classical Functions of Mathematical Physics\/}. New York: Wiley.

\bibitem[{\scshape Highcock {\scshape et~al.}}(2011){\scshape Highcock, Barnes,
  Parra, Schekochihin, Roach \& Cowley}]{Highcock11}
{\scshape Highcock, E.~G., Barnes, M., Parra, F.~I., Schekochihin, A.~A.,
  Roach, C.~M. \& Cowley, S.~C.} (2011) Transport bifurcation induced by
  sheared toroidal flow in tokamak plasmas. {\em \pop\/} {\bfseries 18},
  102304.

\bibitem[{\scshape Dannert \& Jenko}(2004){\scshape Dannert \&
  Jenko}]{Dannert04}
{\scshape Dannert, T. \& Jenko, F.} (2004) {Vlasov simulation of kinetic shear
  Alfv\'en waves}. {\em Computer Physics Communications\/} {\bfseries 163},
  67--78.

\bibitem[{\scshape Kammerer {\scshape et~al.}}(2008){\scshape Kammerer, Merz \&
  Jenko}]{Kammerer08}
{\scshape Kammerer, M., Merz, F. \& Jenko, F.} (2008) Exceptional points in
  linear gyrokinetics. {\em \pop\/} {\bfseries 15}, 052102.

\bibitem[{\scshape Peeters {\scshape et~al.}}(2009){\scshape Peeters, Camenen,
  Casson, Hornsby, Snodin, Strintzi \& Szepesi}]{Peeters09}
{\scshape Peeters, A., Camenen, Y., Casson, F., Hornsby, W., Snodin, A.,
  Strintzi, D. \& Szepesi, G.} (2009) The nonlinear gyro-kinetic flux tube code
  {GKW}. {\em \cpc\/} {\bfseries 180}, 2650--2672.

\bibitem[{\scshape Pueschel {\scshape et~al.}}(2010){\scshape Pueschel, Dannert
  \& Jenko}]{PueschelDannertJenko10}
{\scshape Pueschel, M.~J., Dannert, T. \& Jenko, F.} (2010) {On the role of
  numerical dissipation in gyrokinetic Vlasov simulations of plasma
  microturbulence}. {\em Computer Physics Communications\/} {\bfseries 181},
  1428--1437.

\bibitem[{\scshape van Kampen \& Felderhof}(1967)]{VanKampenBook}
{\scshape van Kampen, N.~G. \& Felderhof, B.~U.} (1967) {\em {Theoretical
  Methods in Plasma Physics}\/}. Amsterdam: North-Holland Publishing Company.

\bibitem[{\scshape Fried \& Conte}(1961)]{Fried61}
{\scshape Fried, B. \& Conte, S.} (1961) {\em The Plasma Dispersion Function:
  The Hilbert Transform of the Gaussian\/}. New York and London: Academic
  Press.

\bibitem[{\scshape Huba}(1994)]{NRL}
{\scshape Huba, J.~D.} (1994) {\em {NRL Plasma Formulary}\/}. {\em NRL
  publication\/} \! Naval Research Laboratory.

\bibitem[{\scshape Lenard \& Bernstein}(1958){\scshape Lenard \&
  Bernstein}]{LenardBernstein}
{\scshape Lenard, A. \& Bernstein, I.~B.} (1958) Plasma oscillations with
  diffusion in velocity space. {\em \pr\/} {\bfseries 112}, 1456--1459.

\bibitem[{\scshape Ng {\scshape et~al.}}(1999){\scshape Ng, Bhattacharjee \&
  Skiff}]{Ng99}
{\scshape Ng, C.~S., Bhattacharjee, A. \& Skiff, F.} (1999) Kinetic eigenmodes
  and discrete spectrum of plasma oscillations in a weakly collisional plasma.
  {\em \prl\/} {\bfseries 83}, 1974--1977.

\bibitem[{\scshape Ng {\scshape et~al.}}(2004){\scshape Ng, Bhattacharjee \&
  Skiff}]{Ng04}
{\scshape Ng, C.~S., Bhattacharjee, A. \& Skiff, F.} (2004) Complete spectrum
  of kinetic eigenmodes for plasma oscillations in a weakly collisional plasma.
  {\em \prl\/} {\bfseries 92}, 065002.

\bibitem[{\scshape Ng {\scshape et~al.}}(2006){\scshape Ng, Bhattacharjee \&
  Skiff}]{Ng06}
{\scshape Ng, C.~S., Bhattacharjee, A. \& Skiff, F.} (2006) {Weakly collisional
  Landau damping and three-dimensional Bernstein--Greene--Kruskal modes: New
  results on old problems}. {\em \pop\/} {\bfseries 13}, 055903.

\bibitem[{\scshape Maxwell}(1867){\scshape Maxwell}]{Maxwell67}
{\scshape Maxwell, J.~C.} (1867) On the dynamical theory of gases. {\em
  \ptrsl\/} {\bfseries 157}, 49--88.

\bibitem[{\scshape Abramowitz \& Stegun}(1972)]{AbramowitzStegun}
{\scshape Abramowitz, M. \& Stegun, I.~A.} (1972) {\em Handbook of Mathematical
  Functions: With Formulas, Graphs, and Mathematical Tables\/}, 10th edn. New
  York: Dover.

\bibitem[{\scshape Grad}(1949{\natexlab{{\em a\/}}}){\scshape
  Grad}]{Grad49Note}
{\scshape Grad, H.} (1949{\natexlab{{\em a\/}}}) Note on {$N$}-dimensional
  {H}ermite polynomials. {\em Communications on Pure and Applied Mathematics\/}
  {\bfseries 2}, 325--330.

\bibitem[{\scshape Grad}(1949{\natexlab{{\em b\/}}}){\scshape
  Grad}]{Grad49Kinetic}
{\scshape Grad, H.} (1949{\natexlab{{\em b\/}}}) On the kinetic theory of
  rarefied gases. {\em Communications on Pure and Applied Mathematics\/}
  {\bfseries 2}, 331--407.

\bibitem[{\scshape Grad}(1958){\scshape Grad}]{GradHandbuch}
{\scshape Grad, H.} (1958) Principles of the kinetic theory of gases. In {\em
  Thermodynamik der Gase\/} (ed. S.~Fl\"ugge), {\em Handbuch der Physik\/},
  vol.~12, pp. 205--294. Berlin: Springer.

\bibitem[{\scshape Armstrong}(1967){\scshape Armstrong}]{Armstrong67}
{\scshape Armstrong, T.~P.} (1967) Numerical studies of the nonlinear {V}lasov
  equation. {\em Physics of Fluids\/} {\bfseries 10}, 1269--1280.

\bibitem[{\scshape Grant \& Feix}(1967){\scshape Grant \& Feix}]{Grant67}
{\scshape Grant, F.~C. \& Feix, M.~R.} (1967) Fourier--{H}ermite solutions of
  the {V}lasov equations in the linearized limit. {\em Physics of Fluids\/}
  {\bfseries 10}, 696--702.

\bibitem[{\scshape Joyce {\scshape et~al.}}(1971){\scshape Joyce, Knorr \&
  Meier}]{Joyce71}
{\scshape Joyce, G., Knorr, G. \& Meier, H.~K.} (1971) Numerical integration
  methods of the {V}lasov equation. {\em \jcp\/} {\bfseries 8}, 53--63.

\bibitem[{\scshape Gagn\'e \& Shoucri}(1977){\scshape Gagn\'e \&
  Shoucri}]{Gagne77}
{\scshape Gagn\'e, R. R.~J. \& Shoucri, M.~M.} (1977) A splitting scheme for
  the numerical solution of a one-dimensional {V}lasov equation. {\em
  \JCompPhys\/} {\bfseries 24}, 445--449.

\bibitem[{\scshape Dawson}(1983){\scshape Dawson}]{Dawson83}
{\scshape Dawson, J.~M.} (1983) Particle simulation of plasmas. {\em Reviews of
  Modern Physics\/} {\bfseries 55}, 403--447.

\bibitem[{\scshape Hockney \& Eastwood}(1981)]{Hockney88}
{\scshape Hockney, R.~W. \& Eastwood, J.~W.} (1981) {\em Computer Simulation
  using Particles\/}. New York; London: McGraw--Hill.

\bibitem[{\scshape Birdsall \& Langdon}(2005)]{Birdsall04}
{\scshape Birdsall, C.~K. \& Langdon, A.~B.} (2005) {\em Plasma Physics via
  Computer Simulation\/}. Bristol: Institute of Physics.

\bibitem[{\scshape Fan \& Shen}(2001){\scshape Fan \& Shen}]{FanShen01}
{\scshape Fan, J. \& Shen, C.} (2001) Statistical simulation of low-speed
  rarefied gas flows. {\em Journal of Computational Physics\/} {\bfseries 167},
  393--412.

\bibitem[{\scshape Peeters {\scshape et~al.}}(2009){\scshape Peeters, Camenen,
  Casson, Hornsby, Snodin, Strintzi \& Szepesi}]{GKW}
{\scshape Peeters, A.~G., Camenen, Y., Casson, F.~J., Hornsby, W.~A., Snodin,
  A.~P., Strintzi, D. \& Szepesi, G.} (2009) The nonlinear gyro-kinetic flux
  tube code {GKW}. {\em Computer Physics Communications\/} {\bfseries 180},
  2650--2672.

\bibitem[{\scshape Fahey \& Candy}(2004){\scshape Fahey \& Candy}]{GYRO}
{\scshape Fahey, M.~R. \& Candy, J.} (2004) Gyro: A {5-D} gyrokinetic-{M}axwell
  solver. In {\em Proceedings of the 2004 ACM/IEEE conference on
  Supercomputing\/}, pp. 26--33. Washington, DC, USA: IEEE Computer Society.

\bibitem[{\scshape Zocco \& Schekochihin}(2011){\scshape Zocco \&
  Schekochihin}]{Zocco11}
{\scshape Zocco, A. \& Schekochihin, A.~A.} (2011) Reduced fluid-kinetic
  equations for low-frequency dynamics, magnetic reconnection, and electron
  heating in low-beta plasmas. {\em Physics of Plasmas\/} {\bfseries 18},
  102309.

\bibitem[{\scshape Hammett {\scshape et~al.}}(1993){\scshape Hammett, Beer,
  Dorland, Cowley \& Smith}]{Hammett93}
{\scshape Hammett, G.~W., Beer, M.~A., Dorland, W., Cowley, S.~C. \& Smith,
  S.~A.} (1993) Developments in the gyrofluid approach to tokamak turbulence
  simulations. {\em Plasma Physics and Controlled Fusion\/} {\bfseries 35},
  973.

\bibitem[{\scshape Parker \& Carati}(1995){\scshape Parker \&
  Carati}]{Parker95}
{\scshape Parker, S.~E. \& Carati, D.} (1995) Renormalized dissipation in
  plasmas with finite collisionality. {\em Physical Review Letters\/}
  {\bfseries 75}, 441--444.

\bibitem[{\scshape Schekochihin {\scshape et~al.}}(2015){\scshape Schekochihin,
  Kanekar, Hammett, Dorland \& Loureiro}]{Schekochihin14}
{\scshape Schekochihin, A.~A., Kanekar, A., Hammett, G.~W., Dorland, W. \&
  Loureiro, N.~F.} (2015) Stochastic advection and phase mixing in a
  collisionless plasma. In preparation.

\bibitem[{\scshape Kanekar {\scshape et~al.}}(2015){\scshape Kanekar,
  Schekochihin, Dorland \& Loureiro}]{Kanekar14}
{\scshape Kanekar, A., Schekochihin, A.~A., Dorland, W. \& Loureiro, N.~F.}
  (2015) Fluctuation-dissipation theorems for a plasma-kinetic {L}angevin
  equation. {\em Journal of Plasma Physics\/} {\bfseries 81}, 305810104.

\bibitem[{\scshape Plunk \& Parker}(2014){\scshape Plunk \& Parker}]{Plunk14}
{\scshape Plunk, G.~G. \& Parker, J.~T.} (2014) Irreversible energy flow in
  forced {V}lasov dynamics. {\em European Physical Journal D\/} {\bfseries 68},
  296, reproduced in Appendix B.

\bibitem[{\scshape Kolmogorov}(1941){\scshape Kolmogorov}]{Kolmogorov41}
{\scshape Kolmogorov, A.~N.} (1941) Dissipation of energy in locally isotropic
  turbulence. In {\em Akademiia Nauk SSSR Doklady\/}, , vol.~32, p.~16.

\bibitem[{\scshape Hatch {\scshape et~al.}}(2013){\scshape Hatch, Jenko,
  Ba{\~n}{\'o}n~Navarro \& Bratanov}]{Hatch13}
{\scshape Hatch, D.~R., Jenko, F., Ba{\~n}{\'o}n~Navarro, A. \& Bratanov, V.}
  (2013) Transition between saturation regimes of gyrokinetic turbulence. {\em
  Physical Review Letters\/} {\bfseries 111}, 175001.

\bibitem[{\scshape Loureiro {\scshape et~al.}}(2013){\scshape Loureiro,
  Schekochihin \& Zocco}]{Viriato}
{\scshape Loureiro, N.~F., Schekochihin, A.~A. \& Zocco, A.} (2013) Fast
  collisionless reconnection and electron heating in strongly magnetized
  plasmas. {\em Physical Review Letters\/} {\bfseries 111}, 025002.

\bibitem[{\scshape Camporeale {\scshape et~al.}}(2013){\scshape Camporeale,
  Delzanno, Bergen \& Moulton}]{CamporealeDelzannoBergenMoulton13}
{\scshape Camporeale, E., Delzanno, G.~L., Bergen, B.~K. \& Moulton, J.~D.}
  (2013) {On the velocity space discretization for the Vlasov--Poisson system:
  comparison between Hermite spectral and Particle-in-Cell methods. Part 1:
  semi-implicit scheme}. arXiv:1311.2098v2.

\bibitem[{\scshape Holloway}(1996){\scshape Holloway}]{Holloway96}
{\scshape Holloway, J.~P.} (1996) {Spectral velocity discretizations for the
  Vlasov--Maxwell equations}. {\em Transport Theory and Statistical Physics\/}
  {\bfseries 25}, 1--32.

\bibitem[{\scshape Boyd}(2001)]{Boyd01}
{\scshape Boyd, J.~P.} (2001) {\em {Chebyshev and Fourier Spectral Methods}\/}.
  New York: Dover.

\bibitem[{\scshape Tang}(1993){\scshape Tang}]{Tang93}
{\scshape Tang, T.} (1993) The {Hermite} spectral method for {Gaussian}-type
  functions. {\em Society for Industrial and Applied Mathematics Journal on
  Scientific Computing\/} {\bfseries 14}, 594--606.

\bibitem[{\scshape Schumer \& Holloway}(1998){\scshape Schumer \&
  Holloway}]{Schumer98}
{\scshape Schumer, J.~W. \& Holloway, J.~P.} (1998) Vlasov simulations using
  velocity-scaled {H}ermite representations. {\em Journal of Computational
  Physics\/} {\bfseries 144}, 626--661.

\bibitem[{\scshape {Le Bourdiec} {\scshape et~al.}}(2006){\scshape {Le
  Bourdiec}, {de Vuyst} \& Jacquet}]{LeBourdiecEtAl06}
{\scshape {Le Bourdiec}, S., {de Vuyst}, F. \& Jacquet, L.} (2006) Numerical
  solution of the {V}lasov--{P}oisson system using generalized hermite
  functions. {\em Computer Physics Communications\/} {\bfseries 175}, 528--544.

\bibitem[{\scshape Olver {\scshape et~al.}}(2010)Olver, Lozier, Boisvert \&
  Clark]{Olver10:18117}
{\scshape Olver, F.~W.~J., Lozier, D.~W., Boisvert, R.~F. \& Clark, C.~W.}, ed.
  (2010) {\em {NIST Handbook of Mathematical Functions}\/}. New York, NY:
  Cambridge University Press, print companion to NIST Digital Library of
  Mathematical Functions, \url{http://dlmf.nist.gov/}. Equations 18.11.7,
  18.11.8.

\bibitem[{\scshape Gil {\scshape et~al.}}(2007)Gil, Segura \& Temme]{Gil07}
{\scshape Gil, A., Segura, J. \& Temme, N.} (2007) {\em Numerical Methods for
  Special Functions\/}. Philadelphia: Society for Industrial and Applied
  Mathematics.

\bibitem[{\scshape Stoer \& Bulirsch}(2002)]{StoerBulirsch}
{\scshape Stoer, J. \& Bulirsch, R.} (2002) {\em {Introduction to Numerical
  Analysis}\/}. New York: Springer.

\bibitem[{\scshape Whittaker}(1928){\scshape Whittaker}]{Whittaker28}
{\scshape Whittaker, J.~M.} (1928) {The ``Fourier'' theory of the cardinal
  function}. {\em Proceedings of the Edinburgh Mathematical Society\/}
  {\bfseries 1}, 169--176.

\bibitem[{\scshape McNamee {\scshape et~al.}}(1971){\scshape McNamee, Stenger
  \& Whitney}]{McNameeStengerWhitney71}
{\scshape McNamee, J., Stenger, F. \& Whitney, E.~L.} (1971) Whittaker's
  cardinal function in retrospect. {\em \mathcomp\/} {\bfseries 25}, 141--154.

\bibitem[{\scshape Stenger}(1981){\scshape Stenger}]{Stenger81}
{\scshape Stenger, F.} (1981) {Numerical methods based on Whittaker cardinal,
  or sinc functions}. {\em Society for Industrial and Applied Mathematics
  Review\/} {\bfseries 23}, 165--224.

\bibitem[{\scshape Gautschi}(2004)]{GautschiBook}
{\scshape Gautschi, W.} (2004) {\em {Orthogonal Polynomials: Computation and
  Approximation}\/}. Oxford: Oxford University Press.

\bibitem[{\scshape Totik}(2005){\scshape Totik}]{Totik05}
{\scshape Totik, V.} (2005) Orthogonal polynomials. {\em Surveys in
  Approximation Theory\/} {\bfseries 1}, 70--125.

\bibitem[{\scshape Trefethen \& Bau}(1997)]{TrefethenBau}
{\scshape Trefethen, L.~N. \& Bau, D.} (1997) {\em Numerical Linear Algebra\/}.
  Philadelphia: Society for Industrial and Applied Mathematics.

\bibitem[{\scshape Sherman \& Morrison}(1949){\scshape Sherman \&
  Morrison}]{ShermanMorrison49}
{\scshape Sherman, J. \& Morrison, W.~J.} (1949) Adjustment of an inverse
  matrix corresponding to changes in the elements of a given column or a given
  row of the original matrix. {\em The Annals of Mathematical Statistics\/}
  {\bfseries 20}, 620--624.

\bibitem[{\scshape Bunch {\scshape et~al.}}(1978){\scshape Bunch, Nielsen \&
  Sorensen}]{Bunch78}
{\scshape Bunch, J.~R., Nielsen, C.~P. \& Sorensen, D.~C.} (1978) Rank-one
  modification of the symmetric eigenproblem. {\em Numerische Mathematik\/}
  {\bfseries 31}, 31--48.

\bibitem[{\scshape Zhou}(2011){\scshape Zhou}]{Zhou11}
{\scshape Zhou, Y.} (2011) On the eigenvalues of specially low-rank perturbed
  matrices. {\em Applied Mathematics and Computation\/} {\bfseries 217},
  10267--10270.

\bibitem[{\scshape Smith}(1997){\scshape Smith}]{Smith}
{\scshape Smith, S.~A.} (1997) Dissipative closures for statistical moments,
  fluid moments, and subgrid scales in plasma turbulence. PhD thesis, Princeton
  University.

\bibitem[{\scshape Morrison \& Shadwick}(1994){\scshape Morrison \&
  Shadwick}]{MorrisonShadwick94}
{\scshape Morrison, P.~J. \& Shadwick, B.~A.} (1994) Canonization and
  diagonalization of an infinite dimensional {H}amiltonian system: Linear
  {V}lasov theory. {\em \appa\/} {\bfseries 85}, 759--769.

\bibitem[{\scshape Boyd}(2001)]{BoydBook}
{\scshape Boyd, J.~P.} (2001) {\em {Chebyshev and Fourier Spectral Methods}\/}.
  New York: Dover.

\bibitem[{\scshape Landau}(1936){\scshape Landau}]{Landau36}
{\scshape Landau, L.~D.} (1936) {The Transport Equation in the case of Coulomb
  Interactions}. {\em \JPhysUSSR\/} {\bfseries 10}.

\bibitem[{\scshape Cercignani}(1966){\scshape Cercignani}]{Cercignani66}
{\scshape Cercignani, C.} (1966) The method of elementary solutions for kinetic
  models with velocity-dependent collision frequency. {\em Annals of Physics\/}
  {\bfseries 40}, 469--481.

\bibitem[{\scshape Struchtrup}(1997){\scshape Struchtrup}]{Struchtrup97}
{\scshape Struchtrup, H.} (1997) {The BGK-model with velocity-dependent
  collision frequency}. {\em \cmt\/} {\bfseries 9}, 23--31.

\bibitem[{\scshape Passot \& Pouquet}(1988){\scshape Passot \&
  Pouquet}]{PassotPouquet88}
{\scshape Passot, T. \& Pouquet, A.} (1988) Hyperviscosity for compressible
  flows using spectral methods. {\em \jcp\/} {\bfseries 75}, 300--313.

\bibitem[{\scshape Cerutti {\scshape et~al.}}(2000){\scshape Cerutti, Meneveau
  \& Knio}]{CeruttiMeneveauKnio00}
{\scshape Cerutti, S., Meneveau, C. \& Knio, O.~M.} (2000) {Spectral and hyper
  eddy viscosity in high-Reynolds-number turbulence}. {\em Journal of Fluid
  Mechanics\/} {\bfseries 421}, 307--338.

\bibitem[{\scshape Knorr \& Shoucri}(1974){\scshape Knorr \& Shoucri}]{Knorr74}
{\scshape Knorr, G. \& Shoucri, M.~M.} (1974) Plasma simulation as eigenvalue
  problem. {\em \jcp\/} {\bfseries 14}, 1--7.

\bibitem[{\scshape Shoucri \& Gagn\'e}(1977){\scshape Shoucri \&
  Gagn\'e}]{Shoucri77}
{\scshape Shoucri, M.~M. \& Gagn\'e, R. R.~J.} (1977) Numerical solution of a
  two-dimensional {V}lasov equation. {\em \JCompPhys\/} {\bfseries 25},
  94--103.

\bibitem[{\scshape Hazeltine \& Meiss}(2003)]{Hazeltine03}
{\scshape Hazeltine, R. \& Meiss, J.} (2003) {\em {Plasma Confinement}\/}. New
  York: Dover.

\bibitem[{\scshape Cercignani}(1975)]{Cercignani}
{\scshape Cercignani, C.} (1975) {\em {Theory and Application of the Boltzmann
  Equation}\/}. Edinburgh: Scottish Academic Press.

\bibitem[{\scshape Anderson \& O'Neil}(2007){\scshape Anderson \&
  O'Neil}]{Anderson07}
{\scshape Anderson, M.~W. \& O'Neil, T.~M.} (2007) {Eigenfunctions and
  eigenvalues of the Dougherty collision operator}. {\em \pop\/} {\bfseries
  14}, 052103.

\bibitem[{\scshape Pauli}(2000)]{Pauli00}
{\scshape Pauli, W.} (2000) {\em Statistical Mechanics\/}. New York: Dover.

\bibitem[{\scshape Bardos {\scshape et~al.}}(1993){\scshape Bardos, Golse \&
  Levermore}]{Bardos93}
{\scshape Bardos, C., Golse, F. \& Levermore, C.~D.} (1993) Fluid dynamic
  limits of kinetic equations {II} convergence proofs for the {B}oltzmann
  equation. {\em Communications on Pure and Applied Mathematics\/} {\bfseries
  46}, 667--753.

\bibitem[{\scshape Lions \& Masmoudi}(2001){\scshape Lions \&
  Masmoudi}]{Lions01}
{\scshape Lions, P.-L. \& Masmoudi, N.} (2001) {From the Boltzmann Equations to
  the Equations of Incompressible Fluid Mechanics, I}. {\em
  \archiverationalmech\/} {\bfseries 158}, 173--193.

\bibitem[{\scshape Golse \& Saint-Raymond}(2004){\scshape Golse \&
  Saint-Raymond}]{Golse04}
{\scshape Golse, F. \& Saint-Raymond, L.} (2004) {The Navier--Stokes limit of
  the Boltzmann equation for bounded collision kernels}. {\em Inventiones
  Mathematicae\/} {\bfseries 155}, 81--161.

\bibitem[{\scshape Bouchut}(1993){\scshape Bouchut}]{Bouchut93}
{\scshape Bouchut, F.} (1993) {Existence and uniqueness of a global smooth
  solution for the Vlasov--Poisson--Fokker--Planck system in three dimensions}.
  {\em Journal of Functional Analysis\/} {\bfseries 111}, 239--258.

\bibitem[{\scshape Dolbeault}(1999){\scshape Dolbeault}]{Dolbeault99}
{\scshape Dolbeault, J.} (1999) {Free energy and solutions of the
  Vlasov--Poisson--Fokker--Planck system: external potential and confinement
  (Large time behavior and steady states)}. {\em Journal de Math\'ematiques
  Pures et Appliqu\'ees\/} {\bfseries 78}, 121--157.

\bibitem[{\scshape Krommes \& Hu}(1994){\scshape Krommes \& Hu}]{Krommes94}
{\scshape Krommes, J.~A. \& Hu, G.} (1994) The role of dissipation in the
  theory and simulations of homogeneous plasma turbulence, and resolution of
  the entropy paradox. {\em Physics of Plasmas\/} {\bfseries 1}, 3211--3238.

\bibitem[{\scshape Hallatschek}(2004){\scshape Hallatschek}]{Hallatschek04}
{\scshape Hallatschek, K.} (2004) Thermodynamic potential in local turbulence
  simulations. {\em \prl\/} {\bfseries 93}, 125001.

\bibitem[{\scshape Desvillettes \& Villani}(2001){\scshape Desvillettes \&
  Villani}]{Desvillettes01}
{\scshape Desvillettes, L. \& Villani, C.} (2001) On the trend to global
  equilibrium in spatially inhomogeneous entropy-dissipating systems: the
  linear {F}okker--{P}lanck equation. {\em Communications on Pure and Applied
  Mathematics\/} {\bfseries 54}, 1--42.

\bibitem[{\scshape Kirkwood}(1946){\scshape Kirkwood}]{Kirkwood46}
{\scshape Kirkwood, J.~G.} (1946) The statistical mechanics theory of transport
  processes {I}. general theory. {\em Journal of Chemical Physics\/} {\bfseries
  14}, 180--201.

\bibitem[{\scshape Helander \& Sigmar}(2002)]{HelanderSigmar}
{\scshape Helander, P. \& Sigmar, D.~J.} (2002) {\em Collisional Transport in
  Magnetized Plasmas\/}. Cambridge: Cambridge University Press.

\bibitem[{\scshape Watanabe \& Sugama}(2004){\scshape Watanabe \&
  Sugama}]{Watanabe04}
{\scshape Watanabe, T.-H. \& Sugama, H.} (2004) Kinetic simulation of steady
  states of ion temperature gradient driven turbulence with weak
  collisionality. {\em Physics of Plasmas\/} {\bfseries 11}, 1476--1483.

\bibitem[{\scshape Mesinger \& Arakawa}(1976)]{MesingerArakawa76}
{\scshape Mesinger, F. \& Arakawa, A.} (1976) {\em Numerical methods used in
  atmospheric models\/}, {\em Global Atmospheric Research Program\/}, vol.~1.
  Geneva: World Meteorological Organization.

\bibitem[{\scshape Strikwerda}(2004)]{Strikwerda04}
{\scshape Strikwerda, J.~C.} (2004) {\em Finite Difference Schemes and Partial
  Differential Equations\/}, 2nd edn. Philadelphia: Society for Industrial and
  Applied Mathematics.

\bibitem[{\scshape Hinch}(1991)]{Hinch91}
{\scshape Hinch, E.~J.} (1991) {\em Perturbation {M}ethods\/}. Cambridge: \CUP.

\bibitem[{\scshape Hou \& Li}(2007){\scshape Hou \& Li}]{Hou07}
{\scshape Hou, T.~Y. \& Li, R.} (2007) Computing nearly singular solutions
  using pseudo-spectral methods. {\em \JCompPhys\/} {\bfseries 226}, 379--397.

\bibitem[{\scshape Parker \& Dellar}(2015){\scshape Parker \&
  Dellar}]{Parker14}
{\scshape Parker, J.~T. \& Dellar, P.~J.} (2015) Fourier--{H}ermite spectral
  representation for the {V}lasov--{P}oisson system in the weakly collisional
  limit. {\em Journal of Plasma Physics\/} {d}oi:10.1017/S0022377814001287,
  reproduced in Appendix A.

\bibitem[{\scshape Trefethen}(1982){\scshape Trefethen}]{Trefethen82}
{\scshape Trefethen, L.~N.} (1982) Group velocity in finite difference schemes.
  {\em Society for Industrial and Applied Mathematics {R}eview\/} {\bfseries
  24}, 113--136.

\bibitem[{\scshape Whitham}(2011)]{Whitham11}
{\scshape Whitham, G.~B.} (2011) {\em Linear and {N}onlinear {W}aves\/}. New
  York: Wiley-{I}nterscience.

\bibitem[{\scshape Davies}(2002)]{Davies02}
{\scshape Davies, B.} (2002) {\em Integral Transforms and Their
  Applications\/}. New York: Springer.

\bibitem[{\scshape Johnson}(1987){\scshape Johnson}]{FiskJohnson87}
{\scshape Johnson, H.~F.} (1987) An improved method for computing a discrete
  {H}ankel transform. {\em \cpc\/} {\bfseries 43}, 181--202.

\bibitem[{\scshape Piessens}(2000){\scshape Piessens}]{Piessens00}
{\scshape Piessens, R.} (2000) The {H}ankel transform. In {\em The Transforms
  and Applications Handbook\/}, 2nd edn. (ed. A.~D. Poularikas), chap.~9. Boca
  Raton, Fla.: CRC Press.

\bibitem[{\scshape Frappier \& Olivier}(1993){\scshape Frappier \&
  Olivier}]{Frappier93}
{\scshape Frappier, C. \& Olivier, P.} (1993) A quadrature formula involving
  zeros of {B}essel functions. {\em Mathematics of Computation\/} {\bfseries
  60}, 303--316.

\bibitem[{\scshape Tatsuno {\scshape et~al.}}(2009){\scshape Tatsuno, Dorland,
  Schekochihin, Plunk, Barnes, Cowley \& Howes}]{Tatsuno09}
{\scshape Tatsuno, T., Dorland, W., Schekochihin, A.~A., Plunk, G.~G., Barnes,
  M., Cowley, S.~C. \& Howes, G.~G.} (2009) Nonlinear phase mixing and
  phase-space cascade of entropy in gyrokinetic plasma turbulence. {\em
  Physical Review Letters\/} {\bfseries 103}, 015003.

\bibitem[{\scshape {Tatsuno} {\scshape et~al.}}(2010){\scshape {Tatsuno},
  {Barnes}, {Cowley}, {Dorland}, {Howes}, {Numata}, {Plunk} \&
  {Schekochihin}}]{Tatsuno10}
{\scshape {Tatsuno}, T., {Barnes}, M., {Cowley}, S.~C., {Dorland}, W., {Howes},
  G.~G., {Numata}, R., {Plunk}, G.~G. \& {Schekochihin}, A.~A.} (2010)
  {Gyrokinetic simulation of entropy cascade in two-dimensional electrostatic
  turbulence}. {\em Journal of Plasma and Fusion Research\/} {\bfseries 9},
  509--516.

\bibitem[{\scshape Plunk {\scshape et~al.}}(2010){\scshape Plunk, Cowley,
  Schekochihin \& Tatsuno}]{Plunk10}
{\scshape Plunk, G.~G., Cowley, S.~C., Schekochihin, A.~A. \& Tatsuno, T.}
  (2010) {Two-dimensional gyrokinetic turbulence}. {\em Journal of Fluid
  Mechanics\/} {\bfseries 664}, 407--435.

\bibitem[{\scshape Plunk \& Tatsuno}(2011){\scshape Plunk \&
  Tatsuno}]{PlunkTatsuno11}
{\scshape Plunk, G.~G. \& Tatsuno, T.} (2011) Energy transfer and dual cascade
  in kinetic magnetized plasma turbulence. {\em Physical Review Letters\/}
  {\bfseries 106}, 165003.

\bibitem[{\scshape Cooley \& Tukey}(1965){\scshape Cooley \& Tukey}]{Cooley65}
{\scshape Cooley, J.~W. \& Tukey, J.~W.} (1965) An algorithm for the machine
  calculation of complex {F}ourier series. {\em Mathematics of Computation\/}
  {\bfseries 19}, 297--297.

\bibitem[{\scshape Orszag}(1970){\scshape Orszag}]{Orszag70}
{\scshape Orszag, S.~A.} (1970) Transform method for the calculation of
  vector-coupled sums: Application to the spectral form of the vorticity
  equation. {\em Journal of the Atmospheric Sciences\/} {\bfseries 27},
  890--895.

\bibitem[{\scshape Averbuch {\scshape et~al.}}(2006){\scshape Averbuch,
  Coifman, Donoho, Elad \& Israeli}]{Averbuch06}
{\scshape Averbuch, A., Coifman, R.~R., Donoho, D.~L., Elad, M. \& Israeli, M.}
  (2006) Fast and accurate polar {F}ourier transform. {\em Applied and
  Computational Harmonic Analysis\/} {\bfseries 21}, 145--167.

\bibitem[{\scshape Siegman}(1977){\scshape Siegman}]{Siegman77}
{\scshape Siegman, A.~E.} (1977) Quasi fast {H}ankel transform. {\em Optical
  Letters\/} {\bfseries 1}, 13--15.

\bibitem[{\scshape Ogata}(2005){\scshape Ogata}]{Ogata05}
{\scshape Ogata, H.} (2005) {A numerical integration formula based on the
  Bessel functions}. {\em \prims\/} {\bfseries 41}, 949--970.

\bibitem[{\scshape Lawson {\scshape et~al.}}(1979){\scshape Lawson, Hanson,
  Kincaid \& Krogh}]{Lawson79}
{\scshape Lawson, C.~L., Hanson, R.~J., Kincaid, D.~R. \& Krogh, F.~T.} (1979)
  Basic linear algebra subprograms for {F}ortran usage. {\em ACM Transactions
  of Mathematical Software\/} {\bfseries 5}, 308--323.

\bibitem[{\scshape Gradshteyn \& Ryzhik}(2007)]{Gradshteyn07}
{\scshape Gradshteyn, I.~S. \& Ryzhik, I.~M.} (2007) {\em Table of Integrals,
  Series and Products\/}. Amsterdam: Academic Press, \emph{Editors: A. Jeffrey
  and D. Zwillinger}.

\bibitem[{\scshape Howes {\scshape et~al.}}(2008){\scshape Howes, Dorland,
  Cowley, Hammett, Quataert, Schekochihin \& Tatsuno}]{Howes08PRL}
{\scshape Howes, G.~G., Dorland, W., Cowley, S.~C., Hammett, G.~W., Quataert,
  E., Schekochihin, A.~A. \& Tatsuno, T.} (2008) Kinetic simulations of
  magnetized turbulence in astrophysical plasmas. {\em Physical Review
  Letters\/} {\bfseries 100}, 065004.

\bibitem[{\scshape TenBarge \& Howes}(2013){\scshape TenBarge \&
  Howes}]{TenBarge13}
{\scshape TenBarge, J.~M. \& Howes, G.~G.} (2013) Current sheets and
  collisionless damping in kinetic plasma turbulence. {\em Astrophysical
  Journal\/} {\bfseries 771}, L27.

\bibitem[{\scshape Barnes {\scshape et~al.}}(2011){\scshape Barnes, Parra \&
  Schekochihin}]{BarnesEtal11}
{\scshape Barnes, M., Parra, F. \& Schekochihin, A.} (2011) Critically balanced
  ion temperature gradient turbulence in fusion plasmas. {\em Physical Review
  Letters\/} {\bfseries 107}, 115003.

\bibitem[{\scshape Kotschenreuther {\scshape et~al.}}(1995){\scshape
  Kotschenreuther, Rewoldt \& Tang}]{Kotschenreuther95}
{\scshape Kotschenreuther, M., Rewoldt, G. \& Tang, W.~M.} (1995) Comparison of
  initial value and eigenvalue codes for kinetic toroidal plasma instabilities.
  {\em Computer Physics Communications\/} {\bfseries 88}, 128--140.

\bibitem[{\scshape Durran}(1991){\scshape Durran}]{Durran91}
{\scshape Durran, D.~R.} (1991) The third-order {A}dams--{B}ashforth method: an
  attractive alternative to leapfrog time-differencing. {\em Monthly Weather
  Review\/} {\bfseries 119}, 702--720.

\bibitem[{\scshape Durran}(1999)]{Durran99}
{\scshape Durran, D.~R.} (1999) {\em Numerical Methods for Wave Equations in
  Geophysical Fluid Dynamics\/}. New York: Springer.

\bibitem[{\scshape Frigo \& Johnson}(2005){\scshape Frigo \& Johnson}]{FFTW3}
{\scshape Frigo, M. \& Johnson, S.~G.} (2005) The design and implementation of
  {FFTW3}. {\em Proceedings of the IEEE\/} {\bfseries 93}, 216--231.

\bibitem[{\scshape Orszag}(1971){\scshape Orszag}]{Orszag71}
{\scshape Orszag, S.~A.} (1971) On the elimination of aliasing in
  finite-difference schemes by filtering high-wavenumber components. {\em
  Journal of the Atmospheric Sciences\/} {\bfseries 28}, 1074.

\bibitem[{\scshape Belli}(2006){\scshape Belli}]{BelliThesis}
{\scshape Belli, E, A.} (2006) {Studies of numerical algorithms for
  gyrokinetics and the effects of shaping on plasma turbulence}. Doctoral
  thesis, Princeton University.

\bibitem[{\scshape Orszag \& Tang}(1979){\scshape Orszag \&
  Tang}]{OrszagTang79}
{\scshape Orszag, S.~A. \& Tang, C.-M.} (1979) Small-scale structure of
  two-dimensional magnetohydrodynamic turbulence. {\em Journal of Fluid
  Mechanics\/} {\bfseries 90}, 129--143.

\bibitem[{\scshape Parker {\scshape et~al.}}(2015){\scshape Parker, Highcock,
  Schekochihin \& Dellar}]{Parker15ITG}
{\scshape Parker, J.~T., Highcock, E.~G., Schekochihin, A.~A. \& Dellar, P.~J.}
  (2015) Free energy flow and dissipation in ion temperature gradient driven
  turbulence. \emph{In preparation.}

\bibitem[{\scshape Conner \& Wilson}(1994){\scshape Conner \&
  Wilson}]{Conner94}
{\scshape Conner, J.~W. \& Wilson, H.~R.} (1994) Survey of theories of
  anomalous transport. {\em Plasma Physics and Controlled Fusion\/} {\bfseries
  36}, 719--795.

\bibitem[{\scshape Doyle {\scshape et~al.}}(2007){\scshape Doyle
  et~al.others}]{Doyle07}
{\scshape Doyle, E.~J. et~al.} (2007) Chapter 2: Plasma confinement and
  transport. {\em Nuclear Fusion\/} {\bfseries 47}, S18.

\bibitem[{\scshape Watanabe \& Sugama}(2006){\scshape Watanabe \&
  Sugama}]{Watanabe06}
{\scshape Watanabe, T.-H. \& Sugama, H.} (2006) Velocity-space structures of
  distribution function in toroidal ion temperature gradient turbulence. {\em
  Nuclear Fusion\/} {\bfseries 46}, 24--32.

\bibitem[{\scshape Howes {\scshape et~al.}}(2008){\scshape Howes, Cowley,
  Dorland, Hammett, Quataert \& Schekochihin}]{Howes08}
{\scshape Howes, G.~G., Cowley, S.~C., Dorland, W., Hammett, G.~W., Quataert,
  E. \& Schekochihin, A.~A.} (2008) A model of turbulence in magnetized
  plasmas: Implications for the dissipation range in the solar wind. {\em
  Journal of Geophysical Research: Space Physics\/} {\bfseries 113},
  doi:10.1029/2007JA012665.

\bibitem[{\scshape Podesta {\scshape et~al.}}(2010){\scshape Podesta, Borovsky
  \& Gary}]{Podesta10}
{\scshape Podesta, J.~J., Borovsky, J.~E. \& Gary, S.~P.} (2010) A kinetic
  {A}lfv{\'e}n wave cascade subject to collisionless damping cannot reach
  electron scales in the solar wind at 1 {AU}. {\em Astrophysical Journal\/}
  {\bfseries 712}, 685--691.

\bibitem[{\scshape Gould {\scshape et~al.}}(1967){\scshape Gould, O'Neil \&
  Malmberg}]{Gould67}
{\scshape Gould, R.~W., O'Neil, T.~M. \& Malmberg, J.~H.} (1967) Plasma wave
  echo. {\em Physical Review Letters\/} {\bfseries 19}, 219--222.

\bibitem[{\scshape Malmberg {\scshape et~al.}}(1968){\scshape Malmberg,
  Wharton, Gould \& O'Neil}]{Malmberg68}
{\scshape Malmberg, J.~H., Wharton, C.~B., Gould, R.~W. \& O'Neil, T.~M.}
  (1968) Plasma wave echo experiment. {\em Physical Review Letters\/}
  {\bfseries 20}, 95.

\bibitem[{\scshape Schekochihin {\scshape et~al.}}(2015){\scshape Schekochihin,
  Parker, Highcock, Dellar, Dorland \& Hammett}]{Schekochihin15}
{\scshape Schekochihin, A.~A., Parker, J.~T., Highcock, E.~G., Dellar, P.~J.,
  Dorland, W. \& Hammett, G.~W.} (2015) Phase mixing vs.\ nonlinear advection
  in drift-kinetic plasma turbulence. ArXiv:1508.05988.

\bibitem[{\scshape Rogers {\scshape et~al.}}(2000){\scshape Rogers, Dorland \&
  Kotschenreuther}]{Rogers00}
{\scshape Rogers, B.~N., Dorland, W. \& Kotschenreuther, M.} (2000) Generation
  and stability of zonal flows in ion-temperature-gradient mode turbulence.
  {\em Physical Review Letters\/} {\bfseries 85}, 5336--5339.

\bibitem[{\scshape {Goldreich} \& {Sridhar}}(1995){\scshape {Goldreich} \&
  {Sridhar}}]{Goldreich95}
{\scshape {Goldreich}, P. \& {Sridhar}, S.} (1995) {Toward a theory of
  interstellar turbulence. 2: Strong {A}lfv\'enic turbulence}. {\em
  Astrophysical Journal\/} {\bfseries 438}, 763--775.

\bibitem[{\scshape Goldreich \& Sridhar}(1997){\scshape Goldreich \&
  Sridhar}]{Goldreich97}
{\scshape Goldreich, P. \& Sridhar, S.} (1997) Magnetohydrodynamic turbulence
  revisited. {\em Astrophysical Journal\/} {\bfseries 485}, 680.

\bibitem[{\scshape Boldyrev}(2005){\scshape Boldyrev}]{Boldyrev05}
{\scshape Boldyrev, S.} (2005) On the spectrum of magnetohydrodynamic
  turbulence. {\em Astrophysical Journal Letters\/} {\bfseries 626}, L37.

\bibitem[{\scshape Cho \& Lazarian}(2004){\scshape Cho \& Lazarian}]{Cho04}
{\scshape Cho, J. \& Lazarian, A.} (2004) The anisotropy of electron
  magnetohydrodynamic turbulence. {\em Astrophysical Journal Letters\/}
  {\bfseries 615}, L41.

\bibitem[{\scshape Nazarenko \& Schekochihin}(2011){\scshape Nazarenko \&
  Schekochihin}]{Nazarenko11}
{\scshape Nazarenko, S.~V. \& Schekochihin, A.~A.} (2011) Critical balance in
  magnetohydrodynamic, rotating and stratified turbulence: towards a universal
  scaling conjecture. {\em Journal of Fluid Mechanics\/} {\bfseries 677},
  134--153.

\end{thebibliography}
\bibliographystyle{jfmbracket3}

\end{document}